\documentclass[11pt,letterpaper]{report}          
\usepackage[letterpaper,margin=1in]{geometry}
\usepackage[intlimits]{amsmath}
\usepackage{amsfonts}
\DeclareSymbolFontAlphabet{\mathbb}{AMSb}
\usepackage{apalike}
\usepackage[bf]{caption}       
\usepackage{fancyhdr}
\usepackage{fancybox}
\usepackage{ifthen}
\usepackage{url}
\usepackage{lscape,afterpage}
\usepackage{xspace}
\usepackage{epstopdf} 
\usepackage{subfig}
\usepackage{graphicx}
\usepackage{appendix}
\usepackage[utf8]{inputenc}
\usepackage[T1]{fontenc}
\usepackage{microtype}
\usepackage{braket}
\usepackage{qcircuit}
\usepackage{amsthm}
\usepackage{amsmath,amssymb}
\usepackage{mathrsfs}
\usepackage{eufrak}
\usepackage{array}
\usepackage{makecell}
\usepackage{mdframed}
\usepackage{stmaryrd}
\usepackage{hyperref}
\usepackage[bottom]{footmisc}
\usepackage{tikz}
\usepackage{tikz-cd}
\usetikzlibrary{calc,arrows,decorations.pathreplacing,decorations.pathmorphing,intersections}
\usetikzlibrary{positioning}
\usetikzlibrary{fit}
\theoremstyle{plain}
\newtheorem{thm}{Theorem}[section]
\newtheorem{lem}[thm]{Lemma}
\newtheorem{tecq}[thm]{Technique}
\newtheorem{prop}[thm]{Proposition}
\newtheorem{prtl}{Protocol}

\newtheorem{cor}{Corollary}

\newtheorem{claim}[thm]{Claim}
\newtheorem{fact}{Fact}

\newtheorem{otl}{Outline}
\newtheorem{foc}{Flow of Construction}
\theoremstyle{definition}
\newtheorem{defn}{Definition}[section]

\newtheorem{exmp}{Example}[section]
\newtheorem{nota}{Notation}[section]
\theoremstyle{remark}

\DeclareMathOperator{\tr}{tr}

\newcommand{\cC}{{\mathcal{C}}}
\newcommand{\cA}{{\mathcal{A}}}

\newcommand{\cD}{{\mathcal{D}}}

\newcommand{\cF}{{\mathcal{F}}}

\newcommand{\cH}{{\mathcal{H}}}
\newcommand{\cK}{{\mathcal{K}}}

\newcommand{\cU}{{\mathcal{U}}}
\newcommand{\cP}{{\mathcal{P}}}
\newcommand{\cWBS}{{\mathcal{WBS}}}

\newcommand{\mi}{{\mathrm{i}}}
\DeclareMathOperator{\bE}{\mathbb{E}}

\newcommand{\fH}{{\sf H}}

\newcommand{\fI}{{\sf I}}

\newcommand{\fCN}{{\sf CNOT}}

\newcommand{\fEn}{{\mathsf{Enc}}}
\newcommand{\fGdgPrep}{{\mathsf{GdgPrep}}}
\newcommand{\fPadHadamard}{{\mathsf{PadHadamard}}}
\newcommand{\fBasisTest}{{\mathsf{BasisTest}}}
\newcommand{\fSuccUBQC}{{\mathsf{SuccUBQC}}}
\newcommand{\fGAUBQC}{{\mathsf{GAUBQC}}}
\newcommand{\fQFac}{{\mathsf{QFac8}}}
\newcommand{\fLT}{{\mathsf{LT}}}
\newcommand{\fRevLT}{{\mathsf{RevLT}}}
\newcommand{\fPhaseLT}{{\mathsf{PhaseLT}}}
\newcommand{\fRobustRLT}{{\mathsf{RobustRLT}}}
\newcommand{\fSecurityRefreshing}{{\mathsf{SecurityRefreshing}}}
\newcommand{\fPrtl}{{\mathsf{Prtl}}}
\newcommand{\fsubPrtl}{{\mathsf{subPrtl}}}
\newcommand{\fTempPrtl}{{\mathsf{TempPrtl}}}
\newcommand{\fCombine}{{\mathsf{Combine}}}
\newcommand{\fAuxInf}{{\mathsf{AuxInf}}}

\newcommand{\fneg}{{\mathsf{negl}}}
\newcommand{\fpoly}{{\mathsf{poly}}}
\newcommand{\fTHRESHOLD}{{\mathsf{threshold}}}

\newcommand{\fsubexp}{{\mathsf{subexp}}}

\newcommand{\fAdv}{{\mathsf{Adv}}}

\newcommand{\weakparam}[5]{$(2^{#1},#2)\rightarrow \Big\{#3\Bigm\vert(2^{#4},#5)\Big\}$}
\newcommand{\weakparams}[4]{$(2^{#1},#2)\rightarrow (2^{#3},#4)$}
\mdfdefinestyle{figstyle}{ 
  linecolor=black!7, 
  backgroundcolor=black!7
}
\newcommand{\weakparamm}[5]{$(2^{#1},#2)\xrightarrow{#3} (2^{#4},#5)$}
\newcommand{\weakparamstar}[5]{$(2^{#1},#2)^*\xrightarrow{#3} (2^{#4},#5)$}
\newcommand{\weakparaml}[7]{$[(2^{#1},#2),(2^{#3},#4)]\rightarrow \Big\{{#5} \Bigm\vert (2^{#6},#7)\Big\}$}
\mdfdefinestyle{figstyle}{ 
  linecolor=black!7, 
  backgroundcolor=black!7
}
\usepackage{float}
\usetikzlibrary{crypto.symbols}
\begin{document}

\title{Succinct Blind Quantum Computation Using a Random Oracle}

\author{Jiayu Zhang\footnote{Boston University \& California Institute and Technology, jiayu@caltech.edu. This work was supported by NSF award 1763786.}}

\maketitle

\begin{abstract}
	In the universal blind quantum computation problem, a client wants to make use of a single quantum server to evaluate $C\ket{0}$ where $C$ is an arbitrary quantum circuit while keeping $C$ secret. The client's goal is to use as few resources as possible. This problem, with a representative protocol by Broadbent, Fitzsimons and Kashefi\cite{UBQC}, has become fundamental to the study of quantum cryptography, not only because of its own importance, but also because it provides a testbed for new techniques that can be later applied to related problems (for example, quantum computation verification). Known protocols on this problem are mainly either information-theoretically (IT) secure or based on trapdoor assumptions (public key encryptions).\par
	 In this paper we study how the availability of symmetric-key primitives, modeled by a random oracle, changes the complexity of universal blind quantum computation. 
	We give a new universal blind quantum computation protocol. 
	Similar to previous works on IT-secure protocols (for example, BFK\cite{UBQC}), our protocol can be divided into two phases. In the first phase the client prepares some quantum gadgets with relatively simple quantum gates and sends them to the server, and in the second phase the client is entirely classical --- it does not even need quantum storage. Crucially, the protocol's first phase is \emph{succinct}, that is, its complexity is independent of the circuit size. Given the security parameter $\kappa$, its complexity is only a fixed polynomial of $\kappa$, and can be used to evaluate any circuit (or several circuits) of size up to a subexponential of $\kappa$. In contrast, known schemes either require the client to perform quantum computations that scale with the size of the circuit \cite{UBQC}, or require trapdoor assumptions \cite{Mahadev2017ClassicalHE}.\par
\end{abstract}






%


\tableofcontents
\cleardoublepage

%

%
%
%
%


\newpage
        
\cleardoublepage
\chapter{Introduction}\label{cht:1}

\section{Problem Background}\label{sec:1.1}
In the universal blind quantum computation problem, a client wants to make use of a single quantum server to evaluate a  quantum circuit $C$ secretly, where $C$ can be arbitrary (up to a subexponential size). The protocol should at least satisfy the following requirements:
\begin{enumerate}
	\item (Correctness) When the server is honest, the client holds $C\ket{0}$ in the end of the protocol with probability negligibly close to 1.\footnote{A more general form is to consider the evaluation of $C\ket{\varphi}$, and in the security requirement both the circuit $C$ and the input state $\ket{\varphi}$ should be hidden. Although this paper considers $C$ as the only input of the protocol, it could also handle the slightly more general case.}
	\item (Security) For any adversarial server, which might be unbounded, polynomial, etc, depending on the setting, it cannot distinguish whether the current protocol is run on input $C$, or run on input $0^{|C|}$.
	\item (Efficiency) When the protocol is run honestly, the client and the server should be in polynomial time.
\end{enumerate}\par
This problem is important for two reasons.\par
First, the problem itself is very important. The related problems in classical world, like the delegation of computation, multiparty computation or homomorphic encryption, are all very famous and fundamental problems and have a very long history. The blind quantum computation problem is important for the same reasons, and in quantum world there is one more reason to study this problem: It's very possible that the quantum computers will mainly be used as a cloud service. So if a client wants to use the power of a remote quantum server, and simultaneously, wants to keep its data or circuits secret, a blind quantum computation protocol is needed.\par
Second, the blind quantum computation problem is the testbed for new techniques. Empirically, once a new technique for the blind quantum computation problem is developed, it may be also useful in many other problems, including the quantum computation verification, multiparty quantum computation, certifiable randomness, zero knowledge proof for QMA and so on. For example, the MBQC-based techniques started with the UBQC protocol\cite{UBQC} for blind quantum computation, and that work becomes the foundation of the UVBQC protocol for quantum computation verification \cite{UVBQC}; another example of empirical relation between these problems is the trapdoor claw-free function techniques \cite{BCMVV}, which led to a series of works for quantum fully homomorphic encryption \cite{Mahadev2017ClassicalHE}, quantum computation verification\cite{MahadevVerification}, zero-knowledge arguments\cite{VidickZ19} and so on.\par
In classical world, this problem is studied for a long time under the names of two party computation and fully homomorphic encryption. We note that these concepts are not the same, but they are closely related and aiming at the same goal: to delegate the computation while keep it (or the data) secure. There are two fundamental constructions in classical world: one is the garbled circuit, or garbled table, raised by Yao\cite{YaoGCOrigin}; another construction is the fully homomorphic encryption\cite{GentryFHE}, or FHE.\par
\section{Previous Works and Motivating Questions}\label{sec:1.2}
Previous protocols for universal blind quantum computation require either the execution of many quantum gates --- proportional to the size of the circuit \cite{UBQC} --- but not computational assumptions; or the existence of trapdoor cryptographic primitives, such as the quantum hardness of learning with errors (LWE \cite{regevLWE}) \cite{Mahadev2017ClassicalHE}. (See Table \ref{tb:tradeoff} for some existing works.) There are also some protocols that use two separated quantum servers\cite{UBQC} and some protocols that are not universal\cite{AnnesQHELowT,revgt,statisticallyQHEforIQP,QHEcode}; in this paper we focus on the universal protocols using a single quantum server.
\begin{itemize}
\item One representative of information-theoretically (IT) secure protocols is the BFK's UBQC protocol \cite{UBQC}. This protocol is based on the measurement-based quantum computation, and it contains an offline phase and an online phase. In the offline phase the client sends many quantum gadgets to the server. These quantum gadgets can be prepared using single-qubit gates, but the total number of gadgets is linear to the size of the circuit to be evaluated, which is prohibitive. This protocol becomes the basis of many later works.
\item Some earlier representatives of computationally-secure protocols include \cite{QFHEPoly}, which is based on the LWE assumption. 
Then in \cite{Mahadev2017ClassicalHE} a classical fully homomorphic encryption for quantum circuits was constructed. That protocol is based on the classical FHE and a new technique based on a primitive called ``trapdoor claw-free functions'' \cite{BCMVV}, and both primitives were constructed from the LWE assumption. These techniques were applied to many related problems like the quantum computation verification \cite{MahadevVerification}. \end{itemize}
One way to classify these assumptions are through the ``Impagliazzo's Five Worlds'' \cite{FiveWorlds}. IT-secure protocols remain secure in all of these worlds, since it does not rely on any computational assumption; and the trapdoor assumptions, FHE, LWE and many ``fancy'' cryptographic schemes and protocols are secure only in Cryptomania (a world in which trapdoor primitives exist).\par
Minicrypt, intuitively, is the world where ``symmetric cryptography'' (for example, pseudorandom generators) exists but ``public key cryptography'' (trapdoor functions) is not possible. Our motivation is to understand what sorts of cryptography are possible in the quantum analogue of Minicrypt. We work with an abstraction, the QROM, which assumes (1) all parties have oracle access to a common function which is chosen uniformly at random; (2) the adversary is unbounded, but can only makes polynomial number of quantum random oracle queries. This setting allows for symmetric-type primitives (one-way functions, pseudorandom generators, collision-free hash functions), but excludes ``public key primitives''. By the ``Random Oracle Methodology'', once we design a protocol in this setting, we can replace the random oracle by an appropriate hash function or symmetric key encryption scheme. (We note that although this setting itself is formal, the instantiation of protocols proved secure in this setting can be subtle: there do exist some constructions that are not possible to be instantiated\cite{rorevisited}. However, the usage of the random oracle as an ideal model of hash functions or symmetric key encryption schemes is wide-spread, and has greatly helped the design of cryptographic protocols in the past three decades\cite{roreliable}.)\par
 Besides the theoretical motivation, the protocols in this setting have the following advantage: currently there are few choices for post-quantum secure public key encryption schemes\cite{regevLWE,sikd}. If we want to instantiate some more specific and stronger primitives, like trapdoor claw-free functions or FHE (which are much stronger than the existence of trapdoor functions), currently the only known way is through the lattice-based cryptography (for example, LWE). On the other hand, there are many choices for symmetric key primitives, and the protocols can remain to be sound even if lattice-based cryptography is broken.\par
 The design of delegation-style quantum protocols in this setting is not well-understood. As far as we know, except the works on IT-secure protocols, the only work is \cite{revgt}, which designed a quantum delegation (blind quantum computation) protocol for a useful but specific circuit family. Thus we ask the following question:
\begin{center}\emph{ How does the availability of symmetric-key cryptographic primitives (modeled by a random oracle) change the complexity of universal blind quantum computation?}\end{center}
Another factor that we will consider is the ``client side quantum computation''. Existing works assume either the client side quantum gates can be linear to the circuit size (\emph{during the whole protocol}), or the client is classical; little is known for the setting between them, which is, to allow the client to run succinct quantum operations, which can depend on the security parameter, but should be independent of the circuit size. Thus, we can ask the following question:
\begin{center}\emph{ How can we design a universal blind quantum computation protocol in which the client side quantum operations are ``succinct'' (that is, independent of the size of the circuit to be evaluated)?}\end{center}
Thus we want to design a protocol that is more efficient than the IT-secure protocol in terms of the client side quantum operations (here we do not care about the classical computation and communication as long as they are polynomial size), and does not use any public key primitives. None of the existing techniques works for this setting and we need to develop new techniques and a new protocol.
\section{Our Results}
In this paper we prove the following:
\begin{thm}\label{thm:1}
	There exists a universal blind quantum computation protocol (Protocol \ref{prtl:17} in Section \ref{sec:12.3}) for circuits of size up to a fixed subexponential function of the security parameter such that \begin{itemize}\item It contains an offline phase and an online phase. In the offline phase the client prepares and sends some quantum gadgets to the server, and in the online phase the client is completely classical. \item The total number of quantum gates to prepare these quantum gadgets is at most a fixed polynomial of the security parameter, independent to the size of the circuit to be evaluated. \item The classical computation, communication and the server-side quantum computation are bounded by a fixed polynomial of the security parameter and the size of the circuit to be evaluated. \item The protocol is secure in the quantum random oracle model against any unbounded malicious server whose number of queries to the random oracle is bounded by a fixed subexponential function of the security parameter.\end{itemize}
\end{thm}
Thus, based on our work, together with previous works\cite{UBQC,Mahadev2017ClassicalHE} we can complete the following table (Table \ref{tb:tradeoff}) about the different tradeoffs between client side quantum resources and assumptions. Thus we claim our result reveals a more complete cryptographic picture for single-server quantum (blind) delegation problem.\par

Our result required the development of a set of new techniques for protocol design and security proof. Section \ref{sec:1.4} provides a brief technical overview. As discussed before, new techniques in blind quantum computation often led to protocols for many related problems. We hope the techniques and protocols developed here will also lead to advances on a range of related problems. 
{\footnotesize{
\begin{table*}\small
	\begin{tabular}{|c||c|c|c|}\hline
		Client side  &                                & QROM       & LWE \\
		 quantum&IT-secure&(Idealized symmetric &(Public key                         encryption \\
		 computation&&key primitives)&with functionalities)\\\hline
		Classical                         & \begin{tabular}{c}
		May be impossible \\
		\cite{AaronsonLimit}\\
		\end{tabular} & Unknown    & \cite{Mahadev2017ClassicalHE} \\\hline
		Succinct                        & Unknown                                 & \textbf{This paper} &                               \\\hline
		Linear                          & \cite{UBQC}                             &            &  \cite{QFHEPoly}                             \\\hline
	\end{tabular}\\
	``Succinct'' means it's at most a fixed polynomial of the security parameter; and ``Linear'' means it's linear to the size of the circuit to be evaluated.
	\caption{Different tradeoff between client side quantum operations and assumptions in quantum computation delegation problem. }\label{tb:tradeoff}
\end{table*}}}

\section{A Top-down Overview of Our Techniques}\label{sec:1.3}

\subsection{Two-step construction via remote gadget preparation}
How can the client allow the server to evaluate $O(|C|)$ gates using only succinct quantum computation? In our protocol, the client will first prepare $\fpoly(\kappa)$ \emph{gadgets} ($\fpoly$ is a fixed polynomial), then use classical interactions to allow the server to \emph{expand} them to $O(|C|)$ gadgets ``securely''. Here the \emph{gadget} is defined to be the states in the form of $\ket{y_0}+\ket{y_1}$, where $y_0,y_1$ are random different strings, or \emph{keys}. The client holds the keys, and the server should hold the state.\footnote{This form of states was also previously used in several papers like \cite{BCMVV}.}\par
This step --- the preparation and expansion of gadgets --- is called \emph{remote gadget preparation}. Let's give the correctness and security definition informally below (for the formal definition, see Section \ref{sec:3.1}):
\begin{defn}[Correctness of the remote gadget preparation, informal] A protocol is called a remote gadget preparation protocol of output number $L$ and output length $\kappa_{out}$ if: if the server behaves honestly, it passes the protocol and  in the end of the protocol (1)the client gets key set $\{y_b^{(i)}\}_{i\in [L],b\in \{0,1\}}$ where for any $i$, $y_b^{(i)}$ is a string of length $\kappa_{out}$, $y^{(i)}_0\neq y^{(i)}_1$; (2)the server holds the state $\otimes_{i=1}^L (\ket{y^{(i)}_0}+\ket{y^{(i)}_1})$.\par
And we say this protocol has input number $N$ if initially the server holds (or equivalently, the client prepares and sends) $\otimes_{i=1}^N (\ket{x^{(i)}_0}+\ket{x^{(i)}_1})$ and the client holds all the keys. We use $N\rightarrow L$ to denote the honest behavior of a protocol of input number $N$ and output number $L$.
\end{defn}
Note that the protocols in this paper do not require quantum communication in the middle of the protocol.
And the security is defined based on the concept of \emph{SC-security}, which describes the adversary's ability to compute both keys simultaneously.
\begin{defn}[SC-security, informal]
	Suppose the client holds a key pair $K$. Suppose the joint state of all parties, after purification, is described by $\ket{\varphi}$. We say this state is $(2^\eta,C|\ket{\varphi}|)$-SC-secure for $K$ given some auxiliary information if for any adversary with query number at most $ 2^\eta$, the norm of outputting both keys in $K$, with access to the auxiliary information and the hash tags of $K$, is at most $ C|\ket{\varphi}|$.
\end{defn}
\begin{defn}[Security definition of the remote gadget preparation, informal]\label{def:1.2} For an $N\rightarrow L$ protocol, suppose the output key set is $K_{out}$. For any unbounded malicious server that makes at most $\fsubexp(\kappa)$ quantum random oracle queries, for any $i\in [L]$, the final state is exponentially SC-secure (which means, $(2^\eta,2^{-\eta})$-SC-secure, for some $\eta$) for $K_{out}^{(i)}$ (the $i$-th output key pair) given auxiliary information $K_{out}-K_{out}^{(i)}$.
\end{defn}
The auxiliary information here is necessary to get rid of potential correlations among keys. And it may be surprising that the security condition is about unpredictability (as opposed to simulation). However, it is sufficient for the final indistinguishability-based security of the blind quantum computation protocol.\par
As an example, we can see the honest final state $\otimes_{i=1}^L (\ket{y^{(i)}_0}+\ket{y^{(i)}_1})$ satisfies the security definition since the server cannot output $y^{(i)}_0||y^{(i)}_1$ from it with high probability, even if $\{y_b^{(j)}\}_{j\neq i,b\in \{0,1\}}$ and the hash values of $\{y_0^{(i)},y_1^{(i)}\}$ are provided.\par
Then we will construct our universal blind quantum computation protocol that satisfies Theorem \ref{thm:1} (denoted $\fSuccUBQC$) as follows:
\begin{otl}\label{ppl:1a} Design of the $\fSuccUBQC$ protocol:
	\begin{enumerate}
		\item Remote gadget preparation
		: (1) the client sends some initial gadgets to the server, whose size and length are succinct; (2) the client uses classical interactions to allow the server to expand the number of gadgets securely. 
		\item Blind quantum computation execution: using the gadgets output from the previous step, the client and the server evaluate $C$ using only classical interactions.
	\end{enumerate}
\end{otl}

The construction of the secure remote gadget preparation protocol (the first step in Outline \ref{ppl:1a}) is the most difficult step. The second step is relatively easier but still non-trivial.
\subsection{Remote gadget preparation via weak security}\label{sec:1.4.2a}
The first step of Outline \ref{ppl:1a} are achieved as follows: we will define the \emph{weak security} of the remote gadget preparation. We will first construct a weakly-secure protocol, then \emph{amplify} it to a fully secure one. The first idea of our construction is to develop a framework for the design of different subprotocols. This is (mainly) captured by the notion of \emph{weak security} and \emph{weak security transform parameter}. Let's give an informal introduction here, and we will revisit this concept in Section \ref{sec:3.2} and give the formal definition in Definition \ref{def:3.12}.
\begin{defn}[Weak security of remote gadget preparation, informal]\label{def:infws} We say a remote gadget preparation protocol run on key set $K$ is weakly secure with \emph{weak security transform parameter} \weakparam{\eta}{C}{p}{\eta^\prime}{C^\prime} if a statement in the following form holds:\par
Suppose the initial (purified joint) state $\ket{\varphi}$ satisfies: \begin{itemize}\item It is $(2^\eta,C|\ket{\varphi}|)$-SC-secure for any input key pair $K^\prime$ in $K$ given $K-K^\prime$ (all the other key pairs) as the auxiliary information; \item The state is not too ``ill-behaved''. (Informally, the state can be written as a sum of several terms with bounded number of RO queries. Formalized in Definition \ref{def:rep}, Notation \ref{nota:wbsn}.)\end{itemize}
 Then for any unbounded adversary with up to $2^\kappa$ random oracle queries during the protocol, at least one of the following is true: \begin{itemize}\item The client accepts with at most norm $p|\ket{\varphi}|$; \item For the output state, for any\footnote{The order of quantifiers is good here: it's not ``the server can output an index $i$ and the corresponding output keys''. This definition seems weak but is on the other hand general enough to be used as the framework.} output key pair, the state is $(2^{\eta^\prime},C^\prime|\ket{\varphi}|)$-SC-secure for this output key pair given all the other output keys as the auxiliary information.
 \end{itemize}
\end{defn}
Then the overall flow of the construction of the secure remote gadget preparation protocol (step 1 of Outline \ref{ppl:1a}) is as follows \footnote{This is only a construction flow, we don't mean there is a two-step protocol.}:
\begin{foc}\label{otl:2a} Protocol construction for the step 1 of Outline \ref{ppl:1a}:
	\begin{enumerate}
		\item Construct a weakly-secure remote gadget preparation protocol such that it can (asymptotically multiplicatively) generate more gadgets than it consumes.
		\item Using some amplification techniques to amplify it to a secure remote gadget preparation protocol.
	\end{enumerate}
\end{foc}
Our work can be seen as the design of a series of subprotocols with different tradeoffs for correctness, (weak) security, etc, and these subprotocols, when combined together, can achieve what we want.\par

\paragraph{Weakly Secure Protocol Step} The goal in this step is to create more gadgets (possibly with weak security) from some input gadgets. First, we consider the simplest case, generating two gadgets using one input gadget, remotely:
$$\ket{x_0}+\ket{x_1}\rightarrow (\ket{y_0}+\ket{y_1})\otimes (\ket{y_0^\prime}+\ket{y_1^\prime})$$
And we want the outputs to have some (possibly weak) security in the malicious setting. This quantum-to-quantum transformation can be enabled using a classical primitive called \emph{reversible look-up tables} (or \emph{reversible garbled tables}  in \cite{revgt}).\footnote{The reversible garbled table in this paper does not carry computation so we name it as the lookup table. We use it for gadget-to-gadget mappings.} But it's not possible to do that directly (See Section \ref{sec:7.1} for a discussion). The first key step is, instead of implementing this transformation directly, we seek for a transformation to the following state as an intermediate step:
$$\ket{x_0}+\ket{x_1}\rightarrow perm((\ket{y_0}+\ket{y_1})\otimes (\ket{y_0^\prime}+\ket{y_1^\prime}))$$
where $perm$ is a random bit-wise permutation sampled by the client, kept secret from the adversary. The secrecy of $perm$ will be a key ingredient on implementing this mapping securely. But the client still needs to reveal it to the server to allow it to de-permute the gadgets in the end, which seems to be a dilemma. Now we introduce the second idea: the realization of the mapping above will make use of an extra helper gadget, and a subprotocol called \emph{padded Hadamard test}. This padded Hadamard test is a padded variant of the Hadamard test in \cite{BCMVV}. We observe that, the padded Hadamard test has several powerful properties, one of which informally say, if such a test is executed between the client and the server, if the server wants to pass the test with high probability, it loses the ability to predict the keys from the post-test state --- a property that we call \emph{unpredictability restriction}. With this property in mind, we can use this subprotocol as a switch that controls when it's safe to reveal the permutation. Now the transformation goes as follows, where we use $\ket{x_0^{\text{helper}}}+\ket{x_1^{\text{helper}}}$ to denote the helper gadget:
{\footnotesize \begin{align}
	&(\ket{x_0^{\text{helper}}}+\ket{x_1^{\text{helper}}})\otimes (\ket{x_0}+\ket{x_1})\label{eq:des1a}\\
	\rightarrow & (\ket{x_0^{\text{helper}}}+\ket{x_1^{\text{helper}}})\otimes perm((\ket{y_0}+\ket{y_1})\otimes (\ket{y_0^\prime}+\ket{y_1^\prime}))\label{eq:des2a}\\ 
	\text{(Hadamard test on $\ket{x_0^{\text{helper}}}+\ket{x_1^{\text{helper}}}$)}\rightarrow &perm((\ket{y_0}+\ket{y_1})\otimes (\ket{y_0^\prime}+\ket{y_1^\prime}))\\
	\text{(Client reveals $perm$ if test passes)}\rightarrow &(\ket{y_0}+\ket{y_1})\otimes (\ket{y_0^\prime}+\ket{y_1^\prime})
\end{align}}
Then in each time step something is secret in the adversary's viewpoint. Before the test the bit-wise permutation is hidden, and after the test, the adversary is not able to have good predictability on $\{x^{\text{helper}}_0,x^{\text{helper}}_1\}$ anymore. And the design of the mapping from (\ref{eq:des1a}) to (\ref{eq:des2a}) will make use of it.\par
Now we go to the construction of (\ref{eq:des1a})$\rightarrow$(\ref{eq:des2a}). Now we can create a \emph{reversible lookup table} for it. The first problem here is the mismatch of the input number and output number, but this can be solved by introducing an auxiliary wire, where the two keys are simply provided classically to the server.\footnote{It's possible to encode the mapping as $1\leftrightarrow 2$ mapping directly, but we choose to encode it in this way for nicer honest setting behavior.} Now (\ref{eq:des1a})$\rightarrow$(\ref{eq:des2a}) is replaced by 
{\small \begin{align}&(\ket{x_0^{\text{helper}}}+\ket{x_1^{\text{helper}}})\otimes \ket{\{x_0^{(2)},x_1^{(2)}\}}\otimes(\ket{x_0^{(3)}}+\ket{x_1^{(3)}})\label{eq:des5ina}\\
\xrightarrow{\substack{\text{table encoding }\\ x^{\text{helper}}x^{(2)}x^{(3)}\leftrightarrow x^{\text{helper}}perm(y^{(2)}y^{(3)})}}& (\ket{x_0^{\text{helper}}}+\ket{x_1^{\text{helper}}})\otimes \underbrace{perm((\ket{y_0^{(2)}}+\ket{y_1^{(2)}})\otimes(\ket{y_0^{(3)}}+\ket{y_1^{(3)}}))}_{\text{reversibly encoded part}}\end{align}}
 However, this is still not sufficient to guarantee the weak security on both of the output keys. Here the final ingredient is, when we encode the mapping above, we design the underlying mapping for the reversibly encoded part carefully. Note that the reversible encoding of two gadgets to two gadgets have multiple ways of encoding. We consider two different encodings for the reversibly encoded part: the \emph{CNOT-style mapping} and the \emph{identity-style mapping}, and associate them to different \emph{branches} of the helper gadget: we encrypt the CNOT-style mapping under $x_1^{\text{help}}$ and the identity-style mapping under $x_0^{\text{help}}$. For the honest setting behavior, the server could still implement the honest mapping coherently on both branches. For the security, we explore a powerful property of the padded Hadamard test called \emph{coherency restriction}, which put a restriction that the behavior of the adversary on two branches should not be too different. But on the other hand, these two branches restrict the adversary's behavior in different ways, which restricts a cheating server's behavior powerfully.\par
In this way we can construct a weakly secure protocol, which is a basic subroutine in our paper. We leave the details and a more technical overview to Section \ref{sec:7}. Overall speaking, what we have achieved could be understood as follows. With the cost of one gadget (the helper gadget), the input gadget on the third wire in (\ref{eq:des5ina}) is ``technically teared up'' into two gadgets, with  securities weaker than the input. (The recovery of the security will be done in the later amplification step.)\par
  But the simple weakly secure protocol above is still not gadget-increasing. The reason is when we save one gadget, we also consume one. But this problem can be solved through a parallel-repetition-style step, and note the helper gadget (consumed in the padded Hadamard test) can be shared in each table. Then we get an $1+n\rightarrow 2n$ protocol (in Section \ref{sec:8l}), which is provable to be both gadget-increasing and weakly-secure.\par
\section{Discussions}\label{sec:1.4}


This result naturally gives rise to the following questions:
\begin{enumerate}
	\item How can we use these techniques on other problems? (for example, quantum computation verification, or zero-knowledge proof.)
	\item Is it possible to do universal blind quantum computation using completely classical client and quantum random oracle model, and make it secure again any unbounded adversary which only makes polynomial number of queries? Is it possible to do universal blind quantum computation using succinct client side quantum computation without relying on any assumptions? We conjecture the answer is no, but we need a formal proof for it.
	\item Is it possible to directly base the protocol on standard model assumptions (quantum-secure oneway functions, or hash functions?)
\end{enumerate}
One intuitive way to think about the future direction is through Table \ref{tb:tradeoff}. There are many unknown cells in this table, and the completion of this table will be interesting. And one interesting thing is: similar (although not the same) tradeoffs also appear in many other problems, not restricted to the delegation-style quantum protocols. For example, in the classical world, for the ``secure key agreement'' problem, symmetric key encryption scheme allows two parties to expand succinct size of pre-shared keys; to achieve key agreement without pre-shared keys, public key encryption is necessary. Thus we wonder how fundamental it is in quantum (or not only quantum) cryptography.
\section*{Acknowledgement}
The author would like to give sincere thanks to Prof. Adam Smith for the advisory. And the author would like to thank anonymous reviewers, Hezi Zhang, Thomas Vidick and Tomoyuki Morimae for useful comments.
\section{Paper Organizations}
This paper is organized as follows.
\begin{enumerate}
\item In Chapter \ref{cht:1} we introduce our work and results, and give a simple overview for the ideas.
\item In Chapter \ref{cht:2}  we give the preliminaries in this paper.
\item In Chapter \ref{cht:4} we give a more detailed overview for the protocol construction and our ideas in Theorem \ref{thm:1}.
\item In Chapter \ref{cht:5} to \ref{cht:8} we formally prove our main result, a universal blind quantum computation protocol where the client only uses $\fpoly(\kappa)$ quantum gates and the assumption is the quantum random oracle model. This proves Theorem \ref{thm:1}.\par
We put the missing proofs of this part in the appendix.\par
In more details:
\begin{itemize}
\item Chapter \ref{cht:5} is the preparation for the protocol design: we define several notions and notations and give a protocol design framework.
\item In Chapter \ref{cht:6} we give a weakly secure gadget increasing protocol.
\item In Chapter \ref{cht:7} we amplify the weakly secure protocol to a fully secure protocol.
\item Finally in Chapter \ref{cht:8} we give a complete universal blind quantum computation protocol (Protocol \ref{prtl:17}).
\end{itemize}	
\end{enumerate}

\cleardoublepage
\chapter{Preliminaries}\label{cht:2}
\section{Basics of Quantum Computation}\label{sec:2.1}
We refer to \cite{NielsenChuangs} for the introduction of quantum computation. And we refer to \cite{AnnesQHELowT} and the preliminary section of \cite{revgt} for an introduction of quantum cryptography. We first clarify some notation here.\par
\begin{nota}\label{nota:2.1}
	We write $\ket{\varphi}\approx_\epsilon\ket{\phi}$ if $|\ket{\varphi}-\ket{\phi}|\leq \epsilon$, where $|\cdots|$ means the norm of a complex vector.
\end{nota}
Note that in the sections later we will give many similar notations, including $\approx_\epsilon^{\fAdv\in \cA}$, $\approx^{st-ind}_\epsilon$ and so on. The notation given in Notation \ref{nota:2.1} is the strongest one.\par
\section{Security Formalization against Quantum Adversaries}\label{sec:2.2}

The security of the blind quantum computation can be formalized as the \emph{qIND-CPA} security, as discussed in \cite{AnnesQHELowT}.
\begin{defn}[qIND-CPA game]\label{def:2.1rr}
	Suppose the protocol that we consider is called $\fSuccUBQC$, which takes the security parameter $\kappa$ and a quantum circuit $C$ as the input. Consider the following game between a challenger and an adversary:
	\begin{itemize}
		\item The adversary chooses a quantum circuit $C$ (whose size is at most a fixed subexponential function of $\kappa$).
		\item The challenger samples $b\in_r\{0,1\}$. If $b=1$, it runs $\fSuccUBQC(\kappa,C)$ with the adversary. If $b=0$, it runs $\fSuccUBQC(\kappa,0^{|C|})$ with the adversary.
		\item The adversary tries to guess $b$.
	\end{itemize}
	The distinguishing advantage is defined to be
	\begin{equation}
		|Pr(\fAdv^{\fSuccUBQC}(\kappa,C)=0)-Pr(\fAdv^{\fSuccUBQC}(\kappa,0^{|C|})=0)|
	\end{equation}
	where the first term inside is the probability that the adversary outputs $0$ in the case of $b=1$, and the second term inside is the probability that the adversary outputs $0$ in the case of $b=0$.

\end{defn}
In this paper we describe the security using the \emph{qIND-CPA security in the quantum random oracle model against unbounded adversaries with up-to subexponential random oracle queries}, which is defined as Definition \ref{def:2.2} below. 
\begin{defn}[qIND-CPA security in QROM]\label{def:2.2}
	We say a protocol $\fSuccUBQC$ is qIND-CPA secure in the quantum random oracle model against unbounded adversary with up-to subexponential RO queries if for any $\fAdv$ with number of RO queries at most $\fsubexp_1(\kappa)$, for any quantum circuit $C$ of size at most $\fsubexp_2(\kappa)$, there is
	\begin{equation}
		|Pr(\fAdv^{\fSuccUBQC}(\kappa,C)=0)-Pr(\fAdv^{\fSuccUBQC}(\kappa,0^{|C|})=0)|\leq 1/\fsubexp_3(\kappa)
	\end{equation}
	where $\fsubexp_1,\fsubexp_2,\fsubexp_3$ are some fixed subexponential functions.
\footnote{We often use subexponential functions in this paper because subexponential functions can upper-bound arbitrary polynomial functions. In cryptography we often discuss security by saying ``for any polynomial something, there exists a negligible function, something''. But this can lead to complicated order or dependency of quantifiers. We will instead say ``for any something less than $\fsubexp_1(\kappa)$, the distinguishing advantage (or something) is less than $1/\fsubexp_2(\kappa)$, where $\fsubexp_1,\fsubexp_2$ are fixed subexponential functions and $\kappa$ is the security parameter''. By describing the statement in this way we avoid the complicated dependency-of-quantifiers, since many things are fixed.}
\end{defn}
\section{Quantum Random Oracle Model}\label{sec:2.3}
A classical random oracle is an oracle of a random function $\cH:\{0,1\}^\kappa\rightarrow \{0,1\}^\kappa$ which all parties can query with classical inputs. It returns independent random value for different inputs, and returns fixed value for the same input. In practice, a random oracle is usually replaced by a hash function.\par
A quantum random oracle allows the users to query it with quantum states: the users can apply the map $\cH: \ket{a}\ket{b}\rightarrow \ket{a}\ket{\cH(a)\oplus b}$ on its state. The quantum random oracle was raised in \cite{QRO}. It becomes the security proof model for many post-quantum cryptographic scheme \cite{roreliable}. On the other hand, the application of the quantum random oracle in quantum cryptographic problems is not very common, and as far as we know, our work is the first application of it in the delegation-stype problems.\par
The security definitions in the quantum random oracle model is given in the last section. Then by the ``Random Oracle Methodology'' we can conjecture the protocol is also secure in the standard model, when the random oracle is replaced by a hash function in practice. As with proofs in the classical random oracle model, interpreting these security claims is subtle, since there exist protocols that are secure in the random oracle model but insecure in any concrete initialization of hash function.\cite{rorevisited}\par
This paper focuses on the quantum cryptographic protocols in the quantum random oracle model. As far as we know, the assumption of a quantum random oracle is incomparable to any trapdoor assumption. We do not know any construction of public key encryption based on solely quantum random oracle. 

\section{Lookup Tables}\label{sec:4.2.3}
	Now we formalize the notation for \emph{lookup tables}, which can be seen as the non-shuffling version of garbled tables discussed in \cite{YaoGCOrigin}. Note that in this work we do not use the underlying gate in the table to encode the computation; we use table to do key transformations.\par
	First, we need to formalize the underlying encryption scheme more concretely:
\begin{defn}[Underlying encryption scheme $\fEn$ used in this work]\label{def:2.13}
	$$\fEn_k(p;\underbrace{ \ell}_{\substack{\text{padding} \\ \text{length}}},
\underbrace{ \kappa_{\text{tag}}}_{\substack{\text{tag} \\ \text{length}}})$$, or $\fEn_k(p)$ if the parameters are implicit, is defined as follows:\par The client samples $R_1\leftarrow \{0,1\}^l$, $R_2\leftarrow \{0,1\}^l$, output $$((R_1,H(R_1||k)\oplus p), (R_2,H(R_2||k)))$$. The first part is the ciphertext and the second part is the key tag. The length of the random oracle output in the first part is the same as the length of $p$ and the length of the random oracle output of the second part is $\kappa_{tag}$.
\end{defn}
Now we introduce the notations for the classical lookup tables. 
	\begin{defn}[Notation for classical lookup tables]\label{def:2.11}
	$$\fLT(\forall b:x_b\rightarrow y_{g(b)}; \underbrace{ \ell}_{\substack{\text{padding} \\ \text{length}}},
\underbrace{ \kappa_{\text{tag}}}_{\substack{\text{tag} \\ \text{length}}})$$ is defined as the lookup table that maps $x_b$ to $y_{g(b)}$, where $\{x_b\}$ and $\{y_b\}$ are two sets of keys (here two symbols whose only difference is the subscript have the same string length), which means, a table of many rows where each row is $\fEn_{x_b}(y_{g(b)};\underbrace{ \ell}_{\substack{\text{padding} \\ \text{length}}},
\underbrace{ \kappa_{\text{tag}}}_{\substack{\text{tag} \\ \text{length}}})$.\par
	And we also use this notation for multi-input multi-output gates: for example, for a Toffoli gate where the input keys are $K=\{x_b^{(1)},x_b^{(2)},x_b^{(3)}\}_{b\in \{0,1\}}$ and the output keys are\\ $K_{out}=\{y_b^{(1)},y_b^{(2)},y_b^{(3)}\}_{b\in \{0,1\}}$, the notation for the lookup table is
	$$\fLT(\forall b_1,b_2,b_3\in\{0,1\}^3:$$ $$x_{b_1}^{(1)},x_{b_2}^{(2)},x_{b_3}^{(3)}\rightarrow y_{b_1}^{(1)},y_{b_2}^{(2)},y_{b_1b_2\oplus b_3}^{(3)}; \underbrace{ \ell}_{\substack{\text{padding} \\ \text{length}}},
\underbrace{ \kappa_{\text{tag}}}_{\substack{\text{tag} \\ \text{length}}})$$
	, and each row in this lookup table is
	$$\fEn_{x_{b_1}^{(1)}||x_{b_2}^{(2)}||x_{b_3}^{(3)}}(y_{b_1}^{(1)}||y_{b_2}^{(2)}||y_{b_1b_2\oplus b_3}^{(3)};\ell, \kappa_{tag})$$
\end{defn}

\section{Basic Notations Needed}
\begin{nota}[Random oracle query number]\label{nota:4.2.3a}
The oracle query number of a quantum operation $\cU$ is denoted as $|\cU|$.	
\end{nota}
Finally we introduce the following notation to simplify the operation of picking a pair of keys from a key set:
\begin{nota}\label{defn:2.17}
	For $K:=\{x_b^{(i)}\}_{i\in [L],b\in\{0,1\}}$, the following notation extracts the key pair with a specific superscript: $K^{(i)}=\{x_b^{(i)}\}_{b\in\{0,1\}}$.\end{nota}
	And the following notation for concatenating keys in several key sets:
	\begin{nota}\label{nota:2.2}
	Suppose $Set_1,Set_2,Set_3$ are three sets where each set contains strings of some fixed length. We use $Set_1||Set_2||Set_3$ to denote the set \\$S=\{x||y||z\}_{x\in Set_1,y\in Set_2,z\in Set_3}$.
\end{nota}
	\subsection{Blinded Oracle}\label{sec:2.2.2r}
One common operation when we study the random oracle is the \emph{blinded oracle}.
\begin{defn}[Blinded Oracle]\label{def:2.3rr}
	Suppose $Set$ is a set of inputs, we say $H^\prime$ is a blinded oracle of $H$ where inputs in $Set$ are blinded to mean (1)$H^\prime (x),x\in Set$ are independently random to the content of $H$ and each other; (2)$H^\prime (x),x\not\in Set$ are the same as $H(x)$.
\end{defn}
For example, we can say $H(Set||K_{out}||\cdots)$ is blinded to mean we blind all the entries $H(pad||y_b||other)$ where $pad\in Set$, $y_b\in K_{out}$, $other$ can be any string (of some fixed length).
\cleardoublepage
\chapter{An Extended Technical Overview}\label{cht:4}

As said in Chapter \ref{cht:1}, in the remaining chapters we prove the following:
\begin{thm}\label{thm:4.1.1}
	There exists a universal blind quantum computation protocol for circuits of size up to a fixed subexponential function of the security parameter such that \begin{itemize}\item It contains an offline phase and an online phase. In the offline phase the client prepares and sends some quantum gadgets to the server, and in the online phase the client is completely classical. \item The total number of quantum gates to prepare these quantum gadgets is at most a fixed polynomial of the security parameter, independent to the size of the circuit to be evaluated. \item The classical computation, communication and the server-side quantum computation are bounded by a fixed polynomial of the security parameter and the size of the circuit to be evaluated. \item The protocol is secure in the quantum random oracle model against any unbounded malicious server whose number of queries to the random oracle is bounded by a fixed subexponential function of the security parameter.\end{itemize}
\end{thm}

\section{An Extended Top-down Overview of Our Techniques}\label{sec:4.1.1}
In this section we give a more formal top-down overview of our protocol than Chapter \ref{cht:1}. It will still be slightly informal, but will be close to the formal definitions given in the remaining chapter.
\subsection{Two-Step Construction via Remote Gadget Preparation}
How can the client allow the server to evaluate $O(|C|)$ gates using only succinct quantum computation? In our protocol, the client will first prepare $\fpoly(\kappa)$ \emph{gadgets} ($\fpoly$ is a fixed polynomial), then use classical interactions to allow the server to \emph{expand} them to $O(|C|)$ gadgets ``securely''. Here the \emph{gadget} is defined to be the states in the form of $\ket{y_0}+\ket{y_1}$, where $y_0,y_1$ are random different strings, or \emph{keys}. The client holds the keys, and the server should hold the state.\footnote{This form of states was also previously used in several papers like \cite{BCMVV}.}\par
This step --- the preparation and expansion of gadgets --- is called \emph{remote gadget preparation}. Let's give the correctness and security definition informally below (see the Chapter \ref{cht:5} for the formal definition):
\begin{defn}[Remote gadget preparation, informal]\label{def:1.1} The correctness and security of remote gadget preparation are defined as follows.\par
\begin{itemize}
\item (Correctness) When the output number is $L$, the protocol should output key set 	$\{y_b^{(i)}\}_{i\in [L],b\in \{0,1\}}$ on the client side and gadgets $\otimes_{i=1}^L (\ket{y^{(i)}_0}+\ket{y^{(i)}_1})$ on the server side.\par
And we say it's an $N\rightarrow L$ protocol if the protocol starts from $N$ gadgets (the server holds $\otimes_{i=1}^N(\ket{x^{(i)}_0}+\ket{x^{(i)}_1})$ and the client knows all the keys) and ends with $L$ gadgets in the honest setting.
\item (Security) For any key pair (say, the $i$-th keys), a malicious server could not output both keys (which means $y^{(i)}_0||y^{(i)}_1$) with non-negligible probability from the protocol's output state, even if the following information is provided additionally (which makes the adversary more powerful):
	\begin{itemize}
	\item All the other keys that are not at index-$i$.
	\item The hash tags of  $y^{(i)}_0$ and $y^{(i)}_1$.
	\end{itemize}
\end{itemize}	
\end{defn}
Which means, after the protocol completes, at least one key in each key pair should be unpredictable, and this unpredictability is not correlated to the unpredictability of the other key pairs.\par
Then we will construct our universal blind quantum computation protocol that satisfies Theorem \ref{thm:4.1.1} (denoted $\fSuccUBQC$) as follows:
\begin{otl}\label{ppl:1} Design of the $\fSuccUBQC$ protocol:
	\begin{enumerate}
		\item Remote gadget preparation
		: (1) the client sends some initial gadgets to the server, whose size and length are succinct; (2) the client uses classical interactions to allow the server to expand the number of gadgets securely to $\Theta(|C|)$. 
		\item Blind quantum computation execution: using the gadgets output from the previous step, the client and the server evaluate $C$ using only classical interactions.
	\end{enumerate}
\end{otl}

The construction of the secure remote gadget preparation protocol (the first step in Outline \ref{ppl:1}) is the most difficult step. The second step is relatively easier but still non-trivial.\par
\subsection{Remote Gadget Preparation via Weak Security}\label{sec:1.4.2}
The first step of Outline \ref{ppl:1} are achieved as follows: we will define the \emph{weak security} of the remote gadget preparation. We will first construct a weakly-secure protocol, then \emph{amplify} it to a fully secure one. First we need to develop a framework for the design of different subprotocols. This is (mainly) captured by the notion of \emph{weak security}. Informally speaking, in weak security the adversary is possible to output both keys in some key pair with bounded (but not necessarily negligible) probability. (For comparison, in the full security, the adversary is not possible to output both keys --- except with negligible probability).
\begin{defn}[Weak security of a state, informal]
	We say a state is $(2^\eta,C)$-SC-secure for a key pair $K$ if for any adversary with query number at most $2^\eta$, with access to the hash tags of keys in $K$, the norm of outputting both keys in $K$ from this state is at most $C$.
\end{defn}
Here ``SC'' means ``simultanously compute'', which is the notion that we define to characterize the weak security.\par
Then a protocol is weakly secure if the output has some reasonable weak security for each out key pair. (Similar to Definition \ref{def:1.1}, the weak security of each output key pair should still hold when the other key pairs are revealed to the adversary, which prevent some potential correlation between weak security of different pairs.)\par
Then the overall flow of the construction of the secure remote gadget preparation protocol (step 1 of Outline \ref{ppl:1}) is as follows \footnote{This is only a construction flow, we don't mean there is a two-step protocol.}:
\begin{otl}\label{otl:2} Protocol construction for the step 1 of Outline \ref{ppl:1}:
	\begin{enumerate}
		\item Construct a weakly-secure remote gadget preparation protocol such that it can (asymptotically multiplicatively) generate more gadgets than it consumes.
		\item Use some amplification techniques to amplify it to a secure remote gadget preparation protocol.
	\end{enumerate}
\end{otl}
Our work can be seen as the design of a series of subprotocols with different tradeoffs for correctness, (weak) security, etc, and these subprotocols, when combined together, can achieve what we want.\par
Below we give a brief introduction to each step above.
\paragraph{Weakly Secure Protocol Step} The goal in this step is to create more gadgets (possibly with weak security) from some input gadgets. First, we consider the simplest case, generating two gadgets using one input gadget, remotely:
$$\ket{x_0}+\ket{x_1}\rightarrow (\ket{y_0}+\ket{y_1})\otimes (\ket{y_0^\prime}+\ket{y_1^\prime})$$
Which stands for a protocol in the following style:
\begin{enumerate}
	\item[(0)] Initially the server holds $\ket{x_0}+\ket{x_1}$.
	\item The client provides some classical information to the server.
	\item The server uses these information to do the transformation above.
\end{enumerate}
And we want the outputs to have some (possibly weak) security in the malicious setting.\par
 First, when the input and output number are the same, a similar form of quantum-to-quantum transformation can be enabled by providing some additional classical information  called \emph{reversible look-up tables}. As the preliminary, we give a simple example on encoding one-to-one gadget transformation and two-to-two gadget transformation:
\begin{exmp}
To transform $\ket{x_0}+\ket{x_1}$ to $\ket{y_0}+\ket{y_1}$ while preserving the security of keys (in the sense of being unable to computing both keys in each key pair), the client can provide the following ciphertexts to the server: $$\text{Forward table: }\fEn_{x_0}(y_0),\fEn_{x_1}(y_1)$$
$$\text{Backward table: }\fEn_{y_0}(x_0),\fEn_{y_1}(x_1)$$
And the server-side operation goes as follows:
\begin{align}
	&\ket{x_0}+\ket{x_1}\\
	\xrightarrow{\text{decrypt forward table in superposition}}&\ket{x_0}\ket{y_0}+\ket{x_1}\ket{y_1}\\
	\xrightarrow{\text{decrypt backward table in superposition}}&\ket{y_0}+\ket{y_1}\\
\end{align}

\end{exmp}
\begin{exmp}\label{exmp:1.2}
To transform $(\ket{x_0}+\ket{x_1})\otimes (\ket{x_0^\prime}+\ket{x_1^\prime})$ to $(\ket{y_0}+\ket{y_1})\otimes (\ket{y_0^\prime}+\ket{y_1^\prime})$ while preserving the security of keys (in the sense of being unable to computing both keys in each key pair), the client can provide the following ciphertexts to the server: \begin{equation}\label{eq:f1}\text{Forward table: }\fEn_{x_bx^\prime_{b^\prime}}(y_by^\prime_{b^\prime})\text{ for each $b,b^\prime$}\end{equation}
\begin{equation}\label{eq:b1}\text{Backward table: }\fEn_{y_by^\prime_{b^\prime}}(x_bx^\prime_{b^\prime})\text{ for each $b,b^\prime$}\end{equation}
And the server-side operation goes similarly:
\begin{align}
	&(\ket{x_0}+\ket{x_1})\otimes (\ket{x_0^\prime}+\ket{x_1^\prime})\\
	\xrightarrow{\text{decrypt forward table}}&\sum_{b,b^\prime}\ket{x_b}\ket{x_{b^\prime}^\prime}\ket{y_b}\ket{y_{b^\prime}^\prime}\\
	\xrightarrow{\text{decrypt backward table}}&(\ket{y_0}+\ket{y_1})\otimes (\ket{y_0^\prime}+\ket{y_1^\prime})\\
\end{align}
Here the input state contains four components in the standard basis, and the output state also has four components. When we do the back-and-forth encryption, we need to find a one-to-one correspondence between the input components and output components. And we note there is some freedom here: in \eqref{eq:f1}\eqref{eq:b1} we match the input component to the output component with the same index; however we can also match them in some other way. This will be useful later.
\end{exmp}
 But it's not easy to achieve a one-to-two gadget transformation directly here. We will introduce several ideas to achieve this transformation securely. First we rewrite this one-to-two mapping to an equivalent form, a two-to-two mapping where one of the input gadget is described classically\footnote{It's possible to encode the mapping as $1\leftrightarrow 2$ mapping directly, but we choose to encode it in this way for nicer honest setting behavior.} (here we relabel the superscripts to make them consistent with later sections):
 \begin{equation}
\underbrace{\ket{\{x_0^{(2)},x_1^{(2)}\}}}_{\text{classical}}\otimes(\ket{x_0^{(3)}}+\ket{x_1^{(3)}})\rightarrow (\ket{y_0^{(2)}}+\ket{y_1^{(2)}})\otimes(\ket{y_0^{(3)}}+\ket{y_1^{(3)}})
 \end{equation}
 Which is an abbreviation of some protocol in the following style:
 \begin{enumerate}
 \item[(0)] Initially the server holds $\ket{x_0^{(3)}}+\ket{x_1^{(3)}}$.
 \item The first gadget $\{x_0^{(2)},x_1^{(2)}\}$ is directly provided to the server in the form of its classical description, and the server can prepare the gadget on its own. Note that this step only uses classical communication. Then the client can send some information to complete this transformation using the previous two-to-two gadget transformation:
 \begin{align}
&\ket{\{x_0^{(2)},x_1^{(2)}\}}\otimes(\ket{x_0^{(3)}}+\ket{x_1^{(3)}})\\
\rightarrow &(\ket{x_0^{(2)}}+\ket{x_1^{(2)}})\otimes(\ket{x_0^{(3)}}+\ket{x_1^{(3)}})\\
\rightarrow & (\ket{y_0^{(2)}}+\ket{y_1^{(2)}})\otimes(\ket{y_0^{(3)}}+\ket{y_1^{(3)}})
 \end{align}
 \end{enumerate}
Now the honest behavior is supported, but the security breaks down: providing both keys to the server will allow the server to get both keys in some output key pair.\par
We will use rescue the security by combining the following two ideas and get the first weakly-secure gadget increasing protocol.
\begin{itemize}
\item A secret bit-wise permutation on the output keys;
\item Usage of (a revised) Hadamard test and two different types of encodings of the two-to-two mapping.	
\end{itemize}
The combination of these two ideas will restrict the adversary's behavior powerfully and give the protocol weak security. This achieves a fundamental and important step in this work.\par
 The first key step is, instead of implementing this transformation directly, we seek for a transformation to the following state as an intermediate step:
$$\ket{\{x_0^{(2)},x_1^{(2)}\}}\otimes(\ket{x_0^{(3)}}+\ket{x_1^{(3)}})\rightarrow \pi((\ket{y_0^{(2)}}+\ket{y_1^{(2)}})\otimes(\ket{y_0^{(3)}}+\ket{y_1^{(3)}}))$$
where $\pi$ is a random bit-wise permutation sampled by the client, kept secret from the adversary. Suppose each output key has length $\kappa_{out}$, then $\pi$ permutes the bit positions of these $2\kappa_{out}$ bits.\par The secrecy of $\pi$ will be a key ingredient on implementing this mapping securely. But the client still needs to reveal it to the server to allow it to de-permute the gadgets in the end, which seems to be a dilemma. Now we introduce the second idea: the realization of the mapping above will make use of an extra helper gadget, and a subprotocol called \emph{padded Hadamard test}. This padded Hadamard test is a padded variant of the Hadamard test in \cite{BCMVV}. We observe that, this revised Hadamard test has several powerful properties, one of which informally say, if such a test is executed between the client and the server, if the server wants to pass the test with high probability, it loses the ability to predict the keys from the post-test state --- a property that we call \emph{unpredictability restriction}. With this property in mind, we can use this subprotocol as a switch that controls when it's safe to reveal the permutation. Now the transformation goes as follows, where we use $\ket{x_0^{\text{helper}}}+\ket{x_1^{\text{helper}}}$ to denote the helper gadget:
\begin{align}
	&(\ket{x_0^{\text{helper}}}+\ket{x_1^{\text{helper}}})\otimes \ket{\{x_0^{(2)},x_1^{(2)}\}}\otimes(\ket{x_0^{(3)}}+\ket{x_1^{(3)}})\label{eq:des1}\\
	\rightarrow & (\ket{x_0^{\text{helper}}}+\ket{x_1^{\text{helper}}})\otimes \pi((\ket{y_0^{(2)}}+\ket{y_1^{(2)}})\otimes(\ket{y_0^{(3)}}+\ket{y_1^{(3)}}))\label{eq:des2}\\ 
	&\text{(Hadamard test on $\ket{x_0^{\text{helper}}}+\ket{x_1^{\text{helper}}}$)}\label{eq:rr17}\\
	\rightarrow &\pi((\ket{y_0^{(2)}}+\ket{y_1^{(2)}})\otimes(\ket{y_0^{(3)}}+\ket{y_1^{(3)}}))\\
	&\text{(Client reveals $\pi$ if test passes)}\\\rightarrow &(\ket{y_0^{(2)}}+\ket{y_1^{(2)}})\otimes(\ket{y_0^{(3)}}+\ket{y_1^{(3)}})
\end{align}
Then in each time step something is secret in the adversary's viewpoint. Before the test the bit-wise permutation is hidden, and after the test, the adversary is not able to have good predictability on $\{x^{\text{helper}}_0,x^{\text{helper}}_1\}$ anymore. The security of the protocol relies on this fact.\par
Now we go to the construction of (\ref{eq:des1})$\rightarrow$(\ref{eq:des2}). Expanding it a little bit, it is
\begin{align}&(\ket{x_0^{\text{helper}}}+\ket{x_1^{\text{helper}}})\otimes \underbrace{\ket{\{x_0^{(2)},x_1^{(2)}\}}}_{\text{given classically}}\otimes(\ket{x_0^{(3)}}+\ket{x_1^{(3)}})\label{eq:des5in}\\
&\xrightarrow{\substack{\text{backward table encoding }\quad x^{(2)}x^{(3)}\text{ under }\pi(y^{(2)}y^{(3)})\\\text{forward table encoding }\quad \pi(y^{(2)}y^{(3)})\text{ under }x^{(2)}x^{(3)}}}\label{eq:22}\\
&(\ket{x_0^{\text{helper}}}+\ket{x_1^{\text{helper}}})\otimes \underbrace{\pi((\ket{y_0^{(2)}}+\ket{y_1^{(2)}})\otimes(\ket{y_0^{(3)}}+\ket{y_1^{(3)}}))}_{\text{reversibly encoded part}}\end{align}
Which follows Example \ref{exmp:1.2} and adds the permutation $\pi$ on the output keys part. However, this is still not sufficient to guarantee the weak security on both of the output keys. But there is still some freedom on the design of mapping in \eqref{eq:22}, and we will make use of it. Here the final ingredient is, when we encode the mapping above, we design the underlying mapping for the reversibly encoded part carefully. Note that the reversible encoding of two gadgets to two gadgets have multiple ways of encoding. We consider two different encodings for the reversibly encoded part: the \emph{CNOT-style mapping} and the \emph{identity-style mapping}:\par
 Identity-style mapping: for each $b,c\in \{0,1\}$,
 $$e_0:=[\quad\substack{\text{backward table encoding }\quad x_b^{(2)}x_c^{(3)}\text{ under }\pi(y_b^{(2)}y_c^{(3)})\\\text{forward table encoding }\quad \pi(y_b^{(2)}y_c^{(3)})\text{ under }x_b^{(2)}x_c^{(3)}}\quad]$$
 CNOT-style mapping: $b,c\in \{0,1\}$
 $$e_1:=[\quad\substack{\text{backward table encoding }\quad x_b^{(2)}x_c^{(3)}\text{ under }\pi(y_b^{(2)}y_{b+c}^{(3)})\\\text{forward table encoding }\quad \pi(y_b^{(2)}y_{b+c}^{(3)})\text{ under }x_b^{(2)}x_c^{(3)}}\quad]$$ 
  Here we use $e_0$ and $e_1$ to denote the encryption results of these mappings. We note both encodings support (\ref{eq:des1})$\rightarrow$(\ref{eq:des2}), but neither can make it secure when used alone. The idea is to use both, and associate them to different \emph{branches} of the helper gadget. That is, the client sends the following to the server for step (\ref{eq:des1})$\rightarrow$(\ref{eq:des2}):
  \begin{equation}\label{eq:24}\fEn_{x^{\text{helper}}_0}(e_0)||\fEn_{x^{\text{helper}}_1}(e_1)\end{equation}
  And this can be analyzed as follows:
  \begin{itemize}
  	\item The honest behavior is supported. The honest server can first decrypt $e_0$ and $e_1$ coherently using the helper gadget and \eqref{eq:24}:
  	\begin{equation}\label{eq:4.25}\ket{x_0^{\text{helper}}}+\ket{x_1^{\text{helper}}}\xrightarrow{\text{decrypt \eqref{eq:24}}}\ket{x_0^{\text{helper}}}\ket{e_0}+\ket{x_1^{\text{helper}}}\ket{e_1}\end{equation}
  	Then since both $e_0$ and $e_1$ support the two-to-two mapping on the reversibly encoded part, server could still implement the honest mapping coherently on keys with superscript $(2)(3)$.\par
  	Finally \eqref{eq:4.25} is applied again to erase the $\ket{e}$ register and the helper gadget is recovered.
  	\item For the security, we explore a powerful property of our Hadamard test called \emph{coherency restriction}, which put a restriction that the behavior of the adversary on two branches of the helper gadget should not be too different. (Here ``branch'' means either $\ket{x_b^{help}}$ tensoring the other parts for each $b=0,1$.) But on the other hand, these two branches restrict the adversary's behavior in different ways (and here the bit-wise permutation also comes in to restrict the adversary's operation), which restricts a cheating server's behavior powerfully.\par
  	Putting it in a more intuitive way, what the protocol achieved can be explained as follows. Without the helper gadget, the adversary can get either $e_0$ or $e_1$ in the clear. However, after we introduce the helper gadget and encrypt these two reversible encodings in the form of \eqref{eq:24}, the server can only use them in the form of the right hand side of \eqref{eq:4.25}. And any meaningful attack of the adversary will collapse of the two branches of \eqref{eq:4.25} --- which means the adversary will not be able to recover the helper gadget and will not be able to pass the Hadamard test. We will give a more formal proof in Section 3. 
  \end{itemize}
    
In this way we can construct a weakly secure protocol, which is a basic subroutine in our paper. Overall speaking, what we have achieved could be understood as follows. With the cost of one gadget (the helper gadget), the input gadget on the third wire in (\ref{eq:des5in}) is ``technically teared up'' into two gadgets, with  securities weaker than the input. (The recovery of the security will be done in the later amplification step.)\par
  But the simple weakly secure protocol above is still not gadget-increasing. The reason is when we save one gadget, we also consume one. But this problem can be solved through a parallel-repetition-style step, and note the helper gadget (consumed in the padded Hadamard test) can be shared in each table. Then we get an $1+n\rightarrow 2n$ protocol, which is provable to be both gadget-increasing and weakly-secure.\par 
\paragraph{Amplification Step} After we complete the first step in the Flow of Construction 1, we move to the amplification part. We give it an overview here and put the full description in the remaining chapter.\par
 Roughly speaking, we call our technique \emph{repeat-and-combine}. 
 The \emph{repeat} part is a parallel repetition of the weakly secure protocol on many different blocks, and the server is required to pass on all the blocks. The overall protocol still asymptotically doubles the number of gadgets. And this gives the upper-level protocol better weak security than the underlying protocol.\par
  The main challenge is to go from weak security to normal (exponential) security. The \emph{combine} technique combines multiple gadgets into one gadget to reduce the server's ability to compute both keys in a key pair. Let's give a minimal example of our \emph{combine} technique, where only two key pairs are combined.\par
Suppose the server holds $(\ket{x_0}+\ket{x_1})\otimes (\ket{x_0^\prime}+\ket{x_1^\prime})$, while the client knows all the keys. Additionally suppose the server knows the hash tags of all the keys. Then it can make a measurement on the xor of the indexes:
$$(\ket{x_0}+\ket{x_1})\otimes (\ket{x_0^\prime}+\ket{x_1^\prime})\rightarrow$$
$$ (output=0)(\ket{x_0}\ket{x^{\prime}_0}+\ket{x_1}\ket{x^{\prime}_1})\quad (output=1)(\ket{x_0}\ket{x^{\prime}_1}+\ket{x_1}\ket{x^{\prime}_0})$$
then it reports the measurement result to the client, and the client can compute and store the new output key pairs ($\{x_0x_0^\prime,x_1x_1^\prime\}$ or $\{x_0x_1^\prime,x_1x_0^\prime\}$). Intuitively, if the server can output both keys in the output key pair, intuitively it means it not only know both $x_0$ and $x_1$, but also know $x_0^\prime$ and $x_1^\prime$. Thus we can hope the adversary's ability of computing both keys for the new key pair is proportional to the multiplication of the corresponding security bounds for the two input key pairs. And if we continue this combination sequentially and combine $\sqrt{\kappa}$ key pairs one-by-one, we can hope this parameter goes down to an exponentially small value.\par
However the story is not that simple. As far as we know, such a simple combination does not always imply the multiplicativity of the bounds of the adversary's ability of computing both keys. To solve this problem, we add an additional layer --- called $\fSecurityRefreshing$ layer --- in the middle of each round of the combination process. This additional layer can be used to ``strengthen the security'' in each round. It makes use of (and consumes) some ``freshly secure'' gadgets, but the consumption is small and it can refresh the security of a large number of key pairs simultaneously. And we can prove, the new protocol, with the \emph{combine} technique and this additional layer, is exponentially secure (in the sense of Definition \ref{def:1.1}).\par
 This is still not the end of the story. The combination part decreases the number of gadgets multiplicatively by a factor of $\kappa^{O(1)}$ and thus we need to do more to make it gadget-increasing again! The solution is, before we do this \emph{repeat} and \emph{combine}, we need to first self-compose the $1+n\rightarrow 2n$ protocol to boost the gadget-expansion ratio (defined by output gadget number divided by input gadget number) from $2$ to $\tilde\Theta(\kappa)$. Then since we can only combine $\sqrt{\kappa}$ gadgets in the combine process into one gadget the overall gadget expansion ratio is still $\tilde\Theta(\sqrt{\kappa})>2$.\par
 Finally we get a remote gadget preparation protocol that is gadget-increasing (with gadget expansion ratio $>2$) and secure (not just weakly-secure). Intuitively we can simply run this protocol again and again to increase the number of gadgets until we get enough gadgets. Again, we make use of the $\fSecurityRefreshing$ layer to bypass the obstacles in the security proof.
\subsection{Security Proof Techniques}
   Formally proving the security, especially for the \emph{combine} technique part, turns out to be technically challenging. One of the most important idea is a series of \emph{state decomposition lemmas} for the quantum random oracle model, which might be of independent interest. These lemmas serve as a bridge from weak security to normal security in our setting.\par
     Let's give a simple, informal example for that. See Section \ref{sec:2.1} for more detailed definitions for notations. Assume the joint purified state of all the parties is described by a normalized state $\ket{\varphi}$ (which means, we first use a cq-state to describe the state of all the parties, where the server's inner state is the quantum part; and then we use the environment to purify all the randomness in the client side, random oracle side, etc). And we want to study the server's ability to predict a single key stored in some client-side register, denoted as $x_0$. The condition is, assume for any server-side operation $\cU$ which makes at most $2^\kappa$ queries to the random oracle, there is
     \begin{equation}\label{eq:new1}|P_{x_0}\cU\ket{\varphi}|\leq A\end{equation} (Again, the server should know some hash tag of $x_0$.)\par
      Then we can prove, (technically nontrivially,) the state, together with some server-side ancillas, can be decomposed into the linear sum of two states $\ket{\phi}+\ket{\chi}$ where \begin{itemize}\item $\ket{\phi}$ is $(2^{O(\kappa)},2^{-O(\kappa)})$-unpredictable for $x_0$ --- which means, for any server-side operation $\cU$ which makes at most $2^{O(\kappa)}$ queries to the random oracle, there is $|P_{x_0}\cU\ket{\phi}|\leq 2^{-O(\kappa)}$. Compare to \eqref{eq:new1}, the right hand side of (\ref{eq:new1}) becomes exponentially small. \item The norm of $\ket{\chi}$ can be bounded: $|\ket{\chi}|\leq (\sqrt{2}+1)A$.
      \item Both states can be written in a well-behaved form using $\ket{\varphi}$.
      \end{itemize}
     Furthermore, this decomposition, in some cases, can be iterated. This will be useful in the security proof of the \emph{combine} technique. 
\subsection{From Remote Gadget Preparation to $\fSuccUBQC$}\label{sec:1.4.3}
Now we give an informal overview of how to reduce the universal blind quantum computation problem to the remote gadget preparation problem.\par
Recall that in BFK's UBQC protocol \cite{UBQC}, to delegate a circuit $C$, the client needs to prepare the state $\ket{+_{\theta^i}}$, $\theta^i\in_r \{n\pi/4: n=0,\cdots 7\}$ for each gate $g^i$ in $C$. Thus the total number of client side quantum computation is linear in $|C|$. One natural idea is to delegate the preparation of these states further; such a primitive for preparing secret single qubit states is abstracted and formalized into a concept called \emph{8-basis qfactory} \cite{qfactory}. However there is an important difference of our setting from \cite{qfactory} here: we cannot delegate the preparation of these single qubit state ``from scratch''; instead, we make use of the output of the remote gadget preparation protocol.\par
Here we only informally describe our protocol, which simplifies it a lot. First, the client and the server run a remote gadget preparation protocol, with output number $L:=\Theta(|C|)$. The honest server will get the gadgets $\otimes_{i=1}^L(\ket{y_0^{(i)}}+\ket{y_1^{(i)}})$. Then our qfactory protocol (simplified due to the assumption that the gadgets are in the honest form) works as follows, which transforms $\ket{y_0^{(i)}}+\ket{y_1^{(i)}}$ into a single qubit state (which can then be used in the BFK protocol). One interesting trick is the usage of phase lookup table, which is given in \cite{revgt} and it's convenient to use it for 8-basis qfactory.\\
\begin{align}
	&\ket{y_0^{(i)}}+\ket{y_1^{(i)}}\\
	\text{(\emph{Phase lookup table})}\Rightarrow &\ket{0}\ket{y_0^{(i)}}+e^{\mi\tilde\theta^i}\ket{1}\ket{y_1^{(i)}}\\&\text{($\tilde\theta^i=\theta^i_2\pi/2+\theta^i_3\pi/4$, $\theta_2^i,\theta_3^i\in_r \{0,1\}^2$)}\\
	(\text{\emph{Hadamard }\cite{MahadevVerification}})\Rightarrow &\ket{+_{\theta^i}} (\theta^i=\theta^i_1\pi+\theta^i_2\pi/2+\theta^i_3\pi/4)
\end{align}
\section{An Overview of the Protocol Design Framework}
In this section we give an overview of the framework for constructing and studying different subprotocols.
\subsection{Purified Joint States and Basic Notations}\label{sec:4.2.1}
The protocol in our work contains many parties including the client, the server, the random oracle, and the environment. Some parties are classical and some party is quantum. Thus the joint state of all the parties can always be described as a cq-state. Then we introduce the notion of ``purified joint state'', which purifies this state and provides a brief way to describe the joint state:
\begin{defn}[Purified joint state]\label{defn:4.2.1}
We say ``purified joint state'' of many parties to mean a pure state defined as follows. Consider	the inner state of all the parties in the protocol we consider. (Including client, server, random oracle.) This can be described as a cq-state where the server is the quantum party and the client and the random oracle are classical. However, we can purify the classical randomness in this cq-state with the environment and it becomes a highly entangled pure state among client, server, random oracle, and the environment. This allows us to use Dirac symbol to describe the inner state of all the parties briefly.
\end{defn}
Thus when we describe the joint state using this purified notation, when we discuss some keys known by the client, like $K=\{x_0,x_1\}$, we are actually discussing client side registers that store these keys --- which are in quantum state after the purification. So there is an abuse of notation: in the purified notation $x_0,x_1$ means the client side key register, while in the usual notation (like when we discuss the honest behavior) they are just key values.\par
When we apply $P_{x_0}$ on a purified joint state where $x_0$ 	denote a client's key system, it means the projection of some system onto the space that is equal to the content of $x_0$ register.\par
In this purified notion, some common operation like ``the client sends a message to the server'' should be seen as a quantum operation on the state. We introduce a notation for it.
\begin{nota}\label{nota:4.2.1}
Suppose $\ket{\varphi}$ is a purified joint state of a client, a server, and some other parties. $X$ is either a client side system or a classical algorithm that takes some client side systems as inputs. We use $\ket{\varphi}\odot X$ to denote the following operation: The $X$ register (if $X$ is an algorithm, we compute it and put the result in a register of the same name) is copied to some server side unused register.
\end{nota}
\subsection{Basic Notations of Quantum Cryptography}
\begin{nota}[Review: random oracle query number]\label{nota:4.2.3}
The oracle query number of a quantum operation $\cU$ is denoted as $|\cU|$.	
\end{nota}
Then we separate an independent part of the random oracle, which is equivalent to the usual random oracle model, but will be useful in our construction since it provides a standard way for defining the \emph{global tag} of some keys.
\begin{nota}\label{nota:4.2.4}
The hash tag of a key $x$ is defined as $H(\mathfrak{tag}||x)$, where $\mathfrak{tag}$ is an unsued special symbol in the alphabet.
\end{nota}

\subsection{SC-security and Well-behaveness}
Below we formalize the ``SC-security'', which describes the adversary's ability to compute both keys from some joint state of all the parties. This will be the foundation of our security framework.\par
See Definition \ref{defn:4.2.1}, Notation \ref{nota:4.2.1}, \ref{nota:4.2.3} for the notions and symbols appeared in the definition.
\begin{defn}[SC-security]\label{defn:3.4}
	Suppose the purified joint state of all the parties is $\ket{\varphi}$. We say $\ket{\varphi}$ is $(2^\kappa, A)$-SC-secure for keys $K=\{x_0,x_1\}$\quad\footnote{Here $x_0,x_1$ should be considered as client side systems that has already been purified.} given $Z$ (which is either a client side system or a classical algorithm that takes some client side systems as inputs) if:\par
	 For any server side operation $\cU$ with query number $|\cU|\leq 2^\kappa$, (note that $\cU$ can introduce server-side ancilla qubits in the zero state, which is inherent in the expression below,)
	$$|P_{x_0||x_1}\cU(\ket{\varphi}\odot Z\odot Tag)|\leq A$$
	where:
	\begin{itemize}
	 \item $Tag$ denotes the set of hash tags of the keys in $K$.
	 \item $P_{x_0||x_1}$ projects onto the space where $\cU$'s outcome register is equal to the concatenation of client side register $x_0$ and $x_1$.
	 \end{itemize}
\end{defn}
Below we introduce the notion of the well-behaveness of a state. This helps us rules out some ``ill-behaved cases'' like a state that contains the xor of all the random oracle content.
\begin{nota}\label{nota:4.3.1}
Define the well-behaved state $\mathcal{WBS}(D)$ to be the set of joint purified states that can be written as the linear sum of at most $2^D$ states where each state can be prepared with at most $2^D$ random oracle queries.
\end{nota}
\subsection{Complete Definition of the Weak Security of Remote Gadget Preparation}\label{sec:4.3.2}
Since we have defined the SC-security and well-behaveness property, we are prepared to introduce the full definition of the weak security. Again we refer to Section \ref{sec:2.1} for the notations inside it.
\begin{defn}\label{def:4.3.3}
	We say an $N\rightarrow L$ remote gadget preparation protocol run on security parameter $\kappa$ has \emph{weak security transform parameter} \weakparam{\eta}{C}{p}{\eta^\prime}{C^\prime} for input state in $\cWBS(D)$ against adversaries\footnote{Note that this is the adversary during the protocol. When we discuss the output security, there is another adversary hidden in the SC-security definition. These two adversaries can be different.} of query number $\leq 2^\kappa$ if a statement in the following form holds for the protocol:\\
	\begin{mdframed}
			Suppose the input keys are $K=\{x_b^{(i)}\}_{i\in [N],b\in\{0,1\}}$. Suppose the initial state, described by the normalized purified joint state $\ket{\varphi}$,  satisfies the following conditions:
			\begin{itemize}
				\item (Input security) $\forall i\in [N]$, $\ket{\varphi}$  is $(2^\eta,C)$-SC-secure for $K^{(i)}$ given $K-K^{(i)}$
				\item (Input well-behavenss) $\ket{\varphi}\in \cWBS(D)$ 
			\end{itemize}
			For any adversary $\fAdv$ of query number $|\fAdv|\leq 2^\kappa$, the final state when the protocol completes, denoted as
			$$\ket{\varphi^\prime}=ProtocolName_\fAdv(K;Parameters)\circ\ket{\varphi}$$
			, (and correspondingly, output keys are $K_{out}=\{y_b^{(i)}\}_{i\in [L],b\in\{0,1\}}$) at least one of the followings is true:
			\begin{itemize}
				\item (Small Passing probability)	$|P_{pass}\ket{\varphi^\prime}|\leq p$
				\item (Good Output security) $\forall i\in [L], P_{pass}\ket{\varphi^\prime}$ is $(2^{\eta^\prime},C^\prime)$-SC-secure for $K_{out}^{(i)}$ given $K_{out}-K_{out}^{(i)}$.
			\end{itemize}\end{mdframed}
\end{defn}
We can see this definition captures how the security, in terms of the SC-security (Definition \ref{defn:3.4}), evolves during the protocol.\par
And it naturally generalizes to the multi-input-key setting:
\begin{defn}\label{def:4.3.4}
Suppose the remote gadget preparation protocol is denoted by $$ProtocolName_\fAdv(K_1,K_2;Parameters)$$. We say the protocol has weak security transform parameter \\\weakparaml{\eta}{C}{\eta_2}{C_2}{p}{\eta^\prime}{C^\prime} for input state in $\cWBS(D)$ against adversaries of query number $\leq 2^\kappa$ if a statement similar to the one shown in Definition \ref{def:4.3.3} holds, with the following differences: the first condition is replaced by the following conditions:\par
Suppose $K_1$ has $N_1$ pairs of keys and $K_2$ has $N_2$ pairs of keys. $\forall i\in [N_1]$, $\ket{\varphi}$ is $(2^\eta,C)$-SC-secure for $K_1^{(i)}$ given $(K_1-K_1^{(i)})\cup K_2$; and $\forall i\in [N_2]$, $\ket{\varphi}$ is $(2^{\eta_2},C_2)$-SC-secure for $K_2^{(i)}$ given $(K_2-K_2^{(i)})\cup K_1$.
\end{defn}

\section{Overview of the Security Proofs of the Weakly Secure Protocols}
In the introduction we already give an informal introduction to the construction of weakly secure protocol. However, the intuition behind its security has not been explained yet. Here we first formalize this part and give a complete protocol. Then we will explain its intuition for the security proof. We will not discuss the remaining steps and the amplification part in details.
\subsubsection{$1+1\rightarrow 2$ Protocol with Weak Security (When the Input State is Honest)}
In this subsection we formalize the protocol in the introduction, and discuss its security. \par 
There are two blocks of it that need to be formalized: the reversible encoding used in (\ref{eq:des1})$\rightarrow$(\ref{eq:des2}) and the Hadamard test in \eqref{eq:rr17}.
First, we formalize the reversible lookup table used in the protocol:
\subsubsection{Reversible Encoding}
As the preparation, we introduce a notation for the reversible lookup table. Recall that the classical lookup table is discussed in Section \ref{sec:4.2.3}.

\begin{defn}[Definition and notation for reversible lookup tables]\label{defn:2.14}
	$$\fRevLT(\forall b:x_b\leftrightarrow y_{g(b)}; \underbrace{ \ell}_{\substack{\text{padding} \\ \text{length}}})$$ is defined as the reversible lookup table that maps $x_b$ to $y_{g(b)}$ and vice versa, which is, the combination of the following two lookup tables:
	\begin{itemize}
		\item Forward table: $\fLT(\forall b:x_b\rightarrow y_{g(b)}; \ell)$, where the tag length is the same as the length of keys $y_{g(b)}$.
		\item Backward table:  $\fLT(\forall b:y_{g(b)}\rightarrow x_b; \ell)$, where the tag length is the same as the length of keys $x_{b}$.
	\end{itemize}
	As Definition \ref{def:2.11}, this notation can be applied in the multi-key case and the keys in different wires are concatenated before feeding into the $\fEn$ operation. The tag length is the same as the total output length.
\end{defn}
\begin{defn}[Simplified notations for some reversible lookup tables]\label{defn:2.15}
	For a reversible lookup table on input key set $K$ and the output key set $K^\prime$, if the type of the gate is implicit, when there is no ambiguity, we can use $\fRevLT(K\leftrightarrow K_{out};\ell)$ to denote the reversible lookup table that maps the keys in $K$ to the corresponding keys in $K_{out}$, and back.
\end{defn}

Then we can construct the lookup table we construct in the introduction:
\begin{defn}\label{def:4.4.3}$\fRobustRLT(K_{\text{help}},K_{in}\leftrightarrow K_{out}, \pi; \ell)$, where \begin{itemize}\item $K_{\text{help}}=\{x_b^{\text{help}}\}_{b\in \{0,1\}}$, $K_{in}=\{x_b^{(2)},x_b^{(3)}\}_{b\in \{0,1\}}$,$K_{out}=\{y_b^{(2)},y_b^{(3)}\}_{b\in \{0,1\}}$, and the keys with the same symbol and superscript have the same length; $y_b^{(2)}$ and $y_b^{(3)}$ have the same length. \item $\pi$ is a bit-wise permutation on the strings of length $2\kappa_{out}$, $\kappa_{out}$ is the key length of $y_b^{(2)}$. \item $\ell$ is the padding length.\end{itemize}

	is defined as follows.\par
First consider the two reversible encoding between $K_{in}^{(2,3)}$ and $K_{out}$: (Below is their encryption structure.)
	$$\text{(Identity-style) }\fRevLT_{b_1=0}: \fRevLT(\forall b_2,b_3\in \{0,1\}^2:$$ $$ x^{(2)}_{b_2}||x_{b_3}^{(3)}\leftrightarrow \pi(y^{(2)}_{b_2}||y_{b_3}^{(3)});\underbrace{ \ell}_{\substack{\text{padding} \\ \text{length}}})$$
	$$\text{(CNOT-style) }\fRevLT_{b_1=1}:\fRevLT(\forall b_2,b_3\in \{0,1\}^2:$$ $$ x^{(2)}_{b_2}||x_{b_3}^{(3)}\leftrightarrow \pi(y^{(2)}_{b_2}||y_{b_2+b_3}^{(3)});\underbrace{ \ell}_{\substack{\text{padding} \\ \text{length}}})$$
	Then the $\fRobustRLT$ is defined as \begin{equation}\label{eq:4.32}\fEn_{x_0^{\text{help}}}(\fRevLT_{b_1=0};\underbrace{ \ell}_{\substack{\text{padding} \\ \text{length}}},\underbrace{ \ell}_{\substack{\text{tag} \\ \text{length}}})\end{equation}\begin{equation*}||\fEn_{x_1^{\text{help}}}(\fRevLT_{b_1=1};\underbrace{ \ell}_{\substack{\text{padding} \\ \text{length}}},\underbrace{ \ell}_{\substack{\text{tag} \\ \text{length}}})\end{equation*}
	\end{defn}

\subsubsection{Hadamard Test}
Then the (padded) Hadamard test is formalized as follows:
\begin{mdframed}[style=figstyle]
\begin{defn}[Padded Hadamard test]\label{def:4.4.4}
	The padded Hadamard test $$\fPadHadamard(K;\underbrace{ \ell}_{\substack{\text{padding} \\ \text{length}}},
\underbrace{ \kappa_{\text{out}}}_{\substack{\text{output} \\ \text{length}}})$$ on $K=\{x_0,x_1\}$ is defined as follows:
	\begin{enumerate}
		\item The client samples $pad\leftarrow_r\{0,1\}^l$ and sends $R$ to the server.
		\item The server returns $d$ such that $d\cdot (x_0||H(pad||x_0))=d\cdot (x_1||H(pad||x_1))$ where $H(pad||x_b)$ has length $\kappa_{out}$, and $d$ is not all zero on the last $\kappa_{out}$ bits. The client checks the server's response.
	\end{enumerate}
	The honest server can pass this test by making Hadamard measurement on $\ket{x_0}\ket{H(pad||x_0)}+\ket{x_1}\ket{H(pad||x_1)}$.
\end{defn}\end{mdframed}
Let's first show some intuition behind our technique. Consider an adversary that can pass this test with high probability. And we want to understand what this fact can tell us about the server's state. It seems to be a hard problem: different from the standard basis measurement in \cite{MahadevVerification}, which directly tells us the state of the server before the measurement, Hadamard basis measurement does not give us anything like that. However, it tells us what the server \textbf{can not do} after the test:
\begin{lem}[Unpredictability restriction, informal]\label{lem:4.4.1} For $K=\{x_0,x_1\}$, suppose (1)the initial state is sufficiently SC-secure for $K$; (2)the adversary can pass the padded Hadamard test with high probability, then for any $b\in \{0,1\}$, the adversary can only compute $x_b$ from the post-test state with small probability. 
\end{lem}
The main idea of the proof is a trick on the time-order-switching of two measurements.
\begin{proof}[Idea of the proof]
	Suppose the adversary can pass the padded Hadamard test with high probability. Then it proceeds and try to compute $x_0$ or $x_1$. Since these two measurements commute (the measurement of getting the output for the Hadamard test, and the measurement that tries to compute $x_b$), imagine that this adversary first measures and gets one of $x_0$ and $x_1$ and then sends out the result for the padded Hadamard test, the probability of passing the test should not change. On the other hand, if the adversary gets one of $x_0$ or $x_1$, by the SC-security of the initial state it cannot get the other key, which implies it should be very hard to pass the padded Hadamard test. (Note that the unpadded Hadamard test does not guarantee this!)
\end{proof}
We note the Hadamard test does not guarantee that the server throws away the keys in $K^{\text{help}}$ completely: the server can cheat with some probability. But we can see this subprotocol does provide some level of security, which is sufficient for our purpose.\par
Another property that we will use is the \emph{coherency restriction}: starting from a initial state that has a two-branch form, the behavior of the corresponding two outcome branches should behave coherently on any efficient adversarial attack:
\begin{lem}[Coherency restriction, informal]\label{lem:4.4.3} For $K=\{x_0,x_1\}$, suppose (1)the initial state is sufficiently SC-secure for $K$; (2) the initial state has the form of $\ket{x_0}\ket{\cdots}+\ket{x_1}\ket{\cdots}$; (3)the adversary can pass the padded Hadamard test with high probability, then denote $\ket{x_0}\ket{\cdots}$ and $\ket{x_1}\ket{\cdots}$ as two \emph{branches} of the state, and denote $\ket{\varphi_0^\prime}$ and $\ket{\varphi_1^\prime}$ as the corresponding outcome state of these two branches. Then for any adversarial operation on the post-test state that ends with a projective measurement, for any measurement result, the probability of getting this result on these two branches should not be too far away from each other.

\end{lem}
This gives the adversary a strong restriction since this lemma intuitively tells us its behavior on these two branches should be ``coherent''. This lemma will be a key property on proving the security of our basic weakly-secure gadget-increasing protocol. The proof is similar to the unpredictability property: assuming a distinguisher can break the coherency restriction, then we can postpone the measurement of Hadamard test output $d$ behind the attack of the distinguisher, and show that the probability of passing the Hadamard test will not be too high.
\subsubsection{Protocol Design}
Now we can formalize our first remote gadget preparation protocol described in the introduction. We name it as $\fGdgPrep^{Basic}$.
\begin{mdframed}[style=figstyle]
\begin{prtl}\label{prtl:4.6}$\fGdgPrep^{Basic}(K^{\text{help}},K^{(3)};\ell,\kappa_{out})$, where $K^{\text{help}}=\{x_b^{\text{help}}\}_{b\in \{0,1\}}$, $K^{(3)}=\{x_b^{(3)}\}_{b\in \{0,1\}}$. $\ell$ is the padding length and $\kappa_{out}$ is the output key length:\\
	For an honest server, the initial state is $\ket{\varphi}=(\ket{x_0^{\text{help}}}+\ket{x_1^{\text{help}}})\otimes (\ket{x_0^{(3)}}+\ket{x_1^{(3)}})$. \par
	\begin{enumerate}
		\item the client samples \begin{itemize}\item A permutation $\pi$ of $[2\kappa_{out}]$. We will view this as acting bit-wisely on $\{0,1\}^{2\kappa_{out}}$.\item a pair of different (input) keys $K^{(2)}=\{x_0^{(2)},x_1^{(2)}\}$ with the same length as $x_b^{(3)}$;\item $2$ pairs of different (output) keys \\$K_{out}=\{y_b^{(2)},y_b^{(3)}\}_{b\in \{0,1\}}$ with key length $\kappa_{out}$.\end{itemize}
		\item The client computes $$\fRobustRLT(K^{\text{help}},K^{(2,3)}\leftrightarrow K_{out}, \pi;\underbrace{ \ell}_{\substack{\text{padding} \\ \text{length}}})$$ and sends it together with $K^{(2)}$ to the server.
		\item An honest server should implement the following mapping:
		      \begin{align}\ket{\varphi}=&(\ket{x_0^{\text{help}}}+\ket{x_1^{\text{help}}})\otimes(\ket{x_0^{(3)}}+\ket{x_1^{(3)}})\\\rightarrow &(\ket{x_0^{\text{help}}}+\ket{x_1^{\text{help}}})\otimes \pi((\ket{y_0^{(2)}}+\ket{y_1^{(2)}})\otimes(\ket{y_0^{(3)}}+\ket{y_1^{(3)}}))\end{align}
		\item The client and the server run the padded Hadamard test on $K^{\text{help}}$. The server can use $\ket{x^{\text{help}}_0}+\ket{x_1^{\text{help}}}$ to pass the test, as described in Definition \ref{def:4.4.4}. Reject if the server does not pass this test.
		\item The client sends out $\pi$.
		\item The server removes the permutation and gets $(\ket{y_0^{(2)}}+\ket{y_1^{(2)}})\otimes (\ket{y_0^{(3)}}+\ket{y_1^{(3)}})$.
	\end{enumerate}
\end{prtl}\end{mdframed}
\paragraph{Correctness} This protocol transforms $2$ gadgets to $2$ gadgets.
\paragraph{Efficiency} Both the client and the honest server run in polynomial time (on the key size and the parameters).\par
The security statement for $\fGdgPrep^{Basic}$ is given below. To focus on the most simplified cases, we will only care about the case where the input state is fully honest. This also allows us to omit the well-behaveness requirement in the full weak security formalism (see Definition \ref{def:4.3.3}). We the general case will be handled in the remaining chapter.\par
See Section \ref{sec:2.1} and Definition \ref{def:4.3.4} for the notations.
\begin{lem}\label{lem:4.4.4}
	There exist constants $A,B\geq 1$ such that the following statement is true for sufficiently large security parameter $\kappa$:\par
	For keys $K^{\text{help}}=\{x_b^{\text{help}}\}_{b\in \{0,1\}},K^{(3)}=\{x_b^{(3)}\}_{b\in \{0,1\}}$ which are both a pair of keys, protocol 
	$$\fGdgPrep^{Basic}(K^{\text{help}},K^{(3)};\underbrace{ \ell}_{\substack{\text{padding} \\ \text{length}}},
\underbrace{ \kappa_{\text{out}}}_{\substack{\text{output} \\ \text{length}}})$$
	has weak security transform parameter \begin{center}\weakparaml{\eta}{2^{-\eta}}{\eta}{4C}{(1-C^2)}{\eta/B}{AC}\end{center} for states defined below against adversaries of query number $\leq 2^\kappa$ when the following input state form and inequalities are satisfied.
	\begin{equation}\label{eq:4.44}\ket{\varphi}=\text{ purified joint state of }(\ket{x^{\text{help}}_{0}}+\ket{x^{\text{help}}_{1}})\otimes (\ket{x^{\text{(3)}}_{0}}+\ket{x^{\text{(3)}}_{1}})\end{equation}

	The inequalities are as follows:
	\begin{enumerate}
		\item (Sufficient security on the inputs) $\eta\geq \kappa\cdot B$. 
		\item (Sufficient padding length, output key length) $\ell\geq 4\eta$, $\kappa_{out}>\ell+4\eta$
		\item For simplicity, additionally assume $\frac{1}{9}>C>2^{-\sqrt{\kappa}}$
	\end{enumerate}
	\end{lem}
	\subsection{Intuition for the Security Proof of Lemma \ref{lem:4.4.4}}\label{sec:4.4.1}
	\subsubsection{How the bit-wise permutation protect the security}\label{sec:7.1}
	Here we use a simple, quick explanation to informally explain how the bitwise permutation takes effect in more details. we analyze the behavior of our protocol, and the behavior of our protocol with the bit-wise permutation removed, against a specific class of adversaries --- adversaries that tries to decrypt the table classically.\par 
Let's see what happens when we remove the bit-wise permutation.
\begin{prtl}[Free Lunch Protocol]
	$\fGdgPrep^{FreeLunch}(K^{\text{help}},K^{(3)})$, where $K^{\text{help}}=\{x_b^{\text{help}}\}_{b\in\{0,1\}},K^{(3)}=\{x_b^{(3)}\}_{b\in\{0,1\}}$\par	The honest initial state is $\ket{\varphi}=(\ket{x_0^{\text{help}}}+\ket{x_1^{\text{help}}})\otimes (\ket{x_0^{(3)}}+\ket{x_1^{(3)}})$. \par
	\begin{enumerate}
		\item The client samples a pair of different keys $K^{(2)}=\{x_0^{(2)},x_1^{(2)}\}$ and samples 2 pairs of different keys $K_{out}=\{y_b^{(2)},y_b^{(3)}\}_{b\in\{0,1\}}$ as the output keys.
		\item The client computes $$\fRevLT(K^{\text{help}},K^{(2,3)}, K_{out};\underbrace{ \ell}_{\substack{\text{padding} \\ \text{length}}})$$ and sends it together with $K^{(2)}$ to the server. (Here we abuse the notation of $\fRevLT$ to mean we remove the permutation in (\ref{eq:rrlt}))
		\item An honest server should implement the following mapping:
		      \begin{equation}\scriptstyle\ket{\varphi}=(\ket{x_0^{\text{help}}}+\ket{x_1^{\text{help}}})\otimes(\ket{x_0^{(3)}}+\ket{x_1^{(3)}})\rightarrow (\ket{x_0^{\text{help}}}+\ket{x_1^{\text{help}}})\otimes (\ket{y_0^{(2)}}+\ket{y_1^{(2)}})\otimes(\ket{y_0^{(3)}}+\ket{y_1^{(3)}})\end{equation}
	\end{enumerate}
\end{prtl}

Let's understand why this idea based on normal reversible lookup table does not work. 
\paragraph{Attack to the ``Free lunch protocol''} The server can simply break the lookup table as follows: 
\begin{enumerate}\item It first makes measurements on the input states that it holds, and gets $x_{b_1}^{\text{help}}$, $x_{b_3}^{(3)}$. Then when $b_1=1$ (CNOT-style branch) together with $x_{0}^{(2)}, x_{1}^{(2)}$ it can decrypt two rows in the forward part of the reversible lookup table:
\begin{equation}\label{eq:53pw}x_{0}^{(2)}x_{0}^{(3)}\leftrightarrow y_{0}^{(2)}y_{0}^{(3)}\end{equation}
\begin{equation}\label{eq:54pw}x_{1}^{(2)}x_{0}^{(3)}\leftrightarrow y_{1}^{(2)}y_{1}^{(3)}\end{equation}
 \item Then it slices out the different blocks on the right side of (\ref{eq:53pw})(\ref{eq:54pw}), and recombines them, and it can decrypt the other rows (that are shown above) in the backward lookup table:
 $$ y_{0}^{(2)}y_{1}^{(3)}\leftrightarrow \cdots$$
$$ y_{1}^{(2)}y_{0}^{(3)}\leftrightarrow\cdots$$
  and then also break the forward table.\end{enumerate}
Now as a comparison, an adversary that attacks the Protocol \ref{prtl:6}, will encounter mappings in the following form:
\begin{equation}\label{eq:53pwn}x_{0}^{(2)}x_{0}^{(3)}\leftrightarrow perm(y_{0}^{(2)}y_{0}^{(3)})\end{equation}
\begin{equation}\label{eq:54pwn}x_{1}^{(2)}x_{0}^{(3)}\leftrightarrow perm(y_{1}^{(2)}y_{1}^{(3)})\end{equation}
but it is not aware of $perm$. Now the right hand side of (\ref{eq:53pwn})(\ref{eq:54pwn}) seems like two independently random strings, and it's not possible to slice out the correct bits and recombine to get other key combinations. Thus this type of attack is avoided.
\subsubsection{Understanding the Protocol in the Padded Hadamard Test Viewpoint}\label{sec:7.2} Here we view the Protocol \ref{prtl:6} starting from the padded Hadamard test. Compared to the ``classical adversary viewpoint'', this viewpoint has the advantage of being more formal and can be used as the guideline for the formal proof.\par
We can view the lookup table with key $x_0^{\text{help}}$ and with key $x_1^{\text{help}}$ as different \emph{branches} of this test. 
Correspondingly, the look-up table can also be divided into two smaller reversible lookup tables: $\fRevLT_{b_1=0}$ and $\fRevLT_{b_1=1}$. 
 On both the $b_1=0$ (identity-style) branch and $b_1=1$ (CNOT-style) branch, the following mapping in the honest setting is encoded (protected under bitwise permutation, which we ignore temporarily):
$$\ket{x_0^{(3)}}+\ket{x_1^{(3)}}\rightarrow(\ket{y_0^{(2)}}+\ket{y_1^{(2)}})\otimes (\ket{y_0^{(3)}}+\ket{y_1^{(3)}})$$
But it is encoded in different ways in $\fRevLT_{b_1=0}$ and $\fRevLT_{b_1=1}$. (See (\ref{eq:b1gc})-(\ref{eq:b11gc2}) below for details.) We can see the structure of these two branches are actually quite different (and actually even distinguishable). Consider an adversary that tried to cheat, then we can consider its behavior on these two branches. Then it is ``driven'' differently by the different table structure on two branches. 
	 On the other hand, the padded Hadamard test is a very powerful tool on controlling the adversary's behavior. Its untestability property intuitively tells us
	 	 \begin{center}
	 \emph{The predictability of keys, if significant, could not behave too differently in these two branches.}	
	 \end{center}
Or in other words, the padded Hadamard test forces the adversary to behave coherently (to some extent) on two branches. And these properties, when combined together, will lead us to the proof of weak security. (Either the adversary could not pass the protocol with very high probability, or it loses the ability to predict the output keys to some extent.)\par
On the other hand, the possible behaviors on these two branches, although quite different, do have an intersection: the honest mapping. So an honest server can choose to follow the honest mapping, and a malicious server which does not following the honest mapping will be caught by the test. (How could the malicious behavior after the test be caught by a test that happens before it? But it could, and that's part of the reason that we say the padded Hadamard test is powerful!)\par
Below we give a more detailed intuition, which is also the intuition behind our formal proof. As a preparation, let's imagine the server gets one of $x_{b_1}^{\text{help}}$ classically. Now the lookup table behaves as follows (and let's ignore the helper gadget; and these two rows correspond to two choices of the pad input gadget):\par
Identity-style (and without loss of generality, assume $b_3=0$):
	      \begin{align}
		      x_0^{(2)},x_0^{(3)} & \leftrightarrow perm(y_0^{(2)}||y_0^{(3)}) \label{eq:b1gc}\\
		      x_1^{(2)},x_0^{(3)} & \leftrightarrow perm(y_1^{(2)}||y_0^{(3)})\label{eq:b1gc2}
	      \end{align}
	      CNOT-style (and without loss of generality, assume $b_3=0$):
	      \begin{align}
		      x_0^{(2)},x_0^{(3)} & \leftrightarrow perm(y_0^{(2)}||y_0^{(3)}) \label{eq:b11gc1}\\
		      x_1^{(2)},x_0^{(3)} & \leftrightarrow perm(y_1^{(2)}||y_1^{(3)})\label{eq:b11gc2}
	      \end{align}
	 
Let's first study the security of output keys on the wires with index ``$(3)$''. Assuming the adversary can pass the protocol with high probability, the security is guaranteed by the following sequence of informal properties:
\begin{align}
&\text{(LT structure on identity-style branch (see (\ref{eq:b1gc})(\ref{eq:b1gc2}))}\\
\Rightarrow &K_{out}^{(3)} \text{ is SC-secure on identity-style branch}\\
&\text{(then by padded Hadamard test property)}\\\Rightarrow &K_{out}^{(3)} \text{ is SC-secure on both branches}
\end{align}
The security on keys with index ``$(2)$'' is more subtle. Very informally, the security on wire ``$(3)$'' is ``partly diffused'' to wire ``$(2)$'' by the permutation. Now the sequence of properties that guarantee the security is as follows:
\begin{align*}
&\text{Proof by contradiction: $K_{out}^{(2)}$ could be computed out}\\
&\text{(By Hadamard test property)}\\\Rightarrow &\text{$K_{out}^{(2)}$ could be computed out on CNOT branch}\\
&\text{(Then by LT structure on CNOT branch }\text{(see (\ref{eq:b11gc1})(\ref{eq:b11gc2}),}\\&\text{ permuted terms in RHS are independently random bits)}\\
&\Rightarrow\text{when provided a \emph{fake}}\text{ (independently sampled) permutation,}\\ &\text{ many bits in (both keys of )}\text{ $K_{out}^{(2)}$ could be computed out on CNOT branch }\\
&\text{(Then by Hadamard test property)}\\\Rightarrow &\text{Many bits in (both keys of )}\\&\text{ $K_{out}^{(2)}$ could be computed out on identity branch}\\
&\text{(LT structure on identity branch (see (\ref{eq:b1gc})(\ref{eq:b1gc2}))}\\\Rightarrow &\text{ contradiction}
\end{align*}
As we can see, we use the padded Hadamard test property twice and use the lookup table properties twice. The lookup table structure on two branches gives restrictions on the adversary in different ways, which, when combined with the padded Hadamard test property, gives what we want to prove.\par
Finally note that the proof-by-contradiction-based intuition above is not how we organize the formal proof. The formal proof will use a forward quantitative analysis structure.
\section{Organizations of the Remaining Sections, and Protocol Diagrams}
We note there are different illustrations of protocol  structure. In section \ref{sec:4.5.1} we give a diagram for the logical dependency of different subprotocols in this paper, and show the honest behavior security property that they guarantee. But this diagram does not show how the protocols are really executed. Then in section \ref{sec:4.5.2} we give a series of diagrams that shows the execution process of these subprotocols. Then in section \ref{sec:4.5.3} we give an overview of different sections of the remaining paper.
\subsection{A Diagram of Flow of Constructions}\label{sec:4.5.1}
Below in Figure \ref{fig:flowofc} we give a diagram which shows the flow of protocol construction. The arrow shows the dependency among subprotocols.  Each node shows the honest behavior (using notation ``input number $\rightarrow$ output number'') and security property. The security property is mainly described by the \emph{weak security transform parameter} but not necessarily in the standard form described above, and we will meet different variants of weak security transform parameter definition in the paper. From above to below, the security of the first three protocols use the multi-input-key-set version of the weak security transform parameter (Definition \ref{def:4.3.4}); Then for Protocol \ref{prtl:9} we use the most basic form of the weak security transform parameter (Definition \ref{def:4.3.3}) since the input keys all have the same security requirement. Then after the repeat technique the $p$ term is removed (Definition \ref{def:3.12b}) and after the combine technique we move to a version that deals with two key sets in a slightly differently way (Definition \ref{def:newdef}). And in the $\fSecurityRefreshing$ layer the secuity conditions on the input keys are weaker than normal thus we add a star symbol there (see Section \ref{sec:10.2}). Finally in Protocol \ref{prtl:13} we don't need to talk about weak security transform parameter anymore.\par 

\begin{figure}
\centering
\begin{tikzpicture}
\node (1a0) {};
\node (1a1) [below=of 1a0] {\parbox{20em}{\scriptsize\centering Protocol \ref{prtl:4.6}, $1+1\rightarrow 2$,\\ \weakparaml{\eta}{2^{-\eta}}{\eta}{C}{(1-\kappa^{-O(1)})}{O(\eta)}{O(C)}\\ assuming specific input form, \\Lemma \ref{lem:7.6}}};
\node (1a05) [below right=-3pt and 0pt of 1a0] {\parbox{17em}{\scriptsize{ reversible lookup table,\\ padded Hadamard test}}} ;
\node (1a2) [below=of 1a1] {\parbox{20em}{\scriptsize\centering Protocol \ref{prtl:7}, $1+1\rightarrow 2$,\\ \weakparaml{\eta}{2^{-\eta}}{\eta}{C}{(1-\kappa^{-O(1)})}{O(\eta)}{O(C)},\\Lemma \ref{lem:7.9}}};
\node (1a3) [below=of 1a2] {\parbox{30em}{\scriptsize\centering Protocol \ref{prtl:n12n}, $1+n\rightarrow 2n$,\\ (when $n\leq \kappa$) \weakparaml{\eta}{2^{-\eta}}{\eta}{C}{(1-\kappa^{-O(1)})}{O(\eta)}{O(C)},\\ Lemma \ref{lem:7.10}}};
\node (1a4) [below=of 1a3]{\parbox{20em}{\scriptsize\centering Protocol \ref{prtl:8}, $\log\kappa+1\rightarrow \kappa$,\\ \weakparam{\eta}{2^{-\eta}}{(1-\kappa^{-O(1)})}{\eta/\kappa^{O(1)}}{1/10},\\Lemma \ref{lem:9.1}}};
\node (1a5) [below=of 1a4] {\parbox{29em}{\scriptsize\centering Protocol \ref{prtl:9}, $M(\log\kappa+1)\rightarrow M\kappa$,\\ (when $M\in [\fpoly(\kappa),\fsubexp(\kappa)]$) \weakparams{\eta}{2^{-\eta}}{\eta/\kappa^{O(1)}}{1/3},\\Lemma \ref{lem:9.2}}};
\node (1a6) [below=of 1a5] {\parbox{13em}{\scriptsize\centering Protocol \ref{prtl:14}, $N{+\fpoly(\kappa)}\rightarrow 2N$,\\ \weakparamm{\eta_1}{2^{-\eta_1}}{\eta_2}{\eta_2/O(\kappa)}{2^{-O(\sqrt{\kappa})}}\\ (when $N\in [\fpoly(\kappa),\fsubexp(\kappa)]$),\\Lemma \ref{lem:10.2}}};
\node (1a7) [below=of 1a6] {\parbox{20em}{\scriptsize Protocol \ref{prtl:13}, \centering $N{+\fpoly(\kappa)}\rightarrow L$ \\ \quad \quad fully secure, output security $O(\kappa^{1/4})$ (when $L\leq\fsubexp(\kappa)$)\\ Theorem \ref{thm:10.3}}};
\node (1a8) [below=of 1a7] {\parbox{20em}{\footnotesize Protocol \ref{prtl:24}, \centering 8-basis qfactory, \\ Lemma \ref{lem:11.1}}};
\node (1a9) [below=of 1a8]{\parbox{15em}{\scriptsize\centering Protocol \ref{prtl:17},  Full Protocol, \\ Theorem \ref{thm:11.2}}};
\node (2a1) [left=of 1a2] {{\parbox{7em}{\scriptsize{ Protocol \ref{prtl:6.2}, \ref{prtl:r9}\\\centering Non-collapsing basis test,\\ Lemma \ref{lem:6.3}, \ref{lem:6.4}}}}};
\node (3a1) [right=10pt of 1a6] {\parbox{10em}{\scriptsize Protocol \ref{prtl:11},\\\centering ``$\fSecurityRefreshing$'',\\ $N+\fpoly(\kappa)\rightarrow N$\\\weakparamstar{\eta_1}{2^{-\eta_1}}{\eta_2}{\eta_2/O(\kappa)}{2^{-O(\eta_1)+O(\kappa)}}\\ Lemma \ref{lem:10.1}}};
\draw[->] (1a0)->(1a1);
\draw[->] (1a1)->(1a2);
\path[->] (1a2) edge node[anchor=west] {\scriptsize{ parallel repetition with gadget sharing}} (1a3);
\path[->] (1a3) edge node[anchor=west] {\scriptsize{ self-composition}}(1a4);
\path[->] (1a4)edge node[anchor=west] {\scriptsize{ repeat technique (parallel repetition)}} (1a5);
\path[->] (1a5) edge node[anchor=west] {\scriptsize{combine technique}} (1a6);
\path[->] (1a6) edge node[anchor=east] {\scriptsize{ self-composition}}(1a7);
\draw (1a7)->(1a8);
\draw[->] (1a8)->(1a9);
\draw[->] (2a1)->(1a2);
\draw[->] (2a1) edge[bend right] (1a8);
\draw[->] (3a1)->(1a7);
\draw[->] (3a1)->(1a6);
\draw[dashed] (-6,-12.5)--(-2.5,-12.5);
\node (abv) [below right= 195pt and -24pt of 2a1]{\parbox{25em}{\scriptsize above: weakly secure}};
\node (bel) [below right= 206pt and -26pt of 2a1]{\parbox{25em}{\scriptsize below: exponentially secure}};
\draw[decorate,decoration={brace,amplitude=10pt},anchor=east] (4.5,0) -- (4.5,-7.5) node [black,midway,anchor=west,xshift=6pt] {\parbox{5em}{\scriptsize Step 1 of Outline \ref{otl:2}\\Weakly secure protocol}};
\draw[decorate,decoration={brace,amplitude=10pt},anchor=east] (-6,-17.5) -- (-6,-7.5) node [black,midway,anchor=east,xshift=-3pt] {\parbox{4.5em}{\scriptsize Step 2 of Outline \ref{otl:2}\\Amplification}};
\draw[decorate,decoration={brace,amplitude=10pt}] (3.5,-17.5) -- (3.5,-21) node [black,midway,anchor=west, xshift=6pt] {\parbox{5em}{\scriptsize Step 2 of Outline \ref{ppl:1}\\Gadget-assisted\\ secure computation}};
\end{tikzpicture}
\caption{Flow of Construction}\label{fig:flowofc}
\end{figure}
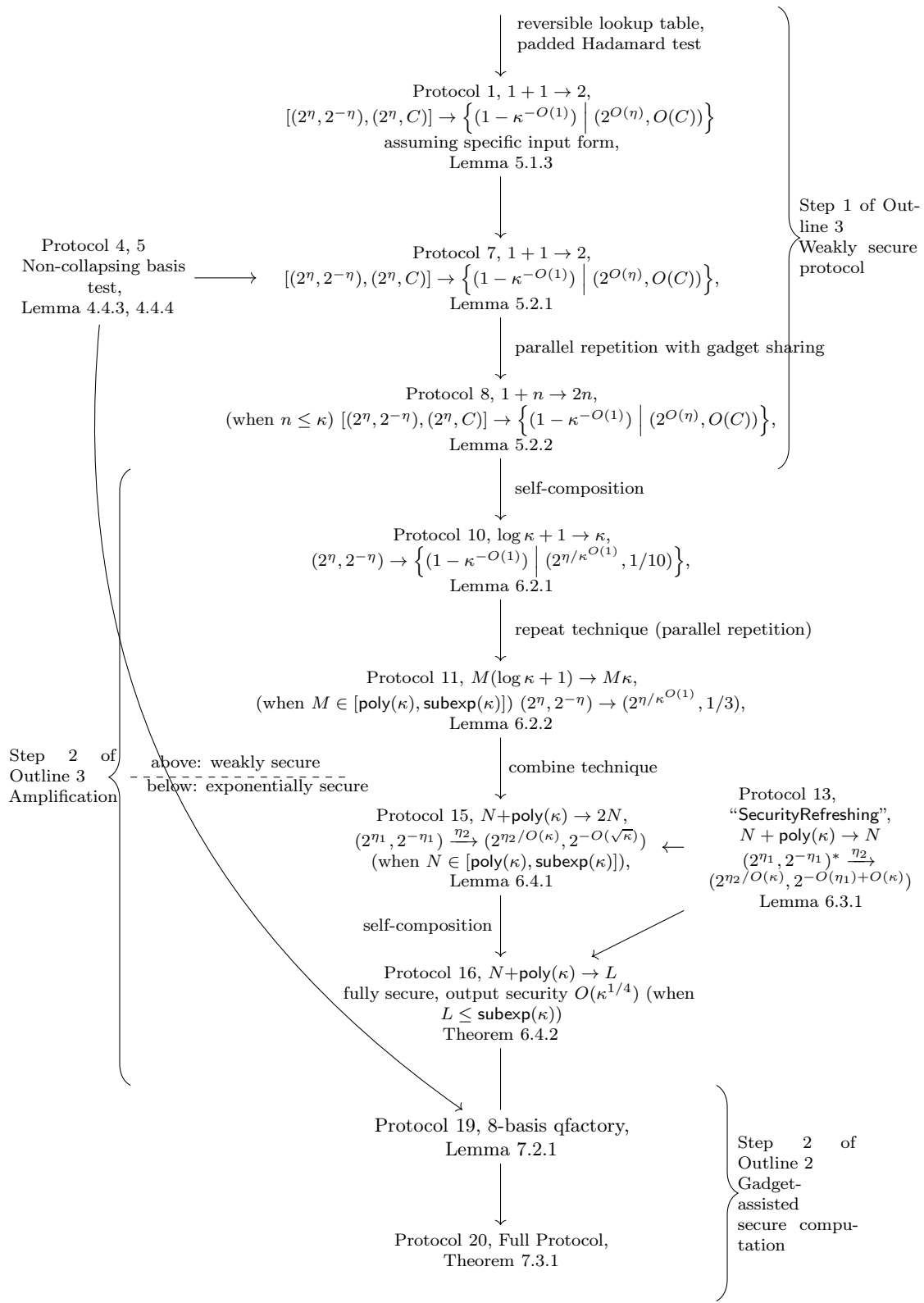

We note that this is not the protocol execution flow. We illustrate the protocol execution flow in the time order in the following section.
\subsection{Diagrams of Protocol Running Structure}\label{sec:4.5.2}
\begin{figure}
\centering
\begin{tikzpicture}
	\node (1a0) {\footnotesize Initial: $s_0$ gadgets};
	\node (1a1) [right=74pt of 1a0] {\footnotesize polynomial new gadgets for security refreshing};
	\node (2a2) [draw, below=70pt of 1a1] {\parbox{6em}{\centering\scriptsize Protocol \ref{prtl:14},\\SR (Protocol \ref{prtl:11})}};
	\node (2a1) [draw, left=33pt of 2a2] {\parbox{6em}{\centering\scriptsize Protocol \ref{prtl:14},\\SR (Protocol \ref{prtl:11})}};
	\node (2a0) [draw, left=33pt of 2a1] {\parbox{6em}{\centering\scriptsize Protocol \ref{prtl:14},\\SR (Protocol \ref{prtl:11})}};
	
	\node (2a3) [right=15pt of 2a2] {\footnotesize $\cdots$};
	\node (2a4) [draw, right=15pt of 2a3] {\parbox{6em}{\centering\scriptsize Protocol \ref{prtl:14},\\SR (Protocol \ref{prtl:11})}};	
	\node (3a2) [draw, below = 70pt of 2a2] {\footnotesize QFac8 (Protocol \ref{prtl:24})};
	\node (4a2) [draw, below = 70pt of 3a2] {\footnotesize Gadget-assisted UBQC (Protocol \ref{prtl:gaubqc})};
	\path[->,anchor=east] (1a0) edge  (2a0) ;
	\path[->,anchor=north] (2a0) edge node {\scriptsize $s_1\geq 2s_0$} (2a1) ;
	\path[->,anchor=north] (2a1) edge node {\scriptsize $s_2\geq 2s_1$} (2a2) ;
	\path[->,anchor=north] (2a2) edge (2a3) ;
	\path[->,anchor=north] (2a3) edge (2a4) ;
	\path[->,anchor=north] (2a4) edge node[anchor=north west] {$L$ gadgets} (3a2) ;
	\path[->,anchor=north] (3a2) edge node[anchor=west] {$L$ $\ket{+_\theta}$ gadgets} (4a2) ;
	\path[->] (1a1) edge node[anchor=south east] {$+\fpoly(\kappa)$} (2a0);
	\path[->] (1a1) edge node[anchor=north west,xshift=-6pt] {$+\fpoly(\kappa)$} (2a1);
	\path[->] (1a1) edge node[anchor=north west] {$+\fpoly(\kappa)$} (2a2);
	\path[->] (1a1) edge node[anchor=south west] {$+\fpoly(\kappa)$} (2a4);
\end{tikzpicture}
\caption{Overall Structure of Protocol Execution}\label{fig:ospe}
\end{figure}
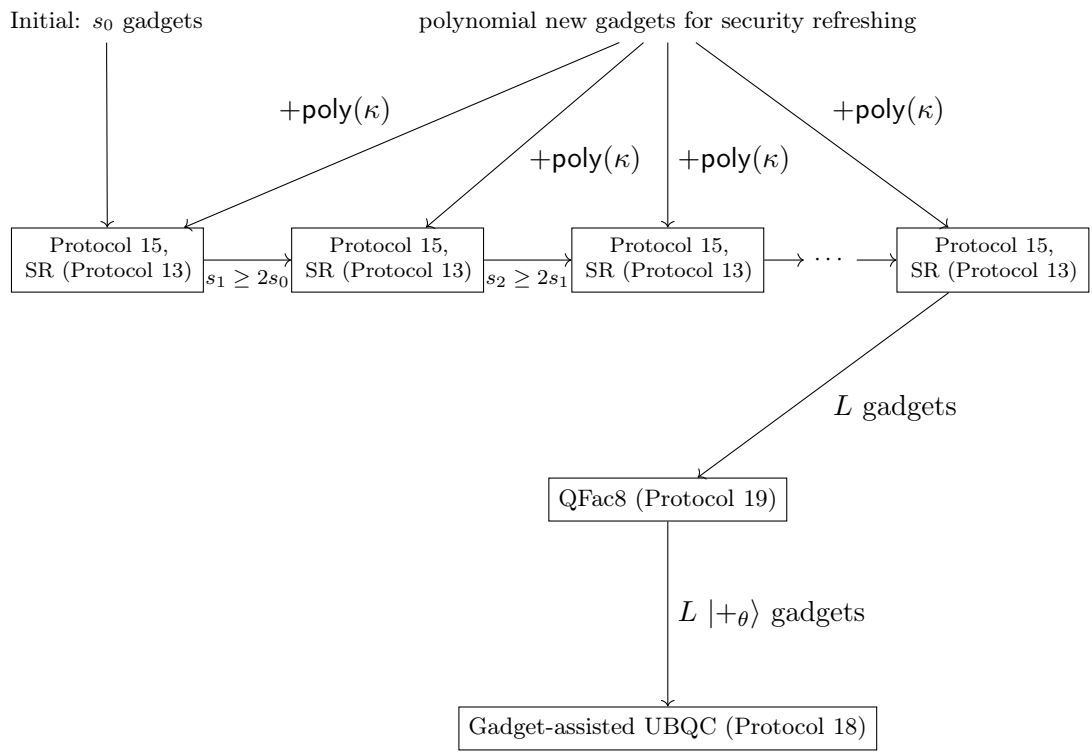

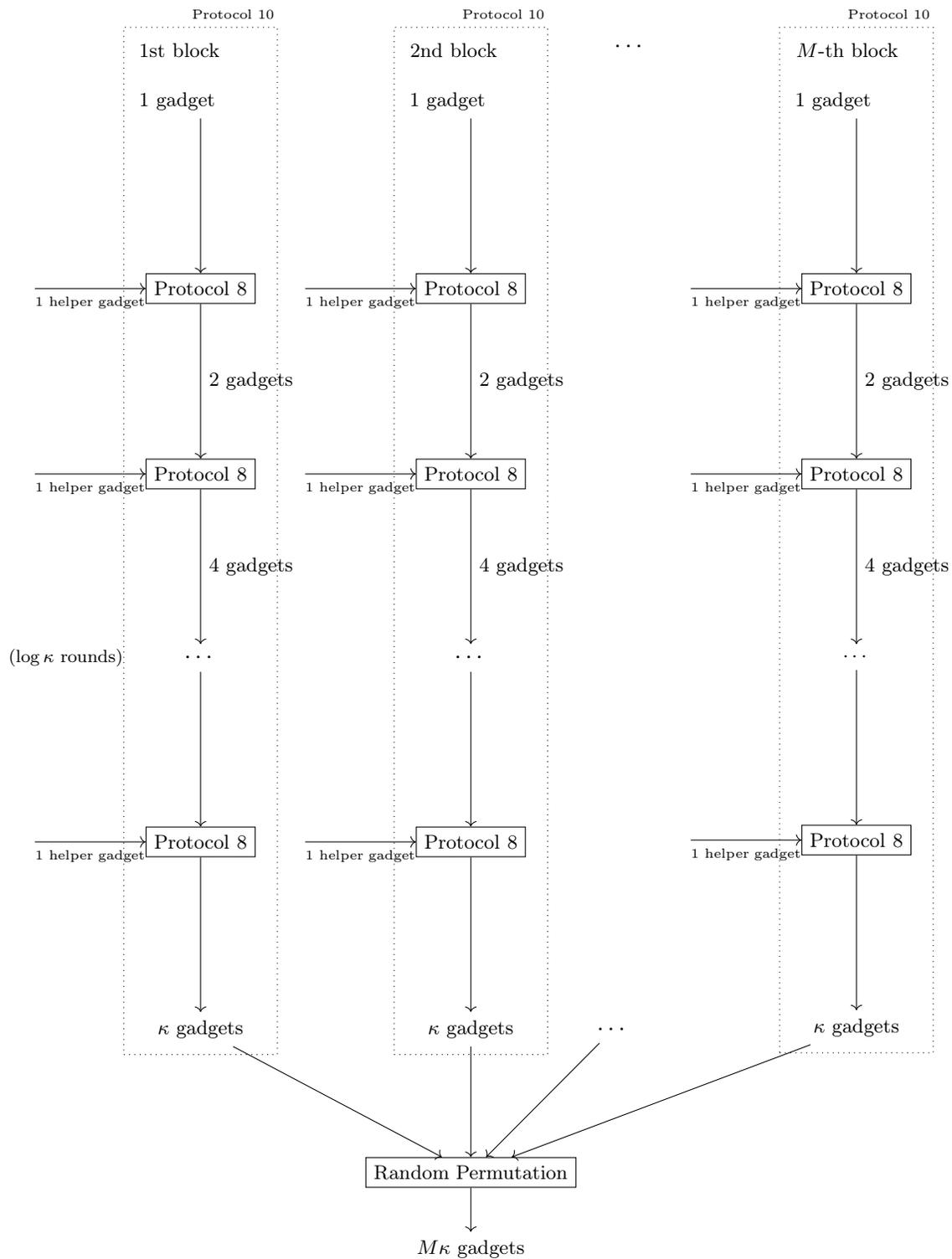
\begin{figure}
\centering
\begin{tikzpicture}
	\node (1a0) {\parbox{5em}{\footnotesize 1st block\\ \\1 gadget}};
	\node (2a0) [draw, below=70pt of 1a0] {\footnotesize Protocol \ref{prtl:n12n}};
	\node (2a0x) [left=50pt of 2a0] {};
	\path[->] (2a0x) edge node[anchor=north] {\tiny $1$ helper gadget} (2a0);
	\node (3a0) [draw, below=70pt of 2a0] {\footnotesize Protocol \ref{prtl:n12n}};
	\node (3a0x) [left=50pt of 3a0] {};
	\path[->] (3a0x) edge node[anchor=north] {\tiny $1$ helper gadget} (3a0);
	\node (4a0) [below=70pt of 3a0] {$\cdots$};
	\node (4a3x) [left=20pt of 4a0] {\scriptsize ($\log\kappa$ rounds)};
	\node (5a0) [draw, below=70pt of 4a0] {\footnotesize Protocol \ref{prtl:n12n}};
	\node (5a0x) [left=50pt of 5a0] {};
	\path[->] (5a0x) edge node[anchor=north] {\tiny $1$ helper gadget} (5a0);
	\node (6a0) [below=70pt of 5a0] {\footnotesize $\kappa$ gadgets};
	\path[->] (1a0) edge (2a0);
	\path[->] (2a0) edge node[anchor=west] {\footnotesize $2$ gadgets} (3a0);
	\path[->] (3a0) edge node[anchor=west] {\footnotesize $4$ gadgets} (4a0);
	\path[->] (4a0) edge (5a0);
	\path[->] (5a0) edge (6a0);
	
	\node (1a1) [right=60pt of 1a0] {\parbox{5em}{\footnotesize 2nd block\\ \\1 gadget}};
	\node (2a1) [draw, below=70pt of 1a1] {\footnotesize Protocol \ref{prtl:n12n}};
	\node (2a1x) [left=50pt of 2a1] {};
	\path[->] (2a1x) edge node[anchor=north] {\tiny $1$ helper gadget} (2a1);
	\node (3a1) [draw, below=70pt of 2a1] {\footnotesize Protocol \ref{prtl:n12n}};
	\node (3a1x) [left=50pt of 3a1] {};
	\path[->] (3a1x) edge node[anchor=north] {\tiny $1$ helper gadget} (3a1);
	\node (4a1) [below=70pt of 3a1] {$\cdots$};
	\node (5a1) [draw, below=70pt of 4a1] {\footnotesize Protocol \ref{prtl:n12n}};
	\node (5a1x) [left=50pt of 5a1] {};
	\path[->] (5a1x) edge node[anchor=north] {\tiny $1$ helper gadget} (5a1);
	\node (6a1) [below=70pt of 5a1] {\footnotesize $\kappa$ gadgets};
	\path[->] (1a1) edge (2a1);
	\path[->] (2a1) edge node[anchor=west] {\footnotesize $2$ gadgets} (3a1);
	\path[->] (3a1) edge node[anchor=west] {\footnotesize $4$ gadgets} (4a1);
	\path[->] (4a1) edge (5a1);
	\path[->] (5a1) edge (6a1);
	
	\node (1a2) [right=30pt of 1a1] {\parbox{5em}{$\cdots$\\ \\ }};
	\node (6a2) [right=30pt of 6a1] {$\cdots$};
	
	\node (1a3) [right=20pt of 1a2] {\parbox{5em}{\footnotesize $M$-th block\\ \\1 gadget}};
	\node (2a3) [draw, below=70pt of 1a3] {\footnotesize Protocol \ref{prtl:n12n}};
	\node (2a3x) [left=50pt of 2a3] {};
	\path[->] (2a3x) edge node[anchor=north] {\tiny $1$ helper gadget} (2a3);
	\node (3a3) [draw, below=70pt of 2a3] {\footnotesize Protocol \ref{prtl:n12n}};
	\node (3a3x) [left=50pt of 3a3] {};
	\path[->] (3a3x) edge node[anchor=north] {\tiny $1$ helper gadget} (3a3);
	\node (4a3) [below=70pt of 3a3] {\footnotesize $\cdots$};
	
	\node (5a3) [draw, below=70pt of 4a3] {\footnotesize Protocol \ref{prtl:n12n}};
	\node (5a3x) [left=50pt of 5a3] {};
	\path[->] (5a3x) edge node[anchor=north] {\tiny $1$ helper gadget} (5a3);
	\node (6a3) [below=70pt of 5a3] {\footnotesize $\kappa$ gadgets};
	\path[->] (1a3) edge (2a3);
	\path[->] (2a3) edge node[anchor=west] {\footnotesize $2$ gadgets} (3a3);
	\path[->] (3a3) edge node[anchor=west] {\footnotesize $4$ gadgets} (4a3);
	\path[->] (4a3) edge (5a3);
	\path[->] (5a3) edge (6a3);
	
	\node (perm) [draw,below=50pt of 6a1] {\footnotesize Random Permutation};
	\path[->] (6a0) edge (perm);
	\path[->] (6a1) edge (perm);
	\path[->] (6a2) edge (perm);
	\path[->] (6a3) edge (perm);
	\node (final) [below=20pt of perm] {\footnotesize $M\kappa$ gadgets};
	\path[->] (perm) edge (final);
	\coordinate (1dtl) [left=of 1a0];
	\coordinate (1dbr) [right=of 6a0];
	\node (1dt) [draw, dotted, fit=(1a0)(6a0),label={north:\tiny \qquad\qquad Protocol \ref{prtl:8}}] {};
	\node (2dt) [draw, dotted, fit=(1a1)(6a1),label={north:\tiny \qquad\qquad Protocol \ref{prtl:8}}] {};
	\node (3dt) [draw, dotted, fit=(1a3)(6a3), label={north:\tiny \qquad\qquad Protocol \ref{prtl:8}}] {};
\end{tikzpicture}
\caption{Protocol \ref{prtl:9}: weak gadget expansion}\label{fig:18}
\end{figure}

First in Figure \ref{fig:ospe} we show how the upper level protocol (Protocol \ref{prtl:17}) works, by encapsulating Protocol \ref{prtl:14} as a single node.
Then Figure \ref{fig:22} shows how the Protocol \ref{prtl:14} works by encapsulating Protocol \ref{prtl:9} and Protocol \ref{prtl:11} (Security Refreshing, abbreviated as SR) as a single node.
\begin{figure}
\centering
\begin{tikzpicture}
	\node (1a0) {\footnotesize (Round 1)};
	\node (1a1) [draw, right= 70pt of 1a0] {Protocol \ref{prtl:9}};
	\node (1a2) [draw, right= 60pt of 1a1] {SR};
	\node (1ax) [above=20pt of 1a2] {};
	\path[->] (1ax) edge node[anchor=west] {+$\fpoly(\kappa)$} (1a2);
	\node (2a0) [below = 70pt of 1a0] {\footnotesize (Round 2)};
	\node (2a1) [draw, right= 70pt of 2a0] {Protocol \ref{prtl:9}};
	\node (2a2) [draw, right= 60pt of 2a1] {SR};
	\node (2ax) [above=20pt of 2a2] {};
	\path[->] (2ax) edge node[anchor=west] {+$\fpoly(\kappa)$} (2a2);
	\node (2a3) [draw, right= 60pt of 2a2] {Combine};
		\node (3a0) [below = 70pt  of 2a0] {\footnotesize ($\cdots$)};
	\node (3a1) [below= 70pt of 2a1] {$\cdots$};
	\node (3a2) [below= 70pt of 2a2] {$\cdots$};
	\node (3a3) [below= 70pt of 2a3] {$\cdots$};
	\node (4a0) [below = 61pt of 3a0] {\footnotesize (Round $\sqrt{\kappa}$))};
	\node (4a1) [draw, below= 70pt of 3a1] {Protocol \ref{prtl:9}};
	\node (4a2) [draw, below= 70pt of 3a2] {SR};
	\node (4ax) [above=20pt of 4a2] {};
	\path[->] (4ax) edge node[anchor=west] {+$\fpoly(\kappa)$} (4a2);
	\node (4a3) [draw, below= 70pt of 3a3] {Combine};
	\node (5a3) [below = 70pt of 4a3] {\footnotesize $M\kappa$ gadgets};
	\path[->] (1a0) edge node[anchor=north] {\tiny $M(1+\log\kappa)$ gadgets} (1a1);
	\path[->] (1a1) edge node[anchor=north] {\footnotesize $M\kappa$ gadgets} (1a2);
	\path[->] (2a0) edge node[anchor=north] {\tiny $M(1+\log\kappa)$ gadgets} (2a1);
	\path[->] (2a1) edge node[anchor=north] {\footnotesize $M\kappa$ gadgets} (2a2);
	\path[->] (2a2) edge node[anchor=north] {\footnotesize $M\kappa$ gadgets} (2a3);
	\path[->] (4a0) edge node[anchor=north] {\tiny $M(1+\log\kappa)$ gadgets} (4a1);
	\path[->] (4a1) edge node[anchor=north] {\footnotesize $M\kappa$ gadgets} (4a2);
	\path[->] (4a2) edge node[anchor=north] {\footnotesize $M\kappa$ gadgets} (4a3);
	\path[->] (1a2) edge node[anchor=south west] {\scriptsize $M\kappa$ gadgets} (2a3);
	\path[->] (2a3) edge node[anchor=west] {\scriptsize $M\kappa$ gadgets} (3a3);
	\path[->] (3a3) edge node[anchor=west] {\scriptsize $M\kappa$ gadgets} (4a3);
	\path[->] (4a3) edge node[anchor=west] {\scriptsize $M\kappa$ gadgets} (5a3);
\end{tikzpicture}
\caption{Structure of Protocol \ref{prtl:14}}\label{fig:22}	
\end{figure}
The Figure \ref{fig:18} shows how the Protocol \ref{prtl:9} works by treating Protocol \ref{prtl:n12n} as a single block.

The Figure \ref{fig:15} shows how the Protocol  \ref{prtl:n12n} goes.\par
\begin{figure}
\centering
	\begin{tikzpicture}
\node (1a0) {\footnotesize {\footnotesize $1$ gadget}};
\node (2a0) [right=60pt of 1a0] {\footnotesize {$1$ gadget}};
\node (0a0) [left= 40pt of 1a0] {\footnotesize $n$ input gadgets:};
\node (3a0) [right=50pt of 2a0] {\footnotesize {$\cdots$}};
\node (4a0) [right=50pt of 3a0] {\footnotesize {$1$ gadget}};
\node (1a1) [draw, below=of 1a0] {\footnotesize Protocol \ref{prtl:r9}};
\node (0a1) [left=of 1a1] {\footnotesize $1$ helper gadget};
\node (2a1) [draw, below=of 2a0] {\footnotesize Protocol \ref{prtl:r9}};
\node (3a1) [below=of 3a0] {\footnotesize $\cdots$};
\node (4a1) [draw, below=of 4a0] {\footnotesize Protocol \ref{prtl:r9}};
\node (1a2) [below=70pt of 2a1] {\footnotesize $\forall i\in [n]$ in parallel: $\fGdgPrep^{basic}$ (Protocol \ref{prtl:4.6}) between helper gadget and $i$-th input gadget};
\node (1a3) [below=70pt of 1a2] {\footnotesize $2n$ gadgets};
\draw[->] (1a0)->(1a1);
\draw[->] (2a0)->(2a1);
\draw[->] (4a0)->(4a1);
\draw[->] (0a1)->(1a1);
\path[->] (1a1) edge node[anchor=west] {\footnotesize 1 gadget} (1a2);
\path[->] (2a1) edge node[anchor=west] {\footnotesize 1 gadget} (1a2);
\path[->] (4a1) edge node[anchor=west] {\footnotesize 1 gadget} (1a2);
\path[->] (1a1) edge node[anchor=north] {\tiny helper gadget} (2a1);
\path[->] (2a1) edge node[anchor=north] {\tiny helper gadget} (3a1);
\path[->] (3a1) edge node[anchor=north] {\tiny helper gadget} (4a1);
\draw[->] (1a2)->(1a3);
\end{tikzpicture}
\caption{Structure of Protocol \ref{prtl:n12n}}\label{fig:15}
\end{figure}
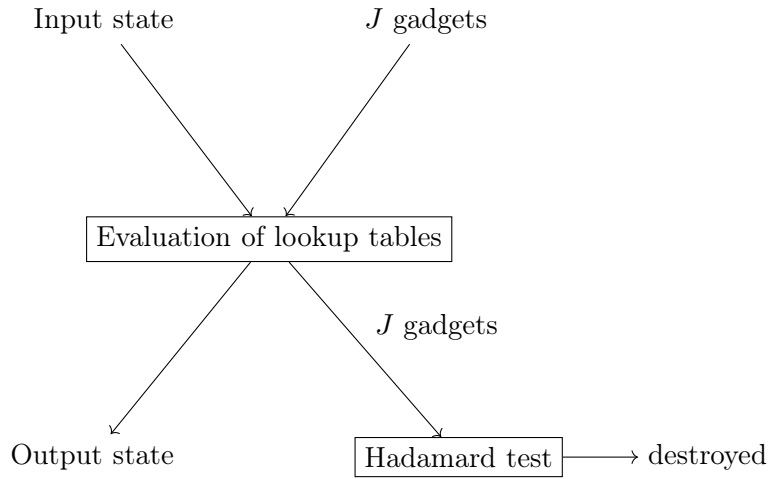
\begin{figure}
\centering
\begin{tikzpicture}
	\node (1a0) {Input state};
	\node (mida0) [right=of 1a0] {};
	\node (2a0) [right=of mida0] {$J$ gadgets};
	\node (mid) [draw, below=70pt of mida0] {Evaluation of lookup tables};
	\node (mida1) [below=70pt of mid] {};
	\node (1a1) [left=of mida1] {Output state};
	\node (2a1) [draw, right=of mida1] {Hadamard test};
	\path[->] (1a0) edge (mid);
	\path[->] (2a0) edge (mid);
	\path[->] (mid) edge (1a1);
	\path[->] (mid) edge node[anchor=south west] {$J$ gadgets} (2a1);
	\node (3a1) [right=of 2a1] {destroyed};
	\path[->] (2a1) edge (3a1);
\end{tikzpicture}
\caption{Structure of SR, Protocol \ref{prtl:11}}\label{fig:sr}
\end{figure}

Finally we give Figure \ref{fig:sr} which shows the structure of SR (Protocol \ref{prtl:11}), security refreshing layer).

\subsection{Section Outline}\label{sec:4.5.3}
The remaining sections are organized as follows. 
 \begin{itemize}\item Chapter \ref{cht:5} is for the preparation of the protocol design.\begin{itemize}\item The setting and notation systems that we use during the whole paper are formalized in Section \ref{sec:5.1}. \item In Section \ref{sec:3} we design a modular framework for tracking the correctness and security properties of different subprotocols.
\item In Section \ref{sec:4} we give some basic lemmas and techniques for the security proofs later. Thus Section \ref{sec:5.1} to \ref{sec:4} are mainly summaries and preparations before we move to the formal construction. The protocol design begins at Section \ref{sec:6}. \item In Section \ref{sec:6} we study a class of subprotocols which we call \emph{non-collapsing basis test}. This class of subprotocols will be useful later, mainly in Section \ref{sec:7} and Section \ref{sec:11.2}.\end{itemize}\item In Chapter \ref{cht:6} we construct the first weakly secure gadget increasing protocol.\begin{itemize} 
\item Section \ref{sec:7} we design a weakly secure  protocol on some specific class of input state, and it's not gadget-increasing. \item In the end of the Section \ref{sec:8l} we give a remote gadget preparation protocol with weak security, that is gadget increasing (Protocol \ref{prtl:n12n}).\end{itemize}\item Chapter \ref{cht:7} amplify the protocol in the last chapter to get a fully secure protocol, and complete the construction of the whole blind quantum computation protocol.\begin{itemize} \item In Section \ref{sec:8} we give an overview of how to amplify the security to normal security. 
\item In Section \ref{sec:9} we give part of the proofs for the amplified protocol (which is the \emph{repeat} part).\item In Section \ref{sec:10.1} we give the $\fSecurityRefreshing$ protocol that we use to overcome the obstacles in the security proof and make the \emph{combine} technique really works. \item In Section \ref{sec:10} we give the formal protocol for the \emph{combine} part and complete the security proof, and complete the amplification and give a fully secure remote gadget preparation protocol.  By this time step 1 of Outline \ref{ppl:1} is completed.  
\end{itemize}
\item Finally in Chapter \ref{cht:8} we give a universal blind quantum computation protocol and complete the proof of Theorem \ref{thm:1}.
\end{itemize}
\subsection{Writing Conventions}
For protocols in the mainline, the descriptions of protocols are put into boxes with grey background. Then simple discussions of correctness and efficiency follows them. Then the security statements are given (either in a new subsubsection or not), which are usually in the form of weak security transform parameter. Then security proofs or discussions and pointers of security proofs follows them.\par
As said before, security statements of subprotocols are usually in the form of \emph{weak security transform parameter}. But some protocol statements use variants of this notion so it's not necessarily in a form that can be sufficiently encapsulated. The security is discussed using the \emph{purified joint state} discussed in Section \ref{sec:5.1}, where the randomness is always purified by the environment. And we use $\ket{\varphi}\odot\cdots$ symbol to mean ``the client sends some messages to the server (actually the read-only buffer)''. This is often used accompanied with the ``$\llbracket\text{Alg}\rrbracket$'' symbol, which means the output of some algorithm $\text{Alg}$.
\cleardoublepage
\chapter{Preparation for the Design of Remote Gadget Preparation Protocol}\label{cht:5}
This chapter is about the preparation before the protocol construction. We will formalize the problem itself, develop a notation system, give a protocol design framework, and discuss a basic protocol that will be useful later --- non-collapsing basis test (in Section \ref{sec:6}).
\section{The Problem Setting and Notation System}\label{sec:5.1}
\subsection{The Problem Setting, and its Notation System}\label{sec:5.1.1}

\subsubsection{CQ-states, purified notation and natural notation}\label{sec:5.1.1.1}

In our protocol, if we temporarily ignore the ``read-only buffer'' system that will be introduced later, there are four parties: the random oracle, the client, the server and the environment. The first two parties are classical during the protocol, while the last two parties can be in quantum state. Thus at any time during the execution of the protocol, the state of the whole system can be described as a cq-state \cite{NielsenChuangs}.
\begin{defn}\label{def:2.1}\cite{NielsenChuangs}
	A cq-state is described as the state ensemble $\{p_c,c,\ket{\varphi_c}\}$, $c\in \cC$, $\sum_{c}p_c=1$. Note that the $\ket{\varphi_c}$ is not necessarily normalized. And it can be equivalently and simply described as $\{c,\frac{1}{\sqrt{p_c}}\ket{\varphi_c}\},c\in \cC$.
\end{defn}
It's convenient to study a cq-state by studying its purified state:
\begin{defn}\label{def:r2.4}\cite{NielsenChuangs}
	The purification of a cq-state given in Definition \ref{def:2.1} is defined to be $\ket{\varphi}=\sum_{c\in \cC}\frac{1}{\sqrt{p_c}} \ket{c}\ket{\varphi_c}\ket{c}$, where the first $\ket{c}$ is held by the parties that are originally considered to be classical (which are the client and the random oracle in our protocols), and the last $\ket{c}$ is in the environment. And its norm is
	$$|\ket{\varphi}|=|\sum_{c\in \cC}\frac{1}{\sqrt{p_c}} \ket{c}\ket{\varphi_c}\ket{c}|=|\sum_{c\in \cC}\frac{1}{\sqrt{p_c}} \ket{c}\ket{\varphi_c}|=\sqrt{\sum_{c\in \cC}\frac{1}{p_c}|\ket{\varphi_c}|^2}$$
\end{defn}

In our protocol, the client will be completely classical after the first step of the protocol (which is, to send a quantum gadget to the server), and the server will hold a quantum state. So the joint state of the client, the random oracle, the server and the environment will form a cq-state. Temporarily denote such a state as $\ket{\varphi^{HCSE}}$, where $H$ is the inner content of the random oracle, $C$ is the memory on the client side, $S$ is the server's system and $E$ is the environment that the state is entangled with. The $HC$ systems are the classical part and the $SE$ systems are the quantum part.\par 
This state can be purified by imagining these classical parts are entangled with the environment. 
As a simplified example, if we assume the client side memory contains key register $K$ and randomness register $R$, which are independently random, the purified state will be:
$$\ket{\varphi^{HKRSE}}:=\sum_{hkr}\underbrace{\frac{1}{\sqrt{\#h}}\ket{h}}_{\text{Random Oracle}}\otimes\underbrace{\frac{1}{\sqrt{\#k}}\ket{k}\otimes\frac{1}{\sqrt{\#r}}\ket{r}}_{\text{Client side}}\otimes\underbrace{\ket{\varphi_{H=h,K=k,R=r}^{SE}}\otimes \ket{hkr}}_{\text{Server+Environment}}$$
where $h$ is the content of the random oracle, $k$ is the client's keys, and $r$ is the client's randomness, $\#$ means the number of possible choices. $\ket{\varphi_{H=h,K=k,R=r}^{SE}}$ is the server-plus-environment's state when the choice of $H,K,R$ is correspondingly $h,k,r$.
\paragraph{Natural notation and purified notation} We note that in this work we use two types of notations to describe the state of the whole system during the protocol. The notation described above, where everything is purified, is called \emph{purified notation}. In this notation the state of the whole system can be simply described as a pure state $\ket{\varphi}$. We will call it the \textbf{\emph{purified joint state}} of the state. This allows us to describe everything using a simple Dirac symbol and is more convenient for the security proofs. On the other hand, in the honest setting, we can use a more natural notation: for example, we can simply write $\ket{x_0}+\ket{x_1}$, to means the server's state is $\ket{x_0}+\ket{x_1}$ when the client side keys are $\{x_0,x_1\}$; if we insist on using the purified notation the expression will be quite long (something like $\sum_{x_0x_1}\frac{1}{\sqrt{|\cK|}}\ket{x_0}\ket{x_1}(\ket{x_0}+\ket{x_1})\ket{x_0}\ket{x_1}$). \textbf{In this work we will use the natural notation in the honest setting and use the purified notation when we prove the security, except explicit stated exceptions.} And we further emphasize that \textbf{using purified notation in the security proof is only a security proof technique and does not mean the client has to hold quantum states.}\par
And we note under the purified notation, the key set symbol, like $K=\{x_0,x_1\}$, does not carry concrete values, but should be considered as the symbol for the client-side systems that stores these keys. And the projection operators discussed in Section \ref{sec:4.2.1} also becomes some highly-entangled projections onto the system of the purified joint state $\ket{\varphi}$.\par
Finally we note that by now we assume in the protocols there are four parties, $HCSE$. In the subsubsections below we will introduce the fifth register. Thus there are five parties ($HCSE$ plus the read-only buffer introduced below).

\subsubsection{Different steps of a protocol}\label{sec:2.2.2}
Let's clarify some details on the formalization of \emph{protocol}. 
Each step of a protocol can be one of the following: \begin{itemize}\item The client does some local computation.\item The client sends some message to the server. This further contains two possibilities: (1) the client sends some classical messages to the server; and (2) the client sends some quantum gadgets to the server. \item The server does some local computation. \item The server sends some message to the client.\end{itemize} 

And let's further make some clarifications. For the server-side operation phase, we note that \begin{itemize}
\item The attacker can introduce server-side ancilla qubits. In the later proofs we might make this part implicit and simply write, for example, $\cU\ket{\varphi}$ where $\cU$ is a server-side operation.
\item We assume the server could only do server-side operations from the time it receives some client side messages to the time it sends out the result. (In other words, we assume the client side computation ``takes no time'', and the adversary could not do anything from the time it sends out the result to the time the client sends the messages in the next round.) The reason for it is, the adversary can always ``postpone'' the attack until it receives the client's message in the next round thus this assumption does not make it weaker.\item When we discuss the adversary \emph{during a protocol}, we do not take the server-side operation after the adversary receives the last message into consideration. Instead, we will consider it separately. In the security statement of our protocol, there are usually two adversaries: $\fAdv$ and $\cD$, where $\fAdv$ is the adversary \emph{during the protocol}, and $\cD$ is the adversary's operation \emph{after the protocol completes}. $\cD$ is also a server-side operation. And we note that $\cD$ can be implicit within the ``SC-security/ANY-security/unpredictability'' (which will be introduced later).\end{itemize}
For later convenience, in the next subsubsection we will further formalize the operation of ``the client sends some classical messages to the server''. We will introduce the \emph{read-only buffer} system, and formalize ``the client sends some classical messages to the server'' as the operation that the client copies the content of some registers in its inner system to the read-only buffer. What's more, in the server-side operation phase, the adversary can read the content of the read-only buffer but could not revise it.
\subsubsection{Introducing the \emph{read-only buffer}, and the ``$\odot$'' symbol for messages and auxiliary information}\label{sec:2.2.3}
\begin{defn}\label{defn:2.6}
	The \emph{read-only buffer} is a standalone system, where the client can write to it using classical gates but the server can only read it.
\end{defn}
\begin{defn}
	In the server-side operation phase, the adversary can read the read-only buffer and the server-side system and can query the random oracle but can only write on the server-side system. Which means the adversary can do a control-operation controlled by the content of the read-only buffer but not the other operations on the read-only buffer.
\end{defn}
Note that, if we purify all the randomness, the read-only buffer will also be entangled with the environment. But the definition above still makes sense.\par
So why do we want to introduce it? The reason is, assuming some information cannot be erased can make the proof much easier. These information allows us to understand what the server can do more clearly.\par
As we said in Section \ref{sec:2.2.2}, the read-only buffer will be used to store the client-to-server classical messages of a protocol. But it's not limited to that. In the security proof we will need to add some more information into it and give the server some \emph{auxiliary information}, or intuitively, assume the adversary can know more than what is given in the honest execution.
\begin{defn}\label{def:2.3}
	We write $\ket{\varphi}\odot Z$ to denote the state that starts from $\ket{\varphi}$, the client sends $Z$ to the \emph{read-only buffer}. 
\end{defn}
Here $Z$ might be the client's messages during a protocol, part of the client's stored key set $K$, or the hash values of part of the keys, or the output of some algorithm run by the client. In the security proof we need to frequently assume the client sends some additional information to the buffer (which does not happen in the protocol, but only in the security proof). \par

Finally we note that introducing the read-only buffer does not make the adversary weaker, thus proving the security in the ``$HCSE+\text{read-only buffer}$'' model implies the security in the usual $HCSE$ model.
\subsubsection{Global tags}\label{sec:global}
In the security proof, one setting that we will use is to introduce the existence of \emph{global tags}:
\begin{defn}
	The global tag of $x$, which is $Tag(x)$, is defined to be $H(\mathfrak{tag}||x)$, where $\mathfrak{tag}$ is a special symbol (which is never used in the honest setting). Note that $\mathfrak{tag}$ is considered to be a single character, and it can only be used in the beginning of the input of the random oracle. Thus $Tag$ can be seen as a new, independent oracle other than $H$.
\end{defn}
Why do we call it ``global tags''? We will see there are a lot of tags (which mean ``hash values with paddings'') for the keys in the protocol construction, for example, in $\fEn_k(x)$ (Definition \ref{def:2.13}), the output contains two parts, the ``ciphertext'' and the ``key tags''. These ``ciphertexts'' can be further used as the inputs of the computation of some other messages, to achieve some advanced functionalities; however, the ``key tags'' part is usually simply given to the server as key tags, as the name suggests, and does not carry advanced functionalities. It's simply used and (intuitively) it can only be used on the server side to verify ``this is the key''.\par
 The global tags also have such properties: it's used in the security proof, but not in the honest setting. In the security proof we frequently assume the client additionally provides the global tags of some key sets to the server as auxiliary information. In most cases, it will not be used in some complicated way.\par 
Why do we need to introduce this concept? The answer is to \emph{make the proof simpler}. We frequently need to assume the adversary holds a version of key tags for some key sets. Lots of key tags appear in the honest execution of our protocol, but in the security proof it's convenient to have a version of key tags that is consistent during the whole proof. It's not ``fair'' to choose any of these key tags that appear during the honest execution, and this will make the proof less readable. Thus we choose to introduce the \emph{global tag} for the security proof.\par
We note that in this definition we expand the input character tables from $\{0,1\}$ to $\{0,1,\mathfrak{tag}\}$. This does not make the adversary weaker.\par
We emphasize that \emph{these global tags only appear during the security proof, and the client does not give the server such message in the real protocol}. So why could we assume this and change the protocol to a different protocol? The reason is simple: ``\textbf{it does not make the adversary weaker}''. If an adversary cannot break the protocol even if the client provides the global tags, the protocol is certainly secure in its original form, since the adversary with global tags can do anything it can do when it does not have them.\par
The next question is: what is the length of the output of global tags? For usual random oracle outputs the output length is either described explicitly, or implicit in the protocol description; but for global tags we hardcode it into the definition, which depends on the length of the input.
\begin{defn}
	The output length of $Tag(x)$ is $2^{2^{2^{|x|}}}$
\end{defn}
(Why is such an ill-looked definition reasonable? The answer is still ``\textbf{it does not make the adversary weaker}''! In fact we believe some much smaller functions like $2^{|x|}$ or even polynomial functions are enough, but we choose to write it in this way to (1) emphasize that the actual length of it does not matter (as long as it's long enough) and (2) it makes the descriptions and proofs of some lemmas later a little bit easier. And we note that since it does not appear in the real protocol and is only used as a security proof tools, it does not make the protocol inefficient.\par
\textbf{But we do need to make it big enough}. The reason is to (approximately) rule out the possibility of collisions of global tags. By making the output length very big $Tag$ becomes almost-impossible to be non-injective. We will formalize this fact in Section \ref{sec:2.4.5}.\par

Sometimes the client needs to provide the global tags for all the keys in a set to the server. We introduce a simplified notation for it.
\begin{nota}
	$Tag(K)$ is defined to be the tuple of the global tags of \emph{all} the keys in $K$. 
\end{nota}
\subsubsection{The initial state in the beginning of the whole protocol}\label{sec:2.4.5}
Finally let's discuss the initial state of our protocols. Our protocol is composed of several subprotocols and the initial state of these subprotocols might vary. But what we are going to discuss, is the initial state in the very beginning, defined as follows:
\begin{defn}
Define $\ket{\mathfrak{init}}$ as the state where the systems of the client and the server are all zeros, and the random oracle is not queried, the read-only buffer is empty, and all the parties haven't done any operations. All the randomness are purified by the environment. 
\end{defn}

Recall that in Section \ref{sec:global} we say $Tag$ is almost-impossible to be non-injective. Now we can formalize this fact.
\begin{fact}\label{fact:injtag}We have, when $\kappa$ is bigger than some constant,
\begin{align}&P_{Tag\text{ is injective on }\{0,1\}^\kappa}\ket{\mathfrak{init}}
\approx_{2^{-2^{2^{\kappa/2}}}}\ket{\mathfrak{init}} 
\end{align}\label{eq:10newnew}
(where $P$ is the projection onto the corresponding space.) \par
And if a state $\ket{\varphi}$ can be written as
$$\ket{\varphi}=\sum_{i\in [2^{\alpha_1}]}\cP_i\ket{\mathfrak{init}}$$

 where each $\cP_i$ is some sequence of (client or server side) unitaries, projections and oracle queries, we have
\begin{equation}P_{Tag\text{ is injective on }\{0,1\}^\kappa}\ket{\varphi}
\approx_{2^{-2^{2^{\kappa/2}}+\alpha_1}}\ket{\varphi} 	
\end{equation}
\end{fact}
\subsubsection{Blinded oracles, revisited}
Let's recall the notation for blinded oracles in Section \ref{sec:2.2.2r}. 
Note that in Chapter \ref{cht:4} we haven't introduce the purified notation and this definition is given under the natural notation. Here $Set$ can also come from some probability distribution and the construction of the blinded oracle actually depends on the actual value of $Set$.\par
Let's understand what this definition means in the purified notation. After we purify the system that stores $Set$, the symbol $Set$ does not refer to a concrete set anymore; it's used as the symbol to describe the content of the system that stores the superposition of different possible $Set$. However, the definition above still works, in the following sense: suppose we are currently studying the following state
$$\ket{\varphi}=\sum_{\text{$s\in$ all the possible choices of $Set$}}\underbrace{\ket{s}}_{\text{system that stores $Set$}}\otimes \ket{\varphi_s}$$
for different basis states of the system that stores $Set$, the blinded oracle constructed out can be different. To make the definition well-behaved, we need to additionally assume the system that stores $Set$ is \emph{read-only} in the protocol later, defined below:
\begin{defn}
We say a system is read-only after some time if no party writes on it in the remaining protocol. In other words, all the possible operations starting from this time that contain this system are control gates controlled by this system.
\end{defn}
\begin{defn}\label{def:3.10rr}Then the ``blinded oracle of $H$ where $Set$ is blinded'' where $Set$ is stored in superposition in some read-only system is interpreted as follows: first for each basis $s$ of the system that stores $Set$, (as we said before, $s$ is some set of inputs), define a blinded oracle following Definition \ref{def:2.3rr}. For each application of this blinded oracle, the oracle query is made as follows: controlled on the content of the system $Set$, apply the corresponding oracle.\end{defn}
Another way is to imagine that the system that stores $Set$ is actually already measured thus Definition \ref{def:2.3rr} makes sense. Since this system is read-only whether it's measured does not affect the protocol execution. In Definition \ref{def:3.10rr} we purify this system and the construction of the blinded oracle takes this system as inputs, which is an equivalent definition.
\subsection{Notations for Describing Protocols, and the $\llbracket\cdots\rrbracket$ Symbol}\label{sec:2.5} 
We introduce the following notations to simplify our security analysis. First, since we are considering the unbounded adversary with limited random oracle queries, the ``power'' of the adversary is characterized by the number of queries it can make. Thus we introduce the following notations:
\begin{defn}\label{def:2.4}
	We use $\cU$ to denote the server-side operation that can be written as a sequence of unitary operations and random oracle queries: $\cU=H_tU_tH_{t-1}\cdots H_1U_1$ where each $U_i$ can be applied without RO queries. What's more, we denote $|\cU|=t$ as the number of oracle queries in $\cU$.\par
	We use $\cP$ to denote server-side operation that contains random oracle queries and projection on some subspace: if $\cP=H_tP_tH_{t-1}\cdots H_1P_1$ where each $P_i$ can be applied without RO queries, and similarly write $|\cP|=t$ for the number of RO queries.\par
	We write $|\fAdv|\leq t$ to mean the total number of random oracle queries in the execution of the adversary $\fAdv$ during the whole protocol is at most $t$.\footnote{We note that $|\cdots|$ symbol is used for both the number of RO queries in an operation, and the norm of a state. But it should be easy to distinguish them.}
\end{defn}
Recall that each of $U_i,P_i$ can read the read-only buffer.\par
And recall that $\cU$ can introduce server-side ancilla qubits that are at the zero state. This part is implicit in ``$\cU\ket{\varphi}$''\par

We then note that, if the initial state $\ket{\varphi}$ (we purify all the randomness as discussed in Section \ref{sec:5.1.1.1}), the protocol itself, the parameters of the protocol and the code of the adversary is fixed, the post-execution state can be determined uniquely. Thus we can write the post-execution state of the protocol concisely:
\begin{nota}
	We use
	\begin{equation}\label{eq:5r}ProtocolName(Keys;Parameters)\end{equation}
	to denote a protocol which takes the description of a set of keys $Keys$ and parameters $Parameters$ as the input.\par
	We use
	$$\ket{\varphi^\prime}:=ProtocolName_\fAdv(Keys;Parameters)\circ\ket{\varphi}$$
	to denote the post-execution state of running protocol \\$ProtocolName(Keys;parameters)$ on $\ket{\varphi}$ against adversary $\fAdv$. Note that the state contains both the passing space and the failing space. And we use
	$$P_{pass}\ket{\varphi^\prime}=P_{pass}ProtocolName_\fAdv(Keys;Parameters)\circ\ket{\varphi}$$
	to denote the state projected onto the passing space.
\end{nota}
We note that if we do not explicitly add the $P_{pass}$ operation, the state should contain both the passing part and the failing part. And we note that the initial state is not necessarily normalized, and after the projection $P_{pass}$, the state is not automatically normalized either.\par
Finally we introduce a notation that means the output of some algorithm:
\begin{nota}
The output of an algorithm $\text{Alg}$ is denoted as $\llbracket\text{Alg}\rrbracket$.	
\end{nota}
This is often used in company with the $\odot$ symbol to mean the client computes something and sends to the server.
\section{The Modular Framework for Protocol Design}\label{sec:3}
\subsection{Remote Gadget Preparation: the Stepping Stone towards UBQC}\label{sec:3.1}

To construct a protocol that satisfies Theorem \ref{thm:1}, our idea is to first construct a protocol for an intermediate problem, which we call the \emph{remote gadget preparation} protocol. We have already discussed this concept informally in the introduction, and here we give a formal definition for it. The correctness and security of the remote gadget preparation protocol are defined as follows.
\subsubsection{What is the gadget?}
We will see, in our construction of the remote gadget preparation protocol, the client needs to first sample a set of initial keys $K=\{x^{(i)}_b\}_{i\in [N],b\in\{0,1\}}$, and prepare the gadgets in the form of $\otimes_{i=1}^N (\ket{x^{(i)}_0}+\ket{x^{(i)}_1})$, and send them to the server; but the initial gadget number $N$, and the initial key length are succinct. Let's introduce a convenient notation to describe such form of states.
\begin{nota}\label{nota:3.1}
	If $K=\{x^{(i)}_0,x^{(i)}_1\}_{i\in [N]}$, $\forall i,x^{(i)}_0\neq x^{(i)}_1$, we define $Gadget(K)$ to be $\otimes_{i=1}^N (\ket{x^{(i)}_0}+\ket{x^{(i)}_1})$.
\end{nota}
 The goal of the client is to allow the server to prepare the gadget $\otimes_{i=1}^L(\ket{y^{(i)}_0}+\ket{y^{(i)}_1})$, where $L$ is an arbitrary polynomial (or even subexponential), while keeps $K_{out}=\{y^{(i)}_b\}_{i\in [L],b\in\{0,1\}}$ ``secure'' (in some sense we are going to define). Notice that these initial states have the same form as the final gadgets to be prepared; the only difference is the number of gadgets. (The key lengths are both succinct, although might be different.) Thus our protocol can be seen as a protocol that securely ``reproduce'' many gadgets on the server side from $\fpoly(\kappa)$ gadgets. ($\fpoly$ is fixed.)\par


\subsubsection{Correctness of remote state preparation}
\begin{defn}[Correctness]\label{defn:3.1}
	Suppose $\kappa$ is the security parameter. We call a protocol a remote gadget preparation protocol with output number $L(\kappa)$ and output length $\kappa_{out}(\kappa)$ if: taking $\kappa$, $L=L(\kappa)$ and $\kappa_{out}=\kappa_{out}(\kappa)$ as the input parameters,  
	in the end of the protocol, an honest server can pass the protocol with probability $\geq 1-\fneg (\kappa)$ and hold the state $\otimes_{i=1}^L(\ket{y_0^{(i)}}+\ket{y_1^{(i)}})$ in the end, where $y_b^{(i)}\in\{0,1\}^{\kappa_{out}}$ such that $y_0^{(i)}\neq y_1^{(i)}$; and the client gets the description of the output keys $K_{out}:=\{y_b^{(i)}\}_{i\in [L],b\in \{0,1\}}$.\par
	And we say a protocol is a correct remote gadget preparation protocol if there exist a \emph{fixed} subexponential function $\fsubexp$ and a \emph{fixed} polynomial $\fpoly$ such that for any $0\leq L(\kappa)< \fsubexp(\kappa)$ and $\fpoly(\kappa)<\kappa_{out}(\kappa)<\fsubexp(\kappa)$, the protocol is a remote gadget preparation protocol with output number $L(\kappa)$ and output length $\kappa_{out}(\kappa)$. 
\end{defn}
\begin{defn}[Correctness, with explicit input gadget number]\label{defn:3.7}
	Suppose $N,L$ are functions of $\kappa$. A remote gadget preparation protocol is an $N\rightarrow L$ remote gadget preparation protocol if it has output number $L$ and in the honest setting, initially the server holds $\otimes_{i=1}^N(\ket{x^{(i)}_0}+\ket{x^{(i)}_1})$ and the client knows all the keys.\par
	We call $L/N$ the \emph{gadget expansion ratio}. If $L/N>1$ we say it's gadget-inceasing.
\end{defn}
And we can define the efficiency requirement for the remote gadget preparation:
\begin{defn}[Efficiency]\label{defn:3.2}
	We call a remote gadget preparation protocol with output number $L(\kappa)$ and output length $\kappa_{out}(\kappa)$ efficient if the client and the honest server runs in time $\fpoly(\kappa,L,\kappa_{out})$, where $\fpoly$ is a \emph{fixed} polynomial.
\end{defn}
And we additionally assume there is no quantum communication in the middle of the protocol. (The quantum communication can only happen in the very beginning of the protocol, where the client sends the initial gadgets to the server. Or equivalently, we assume the server already holds the gadgets before the protocol, then there is no quantum communication.)
\subsubsection{Security Definition: introducing the \emph{SC-security}}
To describe this concept more concisely, we will first define the \emph{SC-security}, which means,  two keys can't be simultaneously computed with norm bigger than some value. This concept will be used frequently in the following sections.
\begin{defn}[Review of Definition \ref{defn:3.4}]\label{defn:5.2.4}
	We say $\ket{\varphi}$ is $(2^\kappa, A)$-SC-secure for keys $K=\{x_0,x_1\}$ given $Z$ if for any server side operation $\cU$ with query number $|\cU|\leq 2^\kappa$, (note that $\cU$ can introduce server-side ancilla qubits in the zero state, which is inherent in the expression below,)
	$$|P_{x_0||x_1}\cU(\ket{\varphi}\odot Z\odot Tag(K))|\leq A$$
\end{defn}
Recall that we are using the notation in Section \ref{sec:2.2.3}.\par
The message $Z$ here can be anything: it might be some other keys, some tags of the keys, or empty.\par
The auxiliary information $Z$ here is mainly to get rid of some strange correlations among different key pairs. For example, assume $\{y_0,y_1\}$ is only one key pair in many different key pairs, if we take $Z$ to be the description of all the other key pairs (for example, $\{y_0^\prime,y_1^\prime\}$), the output state will not contain something like $Enc_{y_0||y_1}(y_0^\prime||y_1^\prime)$. Such form of ``disconnection'' is important and will be used frequently in the security proof.\par

Using this definition, we can re-formalize the security of the remote gadget preparation problem as follows:\par

\begin{defn}[Security, with concise notation]\label{defn:3.5}
	Suppose $\kappa$ is the security parameter. We say a remote gadget preparation protocol is secure against adversaries of query number $\leq 2^\lambda$ with output security $\eta$ ($\lambda,\eta$ are all functions of $\kappa$) if:\par
	For any adversary $\fAdv$ with query number $|\fAdv|\leq 2^{\lambda}$, denote the post-execution state of the protocol projected onto the passing space as $P_{pass}\ket{\varphi^\prime}$ (assume the randomness is purified by the environment) and the output keys as $K_{out}$ which contains $L$ key pairs, then for any index $i\in[L]$, $P_{pass}\ket{\varphi^\prime}$ is $(2^\eta, 2^{-\eta})$-SC-secure for $K_{out}^{(i)}$ given $K_{out}-K_{out}^{(i)}$.\par
	And we say a remote gadget preparation protocol is secure if it is secure against adversaries of query number $\leq 2^\lambda$ with output security $\eta$ where $\lambda,\eta$ are all fixed polynomials of $\kappa$.
\end{defn}
\subsubsection{A summary}
Thus we have the following natural requirements on the remote gadget preparation protocol that we want: (1)correctness (Definition \ref{defn:3.1}); (2) security (Definition \ref{defn:3.5}); (3) efficiency (Definition \ref{defn:3.2}); (4) succinct client-side quantum operations. 

In step 1 of Outline \ref{ppl:1}, we need to design a protocol that satisfies these properties. As we said before, the idea is to first design a weakly-secure protocol, then amplify it to a fully secure one. Let's formalize the concept of weak security.
\subsection{Remote Gadget Preparation with Weak Security: the Stepping Stone towards the Remote Gadget Preparation}\label{sec:5.2.2}
In this section we will give the formal definition of the weak security transform parameter. Before that, let's review the informal definition given in Chapter \ref{cht:4}:
\begin{defn}
	(Review, Incomplete) We say an $N\rightarrow L$ remote gadget preparation protocol run on security parameter $\kappa$ has \emph{weak security transform parameter} \weakparam{\eta}{C}{p}{\eta^\prime}{C^\prime} against adversaries of query number $\leq 2^\kappa$ if assuming the input $\ket{\varphi}$ satisfies 
	\begin{itemize}
		\item For any $i\in [N]$, $\ket{\varphi}$ is $(2^\eta,C|\ket{\varphi}|)$-SC-secure for $K^{(i)}$ given $K-K^{(i)}$;
		\item $\ket{\varphi}$ is not too ``ill-behaved''.
	\end{itemize}
	For any adversary $\fAdv$ of query number $\leq 2^\kappa$, suppose the corresponding post-execution state is
	$$\ket{\varphi^\prime}=ProtocolName_\fAdv(Keys;Parameters)\circ\ket{\varphi}$$
	and the output keys are $K_{out}$, at least one of the followings is true:
	\begin{itemize}
		\item 	$|P_{pass}\ket{\varphi^\prime}|\leq p|\ket{\varphi}|$
		\item For all $i\in [L]$, $P_{pass}\ket{\varphi^\prime}$ is $(2^{\eta^\prime},C^\prime|\ket{\varphi}|)$-SC-secure for $K_{out}^{(i)}$ given $K_{out}-K_{out}^{(i)}$.
	\end{itemize}
\end{defn}

%

The missing part of this definition is what it means by saying the input state is not too ``ill-behaved''. As an example, we don't want the server to know the xor of all the random oracle output. In the next section we will formalize the \emph{representable} property, which is what we want.

\subsection{Ruling out the ill-behaved cases: the representable property, and the behavior of states under padded RO}
Let's start to think about what will happen when the random oracle is padded. The lookup table construction, and many protocols in our paper, have the following procedure: the client samples $pad\leftarrow_r \{0,1\}^l$, and computes something later using $H(pad||\cdots)$. Intuitively, if the adversary's state is ``nice'', if $l$ is long enough, $H(pad||\cdots)$ should ``seems like'' a new random oracle that is not queried by the adversary. On the other hand, if the adversary holds the xor of all the outputs in the random oracle, which is intuitively in the ``ill cases'', such an oracle padding won't work (at least in the simple way). If we want to argue about the security of the protocols abstractly, we need a definition that helps us rule out these ``ill cases''.\par
 We will define a property called \emph{representability}, which intuitively means the state can be ``represented'' from another state via RO queries and linear decomposition:
\begin{defn}[Representability of a state]\label{def:rep}
	We say $\ket{\varphi}$ is $(2^{\alpha_1},2^{\alpha_2})$-representable from $\ket{\varphi_{init}}$ if \begin{equation}\label{eq:represen}\ket{\varphi}=\sum_{i=1}^{2^{\alpha_1}}\cP_i\ket{\varphi_{init}}\end{equation}, and $\forall i$, $\cP_i$ can contain unitaries, projections and RO queries, and the query number $|\cP_i|\leq 2^{\alpha_2}$.\par
	We call (\ref{eq:represen}) the \emph{representation} of $\ket{\varphi}$. (This is not related to the ``representation'' in other fields.)\par
	And we say $\ket{\varphi}$ is $(2^{\alpha_1},2^{\alpha_2})$-server-side-representable from $\ket{\varphi_{init}}$ if (besides on the definition above) all the $\cP_i$ are server side operations.
\end{defn}
We note that the \emph{server-side representable} is defined mainly for other purpose. Here we study the \emph{representable} property, which allows the client and the server cooperate to prepare this state.\par
One common choice of the $\ket{\varphi_{init}}$ is the $\ket{\mathfrak{init}}$ state described in Section \ref{sec:2.4.5}.\par
And we have the following lemmas on its properties:
\begin{lem}[Transitivity of representability]\label{lem:3.2}
	If \begin{itemize}\item $\ket{\varphi}$ is $(2^{\alpha_1},2^{\alpha_2})$-representable from $\ket{\mathfrak{init}}$; \item $\ket{\chi}$ is $(2^{\alpha^\prime_1},2^{\alpha^\prime_2})$-representable from $\ket{\varphi}$\end{itemize}, then $\ket{\chi}$ is $(2^{\alpha_1+\alpha_1^\prime},2^{\alpha_2}+2^{\alpha_2^\prime})$-representable from $\ket{\mathfrak{init}}$.
\end{lem}

Before we introduce the next lemma, let's briefly discuss what it means by saying \emph{a state does not depend on part of the random oracle content}.
\begin{defn}\label{def:ndep}
We say \emph{$\ket{\varphi}$ does not depend on $H(\cdots ||Pads||\cdots)$} (where $\cdots$ means arbitrary strings of a fixed length) if the state components that correspond to different initialization of $H(\cdots ||Pads||\cdots)$ is the same. In more details, we can expand $\ket{\varphi}$:
\begin{equation}\label{eq:10pa}\ket{\varphi}=\sum_{\text{all the possible $pads$ of this size and element length}}\underbrace{\ket{pads}}_{Pads}\otimes \ket{\varphi_{pads}}\otimes \underbrace{\ket{pads}}_{\text{in the environment}}\end{equation}
\begin{equation}
	\ket{\varphi_{pads}}=\sum_h\underbrace{\ket{h}}_{\text{description of $\{H(\cdots ||pad||\cdots)\}_{pad\in pads}$}}\otimes \ket{\varphi_{pads,h}}	\otimes \underbrace{\ket{h}}_{\text{in the environment}}
	\end{equation}
	$\ket{\varphi_{pads,h}}$ is invariant when $h$ varies.
\end{defn}
Now we have the following lemma.

\begin{lem}[Representability implies RO-paddability]\label{lem:3.4r}
	If $\ket{\varphi}$ is \\$(2^{\alpha_1},2^{\alpha_2})$-representable from $\ket{\mathfrak{init}}$, then after the client samples a tuple of strings where each element is sampled independently randomly from $\{0,1\}^l$, consider the post-sampling state
	$$\ket{\varphi^\prime}=\sum_{pads\in \text{all the valid choices}}\frac{1}{\sqrt{2^{l\cdot |Pads|}}}\underbrace{\ket{pads}}_{\text{client side system }Pads}\otimes\ket{\varphi}\otimes \underbrace{\ket{pads}}_{\text{environment}}$$
	$$\text{where $|Pads|$ is the number of strings in the string tuple}$$ Define $\ket{\varphi^{\prime\prime}}$ as the result of the following: expand $\ket{\varphi}$ using its representability property, and replace each oracle query by $H(I-P_{\cdots||Pads||\cdots})$ (for each query input, projecting out the space that have a prefix appeared in $Pads$; the ``$\cdots$'' here has fixed length). Then
	\begin{itemize}
	\item $\ket{\varphi^{\prime\prime}}$ does not depend on $H(\cdots||Pads||\cdots)$. Which means, for each choices of the content of $H(\cdots||Pads||\cdots)$, the corresponding component of $\ket{\varphi^{\prime\prime}}$ on this space is a fixed state.
	\item $|\ket{\varphi^{\prime\prime}}-\ket{\varphi^{\prime}}|\leq 2^{\alpha_1+\alpha_2+\log|Pads|-l/2}$.\end{itemize}
\end{lem}

The proof is by a hybrid method and we put it in Appendix \ref{sec:app1}.
\subsection{Complete definition of the weak security of remote gadget preparation}
Since we have defined the SC-security and representable property, we are prepared to introduce the full definition of the weak security, as discussed in Chapter \ref{cht:4}.
\begin{defn}\label{def:3.12}
	We say an $N\rightarrow L$ remote gadget preparation protocol run on security parameter $\kappa$ has \emph{weak security transform parameter} \weakparam{\eta}{C}{p}{\eta^\prime}{C^\prime} for input state in $\cF$ (which is a set of states) against adversaries of query number $\leq 2^\kappa$ if a statement in the following form holds for the protocol:\\
	\begin{mdframed}
			Suppose the input keys are $K=\{x_b^{(i)}\}_{i\in [N],b\in\{0,1\}}$. Suppose the initial state, described by the purified joint state $\ket{\varphi}$,  satisfies the following conditions:
			\begin{itemize}
				\item (Input security) $\forall i\in [N]$, $\ket{\varphi}$  is $(2^\eta,C|\ket{\varphi}|)$-SC-secure for $K^{(i)}$ given $K-K^{(i)}$
				\item (Input well-behaveness) $\ket{\varphi}\in \cF$ 
			\end{itemize}
			For any adversary $\fAdv$ of query number $|\fAdv|\leq 2^\kappa$, the final state when the protocol completes, denoted as
			$$\ket{\varphi^\prime}=ProtocolName_\fAdv(K;Parameters)\circ\ket{\varphi}$$
			, (and correspondingly, output keys are $K_{out}=\{y_b^{(i)}\}_{i\in [L],b\in\{0,1\}}$) at least one of the followings is true:
			\begin{itemize}
				\item (Passing probability)	$|P_{pass}\ket{\varphi^\prime}|\leq p|\ket{\varphi}|$
				\item (Output security) $\forall i\in [L], P_{pass}\ket{\varphi^\prime}$ is $(2^{\eta^\prime},C^\prime|\ket{\varphi}|)$-SC-secure for $K_{out}^{(i)}$ given $K_{out}-K_{out}^{(i)}$.
			\end{itemize}\end{mdframed}
\end{defn}
Here $p,\eta,C,\eta^\prime,C^\prime$ can all be viewed as functions of $\kappa$. But they can also be constants.
\paragraph{The choice of $\cF$} The introduction of $\cF$ is mainly to restrict the initial state to a set of ``well-behaved'' states. For most protocols later, $\cF$ will be taken to be the set of states of the following form:
$$\text{$\ket{\varphi}$ is $(2^{\alpha_1},2^{\alpha_2})$-representable from $\ket{\mathfrak{init}}$. $|\ket{\varphi}|\leq \text{some upper bound}$.} $$
And sometimes we also make some specific requirements on the form of the input state.\par
Let's introduce the following notation for the set of ``well-behaved states'', which will be used frequently later.
\begin{nota}\label{nota:wbsn}
Define $\mathcal{WBS}(D)$ to be the set of joint purified states (denoted as $\ket{\varphi}$) such that:
$$\text{$\ket{\varphi}$ is $(2^{\alpha_1},2^{\alpha_2})$-representable from $\ket{\mathfrak{init}}$. $\alpha_1,\alpha_2,\log(1/|\ket{\varphi}|)\leq D$.} $$
\end{nota}

Now let's introduce some variants of Definition \ref{def:3.12}. We introduce a simplified variant of Definition \ref{def:3.12}, where the $p$ variable is small enough to be omitted:
\begin{defn}\label{def:3.12b}
	We say an $N\rightarrow L$ remote gadget preparation protocol run on security parameter $\kappa$ has \emph{weak security transform parameter} \weakparams{\eta}{C}{\eta^\prime}{C^\prime} for input state in $\cF$ (which is a set of states) against adversaries of query number $\leq 2^\kappa$ if a statement in the similar form as Definition \ref{def:3.12} holds for the protocol, and the only difference is the first case ($|P_{pass}\ket{\varphi^\prime}|\leq p|\ket{\varphi}|$) is removed.
\end{defn}
And sometimes the protocol is designed to be run on different sets of keys, and the initial conditions might be different for different key sets. So let's introduce a multi-key-set version of Definition \ref{def:3.12}:
\begin{defn}\label{def:3.12c}
Suppose the remote gadget preparation protocol is denoted by $$ProtocolName_\fAdv(K_1,K_2;Parameters)$$. We say the protocol has weak security transform parameter \weakparaml{\eta}{C}{\eta_2}{C_2}{p}{\eta^\prime}{C^\prime} for input state in $\cF$ (which is a set of states) against adversaries of query number $\leq 2^\kappa$ if a statement similar to the one shown in Definition \ref{def:3.12} holds, with the following differences: the first condition is replaced by the following  conditions:\par
Suppose $K_1$ has $N_1$ pairs of keys and $K_2$ has $N_2$ pairs of keys. $\forall i\in [N_1]$, $\ket{\varphi}$ is $(2^\eta,C|\ket{\varphi}|)$-SC-secure for $K_1^{(i)}$ given $(K_1-K_1^{(i)})\cup K_2$; and $\forall i\in [N_2]$, $\ket{\varphi}$ is $(2^{\eta_2},C_2|\ket{\varphi}|)$-SC-secure for $K_2^{(i)}$ given $(K_2-K_2^{(i)})\cup K_1$.
\end{defn}

\subsubsection{Composition of the security statement: From (weak) security of subprotocols to (weak) security of big protocols}\label{sec:4.8.1}
Since our protocol comes from the composition of many small subprotocols, we want the security property of the subprotocols to be ``composable'', too. Thus we can argue about the security of the big protocol with a series of statements in the form of ``the input state of this subprotocol satisfies some properties, so the output state of this subprotocol satisfies some properties''. Our proof basically uses such technique, although the details may be different somewhere.\par
So what kind of properties will we study? As we showed in the definition of the weak security (Definition \ref{def:3.12}), there are three properties that we care about:
\begin{itemize}
	\item (1) the norm of a state, like $|\ket{\varphi}|$, or $|P_{pass}\ket{\varphi}|$, etc, which intuitively describes the ``passing probability'';
	\item (2) the SC-security of a state for some keys;
	\item (3) the state is not too ``ill-behaved'', which is characterized by the \emph{representable} property.\end{itemize}
The first two are already covered by the definition of weak security of protocols; and the third one is easy to deal with: as long as the initial state is not ``ill-behaved'' and $|\fAdv|$ is bounded during the whole protocol, the output state will not be ``ill-behaved'' by Lemma \ref{lem:3.2}. So to study the security of the subprotocols given later, our main work will be to prove the weak security of the protocols, which is, to study how a subprotocol affect the property (1) and (2) of the server's state. We refer to the definition of weak security (Definition \ref{def:3.12b}, and the whole Section \ref{sec:3}) for more details. 
\section{A Collection for Notations, Lemmas and Techniques for Security Proofs}\label{sec:4}
In this section we give several basic concepts, techniques and lemmas that are useful in the later sections. These lemmas are usually intuitive, but play a fundamental role in the security proofs later.
\subsection{More State-related security: ANY-security and unpredictability for key(s)}
%
In Section \ref{sec:3.1} we give the definition of \emph{SC-security}. We will discuss something more in this subsubsection. Let's first repeat the definition of SC-security:
\begin{defn}[Definition of SC-security, repeated]
	We say $\ket{\varphi}$ is $(2^\kappa, A)$-SC-secure for keys $K=\{x_0,x_1\}$ given $Z$ if for any server side operation $\cU$ with query number $|\cU|\leq 2^\kappa$,
	$$|P_{x_0||x_1}\cU(\ket{\varphi}\odot Z\odot Tag(K))|\leq A$$
\end{defn}
The following definition will be useful in the proof, which is similar to the SC-security, but replace $P_{x_0||x_1}$ with $P_{span\{x_0, x_1\}}=P_{x_0}+P_{x_1}$, which is the projection onto the 2-dimensional space spanned by $x_0$ and $x_1$, and the projection is done on some fixed implicit system. (When the randomness of these keys are purified by the environment it's the projection onto the space whose value is equal to the content of the systems that store $x_0$ or $x_1$.) In other words, if the adversary can compute \emph{any} one of the keys, it breaks the ANY-security. For comparison, in the definition of SC-security the adversary has to know both.
\begin{defn}[Definition of ANY-security]\label{def:4.2}
	We say $\ket{\varphi}$ is $(2^\kappa, A)$-ANY-secure for keys $K=\{x_0,x_1\}$ given $Z$ if for any server side operation\footnote{Recall that $\cU$ can introduce server-side ancillas, which is implicit in the expression below.} $\cU$ with at most $2^\kappa$ RO queries,
	$$|P_{span\{x_0, x_1\}}\cU(\ket{\varphi}\odot Z\odot Tag(K))|\leq A$$
\end{defn}
The reason for the existence of auxiliary information $Z$ is similar to the SC-security case discussed in Section \ref{sec:3.1}. And we refer to Section \ref{sec:global} for the \emph{global tag} $Tag$.\par
In the following section we will need the triangle inequality of the SC security, formalized as follows:
\begin{lem}[Triangle Inequality of the SC/ANY-security]\label{lem:4.1}

	\begin{itemize}
		\item If $\ket{\varphi_1}$ is $(2^\kappa,A)$-SC-secure for $K$ given $Z$, 	$\ket{\varphi_2}$  is $(2^\kappa,B)$-SC-secure for $K$ given $Z$, then $\ket{\varphi_1}+\ket{\varphi_2}$ is $(2^\kappa,A+B)$-SC-secure for $K$ given $Z$.
		\item If $\ket{\varphi_1}$ is $(2^\kappa,A)$-ANY-secure for $K$ given $Z$, 	$\ket{\varphi_2}$ is $(2^\kappa,B)$-ANY-secure for $K$ given $Z$, then $\ket{\varphi_1}+\ket{\varphi_2}$ is $(2^\kappa,A+B)$-ANY-secure for $K$ given $Z$.
	\end{itemize}
\end{lem}

SC-security and ANY-security are both for a pair of keys. We can also define the unpredictability for a single key:
\begin{defn}[Definition of unpredictability for a key]\label{def:4.3n}
	We say $\ket{\varphi}$ is $(2^\kappa, A)$-unpredictable for a key $x_b\in K=\{x_0,x_1\}$ given $Z$ if for any server side operation\footnote{Recall that $\cU$ can introduce server-side ancillas, which is implicit in the expression below.} $\cU$ with at most $2^\kappa$ RO queries,
	$$|P_{x_b}\cU(\ket{\varphi}\odot Z\odot Tag(x_b))|\leq A$$
\end{defn}
\subsection{Proving the Security by Adding Auxiliary Information}\label{sec:4.2}
In our security proof, one very common technique is to consider the behavior of the protocol when some auxiliary information is provided to the adversary.\par
The intuition is as follows. If the client provides some auxiliary information to the adversary, it does not make the adversary weaker: an adversary with this auxiliary information can do anything that an adversary without it can do, by simply ignoring it. Thus proving a security statement when such auxiliary information exists implies the same statement when such auxiliary information does not exist.\par
But why do we want to give extra auxiliary information to the adversary? The reason is, in many cases, sending an auxiliary information to the server (by copying it to the read-only buffer) will make the server's state more ``well-behaved''. This allows us to analyze the structure of the state in many different ways. \par
We have the following lemma, which is intuitive, but turns out to be very useful in our security proof:
\begin{tecq}[Auxiliary-Information technique]\label{lem:4.2}
	The following statements are true when ``some property'' is replaced by appropriate concrete statements:
	\begin{enumerate}\item Suppose a protocol $\fPrtl$, initial state $\ket{\varphi}$ and auxiliary information $\llbracket\fAuxInf\rrbracket$ satisfy: for any adversary $\fAdv$ with query number $|\fAdv|\leq 2^\lambda$,
		      $$\fPrtl_\fAdv\circ(\ket{\varphi}\odot \llbracket\fAuxInf\rrbracket)$$
		      satisfies some property, then for any adversary $\fAdv$ with query number $|\fAdv|\leq 2^\lambda$,
		      $$\fPrtl_\fAdv\circ\ket{\varphi}$$
		      satisfies this property.\par
		      This is the most basic form of the \emph{Auxiliary-Information technique}: providing some auxiliary information does not make the adversary weaker.

		\item Suppose a protocol $\fPrtl$, initial state $\ket{\varphi}$ and auxiliary information $\llbracket\fAuxInf\rrbracket$  satisfy: for any adversary $\fAdv$ with query number $|\fAdv|\leq 2^\lambda$,
		      $$\fPrtl_\fAdv\circ(\ket{\varphi}\odot \llbracket\fAuxInf\rrbracket)$$
		      satisfies some property, then for any adversary $\fAdv$ with query number $|\fAdv|\leq 2^\lambda$,
		      $$((\fPrtl_\fAdv\circ\ket{\varphi})\odot \llbracket\fAuxInf\rrbracket)$$
		      satisfies this property.\par
		      Intuitively, it means, if the auxiliary information is provided earlier, it does not make the adversary weaker.
		\item Suppose a protocol $\fPrtl$, initial state $\ket{\varphi}$ and auxiliary information $\llbracket\fAuxInf\rrbracket$ satisfy: for any adversary $\fAdv$ with query number $|\fAdv|\leq 2^\lambda$,
		      $$\fPrtl_\fAdv\circ(\ket{\varphi}\odot \llbracket\fAuxInf\rrbracket)$$
		      satisfies some property, and if $\llbracket\fAuxInf^\prime\rrbracket$ can be computed from $\llbracket\fAuxInf\rrbracket$ using a randomized algorithm that makes at most $Q$ RO-queries, then for any adversary $\fAdv$ with query number $|\fAdv|\leq 2^\lambda-Q$,
		      $$\fPrtl_\fAdv\circ(\ket{\varphi}\odot \llbracket\fAuxInf^\prime\rrbracket)$$
		      satisfies this property.
	\end{enumerate}
\end{tecq}
We note that this is not a formal statement, but a technique. This technique can be seen as a simpler way to write the ``simulation-based proof'' in our setting. Obviously, the ``some property'' cannot be chosen arbitrarily, for example, it cannot be replaced by the size of the state; but the choices are very general, and it can be replaced by one of the follows, which is general enough for our proof:
\begin{itemize}
	\item The SC/ANY-security: $\ket{\varphi}$ is $(B,C)$-SC/ANY secure for some keys given something;
	\item The SC/ANY-security on the passing space: $P_{pass}\ket{\varphi}$ is $(B,C)$-SC/ANY secure for some keys given something;
	\item The weak security of remote gadget preparation: $\ket{\varphi}$ satisfies either $|P_{pass}\ket{\varphi}|\leq A|\ket{\varphi}|$ or $P_{pass}\ket{\varphi}$ is $(B,C)$-SC/ANY secure for some keys given something.
\end{itemize}
The proof of the correctness of the technique is as follows. The proof is similar for all the three cases.
\begin{proof}
	Note that for all the three cases of the technique, the adversaries in the conclusions (the statement after the word ``then'') are no more powerful than the adversaries in the conditions (the ``suppose'' part): For example, in the first case, the adversary in the ``suppose'' part can simply ignore the auxiliary information. Thus if the statement after the word ``then'' does not hold, the adversary in the ``suppose'' part can simply ignore the auxiliary information, run the adversary's code in the ``then'' part, and break the condition.\par
	In the second case of Technique \ref{lem:4.2}, if the auxiliary information is provided in advance, the adversary can choose to keep it in a separate register and use it later.\par
	In the third case, the adversary can compute $\llbracket\fAuxInf^\prime\rrbracket$ from $\llbracket\fAuxInf\rrbracket$ using at most $Q$ queries.
\end{proof}
In the following sections we will need to use this method very frequently. We can apply this technique to reduce the statement we want to prove to a new statement where the adversary: (1) gets the auxiliary information; (2)gets the auxiliary information in advance; (3) gets $\llbracket\fAuxInf\rrbracket$ instead of $\llbracket\fAuxInf^\prime\rrbracket$.\par
Finally we give a convenient lemma that combines the auxiliary-information technique and the SC/ANY security. (The ``SC/ANY'' in the statement below can be replaced by either ``SC'' or ``ANY''.)\par
\begin{lem}\label{lem:basic}
	Suppose $K_1$ is a set of keys and $K_2$ is a pair of keys. The initial purified joint state $\ket{\varphi}$ is $(B,C)$-SC/ANY-secure for $K_2$ given $K_1$.\par A protocol $\fPrtl$ with non-adaptive structure satisfies: at any time of this protocol, the computation of all the client-side messages only uses algorithms that only takes (1) the description of $K_1$; (2) freshly new random coins; and (3) the server's response before this time during this protocol as the inputs, and the number of queries to prepare these messages is at most $Q$.\par $\fAdv$ is an adversary with query number $|\fAdv|$.\par Then $$\text{$\fPrtl_\fAdv\circ\ket{\varphi}$ is $(B-|\fAdv|-Q,C)$-SC/ANY secure for $K_2$ given $K_1$.}$$
	Furthermore, if $\ket{\tilde\varphi^\prime}$ can be written as the sum of $2^{\alpha_1}$ terms where each term has the form of $\fPrtl_\fAdv\circ\ket{\varphi}$ (but for different terms $\fAdv$ can be different), then $$\text{$\ket{\tilde\varphi^\prime}$ is $(B-|\fAdv|-Q,2^{\alpha_1}C)$-SC/ANY secure for $K_2$ given $K_1$.}$$
\end{lem}

\subsection{State Decomposition Lemmas}\label{sec:4.3}
In this subsection we prove a series of \emph{state decomposition lemmas}. These lemmas will be useful in our security proof, and we feel they are both nontrivial and general, and are potentially useful in other random oracle problems.\par
We will first start the state decomposition lemma the SC-security, give a detailed proof. Then we generalize the lemma to ANY-security and unpredictability. Then we give a lemma that has different parameters. Finally we give a multi-key state decomposition lemma.\par
Then in the following two subsubsections we discuss the \emph{linear decomposition technique}, which is based on these decomposition lemmas.
\subsubsection{The lemmas}
Compare the weak security with the usual meaning of security, one source of ``weak'' is: the final state can be $(2^\kappa,A)$-SC-secure for some keys, where $A$ can be a constant or inverse-polynomial instead of exponentially small values. To analyze a protocol that is weakly secure, we need some lemmas to understand the structure of the state $\ket{\varphi}$ that is $(2^\kappa,A)$-SC-secure for $K$. We will show that, such ``non-negligibly-secure SC-security'' can be related to the SC-security with exponentially small ``the second parameter in the SC-security'', through some decomposition lemmas. In this section we will give many decomposition lemmas for this (and related) problem, and these lemmas will be useful in the security proof in the later sections.\par
The intuition is as follows. Suppose a state $\ket{\varphi}$ is $(2^\kappa, A)$-SC-secure for $K=\{x_0,x_1\}$. For simplicity let's temporarily assume $\ket{\varphi}$ is normalized. Imagine $A$ is a constant less than 1. Intuitively, when the adversary makes more and more queries, it should get more and more power on computing the keys in $K$. However, if $\kappa$ is big,  it means the ``information'', or the ``ability of outputting both keys'' that the adversary can get from querying the random oracle for $2^\lambda$ times, $\lambda<\kappa$, is limited. Let's first understand what the state can be: the most natural construction of such a state is to first find a state $\ket{\phi}$ that is $(2^\kappa, 2^{-\kappa})$-SC-secure for $K$ (for example, $\ket{x_0}+\ket{x_1}$, assuming the key length is long enough), then take $|\ket{\chi}|\leq A-\epsilon$ which can be any state(for example, $x_0||x_1$, which simply gives the adversary both keys), and define $\ket{\varphi}=\ket{\phi}+\ket{\chi}$. We can see $\ket{\varphi}$ constructed in this way is $(2^\kappa,A)$-SC-secure for $K$: if the adversary wants to get both keys with non-negligible amplitude with small number of queries, it can only make use of the $\ket{\chi}$ part.\par
The question is: is the inverse also true? Which means, if $\ket{\varphi}$ is $(2^\kappa, A)$-SC-secure for $K$, could we decompose it as $\ket{\phi}+\ket{\chi}$ where $\ket{\phi}$ is SC-secure for $K$ with the second parameter being exponentially small (instead of putting an ``$A$'' there), and $\ket{\chi}$ has bounded norm? In the following lemma we will show such a decomposition always exists.
\begin{lem}\label{lem:4.4}(State decomposition lemma for SC-security)

	The following statement is true for sufficiently large $\kappa$:\par For a pair of keys $K$, suppose the global tag $Tag(K)$ is stored in some fixed place of the read-only buffer, if a purified joint state $\ket{\varphi}$ satisfies:
	\begin{itemize}
		\item $\ket{\varphi}$ is $(2^\kappa, A)$-SC-secure.
		\item $P_{\text{$Tag$ is not injective on inputs with the same length as keys in $K$}}\ket{\varphi}=0$. (Recall $P$ denotes the projection.)

	\end{itemize}
	then we can decompose $\ket{\varphi}$, together with finite number of server-side ancilla qubits (which are all at state zero), into $\ket{\phi}+\ket{\chi}$ where
	\begin{itemize}
		\item $\ket{\phi}$ is $(1,2^\kappa)$-server-side-representable from $\ket{\varphi}$ and is $(2^{\kappa/5},2^{-\kappa/6}A)$-SC-secure for $K$
		\item $|\ket{\chi}|\leq (\sqrt{2}+1)A$ and is $(\kappa/2,2^\kappa)$-server-side-representable from $\ket{\varphi}$.
	\end{itemize}
\end{lem}
\paragraph{A note on the second condition} This condition is not a natural condition on $\ket{\varphi}$ since in the proofs later we will not meet a state $\ket{\varphi}$ that perfectly satisfies this condition. But we can use Fact \ref{fact:injtag} to prove many states that we will meet later is close to this condition. Thus we can use this lemma on the injective subspace and use Fact \ref{fact:injtag} to deal with the error term. Other lemmas in this subsection are also used in this form.\par
\paragraph{A note on the second parameter in the SC-security} Here the condition is ``$\ket{\varphi}$ is $(2^\kappa, A)$-SC-secure''. Note that the second parameter is $A$, not $A|\ket{\varphi}|$. And the initial state is not necessarily normalized. For some lemmas and cases we describe the properties of states in a ``relative'' way (for example, the lemmas in Section \ref{sec:4.4}), and sometimes we describe it in the ``absolute'' way, like this lemma.
\begin{proof}[Proof of Lemma \ref{lem:4.4}]
	We can prove the following strengthened lemma: there exist server-side operators $\cP_0,\cP_1,\cdots \cP_{\kappa/3}$ such that:
	\begin{itemize}
		\item Each $\cP_i$ is a sequence of projections and unitaries and random oracle queries, $\sum_i\cP_i=I$
		\item $\forall i$, the query number $|\cP_i|\leq 2^\kappa$
		\item $\cP_0\ket{\varphi}$ is $(2^{\kappa/5},A(\sqrt{2})^{-\kappa/3})$-SC-secure for $K$
		\item $|\sum_{i\geq 1}\cP_i\ket{\varphi}|\leq (\sqrt{2}+1)A$
	\end{itemize}
	Then we can choose $\ket{\phi}=\cP_0\ket{\varphi}$ and $\ket{\chi}=\sum_{i\geq 1}\cP_i\ket{\varphi}$.\par
	\paragraph{Ideas for the proof} Repeatedly apply an argument to construct a series of server-side operators (which can contain projections) $\cP_1^\prime,\cP_2^\prime\cdots$ one by one. Here $\cP_{i+1}$ is constructed based on the properties of $\ket{\varphi^i}$, during which $\ket{\varphi^{i+1}}$  also gets defined. Each $\ket{\varphi^{\cdots}}$ has a security property of a similar form. And the exponent on the SC-security of $\ket{\varphi^{\cdots}}$ decreases only additively by a constant in each round thus we can afford this decrease for $\kappa/3$ times.\par And $\cP_1,\cP_2,\cdots$ can be represented using $\cP_1^\prime,\cP_2^\prime\cdots$.\par
	We will put the argument in the box, and outside the box, we will first use the first round as an example, and then show how to apply this argument recursively to the end.\par
	To construct $\cP_1^\prime$, we first denote $\ket{\varphi^0}=\ket{\varphi}$ to make the argument more consistent in each round. Which means
	\begin{equation}\label{eq:4rr}\text{$\ket{\varphi^0}$ is $(2^\kappa,A)$-SC-secure for $K$.}\end{equation}
	The construction is given below.
	\begin{mdframed}
	\textbf{Construct $\cP^\prime_{i+1}$ from $\ket{\varphi^i}$}\par
	The condition is: $\ket{\varphi^i}$ is $(2^{\kappa-i\log_25.1},\omega_i)$-SC-secure for $K$. ($\omega_0:=A$)\par
		Discuss by cases:
		\begin{itemize}
			\item (Case 1) If $\ket{\varphi^i}$ is $(2^{\kappa-i\log_25.1}/5,\omega_i/\sqrt{2})$-SC-secure for $K$, choose $\cP_{i+1}^\prime=0$.
			\item (Case 2) Otherwise, there exists a server-side operation $\cU_{i+1}$ with query number $|\cU_{i+1}|\leq 2^{\kappa-i\log_25.1}/5$ such that $|P_{x_0||x_1}\cU_1\ket{\varphi^i}|> \omega_i/\sqrt{2}$. Choose $\cP_{i+1}^\prime=\cU_{i+1}^\dagger P_{x_0||x_1}\cU_{i+1}$. Note here $|\cP^\prime_{i+1}|\leq 2^{\kappa-i\log_25.1}\cdot 2/5+2$.
		\end{itemize}
		In the both cases, we can prove that
		\begin{equation*}\ket{\varphi^{i+1}}:=\ket{\varphi^i}-\cP_{i+1}^\prime\ket{\varphi^i}=\cU_{i+1}^\dagger (I-P_{x_0||x_1})\cU_{i+1}\ket{\varphi^i}\end{equation*}\begin{equation}\label{eq:4n}\text{ is $(2^{\kappa-i\log_25.1}/5-1,\omega_i/\sqrt{2})$-SC-secure for $K$.}\end{equation}
		Equivalently, we can introduce a new variable $\omega_{i+1}$ and express the statement as: $\ket{\varphi^{i+1}}$ is $(2^{\kappa-i\log_25.1}/5-1,\omega_{i+1})$-SC-secure for $K$, for some value $\omega_{i+1}\leq \omega_i/\sqrt{2}$.\par
		The proof of (\ref{eq:4n}) follows.
		\begin{itemize} \item For Case 1 above it's obvious. \item For Case 2, that's because otherwise (here we use the proof-by-contradiction, and assume there is a server-side operation with $\leq 2^{\kappa-i\log_25.1}/5-1$ queries that can compute $x_0||x_1$ from $\ket{\varphi^{i+1}}$ with norm at least $\omega_i/\sqrt{2}$), we can construct the following server-side operation on $\ket{\varphi^i}$ to violate the property that ``$\ket{\varphi^{i}}$ is $(2^{\kappa-i\log_25.1},\omega_i)$-SC-secure for $K$'':
			      \begin{enumerate}\item In $\cP^\prime_{i+1}$, after applying $\cU_{i+1}$, instead of making a projection onto $x_0||x_1$, purify the projection operator as a unitary (that is, taking an auxiliary qubit to store the result of checking whether the state is in $x_0||x_1$.) This can be done with the help of $Tag(K)$. \item Then controlled by the $\ket{0}$ part of the auxiliary qubit (which corresponds to $(I-P_{x_0||x_1})\cU_{i+1}\ket{\varphi^i}$), apply $\cU_{i+1}^\dagger$ and continue to apply the operation from the ``proof-by-contradiction assumption'' that computes both keys from $\ket{\varphi^{i+1}}$ (which compute $x_0||x_1$ with norm at least $\omega_i/\sqrt{2}$).\end{enumerate}
			      The whole operation computes $x_0||x_1$ with norm $>\sqrt{(\omega_i/\sqrt{2})^2+(\omega_i/\sqrt{2})^2}=\omega_i$, and the query number is at most $|\cU_{i+1}|+2+2(|\cU_{i+1}^\dagger|+2^{\kappa-i\log_25.1}/5-1)\leq 2^{\kappa-i\log_25.1}$, which is a contradiction.\end{itemize}\end{mdframed}
	Thus (by taking $i=0$ in (\ref{eq:4n}))
	\begin{equation}\label{eq:5rr}\text{$\ket{\varphi^1}:=(I-\cP^\prime_1)\ket{\varphi^0}$ is $(2^{\kappa-\log_25.1},\omega_1)$-SC-secure for $K$,}\end{equation}\begin{equation}\text{ $\omega_1\leq A/\sqrt{2}$, $|\cP_1^\prime|\leq \frac{2}{5}2^\kappa+2$}\end{equation}
	Note that $I-\cP_1^\prime$ can be implemented using $\frac{2}{5}2^\kappa+2$ queries. And define $$\cP_1=\cP_1^\prime$$
	Thus in the first step $\cP_1$ and $\cP_1^\prime$ are the same. (But later $\cP_i$ and $\cP_i^\prime$ will not be.) Now we complete the construction of $\cP_1^\prime,\cP_1$ by analyzing $\ket{\varphi^0}$, and define $\ket{\varphi^1}$.\par
	Note that (\ref{eq:5rr}) has the same form as the condition on $\ket{\varphi^0}$ (we mean (\ref{eq:4rr})). Thus we can repeat the similar argument on $\ket{\varphi^1}$ (similar to the ``construct $\cP_1^\prime$ from $\ket{\varphi^0}$'', with differences on indexes and parameters) and construct $\cP^\prime_2$ such that \begin{equation}\label{eq:7rr}\text{$\ket{\varphi^2}=(I-\cP^\prime_2)\ket{\varphi^1}$ is $(2^{\kappa-2\log_25.1},\omega_2)$-SC-secure for $K$,}\end{equation}\begin{equation}\text{ $\omega_2\leq \omega_1/\sqrt{2}$, $|\cP^\prime_2|\leq \frac{2}{5}2^{\kappa-\log_25.1}+2$}\end{equation}, and $\cP_2$ comes from expanding $\cP_2^\prime\ket{\varphi^1}$ until reaching $\ket{\varphi}$, which is $$\cP_2=\cP_2^\prime(I-\cP_1^\prime)$$
	Then based on (\ref{eq:7rr}), repeat the argument on $\ket{\varphi^2}$ to get $\cP_3^\prime$, and
	$$\cP_3=\cP_3^\prime(I-\cP_2^\prime)(I-\cP_1^\prime)$$. Repeat this process until all the $\cP_i$ are constructed. Then $\cP_0=I-\sum_{i\geq 1}\cP_i$. Notice for each $\cP_i$, the query number $\leq \sum_{t\in [\kappa/3]}(2^{\kappa-t\log_25.1}+2)\leq 2^\kappa$.\par
	 And for the security of $\cP_0\ket{\varphi}$, we can verify $\kappa-\frac{\kappa}{3}\log_25.1\geq \kappa/5$.\par
	To bound $|\sum_{i\geq 1}\cP_i\ket{\varphi}|$, it's easy to see $|\cP_i\ket{\varphi}|$ can be bounded by a geometric decreasing sequence $A/(\sqrt{2})^t$ thus their sum converges and can be bounded by $A/(1-1/\sqrt{2})$. To get a better bound, we can make use of the relations $$\text{$\omega_0:=A$, $\forall i\in [0,\kappa/3-1]$, $\omega_{i+1}\leq \omega_i/\sqrt{2}$ and $\omega_{i+1}^2+|\cP_{i+1}\ket{\varphi}|^2\leq \omega_i^2$}$$. Some elementary calculation (see Appendix \ref{sec:app1}) gives the bound $|\sum_{i\geq 1}\cP_i\ket{\varphi}|\leq (\sqrt{2}+1)A$.
\end{proof}

Let's explain one subtleness of this lemma. This lemma allows us to decompose a state $\ket{\varphi}$ that is SC-secure into $\ket{\phi}+\ket{\chi}$. However, there is no guarantee that these two states are orthogonal. What's more, the bound on the norm of $\ket{\chi}$ does not come from the orthogonality either.\par
\paragraph{Application Scenario} We will use this lemma heavily in the protocol design and security proof in the next section. The reason is when we design a subprotocol and state its security, the conditions are usually in the form of ``$\ket{\varphi}$ is $(2^{\fpoly(\kappa)},2^{-\fpoly(\kappa)})$-SC-secure for the input keys''. (Let's temporarily only consider the normalized state to simplify the expression.) However the conclusions can be ``$P_{pass}\ket{\varphi^\prime}$ is $(2^{\fpoly(\kappa)},A)$-SC-secure for the output keys''. Thus when we compose the subprotocols together, the security proofs can't be composed directly. Using this lemma, we can decompose a state that is $(2^{\fpoly(\kappa)},A)$-SC-secure into two states $\ket{\phi}$ and $\ket{\chi}$. Since $\ket{\phi}$ is $(2^{\fpoly(\kappa)},2^{-\fpoly(\kappa)})$-SC-secure, when this part is given as the initial state of some other protocols, the security proof goes through. And we can deal with the $\ket{\chi}$ part using some other techniques.\par

Using similar techniques we can also prove a similar lemma for ANY-security:
\begin{lem}\label{lem:4.5}
	(State decomposition for ANY-security)

	The following statement is true for sufficiently large security parameter $\kappa$.\par
	Consider keys denoted as $K=\{x_0,x_1\}$. Suppose the initial state is described by the purified joint state $\ket{\varphi}$, and suppose the global tag  $Tag(K)$ is stored in some fixed place of the read-only buffer, and the followings are satisfied:
	\begin{itemize}
	\item $\ket{\varphi}$ is $(2^\kappa, A)$-ANY-secure	for $K$.
	\item $P_{\text{$Tag$ is not injective on inputs with the same length as keys in $K$}}\ket{\varphi}=0$. (Recall $P$ denotes the projection.)
	\end{itemize}
  we can decompose $\ket{\varphi}$ together with finite number of server-side ancilla qubits (which are all at state zero), into $\ket{\phi}+\ket{\chi}$ where
	\begin{itemize}
		\item 	$\ket{\phi}$ is $(1,2^\kappa)$-server-side-representable from $\ket{\varphi}$ and is $(2^{\kappa/5},2^{-\kappa/6}A)$-ANY-secure for $K$.
		\item $|\ket{\chi}|\leq (\sqrt{2}+1)A$ and is $(\kappa/2,2^\kappa)$-server-side-representable from $\ket{\varphi}$.
	\end{itemize}
\end{lem}

\begin{proof}
	The proof is very similar to the proof above, and the only difference is we need to replace the server-side projection $P_{x_0||x_1}$ with $P_{span\{x_0, x_1\}}$. Both operations can be done given $Tag(K)$.
\end{proof}
And for the unpredictability discussed in Definition \ref{def:4.3n}:
\begin{lem}\label{lem:4.7r}
	(State decomposition for unpredictability)

	The following statement is true when $\kappa$ is bigger than some constant.\par
	Consider keys denoted as $K=\{x_0,x_1\}$. Consider the initial state described by the purified joint state $\ket{\varphi}$, and a bit $b\in \{0,1\}$, the global tag  $Tag(x_b)$ is stored in some fixed place of the read-only buffer, and the followings are satisfied:
	\begin{itemize}
	\item $\ket{\varphi}$ is $(2^\kappa, A)$-unpredictable for key $x_b\in K=\{x_0,x_1\}$;
	\item $P_{\text{$Tag$ is not injective on inputs with the same length as keys in $K$}}\ket{\varphi}=0$. (Recall $P$ denotes the projection.)	
	\end{itemize}
 then  we can decompose $\ket{\varphi}$ together with finite number of server-side ancilla qubits (which are all at state zero), into $\ket{\phi}+\ket{\chi}$ where
	\begin{itemize}
		\item 	$\ket{\phi}$ is $(1,2^\kappa)$-server-side-representable from $\ket{\varphi}$ and is $(2^{\kappa/5},2^{-\kappa/6}A)$-unpredictable for $x_b$.
		\item $|\ket{\chi}|\leq (\sqrt{2}+1)A$ and is $(\kappa/2,2^\kappa)$-server-side-representable from $\ket{\varphi}$.
	\end{itemize}
\end{lem}

\begin{proof}
	The proof is very similar to the proof above, and the only difference is we need to replace $P_{span\{x_0,x_1\}}$ with $P_{x_b}$. This can be done given $Tag(K)$.
\end{proof}
And we also need the following lemma in the later sections, which is similar to the lemma above, but the parameters are different:
\begin{lem}\label{lem:4.6}
	(Another state decomposition for ANY-security)
	The following statement is true when $\kappa$ is bigger than some constant:\par
	Consider keys denoted as $K=\{x_0,x_1\}$. Consider initial state described by purified joint state $\ket{\varphi}$. Suppose the global tag  $Tag(K)$ is stored in some fixed place of the read-only buffer, and the followings are satisfied:
	\begin{itemize}
	\item $\ket{\varphi}$ is $(2^\kappa, (1-\frac{1}{T})|\ket{\varphi}|)$-ANY-secure for $K$. $T>2$.
	\item $P_{\text{$Tag$ is not injective on inputs with the same length as keys in $K$}}\ket{\varphi}=0$. (Recall $P$ denotes the projection.)
	\end{itemize}
	then 
	we can decompose $\ket{\varphi}$ together with finite number of server-side ancilla qubits (which are all at state zero), into $\ket{\phi}+\ket{\chi}$ where
	\begin{itemize}
		\item $\ket{\phi}$ is $(1,2^\kappa)$-server-side-representable from $\ket{\varphi}$ and is \\$(2^{(\kappa-150T^2)/6},2^{-(\kappa-150T^2)/6}|\ket{\varphi}|)$-ANY-secure for $K$.
		\item $\ket{\chi}$ is $(\kappa,2^\kappa)$-server-side-representable from $\ket{\varphi}$, $|\ket{\chi}|\leq (1-\frac{1}{2T})|\ket{\varphi}|$.
	\end{itemize}
\end{lem}
Note that the main difference from the previous lemmas is here the second parameter in the SC-security of the condition, which is $(1-1/T)|\ket{\varphi}|$, is very close to $|\ket{\varphi}|$. If we simply use the previous lemma, $(\sqrt{2}+1)(1-1/T)>1$, and the conclusion will become trivial. But this lemma can help us decompose $\ket{\varphi}$ in this setting. The burden is the input needs to have higher level of security: $(\kappa-150T^2)/6$ should be big to make the conclusion non-trivial.
\begin{proof}
	The proof uses similar ideas as the previous lemma (see the box in the proof of Lemma \ref{lem:4.4}), but we need to use different parameters when we construct $\cU_i$ (and correspondingly, $\cP_i^\prime$) in each step. The proof starts by constructing $\cP_1$ on $\ket{\varphi^0}:=\ket{\varphi}$.\par
	Starting from $\cU_1$, instead of considering whether there exists $\cU_1$ such that\\ $|P_{span\{x_0, x_1\}}\cU_1\ket{\varphi^0}|> (1-1/T)/\sqrt{2}|\ket{\varphi^0}|$ (this is what we did in the previous lemmas), we consider whether there exists a server-side operation $\cU_1$ with query number $|\cU_1|\leq 2^\kappa/5$ such that \begin{equation}\label{eq:5n}|P_{span\{x_0, x_1\}}\cU_1\ket{\varphi^0}|> (1-1/T-1/(50T^2))|\ket{\varphi^0}|\end{equation}
	And we can see the difference to the proof of Lemma \ref{lem:4.4}: In that proof the right hand side of (\ref{eq:5n}) is $\frac{1}{\sqrt{2}}$ times the original norm. But in this proof we subtract $1/(50T^2)$ from it.\par
	We will repeat the argument in the following box step-by-step: 
	\begin{mdframed}
	\textbf{Construct $\cP_{t+1}$ on $\ket{\varphi^{t}}$}\par
	The condition is: $\ket{\varphi^{t}}$ is $(2^{\kappa-t\log_25.1},(1-1/T-t/(50T^2))|\ket{\varphi}|)$-ANY-secure for $K$.\par
	Discuss by whether there exists a server side operation $\cU_{t+1}$ with query number $2^{\kappa-t\log_25.1}/5$ such that $|P_{span\{x_0, x_1\}}\cU_{t+1}\ket{\varphi^t}|> (1-1/T-(t+1)/(50T^2))|\ket{\varphi}|$. So there are two cases:
	\begin{itemize}
		\item Exist: take $\cP^\prime_{t+1}=\cU^\dagger_{t+1}P_{span\{x_0, x_1\}}\cU_{t+1}$, then $\ket{\varphi^{t+1}}:=(I-\cP^\prime_{t+1})\ket{\varphi^t}$ is\\ $(2^{\kappa-(t+1)\log_25.1},\frac{1}{5T}|\ket{\varphi}|)$-ANY-secure for $K$.
		\item Not exist: take $\cP^\prime_{t+1}=0$, then $\ket{\varphi^{t+1}}:=(I-\cP^\prime_{t+1})\ket{\varphi^t}$ is $(2^{\kappa-(t+1)\log_25.1},(1-1/T-(t+1)/(50T^2))|\ket{\varphi}|)$-ANY-secure for $K$.
	\end{itemize}
	Then correspondingly:
	\begin{itemize}
	\item ``Exist'' case: stop this construction, and jump to the argument below starting from ``at this time the remaining state $\cdots$''.\par
	\item ``Not exist'' case, repeat the same argument on $\ket{\varphi^{t+1}}$, and this gives us $\cP^\prime_{t+2}$ and $\ket{\varphi^{t+2}}$ and so on.
	\end{itemize}
	\end{mdframed}
	Which means, for the ``Exist'' case, stop, and jump to the argument below, otherwise continue this process to the next round, with index increased by 1. If we keep meeting the ``not exist'' case, every time we minus $1/50T^2$ in the right hand of (\ref{eq:5n}). We note that this process cannot be continued infinitely, and after at most $50T^2$ steps we can get a non-zero $\cP_i^\prime$ at index $i$.\par
	Now we have got a non-zero $\cP^\prime_i$ and all the $\cP_{\cdots}^\prime$ before it are zero. At this time the ``remaining state'' $\ket{\varphi^i}:=(\fI-\cP^\prime_i)\ket{\varphi^{i-1}}=(\fI-\cP^\prime_i)\ket{\varphi}$ is $(2^{\kappa-150T^2},\frac{1}{5T}|\ket{\varphi}|)$-ANY-secure for $K$. The query number $|\cP^\prime_i|\leq 2^{\kappa-\log_25}$.\par
	 Then we apply Lemma \ref{lem:4.5} and decompose $\ket{\varphi^i}$ we defined just now as $\ket{\phi}+\ket{\chi^\prime}$ where
	\begin{itemize}
		\item 	$\ket{\phi}$ is $(1,2^{\kappa-150T^2})$-server-side-representable from $\ket{\varphi^i}$ and is\\ $(2^{(\kappa-150T^2)/6}, 2^{-(\kappa-150T^2)/6}|\ket{\varphi}|)$-ANY-secure for $K$
		\item $\ket{\chi^\prime}$ is $(\kappa-150T^2,2^{\kappa-150T^2})$-server-side-representable from $\ket{\varphi^i}$ and $|\ket{\chi^\prime}|\leq \frac{\sqrt{2}+1}{5}\frac{1}{T}|\ket{\varphi}|$
	\end{itemize}
	. Combining these two decompositions (which means, use the $\ket{\phi}$ just now and define $\ket{\chi}=\ket{\chi^\prime}+\cP^\prime_i\ket{\varphi}$) gives the decomposition we need. (Note $|\cP^\prime_i\ket{\varphi}|\leq (1-1/T)|\ket{\varphi}|$ and $|\ket{\chi^\prime}|\leq \frac{\sqrt{2}+1}{5}\frac{1}{T}|\ket{\varphi}|$ thus $|\ket{\chi}|\leq (1-\frac{1}{2T})|\ket{\varphi}|$.)
\end{proof}
The previous lemmas only consider the decomposition for a single pair of keys. The following lemma can help us decompose the state for multiple keys simultaneously:
\begin{lem}\label{lem:4.7}(Multi-keys state decomposition lemma)
	The following statement is true for sufficiently large $\kappa$:\par
	Consider a purified joint state $\ket{\varphi}$. Consider a key set denoted as $K=\{x^{(i)}_0,x^{(i)}_1\}_{i\in[L]}$. Suppose the global tag $Tag(K)$ is stored in some fixed place of the read-only buffer, and the following conditions are satisfied:
	\begin{itemize}
	\item $\forall i\in [L]$, $\ket{\varphi}$ is $(2^\kappa, A)$-SC-secure for keys $K^{(i)}$,  $\kappa> 4\log(6L)$.
	\item $P_{\text{$Tag$ is not injective on inputs with the same length as keys in $K$}}\ket{\varphi}=0$. (Recall $P$ denotes the projection.)
	\end{itemize}

	then we can decompose $\ket{\varphi}$ together with finite number of server-side ancilla qubits (which are all at state zero), into $\ket{\phi}+\ket{\chi}$ where:
	\begin{itemize}
		\item $\ket{\phi}$ is $(1,2^\kappa)$-server-side-representable from $\ket{\varphi}$ and $\forall i\in [L]$, $\ket{\phi}$ is \\$(2^{\frac{\kappa}{2\log 6L}}, 2^{-\frac{\kappa}{4\log 6L}}A)$-SC-secure for $K^{(i)}$.
		\item $\ket{\chi}$ is $(\kappa L,2^\kappa)$-server-side-representable from $\ket{\varphi}$, and $|\ket{\chi}|\leq 4AL$.
	\end{itemize}
\end{lem}
We note that after applying this lemma, in the state $\ket{\phi}$ the exponent in the security decreases multiplicatively, but the decrease ratio only depends logarithmically on $L$.\par
And note that, when we use this lemma --- the only application of this lemma is in Section \ref{sec:10} --- $A$ is already exponentially small. And $AL$ is also exponentially small (Note that if $AL>1$ this lemma will become trivial, but luckily $L$ is not big enough to make it big), thus $|\ket{\chi}|$ is small and can be omitted.\par
\begin{proof}
	The overall technique is similar to the proof of the single-key decomposition lemma. We need to repeat an argument multiple times, but here in each round of this repetition we construct up to $L$ operators, instead of $1$ operator. We will construct $\cP^\prime_{t,1},\cdots \cP^\prime_{t,L}$ from $\ket{\varphi^{t-1}}$, and define $\ket{\varphi^t}$ during this process. Thus we can begin at $\ket{\varphi^0}$ and repeat the following argument round-by-round.
	\begin{mdframed}
	\textbf{Construct $\cP^\prime_{t,1},\cdots \cP^\prime_{t,L}$ from $\ket{\varphi^{t-1}}$}\par
	The condition is $\ket{\varphi^{t-1}}$ is $(2^{\kappa-(t-1)\log_2 (6L)},\omega_{t-1})$-SC-secure for $K^{(i)}$.\par
		Repeat the following for at most $L$ times:
		\begin{itemize}
			\item Suppose this is the $l$-th round. ($l\in [1,L]$.)\par
			 Consider whether there exists a server-side operation $\cU_{t,l}$ with query number $|\cU_{t,l}|\leq 2^{\kappa-(t-1)\log_2 6L}/(6L)$ such that
			      \begin{equation}\label{eq:6}\text{For some $i\in [L]$, $|P_{x_0^{(i)}||x_1^{(i)}}\cU_{t,l}\ket{\varphi^{t-1,l-1}}|\geq \frac{\omega_{t-1}}{\sqrt{2}}$.}\end{equation}
			      (Denote $\ket{\varphi^{t-1,0}}:=\ket{\varphi^{t-1}}$.) Discuss by cases:
			      \begin{itemize}
			      \item If it exists, define $\cP^\prime_{t,l}:=\cU^\dagger_{t,l}P_{x_0^{(i)}||x_1^{(i)}}\cU_{t,l}$ and $\ket{\varphi^{t-1,l}}:=(I-\cP^\prime_{t,l})\ket{\varphi^{t-1,l-1}}$, as what we did in the proof of Lemma \ref{lem:4.4}.\par
			      Then continue to the next round, with $l$ increased by $1$. We note that in the next round the right hand side of (\ref{eq:6}) is the same.  
			      \item If we reach a state such that such a server-side operation $\cU_{t,l}$ does not exist, stop the iteration. Define $\cP^\prime_{t,l},\cP^\prime_{t,l+1},\cdots \cP^\prime_{t,L}$ to be all-zero. And define $\ket{\varphi^{t-1,L}}=\ket{\varphi^{t-1,l-1}}$.
			      \end{itemize}
			\end{itemize}      
			      Now we have completed the iterated construction. And we get $\cU_{t,1}\cdots \cU_{t,L}$ and correspondingly $\cP^\prime_{t,1}\cdots \cP^\prime_{t,L}$ and $\cP_{t,1}\cdots \cP_{t,L}$ (similar to the proof of Lemma \ref{lem:4.4}, we define $\cP_{t,l}=\cP_{t,l}^\prime(I-\cP_{t,l-1})\cdots (I-\cP_{t,1})$ ). If the construction above stops because a ``not exist'' case is reached, we already get a decomposition, and we can skip the argument below and go to the ``thus we can decompose ...'' in the end of this box. Otherwise, the iterated construction ends when all the $l\in [1,L]$ rounds have been completed, this means:
			      \begin{equation}\text{$\forall l\in [L], \exists i\in [L]$ such that $|P_{x_0^{(i)}||x_1^{(i)}}\cP_{t,l}\ket{\varphi^{t-1}}|\geq \frac{\omega_{t-1}}{\sqrt{2}}$.}\end{equation}
	If there is still some server-side operation $\cU^\prime$ with query number \\$|\cU^\prime|\leq 2^{\kappa-(t-1)\log_2 (6L)}/(6L)$ such that $|P_{x_0^{(i)}||x_1^{(i)}}\cU^\prime\ket{\varphi^{t-1,L}}|\geq \frac{\omega_{t-1}}{\sqrt{2}}$ holds for some $i\in [L]$, by the pigeonhole principle it's always possible to choose two operations in $\{\cP_1,\cdots \cP_L\}\cup\{\cU^\prime\}$ such that they correspond to the same pair of keys. Then by the same technique as the proof of Lemma \ref{lem:4.4} $\ket{\varphi^{t-1}}$ will not be $(2^{\kappa-(t-1)\log_2 6L}, \omega_{t-1})$-SC-secure for this pair of keys. Thus we get a contradiction and thus we prove that $\forall i\in [L],\ket{\varphi^{t-1,L}}$ is $(2^{\kappa-(t-1)\log_2 (6L)}/(6L), \frac{\omega_{t-1}}{\sqrt{2}})$-SC-secure for $K^{(i)}$.\par
		Thus we can decompose $\ket{\varphi^{t-1}}$ as $\sum_{l\leq L}\cP_{t,l}\ket{\varphi^{t-1}}+\ket{\varphi^t}$ where 
		\begin{itemize}
		\item $\forall l$, the query number $|\cP_{t,l}|\leq 2^{\kappa-(t-1)\log_2 6L}$.
		\item $\ket{\varphi^t}:=\ket{\varphi^{t-1,L}}$ is $(1, 2^{\kappa-(t-1)\log_2 6L})$-server-side-representable from $\ket{\varphi^{t-1}}$, and 
		\item $\forall i\in [L]$, $\ket{\varphi^t}$ is $(2^{\kappa-t\log_2 6L},\omega_{t-1}/\sqrt{2})$-SC-secure for $K^{(i)}$.
		\end{itemize}
	\end{mdframed}

	Thus we can repeat this process for $t\in [1,\kappa/(2\log(6L))]$ and complete the proof. Then we can define $\ket{\phi}:=\ket{\varphi^{\kappa/(2\log (6L))}}$. And the inequality on the norm of $\ket{\chi}:=\sum_t(\sum_{l\leq L}\cP_{t,l}\ket{\varphi^{t-1}})$ holds from the convergence of the sum of geometric sequence.
\end{proof}

\subsubsection{Linear decomposition technique}\label{sec:4.8.2}
Another technique that we will use is the \emph{linear decomposition technique}. This is also why we develop the lemmas in Section \ref{sec:4.3}.\par
For example, suppose the initial state of a subprotocol $\fPrtl$ is $\ket{\varphi}$, and $\fPrtl$ is some remote gadget preparation protocol with some weak security. And we want to know something about the post-execution state $\ket{\varphi^\prime}=\fPrtl\circ \ket{\varphi}$. However, sometimes it's hard or impossible to prove the initial state $\ket{\varphi}$ satisfies the conditions in the security statement of the protocol $\fPrtl$. One example is the security statement of $\fPrtl$ requires the initial state to be $(2^{\eta},2^{-\eta}|\ket{\varphi}|)$-SC-secure for $K$, while we only have $\ket{\varphi}$ is $(2^{\eta},C|\ket{\varphi}|)$-SC-secure for $K$ where $C$ might be inverse-polynomial, thus the security statement of $\fPrtl$ cannot be applied directly.\par
In this case, we can first apply the decomposition lemma (Lemma \ref{lem:4.4}) and do the decomposition: $\ket{\varphi}$, together with finite number of server-side ancilla qubits (which are all at state zero), into $\ket{\phi}+\ket{\chi}$, then we can study $\fPrtl\circ\ket{\varphi}$ through $\fPrtl\circ\ket{\phi}$ and $\fPrtl\circ\ket{\chi}$ separately. Usually we can directly apply the security property of $\fPrtl$ to understand the $\fPrtl\circ\ket{\phi}$ part (since from the decomposition we know $\ket{\phi}$ is $(2^{\Theta(\eta)},2^{-\Theta(\eta)}|\ket{\phi}|)$-SC-secure for $K$), and for the $\fPrtl\circ\ket{\chi}$ part, we can study it in some other ways, for example, we can simply bound the norm $|P_{pass}\fPrtl\ket{\chi}|\leq |\ket{\chi}|$ and add it back by triangle inequality.\par
This technique is used commonly in the sections later. For example, it is used in Section \ref{sec:6}, and its ``advanced'' form is discussed below and used in Section \ref{sec:10}.
\subsubsection{Multi-round linear decomposition technique}\label{sec:4.8.3}
The method in the previous subsubsection can be further generalized to a multi-round protocol. As an example, suppose a protocol $\fPrtl$ is in the following form:
\begin{mdframed}
		\textbf{Structure of $\fPrtl$:}\\
		For $i=1,\cdots \kappa$:\par
		\quad Run subprotocol $\fsubPrtl$.\end{mdframed}
Suppose $\fsubPrtl$ has weak security with security transform parameter \weakparams{\eta}{2^{-\eta}}{{\eta^\prime}}{1/3}, and the input state is $(2^\eta,2^{-\eta}|\ket{\varphi}|)$-SC-secure for some key $K$. Thus we can argue that after the first round the output state (denoted as $P_{pass}\fsubPrtl\ket{\varphi}$) is $(2^{\eta^\prime}, 1/3|\ket{\varphi}|)$-SC-secure for some key. Then we can decompose the state into $\ket{\phi^1}+\ket{\chi^1}$. If the final state of the whole protocol is denoted as $\ket{\varphi^\prime}$, we can write
\begin{align}
	P_{pass}\ket{\varphi^\prime} & :=P_{pass}\fPrtl\ket{\varphi}\label{eq:14}                                  \\
	                             & =P_{pass}\fPrtl_{>1}\fsubPrtl\ket{\varphi}\label{eq:15}                       \\
	                             & =P_{pass}\fPrtl_{>1}\ket{\phi^1}+P_{pass}\fPrtl_{>1}\ket{\chi^1}\label{eq:16}
\end{align}
Where $\fPrtl_{>1}$ is the protocol starting from the second round.\par
If $\eta^\prime$ is big enough, or the initial state satisfies some other conditions, we can prove either $\ket{\phi^1}$ or $\ket{\chi^1}$ still satisfies the conditions in the security statement of $\fsubPrtl$, which means we can apply the same argument on $P_{pass}\fPrtl_{>1}\ket{\chi^1}$. (The exact form of how this happens varies in different protocols and proofs.) This suggests that the same argument can be applied repeatedly and the proof will be ``induction-style''.\par
For example, if for some reason we can continue the same argument on $P_{pass}\fPrtl_{>1}\ket{\chi^1}$, continue this argument we can get
{\small\begin{align}
	P_{pass}\ket{\varphi^\prime} & =P_{pass}\fPrtl\ket{\varphi}                                                                                                                 \\
	                             & :=P_{pass}\fPrtl\ket{\chi^0}\text{(Here we denote $\ket{\chi^0}:=\ket{\varphi}$ to make the notation consistent)}                            \\
	                             & =P_{pass}\fPrtl_{>1}\ket{\phi^1}+P_{pass}\fPrtl_{>1}\ket{\chi^1}                                                                               \\
	                             & =\cdots                                                                                                                                    \\
	                             & =P_{pass}\fPrtl_{>1}\circ\ket{\phi^{1}}+P_{pass}\fPrtl_{>2}\circ\ket{\phi^{2}}+\cdots +P_{pass}\ket{\phi^{\kappa}}+P_{pass}\ket{\chi^{\kappa}}
\end{align}}
Note that since every time we get a new $\ket{\chi^t}$ we need to ensure these states have similar properties thus the same argument can be applied repeatedly. When we use this method, we will write down these properties explicitly.\par
Finally we can argue that each $P_{pass}\fPrtl_{>t}\circ\ket{\phi^{t}}$ satisfies some properties and the last term is exponentially small, (in each round when we apply the decomposition lemmas in Section \ref{sec:4.3} there is $|\ket{\chi^t}|\leq \frac{2.5}{3}|\ket{\chi^{t-1}}|$) and we can prove the properties of $P_{pass}\ket{\varphi^\prime}$ by combining them through triangle inequality.\par
This technique is used frequently in Section \ref{sec:10}.
\subsection{Other Lemmas}\label{sec:techovw}
In Appendix \ref{sec:basiclemmas} we give many other lemmas that we need for later usage. Here we give an overview of them.\par
\begin{enumerate}
\item In Section \ref{sec:4.4} we focus on the following question: how will extra lookup tables affect the SC/ANY-security/unpredictability of a state for some key(s)? Intuitively if a state is secure for some keys, if some extra ciphertexts under this keys is provided to the adversary, it should be ``indistinguishable'' to some random strings. Formalizing this informal intuition in our framework is a little bit technical, but it's possible to do. We will go through different cases and give several lemmas that will be useful in later sections.
\item In Section \ref{sec:4.5} we re-phrase the \emph{oneway-to-hiding} lemma and the \emph{collapsing property} in our framework, into a form that is convenient for later usage.
\item In Section \ref{sec:4.6} we give several lemmas about the blinded oracle, and its interplay with different security notions.
\item Finally in Section \ref{sec:4.7}	we give a lemma about indistinguishability of lookup tables under a state with some security.
\end{enumerate}
We note that later when we use these lemmas we will directly refer to the lemmas in Appendix \ref{sec:basiclemmas}. There is no circular proof here.
\section{Non-collapsing Basis Test}\label{sec:6}
Before we give our weakly secure remote gadget preparation protocol, in this section we will discuss a class of subprotocols called \emph{non-collapsing basis test}, which will be a component and used very frequently in the remaining sections.
\subsection{Non-collapsing Basis Test on a Single Pair of Keys}\label{sec:6.1}
\subsubsection{Problem setting}\label{sec:6.1.1}
Suppose in some subprotocol an honest server is supposed to hold the gadget corresponding to the keys $K=\{x_0,x_1\}$, which is the state $\ket{x_0}+\ket{x_1}$. But the server can cheat. So the client wants to verify that the server really holds such a state.\par
We note that our original problem, the blind quantum computation problem,  is only about ``blindness'', instead of the ``verification''. Informally, here ``blindness'' means the server cannot know the client's input; while ``verification'' means the server has to hold a specific state or messages, otherwise it will not pass the client's verification procedure. (This is only to informally distinguish ``blindness'' and ``verification'' and should not be considered a definition.) Our original problem is only about blindness; but during the construction of our protocols, to simplify the construction and security proofs, having some verifiability property in the middle can be very useful. The \emph{non-collapsing basis test} talks about the verifiability, and will be useful in the construction of the weakly secure remote gadget preparation protocol.\par
In this setting, the strongest form of verification requires the server to hold the state $\ket{x_0}+\ket{x_1}$, otherwise it will make the client reject. However, such a strong form of verification is hard. So we will weaken the problem as follows:\par
Problem setting for a non-collapsing basis test for key $K=\{x_0,x_1\}$:
\begin{itemize}
	\item In this protocol, the client wants to verify that, (or more formally, if the server makes the client accept with some probability, then conditioned on the client's acceptance in this protocol,) the server's state can be unitarily transformed into a state close to
	      \begin{equation}\label{eq:6.1.1}
	      \ket{x_0} \ket{\cdots}+\ket{x_1}\ket{\cdots}
	      \end{equation}
	      Or equivalently, a state that can be unitarily transformed into
	      \begin{equation}\label{eq:6.1.2}
		      \ket{x_0} \ket{\cdots}+\ket{x_1}\ket{\cdots}+\ket{\chi^\prime},\text{ where $|\ket{\chi^\prime}|$ is small.}\end{equation}
	\item What's more, as the ``non-collapsing'' suggests, it should be possible for the honest server (which holds the state $\ket{x_0}+\ket{x_1}$) to pass these tests with probability close to 1 without disturbing the state.
\end{itemize}

We note that if the client asks the server to make a measurement and report the result, (and accept if it is either $x_0$ or $x_1$,) the state will be destroyed and it's not possible to use it for further tasks. Thus we need to add the ``non-collapsing'' requirement.\par
And the description above is under the natural notation instead of the purified notation (where all the randomness are purified, see Section \ref{sec:2.1}). In the purified notation (\ref{eq:6.1.1})(\ref{eq:6.1.2}) are all entangled with the client-side description of $K$. Using the purified notation, what we want is, there exists some server-side efficient unitary operation $\cU$,
\begin{equation}\label{eq:17}
	|(I-P_{K})\cU P_{pass}\ket{\varphi^\prime}|,\text{ or (a different notation) }|(I-P_{span\{x_0, x_1\}})\cU P_{pass}\ket{\varphi^\prime}|	\text{, is small.}
\end{equation}
where $\ket{\varphi^\prime}$ is the state after the execution of the protocol, $P_K$, or $P_{span\{x_0, x_1\}}$, is the projection onto the space where a server-side register holds either $x_0$ or $x_1$, the keys in $K$.\par
We will design some protocols that try to address this question.\par
\paragraph{Note}The protocols given below do not really give us (\ref{eq:17}), but it satisfies (\ref{eq:17}) if the client additionally provides some specific auxiliary information. This is enough for later usage, since the extra auxiliary information can be handled via techniques in Section \ref{sec:4.2}.
\subsubsection{Protocol design, single round}
We first give a single-round protocol for a single pair of keys. To achieve this task, the client sends a lookup table with the same output keys for both input keys. 
\begin{mdframed}[style=figstyle]
\begin{prtl}\label{prtl:6.1}($\fBasisTest(K;\ell,\kappa_{out})$) is defined as follows, where $K=\{x_0,x_1\}$, $\ell$ is the padding length, $\kappa_{out}$ is the output key length:\par
The initial state in the honest setting is $(\ket{x_0}+\ket{x_1})\otimes\ket{\text{other part}}$:
	\begin{enumerate}
		\item  The client picks $r\leftarrow_r\{0,1\}^{\kappa_{out}}$ and computes and sends $$\fLT(\forall b, x_b\rightarrow r;\underbrace{ \ell}_{\substack{\text{padding} \\ \text{length}}},
\underbrace{ \kappa_{\text{out}}}_{\substack{\text{tag} \\ \text{length}}})$$ to the server.
		\item The honest server should implement the mapping by evaluating the lookup table:
		      \begin{equation}\ket{x_0}+\ket{x_1}\rightarrow (\ket{x_0}+\ket{x_1})\otimes\ket{r}\end{equation}
		      and send $r$ to the client.
		\item The client checks if $r$ is correct, if not, reject.
	\end{enumerate}

\end{prtl}\end{mdframed}
We can see in the honest setting, since the lookup table is $\fLT(x_0\rightarrow r, x_1\rightarrow r)$, and the server holds $\ket{x_0}+\ket{x_1}$, it can decrypt the lookup table in superposition and get $r$ with probability 1, without disturbing the gadget.\par
In the adversarial setting, intuitively, we want to argue that if the server's initial state satisfies some conditions, if the server can pass the test with high probability, it should hold the state that can be unitarily transformed to the state (\ref{eq:6.1.2}) (or equivalently, (\ref{eq:17})). However proving this fact is hard. To study its security, first, we will give an improvement to our protocol below; then, when we discuss their security properties in the next subsection, as we discussed in the end of Section \ref{sec:6.1.1}, its verifiability is formalized in a way that an extra auxiliary information is introduced.
\subsubsection{Protocol design, multi-round}
The first attack to consider is, the server can choose to only pass the test only with constant probability, for example, $99/100$, $2/3$, or $1/1000$. So there is a tradeoff between the adversary's ``passing probability'' that the client can stand and the verifiability provided by the protocol (which means, how close the server's state is to the correct state (\ref{eq:6.1.1})).  Below we will first revise Protocol \ref{prtl:6.1} to partially get rid of this tradeoff. We note that we do not mean the original protocol (Protocol \ref{prtl:6.1}) is ``broken''; it's just because the its property is not good enough for some of our later tasks.\par
Intuitively, if the client runs this test multiple times and requires the server to pass the test in all the rounds, this test will become more powerful. Thus we get the following test (non-collapsing basis test for single pair of keys, multiple rounds):
\begin{mdframed}[style=figstyle]
\begin{prtl}\label{prtl:6.2}
	($\fBasisTest(K;T,\ell,\kappa_{out})$), where $K=\{x_0,x_1\}$, $T$ is the number of test rounds, $\ell$ is the padding length, $\kappa_{out}$ is the output key length:\par
	The server's initial state in the honest setting is $(\ket{x_0}+\ket{x_1})\otimes\ket{\text{other part}}$.
	\begin{enumerate}
		\item For $t=1,\cdots T$:
		      \begin{enumerate}
			      \item  The client picks $r^t\leftarrow_r\{0,1\}^{\kappa_{out}}$ and sends $$\fLT(\forall b, x_b\rightarrow r^t;\underbrace{ \ell}_{\substack{\text{padding} \\ \text{length}}},
\underbrace{ \kappa_{\text{out}}}_{\substack{\text{tag} \\ \text{length}}})$$ to the server.
			      \item An honest server should implement the mapping
			            \begin{equation}\ket{x_0}+\ket{x_1}\rightarrow (\ket{x_0}+\ket{x_1})\otimes\ket{r^t}\end{equation}
			            and sends $r^t$ to the client.
			      \item The client checks if the response from the server is correct, if not, reject.
		      \end{enumerate}
	\end{enumerate}

\end{prtl}\end{mdframed}
The following lemma is intuitive. It means, for any adversary, it either can only pass the whole protocol with some probability, or in some of the iteration it has to pass the protocol with high probability:
\begin{lem}\label{lem:6.1}
	In the $\fBasisTest(K;T)$ protocol, (note that we omit some parameters that are not important here) suppose the initial state is described by the purified joint state\footnote{the randomness are purified by the environment, as discussed in Section \ref{sec:5.1.1}} $\ket{\varphi}$  and the post-execution state after the $t$-th round of iterations is $\ket{\varphi^t}$, (additionally define $\ket{\varphi^0}:=\ket{\varphi}$), at least one of the followings is true:
	\begin{itemize}
		\item $|P_{pass}\ket{\varphi^T}|\leq \frac{1}{2}|\ket{\varphi}|$.
		\item There exists a $t$, $1\leq t\leq T$, such that $|P_{pass}\ket{\varphi^t}|\geq (1-\frac{1}{T})|P_{pass}\ket{\varphi^{t-1}}|$
	\end{itemize}
\end{lem}
In the second case of the lemma above, intuitively, $\ket{\varphi^{t-1}}$ should be close to the form that we want ((\ref{eq:6.1.2}) or (\ref{eq:17})). In the next section, we will prove, this is true when some additional auxiliary information is provided to the server. (Note that in Section \ref{sec:4.2} we discussed the auxiliary-information technique. This is one of the examples that adding some auxiliary information can be useful. We further note that since adding this auxiliary information does not affect the SC-security of the state too much, it does not affect the usage of this protocol.)

\subsection{Security of the $\fBasisTest$ Protocol for Single Key Pair}\label{sec:6.2}
We give some protocols in the last section. Although it's hard to prove the state can be unitarily transformed by a server-side operation to (\ref{eq:6.1.2}), we can prove, when some auxiliary information is provided, it satisfies this property. Let's first give the security statement for the single-round basis test:
\subsubsection{Security statement for the single round test}
\begin{lem}\label{lem:6.2}
	The following statement is true for sufficiently large security parameter $\kappa$:\par
	Suppose the keys are denoted as $K=\{x_0,x_1\}$, the initial state is described by the purified joint state $\ket{\varphi}$, and the protocol considered is $$\fBasisTest(K;\underbrace{ \ell}_{\substack{\text{padding} \\ \text{length}}},
\underbrace{ \kappa_{\text{out}}}_{\substack{\text{output} \\ \text{length}}})$$, suppose $Tag(K)$ is in some fixed place of the read-only buffer, suppose the following conditions are satisfied, where $T$ is a positive integer:
	\begin{itemize}
		\item (Security of input state) $\ket{\varphi}$ is $(2^{\eta}, 2^{-\eta}|\ket{\varphi}|)$-SC-secure for $K$, $(\eta/3-150T^2)/24>\kappa$.
		\item (Well-behaveness of input state) $\ket{\varphi}\in \cWBS(D)$, $D\leq 2^\eta$.
		\item $20<T<2^{\sqrt{\kappa}}$
		\item (Sufficient padding length, output key length) $l>6D+4\eta$, $\kappa_{out}>l+\eta$
	\end{itemize}
	then the following conclusion holds:\par
	For all the adversaries $\fAdv$ with query number $ |\fAdv|\leq 2^{\kappa}$, denote the corresponding post-execution state as
	$$\ket{\varphi^\prime}=\fBasisTest_\fAdv(K;\ell,\kappa_{out})\circ \ket{\varphi}$$
	at least one of the followings is true:
	\begin{itemize}
		\item (Passing probability) $|P_{pass}\ket{\varphi^\prime}|\leq (1-\frac{1}{T})|\ket{\varphi}|$.
		\item (Test result) There exists a server side operation $\cU$ (whose form deterministically only depends on the code of $\fAdv$) with query number $|\cU|\leq |\fAdv|+20$ such that:
		      \begin{equation}\label{eq:26r}
			      |P^S_{span\{x_0, x_1\}}\cU  (\ket{\varphi}\odot \llbracket\fAuxInf\rrbracket)|\geq (1-\frac{4}{\sqrt{T}})|\ket{\varphi}|
		      \end{equation}
		      where $S$ is some server-side system, $\fAuxInf$ is defined as follows: The client samples a pair of different keys $K_{out}=\{r_0,r_1\}$ with key length $\kappa_{out}$, and computes and uses the reversible lookup table $\fRevLT(K\leftrightarrow K_{out}; \ell)$ as the $\llbracket\fAuxInf\rrbracket$.
	\end{itemize}
\end{lem}
Note that the conclusion in the second case implies $\ket{\varphi}$ can be written as follows:
\begin{equation}\cU(\ket{\varphi}\odot \llbracket\fAuxInf\rrbracket)=\ket{\psi_0}+\ket{\psi_1}+\ket{\chi^\prime}\end{equation}
where for some server-side system $S$, \begin{equation}\text{$P^S_{x_0}\ket{\psi_0}=\ket{\psi_0}$, $P^S_{x_1}\ket{\psi_1}=\ket{\psi_1}$, $P^{S}_K\ket{\chi^\prime}=0$, $|\ket{\chi^\prime}|\leq 3T^{-1/4}|\ket{\varphi}|$}\end{equation}. This form will be useful in the later sections.\par
\paragraph{How to understand this lemma}First we note the conclusion is what we want in the discussion around equation (\ref{eq:6.1.1}). And we also note that the form of the theorem is a little bit similar to the weak security definition (Definition \ref{def:3.12}) of the remote state preparation, even if it is a different problem. The conditions say: if the SC-security is good enough, and if the state is not ill-behaved, if the number of ``test rounds'' (see below) is big enough (but not extraordinarily big), if the pad length and output key length are long enough (intuitively the longer they are the more difficult the server's attack will be), we get the conclusions.\par
Finally, we point out that in this lemma --- the security for the single-round protocol, $T$ can be chosen arbitrarily, as long as the conditions are all satisfied. In the multi-round protocol, $T$ will become the ``test round''. In this lemma, different $T$ has different influences on the two cases in the conclusion (below ``one of the followings is true''). When we make $T$ bigger, the first case will become a weaker statement but the second case will become a stronger statement.
\paragraph{Intuition behind the lemma}The reader may wonder how the introduction of $\fRevLT$ affects the state. The trick inside it is similar to the introduction of ``trapdoor-claw-free-function and trapdoor-injection-function pair'' in Mahadev's protocol\cite{MahadevVerification}. We note the lookup table in the protocol is a ``2-to-1'' table, while the forward part of $\fRevLT$ is a ``2-to-2'' table, which is also indistinguishable to the original table --- and it enables the server to decrypt coherently, with a similar output norm, without really making the measurement (this step requires some technical proof). Finally the backward part allows the server to map the output back to the input and complete the test.\par
Then based on this lemma, we get the security for the multiple-round basis test for single pair of keys.
\subsubsection{Security statement and proof for multi-round protocol}
Below we give the security statement and proof for Protocol \ref{prtl:6.2}.
\begin{lem}\label{lem:6.3}
	The following statement is true for sufficiently large security parameter $\kappa$:\par
	Consider a pair of keys denoted as $K=\{x_0,x_1\}$ , initial state described by a joint purified state $\ket{\varphi}$, and protocol $$\fBasisTest(K;T,\underbrace{ \ell}_{\substack{\text{padding} \\ \text{length}}},
\underbrace{ \kappa_{\text{out}}}_{\substack{\text{output} \\ \text{length}}})$$,   suppose $Tag(K)$ is stored in some fixed place of the read-only buffer, and 
suppose the following conditions are satisfied:
	\begin{itemize}
		\item (Security of the inputs) $\ket{\varphi}$ is $(2^{\eta}, 2^{-\eta}|\ket{\varphi}|)$-SC-secure for $K$, $(\eta/19-150T^2)/24>\kappa$.
		\item (Well-behaveness of the inputs) $\ket{\varphi}\in \cWBS(D)$, $D\leq 2^{\eta/7}$.
		\item $20<T<2^{\sqrt{\kappa}}$
		\item (Sufficient padding length and output key length) $l>6D+6\eta$, $\kappa_{out}>l+\eta$
	\end{itemize}
	then the following conclusion holds:\par
	For all the adversaries $\fAdv$ with query number $ |\fAdv|\leq 2^{\kappa}$, denote the corresponding post-execution state as
	$$\ket{\varphi^\prime}=\fBasisTest_\fAdv(K;T,\ell,\kappa_{out})\circ\ket{\varphi}$$
	and denote the post-execution state after the $t$-th round as $\ket{\varphi^t}$, $\ket{\varphi^0}:=\ket{\varphi}$, at least one of the followings are true:
	\begin{itemize}
		\item (Passing probability) $|P_{pass}\ket{\varphi^\prime}|\leq \frac{1}{2}|\ket{\varphi}|$.
		\item (Test result) There exists an integer $t\in [0, T)$, a server side operation $\cU$ (whose form deterministically only depends on the code of $\fAdv$) with query number $|\cU|\leq |\fAdv|+20$ such that
		      \begin{equation}\label{eq:44}
			      |P_{span\{x_0, x_1\}}\cU (P_{pass}\ket{\varphi^{t}}\odot \llbracket\fAuxInf\rrbracket)|\geq (1-\frac{4}{\sqrt{T}})|P_{pass}\ket{\varphi^t}|
		      \end{equation}
		      where $\fAuxInf$ is defined as follows: The client samples $K_{out}=\{r_0,r_1\}$ with key length $\kappa_{out}$ such that $r_0\neq r_1$, and computes and adds the reversible lookup table $\fRevLT(K\leftrightarrow K_{out}; \ell)$ into $\fAuxInf$.
	\end{itemize}
\end{lem}
This lemma is very similar to Lemma \ref{lem:6.2}. (And we refer to the ``How to understand this lemma'' under that lemma for an explanation.) One difference is here $T$ is a parameter of the protocol rather than a parameter that can be chosen arbitrarily when we apply the lemma. Again, the second case in the conclusion implies $\cU(P_{pass}\ket{\varphi^t}\odot \llbracket\fAuxInf\rrbracket)$ can be written as
$$\cU(P_{pass}\ket{\varphi^t}\odot \llbracket\fAuxInf\rrbracket)=\ket{\psi_0}+\ket{\psi_1}+\ket{\chi^\prime}$$
where \begin{equation}\text{$P_{x_b}\ket{\psi_b}=\ket{\psi_b}$, $P_K\ket{\chi^\prime}=0$, $|\ket{\chi^\prime}|\leq 3T^{-1/4}|P_{pass}\ket{\varphi^t}|$}\end{equation}
The proof is given below. It's basically a combination of Lemma \ref{lem:6.1} and Lemma \ref{lem:6.2}.
\begin{proof}[Proof of Lemma \ref{lem:6.3}]
	Apply Lemma \ref{lem:6.1} and suppose the first case is not true:
	\begin{equation}\label{eq:45rr}|P_{pass}\ket{\varphi^{T}}|>\frac{1}{2}|\ket{\varphi}|\end{equation}
	(Otherwise the statement already holds.) So there exists an $1\leq t\leq T$ such that
	$$|P_{pass}\ket{\varphi^t}|\geq (1-\frac{1}{T})|P_{pass}\ket{\varphi^{t-1}}|$$
	Each single round of $\fBasisTest(K;T)$ is a $\fBasisTest(K)$ protocol. (We omit some parameters.) Apply Lemma \ref{lem:6.2} on initial state $P_{pass}\ket{\varphi^{t-1}}$ we get the conclusion we want when $t$ (in (\ref{eq:44})) chosen to be $t-1$ here.\par
	Below is a checklist for the conditions of applying Lemma \ref{lem:6.2} on $P_{pass}\ket{\varphi^{t-1}}$ in the last step.\\
	\begin{mdframed}
			\textbf{Checklist for applying Lemma \ref{lem:6.2} on $P_{pass}\ket{\varphi^{t-1}}$}
			\begin{itemize}
				\item $P_{pass}\ket{\varphi^{t-1}}=P_{pass}\fBasisTest_{\fAdv_{1\sim t-1}}(K;t-1)\circ\ket{\varphi}$. First by Lemma \ref{lem:4.8} \begin{equation}\text{$\ket{\varphi}\odot \llbracket\fBasisTest(K,t-1)\rrbracket$ is $(2^{\eta/6},2^{-\eta/6}|\ket{\varphi}|)$-SC-secure for $K$}\end{equation}, thus by Lemma \ref{lem:basic} \begin{equation}
					      \text{$P_{pass}\ket{\varphi^{t-1}}$ is $(2^{\eta/6}-2^\kappa,2^{-\eta/6}|\ket{\varphi}|)$-SC-secure for $K$.}
				      \end{equation}
				      Thus by (\ref{eq:45rr}) $P_{pass}\ket{\varphi^{t-1}}$ is $(2^{\eta/6-1},2^{-\eta/6+1}|P_{pass}\ket{\varphi^{t-1}}|)$-SC-secure for $K$.
				\item The number of RO queries by both parties during $\fBasisTest(K;t-1)$ is $\leq |\fAdv|+O(t)$ thus $P_{pass}\ket{\varphi^{t-1}}$ is $(2^{D},2^{D}+|\fAdv|+O(T))$-representable from $\ket{\mathfrak{init}}$. And $|P_{pass}\ket{\varphi^{t-1}}|>\frac{1}{2}|\ket{\varphi}|$. Thus when we use $|P_{pass}\ket{\varphi^{t-1}}|$ instead of $|\ket{\varphi}|$ to check the inequalities in the 4th condition in Lemma \ref{lem:6.2}, add an additional ``$\eta$'' in the inequality on $l$ is enough.
			\end{itemize}
			We can see when we apply Lemma \ref{lem:6.2} we need to choose $\eta$ (in Lemma \ref{lem:6.2}) to be $\eta/6-1$.
		\end{mdframed}
\end{proof}
\subsection{$\fBasisTest$ for Two Pairs of Keys}\label{sec:6.3}
\subsubsection{Problem setting and protocol}
In the previous sections, we are considering the non-collapsing basis test for a single pair of keys. But in the following sections, we will need a non-collapsing basis test protocol for two pairs of keys simultaneously. In other words, for keys $K^{(1)}=\{x_b^{(1)}\}_{b\in \{0,1\}}$, $K^{(3)}=\{x_b^{(3)}\}_{b\in \{0,1\}}$, we want to verify the adversary's state is close to a state that is unitarily isomorphic to the following state:
\begin{equation}\label{eq:32}\sum_{b_1\in \{0,1\}}\ket{x^{(1)}_{b_1}}\sum_{b_2\in \{0,1\}}\ket{x^{(3)}_{b_2}}\ket{\cdots}\end{equation}
Or, equivalently, a state that is unitarily isomorphic to the state
\begin{equation}\label{eq:4}\sum_{b_1\in \{0,1\}}\ket{x^{(1)}_{b_1}}(\sum_{b_2\in \{0,1\}}\ket{x^{(3)}_{b_2}}\ket{\cdots}+\ket{\chi_{b_1}^\prime})+\ket{\chi^\prime}\end{equation}
where $\ket{\chi_{b_1}}$, $\ket{\chi^\prime}$ are small.
\paragraph{Note}The reason that we choose $1,3$ as the superscripts is to make it consistent with its usage in Section \ref{sec:7}.\\

Note that similar to the previous case the description above is under the \emph{natural notation} instead of the \emph{purified notation}. If we purify all the randomness, the final state form we want can be described as
\begin{equation}
	\ket{\psi_0}+\ket{\psi_1}+\ket{\chi^\prime}
\end{equation}
\begin{equation}
	\forall b_1\in \{0,1\}, \ket{\psi_{b_1}}=\ket{\psi_{b_10}}+\ket{\psi_{b_11}}+\ket{\chi_{b_1}^\prime}
\end{equation}
where for some server-side system $S_1$, $$\text{$P^{S_1}_{x^{(1)}_0}\ket{\psi_0}=\ket{\psi_0}$, $P^{S_1}_{x^{(1)}_1}\ket{\psi_1}=\ket{\psi_1}$, $P^{S_1}_{K^{(1)}}\ket{\chi^\prime}=0$, $|\ket{\chi^\prime}|$ is small}$$
, and furthermore for some server-side system $S_3$, $\forall b_1\in \{0,1\}$,
$$\text{$P^{S_1}_{x^{(1)}_{b_1}}P^{S_3}_{x^{(3)}_0}\ket{\psi_{b_10}}=\ket{\psi_{b_10}}$, $P^{S_1}_{x^{(1)}_{b_1}}P^{S_3}_{x^{(3)}_1}\ket{\psi_{b_11}}=\ket{\psi_{b_11}}$, $P^{S_3}_{K^{(3)}}\ket{\chi_{b_1}^\prime}=0$, $|\ket{\chi_{b_1}^\prime}|$ is small.}$$
Let's first give the $\fBasisTest$ protocol for the two key pairs setting:
\begin{mdframed}[style=figstyle]
\begin{prtl}\label{prtl:r9}
	$\fBasisTest(K^{(1)},K^{(3)};T,\ell,\kappa_{out})$ where $T$ is the round of tests (for $K^{(3)}$), $\ell$ is the padding length, $\kappa_{out}$ is the output key length:
	\begin{enumerate}
		\item The client and the server run $\fBasisTest(K^{(3)};T,\ell,\kappa_{out})$;
		\item The client and the server run $\fBasisTest(K^{(1)};\ell,\kappa_{out})$;
	\end{enumerate}
\end{prtl}\end{mdframed}
The first step is Protocol \ref{prtl:6.2} on $K^{(3)}$ where the number of rounds is $T$, and the second step is Protocol \ref{prtl:6.1} on $K^{(1)}$, which is only a single round test. Although in the original problem $K^{(1)}$ and $K^{(3)}$ are in the equivalent places, in the protocol and the security statement we will handle these two pairs differently. The reason is, in the next section, when we use this protocol, $K^{(1)}$ and $K^{(3)}$ actually play different roles, so here we also need to handle them differently.\par
For the security statement, we need to consider the case where the state is $(2^{\eta_2},C|\ket{\varphi}|)$-SC-secure for $K^{(3)}$ and $C$ may be not exponentially small. This will be needed when we use this protocol in Section \ref{sec:7}.
\subsubsection{Security statement}
\begin{lem}\label{lem:6.4}
	The following statement is true for sufficiently big security parameter $\kappa$:\par
	Consider keys denoted as $K^{(1)}=\{x_b^{(1)}\}_{b\in \{0,1\}},K^{(3)}=\{x_b^{(3)}\}_{b\in \{0,1\}}$, the initial state described by the purified joint state $\ket{\varphi}$, and the protocol $$\fBasisTest(K^{(1)}, K^{(3)};\underbrace{T}_{\substack{\text{test}\\\text{rounds}}},\underbrace{ \ell}_{\substack{\text{padding} \\ \text{length}}},
\underbrace{ \kappa_{\text{out}}}_{\substack{\text{output} \\ \text{length}}})$$, suppose $Tag(K^{(1)},K^{(3)})$ is already stored in some fixed place of the read-only buffer, and 
if the following conditions are satisfied:
	\begin{itemize}
		\item (Security of the inputs) $\ket{\varphi}$ is $(2^{\eta},2^{-\eta}|\ket{\varphi}|)$-SC-secure for $K^{(1)}$ given $K^{(3)}$, $(\eta/19-150T^2)/24>\kappa$.
		\item (Security of the inputs) $\ket{\varphi}$ is $(2^{\eta},C|\ket{\varphi}|)$-SC-secure for $K^{(3)}$ given $K^{(1)}$, $2^{-\sqrt{\kappa}/10}\leq C\leq 1/9$.
		\item (Well-behaveness of the inputs) $\ket{\varphi}\in \cWBS(D)$, $D\leq 2^{\eta/100}$.
		\item (Suitable number of test rounds) $2/C^4>T>1/C^4$
		\item (Sufficient padding length and output key length) $l>6D+6\eta$, $\kappa_{out}>l+4\eta$
	\end{itemize}
	then the following conclusion holds:\par
	For all the adversaries $\fAdv$ with query number $ |\fAdv|\leq 2^\kappa$, denote the corresponding output state as
	$$\ket{\varphi^\prime}=\fBasisTest_\fAdv(K^{(1)}, K^{(3)};T,\ell,\kappa_{out})\circ\ket{\varphi}$$
	at least one of the following two is satisfied:
	\begin{enumerate}
		\item $|P_{pass}\ket{\varphi^\prime}|\leq (1-C^{12})|\ket{\varphi}|$
		\item There exists a server-side operation $\cU$ (whose code depends deterministically on the code of $\fAdv$) with query number $|\cU|\leq 2^{\kappa+3}$ such that
		      \begin{equation}\label{eq:r36}\cU (P_{pass}\ket{\varphi^\prime}\odot \llbracket\fAuxInf\rrbracket)=\ket{\psi_0}+\ket{\psi_1}+\ket{\chi^\prime}\end{equation}
		      \begin{equation}\label{eq:r38}
			      \forall b_1\in \{0,1\},\ket{\psi_{b_1}}=\ket{\psi_{b_10}}+\ket{\psi_{b_11}}+\ket{\chi_{b_1}^\prime}
		      \end{equation}
		      where for some server-side system $S_1$,
		      \begin{equation}\label{eq:55r}
			      \text{$P^{S_1}_{x^{(1)}_0}\ket{\psi_0}=\ket{\psi_0}$, $P^{S_1}_{x^{(1)}_1}\ket{\psi_1}=\ket{\psi_1}$, $P^{S_1}_{K^{(1)}}\ket{\chi^\prime}=0$, $|\ket{\chi^\prime}|\leq \frac{2}{3}C^2|\ket{\varphi}|$}\end{equation}
		      , and furthermore for some server-side system $S_3$, $\forall b_1\in \{0,1\}$,
		      {\small\begin{equation}\label{eq:61rr}\text{$P^{S_1}_{x^{(1)}_{b_1}}P^{S_3}_{x^{(3)}_0}\ket{\psi_{b_10}}=\ket{\psi_{b_10}}$, $P^{S_1}_{x^{(1)}_{b_1}}P^{S_3}_{x^{(3)}_1}\ket{\psi_{b_11}}=\ket{\psi_{b_11}}$, $P^{S_3}_{K^{(3)}}\ket{\chi_{b_1}^\prime}=0$, $|\ket{\chi_{b_1}}|\leq 8C|\ket{\varphi}|$}\end{equation}}
		      , where $\llbracket\fAuxInf\rrbracket$ in (\ref{eq:r36}) is defined as the concatenation of the followings:
		      \begin{itemize}
			      \item $\llbracket\fAuxInf_1\rrbracket$: the client samples $\{r_b^{(3)}\}_{b\in\{0,1\}}$ differently with key length $\kappa_{out}$, and prepares the reversible lookup table $\fRevLT(K^{(3)}\leftrightarrow \{r_0^{(3)},r_1^{(3)}\};\ell)$.
			      \item $\llbracket\fAuxInf_2\rrbracket$: the client samples $\{r_b^{(1)}\}_{b\in\{0,1\}}$ differently with key length $\kappa_{out}$, and prepares the reversible lookup table $\fRevLT(K^{(1)}\leftrightarrow \{r^{(1)}_0,r_1^{(1)}\};\ell)$.
		      \end{itemize}
	\end{enumerate}
\end{lem}
\paragraph{How to understand this lemma}The ``natural notation'' version of \\(\ref{eq:r36})(\ref{eq:r38})(\ref{eq:55r})(\ref{eq:61rr}) is in (\ref{eq:4}), which is more intuitive. And we again refer to ``How to understand this lemma'' under Lemma \ref{lem:6.2}. The lemma says: if the security for the keys is good enough (note that the adversary knows the global tags of the keys), the state is not ill-behaved, the test round is suitable (it could be much larger, but it's not needed), the pad length and output key length are enough, we get what we want: the server either has some non-negligible failing probability, or the final state has the form we want.\par
The proof is given in Appendix \ref{sec:AB}. The proof makes use of the lemmas in the single-key-pair case. Here there are two pairs of keys, we need to apply the previous lemmas (\ref{lem:6.2} and \ref{lem:6.3}) twice, and analyze the form of the state carefully.


\cleardoublepage
\chapter{Construction of Weakly Secure Gadget Increasing Protocol}\label{cht:6}
In this chapter we will construct the first weakly-secure gadget-increasing protocol. In the end of this chapter we will give a protocol (Protocol \ref{prtl:10}) that is both gadget-increasing and weakly-secure, which completes the first step of Outline \ref{ppl:1}.
\section{Remote Gadget Preparation With Weak Security}\label{sec:7}
In this section we give the protocol that achieve remote gadget preparation with weak security. The goal of this section is to construct the first remote gadget preparation protocols ($\fGdgPrep^{basic}$) with weak security on a specific class of input states.\par

\subsection{A Review of the Protocol Overview, and the Formalization of Underlying Encoding}
In Section \ref{sec:1.4.2} we have given an overview of construction of this protocol. Now we repeat its key steps and explain more details.\par
As we said in the introduction, our goal is to generate 2 gadgets from 1 input gadget. And we choose to generate the bitwise-permuted output gadgets as an intermediate step. Then we introduce a helper gadget, and consider an interactive protocol as follows:
{\small\begin{align}
	&(\ket{x_0^{\text{helper}}}+\ket{x_1^{\text{helper}}})\otimes (\ket{x_0}+\ket{x_1})\label{eq:dese1}\\
	\rightarrow & (\ket{x_0^{\text{helper}}}+\ket{x_1^{\text{helper}}})\otimes perm((\ket{y_0}+\ket{y_1})\otimes (\ket{y_0^\prime}+\ket{y_1^\prime}))\label{eq:dese2}\\ 
	\text{(Test on $\ket{x_0^{\text{helper}}}+\ket{x_1^{\text{helper}}}$)}\rightarrow &perm((\ket{y_0}+\ket{y_1})\otimes (\ket{y_0^\prime}+\ket{y_1^\prime}))\\
	\text{(Reveals $perm$ if test passes)}\rightarrow &(\ket{y_0}+\ket{y_1})\otimes (\ket{y_0^\prime}+\ket{y_1^\prime})\label{eq:dese4}
\end{align}}
where the first step is achieved using \emph{reversible garbled table}, or \emph{reversible lookup table}:
\begin{align}&(\ket{x_0^{\text{helper}}}+\ket{x_1^{\text{helper}}})\otimes \ket{\{x_0^{(2)},x_1^{(2)}\}}\otimes(\ket{x_0^{(3)}}+\ket{x_1^{(3)}})\label{eq:des5}\\
\xrightarrow{\substack{\text{table encoding }\\  x^{\text{help}}x^{(2)}x^{(3)}\leftrightarrow x^{\text{help}}perm(y^{(2)}y^{(3)})}}&\footnotesize (\ket{x_0^{\text{helper}}}+\ket{x_1^{\text{helper}}})\otimes \underbrace{perm((\ket{y_0^{(2)}}+\ket{y_1^{(2)}})\otimes(\ket{y_0^{(3)}}+\ket{y_1^{(3)}}))}_{\text{reversibly encoded part}}\label{eq:des51}\end{align}
We refer to Definition \ref{defn:2.15} for an intuitive meaning of the ``$\leftrightarrow$'' notation in a look-up table. And 
we can change the notation in (\ref{eq:dese1}) to (\ref{eq:dese4}) to match (\ref{eq:des5})(\ref{eq:des51}) as follows:
\begin{align}
	&(\ket{x_0^{\text{helper}}}+\ket{x_1^{\text{helper}}})\otimes \ket{\{x_0^{(2)},x_1^{(2)}\}}\otimes (\ket{x^{(3)}_0}+\ket{x^{(3)}_1})\label{eq:rgtm}\\
	\rightarrow & (\ket{x_0^{\text{helper}}}+\ket{x_1^{\text{helper}}})\otimes \nonumber\\&\qquad perm((\ket{y_0^{(2)}}+\ket{y_1^{(2)}})\otimes (\ket{y_0^{(3)}}+\ket{y_1^{(3)}}))\label{eq:rgtm2}\\ 
	\text{\footnotesize (Test on $\ket{x_0^{\text{helper}}}+\ket{x_1^{\text{helper}}}$)}\rightarrow &perm((\ket{y_0^{(2)}}+\ket{y_1^{(2)}})\otimes (\ket{y_0^{(3)}}+\ket{y_1^{(3)}}))\\
	\text{\footnotesize (Reveals $perm$ if passes)}\rightarrow &(\ket{y_0^{(2)}}+\ket{y_1^{(2)}})\otimes (\ket{y_0^{(3)}}+\ket{y_1^{(3)}})
\end{align}
As what we said in the introduction, we need to design the underlying mapping in this reversible encoding carefully, for a reason that will be clear in Section \ref{sec:7.2} (we formalize the \emph{unpredictability restriction} property in Section \ref{sec:fsspht} and give an intuitive explanation in Section \ref{sec:phtvp}). In more details, for the $x_0^{\text{help}}$ branch we implement the mapping with the \emph{identity-style mapping} and for $x_1^{\text{help}}$ branch we implement the mapping with the \emph{$\fCN$-style mapping}. In the next subsubsection we first formalize our special reversible encoding of this mapping.
\subsubsection{Formalization of the reversible encoding}

The lookup table used in (\ref{eq:rgtm})(\ref{eq:rgtm2}) is defined as follows.\footnote{The previous versions put $K^{\text{help}}$ into the reversibility part. In this version we move it out of the reversibility part.}
\begin{defn}\label{def:rrlt}$\fRobustRLT(K_{\text{help}},K_{in}\leftrightarrow K_{out}, perm; \ell)$, where \begin{itemize}\item $K_{\text{help}}=\{x_b^{\text{help}}\}_{b\in \{0,1\}}$, $K_{in}=\{x_b^{(2)},x_b^{(3)}\}_{b\in \{0,1\}}$,$K_{out}=\{y_b^{(2)},y_b^{(3)}\}_{b\in \{0,1\}}$, and the keys with the same symbol and superscript have the same length; $y_b^{(2)}$ and $y_b^{(3)}$ have the same length. \item $perm$ is a bit-wise permutation on the strings of length $2\kappa_{out}$, $\kappa_{out}$ is the key length of $y_b^{(2)}$. \item $\ell$ is the padding length.\end{itemize}

	is defined as follows.\par
	The forward table part is defined as
	$$\fLT(x_{b_1}^{\text{help}}x_{b_2}^{(2)}x_{b_3}^{(3)}\rightarrow perm(y^{(2)}_{b_2}||y_{b_3+b_1b_2}^{(3)});\underbrace{ \ell}_{\substack{\text{padding} \\ \text{length}}},\underbrace{ \ell}_{\substack{\text{tag} \\ \text{length}}})$$
	The backward table is defined as
		$$\fLT(x_{b_1}^{\text{help}}perm(y^{(2)}_{b_2}||y_{b_3+b_1b_2}^{(3)})\rightarrow x_{b_2}^{(2)}x_{b_3}^{(3)};\underbrace{ \ell}_{\substack{\text{padding} \\ \text{length}}},\underbrace{ \ell}_{\substack{\text{tag} \\ \text{length}}})$$
	\end{defn}
	An intuitively equivalent form is as follows, which shows the 2-branch table structure clearly. One difference is on the encryption structure: Definition \ref{def:rrlt} uses key-concatenation encryption, while Definition \ref{def:rrlta} uses cascading encryption. For intuition description we often use the definition below, but the formal definition uses Definition \ref{def:rrlt} since the encryption structure is easier to deal with in our proof.
	\begin{defn}[$\fRobustRLT$, intuitively equivalent form]\label{def:rrlta}
	First consider the two reversible encoding between $K_{in}^{(2,3)}$ and $K_{out}$:
	{\small$$\text{(Identity-style) }\fRevLT_{b_1=0}: \fRevLT(\forall b_2,b_3\in \{0,1\}^2: x^{(2)}_{b_2}||x_{b_3}^{(3)}\leftrightarrow perm(y^{(2)}_{b_2}||y_{b_3}^{(3)});\underbrace{ \ell}_{\substack{\text{padding} \\ \text{length}}})$$
	$$\text{(CNOT-style) }\fRevLT_{b_1=1}:\fRevLT(\forall b_2,b_3\in \{0,1\}^2: x^{(2)}_{b_2}||x_{b_3}^{(3)}\leftrightarrow perm(y^{(2)}_{b_2}||y_{b_2+b_3}^{(3)});\underbrace{ \ell}_{\substack{\text{padding} \\ \text{length}}})$$}
	Then the $\fRobustRLT$ is defined as \begin{equation}\label{eq:rrlt}\fEn_{x_0^{\text{help}}}(\fRevLT_{b_1=0};\underbrace{ \ell}_{\substack{\text{padding} \\ \text{length}}},\underbrace{ \ell}_{\substack{\text{tag} \\ \text{length}}})||\fEn_{x_1^{\text{help}}}(\fRevLT_{b_1=1};\underbrace{ \ell}_{\substack{\text{padding} \\ \text{length}}},\underbrace{ \ell}_{\substack{\text{tag} \\ \text{length}}})\end{equation}
\end{defn}
This construction implements (\ref{eq:des5})$\rightarrow$(\ref{eq:des51}).\par
Now we turn to formalize the padded Hadamard test, another ingredient in our construction.
\subsection{The padded Hadamard Test}\label{sec:712}
Another key tool that we need is an updated version of the Hadamard test in the previous Mahadev-based constructions\cite{Mahadev2017ClassicalHE,BCMVV}. The test comes from the following formula:
$$\fH^{\otimes n}(\ket{x_0}+\ket{x_1})\propto\sum_{d:d\cdot x_0=d\cdot x_1}\ket{d}$$
In this test the client will ask for a non-zero $d$ such that $d\cdot (x_0\oplus x_1)=0$.\par
In these protocols, the standard basis measurement is used to test the form of the server's state, and the Hadamard basis measurement is used to collapse a state of the form $\ket{0}\ket{x_0}+\ket{1}\ket{x_1}$ to a single qubit. In some sense, if we want to use it for ``controlling'' the form of the state, the Hadamard test seems not that powerful. In our protocol, what we need is a protocol called \emph{padded Hadamard test}. It's a revised version of the Hadamard test, and this revision allows us to view the test from a different viewpoint (and allow the corresponding proofs work). We will see, the ability of passing the padded Hadamard test with high probability gives a very strong control on the state of the adversary.\par
The protocol is as follows:
\begin{mdframed}[style=figstyle]
\begin{defn}[Padded Hadamard test]\label{def:pht}
	The padded Hadamard test $$\fPadHadamard(K;\underbrace{ \ell}_{\substack{\text{padding} \\ \text{length}}},
\underbrace{ \kappa_{\text{out}}}_{\substack{\text{output} \\ \text{length}}})$$ on $K=\{x_0,x_1\}$ is defined as follows:
	\begin{enumerate}
		\item The client samples $pad\leftarrow_r\{0,1\}^l$ and sends $R$ to the server.
		\item The server returns $d$ such that $d\cdot (x_0||H(pad||x_0))=d\cdot (x_1||H(pad||x_1))$ where $H(pad||x_b)$ has length $\kappa_{out}$, and $d$ is not all zero on the last $\kappa_{out}$ bits. The client checks the server's response.
	\end{enumerate}
	The honest server can pass this test by making Hadamard measurement on $\ket{x_0}\ket{H(pad||x_0)}+\ket{x_1}\ket{H(pad||x_1)}$.
\end{defn}\end{mdframed}

 We provide some informal discussion of the unpredictability restriction and the coherency restriction in Chapter \ref{cht:4}. We give some additional notes for the coherency restriction. For the coherency restriction, we can even consider the case that some extra auxiliary information is provided, where this auxiliary information can be chosen arbitrarily, which allows us to apply the lemma multiple times on different auxiliary information, even if the post-execution state itself is fixed. This tells us the padded Hadamard test is a powerful tool to test and control the adversary's state.\par
Certainly, formalizing the security and giving a security proof is still tricky. For example, the padded Hadamard test protocol does not guarantee that the server throws away the keys completely: the server can cheat with some probability. But we can see this subprotocol does provide some level of security, which is sufficient for our purpose.\par
A formal treatment is below.
\subsubsection{Formal security statements of the padded Hadamard test}\label{sec:fsspht}

In the \emph{unpredictability restriction} of the padded Hadamard test, we will relate the passing probability with the ANY-security of the post-execution state. 
\begin{lem}[unpredictability restriction of the Padded Hadamard test]\label{lem:7.3}
	The following statement is true for sufficiently big security parameter $\kappa$:\par
	Suppose the initial state is described by the purified joint state $\ket{\varphi}$. Suppose:
	\begin{itemize}
		\item (Security of the state) $\ket{\varphi}$  is $(2^\eta,2^{-\eta}|\ket{\varphi}|)$-SC-secure for $K$. $\eta>10\kappa$.
		\item (Well-behaveness of the state) $\ket{\varphi}\in \cWBS(D)$, $D\leq 2^\eta$.
		\item (Sufficient padding length and output length) $l>6D+2\eta$, $\kappa_{out}>l+\eta$.
	\end{itemize}
	then the following conclusion is true for any $C>2^{-\kappa/10}$:\par
	For any adversary $\fAdv$ with query number $|\fAdv|\leq 2^\kappa$, consider the post execution state, which is
	$$\ket{\varphi^\prime}=\fPadHadamard_\fAdv(K;\underbrace{ \ell}_{\substack{\text{padding} \\ \text{length}}},
\underbrace{ \kappa_{\text{out}}}_{\substack{\text{output} \\ \text{length}}})\circ\ket{\varphi}$$
	one of the following two is true:
	\begin{itemize}
		\item 	$|P_{pass}\ket{\varphi^\prime}|\leq (1-C^2)|\ket{\varphi}|$
		\item $P_{pass}\ket{\varphi^\prime}$ is $(2^{\eta/6},2C|\ket{\varphi}|)$-ANY-secure for $K$.
	\end{itemize}
\end{lem}
\paragraph{How to understand this lemma} We refer to Definition \ref{def:3.12}, which is for a different problem, but has a similar structure. The lemma says: if the state has sufficient security, and is not ill-behaved, and if the pad length and output key length is enough, the post-execution state has some properties that we want.\par
And we get the following corollary by choosing an appropriate $C$, which will be used in some other section (on the $\fSecurityRefreshing$ layer):
\begin{cor}\label{cor:7.3}
	The following statement is true for sufficiently large security parameter $\kappa$:\par
	Consider the initial state, described by the purified joint state $\ket{\varphi}$, satisfies the conditions listed in Lemma \ref{lem:7.3}.\par
	Then for any adversary $\fAdv$ with query number $|\fAdv|\leq 2^\kappa$, consider the post execution state, which is
	$$\ket{\varphi^\prime}=\fPadHadamard_\fAdv(K;\underbrace{ \ell}_{\substack{\text{padding} \\ \text{length}}},
\underbrace{ \kappa_{\text{out}}}_{\substack{\text{output} \\ \text{length}}})\circ\ket{\varphi}$$
	one of the following two is true:
	\begin{itemize}
		\item 	$|P_{pass}\ket{\varphi^\prime}|\leq \frac{35}{36}|\ket{\varphi}|$
		\item $P_{pass}\ket{\varphi^\prime}$ is $(2^{\eta/6},\frac{1}{3}|\ket{\varphi}|)$-ANY-secure for $K$.
	\end{itemize}
\end{cor}

If we require the input to have a two-branch form, we can have better control on server's behavior in the padded Hadamard test --- what we call \emph{coherency restriction}.
\begin{lem}[coherency restriction of the Padded Hadamard test]\label{lem:7.4}
	The following statement is true for sufficiently large security parameter $\kappa$:\par
	Suppose the initial state, described by the purified joint state $\ket{\varphi}$, satisfies the conditions listed in Lemma \ref{lem:7.3}, and additionally, it has the form
	\begin{equation}\label{eq:5i}\ket{\varphi}=\ket{\psi_0}+\ket{\psi_1}\end{equation}
	\begin{equation}\label{eq:5i2}
		\forall b\in \{0,1\},P^{S}_{x_b}\ket{\psi_b}=\ket{\psi_b}
	\end{equation}
	where $S$ is a system on the server side and $P^{S}_{x_b}$ is the projection onto $x_b\in K$ on system S.\par
	And an algorithm $\fAuxInf$ is a client-side algorithm on some read-only system, and it does not require random oracle queries.\par
	Then the following conclusion holds for any $C>2^{-\kappa/10}$:\par
	For any adversary $\fAdv$ of query number $|\fAdv|\leq 2^\kappa$, for the post-execution state, which is
	$$\ket{\varphi^\prime}=\fPadHadamard_\fAdv(K;\underbrace{ \ell}_{\substack{\text{padding} \\ \text{length}}},
\underbrace{ \kappa_{\text{out}}}_{\substack{\text{output} \\ \text{length}}})\circ\ket{\varphi}$$
	at least one of the following is true:
	\begin{itemize}
		\item 	$|P_{pass}\ket{\varphi^\prime}|\leq (1-C^2)|\ket{\varphi}|$.
		\item For any fixed standard basis subspace $S$ on some server-side system (here ``fixed'' means $S$ should not depend on keys in $K$)\footnote{In practice $S$ can actually be chosen to be some keys --- even if the conditions say it should be fixed, we can bypass this problem using hash tags in practice --- which means the server's ability to compute some keys will be small.}, any server-side operation $\cD$ with query number $|\cD|\leq 2^{\eta/4}$, $\forall b\in \{0,1\}$, define $\ket{\varphi^\prime_b}$  as the post-execution state of feeding part of the input to the protocol:
		      $$\ket{\varphi^\prime_b}=\fPadHadamard_\fAdv(K;\ell,\kappa_{out})\circ\ket{\varphi_b}$$
		      Thus $\ket{\varphi^\prime}=\ket{\varphi_0^\prime}+\ket{\varphi_1^\prime}$. And define
		      $$p=|P_S \cD(\ket{\varphi^\prime}\odot \llbracket\fAuxInf\rrbracket)|/|\ket{\varphi}|$$
		      $$p_0=|P_S \cD(\ket{\varphi^\prime_0}\odot  \llbracket\fAuxInf\rrbracket)|/|\ket{\varphi}|$$
		      $$p_1=|P_S \cD(\ket{\varphi^\prime_1}\odot  \llbracket\fAuxInf\rrbracket)|/|\ket{\varphi}|$$
		      where $P_S$ is the projection onto $S$ on some server-side system. \footnote{We note that $S$ is the abbreviation of ``subspace'' (for example, $\ket{0}$) instead of ``subsystem''.}\par
		      Then at least one of the following two is true:
		      \begin{equation}\label{eq:44r}p\leq 5C\end{equation}
		      \begin{equation}\label{eq:45r}\min\{p_0,p_1\}\geq \frac{p}{6}\end{equation}
	\end{itemize}

\end{lem}
\paragraph{How to understand this lemma} This lemma can be understood as follows: first, suppose the adversary can pass the test with some high probability, (correspondingly, the first case in the conclusion is false.) then starting from the post-execution state, the adversary wants to use the operation $P_S\cD$ to ``separate the behavior'' of the two \emph{branches} in the input: note that $p_0$ corresponds to the case where the input state is $\ket{\psi_0}$ and $p_1$ corresponds to the case where the input state is $\ket{\psi_1}$, and the adversary wants to make one of $p_0$ and $p_1$ big and the other one small. The lemma says it won't succeed, in the following sense: if the adversary can pass the test with high probability, then either $p$ is not big enough, or $p_0$ and $p_1$ do not differ too much --- which intuitively means the two branches could not be tested apart.\par
What's more, we allow the adversary to get some auxiliary information. The main condition we need here is $\ket{\varphi}\odot \llbracket\fAuxInf\rrbracket$ is still SC-secure.\par
One important thing is $\llbracket\fAuxInf\rrbracket$ and $\cD$ can be chosen arbitrarily. This is important and is what makes this lemma powerful: in the following sections, when we analyze a protocol that contains the padded Hadamard test, we will apply this lemma multiple times on different auxiliary information and different $\cD$ to get multiple inequalities, assuming $|P_{pass}\ket{\varphi^\prime}|> (1-C^2)|\ket{\varphi}|$. We will see, these inequalities, together with some other tools, will lead to the result we need.\par
Finally, we note that this lemma does not say the two branches of the input state are indistinguishable. Actually, we can see, when we use the padded Hadamard test in our $\fGdgPrep^{basic}$ protocol, the corresponding two branches are actually distinguishable, what this lemma tells us is the adversary's behavior has to be ``coherent'' (at least not in a detectable way) in these two branches.\par
We put the proofs in Appendix \ref{sec:AC}.

\subsection{The first protocol for Remote Gadget Preparation: $\fGdgPrep^{basic}$ (Protocol Design)}\label{sec:7.3}
Now we can describe our first remote gadget preparation protocol formally. We name it as $\fGdgPrep^{basic}$.
\begin{mdframed}[style=figstyle]
\begin{prtl}\label{prtl:6}$\fGdgPrep^{basic}(K^{\text{help}},K^{(3)};\ell,\kappa_{out})$, where $K^{\text{help}}=\{x_b^{\text{help}}\}_{b\in \{0,1\}}$, $K^{(3)}=\{x_b^{(3)}\}_{b\in \{0,1\}}$. $\ell$ is the padding length and $\kappa_{out}$ is the output key length:\\
	For an honest server, the initial state is $\ket{\varphi}=(\ket{x_0^{\text{help}}}+\ket{x_1^{\text{help}}})\otimes (\ket{x_0^{(3)}}+\ket{x_1^{(3)}})$. \par
	\begin{enumerate}
		\item the client samples \begin{itemize}\item The bit-wise permutation $perm$ on strings of length $2\kappa_{out}$;\item a pair of different (input) keys $K^{(2)}=\{x_0^{(2)},x_1^{(2)}\}$ with the same length as $x_b^{(3)}$;\item $2$ pairs of different (output) keys $K_{out}=\{y_b^{(2)},y_b^{(3)}\}_{b\in \{0,1\}}$ with key length $\kappa_{out}$.\end{itemize}
		\item The client computes $$\fRobustRLT(K^{\text{help}},K^{(2,3)}\leftrightarrow K_{out}, perm;\underbrace{ \ell}_{\substack{\text{padding} \\ \text{length}}})$$ and sends it together with $K^{(2)}$ to the server.
		\item An honest server should implement the following mapping:
		      $$\ket{\varphi}=(\ket{x_0^{\text{help}}}+\ket{x_1^{\text{help}}})\otimes(\ket{x_0^{(3)}}+\ket{x_1^{(3)}})$$ $$\rightarrow (\ket{x_0^{\text{help}}}+\ket{x_1^{\text{help}}})\otimes perm((\ket{y_0^{(2)}}+\ket{y_1^{(2)}})\otimes(\ket{y_0^{(3)}}+\ket{y_1^{(3)}}))$$
		\item The client and the server run the padded Hadamard test on $K^{\text{help}}$. The server can use $\ket{x^{\text{help}}_0}+\ket{x_1^{\text{help}}}$ to pass the test, as described in Definition \ref{def:pht}. Reject if the server does not pass this test.
		\item The client sends out $perm$.
		\item The server removes the permutation and gets $(\ket{y_0^{(2)}}+\ket{y_1^{(2)}})\otimes (\ket{y_0^{(3)}}+\ket{y_1^{(3)}})$.
	\end{enumerate}
\end{prtl}\end{mdframed}
which formalizes the informal description in the introduction. 

\subsection{Formal Statement and Security Proof}\label{sec:7.4}
We make use of the \emph{weak security transform parameter} defined in Definition \ref{def:3.12c} to describe the security of Protocol \ref{prtl:6}. We note that:
\begin{itemize}
	\item We need to assume the initial state has a specific form. This form can be verified by the non-collapsing basis test protocol described in Protocol \ref{prtl:r9}. Thus although this is an extra requirement, we can remove it later when we further revise the protocol.
	\item The initial state has different SC-security conditions for $K^{\text{help}}$ and $K^{(3)}$.
	\end{itemize}
\subsubsection{An additional requirement on the initial state: it has to have a specific form}\label{sec:7.4.1}

We can prove the security of this protocol when the state has a specific form. In more details, the initial state should be
\begin{equation}\label{eq:40}\ket{\varphi}=\sum_{b_1\in \{0,1\}}\ket{x^{\text{help}}_{b_1}}(\sum_{b_3\in \{0,1\}}\ket{x^{(3)}_{b_3}}\ket{\cdots}+\ket{\chi_{b_1}})\end{equation}
Where the norm of $\ket{\chi_{b_1}}$ should be small. The states are not necessarily normalized.\par
The honest state has this form. We choose to consider this state because, on the one hand, we can still prove the security on this state; on the other hand, we can test whether a state has such form with the $\fBasisTest$ protocol discussed in Protocol \ref{prtl:r9}, Section \ref{sec:6.3}. Recall that using Protocol \ref{prtl:r9} ($\fBasisTest(K^{\text{help}},K^{(3)};T)$) by its security statement (Lemma \ref{lem:6.4}) the client can verify the state, after given some auxiliary information, can be transformed to the following state using a server-side operation:
\begin{equation}\label{eq:30}\ket{\varphi}=\sum_{b_1}\ket{x^{\text{help}}_{b_1}}(\sum _{b_2}\ket{x^{(3)}_{b_2}}\ket{\cdots}+\ket{\chi_{b_1}})+\ket{\chi^\prime}\end{equation}
In this subsection we will first ignore the $\ket{\chi^\prime}$ part, and deal with it in Section \ref{sec:7.5}. (After that we can easily adding $\ket{\chi^\prime}$ back, and the final conclusion will not be affected too much.) We further note that the extra auxiliary information needed for transforming the state into (\ref{eq:30}) can be easily handled with the auxiliary-information technique (Technique \ref{lem:4.2}).\par
Finally note that (\ref{eq:40})(\ref{eq:30}) are all natural notations, but for the security proof below we need to use the purified notations. It's less intuitive, and the readers can refer to (\ref{eq:40}) to get the intuition.
\subsubsection{Security statement and proof}
The security statement for $\fGdgPrep^{basic}$ is given below.
\begin{lem}\label{lem:7.6}
	There exist constants $A,B\geq 1$ such that the following statement is true for sufficiently large security parameter $\kappa$:\par
	For keys $K^{\text{help}}=\{x_b^{\text{help}}\}_{b\in \{0,1\}},K^{(3)}=\{x_b^{(3)}\}_{b\in \{0,1\}}$ which are both a pair of keys, Protocol 
	$$\fGdgPrep^{basic}(K^{\text{help}},K^{(3)};\underbrace{ \ell}_{\substack{\text{padding} \\ \text{length}}},
\underbrace{ \kappa_{\text{out}}}_{\substack{\text{output} \\ \text{length}}})$$
	has weak security transform parameter \weakparaml{\eta}{2^{-\eta}}{\eta}{4C}{(1-C^2)}{\eta/B}{AC} for states in $\cF$ defined below against adversaries of query number $\leq 2^\kappa$ when the following inequalities are satisfied.\par
	$\cF$ is defined to be the intersection of $\mathcal{WBS}(D),D>0$  and states of the following form (see above for an informal description):
	\begin{equation}\label{eq:36nn}
		\ket{\varphi}=\ket{\varphi_0}+\ket{\varphi_1}, P^{S_1}_{x_0^{\text{help}}}\ket{\varphi_0}=\ket{\varphi_0}, P^{S_1}_{x_1^{\text{help}}}\ket{\varphi_1}=\ket{\varphi_1}, \text{$S_1$ is a server-side system}
	\end{equation}
	\begin{equation}\label{eq:37nn}
		\forall b_1\in \{0,1\},\ket{\varphi_{b_1}}=\ket{\varphi_{b_10}}+\ket{\varphi_{b_11}}+\ket{\chi_{b_1}},\end{equation}\begin{equation*} \forall b_3\in \{0,1\},P^{S_1}_{x_{b_1}^{\text{help}}}P^{S_3}_{x_{b_3}^{(3)}}\ket{\varphi_{b_1b_3}}=\ket{\varphi_{b_1b_3}}, P^{S_3}_{K^{(3)}}\ket{\chi_{b_1}}=0\end{equation*}
	\begin{equation}\text{where $S_3$ is a server-side system}, \forall b_1\in \{0,1\},|\ket{\chi_{b_1}}|\leq 9C|\ket{\varphi}|
	\end{equation}

	The inequalities are as follows:
	\begin{enumerate}
		\item (Sufficient security on the inputs) $\eta\geq \kappa\cdot B$. 
		\item (Well-behaveness of the inputs) $D\leq 2^{\eta^{0.99}}$. 
		\item (Sufficient padding length, output key length) $\ell\geq 6D+4\eta$, $\kappa_{out}>\ell+4\eta$
		\item For simplicity, additionally assume $\frac{1}{9}>C>2^{-\sqrt{\kappa}}$
	\end{enumerate}

\end{lem}
\paragraph{How to understand this lemma}
One way to understand this statement is to  understand in a reverse way. Note that the output properties is parameterized by $C$. If we want the output has the security properties for some $C$, we can ``trace back'' using this lemma and know what conditions we need to require for the initial state. And when we use this subprotocol in an upper-level protocol, we need its pervious steps to be secure enough such that these conditions on the initial states are satisfied.\par
When we use this lemma the choice of $C$ can vary. And it is usually chosen to be some inverse-polynomial of $\kappa$.\par
Note again that do not be confused by the condition ``$C>2^{-\sqrt{\kappa}}$'' and think $\ket{\varphi}$ cannot be $(2^{\eta},2^{-\eta}|\ket{\varphi}|)$-SC-secure for $K^{(3)}$. For an initial state that is $(2^{\eta},2^{-\eta}|\ket{\varphi}|)$-SC-secure, $\eta>>\kappa$, we can definitely say it is  $(2^{\eta},4C|\ket{\varphi}|)$-SC-secure for $K^{(3)}$ thus it satisfies the initial security condition.\par
Let's use $w\in \{2,3\}$ to denote the output keys superscript. We will give a more formal proof for this $w=3$ case in the next subsubsection.\par
Then how about the case for $w=2$? The proof for the $w=2$ case is more challenging. We need to make use of the permutation in the lookup table: we can prove that, the security of the $w=3$ case, in some sense, is ``mixed'' with the $w=2$ case. Which means, if the adversary can compute the keys on output wire $w=2$, the property of this permuted lookup table will imply the adversary can also compute some part of the keys at $w=3$ under some situation. Then we can use a similar argument as the $w=3$ case above. We put a proof overview in Chapter \ref{cht:4}. We will give a more detailed proof overview in Section \ref{sec:7.4.5}.
\subsubsection{Security proofs for the $w=3$ case}\label{sec:7.4.4}
The formal proof of Lemma \ref{lem:7.6} for $w=3$ case is given below. The details are postponed to Appendix \ref{sec:afn}.
\begin{proof}[Proof of Lemma \ref{lem:7.6} for $w=3$ case]
As before, use $\ket{\varphi^\prime}$ to denote the post-execution state ($:=\fGdgPrep^{basic}_\fAdv\circ \ket{\varphi}$). $\ket{\varphi}$ is the initial state.	By the auxiliary-information technique (see Technique \ref{lem:4.2}) we can assume $Tag(K_{out}^{(3)})$ are given to the server.\par
	Suppose $|P_{pass}\ket{\varphi^\prime}|> (1-C^2)|\ket{\varphi}|$.\par
	And we choose $B$ to be a big enough constant to make all the arguments below work. (Note that when we apply the lemmas mentioned below, one condition that often appear is the adversary's query number $2^\kappa$ and the security of the initial state $2^{\eta}$ should satisfy $\eta>\kappa\cdot O(1)$. This can be satisfied by choosing a big enough $B$. We note that choosing $B=10000000$ is enough, but it can also be much smaller, although we didn't estimate the exact threshold constant.)\par
	Define 
		      \begin{equation}\label{eq:70n}
		      \ket{\varphi^\prime_b}:=\fGdgPrep^{basic}_\fAdv(K^{\text{help}},K^{(3)};\underbrace{ \ell}_{\substack{\text{padding} \\ \text{length}}},
\underbrace{ \kappa_{\text{out}}}_{\substack{\text{output} \\ \text{length}}})\circ\ket{\varphi_b}, b\in \{0,1\}
	      \end{equation}
	$\ket{\varphi^\prime}$ can be written as $\ket{\varphi^\prime_0}+\ket{\varphi^\prime_1}$, as defined in (\ref{eq:70n}).\par 
	Furthermore $$\ket{\varphi^\prime_0}\approx_{9C|\ket{\varphi}|}\ket{\varphi^\prime_{00}}+\ket{\varphi^\prime_{01}}$$
	where $\ket{\varphi^\prime_{00}}$ is the output state of using the state $\ket{\varphi_{00}}$ as the initial state (of the overall $\fGdgPrep^{basic}$ protocol) and $\ket{\varphi^\prime_{01}}$ is defined correspondingly. Then first we have
	\begin{itemize}
		\item $\ket{\varphi_{00}}$ is $(2^{\eta}-4,2^{-\eta}|\ket{\varphi}|)$-unpredictable for $x_1^{\text{help}}$.
		\item 	$\ket{\varphi_{00}}$ is $(2^{\eta}-4,4C|\ket{\varphi}|)$-unpredictable for $x_1^{(3)}$.
	\end{itemize}

	By applying Lemma \ref{lem:lrr2}, we can prove $\ket{\varphi_{00}^\prime}$ is $(2^{\eta/37}, 12C|\ket{\varphi}|)$-unpredictable for $y_1^{(3)}$ given $K_{out}^{(2)}$. (Lemma \ref{lem:lrr2} additionally assumes $4C\leq \frac{1}{3}$, but in the case $4C> \frac{1}{3}$, this statement already holds.) And similarly $\ket{\varphi_{01}^\prime}$ is $(2^{\eta/37}, 12C|\ket{\varphi}|)$-unpredictable for $y_0^{(3)}$ given $K_{out}^{(2)}$. Thus
	\begin{equation}\text{$\ket{\varphi_0^\prime}$ is $(2^{\eta/37}, 33C|\ket{\varphi}|)$-SC-secure for $K_{out}^{(3)}$ given $K_{out}^{(2)}$.}
	\end{equation}
	Then applying Lemma \ref{lem:7.4} and taking the projection $P_S$ to be the projection onto $y_0^{(3)}||y_1^{(3)}$ (what we mean is actually first using the global tags to check the values then do a projection onto the ``yes'' space) we know $\ket{\varphi^\prime}$ is $(2^{\eta/37},200C|\ket{\varphi}|)$-SC-secure for $K_{out}^{(3)}$ given $K_{out}^{(2)}$. (The condition for applying this lemma is the SC-security of $$\ket{\varphi}\odot \llbracket\fRobustRLT\rrbracket\odot K^{(2)}\odot perm\odot K_{out}^{(2)}\odot Tag(K_{out}^{(3)})$$ for $K^{\text{help}}$, which comes from Lemma \ref{lem:af3}.) This completes the proof.
\end{proof}
\subsubsection{(An overview of) the security proof of the $w=2$ case}\label{sec:7.4.5}
We already give some upper level intuition of how different steps protect the security of this protocol in Section \ref{sec:7.3} and \ref{sec:phtvp}. But these intuitions are not enough for the formal proof of the security statement. In this subsubsection we will give some intuition of how the formal security proof for the $w=2$ case works. The following proof overview is similar to what is given in Chapter \ref{cht:4} but will add some missing details, thus reflecting what we are really doing in the formal proof. The detailed formal proof is postponed to Appendix \ref{sec:ac2}. The ideas we use in the proof can be informally listed as follows:\par
As before, use $\ket{\varphi^\prime}$ to denote the post-execution state ($:=\fGdgPrep^{basic}_\fAdv\circ \ket{\varphi}$). $\ket{\varphi}$ is the initial state.
\begin{enumerate}
	\item Similar to the $w=2$ case, we assume the first case does not hold (thus $|P_{pass}\ket{\varphi^\prime}|> (1-C^2)|\ket{\varphi}|$) thus we need to prove $P_{pass}\ket{\varphi^\prime}$ is \\$(2^{\eta/O(1)},O(1)C|\ket{\varphi}|)$-SC-secure for $K_{out}^{(2)}$ given $K_{out}^{(3)}$.
	\item Let's first introduce some symbols. (In the formal proof we will also use the same symbols.) If an adversary wants to break this SC-security statement, the adversary needs to compute $y_0^{(2)}||y^{(2)}_1$. Suppose the adversary's operation is $\cU$, and denote:
	      \begin{equation}\label{eq:52}p:=|P_{y_0^{(2)}||y_1^{(2)}}\cU(\ket{\varphi^\prime}\odot K_{out}^{(3)}\odot Tag(K_{out}^{(2)}))|/|\ket{\varphi}|\end{equation}
	      We want to get an upper bound for $p$.\par
	\item (\ref{eq:52}) implies, there exists a server-side operation $\cU^\prime$ with query number $O(1)+|\cU|$ such that
	      \begin{equation}|P_{y_0^{(2)}||y_1^{(2)}}\cU^\prime(\ket{\varphi^\prime}\odot K_{out}^{(3)}\odot TAG)|/|\ket{\varphi}|\geq p/\sqrt{3}\end{equation}
	      $TAG$ is the random shuffling of $Tag(K_{out}^{(2)})$ and two other \emph{fake tags}. We temporarily omit the definition of $TAG$ and the explanation of this step here since we want to move to the explanation of the remaining step faster. But this step is indeed an important preparation for the proof later.
	\item The first step is to make use of the idea that the padded Hadamard test is a \emph{unpredictability restriction} for $K^{\text{help}}$. The lemma is Lemma \ref{lem:7.3}, and the reason for doing this is discussed informally in Section \ref{sec:7.3}. To formalize these intuitions we will consider the \emph{blinded} adversary where the adversary's queries in $\cU^\prime$ are replaced by queries to a blinded oracle where $H(\cdots ||K^{\text{help}}||\cdots)$ part are blinded (the prefix padding has length $l$), and we can prove, the output of using this blinded operation (denoted as $\cU^{blind}$) does not differ too much from the output of using $\cU^\prime$:
	      \begin{equation}\label{eq:47}|P_{y_0^{(2)}||y_1^{(2)}}\cU^{blind}(\ket{\varphi^\prime}\odot K_{out}^{(3)}\odot TAG)|/|\ket{\varphi}|\geq O(1)p-O(1)C\end{equation}
	      where $O(1)$-s are going to be replaced by some fixed constants. Note that this step is not needed in the $w=3$ case but is crucial in the $w=2$ case.
	\item Then, similar to the proof of the $w=3$ case, by applying Lemma \ref{lem:7.4} we know, the adversary should be able to compute $y_0^{(2)}||y_1^{(2)}$ on the $\ket{\varphi_1}$ part of the initial state with a not-too-small norm:
	      \begin{equation}\label{eq:48}|P_{y_0^{(2)}||y_1^{(2)}}\cU^{blind}(\ket{\varphi^\prime_1}\odot K_{out}^{(3)}\odot TAG)|/|\ket{\varphi}|\geq O(1)p-O(1)C\end{equation}
	      Note that different $O(1)$-s can be different constants. $\ket{\varphi^\prime_1}$ is defined as (\ref{eq:70n}), the same as the $w=3$ case.
	      \paragraph{A note on the proof structure} We will see, in our proof, we keep deriving lower bounds in the form of $O(1)p-O(1)C$ for different expressions. Once we reach an expression that can also be upper-bounded by $O(1)C$, we get $O(1)C\geq O(1)p-O(1)C$ thus $p\leq AC$ for some constant $A$. We note that the whole structure might be counter-intuitive, since we are not trying to prove the adversary cannot do something in some settings; instead, we are proving the adversary can do something in different settings, assuming (\ref{eq:52}). (Thus (\ref{eq:47})(\ref{eq:48}) are all ``$\geq$'' inequalities.) Finally we reach something that can be bounded in the other direction and complete the proof.\par
	\item Where can we go from (\ref{eq:48})? The next idea is to consider what will happen if the client replaces the $perm$ in the fifth step of the protocol by a random permutation. We would like to prove, if (\ref{eq:48}) holds, which means the adversary can compute the description of $K_{out}^{(2)}$ when the real permutation is provided in the fifth step, the same operation will also compute a set of \emph{fake keys} when the permutation in the fifth step is a random \emph{fake permutation}.\par
	      Note that in (\ref{eq:48}) we only consider the $\ket{\varphi_1}$ branch of the input, and similar to the $w=3$ case, $\ket{\varphi_1}\approx_{9C|\ket{\varphi}|}\ket{\varphi_{10}}+\ket{\varphi_{11}}$. (See (\ref{eq:36nn})(\ref{eq:37nn}) for the meaning of notations.)\footnote{Recall that $\ket{\varphi_0}$ corresponds to the identity-style branch and $\ket{\varphi_1}$ corresponds to the CNOT-style branch.}\par
	      Informally and without loss of generality let's consider the case where the input is $\ket{\varphi_{10}}$.  The adversary already knows or can decrypt the followings directly from the lookup table:
	      \begin{equation}\label{eq:49}\text{(Given $x_1^{\text{help}}$)}:x_0^{(2)},x_{0}^{(3)}\leftrightarrow perm(y_0^{(2)}||y_{0}^{(3)})\end{equation}
	      \begin{equation}\label{eq:49.2}\text{(Given $x_1^{\text{help}}$)}:x_1^{(2)},x_{0}^{(3)}\leftrightarrow perm(y_1^{(2)}||y_{1}^{(3)})\end{equation}
	      (\ref{eq:49}) is what the adversary can get before the 5th step of the lookup table. Since $perm$ is hidden before the 5th step, the server can't extract $y_b^{(w)},w\in \{2,3\}$ from the right of (\ref{eq:49}), and the right of (\ref{eq:49}) looks (almost) the same as two independently random strings.\par
	      In the 5th step of the protocol the client provides $perm$. Then since the goal of the adversary is to output $y_0^{(2)}||y_1^{(2)}$, it has to extract $y_0^{(2)}$ from (\ref{eq:49}) and extract $y_1^{(2)}$ from (\ref{eq:49.2}):
	      \begin{equation}\label{eq:50}\text{(Given $x_1^{\text{help}}$)}:x^{(2)}_{0},x_{0}^{(3)}\leftrightarrow perm(y_0^{(2)}||y_{0}^{(3)})\xrightarrow{perm} y_0^{(2)}\end{equation}
	      \begin{equation}\label{eq:51}\text{(Given $x_1^{\text{help}}$)}:x^{(2)}_{1},x_{0}^{(3)}\leftrightarrow perm(y_1^{(2)}||y_{1}^{(3)})\xrightarrow{perm} y_1^{(2)}\end{equation}
	      Imagine that in the 5th step, instead of providing the real permutation $perm$, the client provides a different $perm^\prime$. The key observation is, from (\ref{eq:50})(\ref{eq:51}), the server can't distinguish this \emph{fake permutation} $perm^\prime$ from the real $perm$ using only the decrypted plaintext in (\ref{eq:49}). In the server's viewpoint,  \\$perm(y_0^{(2)}||y_0^{(3)}),perm(y_1^{(2)}||y_1^{(3)})$ are (almost) just two strings whose bits are all independently random. Thus if some adversary can do the extraction shown in (\ref{eq:50})(\ref{eq:51}), if the client chooses to provide a different $perm^\prime$, it should also be able to compute the \emph{fake keys} $K^{fake-0-(2)}_{out},K^{fake-0-(3)}_{out}$, defined as the key pairs that have the same length with $K^{(2)}_{out},K^{(3)}_{out}$, and satisfy the following equation:
	      $$\forall w\in \{2,3\},K_{out}^{fake-0-(w)}=\{y_b^{fake-0-(w)}\}_{b\in \{0,1\}}\text{ satisfy}:$$
	      $$\forall b\in \{0,1\},perm^\prime(y_b^{fake-0-(2)}||y_{b}^{fake-0-(3)})=perm (y_b^{(2)}||y_{b}^{(3)})$$
	      Writing it in the form of (\ref{eq:50})(\ref{eq:51}), in the server's viewpoint, it gets the following:
	      $$\text{(Given $x_1^{\text{help}}$)}:x^{(2)}_{0},x_{0}^{(3)}\leftrightarrow perm(y_0^{(2)}||y_{0}^{(3)})$$ $$\xrightarrow{perm^\prime} y_0^{fake-0-(2)}:=\text{first half of }{perm^\prime}^{-1}(perm(y_0^{(2)}||y_{0}^{(3)}))$$
	      $$\text{(Given $x_1^{\text{help}}$)}:x^{(2)}_{1},x_{0}^{(3)}\leftrightarrow perm(y_1^{(2)}||y_{1}^{(3)})$$ $$\xrightarrow{perm^\prime} y_1^{fake-0-(2)}:=\text{first half of }{perm^\prime}^{-1}(perm(y_1^{(2)}||y_{1}^{(3)}))$$
	      Formally speaking, it implies
	     {\small \begin{align}\label{eq:82}q_1^{blind,fake,0}:=&|P_{y_0^{fake-0-(2)}||y_1^{fake-0-(2)}}\cU^{blind}(\ket{\phi_1^\prime}\odot K_{out}^{fake-0-(3)}\odot tag)|/|\ket{\varphi}|\\\geq& O(1)p-O(1)C\end{align}}
	      where $\ket{\phi_1^\prime}$ is the output state when the initial state is $\ket{\varphi_1}$ and the permutation provided is $perm^\prime$.
	\paragraph{Note} There is one more detail missing above: in the argument above we implicitly assume the adversary can only decrypt at most two rows in the forward table and two rows in the backward table ((\ref{eq:50})(\ref{eq:51})). But why is this still true when the $perm$ is provided to the server? The reason is, we have already ``blinded'' the server's operation $\cU$! In other words, before the adversary knows $perm$, this property is guaranteed by the fact that the adversary does not know $perm$; after the $perm$ is provided, the adversary has already been blinded, and this property still holds.
	\item On the other hand, we can prove, these fake-keys cannot be simultaneously computed in the $\ket{\varphi_0}$ part of the input:
	      {\small\begin{equation}\label{eq:83}q_0^{blind,fake,0}:=|P_{y_0^{fake-0-(2)}||y_1^{fake-0-(2)}}\cU^{blind}(\ket{\phi_0^\prime}\odot K_{out}^{fake-0-(3)}\odot tag)|/|\ket{\varphi}|\leq O(1)C\end{equation}}
	       where $\ket{\phi_0^\prime}$ is the output state when the initial state is $\ket{\varphi_0}$ and the permutation provided is $perm^\prime$. Then using (\ref{eq:82})(\ref{eq:83}) and applying Lemma \ref{lem:7.4} again complete the proof.
\end{enumerate}
We note that this is only an overview of the formal proof and there are a lot of details missing here. In the formal proof we will combine all these details.\par
We leave the complete proof to Appendix \ref{sec:ac2}.
\section{Weakly Secure Gadget-increasing Protocol}\label{sec:8l}
\subsection{The $1+1\rightarrow 2$ Gadget Preparation Protocol}\label{sec:7.5}
\subsubsection{Protocol design and security statement}
In the last section we formalized the protocol that generate two output gadgets from two input gadgets, but we need to assume the input state has a specific form. However, if we want to glue the subprotocols together into a big protocol, the only things we want to assume about the initial state are its norm, the SC-security property and the representable property, as discussed in Section \ref{sec:4.8.1}.\par
So in this section we add a $\fBasisTest$ step before it. Then we can get rid of the condition on the form of the initial state: using the property of the $\fBasisTest$ protocol, we can derive that some state in the middle of the execution is close to the state we want, and then we can consider this state as the initial state and apply Lemma \ref{lem:7.6}. (The reader can refer to Section \ref{sec:7.4.1} for some discussions on it.)\par
Then we give the remote gadget preparation protocol of this subsection. Using the notation in Definition \ref{defn:3.7}, this protocol is a $2\rightarrow 2$ protocol. Since the two pairs of keys play different roles, we name it as ``$1+1\rightarrow 2$ protocol'', or $\fGdgPrep^{1+1\rightarrow 2}$.
\begin{mdframed}[style=figstyle]
\begin{prtl}\label{prtl:7}$\fGdgPrep^{1+1\rightarrow 2}(K^{\text{help}},K^{(3)};\ell,\kappa_{out}, T)$ is defined as follows, where $K^{\text{help}},K^{(3)}$ are both single pairs of keys, $\ell$ is the padding length, $\kappa_{out}$ is the output key length, $T$ is the number of test rounds in the basis test step:
	\begin{enumerate}
		\item The client and the server execute $\fBasisTest(K^{\text{help}}, K^{(3)};T,\ell,\kappa_{out})$. (Protocol \ref{prtl:r9})
		\item The client and the server execute $\fGdgPrep^{basic}(K^{\text{help}},K^{(3)};\ell,\kappa_{out})$. The client stores the returned output keys.
	\end{enumerate}
\end{prtl}\end{mdframed}
Now we have the following properties for this protocol:
\paragraph{Correctness} This protocol transforms $2$ gadgets to $2$ gadgets.
\paragraph{Efficiency} Both the client and the server run in polynomial time (on the key size and the parameters).\par
\begin{lem}\label{lem:7.9}
	There exist constants $A_2,B_2>1$ such that the following is true for sufficient large security parameter $\kappa$:\par
	Suppose $K^{\text{help}}=\{x_b^{\text{help}}\}_{b\in \{0,1\}}$, $K^{(3)}=\{x_b^{(3)}\}_{b\in \{0,1\}}$. Protocol 
	$$\fGdgPrep^{1+1\rightarrow 2}(K^{\text{help}},K^{(3)};
\underbrace{ \ell}_{\substack{\text{padding} \\ \text{length}}},
\underbrace{ \kappa_{\text{out}}}_{\substack{\text{output} \\ \text{length}}},
\underbrace{ T}_{\substack{\text{rounds} \\ \text{of test}}}
)$$
	has weak security transform parameter \weakparaml{\eta}{2^{-\eta}}{\eta}{C}{(1-C^{12})}{\eta/B_2}{A_2C} on inputs in $\cWBS(D)$ against adversaries of query number $\leq 2^\kappa$ when the following inequalities are satisfied:
	\begin{itemize}
	\item (Well-behaveness of the inputs) $D\leq 2^{\eta^{0.98}}$.
	\item (Range of $C$) $1/9>C>2^{-\sqrt{\kappa}/10}$
	\item (Suitable rounds of tests) $2/C^4>T>1/C^4$
	\item (Sufficient security of the inputs) $\eta>B_2\kappa $. $\eta>500\kappa+2000T^2$.
	\item (Sufficient padding length and output key length) $\ell>6D+10\eta$, $\kappa_{out}\geq \ell+4\eta$
	\end{itemize}


\end{lem}

We add the following note to help the reader understand the conditions in Lemma \ref{lem:7.9}.\par
\begin{itemize}
	\item The relations between different variables in the statement might seem complicated. One way to understand this statement is to understand it in the ``reverse'' way: we can view $C$ as a value that decides the properties of the output. If we want the post-execution state to satisfy the conclusion (which is parameterized by $C$, $\frac{1}{9}>C>2^{\sqrt{\kappa}/10}$, then the protocol parameter should satisfy the condition 4, 5, and the initial state should satisfy the first three conditions.
	\item In the sections later, $1/C,T$ are all at most chosen to be a fixed polynomial of $\kappa$. Thus $\eta$ only needs to be a fixed polynomial of $\kappa$, thus the size of the initial gadgets is succinct.
	\item The reader might get confused on the fact that we seem to be requiring $C$ to be ``not-too-small'': in the conditions we say $\ket{\varphi}$ is $(2^\eta,C|\ket{\varphi}|)$-SC-secure for $K^{(3)}$ given $K^{\text{help}}$ and we additionally require $C>2^{-\sqrt{\kappa}}$. Intuitively it's better to have $\ket{\varphi}$ to be $(2^\eta,2^{-\eta}|\ket{\varphi}|)$-SC-secure, does that mean we want the initial state to be ``insecure''? The answer is no: for an input state $\ket{\varphi}$, if $\ket{\varphi}$ is $(2^{\eta},2^{-\eta}|\ket{\varphi}|)$-SC-secure, then we can also say it's  $(2^{\eta},C|\ket{\varphi}|)$-SC-secure for $C=\fpoly(\kappa)>2^{-\sqrt{\kappa}}>2^{-\eta}$! Thus the reader should not think the condition 3 means $\ket{\varphi}$ is ``only'' SC-secure with that parameter. 
	\item The requirement $C>2^{-\sqrt{\kappa}/10}$ is a simple way to cover the all cases where $C$ is a fixed inverse-polynomial function of $\kappa$.\par
	\item Similar to Lemma \ref{lem:6.4}, we need a two-sided constraint on $T$. We can definitely choose $T$ in a bigger interval, but it's not needed here.
\end{itemize}
\subsubsection{Security proof}
\begin{proof}
	The problem here is to combine the analysis of the $\fBasisTest$ protocol and the basic gadget preparation protocol. Denote the state after the first ($\fBasisTest$) step of the protocol as $\ket{\varphi^1}$, apply the property of the $\fBasisTest$ protocol (Lemma \ref{lem:6.4}), we know:\par
	Either
	\begin{equation}\label{eq:58r}
		|P_{pass}\ket{\varphi^1}|\leq (1-C^{12})|\ket{\varphi}|
	\end{equation}
	or there exists a server-side operation $\cU$, query number $|\cU|\leq 2^{\kappa+3}$ such that
	\begin{equation}\label{eq:59}\ket{\tilde \varphi^1}:=\cU (P_{pass}\ket{\varphi^1}\odot \llbracket\fAuxInf\rrbracket)=\ket{\psi_0}+\ket{\psi_1}+\ket{\chi^\prime},\end{equation}
	\begin{equation}\label{eq:60r}\text{for some server-side system $S_1$, }\forall b_1\in \{0,1\},P^{S_1}_{x_{b_1}^{\text{help}}}\ket{\psi_{b_1}}=\ket{\psi_{b_1}},P^{S_1}_{K^{\text{help}}}\ket{\chi^\prime}=0\end{equation}
	\begin{equation}\label{eq:61r}\forall b_1\in \{0,1\},\ket{\psi_{b_1}}=\ket{\psi_{b_10}}+\ket{\psi_{b_11}}+\ket{\chi_{b_3}},\end{equation}
	{\small\begin{equation}\label{eq:62r}\text{for some server-side system $S_3$, }\forall b_1,b_3\in \{0,1\}^2, P^{S_1}_{x_{b_1}^{\text{help}}}P^{S_3}_{x_{b_3}^{(3)}}\ket{\psi_{b_1b_3}}=\ket{\psi_{b_1b_3}},P^{S_3}_{K^{(3)}}\ket{\chi_{b_1}}=0\end{equation}}
	\begin{equation}\text{ $\llbracket\fAuxInf\rrbracket$ is defined as in Lemma \ref{lem:6.4},}\end{equation}
	\begin{equation}\label{eq:69}\text{$|\ket{\chi^\prime}|\leq 2C^2/3|\ket{\varphi}|$, $\forall b_3\in \{0,1\},|\ket{\chi_{b_3}}|\leq 8C|\ket{\varphi}|$}\end{equation}
	To prove the output security property in this lemma (the right hand side of the weak security transform parameter), we can instead view the final output state $\ket{\varphi^\prime}$ as the output state of applying the second step of the protocol (which is a $\fGdgPrep^{basic}$ protocol against $\fAdv_{=2}$, the adversary for the second step of the protocol) on initial state $P_{pass}\ket{\varphi^1}$, and $P_{pass}\ket{\varphi^1}$ satisfies (\ref{eq:58r})-(\ref{eq:69}).\par
	Do a small \textbf{discussion by cases.} Note that the first case (we mean (\ref{eq:58r})) already implies the final security property in the weak security transform parameter. So we need to understand what the second case ((\ref{eq:59})-(\ref{eq:69})) gives us.\par
	Then by the auxiliary-information technique (Technique \ref{lem:4.2}) adding the extra $\llbracket\fAuxInf\rrbracket$ can only make the conclusion stronger, thus we only need to prove the final output security property in the lemma  when:
	\begin{enumerate}
		\item the initial state is $\ket{\tilde \varphi^1}$
		\item the bound on the query number of the adversary becomes $|\fAdv^\prime|\leq |\fAdv_{=2}|+|\cU|\leq 2^{\kappa+4}$
		\item the protocol becomes a $\fGdgPrep_{\fAdv^\prime}^{basic}$ protocol with the same parameters (the same key set, pad length and output length, and the ``$TestRound$'' parameter is useless here)
	\end{enumerate}
	We will make use of the security statement of $\fGdgPrep^{simplfied}$ (Lemma \ref{lem:7.6}). Note that in Lemma \ref{lem:7.6} the initial state should have a specific form, and here $\ket{\tilde\varphi}$ is already very close but not exactly the same. Denote
	\begin{equation}\label{eq:63}\ket{\psi}:=\ket{\psi_0}+\ket{\psi_1}\approx_{\frac{2}{3}C^2|\ket{\varphi}|}\ket{\tilde\varphi}\end{equation}
	We will first study what happens when the initial state is $\ket{\psi}$, and the case when $\ket{\tilde\varphi}$ is the initial state will be close to it. We can verify the conditions of applying Lemma \ref{lem:7.6} as follows:\par 
	\begin{enumerate}
		\item Query number of the adversary: $|\fAdv^\prime|\leq 2^{\kappa+4}$
		\item $\ket{\psi}$ has the form we need: (\ref{eq:60r})(\ref{eq:61r})(\ref{eq:62r})(\ref{eq:63}), and $\forall b_3\in \{0,1\}$, $|\ket{\chi_{b_3}}|\leq 8C|\ket{\varphi}|\leq 9C|\ket{\psi}|$.
		\item $\ket{\psi}$ is $(2^{\eta/7},2^{-\eta/7}|\ket{\psi}|)$-SC-secure for $K^{\text{help}}$ given $K^{(3)}$. Its proof is given below.\par
		      To prove it, by (\ref{eq:59})(\ref{eq:63}) this can be reduced to proving
		      $$\text{$\ket{\varphi^1}\odot \llbracket\fAuxInf\rrbracket$ is $(2^{\eta/7}+|\cU|,2^{-\eta/7-1}|\ket{\varphi}|)$-SC-secure for $K^{\text{help}}$ given $K^{(3)}$.}$$
		      By Lemma \ref{lem:basic} this is reduced to proving
		      \begin{center}\emph{$\ket{\varphi}\odot \llbracket\fBasisTest(K^{\text{help}},K^{(3)};T)\rrbracket\odot \llbracket\fAuxInf\rrbracket$ is $(2^{\eta/7}+|\cU|+|\fAdv|,2^{-\eta/7-1}|\ket{\varphi}|)$-SC-secure for $K^{\text{help}}$ given $K^{(3)}$.}\end{center}
		      When $K^{(3)}$ is given beforehand, the $\llbracket\fBasisTest(K^{\text{help}},K^{(3)};T)\rrbracket\odot \llbracket\fAuxInf\rrbracket$ are sets of padded hash values and reversible lookup tables on $K^{\text{help}}$ (with pad length $\ell$). Since $\ket{\varphi}\odot K^{(3)}$ is $(2^\eta,2^{-\eta}|\ket{\varphi}|)$-SC-secure for $K^{\text{help}}$, by Lemma \ref{lem:4.8} $\ket{\varphi}\odot K^{(3)}\odot \llbracket\fBasisTest(K^{\text{help}},K^{(3)};T)\rrbracket\odot \llbracket\fAuxInf\rrbracket$ is $(2^{\eta/6},2^{-\eta/6}|\ket{\varphi}|)$-SC-secure for $K^{\text{help}}$, which completes the proof.
		\item Applying Lemma \ref{lem:4.10} and use a similar argument as above we can prove
		      \begin{center}\emph{$P_{pass}\ket{\varphi^1}\odot \llbracket\fAuxInf\rrbracket$ is $(2^{\eta/37},3C|\ket{\varphi}|)$-SC-secure for $K^{(3)}$ given $K^{\text{help}}$.}\end{center}
		      Then $\ket{\psi}$ is $(2^{\eta/40}, 4C|\ket{\psi}|)$-SC-secure for $K^{(3)}$ given $K^{\text{help}}$.
		\item $\ket{\psi}$ is $(1,|\fAdv|+|\cU|+O(T))$-representable from $\ket{\varphi}$ thus is $(2^{D},2^{D}+2^{\kappa+5})$-representable from $\ket{\mathfrak{init}}$. Here $l$ is chosen to be bigger than the one given in Lemma \ref{lem:7.6} thus the pad length is enough.
	\end{enumerate}
	By the security of the $\fGdgPrep^{basic}$ protocol (Lemma \ref{lem:7.6}) we know $\ket{\psi^\prime}$, defined as the output when $\ket{\psi}$ is the initial state, satisfies either
	\begin{equation}\label{eq:64}|P_{pass}\ket{\psi^\prime}|\leq (1-C^2)|\ket{\psi}|\end{equation}
	or
	\begin{equation}\label{eq:88}
		\text{$P_{pass}\ket{\psi^\prime}$ is $(2^{\eta/40B},4AC|\ket{\psi}|)$-SC-secure for ($\forall w\in \{2,3\}$) $K_{out}^{(w)}$ given $K_{out}-K_{out}^{(w)}$}
	\end{equation}
	Now we can combine this with (\ref{eq:63}) to draw the final conclusion. In the first case (we mean (\ref{eq:64})) we have $1-C^2+2/3C^2<1-C^{12}$ and in the second case (\ref{eq:88}) we only need to choose the constant $A_2$ to be slightly bigger than $4A$.

\end{proof}

\subsection{An $1+n\rightarrow 2n$ Gadget Preparation Protocol}\label{sec:7.6}
In the previous section we designed the $\fGdgPrep^{1+1\rightarrow 2}$ protocol. However, this protocol cannot generate new gadgets, even on an honest server. To solve this problem, we take a parallel repetition of the previous protocol, and make the gadget for the padded Hadamard test being shared:
\subsubsection{Double (asymptotically) the number of gadgets via parallel repetition and gadget sharing}

We note that, the gadget corresponding to $K^{\text{help}}$ (for padded Hadamard test) can be shared among different execution of the protocol. If we run $n$ such protocols simultaneously and let them share the $Gadget(K^{\text{help}})$ gadget, we get a protocol which can generate $2n$ gadgets from $1+n$ gadgets.\par
Similar to (\ref{eq:des5})(\ref{eq:des51}), (the mapping of the $\fGdgPrep^{basic}$ protocol in the honest setting) such a construction allows the server to do the following mapping:
$$(\ket{x_0^{\text{help}}}+\ket{x_1^{\text{help}}})\otimes (\otimes_{i=1}^n(\ket{x_0^{(3)(i)}}+\ket{x_1^{(3)(i)}}))\xrightarrow{n\text{ $\fRobustRLT$ tables, together with }\{K^{(2)(i)}\}_{i\in [n]}} $$
$$(\ket{x_0^{\text{help}}}+\ket{x_1^{\text{help}}})\otimes (\otimes_{i=1}^n perm^{(i)}((\ket{y_0^{(i)(2)}}+\ket{y_1^{(i)(2)}})\otimes (\ket{y_0^{(i)(3)}}+\ket{y_1^{(i)(3)}}))$$
and the corresponding look-up tables are constructed to encode the mapping:
\begin{align}\label{eq:36n}
	\text{$i$-th table, $i\in [n]$, under  $K^{\text{help}}$ : }K^{(2)(i)},K^{(3)(i)}\leftrightarrow perm^{(i)}(K_{out}^{(i)(2)},K_{out}^{(i)(3)})
\end{align}
the keys are correspondingly: $K^{\text{help}}=\{x_0^{\text{help}},x_1^{\text{help}}\}$, $K^{(3)(i)}=\{x_0^{(3)(i)},x_1^{(3)(i)}\}$. $K^{(i)(w)}_{out}=\{y_0^{(i)(w)},y_1^{(i)(w)}\}$\par
Then we need to make one further revision to the idea above: note that the discussion above focuses on running $n$ blocks of $\fGdgPrep^{basic}$ protocol. But this is not enough: what we will do next is to run $n$ blocks of $\fGdgPrep^{1+1\rightarrow 2}$ protocol simultaneously, thus we need to insert the $\fBasisTest$ part suitably. The formal description of the protocol is given below.
\subsubsection{Protocol design}
Note that in the protocol below we use notation $\tilde K_{out}$ to denote the output keys used in the lookup table, since in the final step there a ``change of notation'' step, and we will denote the final key set as $K_{out}$.
\begin{mdframed}[style=figstyle]
\begin{prtl}\label{prtl:n12n}
	$\fGdgPrep^{1+n\rightarrow 2n}(K^{\text{help}},K^{(3)}; \ell,\kappa_{out}, T)$ is defined below, where $$\text{$K^{\text{help}}=\{x_b^{\text{help}}\}_{b\in \{0,1\}}$, $K^{(3)}:=\{K^{(3)(i)}\}_{i\in [n]}$, $K^{(3)(i)}=\{x_b^{(3)(i)}\}_{b\in \{0,1\}}$.}$$, $\ell$ is the padding length, $\kappa_{out}$ is the output key length, $T$ is the number of test round in the basis test step.\par
	In the honest setting the server should hold $$Gadget(K)=(\ket{x_0^{\text{help}}}+\ket{x_1^{\text{help}}})\otimes (\otimes_{i=1}^n(\ket{x_0^{(3)(i)}}+\ket{x_1^{(3)(i)}}))$$ initially.
	\begin{enumerate}
		\item For $i=1,\cdots n$, the client and the server execute Protocol \ref{prtl:r9}:
		      $$\fBasisTest(K^{\text{help}},K^{(3)(i)};T,\ell,\kappa_{out})$$
		\item The client samples a pair of different keys $K^{(2)(i)}$ for each $i\in [n]$ whose length is the same as $K^{(3)(i)}$.\par
		      And it samples $\{\tilde K_{out}^{(i)(2)},\tilde K_{out}^{(i)(3)}\}_{i\in [n]}$, where $$\tilde K_{out}^{(i)(w)}=\{y_0^{(i)(w)},y_1^{(i)(w)}\},i\in [n],w\in \{2,3\}$$
		      and each key pair are sampled differently independently and each key has length $\kappa_{out}$.
		\item For each $i$, the client samples $perm^{(i)}$ from the bit-wise permutations on strings of length $2\kappa_{out}$ and sends
		      $$\fRobustRLT(K^{\text{help}},\{K^{(2)(i)},K^{(3)(i)}\}\leftrightarrow \{\tilde K_{out}^{(i)(2)},\tilde K_{out}^{(i)(3)}\},perm^{(i)};\underbrace{\ell}_{\substack{\text{padding}\\ \text{length}}})$$
		      together with the description of $K^{(2)(i)}$ to the server.\\
		      The honest server can implement the mapping
		      \begin{align}            & (\ket{x_0^{\text{help}}}+\ket{x_1^{\text{help}}})\otimes (\otimes_{i=1}^n(\ket{x_0^{(3)(i)}}+\ket{x_1^{(3)(i)}}))                                                                            \\
			      \rightarrow & (\ket{x_0^{\text{help}}}+\ket{x_1^{\text{help}}})\otimes (\otimes_{i=1}^n (perm^{(i)}((\ket{y_0^{(i)(2)}}+\ket{y_1^{(i)(2)}})\otimes (\ket{y_0^{(i)(3)}}+\ket{y_1^{(i)(3)}})))\label{eq:70r}\end{align}
		\item The client and the server run the padded Hadamard test on $K^{\text{help}}$. The honest server can use $\ket{x_0^{\text{help}}}+\ket{x_1^{\text{help}}}$ in (\ref{eq:70r}) to pass this test. Reject if the server cannot pass the test.
		\item The client sends out all the $perm^{(i)},i\in [n]$. The honest server can remove the permutation and get the state
		      $$Gadget(\{\tilde K_{out}^{(i)(2)},\tilde K_{out}^{(i)(3)}\}_{i\in [n]})=\otimes_{i=1}^n ((\ket{y_0^{(i)(2)}}+\ket{y_1^{(i)(2)}})\otimes (\ket{y_0^{(i)(3)}}+\ket{y_1^{(i)(3)}}))$$
	\end{enumerate}
	The protocol outputs $K_{out}$ on the client side as the returned output keys where $K_{out}$ is just a change of the notation from $\{\tilde K_{out}^{(i)(2)},\tilde K_{out}^{(i)(3)}\}_{i\in [n]}$ to $K_{out}=\{K_{out}^{(i)}\}_{i\in [2n]}$ where $K_{out}^{(i)}$ corresponds to $\tilde K_{out}^{([i/2])(i\mod 2+2)}$.
\end{prtl}\end{mdframed}
%
Now we have the following properties for this protocol:
\paragraph{Correctness} This protocol transforms $1+n$ gadgets to $2n$ gadgets, thus allows an honest server to (asymtotically) double the number of states.
\paragraph{Efficiency} Both the client and the server run in polynomial time (on the key size and the parameters).\par
To describe its (weak) security, the security lemma is given below.
\begin{lem}[Security of Protocol \ref{prtl:n12n}]\label{lem:7.10}
	There exist constants $A_3,B_3>1$ such that the following is true for sufficiently large $\kappa$:\par
	Protocol
$$\fGdgPrep^{1+n\rightarrow 2n}(K^{\text{help}},K^{(3)};
\underbrace{ \ell}_{\substack{\text{padding} \\ \text{length}}},
\underbrace{ \kappa_{\text{out}}}_{\substack{\text{output} \\ \text{length}}},
\underbrace{ T}_{\substack{\text{rounds} \\ \text{of test}}}
)$$
has weak security transform parameter \weakparaml{\eta}{2^{-\eta}}{\eta}{C}{(1-C^{12})}{\eta/B_3}{A_3C} on input states in $\cWBS(D)$ against adversaries of query number $\leq 2^\kappa$ when the following inequalities are satisfied:
\begin{itemize}
\item (Limitations on the domain of the protocol) $n\leq \kappa$
\item (Well-behaveness of the inputs) $D\leq 2^{\eta^{0.97}}$.
\item (Range of $C$) $1/9>C>2^{-\sqrt{\kappa}/10}$
\item (Suitable rounds of tests) $2/C^4>T>1/C^4$
\item (Sufficient security on the input) $\eta\geq \kappa B_3$, $\eta>2000T^2+500\kappa$
\item (Sufficient padding length and output key length) $\ell>6D+12\eta$, $\kappa_{out}>\ell+4\eta$
\end{itemize}


\end{lem}
\subsubsection{Security proof}
\paragraph{Proof overview} For the proof, we make use of the security statement of $\fGdgPrep^{1+1\rightarrow 2}$ (Lemma \ref{lem:7.9}). Before we go to the proof techniques, let's first ``unroll'' the security statement by expanding the definition of weak security transform parameter:
\begin{mdframed}
\textbf{Unrolled version of Lemma \ref{lem:7.10}}\par
There exist constants $A_3,B_3>1$ such that the following is true for sufficiently large $\kappa$:\par
Consider a pair of keys $K^{\text{help}}$ and $n$ pairs of keys $K^{(3)}$. Consider the protocol $$\fGdgPrep^{1+n\rightarrow 2n}(K^{\text{help}},K^{(3)};
\underbrace{ \ell}_{\substack{\text{padding} \\ \text{length}}},
\underbrace{ \kappa_{\text{out}}}_{\substack{\text{output} \\ \text{length}}},
\underbrace{ T}_{\substack{\text{rounds} \\ \text{of test}}}
)$$. Suppose a purified joint state $\ket{\varphi}$ satisfies:
\begin{itemize}
\item $\ket{\varphi}$ is $(2^{\eta},2^{-\eta}|\ket{\varphi}|)$-SC-secure for $K^{\text{help}}$ given $K^{(3)}$;
\item $\forall i\in [n]$, $\ket{\varphi}$ is $(2^{\eta},C|\ket{\varphi}|)$-SC-secure for $K^{(3)(i)}$ given $K^{\text{help}}$ and $K^{(3)}-K^{(3)(i)}$;
\item $\ket{\varphi}\in \cWBS(D)$
\end{itemize}
And suppose the inequalities in Lemma \ref{lem:7.10} are satisfied.\par
Then for any adversary $\fAdv$ with query number $\leq 2^\kappa$, denote the post-execution state:
\begin{equation}\label{eq:37}\ket{\varphi^\prime}=\fGdgPrep^{1+n\rightarrow 2n}_\fAdv(K^{\text{help}},K^{(3)};\ell,\kappa_{out}, T)\circ\ket{\varphi}\end{equation}
at least one of the following two is true:
	\begin{itemize}
		\item $|P_{pass}\ket{\varphi^\prime}|\leq (1-C^{12})|\ket{\varphi}|$
		\item For any $i\in [2n]$, $P_{pass}\ket{\varphi^\prime}$ is $(2^{\eta/B_3},A_3 C|\ket{\varphi}|)$-SC-secure for $K_{out}^{(i)}$ given $K_{out}-K_{out}^{(i)}$.
	\end{itemize}
\end{mdframed}

First, we can prove this statement separately, for each $i\in [2n]$. Use the notation before the re-indexing, the second case of the statement becomes: for any $i\leq n$, $w\in\{2,3\}$, $P_{pass}\ket{\varphi^\prime}$ is $(2^{\eta/B_3},A_3 C|\ket{\varphi}|)$-SC-secure for $\tilde K_{out}^{(i)(w)}$ given $\tilde K_{out}-\tilde K_{out}^{(i)(w)}$. For a specific $i\in [n]$, such a statement can be reduced to the security of $\fGdgPrep^{1+1\rightarrow 2}$ by the auxiliary-information technique (Section \ref{sec:4.2}): recall that the protocol can be seen as a simultaneous running of $n$ blocks of subprotocols, and using this technique we can ``remove'' the part of the subprotocol that is not at index $i$ (by applying Technique \ref{lem:4.2} and Lemma \ref{lem:basic}), and the remaining protocol becomes a $\fGdgPrep^{1+1\rightarrow 2}$ protocol. And the new initial state becomes $\ket{\varphi}\odot \llbracket\fPrtl_{otherpart}\rrbracket$ where $\llbracket\fPrtl_{otherpart}\rrbracket$ comes from applying the auxiliary-information technique.\par
The formal proof is given below. 

\begin{proof}
	Consider a specific $i\leq n$, which corresponds to the input keys $K^{(3)(i)}$ and output keys $\tilde K_{out}^{(i)(w)}$ ($w\in \{2,3\}$). Thus to prove the conclusion (the statement below ``then the following conclusion holds''), proving it for $i$ is enough.\par
	We only care about the input, protocol and the output that are applied on the keys at index $i\in [n]$. (Note that ``$i\in [n]$'' and ``$i\in [2n]$'' represent different things. The former one is the index before the change-of-notation, and the later one is the index after the change-of-notation, and will not be used in this proof. See the last step of the protocol.) Applying the auxiliary-information technique (Technique \ref{lem:4.2}, Lemma \ref{lem:basic}), proving this is reduced to proving a new statement where:
	\begin{itemize}
		\item 	The initial state (the right side of equation (\ref{eq:37})) is replaced by $\ket{\varphi}\odot X\odot Y$ where $X:=K^{(3)}-K^{(3)(i)}$. $Y$ is the client-side messages of the first, third and fifth steps of the protocol, except the index $i$. In more details, $Y$ contains
		      \begin{enumerate}
			      \item  $\llbracket\fBasisTest(K^{\text{help}},K^{(3)(i^\prime)};\underbrace{T}_{\substack{\text{test}\\\text{round}}},\underbrace{ \ell}_{\substack{\text{padding} \\ \text{length}}},
\underbrace{ \kappa_{\text{out}}}_{\substack{\text{output} \\ \text{length}}})\rrbracket$ for all $i^\prime\neq i$;
			      \item For all $i^\prime\neq i$, the description of $K^{(2)(i^\prime)}$, and $$\fRobustRLT(K^{\text{help}},\{K^{(2)(i^\prime)},K^{(3)(i^\prime)}\}\leftrightarrow \{\tilde K_{out}^{(i^\prime)(2)},\tilde K_{out}^{(i^\prime)(3)}\},perm^{(i^\prime)};\underbrace{ \ell}_{\substack{\text{padding} \\ \text{length}}})$$
			      \item $perm^{(i^\prime)}$ for all $i^\prime\neq i$.
		      \end{enumerate}
		\item Correspondingly, the messages in the protocol that are not on the index $i$ in the first, third and fifth step are removed. And the remaining protocol is just an execution of $\fGdgPrep^{1+1\rightarrow 2}$ protocol applied on $K^{\text{help}}$, $K^{(3)(i)}$.
		\item The adversaries are of query number $\leq 2|\fAdv|+O(\kappa)$
	\end{itemize}
	In summary, to prove the original conclusion in Lemma \ref{lem:7.10}, we only need to prove the following under the same conditions of Lemma \ref{lem:7.10}:\\
\begin{mdframed}
			For any adversary $\fAdv^\prime$ with query number $|\fAdv^\prime|\leq 2|\fAdv|+O(\kappa)\leq 2^{\kappa+2}$, denote the post-execution state as (where $X$, $Y$ are defined above)
			\begin{equation}\label{eq:46}\ket{\tilde\varphi^\prime}=\fGdgPrep^{1+1\rightarrow 2}_{\fAdv^\prime}(\{K^{\text{help}},K^{(3)(i)}\};\ell,\kappa_{out}, T)\circ(\ket{\varphi}\odot X\odot Y)\end{equation}
			, denote the output keys as $\{\tilde K_{out}^{(i)(w)}\}_{w\in \{2,3\}}$, at least one of the following two is true:
			\begin{itemize}
				\item $|P_{pass}\ket{\tilde\varphi^\prime}|\leq (1-C^{12})|\ket{\varphi}|$
				\item $\forall w\in \{2,3\}$, $P_{pass}\ket{\tilde\varphi^\prime}$ is $(2^{\eta/B_3},A_3 C|\ket{\varphi}|)$-SC-secure for $\tilde K_{out}^{(i)}$ given $\tilde K^{(i)}_{out}-\tilde K_{out}^{(i)(w)}$.
			\end{itemize}
		\end{mdframed}
	Note that $i$ is already fixed.\par
	We are almost at the place to apply the property of $\fGdgPrep^{1+1\rightarrow 2}$ to draw the conclusion. We check the conditions as follows:
	\begin{itemize}
		\item SC-security of $\ket{\varphi}\odot X\odot Y$ for $K^{\text{help}}$ given $K^{(3)(i)}$: After $K^{(3)(i)}$ is given to the server, together with $X=K^{(3)}-K^{(3)(i)}$, we can simply assume the adversary gets $K^{(3)}$ in the beginning. And everything in $Y$ can be simulated using some tables or reversible tables encrypted under $K^{\text{help}}$, with some extra paddings. The total number of (reversible) lookup tables is at most $O(nT)$. Since $\ket{\varphi}\odot K^{(3)}$ is $(2^\eta,2^{-\eta}|\ket{\varphi}|)$-SC-secure for $K^{\text{help}}$, by Lemma \ref{lem:4.8} we know
		      \begin{equation}\text{$\ket{\varphi}\odot X\odot Y$ is $(2^{\eta/6},2^{-\eta/6}|\ket{\varphi}|)$-SC-secure for $K^{\text{help}}$ given $K^{(3)(i)}$.}\end{equation}
		\item After $K^{\text{help}}$ is provided, since $Y$ can be simulated from $X$ and $K^{\text{help}}$ with $O(nT)$ queries, and we know $\ket{\varphi}\odot X\odot K^{\text{help}}$ is $(2^\eta,C|\ket{\varphi}|)$-SC-secure for $K^{(3)(i)}$,  by Technique \ref{lem:4.2}
		      \begin{equation}\text{$\ket{\varphi}\odot X\odot Y$ is $(2^{\eta}-O(nT),C|\ket{\varphi}|)$-SC-secure for $K^{(3)(i)}$ given $K^{\text{help}}$.}\end{equation}
		\item Since $\ket{\varphi}$ is $(2^{D},2^{D})$-representable from $\ket{\mathfrak{init}}$, and since $X$ and $Y$ can be prepared using $O(nT)$ queries,
		      \begin{equation}
			      \text{$\ket{\varphi}$ is $(2^{D},2^{D}+O(nT))$-representable from $\ket{\mathfrak{init}}$.}
		      \end{equation}  Note that the length of $l$ in this lemma is bigger than the one in Lemma \ref{lem:7.9} by $\kappa$. Thus the inequalities on the parameters in Lemma \ref{lem:7.9} (the 5th condition) are satisfied.
	\end{itemize}
	Thus we can apply Lemma \ref{lem:7.9} and conclude that the final state $\ket{\tilde\varphi^\prime}$ in (\ref{eq:46}) satisfies either $|P_{pass}\ket{\tilde\varphi^\prime}|\leq (1-C^{12})|\ket{\varphi}|$ or $P_{pass}\ket{\tilde\varphi^\prime}$ is $(2^{\eta/6B_2},A_2C|\ket{\varphi}|)$-SC-secure for $\tilde K_{out}^{(i)(w)}$ given $\tilde K_{out}^{(i)}-\tilde K_{out}^{(i)(w)}$ ($\forall w\in \{2,3\}$). Thus we complete the proof. 
\end{proof}
And in the next section we will see how to amplify it to a fully secure protocol.
\cleardoublepage
\chapter{Amplification of Gadget Security}\label{cht:7}
In this chapter we will amplify the weakly-secure protocol in the last chapter to a fully secure remote gadget preparation protocol (Protocol \ref{prtl:13}).
\section{Overview of the Amplification Techniques}\label{sec:8}
In the previous section we get an $1+n\rightarrow 2n$	protocol with weak security of transform parameter \weakparam{\eta}{C}{(1-\kappa^{-O(1)})}{ O(\eta)}{O(C)}. (Recall the definition of the weak security (Definition \ref{def:3.12}).) In this section we give an overview of how to \emph{amplify} the weak security to normal security for the remote state preparation protocol. The formal protocols are given in Sections \ref{sec:9} and \ref{sec:10}.
\subsection{Simplifying Things to Get Intuitions: General Setting, simplified Adversary Setting and i.i.d Adversary Setting}\label{sec:8.2}
\subsubsection{Motivations}
We will consider a specific class of adversary to gain intuitions. Recall that, as discussed in Section \ref{sec:4.8.1}, in the security proof, we will focus on (1)the norm of the state (which reflects the ``passing probability'') and (2)the SC-security of a state. 
This leads us to the definition of the i.i.d adversaries, as follows. And we emphasize these definitions are just for intuitions and some intuitively quantitive analysis, and the formal proof does not require it.\subsubsection{Settings}
We first define the \emph{simplified adversary}, on a single execution of some remote gadget preparation protocol, and then define the \emph{i.i.d adversary} for multi-round execution of some remote gadget preparation protocol:
\begin{defn}[Simplified adversary setting]\label{def:8.1}
	For initial state $\ket{\varphi}$, and an $N\rightarrow L$ remote gadget preparation protocol $\fPrtl$ that has weak security of transform parameter \weakparam{\eta}{C}{p}{\eta^\prime}{C^\prime}, we define the simplified adversary setting as follows:\par
	Suppose the input state $\ket{\varphi}$ is honest\footnote{We use the natural notation instead of the jointly purified state.}: $\ket{\varphi}=\otimes_{i=1}^N (\ket{x_0^{(i)}}+\ket{x_1^{(i)}})$, where \\$\{x_b^{(i)}\}_{i\in [N],b\in \{0,1\}}$ are the input keys, and the adversary chooses to do one of the following two:
	\begin{itemize}
		\item Make the client reject with probability $1-p^2$. Then all the keys are revealed to the adversary, no matter the client accepts or rejects.
		\item With probability $C^2$, all the keys are revealed to the adversary.\\
		      Otherwise (with probability $1-C^2$), the adversary behaves honestly and generate $\otimes_{i=1}^L(\ket{y_0^{(i)}}+\ket{y_1^{(i)}})$, where $\{y_b^{(i)}\}_{i\in [L],b\in \{0,1\}}$ are the output keys. The client gets the output keys.
	\end{itemize}
\end{defn}
The simplified adversary captures the case where a single round of some protocol $\fPrtl$ is executed. We also need to consider the case where multiple rounds of a subprotocol is executed on different blocks of the input. Thus we define the \emph{i.i.d} adversary setting as a generalization of the \emph{simplified adversary setting}:
\begin{defn}[i.i.d adversary setting]\label{defn:8.2}
	The i.i.d adversary setting is defined as follows. Suppose the initial state is $\ket{\varphi}=\otimes_{m=1}^M (\ket{Block_m})$, and a remote gadget preparation protocol $\fPrtl$ is applied separately (which might be parallelly or sequentially) on each block of the input. In the i.i.d adversary setting, each block is an honest state (for example, if each block has $\kappa$ gadgets, it is of the form $\ket{Block_m}=\otimes_{i=1}^\kappa (\ket{x_0^{(m)(i)}}+\ket{x_1^{(m)(i)}})$) the adversary behaves \emph{independently as a simplified adversary} on each block.
\end{defn}

These definitions might be counter-intuitive, since the adversary seems to be over-powerful (it can even ``control'' the client to do something, for example, to reveal the keys), and there is no explicit way to achieve what we assume it can do in practice. Why do we still want to consider such a setting? We argue below that it's a suitable simplification for gaining intuitions.\par
 Recall that in the security proof we focus on the norm (or the ``passing probability'') of the state, and the SC-security of the output state. Thus in the security proof we only have very limited control on what the adversary actually can do. So the principle behind the design of the simplified adversary setting and the i.i.d setting is: \textbf{Once during the security proof we lose the control on the SC-security (which reflects how much norm the adversary can compute the keys with), intuitively we can simply imagine the adversary already knows everything, because our security proof technique does not distinguish them.}\par
Correspondingly:
\begin{itemize} \item The first case in the conclusion part of the weak security (Definition \ref{def:3.12}), which is ``$|P_{pass}\ket{\varphi^\prime}|\leq p|\ket{\varphi}|$'', we do not know any other thing about the adversary's state except the fact that it at most passes the protocol with probability $p^2$, so we simply assume the adversary knows everything.
\item For the second case, we further simplify the state with SC-security by a mixture of honest state and a state where the adversary knows everything:	When we encounter a state $\ket{\varphi}$ that is $(2^{\eta},C|\ket{\varphi}|)$-SC-secure for $\{x_0,x_1\}$, we can simply assume the server's state is the mixture of the honest state ($\ket{x_0}+\ket{x_1}$, with probability $1-C^2$) and a state where the adversary already knows everything ($x_0||x_1$, with probability $C^2$).
\end{itemize}

And an advantage of considering such i.i.d setting is: we only need to consider the mixtures of adversaries at the two extreme cases: the adversary that is honest and the adversaries that already break everything for a given block of the protocol.\par

\subsection{Amplification: the Repeat-and-combine Technique}\label{sec:8.3}
Now we discuss our amplification technique using the i.i.d adversary setting.\par
\subsubsection{Repeat technique: amplification on the ``(square root of the) passing probability''}\label{sec:8.2.1}
First, as we showed in the end of Section \ref{sec:7}, there exists a $1+n\rightarrow 2n$ protocol with weak security transform parameters \weakparam{\eta}{C}{p}{\eta^\prime}{C^\prime} where $p=1-\kappa^{-O(1)}$, $\eta^\prime=\eta/O(1)$, $C^\prime=O(1)C$ ($O(1)$-s in the statement should be replaced by constants).\par
This $1+n\rightarrow 2n$ protocol is only a weak protocol since $p,C^\prime$ are all bounded by an inverse polynomial, but we want them to be negligible. First, let's see what we can do for the ``(square root of) passing probability'' $p$. We will show a technique to reduce the parameter $p$ in the weak security from $1-\kappa^{-O(1)}$ to a constant, and give a protocol with weak security \weakparams{\eta}{C}{\eta^\prime}{C^\prime}, where $\eta^\prime$ is not too small and $C^\prime$ is not too big. (See Definition \ref{def:3.12b} for the notation without $p$.)\par
The technique is simple: \textbf{(Step 1 of this technique) The client and the server run $M$ blocks of the original protocol on (if the input is honest) $M$ blocks of states, and require that the server passes in all the blocks of protocol.}\par 
The correctness is obvious. To analyze the security, we temporarily use the i.i.d adversary setting and assume the input is honest and the adversary behaves independently on each block. We will see when the number of blocks is bigger than some \emph{threshold polynomial} (which is $M>40\kappa(1-p)^{-1}$), such a technique amplifies a protocol with weak security of transform parameter \weakparam{\cdots}{\cdots}{p}{\cdots}{\cdots} to a protocol with weak security of transform parameter \weakparams{\cdots}{\cdots}{\cdots}{\cdots}.\par
We will refer to this technique as the \emph{repeat} technique.\par
\textbf{Why it works (in the i.i.d setting):} Notice that in the i.i.d setting the adversary needs to choose a choice for each block (see Definition \ref{def:8.1}), and here the server can only choose the first choice in less than $\kappa (1-p)^{-1}$ number of blocks, otherwise the probability of passing the verification will be negligible! (because $p^{(1-p)^{-1}}\leq \frac{1}{2}$.) So if the attacker wants to pass this protocol with non-negligible probability, it has to choose the second choice (which means it should behave ``partially honest'') on at least $M-\kappa(1-p)^{-1}$ blocks. Since we already assume $M>10\kappa(1-p)^{-1}$ this means the server has to be partially honest on most of the blocks!\par

If the client can know in which block the adversary will choose the second choice, and throws away the blocks where the server choose the first choice, this will be a remote gadget preparation protocol with weak security of transform parameter \weakparam{\eta}{C}{2^{-\Theta(\kappa)}}{ \eta^\prime}{C^\prime}. When $C^\prime$ is not too small we can omit the $p$ term and say it has weak security \weakparams{\eta}{C}{\eta^\prime}{C^\prime}. However, the client doesn't know on which blocks the server will choose so it can't drop the corresponding blocks. The solution is: \textbf{(Step 2 of this technique) The client does a random shuffling on all the blocks of keys, and asks the server to do the shuffling on the gadgets too}. Then for any fixed index, with $>\frac{9}{10}$ probability the block after the shuffling is an ``honest block''. (Which means, the adversary behaves honestly here.) Thus this is a protocol with weak security of transform parameter \weakparams{\eta}{C}{\eta^\prime}{O(1)}.\par
The next problem is, even if we already make $p$ negligibly small and make the protocol weakly-secure under Definition \ref{def:3.12b} instead of \ref{def:3.12}, $C^\prime$ is still inverse-polynomial (actually, becomes a constant). To solve these problems, we will describe the \emph{combine} technique in the next subsubsection.\par
\subsubsection{Combine technique: constant-to-negligible amplification on the SC-security}
In Section \ref{sec:8.2.1} we describe the \emph{repeat} technique for amplifying the ``passing probability'' parameter, and raise a problem.\par
Let's first consider a simplified case. Assume there are only two pairs of keys: $K_1=\{x_b\}_{b\in\{0,1\}}$ and $K_2=\{x_b^\prime\}_{b\in\{0,1\}}$. The initial state is $(\ket{x_0}+\ket{x_1})\otimes (\ket{x^\prime_0}+\ket{x^\prime_1})$. Now a malicious server can choose to break one block and get both keys on this block, and it has to stay honest on the other block. In other words, the malicious server can choose to get one of the followings:
$$(\ket{x_0}+\ket{x_1})\otimes (\ket{x^\prime_0}+\ket{x^\prime_1})\otimes x_0||x_1,\quad (\ket{x_0}+\ket{x_1})\otimes (\ket{x^\prime_0}+\ket{x^\prime_1})\otimes x^\prime_0||x^\prime_1$$
the client knows all the keys, but doesn't know which state the server holds.\par
The client wants to transform it into a state with good SC-security for some output keys, without knowing which block the server chooses to break.  To solve this problem, we use a \emph{combine} technique, as follows: the client can ask the server to make a measurement on the xor of the subscripts of the keys; in other words, for the honest server, it's a projection onto states $\ket{x_0x^\prime_0}+\ket{x_1x^\prime_1}$ and $\ket{x_0x^\prime_1}+\ket{x_1x^\prime_0}$. This measurement can be done easily given the hash value of these keys. On the one hand, for an honest server, after it reports the measurement result, the client can update the output keys $K_{out}$ as $\{x_0x^\prime_0,x_1x^\prime_1\}$ (if output is 0) or $\{x_0x^\prime_0,x_1x^\prime_1\}$ (if output is 1). On the other hand, for a malicious party, if it can only break one of the blocks, it can't get both keys in the updated keys (in other words, output $x_0x^\prime_0x_1x^\prime_1$ or $x_0x^\prime_0x_1x^\prime_1$, depending on the measurement output), even if it can report the measurement output maliciously.\par
A more formal description is given below. (This will not be used in the formal protocol; we will use a further revised version there.)
\begin{prtl}\label{prtl:15}
	$\fCombine(K_1,K_2)$, where $K_1=\{x_0,x_1\}$, $K_2=\{x_0^\prime,x_1^\prime\}$\\
	The honest server should hold $Gadget(K_1)\otimes Gadget(K_2)$. Suppose the server knows some hash values of these keys.\\
	The server makes a measurement on the xor of the indexes:
	$$(\ket{x_0}+\ket{x_1})\otimes (\ket{x^{\prime}_0}+\ket{x^{\prime}_1})\rightarrow$$
	$$ (output=0)(\ket{x_0}\ket{x^{\prime}_0}+\ket{x_1}\ket{x^{\prime}_1})\qquad (output=1)(\ket{x_0}\ket{x^{\prime}_1}+\ket{x_1}\ket{x^{\prime}_0})$$
	and sends the output to the client.	The client updates the keys based on the concatenation of the keys in $ K_{1}$ and $K_{2}$: if $output=0$, the client stores $\{x_0||x^\prime_0,x_1||x^{\prime}_1\}$; if $output=1$, the client stores $\{x_0||x^{\prime}_1,x_1||x^{\prime}_0\}$.
\end{prtl}

Generalize the technique above, suppose the initial state is $\otimes_{i=1}^{\kappa}(\ket{Block_i})$ where $\ket{Block_i}=\otimes_{m=1}^{M}(\ket{x_0^{(i)(m)}}+\ket{x_1^{(i)(m)}})$. The server can break each block independently, and in each block, the server can break half of the indexes and get both keys at these places. For a fixed $m\in [M]$, the client will run the $Combine$ protocol step by step, as follows: First combine the states in $\ket{Block_1}$ and $\ket{Block_2}$ into a new state, then combine this new state with the randomly-chosen state in $\ket{Block_3}$ into a new state, etc. If the adversary wants to get both keys in the combined key pair, it has to know all the key pairs that the client chose when they do the combination. For each block this probability is at most $1/2$ from the original assumption, thus the total probability is exponentially small (in the i.i.d setting).\par
\paragraph{Another way to understand this \emph{repeat-and-combine} process}. Note that  in the end of the \emph{repeat} technique, the client does a random shuffling on the indexes within each block, and provides the permutation to the server thus the server can also permute correspondingly. Then both parties run the \emph{combine} technique as described above. Then the random shuffling, together with this \emph{combine} technique, can be seen as a process as follows: the client picks a random subset of all the keys, and ask the server to combine them into one pair of keys.\par

\subsubsection{Putting everything together, dealing with both the honest setting and the malicious setting}
We have described the \emph{repeat} and \emph{combine} technique separately. Now return to the original problem and see how this technique works.\par
An overview of our technique by now is as follows. First the \emph{repeat} technique gives us a ''$M\times (1+n)\rightarrow 2n(M-\kappa (1-p)^{-1})$ remote gadget preparation'', which aympototically doubles the number of gadgets. On the other hand, the protocol has weak security transfer parameter $(2^{\eta},2^{-\eta}\rightarrow 2^{\eta/{O(1)}},1/3)$.\par
Then apply the \emph{combine} technique described above to amplify the constant $1/3$ to an exponentially small value. The post-measurement state will only be broken if the client is so unlucky and the corresponding indexes in all the blocks are all broken by the server initially. This probability is exponentially small.\par
This is still not the end of the story: the honest behavior is affected! The number of gadgets become fewer, since in the \emph{repeat} step the protocol only doubles the number of gadgets and in the \emph{combine} step it decreases the number of gadgets by a factor $\kappa$. But this can be solved by revising the lower level protocol as follows: \begin{itemize}\item Previously we are doing the \emph{repeat} technique based on the $1+n\rightarrow 2n$ protocol, but it's actually possible to first self-compose the protocol to get a $\log \kappa+1\rightarrow \kappa$ protocol, which asymptotically increase the number of gadgets by a factor of $\tilde\Theta(\kappa)$; \item On the other hand, in the \emph{combine} technique, we do not need so many blocks: in the formal protocol we will use $\sqrt{\kappa}$ blocks, which is still enough.\end{itemize} Thus the whole protocol will be gadget-increasing and still preserve security (in the i.i.d setting).\par
A good reference is the diagram in Section \ref{sec:4.5.1}: we are discussing the ``self-composition --- repeat --- combine'' step in the ``amplification'' part of it.\par
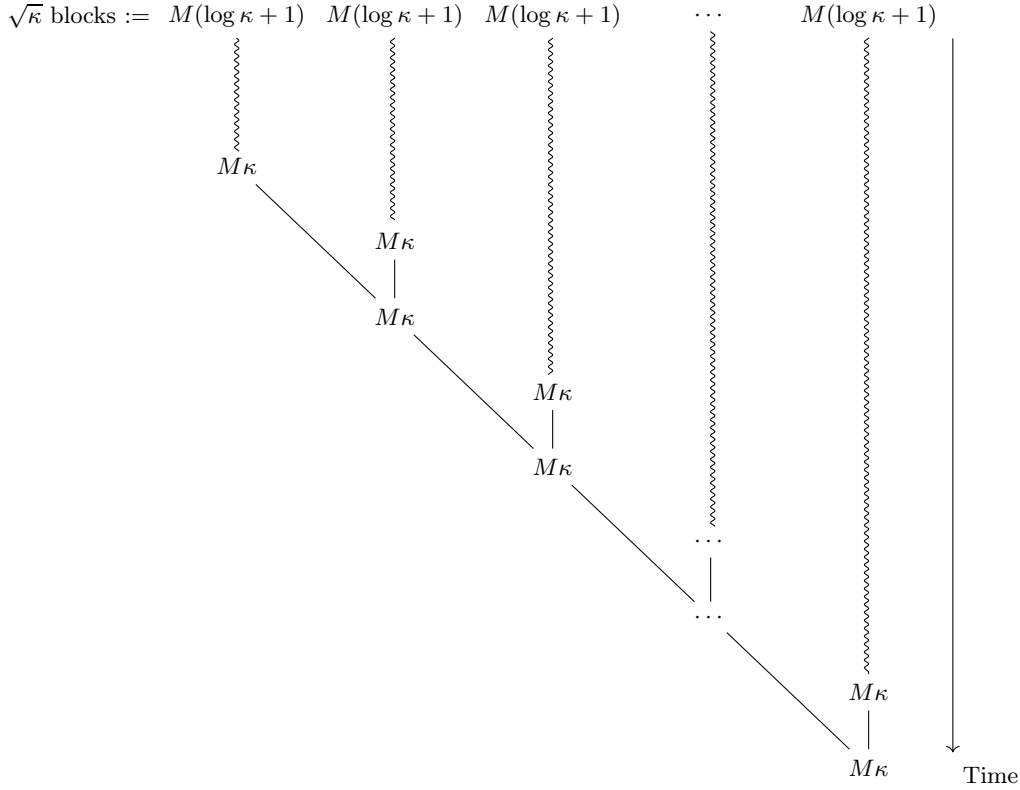
\begin{tikzpicture}[xscale=1.4,yscale=2,radiative/.style={decorate,decoration={snake,segment length=3,amplitude=0.8}}]
\path
(-4.5,0)     node    (1a1) {\footnotesize{$\sqrt{\kappa}$ blocks $:=$}}
(-3,0)     node    (1a1) {\footnotesize{$M(\log\kappa+1)$}}
+(0,-1)     node    (3a1) {\footnotesize{$M\kappa$}}
(-1.5,0)     node    (1a2) {\footnotesize{$M(\log\kappa+1)$}}
+(0,-1.5)     node    (2a2) {\footnotesize{$M\kappa$}}
+(0,-2)     node    (3a2) {\footnotesize{$M\kappa$}}

(0,0)     node    (1a3) {\footnotesize{$M(\log\kappa+1)$}}
+(0,-2.5)     node    (2a3) {\footnotesize{$M\kappa$}}
+(0,-3)     node    (3a3) {\footnotesize{$M\kappa$}}
(1.5,0)     node    (1a4) {\footnotesize{$\cdots$}}
+(0,-3.5)     node    (2a4) {\footnotesize{$\cdots$}}
+(0,-4)     node    (3a4) {\footnotesize{$\cdots$}}
(3,0)     node    (1a5) {\footnotesize{$M(\log\kappa+1)$}}
+(0,-4.5)     node    (2a5) {\footnotesize{$M\kappa$}}
+(0,-5)     node    (3a5) {\footnotesize{$M\kappa$}};

\draw[->]
(3.8,-0.15)--(3.8,-4.9) 
node[right,shift={(-90:.3)}]{\footnotesize{Time}};
\draw 
 (3a1)--(3a2) (2a2)--(3a2) (3a2)--(3a3) (2a3)--(3a3) (3a3)--(3a4)  (2a4)--(3a4)  (2a5)--(3a5) (3a4)--(3a5);
\draw[radiative] (1a1)--(3a1) (1a2)--(2a2) (1a3)--(2a3) (1a4)--(2a4) (1a5)--(2a5);
\end{tikzpicture}\par
A picture for the structure of the \emph{combine} technique by this time, which works in the simplified/i.i.d setting. The number in each node represents the number of gadgets. Initially there are $\sqrt{\kappa}$ blocks of gadgets where each block contains $M(\log\kappa+1)$ gadgets. The snake line represents the protocol we get after we self-compose the weakly secure protocol and use the \emph{repeat} technique. The straight line represents the gadgets in two nodes are combined together. Each node contains $M\kappa$ gadgets and when they are combined together correspondingly we get $M\kappa$ output gadgets. And from the time arrow we can see the execution process is: Generate new gadgets using the 1st block (snake line) --- Generate new gadgets using the 2nd block (snake line) --- Combine them to the old gadgets --- Generate new gadgets using the 3rd block (snake line) --- Combine them --- $\cdots$.
\subsubsection{Overcoming the obstacles in the security proof of the \emph{combine} technique and really get a secure protocol: the introduction of the ``$\fSecurityRefreshing$'' layer}
As we said before, we use the simplified/i.i.d setting to describe our intuition for the protocol design. But this protocol could not be proven secure in the actual setting. However, we can overcome these obstacles by further revising the protocols: we will design a new protocol, named $\fSecurityRefreshing$ (Protocol \ref{prtl:11}), and use it to bypass the obstacles: we will use it before each round of the combine technique. We will explain it in details in Section \ref{sec:10.1}. Let's informally discuss its properties.\par
If we go through the security properties of the previous protocols, we will see, when we say they have the weak security transform parameters \weakparam{\eta}{C}{p}{\eta^\prime}{C^\prime}, there is always $\eta^\prime<\eta$. When we compose the subprotocols together such decrease will accumulate and make the upper-level protocol insecure. Thus we need to design a ``$\fSecurityRefreshing$'' layer, which is a specially-designed $N+\fpoly(\kappa)\rightarrow N$ protocol. We will use it as an extra layer when we revise Protocol \ref{prtl:10} to overcome the obstacles. This protocol also has some additional properties that helps us do the security proof.\par
 On the one hand, in the honest setting, this $\fSecurityRefreshing$ layer only uses succinct extra client side quantum computation; (note that this protocol does not generate new gadget; it even consumes gadgets, but the consumption is succinct.) On the other hand, (informally speaking,) it has weak security transform parameters  \weakparam{\eta}{C}{p}{\eta^\prime}{C^\prime} where $p,C^\prime$ are exponentially small, and $\eta^\prime$ can be much bigger than $\eta$. This helps us overcome the obstacles.\par
 \begin{tikzpicture}[xscale=1.5,yscale=2,radiative/.style={decorate,decoration={snake,segment length=3,amplitude=0.8}}]
\path
(-4.5,0)     node    (1a1) {\footnotesize{$\sqrt{\kappa}$ blocks $:=$}}
(-3,0)     node    (1a1) {\footnotesize{$M(\log\kappa+1)$}}
+(-0.641,-0.65) node {\tiny{$SecurityRefresh$}}
+(0,-1)     node    (3a1) {\footnotesize{$M\kappa$}}
(-1.5,0)     node    (1a2) {\footnotesize{$M(\log\kappa+1)$}}
+(0,-1.5)     node    (2a2) {\footnotesize{$M\kappa$}}
+(-0.641,-1.15) node {\tiny{$SecurityRefresh$}}
+(0,-2)     node    (3a2) {\footnotesize{$M\kappa$}}

(0,0)     node    (1a3) {\footnotesize{$M(\log\kappa+1)$}}
+(0,-2.5)     node    (2a3) {\footnotesize{$M\kappa$}}
+(-0.641,-2.15) node {\tiny{$SecurityRefresh$}}
+(0,-3)     node    (3a3) {\footnotesize{$M\kappa$}}
(1.5,0)     node    (1a4) {\footnotesize{$\cdots$}}
+(0,-3.5)     node    (2a4) {\footnotesize{$\cdots$}}
+(-0.641,-3.15) node {\tiny{$SecurityRefresh$}}
+(0,-4)     node    (3a4) {\footnotesize{$\cdots$}}
(3,0)     node    (1a5) {\footnotesize{$M(\log\kappa+1)$}}
+(0,-4.5)     node    (2a5) {\footnotesize{$M\kappa$}}
+(-0.641,-4.15) node {\tiny{$SecurityRefresh$}}
+(0,-5)     node    (3a5) {\footnotesize{$M\kappa$}};
\draw[->]
(-3.8,-0.75)--(-3.1,-0.75);
\path
(-3.4,-0.85) node {\tiny{$poly(\kappa)$}};
\draw[->]
(-2.3,-1.25)--(-1.6,-1.25);
\path
(-1.9,-1.35) node {\tiny{$poly(\kappa)$}};
\draw[->]
(-0.8,-2.25)--(-0.1,-2.25);
\path
(-0.4,-2.35) node {\tiny{$poly(\kappa)$}};
\draw[->]
(0.7,-3.25)--(1.4,-3.25);
\draw[->]
(2.2,-4.25)--(2.9,-4.25);
\path
(2.6,-4.35) node {\tiny{$poly(\kappa)$}};
\path
(1.1,-3.35) node {\tiny{$poly(\kappa)$}};
\draw[->]
(3.8,-0.15)--(3.8,-4.9) 
node[right,shift={(-90:.3)}]{\footnotesize{Time}};
\draw 
 (3a1)--(3a2) (2a2)--(3a2) (3a2)--(3a3) (2a3)--(3a3) (3a3)--(3a4)  (2a4)--(3a4)  (2a5)--(3a5) (3a4)--(3a5);
\draw[radiative] (1a1)--(3a1) (1a2)--(2a2) (1a3)--(2a3) (1a4)--(2a4) (1a5)--(2a5);
\end{tikzpicture}
A diagram for the protocol execution after we add the $\fSecurityRefreshing$ layer. In the end we get $M\kappa$ gadgets from $\sqrt{\kappa}M(\log\kappa+1)+\sqrt{\kappa}\fpoly(\kappa)$ gadgets. Here $\fpoly$ is a fixed polynomial but $M$ can be very big thus the whole protocol is gadget-increasing. We can see the execution process is: Generate new gadgets using the 1st block (snake line) --- $\fSecurityRefreshing$ for this part --- Generate new gadgets using the 2nd block (snake line) --- $\fSecurityRefreshing$ for this part --- Combine them to the old gadgets --- Generate new gadgets using the 3rd block (snake line) --- $\fSecurityRefreshing$ for this part --- Combine them --- $\cdots$.
\subsection{A summary of the whole amplification techniques, and the organizations of the next two sections}
Thus the design of the whole amplification part is as follows (and we refer to the diagram in Section \ref{sec:4.5.1}):
\begin{otl}[Outline of the Amplification Part]\label{otl:4r}\quad
	\begin{enumerate}
		\item As the discussion in the previous subsection, we design an $M\times (\log\kappa+1)\rightarrow M\times \kappa$ protocol by applying the \emph{repeat} technique on the self-composition of the $1+n\rightarrow 2n$ protocol (given in Protocol \ref{prtl:n12n}). This step is given in Section \ref{sec:9}.
		\item Use the \emph{combine} part of the repeat-and-combine technique to design the protocol $\fGdgPrep^{OneRound}$, and after each iteration within the \emph{combine} protocol, add a $\fSecurityRefreshing$ layer to strengthen and refresh the security. We first discuss the $\fSecurityRefreshing$ layer in Section \ref{sec:10.1}, then discuss the $\fGdgPrep^{OneRound}$ protocol in Section \ref{sec:10.3}.
		\item To get an $N\rightarrow L$ protocol, both parties repeat the $\fGdgPrep^{OneRound}$ protocol for $\log (L/N)$ times and after each round the number of gadgets doubles (in the honest setting.) And to overcome the difficulties in the security proof, after each iteration, again we add a $\fSecurityRefreshing$ layer into it. This is given in Section \ref{sec:10.4}.
	\end{enumerate}\end{otl}
	\subsection{Subtleness in the Security Proof}
	Let's first list the weak security transform parameter for the different protocols in this amplification process, and discuss the subtleness.\par
	\begin{enumerate}
		\item $\fGdgPrep^{\log\kappa+1\rightarrow\kappa}$ has weak security transform parameter \weakparam{\eta}{2^{-\eta}}{(1-\kappa^{-O(1)})}{\eta/\kappa^{O(1)}}{\frac{1}{10}}
		\item $\fGdgPrep^{M\times (\log\kappa+1)\rightarrow M\kappa}$ has weak security transform parameter \weakparams{{\eta}}{2^{-\eta}}{{\eta/\kappa^{B_4}}}{1/3}
		\item $\fGdgPrep^{OneRound}$ has weak security transform parameter \\ \weakparamm{\eta_1}{2^{-\eta_1}}{\Lambda:\eta_2}{\eta_2/350\kappa}{2^{-\sqrt{\kappa}/10}} 
		\item For the security of the final protocol, we directly use the full security: it has output security  $\kappa^{1/4}/10$.
	\end{enumerate}
	On the technique aspect, the security proofs for $2\rightarrow 3$ and $3\rightarrow 4$ above mainly makes use of the multi-round linear decomposition method described in Section \ref{sec:4.8.3}. But this is not without cost. One cost is, the security is usually described using SC-security for some key pair \emph{given all the other keys}. When we apply the decomposition lemmas in $2\rightarrow 3$ above, in the middle of the proof, security properties do not always have this important auxiliary information. That's part of the reason that we need the security refreshing layer in $2\rightarrow 3$, and with this layer, the security can be recovered.\par
	$3\rightarrow 4$ also has similar subtleness. Additionally, we discuss the following question: why isn't the final protocol a simple self-composition of $\fGdgPrep^{OneRound}$? The reason is, in the security statement of $\fGdgPrep^{OneRound}$, for the output part, it's in the form of $(2^{\eta_2/O(\kappa)},2^{-\sqrt{\kappa}/10})$, which is both very secure in terms of the adversary's query bound, and not-that-secure in terms of the adversary's outputting norm. It's not as ideal as $(2^{O(\eta_2)},2^{-O(\eta_2)})$. But we can bypass it by applying the multi-key state decomposition lemma (Lemma \ref{lem:4.7}) and the properties of the $\fSecurityRefreshing$ layer.
\section{Amplification, Part I (The Self-composition and \emph{Repeat} Part)}\label{sec:9}
\subsection{The $\log\kappa+1\rightarrow \kappa$ Remote Gadget Preparation Protocol (the Self-composition Step)}\label{sec:9.1}
In Section \ref{sec:7} we get an $1+n\rightarrow 2n$ protocol, whose gadget-increasing ratio is approximately $2$. As described in Section \ref{sec:8.3}, this is still not enough for later use, since later we will encounter a protocol that combines $\sqrt{\kappa}$ gadgets into one gadgets.\par
In Protocol \ref{prtl:8} we give a $\log\kappa+1\rightarrow \kappa$ protocol by self-composing the $1+n\rightarrow 2n$ protocol for $\log\kappa$ times, whose gadget-increasing ratio is approximately $\tilde\Theta(\kappa)$, and is enough for later use.
\subsubsection{Protocol and statement}
\begin{mdframed}[style=figstyle]
\begin{prtl}\label{prtl:8}
	$\fGdgPrep^{\log\kappa+1\rightarrow\kappa}(K; \ell,\kappa_{out})$ is defined as follows, where \\$K=\{K_1,K_2\}$, $K_1=\{K_1^{(i)}\}_{i\in [\log \kappa ]}$, where $K_1^{(i)}$ (for each $i$) and $K_2$ are all single pairs of keys; $\ell$ is the padding length, $\kappa_{out}$ is the output key length:\par
	The honest server should hold the state $Gadget(K)$ initially.
	\begin{enumerate}
		\item For $t=1,\cdots \log\kappa$:
		      \begin{enumerate}
			      \item Client and server execute $$\fGdgPrep^{1+n\rightarrow 2n}(K_1^{(t)},K_2;\ell,\kappa_{out}, \underbrace{1.5\times 10^4A_3^{8(\log\kappa-(t-1))}}_{\text{test round}})$$
			            where $A_3$ is the constant in Lemma \ref{lem:7.10} (the security of the $1+n\rightarrow 2n$ protocol). The client updates $K_2$ as the returned keys of this protocol call.
		      \end{enumerate}
	\end{enumerate}
	The client stores the final $K_2$ as the returned keys of this protocol.
\end{prtl}\end{mdframed}
\paragraph{Correctness} This is a $\log\kappa+1\rightarrow\kappa$ protocol. Thus its gadget expansion ratio is asymptotically $\tilde\Theta(\kappa)$.
\paragraph{Efficiency} Both parties run in polynomial time (in key size and the parameters).\par
For this protocol, we have the following security statement:
\begin{lem}[Security of Protocol \ref{prtl:8}]\label{lem:9.1}
	There exist constants $A_4>1,B_4>1$ such that the following statement is true for sufficiently large security parameter $\kappa$:\par
	Protocol $$\fGdgPrep^{\log\kappa+1\rightarrow \kappa}(K;\underbrace{ \ell}_{\substack{\text{padding} \\ \text{length}}},\underbrace{ \kappa_{\text{out}}}_{\substack{\text{output} \\ \text{length}}})$$ has weak security transform parameter \weakparam{\eta}{2^{-\eta}}{(1-\kappa^{-A_4})}{\eta/\kappa^{B_4}}{\frac{1}{10}} for input states in $\cWBS(D),D>0$ against adversaries of query number $\leq 2^\kappa$ when the following inequalities are satisfied: 
	\begin{itemize}
	\item (Well-behaveness of the inputs) $D\leq 2^{\kappa^{0.95}}$
		\item (Sufficient security on the inputs) $2^\kappa> \eta > \kappa^{B_4+2}$
		\item (Sufficient pad length and output key length) $\ell\geq 6D+16\eta+8$, $\kappa_{out}\geq \ell+4\eta$
	\end{itemize}


\end{lem}
Intuitively, since within the protocol the composition of the $1+n\rightarrow 2n$ protocol is only repeated by $\log\kappa$ times, the ``exponential blows-up'' is actually a polynomial: $B_3^{\log\kappa}=\kappa^{\log B_3}$, $A_3^{\log\kappa}=\kappa^{\log A_3}$, where $A_3,B_3$ are the constants appeared in the security of the $1+n\rightarrow 2n$ protocol. And the test round in each iteration is also succinct (we will see the $C$ we use when we apply Lemma \ref{lem:7.10} can be lower-bounded by a reciprocal of a fixed polynomial of $\kappa$), thus the $\eta>2000T^2+500\kappa$ condition in the Lemma \ref{lem:7.10} can be satisfied by choosing $\eta$ to be a big enough \textbf{fixed} polynomial function. Thus it's still succinct we can stand it.
\subsubsection{Proof}
The $\log\kappa+1\rightarrow \kappa$ protocol is a self-composition of the $\fGdgPrep^{1+n\rightarrow 2n}$ protocol, and the proof also uses the security of $\fGdgPrep^{1+n\rightarrow 2n}$ (Lemma \ref{lem:7.10}) inductively. 
\begin{proof}[Proof of Lemma \ref{lem:9.1}]
	Suppose the initial purified joint state is $\ket{\varphi}$, which is in $\cWBS(D)$ and is $(2^\eta,2^{-\eta}|\ket{\varphi}|)$-SC-secure for each key pair in $K$ given the other key pairs. And denote the post-execution state (for adversary $\fAdv$, $|\fAdv|\leq 2^\kappa$) as $\ket{\varphi^\prime}$.\par
	Define $A_3,B_3$ as in Lemma \ref{lem:7.10}. Denote $C_{min}=\frac{1}{10}A_3^{-2\log\kappa}$. Assume \begin{equation}\label{eq:26}|P_{pass}\ket{\varphi^\prime}|> (1-C_{min}^{12})|\ket{\varphi}|\end{equation}
	otherwise the first case (which corresponds to the parameter ``$(1-\kappa^{O(1)})$'') on the output part is already true.\par
	Denote the post-execution state of all the parties' systems after the $t$-th round of the iteration as $\ket{\varphi^t}$. Thus $\ket{\varphi^{\log\kappa}}=\ket{\varphi^\prime}$. Additionally define $\ket{\varphi^0}:=\ket{\varphi}$.\par
	First we have $|P_{pass}\ket{\varphi^{t}}|> (1-C_{min}^{12})|P_{pass}\ket{\varphi^{t-1}}|$ holds for any $t\in [\log\kappa]$.\par
	Denote the set of output keys in the $t$-th round as $K_{temp\_t}$. $K_{temp\_0}$ is defined to be the initial $K_2$. We will avoid using $K_2$ below since it has different meaning in each round.\par
	Each round of the protocol is an execution of the $1+n\rightarrow 2n$ protocol. We will argue about the SC-security of the state for the output keys in each round inductively. The problem is how to write down the argument for the inductive proof. We will show that, inductively for $\forall t=0,1,\cdots \log\kappa$:
	\begin{center}\emph{$\forall i$, $P_{pass}\ket{\varphi^t}$ is $(2^{\eta/B_3^{t}},\frac{1}{10}A_3^{-2(\log\kappa-t)}|P_{pass}\ket{\varphi^t}|)$-SC-secure for $K_{temp\_t}^{(i)}$ given $K_{temp\_t}-K_{temp\_t}^{(i)}$ and $\{K_1^{(t^\prime)}\}_{t^\prime>t}$.}\end{center}
	The statement is already true for $t=0$ by the conditions. (Note that $\frac{1}{10}A_3^{-2(\log\kappa-t)}$ is only inverse-polynomial in $\kappa$.) Assume the statement is true for time $t$. The protocol in the $(t+1)$-th round is just an $1+n\rightarrow 2n$ protocol. To apply the security of $\fGdgPrep^{1+n\rightarrow 2n}$ (Lemma \ref{lem:7.10}) on $P_{pass}\ket{\varphi^t}$ and argue about the property of the state $P_{pass}\ket{\varphi^{t+1}}$, let's first verify that the conditions for applying the lemma are satisfied by the state $P_{pass}\ket{\varphi^t}$:
	\begin{itemize}
		\item First we will prove $P_{pass}\ket{\varphi^t}$ is $(2^{\eta}-|\fAdv|-\fpoly(\kappa),2^{-\eta}|\ket{\varphi}|)$-SC-secure for $K_1^{(t+1)}$ given $K_{temp\_t}$ and $\{K_1^{(t^\prime)}\}_{t^\prime>t+1}$. This is because \begin{enumerate}\item $\ket{\varphi}$ is $(2^{\eta},2^{-\eta}|\ket{\varphi}|)$-SC-secure for $K_1^{(t+1)}$ given $(K_1-K^{(t+1)}_1)\cup K_2$;\item The client's messages by the completion of the $t$-th round, together with $K_{temp\_t}$, can be computed from $(K-K^{(t+1)}_1)\cup K_2$ and random coins with RO queries at most $\fpoly(\kappa)$.\end{enumerate} By Lemma \ref{lem:basic} we get the conclusion.\\
		      Then note the fact that $|P_{pass}\ket{\varphi^t}|$ and $\ket{\varphi}|$ are almost the same (equation (\ref{eq:26})). Thus \begin{center}\emph{$P_{pass}\ket{\varphi^t}$ is $(2^{\eta-1},2^{-\eta+1}|P_{pass}\ket{\varphi^t}|)$-SC-secure for $K_1^{(t+1)}$ given $K_{temp\_t}$ and $\{K_1^{(t^\prime)}\}_{t^\prime>t+1}$.}\end{center}
		\item $\forall i$, $P_{pass}\ket{\varphi^t}$ is $(2^{\eta/B_3^t},\frac{1}{10}A_3^{-2(\log\kappa-t)}|P_{pass}\ket{\varphi^t}|)$-SC-secure for $K_{temp\_t}^{(i)}$ given \\$K_{temp\_t}-K_{temp\_t}^{(i)}$ and $\{K_1^{(t^\prime)}\}_{t^\prime>t}$ by the inductive hypothesis.
		\item $P_{pass}\ket{\varphi^t}$ is $(1,|\fAdv|+\fpoly(\kappa)\cdot t)$-representable from $\ket{\varphi}\in \cWBS(D)$. We choose the lowerbound of $\ell$ to be bigger than the bound in Lemma \ref{lem:7.10} by $\eta+1$ then the pad length is enough.
	\end{itemize}
	Thus we can apply Lemma \ref{lem:7.10}. Since the first case in the conclusion of Lemma \ref{lem:7.10} is already ruled out, applying this lemma with $C$ (we mean $C$ in Lemma \ref{lem:7.10}) chosen to be $\frac{1}{10}A_3^{-2(\log\kappa-t)}$, and $\eta$ (we mean $\eta$ in Lemma \ref{lem:7.10}) chosen to be $\eta/B_3^t$, we prove that
	\begin{equation}\label{eq:104}
		\text{$\forall i$, $P_{pass}\ket{\varphi^{t+1}}$ is $(2^{\eta/B_3^{t+1}},\frac{1}{10}A_3^{-2(\log\kappa-t)+1}|P_{pass}\ket{\varphi^t}|)$-SC-secure for $K_{temp\_(t+1)}^{(i)}$ given}\end{equation}\begin{equation*}\text{ $K_{temp\_(t+1)}-K_{temp\_(t+1)}^{(i)}$ and $\{K_1^{(t^\prime)}\}_{t^\prime>t+1}$.}\end{equation*}
	Note that we implicitly use the \emph{auxiliary-information technique} (Technique \ref{lem:4.2}) here: the conclusion of Lemma \ref{lem:7.10} does not have the ``and $\{K_1^{(t^\prime)}\}_{t^\prime>t+1}$'' term. But Lemma \ref{lem:7.10} allows us to prove the property of $P_{pass}\ket{\varphi^{t+1}}$ when $\{K_1^{(t^\prime)}\}_{t^\prime>t+1}$ is provided in advance (which means, the initial state is $P_{pass}\ket{\varphi^{t}}\odot \{K_1^{(t^\prime)}\}_{t^\prime>t+1}$). Thus by Technique \ref{lem:4.2} we can conclude about the property of $P_{pass}\ket{\varphi^{t+1}}$ when $\{K_1^{(t^\prime)}\}_{t^\prime>t+1}$ is provided after the $(t+1)$-th round of the protocol completes, as described here.\par
	Finally $|P_{pass}\ket{\varphi^t}|$ and $|P_{pass}\ket{\varphi^{t+1}}|$ are very close by equation (\ref{eq:26}): in the statement above we have (note that we can choose $\kappa$ to be bigger than some constant to make it true) $$\frac{1}{10}A_3^{-2(\log\kappa-t)+1}|P_{pass}\ket{\varphi^t}|\leq \frac{1}{10}A_3^{-2(\log\kappa-(t+1))}|P_{pass}\ket{\varphi^{t+1}}|$$
	substituting it into (\ref{eq:104}) proves the inductive hypothesis thus completes the inductive proof.\par
	Finally choose $t=\log\kappa$ completes the proof. 
\end{proof}

\subsection{The \emph{Repeat} Technique}\label{sec:9.2}
In this section we complete the security proof the \emph{repeat} part of the \emph{repeat-and-combine} technique, described in Section \ref{sec:8.3}. 
\subsubsection{Protocol and statement}
In the following protocol the client and the server runs many rounds of the previous protocol and do a random permutation on the returned keys. We call the protocol as $\fGdgPrep^{M\times (\log\kappa+1)\rightarrow M\kappa}$:
\begin{mdframed}[style=figstyle]
\begin{prtl}\label{prtl:9}
	$\fGdgPrep^{M\times (\log\kappa+1)\rightarrow M\kappa}(K;\ell,\kappa_{out})$ is defined as follows, where $K:=\{K^{(m)}\}_{m\in [M]}$, and $\forall m\in [M]$, the form of $K^{(m)}$ is compatible with the input of $\fGdgPrep^{\log\kappa+1\rightarrow\kappa}$ (thus each $K^{(i)}$ contains $(\log\kappa+1)$ pairs of keys); $\ell$ is the padding length, $\kappa_{out}$ is the output key length:\par
	The honest server should hold the state $Gadget(K)$ initially.
	\begin{enumerate}
		\item For $m=1$ to $M$:
		      \begin{enumerate}
			      \item The client and the server execute $\fGdgPrep^{\log\kappa+1\rightarrow\kappa}(K^{(m)};\ell,\kappa_{out})$. The client stores the output keys of this step as $\tilde K_{out}^{(m)}$. (Note that $\tilde K_{out}^{(m)}$ contains $\kappa$ pairs of keys.)
		      \end{enumerate}

		\item The client chooses a random permutation $perm$ on $[M]$, permutes the key sets $\tilde K_{out}^{(m)},m\in [M]$ to $K_{out}^{(m)},m\in [M]$, based on the following relation: $$K_{out}^{(m)}=\tilde K_{out}^{(perm(m))}$$. The client sends $perm$ to the server so that the honest server can permute the position of different gadgets correspondingly.
		\item Change the notation for the output keys from $ K_{out}^{(m)},m\in [M]$ (where each $ K_{out}^{(m)}$ contains $\kappa$ pairs of keys) to the ``flatten notation'' $K_{out}^{(i)},i\in [M\kappa]$ (where each $K_{out}^{(i)}$ is a single pair of keys).
	\end{enumerate}
\end{prtl}\end{mdframed}
\paragraph{Correctness} This protocol transforms $M(\log\kappa+1)$ gadgets to $M\kappa$ gadgets, thus achieves gadget expansion ration $\tilde\Theta(\kappa)$.
\paragraph{Efficiency} Both the client and the server run in polynomial time (on the key size and the parameters).\par
We can prove, in Protocol \ref{prtl:9}, when $M$ is chosen to be bigger than a fixed polynomial, the $p$ parameter in the weak security will be at most $\frac{1}{3}$ --- thus we can switch the security definition from Definition \ref{def:3.12} to Definition \ref{def:3.12b}. (This protocol has weak security transform parameters \weakparams{\eta}{C}{{\eta/\kappa^{O(1)}}}{\frac{1}{3}}.) For comparison, in Protocol \ref{prtl:8}, this parameter is $1-\kappa^{-O(1)}$, which is very close to $1$.\par
The security statement of Protocol \ref{prtl:9} is given below. 
\begin{lem}\label{lem:9.2}
	There exists a fixed polynomial $\fTHRESHOLD(\kappa)$, a constant $B_4>1$ such that the following is true for large enough security parameter $\kappa$:\par
	Protocol $$\fGdgPrep^{M\times (\log\kappa+1)\rightarrow M\kappa}(K;\underbrace{ \ell}_{\substack{\text{padding} \\ \text{length}}},
\underbrace{ \kappa_{\text{out}}}_{\substack{\text{output} \\ \text{length}}}),\quad 2^{\sqrt{\kappa}}>M>\fTHRESHOLD(\kappa)$$ has weak security transform parameter \weakparams{{\eta}}{2^{-\eta}}{{\eta/\kappa^{B_4}}}{1/3} for input states in $\cWBS(D)$ against adversaries of query number $\leq 2^\kappa$ when the following inequalities are satisfied:
	\begin{itemize}
	\item (Well-behaveness of the inputs) $D\leq 2^{\kappa^{0.9}}$
	\item (Sufficient security on the inputs)	$2^\kappa>\eta> 2\kappa^{B_4+2}$
	\item (Sufficient pad length and output key length) $\ell>6D+20\eta+12$, $\kappa_{out}>\ell+4\eta$
	\end{itemize}

\end{lem}
\subsubsection{Proof}
The proof is given below.
\begin{proof}[Proof of Lemma \ref{lem:9.2}]
Suppose the initial purified joint state is $\ket{\varphi}$, which is in $\cWBS(D)$ and is $(2^\eta,2^{-\eta}|\ket{\varphi}|)$-SC-secure for each key pair in $K$ given the other key pairs. And denote the post-execution state (for adversary $\fAdv$, $|\fAdv|\leq 2^\kappa$) as $\ket{\varphi^\prime}$.\par

	Suppose
	\begin{equation}\label{eq:42}|P_{pass}\ket{\varphi^\prime}|> \frac{1}{3}|\ket{\varphi}|\end{equation}
	(otherwise the conclusion is already true).\par
	Then suppose the state after the $m$-th round of the first step is $\ket{\varphi^m}$. $\ket{\varphi^0}:=\ket{\varphi}$ and $\ket{\varphi^{M}}$ is the state when the first step completes.\par
	$A_4,B_4$ are the same as the constants in Lemma \ref{lem:9.1}.\par
	Consider the quotient $|P_{pass}\ket{\varphi^{m}}|/|P_{pass}\ket{\varphi^{m-1}}|$, the ``square-root of the passing probability'' in the $m$-th round. Denote $S$ as the set of time $m$ such that this quotient is $\leq 1-\kappa^{-A_4}$. Then $$|S|\leq 4\kappa^{A_4}\leq M/10$$ (otherwise $|P_{pass}\ket{\varphi^\prime}|$ will be too small), and $$\forall m\not\in S,\
		|P_{pass}\ket{\varphi^{m}}|> (1-\kappa^{-A_4})|P_{pass}\ket{\varphi^{m-1}}|.$$
	Consider an arbitrary $m\not\in S$, and we are going to prove the keys generated in the $m$-th round (for example, $\tilde K_{out}^{(m)(i)}$) is SC-secure with some parameters given all the other keys (which are $\tilde K_{out}-\tilde K_{out}^{(m)(i)}$).\par
	We are going to prove the following statement:
	\begin{center}
		\emph{(Statement 1)For any $m\in [M]$, if $|P_{pass}\ket{\varphi^m}|> (1-\kappa^{-A_4})|P_{pass}\ket{\varphi^{m-1}}| $, then $\forall i\in [\kappa]$, $P_{pass}\ket{\varphi^M}$ is $(2^{\eta/\kappa^{B_4}},\frac{1}{10}|\ket{\varphi}|)$-SC-secure for $\tilde K_{out}^{(m)(i)}$ given $\tilde K_{out}-\tilde K_{out}^{(m)(i)}$}
	\end{center}
	note that $P_{pass}\ket{\varphi^M}$ is the state by the beginning of the second step of the protocol onto the passing space. (That is, before the random permutation.)\par
	Let's divide $\tilde K_{out}-\tilde K_{out}^{(m)(i)}$ into three parts: $$Y_<=\{\tilde K_{out}^{(m^\prime)}\}_{m^\prime<m},\qquad\tilde K_{out}^{(m)}-\tilde K_{out}^{(m)(i)},\qquad Y_>=\{\tilde K_{out}^{(m^\prime)}\}_{m^\prime>m}$$ Thus $\tilde K_{out}-\tilde K_{out}^{(m)(i)}=Y_<\cup (K_{out}^{(m)}-\tilde K_{out}^{(m)(i)})\cup Y_>$.\par
	Define $X_{>}=\{K^{(j)}\}_{j\in[M],j> m}$, the subset of the initial keys after the index $m$. Note that the client's messages on the rounds after the $m$-th round, together with $Y_>$, can all be simulated from $X_>$ using $O(\fpoly(\kappa) M)$ RO queries, $\fpoly$ is a fixed polynomial. So by the auxiliary-information technique (Technique \ref{lem:4.2}) and Lemma \ref{lem:basic} we only need to prove the following statement:\\
	\begin{mdframed}For any adversary $\fAdv^\prime$ with query number $|\fAdv^\prime|\leq 2^{\kappa+2}$, the state
			$$\ket{\tilde\varphi}=\fGdgPrep^{\log\kappa+1\rightarrow\kappa}_{\fAdv^\prime}(K^{(m)};\ell,\kappa_{out})\circ (P_{pass}\ket{\varphi^{m-1}}\odot X_{>}\odot Y_{<})$$
			satisfies one of the following two:
			\begin{itemize}\item $|P_{pass}\ket{\tilde\varphi}|\leq  (1-\kappa^{-A_4})|P_{pass}\ket{\varphi^{m-1}}|$
				\item $P_{pass}\ket{\tilde\varphi}$ is $(2^{\eta/\kappa^{B_4}},\frac{1}{10}|\ket{\varphi}|)$-SC-secure for $\tilde K_{out}^{(m)(i)}$ given $\tilde K_{out}^{(m)}-\tilde K_{out}^{(m)(i)}$\end{itemize}\end{mdframed}
	Once we prove it, we prove Statement 1.\par
	To prove it, first notice that if $|P_{pass}\ket{\varphi^{m-1}}|\leq \frac{1}{10}|\ket{\varphi}|$ the statement is already true. Otherwise we can apply the security property of the $\log\kappa+1\rightarrow\kappa$ protocol (Lemma \ref{lem:9.1}). 
	We need to verify the conditions for applying this lemma.
	\begin{mdframed}
		\textbf{Checklist for conditions for applying Lemma \ref{lem:9.1}}
		\begin{enumerate}
			\item The main condition that we need to prove is:
			      \begin{center}\emph{ For any single pair of keys $K^\prime$ in $K^{(m)}$, $P_{pass}\ket{\varphi^{m-1}}\odot X_>\odot Y_<$ is $(2^\eta-2^\kappa-O(M\fpoly(\kappa)), 2^{-\eta+4}|P_{pass}\ket{\varphi^{m-1}}|)$-SC-secure for $K^\prime$ given $K^{(m)}-K^\prime$}\end{center}
			      This is true since first we know \begin{center}\emph{$\ket{\varphi}$ is $(2^\eta,2^{-\eta}|\ket{\varphi}|)$-SC-secure for $K^\prime$ given $X_<$, $X_>$ and $K^{(m)}-K^\prime$.}\end{center} ($X_{<}=\{K^{(j)}\}_{j\in[M],j< m}$. Thus $X_<\cup X_>\cup (K^{(m)}-K^\prime)=K-K^\prime$.)\par Then since the client's messages on the rounds before the $m$-th round, together with $Y_<$, can all be computed from $X_<$ and random coins using $O(\fpoly(\kappa) M)$ RO queries, and the number of adversary's queries is at most $2^\kappa$, applying Lemma \ref{lem:basic} we know \begin{center}\emph{$P_{pass}\ket{\varphi^{m-1}}\odot X_>\odot Y_<$ is $(2^\eta-2^\kappa-O(M\fpoly(\kappa)), 2^{-\eta}|\ket{\varphi}|)$-SC-secure for $K^\prime$ given $K^{(m)}-K^\prime$.}\end{center} Finally using $|P_{pass}\ket{\varphi^{m-1}}|> \frac{1}{10}|\ket{\varphi}|$ completes the proof of this condition.
			\item $P_{pass}\ket{\varphi^{m-1}}$ is $(1,2^\kappa+O(\fpoly(\kappa)M))$-representable from $\ket{\varphi}\in \cWBS(D)$. Thus if we choose the pad length lowerbound to be longer than the bound given in Lemma \ref{lem:9.1} by $2\eta$ the pad length is enough.
		\end{enumerate}
	\end{mdframed}
	Thus we complete the proof of Statement 1 above.\par
	Now we can return to the proof of Lemma \ref{lem:9.2}. The conclusion of Statement 1 holds for any $m\not\in S$. Since $M\geq 10|S|$, the number of $m$ that are not in $S$ will occupy at least $9/10$ of $[M]$. Since $\sqrt{(\sqrt{9/10}\times 1/10)^2+1/10}<1/3$, after the permutation of the index, for any index $m\in [M]$, $i\in [\kappa]$, the final state is $(2^{\eta/\kappa^{B_4}},\frac{1}{3}|\ket{\varphi}|)$-SC-secure for $K^{(m)(i)}_{out}$ given $K_{out}-K_{out}^{(m)(i)}$. Then after flattening the notation we complete the proof.
\end{proof}

\subsection{A Brief Discussion of the Security of the \emph{Combine} Technique}
It's premature to discuss the formal version of the \emph{combine} technique, since we haven't introduce the ``$\fSecurityRefreshing$'' layer and could not give a formal description. But we can give a toy version, as discussed in Section \ref{sec:8.3}.\par
\begin{prtl}\label{prtl:10}
	$\fGdgPrep^{OneRound-toy}(K;\ell,\kappa_{out})$ is as follows, where $K:=\{K^{(t)}\}_{t\in [\kappa]}$, where each $K^{(t)}$ is compatible with the input of \\$\fGdgPrep^{M\times (\log\kappa+1)\rightarrow M\kappa}$ (which means, $K^{(t)}$ contains $M$ sets of keys, where each set is compatible with the input of $\fGdgPrep^{\log\kappa+1\rightarrow \kappa}$. Note that each $K^{(t)}$ contains the same number of sets of keys.); $\ell$ is the padding length, $\kappa_{out}$ is the output key length.\par
	The honest server should hold the state $Gadget(K)$ initially.\\
	For $t=1$ to $\kappa$:
	\begin{enumerate}
		\item The client and the server execute $\fGdgPrep^{M\times (\log\kappa+1)\rightarrow M\kappa}(K^{(t)};\ell,\kappa_{out})$. Denote the returned output keys as $ K_{temp\_t}$.
		\item If $t=1$, skip this step and simply let $ K_{out\_1}= K_{temp\_1}$. Otherwise do the following:\\  For each $m\in [M]$:\begin{enumerate}\item The client and the server run $\fCombine(K_{temp\_t}^{(m)},K_{out\_(t-1)}^{(m)})$ (Protocol \ref{prtl:15}).\\ Store the result as $K_{out\_t}^{(m)}$.\end{enumerate}

	\end{enumerate}
	Denote the returned keys in the last step as $K_{out}$.
\end{prtl}

As we said before, Protocol \ref{prtl:10} is secure in the i.i.d setting, but we don't know how to prove the security in the general setting. But in the next section we will overcome this obstacle by making use of the ``$\fSecurityRefreshing$'' layer. Next we discuss some difficulties of proving the security of Protocol \ref{prtl:10} directly.\par
From the security lemmas in the previous section, we can observe that, in the statement of the weak security of the protocols, suppose the weak security has transform parameter \weakparam{\eta}{C}{p}{\eta^\prime}{C^\prime}, there is always $\eta^\prime<\eta$, or even multiplicative decreases. And the other tools that we have now do not help either: for example, we can't improve the $\eta$ using the decomposition lemmas in Section \ref{sec:4.3}. Thus, when we try to glue the subprotocols together, this parameter will decrease multiplicatively in each composition and finally ``blows up'' (to a very small value). This problem does not exist in the i.i.d setting since such a decay is not obvious. We have no idea whether this phenomenon is inherent in our protocol, or it's just because our proof techniques are not powerful enough. However, we can get rid of this problem by revising the protocol as follows: we design a ``$\fSecurityRefreshing$'' layer (Protocol \ref{prtl:11}), which can help us recover this parameter before it accumulates and blows up. And it also, in some sense, ``de-correlates'' the keys in different places. We will insert this layer to each round of Protocol \ref{prtl:10}, and get a secure protocol (Protocol \ref{prtl:14}).\par
\section{The ``$\fSecurityRefreshing$'' Layer}\label{sec:10.1}
In the last section, we have already discuss why we need to design a ``$\fSecurityRefreshing$'' layer in our protocol. In this subsection we will design such a protocol and in the next subsection we will show how to use it to overcome the obstacle in the \emph{combine} technique (Protocol \ref{prtl:10}).\par
Briefly speaking, in this layer, we need to ``refresh '' and strengthen the security of many key pairs. We will make use of a small number of extra freshly secure gadgets to ``refresh'' the security of many gadgets.\par 
In more details, this ``$\fSecurityRefreshing$'' layer will satisfy: \begin{itemize}\item It is a remote gadget preparation protocol with security transform parameter \weakparams{\eta}{2^{-\eta}}{\eta^\prime}{2^{-\Theta(\eta)}} where $\eta^\prime$ can be bigger than $\eta$. (This is informal, and we will explain it later.) \item What's more, we need this protocol to only use succinct extra client side quantum operations, which means, this protocol will be an $L+\fpoly(\kappa)\rightarrow L$ protocol: with the help of a gadget that can be prepared using succinct extra client side quantum computation, the protocol can map $L$ input gadgets to $L$ output gadgets, where $L$ can be very big (an arbitrary polynomial of $\kappa$, or even sub-exponential).\end{itemize}
The protocol is given below. We will first give the formal definition, then discuss some intuition behind it. We note that it has some additional features (other than improving $\eta$): it helps us ``disconnect'' the output keys on different indexes. The exact meaning of these will become clear later.
\begin{mdframed}[style=figstyle]
\begin{prtl}\label{prtl:11} The protocol $\fSecurityRefreshing(K,\Lambda;\ell,\kappa_{out})$ is defined as follows, where $\ell$ is the padding length, $\kappa_{out}$ is the output key length. 
	Suppose $K=\{x_{b}^{(i)}\}_{i\in [N],b\in \{0,1\}}$, $\Lambda=\{r_{b}^{(j)}\}_{j\in [J],b\in \{0,1\}}$.\par
	The honest server should hold the state $(\otimes_{i=1}^N(\ket{x_0^{(i)}}+\ket{x_1^{(i)}}))\otimes (\otimes_{j=1}^J(\ket{r_0^{(i)}}+\ket{r_1^{(i)}}))$.\\
	For $j=1,\cdots J$:
	\begin{enumerate}
		\item For all $i\in [N]$, the client samples $K_{temp}^{(i)(j)}=\{y_0^{(i)(j)},y_1^{(i)(j)}\}$, which are $N$ pairs of different keys with key length $\kappa_{out}$. 
		      Then for all $i\in [N]$, the client computes \begin{equation}\label{eq:117zz}\fLT(\forall b,b_2\in \{0,1\}^2:(x_{b}^{(i)},r_{b_2}^{(j)})\rightarrow y_b^{(i)(j)} ;\underbrace{ \ell}_{\substack{\text{padding} \\ \text{length}}},
\underbrace{ \kappa_{\text{out}}}_{\substack{\text{tag} \\ \text{length}}})\end{equation} and sends them to the server.\par
		      Note that in (\ref{eq:117zz}) different $b$ corresponds to different output keys, but different $b_2$ corresponds to the same output key.
		\item Denote \begin{equation}\label{eq:76r}\forall b\in \{0,1\}, i\in [N], \qquad x_{b}^{(i)(0)}:=x_{b}^{(i)},\qquad x_{b}^{(i)(j)}:=x_{b}^{(i)}||y_{b}^{(i)(1)}||y_{b}^{(i)(2)}\cdots y_{b}^{(i)(j)}\end{equation}
		      The honest server can use the lookup tables from the last step to implement the following mapping
		      {\small\begin{align*}&(\otimes_{i=1}^N(\ket{x_0^{(i)(j-1)}}+\ket{x_1^{(i)(j-1)}}))\otimes (\ket{r_0^{(j)}}+\ket{r_1^{(j)}})\\\rightarrow& (\otimes_{i=1}^N(\ket{x_0^{(i)(j)}}+\ket{x_1^{(i)(j)}}))\otimes (\ket{r_0^{(j)}}+\ket{r_1^{(j)}})\end{align*}}
		\item The client and the server run the padded Hadamard test on $\Lambda^{(j)}$  with pad length $\ell$ and output length $\kappa_{out}$. The honest server uses and measures the $\ket{r_0^{(j)}}+\ket{r_1^{(j)}}$ state as the initial state of this step.
	\end{enumerate}
	After all the $J$ rounds completes, the client and the server do the following:\par
	The client samples $\{pad^i\}_{i\in [N]}$ where each $pad^i$ is sampled independently randomly from $\{0,1\}^l$, and sends it to the server. The client stores \\$\{pad^i||x_b^{(i)(J)}\}_{i\in [N],b\in \{0,1\}}$ as the returned output keys $K_{out}$. (Note that there are $N$ pairs of keys.) The honest server can add the random pads to the output states and prepare $$Gadget(K_{out})=(\otimes_{i=1}^N(\ket{pad^i||x_0^{(i)(J)}}+\ket{pad^i||x_1^{(i)(J)}}))$$
\end{prtl}\end{mdframed}
\paragraph{Correctness}We can see the honest server does the following mapping in the whole protocol:
\begin{equation}(\otimes_{i=1}^N(\ket{x_0^{(i)}}+\ket{x_1^{(i)}}))\otimes (\otimes_{j=1}^J(\ket{r_0^{(i)}}+\ket{r_1^{(i)}}))\rightarrow \otimes_{i=1}^N\ket{Gadget(K_{out}^{(i)})}\end{equation}
where, as defined in (\ref{eq:76r}) and what we discussed above,
\begin{equation}\label{eq:outkeys}K_{out}^{(i)}=\{pad^i||x_b^{(i)(J)}\}_{b\in \{0,1\}}, x_{b}^{(i)(J)}:=x_{b}^{(i)}||y_{b}^{(i)(1)}||y_{b}^{(i)(2)}\cdots y_{b}^{(i)(J)}\end{equation}
In the beginning of this subsection we say this protocol is an $L+\fpoly(\kappa)\rightarrow L$ protocol. Here we use a different symbol $N$ which plays the role of $L$, which can be an arbitrary polynomial (up to a fixed subexponential). The parameter related to the ``$\fpoly(\kappa)$'' part is the number of ``extra'' gadgets $J$ and the keys in $\Lambda$, which only need to be succinct.\par
\paragraph{Security}For the security, in the beginning of this subsection we say this protocol has security transform parameters \weakparams{\eta}{2^{-\eta}}{\eta^\prime}{2^{-\Theta(\eta)}}. The left hand side is about the security for $K$ part. And $\eta^\prime$ will depend on the security for $\Lambda$. Making the security on $\Lambda$ explicit in the notation, we can say this protocol has weak security transform parameter \weakparamstar{\eta_1}{2^{-\eta_1}}{\Lambda:\eta_2}{\eta_2/O(\kappa)}{2^{-\Theta(\eta_1)}} which means: 
\begin{itemize}\item The left hand side and the arrow part mean the conditions for the security statement of the protocol is in the form of (1)$\forall i\in [N]$, the initial state is $(2^{\eta_1},2^{-\eta_1}|\ket{\varphi}|)$-SC-secure for $K^{(i)}$;\footnote{Here the difference on the auxiliary information provided is why we put a star symbol in the weak security above --- and this is still informal.} (2)$\forall j\in [J]$, the initial state is $(2^{\eta_2},2^{-\eta_2}|\ket{\varphi}|)$-SC-secure for $\Lambda^{(j)}$ given the other keys.\footnote{Let's explain why we put the security for $\Lambda$ above the arrow instead of the left hand side of the arrow: although in the protocol the two key sets are both part of the initial keys, their functionalities are different. This can be seen more clearly in the following weakened setting: we allow the quantum communication during the protocol, but the total amount of it should be small. Then $K$ will still be considered as part of the initial keys, but $\Lambda$ can be provided ``on the fly''. In some sense $\Lambda$ are freshly new gadgets that are provided additionally to ``refresh the security'' of $K$.} \item And the right hand side means the conclusion from the security of the protocol is in the form of ``$P_{pass}\ket{\varphi^\prime}$ is $(2^{\eta^\prime},2^{-\Theta(\eta_1)}|\ket{\varphi}|)$-SC-secure for $K_{out}^{(i)}$ given $K_{out}-K_{out}^{(i)}$''.\end{itemize} (And we can see $\eta_2/O(\kappa)$ can be much bigger than $\eta_1$.) \par
\subsubsection{Intuitions and ideas behind Protocol \ref{prtl:11}}
Informally, we summaries the ideas of the protocol as follows:
\begin{enumerate}
	\item Initially, the system is in a state that is not-very-secure: the first parameter in its SC-security is $2^{\eta_1}$. (In other words, the condition on the input is in the form of ``$\ket{\varphi}$ is $(2^{\eta_1},\cdot)$-SC-secure for $K^{(i)}$''.) We want to amplify it to $2^{\eta_2}$ (for some $K_{out}$), $\eta_2>\eta_1$. Recall that the first parameter in the definition of the SC-security describes the query number of the adversary.
	      \footnote{Note that simply extending the key length by providing a reversible lookup table $\fRevLT(K\leftrightarrow K_{out})$ with longer output key length does not work: although it extends the key length, it does not improve the security, since the adversary can break the input keys using $2^{\Theta(\eta_1)}$ queries, then the reversible lookup table will be broken either.}
	\item We note that, in the protocol, in step 1 of each round, the client creates lookup tables that are encrypted under $r_{b_2},b_2\in \{0,1\}$. (They are also encrypted under $x_b$, but we temporarily focus on the $r$ part.) If the server knows one of $r_0$ and $r_1$, together with $x_b$, (or, as in the protocol, the honest server holds the superposition of $\ket{r_0}$ and $\ket{r_1}$,) the server can decrypt the table and get the output. However, if the server doesn't know any of them, it cannot decrypt the table.\par
	      So there seems to be a dilemma: for the correctness the client wants to give the server the values of $r$, but for the security the client does not want to give the server the value of $r$. So what if the client first gives server the knowledge of $r$, then uses some method to ``force'' the server to ``throw away'' $r$? This is what the client does in the protocol! Let's explain it by focusing on a fixed $j\in [J]$, and the corresponding \emph{temporary output keys} $K^{(i)(j)}_{temp}$:
	      \begin{enumerate}\item The honest server initially holds $\ket{r_0^{(j)}}+\ket{r_1^{(j)}}$, and it can use it to do the step $2$, which is to decrypt the table.
		      \item In step 3 a padded Hadamard test on $\Lambda^{(j)}=\{r^{(j)}_0,r^{(j)}_1\}$ is executed. By the discussion in Section \ref{sec:7.2} we know \textbf{padded Hadamard test is a \emph{unpredictability restriction} for the keys being tested.} After the third step of the protocol, intuitively, it has to break the $\Lambda$ part to decrypt the lookup table to get the output keys $y_b^{(i)(j)}$. Then the SC-security for $K_{out}$ will somewhat rely on the security of the initial state for $\Lambda$, which has parameter $\eta_2$, which can be much bigger than $\eta_1$.
	      \end{enumerate}
	\item The fact that (informally) the adversary doesn't know the keys in $\Lambda$ relies on the properties of the padded Hadamard test. As we discussed in Section \ref{sec:7.1}, after the Hadamard test the server shouldn't be able to learn the keys any more, otherwise it will not be able to pass the test with high probability. But the server can still choose to only pass the test with constant probability, thus there is a \emph{random guessing attack}: if we simply take $J=1$, the adversary can try its best to get the keys, and to pass the padded Hadamard test by simply flipping random coins, hoping it can work. Thus we choose $J$ pairs of ``extra'' gadgets and keys $\{\Lambda^{(j)}\}_{j\in [J]}$. We repeat the previous ideas for $J$ times, and if $J$ is a big enough succinct value we avoid this problem. Note that the final output keys are the concatenation of the output keys in each round.
	\item The final paddings are for the ease of the security proof.
\end{enumerate}

Finally, we note the gadgets on the $\Lambda$ part are shared: this protocol can let us refresh and strengthen the security on all the $N$ keys simultaneously, but the extra client side quantum resources (we mean the gadgets corresponding to $\Lambda$) are succinct. In other words, we refresh the security for many keys by introducing only a small number of ``freshly secure'' gadgets. And in the later section we only need to use this protocol for succinct times, each time it consumes succinct number of gadgets.
\subsection{Security of the $\fSecurityRefreshing$ Layer}\label{sec:10.2}
The formal security statement is given below. We note that there are two additional features of this statement: in the conditions, for keys $K$, we only require the state to be SC-secure for $K$ given the tags (which does not offer the ``disconnection'' among the keys), but the output state  is SC-secure for $K^{(i)}_{out}$ given all the other keys ($K_{out}-K_{out}^{(i)}$); (2) an extra $\llbracket\fAuxInf\rrbracket$ can be considered, which will be useful when we use this protocol in some upper level protocols.
\begin{lem}\label{lem:10.1}
	The following statement is true for sufficiently large security parameter $\kappa$:\par
	Consider two sets of keys $K=\{x_{b}^{(i)}\}_{i\in [N],b\in \{0,1\}}$, $\Lambda=\{r_{b}^{(j)}\}_{j\in [J],b\in \{0,1\}}$, the initial state described by a purified joint state $\ket{\varphi}$, and a randomized algorithm $\fAuxInf$ applied on some client-side read-only system whose operation is simply to output some client side storage, and the protocol
	$$\fSecurityRefreshing(K,\Lambda;\underbrace{ \ell}_{\substack{\text{padding} \\ \text{length}}},
\underbrace{ \kappa_{\text{out}}}_{\substack{\text{output} \\ \text{length}}})$$
	, 
	suppose the following conditions are satisfied:
	\begin{itemize}
		\item (Sufficient security for $K$) $\ket{\varphi}$ is $(2^{\eta_1}, 2^{-\eta_1}|\ket{\varphi}|)$-SC-secure for $K^{(i)}$ given $Tag(K)$ and $Tag(\Lambda)$. $2^\kappa>\eta_1>8\kappa$.  $N<2^{\sqrt{\kappa}}$.
		\item (Sufficient security for $\Lambda$) For any $j\in [J]$, $\ket{\varphi}$ is $(2^{\eta_2}, 2^{-\eta_2}|\ket{\varphi}|)$-SC-secure for $\Lambda^{(j)}$ given $K$, $\Lambda-\Lambda^{(j)}$, and $\llbracket\fAuxInf\rrbracket$. $2^\kappa>\eta_2\geq 10000\kappa\eta_1$, $\eta_2^2>J\geq \eta_2$.
		\item (Well-behaveness of the inputs) $\ket{\varphi}\in \cWBS(D)$, $D\leq 2^\kappa$.
		\item (Sufficient padding length, output length) $\ell\geq 6D+8\eta_2$, $\kappa_{out}\geq \ell+4\eta_2$.
	\end{itemize}
	then the following conclusion holds:\par
	For any adversary $\fAdv$ with query number $ |\fAdv|\leq 2^{\kappa}$, denote the post-execution state as $$\ket{\varphi^\prime}=\fSecurityRefreshing_\fAdv(K,\Lambda;\underbrace{ \ell}_{\substack{\text{padding} \\ \text{length}}},
\underbrace{ \kappa_{\text{out}}}_{\substack{\text{output} \\ \text{length}}})\circ\ket{\varphi}$$
	and suppose the output keys are $K_{out}$, then: $P_{pass}\ket{\varphi^\prime}$ is $(2^{\eta_2/100\kappa}, 2^{-\eta_1/4+2\kappa+2}|\ket{\varphi}|)$-SC-secure for $K_{out}^{(i)}$ given $K_{out}-K_{out}^{(i)}$ and $\llbracket\fAuxInf\rrbracket$.
\end{lem}
\paragraph{How to understand this lemma} The parameter inequalities and conditions might seem complicated. We refer to Definition \ref{def:3.12b}. Informally speaking, we are proving the weak security of this protocol, and we explicitly write down the conditions (note that there are two initial key sets, and the conditions are not really the same as what is required in the weak security transform parameter --- one difference is the first condition does not take the other keys as the auxiliary information. This is important and it helps us use this protocol to overcome the difficulty in the security proof of the \emph{combine} technique.). And we note the inequalities in the first two conditions all have the form of $[\fpoly(\kappa),\fsubexp(\kappa)]$. The third condition is to rule out some ``ill-behaved'' cases, and the final condition just means the pad length and output key length should not be too small --- but they don't need to be too big either.\par
\paragraph{Why the lemma and protocol are very useful later} Let's continue the motivation discussion in the beginning of this section. As we said before, one motivation is to ``recover the first parameter in the SC-security'', which means, in the input part, initially the condition only says the state is $(2^{\eta_1},\cdots)$-SC-secure for the keys in $K$, but the output state can have a much better resilience on the query number of the attacker. There are two more additional importances: as we said before, the first condition is even weaker than what is required in the definition of the weak security. Furthermore, we can consider an additional auxiliary information $\llbracket\fAuxInf\rrbracket$. These two points, in some sense, ``de-correlate'' different key pairs, and broaden the application scope of this protocol. These properties are important on the road to a provable-secure \emph{combine} technique. \par
The proof is through a linear decomposition of the adversary's operations. The details are postponed to Appendix \ref{sec:AF}.\par
And we have the following immediate corollary, which deal with multiple single key pairs simultaneously instead of single key pair. The differences are on the first condition and the conclusion: whether $i$ is fixed in advance, or there are ``$\forall i$''. 
\begin{cor}\label{cor:101c2} The statement coming from replacing the first condition in Lemma \ref{lem:10.1} by
$$\text{$\forall i\in [N]$, $\ket{\varphi}$ is $(2^{\eta_1}, 2^{-\eta_1}|\ket{\varphi}|)$-SC-secure for $K^{(i)}$ given $Tag(K)$ and $Tag(\Lambda)$.}$$ $$\text{ $2^\kappa>\eta_1>8\kappa$.  $N<2^{\sqrt{\kappa}}$.}$$
and replacing the result (starting from ``then'') by
$$\text{$\forall i\in [N]$, $P_{pass}\ket{\varphi^\prime}$ is $(2^{\eta_2/100\kappa}, 2^{-\eta_1/4+2\kappa+2}|\ket{\varphi}|)$-SC-secure for $K_{out}^{(i)}$}$$ $$\text{ given $K_{out}-K_{out}^{(i)}$ and $\llbracket\fAuxInf\rrbracket$}$$
is also true.
\end{cor}
Now we can introduce new symbol to describe this statement using the notation of weak security transform parameters. Suppose we say a protocol run on key sets $K,\Lambda$ has weak security transform parameter \weakparamstar{\eta_1}{C}{\Lambda:\eta_2}{\eta_2^\prime}{C^\prime} (for input state in $\cF$ against adversaries of query number $\leq 2^\kappa$) to mean the protocol satisfies a security statement with a form similar to Definition \ref{def:3.12b}, with the following difference:\footnote{See discussion around Definition \ref{def:newdef} for why we treat $K$ and $\Lambda$ differently in the notation.}\par
The input security condition part is (1)$\forall i\in [N]$, the initial state is $(2^{\eta_1},C|\ket{\varphi}|)$-SC-secure for $K^{(i)}$ given $Tag(K,\Lambda)$; (2)$\forall j\in [J]$, the initial state is $(2^{\eta_2},C|\ket{\varphi}|)$-SC-secure for $\Lambda^{(j)}$ given the other keys ($K$, $\Lambda-\Lambda^{(j)}$).\par
Then we can say the $\fSecurityRefreshing$ layer has weak security transform parameter \weakparamstar{\eta_1}{2^{-\eta_1}}{\Lambda:\eta_2}{\eta_2/100\kappa}{2^{-\eta_1/4+2\kappa+2}} (for states in a reasonable range). This is not strong enough to capture everything described in the security statement above, but it characterize part of its properties in a way that is more consistent with the other parts of this work.
\section{Amplification, Part II (the ``combine'' part, and the whole protocol for remote gadget preparation)}\label{sec:10}
\subsection{Formalizing and Completing the ``Combine'' Technique: Remote Gadget Preparation Protocol with Normal Security}\label{sec:10.3}
Finally we can give a protocol that allows to at least double (asymtotically) the number of gadgets with normal security (instead of weak security). This protocol basically follows the idea in Section \ref{sec:8.3} and simplified protocols (Protocol \ref{prtl:10}): we can make the output keys exponentially secure by combining the output keys from different subprotocol calls. Here we add an extra $\fSecurityRefreshing$ layer in it, and that helps us complete the security proof.\par We name it as $\fGdgPrep^{OneRound}$ since the final protocol is from iterations of this protocol. 
\subsubsection{Protocol design}
To design such a protocol, we need to first revise the $Combine$ protocol a little bit to make it better.
\begin{mdframed}[style=figstyle]
\begin{prtl}\label{prtl:14n}
	$\fCombine^{improved}(K_1,K_2;\ell,\kappa_{out})$ is defined as follows, where $K_1=\{x_0,x_1\}$, $K_2=\{x_0^\prime,x_1^\prime\}$; $\ell$ is the pad length, $\kappa_{out}$ is the tag length.\par
	The honest server should hold $Gadget(K_1)\otimes Gadget(K_2)$. 
	\begin{enumerate}
		\item The client samples $R_0,R_1,R_0^\prime,R_1^\prime$ independently randomly from $\{0,1\}^l$. And it sends $H(R_0||x_0)$, $H(R_1||x_1)$, $H(R_0^\prime||x_0^\prime)$, $H(R_1^\prime||x_1^\prime)$ together with these random pads to the server.
		\item The server makes a measurement on the xor of the indexes:
		      $$(\ket{x_0}+\ket{x_1})\otimes (\ket{x^{\prime}_0}+\ket{x^{\prime}_1})\rightarrow$$
		      $$ (output=0)(\ket{x_0}\ket{x^{\prime}_0}+\ket{x_1}\ket{x^{\prime}_1})\qquad (output=1)(\ket{x_0}\ket{x^{\prime}_1}+\ket{x_1}\ket{x^{\prime}_0})$$
		      and sends the output to the client (which is a single bit). The client updates the keys based on the response from the server: if $output=0$, the client temporarily stores $\{x_0||x^\prime_0,x_1||x^{\prime}_1\}$; if $output=1$, the client temporarily stores $\{x_0||x^{\prime}_1,x_1||x^{\prime}_0\}$.
		\item The client samples two random pads $pad_0,pad_1\leftarrow_r\{0,1\}^l$ and sends it to the server. The client stores the output keys $K_{out}$ by concatenating $pad$ to the keys stored temporarily in the last step: it stores $K_{out}$ as $\{pad_0||x_0||x^\prime_0,pad_1||x_1||x^{\prime}_1\}$ or $\{pad_0||x_0||x^{\prime}_1,pad_1||x_1||x^{\prime}_0\}$, depending on the server's response in the last step. The former one in the output key set is considered to have subscript 0 and the later one is considered to have subscript 1.\par
		      The honest server can also pad the gadget and holds $Gadget(K_{out})$.
	\end{enumerate}
\end{prtl}\end{mdframed}
So compare to Protocol \ref{prtl:10}, in the last step the client samples a new random pad and replaces the keys by the padded keys. The reason for doing this is in the security proof of this protocol the client needs to provide a lot of \emph{global tags}. Such a padding allows us to analyze them more easily.\par
Now we can give our protocol.
\begin{mdframed}[style=figstyle]
\begin{prtl}[$\fGdgPrep^{OneRound}$ protocol]\label{prtl:14}
	Suppose the key sets $K,\Lambda$ satisfy:
	\begin{itemize}\item  $K:=\{K^{(t)}\}_{t\in [\sqrt{\kappa}]}$, and each $K^{(t)}$ is compatible with the input of\\ $\fGdgPrep^{M\times (\log\kappa+1)\rightarrow M\kappa}$ and has $M$ sets of keys (thus each $K^{(t)(m)} (t\in[\sqrt{\kappa}],m\in [M])$ contains $(\log\kappa+1)$ pairs of keys);\item $\Lambda:=\{\Lambda^{(t)}\}_{t\in [\sqrt{\kappa}]}$, and each $\Lambda^{(t)}$ has $J$ pairs of keys.
	\end{itemize}
	Then the protocol $\fGdgPrep^{OneRound}(K, \Lambda;\ell,\kappa_{out})$ is defined as follows, where $\ell$ is the pad length, $\kappa_{out}$ is the output key length:\par
	Initially the honest server should hold the state $Gadget(K)\otimes Gadget(\Lambda)$.\\
	For $t=1$ to $\sqrt{\kappa}$:
	\begin{enumerate}
		\item The client and the server execute $\fGdgPrep^{M\times (\log\kappa+1)\rightarrow M\kappa}(K^{(t)};\ell,\kappa_{out})$. Denote the returned output keys as $K_{temp\_t}$.  $K_{temp\_t}$ has $M\kappa$ pairs of keys.
		\item The client and the server execute $\fSecurityRefreshing(K_{temp\_t},\Lambda^{(t)};\ell,\kappa_{out})$. Denote the returned keys as $K_{temp^\prime\_t}$.
		\item If $t=1$, skip this step and simply let $K_{out\_1}=K_{temp^\prime\_1}$. Otherwise do the following:\\
		      For each $m\in [M\kappa]$:\begin{enumerate}
			      \item The client and the server run $\fCombine^{improved}(K^{(m)}_{temp^\prime\_t},K_{out\_(t-1)}^{(m)};\ell, \kappa_{out})$. The client stores the returned keys as $K_{out\_t}^{(m)}$.
		      \end{enumerate}
		\item Thus by the end of this round the client stores $K_{out\_t}$ and the honest server holds $Gadget(K_{out\_t})\otimes Gadget(K^{(>t)})\otimes  Gadget(\Lambda^{(>t)})$.
	\end{enumerate}
	Denote the returned keys in the last step as $K_{out}$. And the honest server holds $Gadget(K_{out})$.
\end{prtl}\end{mdframed}
\paragraph{Correctness} In summary, this protocol is a $\sqrt{\kappa}M(\log\kappa+1)+\sqrt{\kappa}J\rightarrow M\kappa$ remote gadget preparation protocol.
\paragraph{Efficiency} Both parties run in polynomial time (in key size and the parameters)
\paragraph{Gadget Expansion Ratio} Totally the \emph{gadget expansion ratio} is $\tilde\Theta(\sqrt{\kappa})>2$ (when $J$ is succinct). In the first step of each round, the number of gadgets increases by a factor of $\tilde\Theta(\kappa)$; and since the \emph{combine} technique combines $\sqrt{\kappa}$ gadgets into one big gadget, the number of gadgets decrease only by a factor of $\sqrt{\kappa}$ here. 
\paragraph{The ``extra'' gadgets are succinct} So compared to Protocol \ref{prtl:10}, this protocol requires $\sqrt{\kappa}\times J$ number of ``extra'' gadgets for the $\fSecurityRefreshing$ step. (Because in the $t$-th round $Gadget(\Lambda^{(t)})$ is consumed, and $\Lambda^{(t)}$ contains $J$ pairs of keys). The size of $K$ can be as big as an arbitrary polynomial, but $\Lambda$ is succinct, and the gadgets in $Gadget(\Lambda)$ are shared thus the total number of extra client side gadgets is still succinct.
 
\subsubsection{Security statement}
We note that in the security treatment we need to deal with the two key sets $K,\Lambda$ slightly differently. Let's first generalize the definition of weak security transform parameter (originally Definition \ref{def:3.12b}).\footnote{Let's explain why we put the security for $\Lambda$ above the arrow instead of the left hand side of the arrow: although in the protocol the two key sets are both part of the initial keys, their functionalities are different. This can be seen more clearly in the following weakened setting: we allow the quantum communication during the protocol, but the total amount of it should be small. Then $K$ will still be considered as part of the initial keys, but $\Lambda$ can be provided ``on the fly''. In some sense $\Lambda$ are ``freshly new gadgets'' that are provided additionally to ``refresh the security'' of $K$.}
\begin{defn}\label{def:newdef}
	We say an $N\rightarrow L$ remote gadget preparation protocol run on security parameter $\kappa$ has \emph{weak security transform parameter} \weakparamm{\eta}{C}{\Lambda:\eta_2}{\eta^\prime}{C^\prime} for initial state in $\cF$ (which is a set of states) against adversaries of query number $\leq 2^\kappa$ if a statement in the following form holds for the protocol:\\
	\begin{mdframed}
			Suppose the input keys are $K=\{x_b^{(i)}\}_{i\in [N],b\in\{0,1\}}$ and $\Lambda=\{r_b^{(j)}\}_{j\in [J],b\in\{0,1\}}$. Suppose the purified joint initial state $\ket{\varphi}$  satisfies the following conditions:
			\begin{itemize}
				\item $\forall i\in [N]$, $\ket{\varphi}$  is $(2^\eta,C|\ket{\varphi}|)$-SC-secure for $K^{(i)}$ given $K-K^{(i)}$ and $Tag(\Lambda)$.
				\item $\forall j\in [J]$, $\ket{\varphi}$  is $(2^{\eta_2},2^{-\eta_2}|\ket{\varphi}|)$-SC-secure for $\Lambda^{(j)}$ given $K$ and $\Lambda-\Lambda^{(j)}$.
				\item $\ket{\varphi}\in \cF$ 
			\end{itemize}
			For any adversary $\fAdv$ of query number $|\fAdv|\leq 2^\kappa$, the final state when the protocol completes, denoted as
			$$\ket{\varphi^\prime}=ProtocolName_\fAdv(K,\Lambda;Parameters)\circ\ket{\varphi}$$
			, (and correspondingly, output keys are $K_{out}=\{y_b^{(i)}\}_{i\in [L],b\in\{0,1\}}$) there is:
				$$\text{$\forall i\in [L], P_{pass}\ket{\varphi^\prime}$ is $(2^{\eta^\prime},C^\prime|\ket{\varphi}|)$-SC-secure for $K_{out}^{(i)}$ given $K_{out}-K_{out}^{(i)}$.}$$
			\end{mdframed}
\end{defn}

Note that we use \weakparamm{\eta}{C}{\eta_2}{\eta^\prime}{C^\prime} in the diagram in Section \ref{sec:4.5.1}.\par
Below is the security statement for Protocol \ref{prtl:14}.
\begin{lem}[Security of $\fGdgPrep^{OneRound}$]\label{lem:10.2} Define $\fTHRESHOLD(\kappa)$, constant $B_4>1$ as in Lemma \ref{lem:9.2}. The following statement is true for sufficiently large security parameter $\kappa$:\par
Protocol  $$\fGdgPrep^{OneRound}(K,\Lambda; \underbrace{ \ell}_{\substack{\text{padding} \\ \text{length}}},
\underbrace{ \kappa_{\text{out}}}_{\substack{\text{output} \\ \text{length}}})$$ has weak security transform parameter \weakparamm{\eta_1}{2^{-\eta_1}}{\Lambda:\eta_2}{\eta_2/350\kappa}{2^{-\sqrt{\kappa}/10}} for inputs in $\cWBS(D)$ against adversaries of query number $\leq 2^\kappa$ when the following inequalities are satisfied:
\begin{itemize}
\item (Sufficient security on the inputs) $2^{\sqrt{\kappa}}>\eta_1>48\kappa^{B_4+2}$.
\item (Suitable number of input keys) $2^{\sqrt{\kappa}}\geq M\geq \fTHRESHOLD(\kappa)$ (recall that $M$ is the number of key sets in $K^{(i)}$).
\item (Sufficient security on the ``extra gadgets'' $\Lambda$) $2^{\sqrt{\kappa}/5}\geq \eta_2\geq 40000\kappa\eta_1$. $\eta_2^2\geq J\geq \eta_2$.
\item (Well-behaveness of the inputs) $D\leq 2^{\sqrt{\kappa}}$
\item (Sufficient pad length and output key length) $\ell\geq 6D+24\eta_2+12$, $\kappa_{out}\geq \ell+4\eta_2$.
\end{itemize}

\end{lem}
We note that the final security statement has parameter $(2^{\Theta(\eta_2/\kappa)},1/\fsubexp(\kappa))$, while in the security statement of Protocol \ref{prtl:8} and \ref{prtl:9}, the first parameter is $2^{O(\eta_1)}$. Here the introduction of $\fSecurityRefreshing$ layer makes a difference: the first parameter in the SC-security in the security statement depends on $\eta_2$ instead of $\eta_1$. When we use this protocol in an upper level protocol, $\eta_2$ can be much bigger than $\eta_1$.\par
\subsubsection{Security proof}
The proof is given below. Let's first give a summary for different characters used in this proof.
\begin{itemize}
	\item $i$ is used for the index of keys.
	\item $t\in [\sqrt{\kappa}]$ denotes the round count of iterations, and is used as the index of the first dimension of $\Lambda$.
	\item $j\in [J]$ is used as the index of the second dimension in $\Lambda$: for example, $\Lambda^{(t^\prime)(j)}$.
\end{itemize}
\paragraph{Proof overview}We will use the multi-round decomposition method described in Section \ref{sec:4.8.3}. Overall speaking, we will repeat an argument of the same form for many rounds, and decompose the final state step-by-step. Note that in this protocol we iterate a block of subprotocols for $\sqrt{\kappa}$ rounds, so it has the structure described in Section \ref{sec:4.8.3}. But the details are more complicated. Note that in each round --- for example, at round $t$ --- the followings are executed: \begin{enumerate}\item Generate new gadgets; \item Run the security refreshing layer $\fSecurityRefreshing$; \item Combine the new gadgets to the old gadgets.\end{enumerate}
Correspondingly, each round (suppose it's the $t$-th round) of the proof goes as follows:
\begin{otl}[Proof outline of Lemma \ref{lem:10.2}, for round $t$]\label{otl:121}
\begin{enumerate}
\item[0.] The analysis starts with a state denoted by $\ket{\chi^{t}}$.
\item Prove the security properties for the newly-generated gadgets; The tools for this step are the properties of $\ket{\chi^t}$ and the security lemma for subprotocol ($\fGdgPrep^{M\times(\log\kappa+1)\rightarrow M\kappa}$).
\item[1.5.] Apply the decomposition lemma to decompose the state into two terms $\ket{\phi^{t.1}}$ and $\ket{\chi^{t.1}}$, corresponding to $\ket{\phi}$ and $\ket{\chi}$ in the state decomposition lemma. (This idea is discussed in Section \ref{sec:4.8.3}.) Let's discuss these two terms separately:
\begin{itemize}
\item For branch $\ket{\phi^{t.1}}$: note that the remaining protocols, denoted by $\fPrtl_{>t.1}$, contain (1) an execution of $\fSecurityRefreshing$ layer, (2) a $\fCombine^{improved}$ step, (3) several rounds of iterations of the same protocol.
\begin{enumerate}
\item[2.] The $\fSecurityRefreshing$ layer refreshes and strengthens the security property.
\item[3.] Prove the combination of the old gadgets, and the generation and combination of the future gadgets, do not affect the security too much. 
\end{enumerate}
\item For branch $\ket{\chi^{t.1}}$: the main property we need is the norm of this part decrease multiplicatively compared to the state in the beginning of this round. And the $\fSecurityRefreshing$ step and the combination step do not increase the norm. So we can simply go to analyze the next round.
\end{itemize}
\item[...]Finally we combine all the terms together:
\begin{align}P_{pass}\ket{\varphi^\prime}		=                             & P_{pass}\fPrtl_{>1.1}\circ\ket{\phi^{1.1}}+P_{pass}\fPrtl_{>2.1}\circ\ket{\phi^{2.1}}+\cdots                                                        \\
		                              & +P_{pass}\fPrtl_{>{\sqrt{\kappa}}.1}\circ\ket{\phi^{\sqrt{\kappa}.1}}+P_{pass}\fPrtl_{>\sqrt{\kappa}.1}\ket{\chi^{\sqrt{\kappa}.1}}
	\end{align}
 The $\ket{\phi}$ terms all have the security properties we want and can be summed up through triangle inequality; and the $\ket{\chi}$ term in the end is exponentially small.
\end{enumerate}
\end{otl}

 The following proof will be organized in the following way: \emph{we will divide the argument (in round $t$, for some arbitrary $t$) into pieces, and put it into boxes; outside the box, we will first show how the argument goes in the first round, as an explicit example, then show how the argument goes in an arbitrary round, and finally gives the decomposition and completes the proof.}
\begin{proof}[Proof of Lemma \ref{lem:10.2}]
Suppose the initial purified joint state is $\ket{\varphi}$, which is in $\cWBS(D)$ and is $(2^\eta,2^{-\eta}|\ket{\varphi}|)$-SC-secure for each key pair in $K$ given the other key pairs. And denote the post-execution state (for adversary $\fAdv$, $|\fAdv|\leq 2^\kappa$) as $\ket{\varphi^\prime}$.\par

	Without loss of generality, consider the $i$-th output keys $K^{(i)}_{out}$. Thus $i$ will be fixed in the following proof.\par
	By the auxiliary-information technique (Technique \ref{lem:4.2}) we assume the adversary gets $Tag(K,\Lambda)$ in the read-only buffer in advance. Additionally assume after the first step of each round of the iteration the client provides $Tag(K_{temp\_t}^{(i)})$ to the read-only buffer. This can only make the adversary more powerful thus proving the lemma under this assumption is enough to prove the original lemma.\par
	
	To make the symbol consistent in different steps of the induction, we define $\ket{\chi^1}:=\ket{\varphi}$. We will consider $\ket{\chi^{\cdots}}$ as the initial state of each round of the proof. And we know $\ket{\chi^1}$ satisfies the following conditions:\\
	
		\begin{mdframed}
			Properties of $\ket{\chi^1}$, the state before the 1st iteration:
			\begin{itemize}
				\item $\ket{\chi^1}$ is $(2^{\eta_1},2^{-\eta_1}|\ket{\varphi}|)$-SC-secure for any single pair of keys $K^\prime$ given $K-K^\prime$ and $Tag(K,\Lambda)$. 
				\item $\forall t^\prime\in [\sqrt{\kappa}]$, $\forall j\in [J]$, $\ket{\chi^1}$ is $(2^{\eta_2},2^{-\eta_2}|\ket{\varphi}|)$-SC-secure for $\Lambda^{(t^\prime)(j)}$ given $\Lambda-\Lambda^{(t^\prime)(j)}$ and $K$. (The reason that we use $t^\prime$ instead of $t$ is to make it consistent with the conditions for general $\ket{\chi^t}$.) 
				\item $\ket{\chi^1}$ is $(2^{D},2^{D})$-representable from $\ket{\mathfrak{init}}$.
			\end{itemize}\end{mdframed}
	Let's first analyze the first round ($t=1$) of this protocol.\par
	\textbf{Let's first analyze the property of the output state after the first step} of the first round of the protocol, which can be denoted as $\fPrtl_{=1.1}$, an execution of $\fGdgPrep^{M\times(\log\kappa+1)\rightarrow M\kappa}$ protocol. We can apply Lemma \ref{lem:9.2} (the security of $\fGdgPrep^{M\times(\log\kappa+1)\rightarrow M\kappa}$) and prove the following:
	\begin{equation}\label{eq:81}
		\text{$P_{pass}\fPrtl_{=1.1}\ket{\chi^{1}}$ is $(2^{\eta_1/3{\kappa^{B_4}}},\frac{1}{3}|\ket{\varphi}|)$-SC-secure for $K_{temp\_1}^{(i)}$ given $K_{temp\_1}-K_{temp\_1}^{(i)}$}
	\end{equation}
	The details for proving (\ref{eq:81}) is given below. First note that the following proof is written in a way to support general time step $t$. In this step we can simply choose $t=1$. 
	\begin{mdframed}
		\textbf{How to prove $P_{pass}\fPrtl_{=t.1}\ket{\chi^{t}}$ is $(2^{\eta_1/3{\kappa^{B_4}}},(\frac{2.5}{3})^{t-1}\frac{1}{3}|\ket{\varphi}|)$-SC-secure for $K_{temp\_t}^{(i)}$ given $K_{temp\_t}-K_{temp\_t}^{(i)}$:}\\
		If $|\ket{\chi^t}|\leq 2^{-\kappa}|\ket{\varphi}|$, the statement is already true.\\ Otherwise we can get the followings from ``properties of $\ket{\chi^t}$'':
		\begin{itemize}
			\item $(\frac{2.5}{3})^{t-1}|\ket{\varphi}|\geq |\ket{\chi^t}|$. And we know $|\ket{\chi^t}|\geq 2^{-\kappa}|\ket{\varphi}|$
			\item $\ket{\chi^t}$ is $(2^{\eta_1/2},2^{-\eta_1/2}|\ket{\chi^t}|)$-SC-secure for any single pair of keys $K^\prime$ in $K^{(t)}$ given $K^{(t)}-K^\prime$. Note $\eta_1/2>6\kappa^{B_4+2}$, as required in Lemma \ref{lem:9.2}.
			\item $\ket{\chi^t}$ is $(2^{D+\eta_1},2^{D}+2^{\eta_1})$-representable from $\ket{\mathfrak{init}}$. We can see the pad length is enough.
		\end{itemize}
		Note that we relax some parameters to make these  statements invariant in each round of this induction-style proof.\par
		Thus by Lemma \ref{lem:9.2} we know $P_{pass}\fPrtl_{=t.1}\ket{\chi^{t}}$ is is $(2^{\eta_1/3{\kappa^{B_4}}},\frac{1}{3}|\ket{\chi^t}|)$-SC-secure for $K_{temp\_t}^{(i)}$ given $K_{temp\_t}-K_{temp\_t}^{(i)}$ and this completes the proof.
	\end{mdframed}
	Then by applying Lemma \ref{lem:4.4} to \textbf{decompose the state}: $P_{pass}\fPrtl_{=1.1}\ket{\chi^{1}}$ can be decomposed as $\ket{\phi^{1.1}}+\ket{\chi^{1.1}}$, as follows. As before, the decomposition below is written under general round count $t$, and for this step we can simply substitute $t=1$:\\
	\begin{mdframed}
			\textbf{State Decomposition: $P_{pass}\fPrtl_{=t.1}\ket{\chi^t}$ to $\ket{\phi^{t.1}}+\ket{\chi^{t.1}}$:}\par
			If $|P_{pass}\fPrtl_{=t.1}\ket{\chi^t}|\leq 2^{-\kappa}|\ket{\varphi}|$, simply choose $\ket{\phi^{t.1}}=0$. Otherwise by Fact \ref{fact:injtag} and Lemma \ref{lem:4.4}:
			\begin{itemize}
				\item $|\ket{\chi^{t.1}}|\leq (\frac{2.5}{3})|P_{pass}\fPrtl_{=t.1}\ket{\chi^t}|\leq (\frac{2.5}{3})^t|\ket{\varphi}|$ and is $(\eta_1/3{\kappa^{B_4}},2^{\eta_1/3{\kappa^{B_4}}})$-server-side-representable from $P_{pass}\fPrtl_{=t.1}\ket{\chi^{t}}$.
				\item  $\ket{\phi^{t.1}}$ is $(2^{\eta_1/18\kappa^{B_4}},2^{-\eta_1/18\kappa^{B_4}}\frac{1}{3}|\ket{\varphi}|)$-SC-secure for $K_{temp\_t}^{(i)}$, and is $(1,2^{\eta_1/3\kappa^{B_4}})$-server-side-representable from $P_{pass}\fPrtl_{=t.1}\ket{\chi^{t}}$.
			\end{itemize}\end{mdframed}
	Then the final state can be written as:
	\begin{align}
		P_{pass}\ket{\varphi^\prime}= & P_{pass}\fPrtl_{\geq 1}\ket{\varphi}\label{eq:7.19}                                                                                  \\
		=                             & P_{pass}\fPrtl_{>1.1}\circ\ket{\phi^{1.1}}+P_{pass}\fPrtl_{>1.1}\circ\ket{\chi^{1.1}}\label{eq:7.20}                                   \\
		=                             & P_{pass}\fPrtl_{\geq 2}\circ \fPrtl_{1.3}\circ \fPrtl_{1.2}\circ\ket{\phi^{1.1}}+P_{pass}\fPrtl_{\geq 2}\circ\ket{\chi^{2}}\label{eq:7.21} \\
		                              & \text{(where we denote $\ket{\chi^2}:=\fPrtl_{1.3}\circ \fPrtl_{1.2}\circ\ket{\chi^{1.1}}$)}\label{eq:91rr}
	\end{align}
	\textbf{We first analyze the $\ket{\phi^{1.1}}$ term} in (\ref{eq:7.21}). Define
	\begin{equation}\label{eq:92rrr}\ket{\phi^{1.2}}:=P_{pass}\fPrtl_{1.2}\circ\ket{\phi^{1.1}}=P_{pass}\fSecurityRefreshing_{\fAdv_{1.2}}(K_{temp\_1},\Lambda^{(1)})\circ\ket{\phi^{1.1}}\end{equation}.
	We can apply the security property of the $\fSecurityRefreshing$ protocol (Lemma \ref{lem:10.1}) on initial state $\ket{\phi^{1.1}}$: to make the conclusion strong enough for later usage, take
	\begin{equation}\label{eq:92rr}\llbracket\fAuxInf^1\rrbracket=\Lambda^{(\geq 2)}\cup K^{(\geq 2)}\cup (K_{out\_0}-K_{out\_0}^{(i)})\end{equation}
	(the last term is empty, since in the first round we simply define $K_{out\_1}$ to be $K_{temp^\prime\_1}$ and there is no ``output keys from the last round''. but we add this to make its form consistent with the auxiliary information for general round count $t$. 
	This will be discussed again later.)\par
	Then we can apply Lemma \ref{lem:10.1} and conclude that {\footnotesize\begin{equation}\label{eq:93r}\text{$\ket{\phi^{1.2}}$ is $(2^{\eta_2/300\kappa},2^{-\kappa}|\ket{\varphi}|)$-SC-secure for $K_{temp^\prime\_1}^{(i)}$ given $K_{temp^\prime\_1}-K_{temp^\prime\_1}^{(i)}$ and $\llbracket\fAuxInf^1\rrbracket$}\end{equation}}
	The details for this step is given below. The proof is described for general $t$. Some notations for general $t$ will be defined later. Substitute $t=1$ we get the proof we need currently.\\
	\begin{mdframed}
		\textbf{How to prove the property of $\ket{\phi^{t.2}}$}\par
		We want to prove
		\begin{equation}\label{eq:115}\text{$\ket{\phi^{t.2}}$ is $(2^{\eta_2/300\kappa},2^{-\kappa}|\ket{\varphi}|)$-SC-secure for $K_{temp^\prime\_t}^{(i)}$}\end{equation}\begin{equation*}\text{ given $K_{temp^\prime\_t}-K_{temp^\prime\_t}^{(i)}$ and $\llbracket\fAuxInf^t\rrbracket$}\end{equation*}
		where
		\begin{equation}\label{eq:117n}\llbracket\fAuxInf^t\rrbracket:=\Lambda^{(\geq t+1)}\cup K^{(\geq t+1)}\cup K_{out\_(t-1)}\end{equation}
		\begin{equation}\label{eq:116}\ket{\phi^{t.2}}:=P_{pass}\fSecurityRefreshing_{\fAdv_{t.2}}(K_{temp\_t},\Lambda^{(t)})\circ\ket{\phi^{t.1}}\end{equation}
		First, if $|\ket{\phi^{t.1}}|\leq 2^{-\kappa}|\ket{\varphi}|$, the final statement is already true. Otherwise, we can verify that $\ket{\phi^{t.1}}$ satisfies the conditions for applying Lemma \ref{lem:10.1}, as follows:
		\begin{enumerate}
			\item We already proved $\ket{\phi^{t.1}}$ is $(2^{\eta_1/18\kappa^{B_4}},2^{-\eta_1/18\kappa^{B_4}}\frac{1}{3}|\ket{\varphi}|)$-SC-secure for $K_{temp\_t}^{(i)}$ given\\ $Tag(K_{temp\_t},\Lambda^{(t)})$. (See the box of ``State Decomposition''.) Thus
			      \begin{equation}
				      \text{$\ket{\phi^{t.1}}$ is $(2^{\eta_1/18\kappa^{B_4}},2^{-\eta_1/18\kappa^{B_4}+\kappa}|\ket{\phi^{t.1}}|)$-SC-secure for $K_{temp\_t}^{(i)}$}\end{equation}\begin{equation}\text{ given $Tag(K_{temp\_t},\Lambda^{(t)})$.}
			      \end{equation}

			\item We can prove \begin{equation*}\text{$\forall j\in [J]$, $\ket{\phi^{t.1}}$ is $(2^{\eta_2/3},2^{-\eta_2/3}|\ket{\phi^{t.1}}|)$-SC-secure for $\Lambda^{(t)(j)}$ given}\end{equation*}\begin{equation}\label{eq:122} (\Lambda^{(t)}-\Lambda^{(t)(j)})\cup\llbracket\fAuxInf^t\rrbracket\end{equation}. Because:
			      \begin{enumerate}
				      \item From the ``Properties of $\ket{\chi^t}$'' we can know $\ket{\chi^t}$ is \\$(2^{\eta_2}-t2^{\eta_2/3},2^{-\eta_2+t\log\eta_2}|\ket{\varphi}|)$-SC-secure for $\Lambda^{(t)(j)}$ given $\Lambda^{(\geq t)}-\Lambda^{(t)(j)}$ and $K_{out\_(t-1)}$ and $K^{(\geq t)}$.
				      \item The computation of client side messages in $\fPrtl_{=t.1}$ is run only on key set $K^{(t)}$, $K_{temp\_t}$ (sampled randomly in this step) and random coins. The adversary makes queries $\leq 2^\kappa$.
				      \item $\ket{\phi^{t.1}}$ is $(1,2^{\eta_1/\kappa})$-server-side-representable from $P_{pass}\fPrtl_{=t.1}\ket{\chi^{t}}$.
			      \end{enumerate}
			      Applying Lemma \ref{lem:basic} and noting $|\ket{\phi^{t.1}}|\geq 2^{-\kappa}|\ket{\varphi}|$ give (\ref{eq:122}).
			\item Since $\ket{\chi^t}$ is $(2^{D+t\log\eta_1},2^{D}+t2^{\eta_1})$-representable from $\ket{\mathfrak{init}}$, by the same reasons as above, we know $\ket{\phi^{t.1}}$ is $(2^{\eta_1+D},2^{D}+2^{2\eta_1})$-representable from $\ket{\mathfrak{init}}$. Thus the pad length is enough.
		\end{enumerate}
		Thus applying Lemma \ref{lem:10.1} completes the proof.

	\end{mdframed}
	We note since (informally) $K_{temp^\prime\_1}-K_{temp^\prime\_1}^{(i)}$ and $\llbracket\fAuxInf^1\rrbracket$ contain enough information to allow the adversary simulate the protocol after this round, we can apply Lemma \ref{lem:basic} and get
	\begin{equation}\label{eq:95r}P_{pass}\fPrtl_{\geq 2}\circ \fPrtl_{1.3}\ket{\phi^{1.2}}\text{ is $(2^{\eta_2/350\kappa},2^{-0.9\kappa}|\ket{\varphi}|)$-SC-secure for $K_{out}^{(i)}$ given $K_{out}-K_{out}^{(i)}$.}\end{equation}
	The details for this part of the proof are given below. The proof is described for general $t$. Substitute $t=1$ we get the proof we need currently.
	\begin{mdframed}
		\textbf{How to prove the property of $P_{pass}\fPrtl_{>t.1}\ket{\phi^{t.1}}=P_{pass}\fPrtl_{\geq (t+1)}\circ \fPrtl_{t.3}\ket{\phi^{t.2}}$}\par
		What we know is (from (\ref{eq:115})) $\ket{\phi^{t.2}}$ is $(2^{\eta_2/300\kappa},2^{-\kappa}|\ket{\varphi}|)$-SC-secure for $K_{temp^\prime\_t}^{(i)}$ given $K_{temp^\prime\_t}-K_{temp^\prime\_t}^{(i)}$ and $\llbracket\fAuxInf^t\rrbracket$. And we want to prove
		\begin{equation}\label{eq:95rgr}P_{pass}\fPrtl_{\geq t+1}\circ \fPrtl_{t.3}\ket{\phi^{t.2}}\text{ is $(2^{\eta_2/350\kappa},2^{-0.9\kappa}|\ket{\varphi}|)$-SC-secure for $K_{out}^{(i)}$}\end{equation}\begin{equation*}\text{ given $K_{out}-K_{out}^{(i)}$.}\end{equation*}
		Define $\ket{\phi^{t.3}}:=\fPrtl_{t.3}\ket{\phi^{t.2}}$. In $\fPrtl_{t.3}$ (which are $\fCombine^{improved}(K_{temp^\prime\_t}^{(m)}, K_{out\_{(t-1)}}^{(m)})$ for each $m$), for index $i$ (not ``for any index $i$''; recall that $i\in [\kappa M]$ is already fixed in the beginning of the security proof), the followings are executed (see Protocol \ref{prtl:14n}):
		\begin{enumerate}
			\item The client sends some hash values about $K_{temp^\prime\_t}^{(i)}$ and $K_{out\_(t-1)}^{(i)}$.
			\item The server sends back a one-bit response.
			\item The client computes $\tilde K_{out\_t}^{(i)}$, defined to be the key set by the completion of the second step of the $\fCombine^{improved}$ protocol. This is a (deterministic) function of the server's response and $K_{temp^\prime\_t}^{(i)}$ and $K_{out\_(t-1)}^{(i)}$. And the keys in $K_{temp^\prime\_t}^{(i)}$ and $K_{out\_(t-1)}^{(i)}$ can be extracted given $\tilde K_{out\_t}^{(i)}$ and the server's response in the last step.
			\item The client samples $pad_0,pad_1\leftarrow_r\{0,1\}^l$, and $K_{out\_t}^{(i)}$ comes from adding $pad_0,pad_1$ before $\tilde K_{out\_t}^{(i)}$.
		\end{enumerate}
		We proceed in two steps.
		\begin{enumerate}\item Starting from (\ref{eq:115}), we will prove
		\begin{equation*}
			\ket{\phi^{t.3}}\text{ is $(2^{\eta_2/300\kappa-20},2^{-\kappa+20}|\ket{\varphi}|)$-SC-secure for $K_{out\_t}^{(i)}$ given}\end{equation*}\begin{equation}\label{eq:125nn}\text{ $K_{out\_t}-K_{out\_t}^{(i)}$ and $\llbracket\fAuxInf^t\rrbracket$}
		\end{equation}
		Note that there are two things that we need to care about: (1) when we discuss the SC-security of $K_{out\_t}^{(i)}$, $Tag(K_{out\_t}^{(i)})$ needs to be revealed to the server; (2) the protocol in this step is interactive. But these two issues can be handled, and the proof of (\ref{eq:125nn}) is as follows:\par
		Suppose (\ref{eq:125nn}) is not true.
		Notice that no matter what the adversary returns in the $i$-th round of $\fPrtl_{=t.3}$, if in the end it can compute both keys in $K_{out\_t}^{(i)}$, it can also compute both keys in $K_{temp^\prime\_t}^{(i)}$ by ``extracting'' the corresponding bits. In other words, if (\ref{eq:125nn}) does not hold, the adversary can run the following attack on $\ket{\phi^{t.2}}$. when $K_{temp^\prime\_t}-K_{temp^\prime\_t}^{(i)}$ and $\llbracket\fAuxInf^t\rrbracket$ are given as auxiliary information, to break the SC-security for $K_{temp^\prime\_t}^{(i)}$:
		\begin{enumerate}
			\item The adversary goes through the operation of $\fPrtl_{t.3}$ and the adversary's operation step-by-step. For each round of $\fPrtl_{t.3}$ that is not at superscript $(i)$, since $K_{temp^\prime\_t}-K_{temp^\prime\_t}^{(i)}$ and $K_{out\_(t-1)}-K_{out\_(t-1)}^{(i)}$ (in $\llbracket\fAuxInf^t\rrbracket$) are provided as auxiliary information, the adversary can simulate all the client-side messages; thus it can simulate all the steps locally. (When we say ``simulate'' we do not mean the server can create the same state that we consider in the purified notation; we use the usual meaning: the server can create a state that looks the same if we switch to the non-purified notation (by throwing away the environment).)
			\item For the $i$-th round inside $\fPrtl_{t.3}$ (which is\\ $\fCombine^{improved}(K_{temp^\prime\_t}^{(i)}, K_{out\_(t-1)}^{(i)})$, see Protocol \ref{prtl:14n}), the adversary:
			      \begin{enumerate}\item In the first step the adversary asks the client to provide the hash values, as shown in the first step of Protocol \ref{prtl:14n}; \item The adversary simulates the remaining protocol; this is possible using the information that it can read. \item And it asks the client to provide $Tag(K_{out\_t}^{(i)})$.\end{enumerate}
		\end{enumerate}
		So this is not an attack that can be executed completely on the server-side; instead, in this ``attack'' the attacker can ask the client to provide some padded hash values, which corresponds to the client-side messages in the protocol.\par
		 Notice that the client side messages in the step $b$ above include
		 \begin{equation}\label{eq:140qa}
		 \text{Hash tags of $K_{temp^\prime\_t}^{(i)}$ shown in the first step of Protocol \ref{prtl:14n}}
		 \end{equation}
\begin{equation}\label{eq:141qa}
		 		 \text{Hash tags of $K_{out\_(t-1)}^{(i)}$ shown in the first step of Protocol \ref{prtl:14n}}	
		 \end{equation}
		 \begin{equation}\label{eq:142qa}
		 Tag(K_{out\_t}^{(i)})
		 \end{equation}
		 where (\ref{eq:140qa})(\ref{eq:142qa}) can be seen as  hash tags of $K_{temp^\prime\_t}^{(i)}$ (with extra paddings), and (\ref{eq:141qa}) can be computed on the server-side. Thus by applying Lemma \ref{lem:4.9} to handle (\ref{eq:140qa})(\ref{eq:142qa}) we get an upper bound on the ``norm of outputting both keys in $K_{temp^\prime\_t}^{(i)}$'' of this attack. On the other hand if (\ref{eq:125nn}) is not true we give an attack to break it. Thus we complete the proof of (\ref{eq:125nn}).
		\item Then we move to study $\fPrtl_{\geq t+1}\ket{\phi^{t.3}}$. We can study each round in $\fPrtl_{\geq t+1}$ step-by-step. Without loss of generality, suppose this is in the $t^\prime$-th round. $t^\prime\in [t+1,\sqrt{\kappa}]$. We can prove the following statement inductively:
		\begin{center}$P_{pass}\fPrtl_{(t+1)\sim t^\prime}\ket{\phi^{t.3}}$ is $(2^{\eta_2/300\kappa-20-22(t^\prime-t)},2^{-\kappa+20+22(t^\prime-t)}|\ket{\varphi}|)$-SC-secure for $K_{out\_t^\prime}^{(i)}$ given $K_{out\_t^\prime}-K_{out\_t^\prime}^{(i)}$ and $\llbracket\fAuxInf^t\rrbracket$.\end{center} 
		Similar to the proof above, let's assume this is not true for $t^\prime$ and construct an interactive attack for the server and lead to a contradiction.
		\begin{enumerate}\item First notice that the first and second step of $\fPrtl_{=t^\prime}$ can all be simulated using $K^{(\geq t+1)}$ and $\Lambda^{(\geq t+1)}$; \item And within the third step of each round, the $\fCombine^{improved}$ protocol on key sets with superscripts that are not $(i)$ can also be simulated;\par
			      And every time we meet a $\fCombine^{improved}$ protocol on key sets with superscript $(i)$, we apply a similar argument as above (the step (b) of the arguments below (\ref{eq:125nn})).\end{enumerate}
			       And we can construct an interactive attack for the adversary where the client side messages are
			      		 \begin{equation}\label{eq:140qaa}
		 \text{Hash tags of $K_{temp^\prime\_t^\prime}^{(i)}$ shown in the first step of Protocol \ref{prtl:14n}}
		 \end{equation}
\begin{equation}\label{eq:141qaa}
		 		 \text{Hash tags of $K_{out\_(t^\prime-1)}^{(i)}$ shown in the first step of Protocol \ref{prtl:14n}}	
		 \end{equation}
		 \begin{equation}\label{eq:142qaa}
		 Tag(K_{out\_t^\prime}^{(i)})
		 \end{equation}
			       One difference is we should reduce the SC-security for $K_{out\_t^\prime}^{(i)}$ to the SC-security of $K_{out\_(t^\prime-1)}^{(i)}$ (instead of $K_{temp^\prime\_t}^{(i)}$). Here (\ref{eq:140qaa}) could be computed from the auxiliary information but (\ref{eq:141qaa})(\ref{eq:142qaa}) are handled by Lemma \ref{lem:4.9}.\par
		\end{enumerate}
		Finally taking $t^\prime=\sqrt{\kappa}$ completes the proof.
	\end{mdframed}
	\textbf{Then let's analyze the second term in equation (\ref{eq:7.21})}. We can view $\ket{\chi^2}$ as the new initial state and use the same technique above. We only need to verify that $\ket{\chi^2}$ has the necessary properties. (Note that we want to use the argument ``inductively'', we will describe the properties of $\ket{\chi^t}$ directly, which are the states that will appear step-by-step in this proof, and taking $t=2$ gives the conditions we need for this round:)\\
	\begin{mdframed}
		\textbf{Properties of $\ket{\chi^t}$, the state we consider before the $t$-th round of iterations:}\\
		We have
		\begin{itemize}
			\item For any single pair of keys $K^\prime$ in $K^{(\geq t)}$, $\ket{\chi^t}$ is $(2^{\eta_1}-t2^{\eta_1/3},2^{-\eta_1+t\log\eta_1}|\ket{\varphi}|)$-SC-secure for $K^\prime$ given $K^{(\geq t)}-K^\prime$ and $K_{out\_(t-1)}$.\par

			\item $\forall t^\prime\geq t$, $j\in [J]$, $\ket{\chi^t}$ is $(2^{\eta_2}-t2^{\eta_1/3},2^{-\eta_2+t\log\eta_1}|\ket{\varphi}|)$-SC-secure for any $\Lambda^{(t^\prime)(j)}$ given $K^{(\geq t)}$, $\Lambda^{(\geq t)}-\Lambda^{(t^\prime)(j)}$ and $K_{out\_(t-1)}$.\par
			\item $\ket{\chi^t}$ is $(2^{D+t\log\eta_1},2^{D}+t2^{\eta_1})$-representable from $\ket{\mathfrak{init}}$.
			\item $|\ket{\chi^t}|\leq (\frac{2.5}{3})^{t-1}|\ket{\varphi}|$
		\end{itemize}
		The first three properties can all be proved inductively (through Lemma \ref{lem:basic}), since we can notice the following about the protocol:
		\begin{enumerate}
			\item In $\fPrtl_{(t-1).1}$ the computation of all the client-side messages only take $K^{(t-1)}$ and $K_{temp\_(t-1)}$ (sampled randomly in this step) as the input.
			\item When we do the decomposition, $\ket{\chi^{(t-1).1}}$ is $(\eta_1/2\kappa^{B_4},2^{\eta_1/2\kappa^{B_4}})$-server-side-representable from $P_{pass}\fPrtl_{(t-1).1}\ket{\chi^{t-1}}$. 
			\item In $\fPrtl_{(t-1).2}$ the computation of all the client side messages only take $K_{temp\_(t-1)}$, $\Lambda^{(t-1)}$ and $K_{temp^\prime\_(t-1)}$ (sampled randomly in this step) as the input.
			\item In $\fPrtl_{(t-1).3}$, the new keys $K_{out\_(t-1)}$ can be computed on the server-side from its response, $K_{out\_(t-2)}$ and $K_{temp^\prime\_(t-1)}$, and the random pads.
			\item The query number of the adversary is bounded by $2^\kappa$.
		\end{enumerate}
		Then we can complete the proof of the first three properties. The fourth property (norm of $\ket{\chi^t}$) appeared during the argument in the last round (see the ``State Decomposition'' box).
	\end{mdframed}
	Now we have completed the argument for one round. Let's describe how the argument goes in round $t$. We have put pieces of this argument into boxes. And the overall structure of the proof is actually the same as what we described in the $t=1$ case:
	\begin{enumerate}
		\item From ``the conditions on $\ket{\chi^t}$'' we can prove (which is analogous to (\ref{eq:81}), but is for general $t$.)
		      \begin{equation}\label{eq:81g}
			      \text{$P_{pass}\fPrtl_{=t.1}\ket{\chi^{t}}$ is $(2^{\eta_1/3{\kappa^{B_4}}},(\frac{2.5}{3})^{t-1}\frac{1}{3}|\ket{\varphi}|)$-SC-secure for $K_{temp\_t}^{(i)}$}\end{equation}\begin{equation*} \text{given $K_{temp\_t}-K_{temp\_t}^{(i)}$}
		      \end{equation*}
		      This can be proved using the same argument as described in ``How to prove the properties of $P_{pass}\fPrtl_{=t.1}\ket{\chi^t}$'' before. Here we implicitly use the fact that $\eta_1-t\log\eta_1>\eta_1/2$ and $\eta_2-t\log\eta_2>\eta_2/2$ hold when $t\leq \sqrt{\kappa}$. (This is the reason that we cannot do these arguments for $t\rightarrow +\infty$.)\par
		\item Starting from $P_{pass}\fPrtl_{=t.1}\ket{\chi^{t}}$, we can continue to apply the decomposition lemma (as shown in the box ``State decomposition $P_{pass}\fPrtl_{=t.1}\ket{\chi^{t}}$ to $\ket{\phi^{t.1}}+\ket{\chi^{t.1}}$'').
		\item Substitute the decomposition back to the protocol, we have
		      \begin{equation}\label{eq:132rr}P_{pass}\fPrtl_{\geq t}\ket{\chi^t}=P_{pass}\fPrtl_{\geq t+1}\circ \fPrtl_{t.3}\circ \fPrtl_{t.2}\circ\ket{\phi^{t.1}}+P_{pass}\fPrtl_{\geq t+1}\circ\ket{\chi^{t+1}}\end{equation}
		      where we denote $\ket{\chi^{t+1}}:=\fPrtl_{t.3}\circ \fPrtl_{t.2}\circ\ket{\chi^{t.1}}$.
		\item For the $\ket{\phi^{\cdots}}$ part in (\ref{eq:132rr}):\par
		      Define $\ket{\phi^{t.2}}$, $\llbracket\fAuxInf^t\rrbracket$ similarly to (\ref{eq:92rrr})(\ref{eq:92rr}),
		      \begin{equation}\label{eq:92rg}\ket{\phi^{t.2}}:=P_{pass}\fSecurityRefreshing_{\fAdv_{t.2}}(K_{temp\_t},\Lambda^{(t)})\circ\ket{\phi^{t.1}}\end{equation}
		      \begin{equation}\label{eq:92g}\llbracket\fAuxInf^t\rrbracket:=\Lambda^{(\geq t+1)}\cup K^{(\geq t+1)}\cup K_{out\_(t-1)}\end{equation}
		      we can prove the following statement, using the proof described in box ``How to prove the property of $\ket{\phi^{t.2}}$'':
		      \begin{equation}\label{eq:93g}\text{$\ket{\phi^{t.2}}$ is $(2^{\eta_2/300\kappa},2^{-\kappa}|\ket{\varphi}|)$-SC-secure for $K_{temp^\prime\_t}^{(i)}$}\end{equation}\begin{equation*}\text{ given $K_{temp^\prime\_t}-K_{temp^\prime\_t}^{(i)}$ and $\llbracket\fAuxInf^t\rrbracket$}\end{equation*}
		      And similarly to (\ref{eq:95r}), by the argument in the box ``How to prove the properties of $P_{pass}\fPrtl_{>t.1}\ket{\phi^{t.1}}=P_{pass}\fPrtl_{\geq (t+1)}\circ \fPrtl_{t.3}\ket{\phi^{t.2}}$, we can get
		      \begin{equation}P_{pass}\fPrtl_{\geq (t+1)}\circ \fPrtl_{t.3}\ket{\phi^{t.2}}\text{ is $(2^{\eta_2/350\kappa},2^{-0.9\kappa}|\ket{\varphi}|)$-SC-secure for $K_{out}^{(i)}$}\end{equation}\begin{equation*}\text{ given $K_{out}-K_{out}^{(i)}$.}\end{equation*}
		\item Finally we can prove the properties of $\ket{\chi^{t+1}}$ (using the arguments in the box with the same title) and continue to the next round of this argument.
	\end{enumerate}
	Continue these arguments inductively and we get
	\begin{align}P_{pass}\ket{\varphi^\prime}= & P_{pass}\fPrtl_{>1.1}\circ\ket{\phi^{1.1}}+P_{pass}\fPrtl_{>1.1}\circ\ket{\chi^{1.1}}\label{eq:133nn}                                               \\
		=                             & \cdots                                                                                                                                          \\
		=                             & P_{pass}\fPrtl_{>1.1}\circ\ket{\phi^{1.1}}+P_{pass}\fPrtl_{>2.1}\circ\ket{\phi^{2.1}}+\cdots                                                        \\
		                              & +P_{pass}\fPrtl_{>{\sqrt{\kappa}}.1}\circ\ket{\phi^{\sqrt{\kappa}.1}}+P_{pass}\fPrtl_{>\sqrt{\kappa}.1}\ket{\chi^{\sqrt{\kappa}.1}}\label{eq:136nn}
	\end{align}
	And the last term has norm $\leq (\frac{2.5}{3})^{\sqrt{\kappa}}|\ket{\varphi}|$. And each other term is $(2^{\eta_2/350\kappa},2^{-0.9\kappa}|\ket{\varphi}|)$-SC-secure for $K_{out}^{(i)}$ given $K_{out}-K_{out}^{(i)}$. We can see $(\frac{2.5}{3})^{\sqrt{\kappa}}>\sqrt{\kappa} 2^{-0.9\kappa}$. Combining them together using triangle inequality of SC-security (Lemma \ref{lem:4.2}) and doing a slight relaxing using $\log_2(\frac{2.5}{3})<-1/10$ gives us the conclusion.
\end{proof}
\subsection{Full Formal Description of the Secure Remote Gadget Preparation Protocol}\label{sec:10.4}
\subsubsection{Protocol design}
Finally we put all the pieces together and give the final protocol for remote gadget preparation. This will complete Outline \ref{otl:2} and complete the first step in Outline \ref{ppl:1}.\par
Since in the last section we have already got a secure protocol that can double the number of gadgets, intuitively we can get an $N\rightarrow L$ protocol by simply repeating it for $\log L/N$ times: starting from $N$ gadgets, in each step, the number of gadgets becomes $2N,4N,\cdots, L$, with $\sqrt{\kappa} J\log L/N$ extra gadgets.\par
However there is some loss on the second parameter of the SC-security of the output keys. (Which means, in Lemma \ref{lem:10.2} the final state is $(2^{\Theta(\eta_2/\kappa)},2^{-\Theta(\sqrt{\kappa})}|\ket{\varphi}|)$-SC-secure for the output keys, instead of $(2^{\Theta(\eta_2/\kappa)},2^{-\Theta(\eta_2/\kappa)}|\ket{\varphi}|)$-SC-secure for the output keys.) The first idea to solve this problem is to use the decomposition lemmas (in Section \ref{sec:4.3}) and the linear decomposition method (in Section \ref{sec:4.8.2}). However there is still some obstacle when we do the security proof. To solve this problem, we add another $\fSecurityRefreshing$ layer into the protocol. This layer, together with the decomposition lemmas, helps us get rid of the obstacle and we can prove the final protocol is secure.\par
The protocol is given below. Note that different from the previous protocols, in this protocol we hardcode the various parameters inside the protocol and only keep the security $\kappa$ and number of output gadgets $L$ as the parameters of this protocol.
\begin{mdframed}[style=figstyle]
\begin{prtl}\label{prtl:13}(Remote Gadget Preparation Protocol) The protocol $$\fGdgPrep(\underbrace{\kappa}_{\substack{security\\  parameter}},\underbrace{L}_{\substack{output \\ number}})$$ is defined as follows:\par
	Define $$\eta=\kappa^{B_4+6},N=\kappa \cdot \fTHRESHOLD(\kappa),T=\log (L/N), J=\eta$$, where $\fTHRESHOLD(\kappa)$, $B_4$ are the constants given in Lemma \ref{lem:10.2}, and everything is rounded to the ceiling.
	\begin{enumerate}
		\item The client samples $$K=\{K^{(i)}\}_{i\in [N]},\Lambda_1=\{\Lambda_1^{(t)(s)(j)}\}_{t\in [T],s\in [\sqrt{\kappa}],j\in [J]},\Lambda_2=\{\Lambda_2^{(t)(j)}\}_{t\in [T],j\in [J]}$$
		      , where\begin{itemize}\item $K^{(i)}=\{x_0^{(i)},x_1^{(i)}\}$, each of them is a single pair of different keys with key length $\eta$; \item $\Lambda_1^{(t)(s)(j)}=\{r_0^{(t)(s)(j)},r_1^{(t)(s)(j)}\}$, $\Lambda_2^{(t)(j)}=\{g_0^{(t)(j)},g_1^{(t)(j)}\}$, each of them is a single pair of different keys with key length $\eta$.\end{itemize}
		\item The client prepares
		      \begin{align}&Gadget(K)\otimes Gadget(\Lambda_1)\otimes Gadget(\Lambda_2)\label{eq:131}                                                                                                                                                                                       \\
			                     =&	(\otimes_{i=1}^N(\ket{x_0^{(i)}}+\ket{x_1^{(i)}}))\\&\otimes (\otimes_{t=1}^{T}\otimes_{s=1}^{\sqrt{\kappa}}\otimes_{j=1}^{J}(\ket{r_0^{(t)(s)(j)}}+\ket{r_1^{(t)(s)(j)}}))\\&\otimes (\otimes_{t=1}^T\otimes_{j=1}^J(\ket{g_0^{(t)(j)}}+\ket{g_1^{(t)(j)}}))\label{eq:132}
		      \end{align}
		      and sends it to the server.
		\item For $t=1,\cdots T$:\par
		      Take the padding length $\ell^t:=100t\eta$. Output key length $\kappa_{out}^t:=\ell^t$.
		      \begin{enumerate}
			      \item Organize the keys\footnote{Let's briefly explain what the ``organize the keys'' means. The returned keys and the initial keys are organized in the ``flatten notation'', which is, for example, $K=\{K^{(i)}\}_{i\in [N]}$ where each $K^{(i)}$ is a single key pair. However when we call some subprotocols the input key set needs to have some inner structure, for example, in the $\fGdgPrep^{\log\kappa+1\rightarrow \kappa}$ protocol the input key set is $K=\{K_1,K_2\}$ where $K_1$ is a single key pair while $K_2$ contains $\log\kappa$ pairs of keys. Thus we need to ``organize'' these input keys. (The order of the keys doesn't matter.) This is basically a change of notations.} from the last step (which is $K$ initially and $K_{out\_(t-1).b}$ if it's not the first round) into a form that is compatible with the input of $\fGdgPrep^{OneRound}$. Denote it as $K_{in\_t}$. The client and the server run $$\fGdgPrep^{OneRound}(K_{in\_t},\Lambda_1^{(t)};\ell^t;\kappa_{out}^t)$$. Suppose the returned keys are $K_{out\_t.a}$.
			      \item The client and the server run $$\fSecurityRefreshing( K_{out\_t.a},\Lambda_2^{(t)};\ell^t;\kappa_{out}^t)$$. Denote the returned keys as $K_{out\_t.b}$.
		      \end{enumerate}

	\end{enumerate}
	The client stores the keys from the last execution of the protocol above as the returned keys of the whole protocol. Denote it as $K_{out}$. The honest server should get $Gadget(K_{out})$.
\end{prtl}\end{mdframed}
And we note that the initial state in the very beginning is all-zero for both the client and the server. In the purified notation viewpoint, it's $\ket{\mathfrak{init}}$. But in the security proof our analysis starts at the third step, thus $\ket{\varphi}$ (transformed into purified notation) is the initial state.
\subsubsection{Security statement}
The security statement is given below.
\begin{thm}[Security of Protocol \ref{prtl:13}]\label{thm:10.3}
	The following statement is true for sufficiently big security parameter $\kappa$:\par
	Suppose the $\fGdgPrep$ protocol (Protocol \ref{prtl:13}) is run on security parameter $\kappa$ and output number $L$. $\log L\leq \kappa^{1/5}$. The initial state on the server side is all-zero.\par
	Then for any adversary $\fAdv$ with query number $|\fAdv|\leq 2^{\sqrt{\kappa}/2}$, denote the post-execution joint purified state as $\ket{\varphi^\prime}$:
	$$\ket{\varphi^\prime}=\fGdgPrep_\fAdv(\underbrace{\kappa}_{\substack{security\\  parameter}},\underbrace{L}_{\substack{output \\ number}})\circ \ket{\mathfrak{init}}$$
	then $\forall i\in [L]$, $P_{pass}\ket{\varphi^\prime}$ is $(2^\kappa, 2^{-\kappa^{1/4}/10})$-SC-secure for $K_{out}^{(i)}$ given $K_{out}-K_{out}^{(i)}$, where $K_{out}$ is the output key set of this protocol.
\end{thm}
In other words, using the language from Definition \ref{defn:3.7}, Definition \ref{defn:3.5}, we can describe the correctness and security of this protocol as follows:
\begin{thm}[Correctness and Security of Protocol \ref{prtl:13}, concise notation]
	For sufficiently large security parameter $\kappa$, for any $L\leq 2^{\kappa^{1/5}}$, $\fGdgPrep(\kappa,L)$ is an $N\rightarrow L$ remote gadget preparation protocol that is secure against adversaries of query number $\leq 2^{\sqrt{\kappa}/2}$ with output security $\kappa^{1/4}/10$.
\end{thm}
Thus once we can prove Theorem \ref{thm:10.3}, we complete the first step of Outline \ref{ppl:1}.
\subsubsection{Proof of Theorem \ref{thm:10.3}}
The proof is given below. As a comment for the proof, the alphabetical letters are used in the following way:
\begin{itemize}
	\item $i$ is used for indexing the keys in $K$ (and also $K_{in\_t}$, $K_{out\_t.a}$, $K_{out\_t.b}$).
	\item $t$ is used for different round. $t\in [T]$.
	\item For key sets that have multi-dimensional structure: for $\Lambda_1$ we use $t\in [T],s\in [\sqrt{\kappa}],j\in [J]$ as the index, for $\Lambda_2$ we use $t\in [T],j\in [J]$.
\end{itemize}
We will use the multi-round decomposition method described in  Section \ref{sec:4.8.3}. The organization of the proof is similar to the proof of Lemma \ref{lem:10.2}. Overall speaking, \emph{we will repeat the an argument of the same form for many rounds, and decompose the final state step-by-step. We will divide the argument (in round $t$, for some arbitrary $t$) into pieces, and put it into boxes; outside the box, we will first show how the argument goes in the first round, as an explicit example, then show how the argument goes in an arbitrary round, and complete the proof.}\par
But here the detailed structure of arguments in each round is different from the proof of Lemma \ref{lem:10.2}. First note the protocol in each round of this protocol only has two steps: generate new gadgets, and strengthen the security. And the security analysis for round $t$ goes as follows:
\begin{otl}
\begin{enumerate}
\item[0.] The analysis starts with a state denoted by $\ket{\varphi^{t}}$.
\item Prove the security properties for the newly-generated gadgets; The tools for this step are the properties of $\ket{\varphi^t}$ and the security lemma for subprotocol ($\fGdgPrep^{OneRound}$).
\item[1.5.] Apply the decomposition lemma to decompose the state into two terms $\ket{\phi^{t.1}}$ and $\ket{\chi^{t.1}}$, corresponding to $\ket{\phi}$ and $\ket{\chi}$ in the state decomposition lemma. (This idea is discussed in Section \ref{sec:4.8.3}.) Let's discuss these two terms separately:
\begin{itemize}
\item For branch $\ket{\phi^{t.1}}$:
\begin{enumerate}
\item[2.] The $\fSecurityRefreshing$ layer strengthens the security property. And the output state is viewed as $\ket{\varphi^{t+1}}$, which is the initial state for the analysis in the next round.
\end{enumerate}
\item For branch $\ket{\chi^{t.1}}$: it's already exponentially small.
\end{itemize}
\item[...]Finally we combine all the terms together:
\begin{align}P_{pass}\ket{\varphi^\prime} & =P_{pass}\fPrtl_{\geq 1}\ket{\varphi^1}                                                                                                 \\
		                             & =P_{pass}\fPrtl_{>1.1}\circ\ket{\chi^{1.1}}+P_{pass}\fPrtl_{\geq 2}\circ \fSecurityRefreshing\circ\ket{\phi^{1.1}}                             \\
		                             & =P_{pass}\fPrtl_{>1.1}\circ\ket{\chi^{1.1}}+P_{pass}\fPrtl_{\geq 2}\circ \ket{\varphi^2}                                                \\
		                             & =\cdots                                                                                                                             \\
		                             & =P_{pass}\fPrtl_{>1.1}\circ\ket{\chi^{1.1}}+P_{pass}\fPrtl_{>2.1}\circ\ket{\chi^{2.1}}+\cdots+P_{pass}\fPrtl_{>T.1}\ket{\chi^{T.1}}\\&\quad+\ket{\varphi^{T}}\end{align}
		                              The only $\ket{\varphi^{T}}$ term has the security properties we want and the $\ket{\chi^{\cdots}}$ terms in the end are all exponentially small.
\end{enumerate}
\end{otl}
Now let's go to the formal proof.
\begin{proof}[Proof of Theorem \ref{thm:10.3}]
	Assume $|P_{pass}\ket{\varphi^\prime}|>2^{-\kappa^{1/4}}$, otherwise the statement is already true.\par
	Our goal is to analyze the property of $\ket{\varphi^\prime}=\fPrtl\circ\ket{\varphi^1}$, where $\fPrtl$ is an abbreviated notation for the whole protocol starting from the sending of $\ket{\varphi^1}$, $\ket{\varphi^1}$ is defined to be the purified joint state of (\ref{eq:131})(\ref{eq:132}).\par
	By Technique \ref{lem:4.2} we can assume $Tag(\Lambda_1,\Lambda_2)$ are additionally stored in the read-only buffer. Additionally assume $Tag(K_{out\_t.a})$ is revealed to the server after the first step of each round (which will be useful in the arguments later). These auxiliary information will only make the adversary more powerful thus prove the theorem under this assumption will imply the original theorem.\par
	
	We will use $\ket{\varphi^{\cdots}}$ to denote the initial state in each round of the argument. Recall that $\ket{\varphi^1}$ is also previously defined to be the input state \textbf{before} the first round. Then we have the following conditions on $\ket{\varphi^1}$:\\
	\begin{mdframed}
			\textbf{Properties of $\ket{\varphi^1}$:}
			\begin{itemize}
				\item $\forall i$, $\ket{\varphi^1}$ is $(2^{\eta/7},2^{-\eta/7})$-SC-secure for $K^{(i)}$ given $K-K^{(i)}$ and $Tag(\Lambda_1,\Lambda_2)$.

				\item $\forall t^\prime\in [T]$, $\forall s\in [\sqrt{\kappa}],j\in [J]$, $\ket{\varphi^1}$ is $(2^{\eta/7},2^{-\eta/7})$-SC-secure for $\Lambda_1^{(t^\prime)(s)(j)}$ given $K$, $\Lambda_1-\Lambda_1^{(t^\prime)(s)(j)}$, and $\Lambda_2$. (The reason that we use $t^\prime$ instead of $t$ is to make it more consistent with the symbols later.)

				\item $\forall t^\prime\in [T]$, $\forall j\in [J]$, $\ket{\varphi^1}$ is $(2^{\eta/7},2^{-\eta/7})$-SC-secure for $\Lambda_2^{(t^\prime)(j)}$ given $K$, $\Lambda_1$, and $\Lambda_2-\Lambda_2^{(t^\prime)(j)}$.

				\item $\ket{\varphi^1}$ is $(1,2^\kappa)$-representable from $\ket{\mathfrak{init}}$.
			\end{itemize}\end{mdframed}
	The first step of the first round of the protocol is $\fPrtl_{=1.1}$, which is a $\fGdgPrep^{OneRound}$ protocol applied on $K_{in\_1}$, $\Lambda_1^{(1)}$. Then we can apply its security property (Lemma \ref{lem:10.2}) on initial state $\ket{\varphi^1}$. Using this lemma we can know the state after the first round of execution of 3(a)(denoted as $\ket{\varphi^{1.1}}$) satisfies:
	\begin{equation}\label{eq:92r}\text{$\forall i$, $P_{pass}\ket{\varphi^{1.1}}$ is $(2^{\eta/\kappa^2}, 2^{-\sqrt{\kappa}/10})$-SC-secure for $K^{(i)}_{out\_1.a}$ given $K_{out\_1.a}- K_{out\_1.a}^{(i)}$.}\end{equation}
	The details for this step is given below. Note that the following proof is written in a way to support general round count $t$. In the analysis of the first round we can simply choose $t=1$ below.\\
		\begin{mdframed}
			\textbf{Details for arguing about the properties of $P_{pass}\ket{\varphi^{t.1}}$.}\par
			If $|\ket{\varphi^t}|\leq 2^{-\kappa}$, this statement is already true. Otherwise, from ``the properties of $\ket{\varphi^t}$'', we can loosen the parameters and know the followings:
			\begin{itemize}
				\item $\forall i$, $\ket{\varphi^t}$ is $(2^{\eta/\kappa^3},2^{-\eta/\kappa^3}|\ket{\varphi^t}|)$-SC-secure for $K_{in\_t}^{(i)}$ given $K_{in\_t}-K_{in\_t}^{(i)}$ and $Tag(\Lambda_1)$.\\
				      (Note that the exponent $\eta/\kappa^3> 48\kappa^{B_4+2}$.)\par
				\item $\forall s\in [\sqrt{\kappa}],j\in [J]$, $\ket{\varphi^t}$  is $(2^{\eta/20},2^{-\eta/20}|\ket{\varphi^t}|)$-SC-secure for $\Lambda_1^{(t)(s)(j)}$ given $K_{in\_t}$, $\Lambda_1^{(t)}-\Lambda_1^{(t)(s)(j)}$.\par
				\item $\ket{\varphi^t}$ is $(1,t2^\eta)$-representable from $\ket{\mathfrak{init}}$.\par
				      Which implies the pad length is enough.
			\end{itemize}
			Then applying Lemma \ref{lem:10.2} completes the proof.
		\end{mdframed}
	From (\ref{eq:92r}), we can decompose $P_{pass}\ket{\varphi^{1.1}}$ as $\ket{\phi^{1.1}}+\ket{\chi^{1.1}}$, as follows:\\
	\begin{mdframed}
			\textbf{Decomposition: $P_{pass}\ket{\varphi^{t.1}}$ to $\ket{\phi^{t.1}}+\ket{\chi^{t.1}}$:}\par
			If $|P_{pass}\ket{\varphi^{t.1}}|\leq 2^{-\sqrt{\kappa}/20}$, simply take $\ket{\phi^{t.1}}=0$. Otherwise:\par
			Based on the ``properties of $P_{pass}\ket{\varphi^{t.1}}$'', apply the multi-key decomposition lemma (Lemma \ref{lem:4.7}) (together with Fact \ref{fact:injtag}) we can decompose $P_{pass}\ket{\varphi^{t.1}}$ as $\ket{\phi^{t.1}}+\ket{\chi^{t.1}}$ and we have:
			\begin{itemize}
				\item $|\ket{\chi^{t.1}}|\leq 4.5L2^{-\sqrt{\kappa}/20}$
				\item  $\forall i$, $\ket{\phi^{t.1}}$ is $(2^{\eta/(4\kappa^2\log 6L)},2^{-\eta/(4\kappa^2\log 6L)})$-SC-secure for $K_{out\_t.a}^{(i)}$. And $\ket{\phi^{t.1}}$ is $(1, 2^{\eta/\kappa^2})$-server-side-representable from $P_{pass}\ket{\varphi^{t.1}}$.
			\end{itemize}\end{mdframed}
	We can write
	$$P_{pass}\ket{\varphi^\prime}=P_{pass}\fPrtl_{>1.1}\circ\ket{\chi^{1.1}}+P_{pass}\fPrtl_{\geq 2}\circ \fPrtl_{=1.2}\circ\ket{\phi^{1.1}}$$
	Note that $\fPrtl_{=1.2}$ is a $\fSecurityRefreshing$ protocol run on key sets $K_{out\_1.a}$ and $\Lambda_2^{(1)}$. Denote $$\ket{\varphi^2}:=P_{pass}\fPrtl_{=1.2}\circ\ket{\phi^{1.1}}$$
	, by the security of the $\fSecurityRefreshing$ protocol (Corollary \ref{cor:101c2} of Lemma \ref{lem:10.1}) we can prove
	\begin{equation}\label{eq:134}\text{$\forall i$, $\ket{\varphi^2}$ is $(2^{\eta/\kappa^2},2^{-\eta/\kappa^{2.5}})$-SC-secure for $K_{out\_1.b}^{(i)}$ given $K_{out\_1.b}-K_{out\_1.b}^{(i)}$.}\end{equation}
	The details are given below. The details are written for general $t$ to support the induction-style proof. For this round just substitute $t=1$.
	\begin{mdframed}
		\textbf{Details for arguing about the properties of $\ket{\varphi^{t+1}}$ for $K_{out\_t.b}$ (given the other keys)}\par
		Recall that $$\ket{\varphi^{t+1}}:=P_{pass}\fPrtl_{=t.2}\circ\ket{\phi^{t.1}}=P_{pass}\fSecurityRefreshing_{\fAdv_{t.2}}(K_{out\_t.a},\Lambda^{(t)}_2)\circ\ket{\phi^{t.1}}$$
		First we can assume \begin{equation}\label{eq:144rr}|\ket{\phi^{t.1}}|\geq 2^{-\eta/\kappa^{2.5}}\end{equation}, otherwise the statement is already true.\par
		We will apply Corollary \ref{cor:101c2} of Lemma \ref{lem:10.1}. The checklist for applying Corollary \ref{cor:101c2} of Lemma \ref{lem:10.1} on $\ket{\phi^{t.1}}$:
		\begin{itemize}
			\item From the result in the last step and (\ref{eq:144rr}) we know $\forall i$, $\ket{\phi^{t.1}}$ is\\ $(2^{\eta/4\kappa^2\log 6L},2^{-\eta/8\kappa^2\log 6L}|\ket{\phi^{t.1}}|)$-SC-secure for $ K_{out\_t.a}^{(i)}$. $Tag( K_{out\_t.a})$ and $Tag(\Lambda_2)$ are already stored in the read-only buffer.
			\item  $\forall j\in [J]$, $\ket{\phi^{t.1}}$ is $(2^{\eta/30},2^{-\eta/30}|\ket{\phi^{t.1}}|)$-SC-secure for $\Lambda_2^{(t)(j)}$ given $\Lambda_2^{(t)}-\Lambda_2^{(t)(j)}$ and $K_{out\_t.a}$. This is because:
			      \begin{enumerate}
				      \item $\ket{\phi^{t.1}}$ is $(1, 2^{\eta/\kappa^2})$-server-side-representable from $P_{pass}\ket{\varphi^{t.1}}$.
				      \item $P_{pass}\ket{\varphi^{t.1}}=P_{pass}\fGdgPrep^{OneRound}_{\fAdv_{t.1}}(K_{in\_t},\Lambda^{(t)}_1)\circ \ket{\varphi^t}$. Note that in this part the computation of client side messages does not take $\Lambda_2^{(t)}$ as the inputs, and all the messages sent to the server in this round can be computed from $K_{in\_t}$, $\Lambda_1^{(t)}$, $K_{out\_t.a}$ (sampled randomly in this round), the server's response and random coins. 
				      \item $\forall j\in [J]$, $\ket{\varphi^t}$ is $(2^{\eta/25},2^{-\eta/25})$-SC-secure for $\Lambda_2^{(t)(j)}$ given $K_{in\_t}$, $\Lambda_1^{(\geq t)}$, $\Lambda_2^{(\geq t)}-\Lambda_2^{(t)(j)}$. This comes from ``properties on $\ket{\varphi^t}$''. (We will see it later. Since we are doing this induction-style proof on round $t$ this is not a circular-proof.)
				      \item We already assume $|\ket{\phi^{t.1}}|> 2^{-\eta/\kappa^{2.5}}$
				      \item The query number of the adversary is at most $2^{\sqrt{\kappa}}$.
			      \end{enumerate}
			      Then applying Lemma \ref{lem:basic} proves it.
			\item For the same reason described above and ``the properties on $\ket{\varphi^t}$'', we know $\ket{\phi^{t.1}}$ is $(1,2^{\eta})$-representable from $\ket{\mathfrak{init}}$. Thus the pad length and output key length is enough.

		\end{itemize}\end{mdframed}
	Then we can repeat the same argument on $\fPrtl_{\geq 2}\circ\ket{\varphi^2}$, as what we did from the beginning of this proof to analyze $\fPrtl\circ\ket{\varphi^1}$. Note that $\ket{\varphi^2}$ defined above satisfies very similar conditions as what we showed for $\ket{\varphi^1}$. Thus we can repeat the same argument above and construct $\ket{\varphi^t}$ inductively. The following is the properties for $\ket{\varphi^t}$, and substituting $t=2$ gives the conditions we need currently:
	\begin{mdframed}
		\textbf{The Properties on $\ket{\varphi^t}$:}
		\begin{itemize}
			\item  We have already proved in the last round that
			      \begin{equation}\label{eq:135r}\text{$\forall i$, $\ket{\varphi^t}$ is $(2^{\eta/\kappa^2},2^{-\eta/\kappa^{2.5}})$-SC-secure for $K_{out\_(t-1).b}^{(i)}$}\end{equation}\begin{equation*}\text{ given $K_{out\_(t-1).b}-K_{out\_(t-1).b}^{(i)}$.}\end{equation*}
			      (Note that (\ref{eq:134}) is a special case for this statement when $t=2$.)\par
			      In other words, \begin{equation}\label{eq:136}\text{$\forall i$, $\ket{\varphi^t}$ is $(2^{\eta/\kappa^2},2^{-\eta/\kappa^{2.5}})$-SC-secure for $K_{in\_t}^{(i)}$ given $K_{in\_t}-K_{in\_t}^{(i)}$.}\end{equation} 
			\item $\forall t^\prime \geq t$, $\forall s\in [\sqrt{\kappa}],j\in [J]$, $\ket{\varphi^t}$ is $(2^{\eta/7}-t2^{\eta/100},2^{-\eta/7})$-SC-secure for $\Lambda_1^{(t^\prime)(s)(j)}$ given $K_{in\_t}$, $\Lambda_1^{(\geq t)}-\Lambda_1^{(t^\prime)(s)(j)}$ and $\Lambda_2^{(\geq t)}$.\par
			      This comes from the fact that
			      \begin{enumerate}
				      \item $K_{in\_t}$ is a change of notation of $K_{out\_(t-1).b}$.
				      \item $\ket{\varphi^t}:=P_{pass}\fSecurityRefreshing_{\fAdv_{(t-1).2}}(K_{out\_(t-1).a},\Lambda_2^{(t-1)})\circ\ket{\phi^{(t-1).1}}$, where:\\ The query number of the adversary in this round (which is $\fAdv_{(t-1).2}$) is at most $2^{\sqrt{\kappa}}$;\\
				            In the protocol the computation of the client side messages only use $K_{out\_(t-1).a}$, $K_{out\_(t-1).b}$ (sampled randomly in this round) and $\Lambda_2^{(t-1)}$ and random coins. (Especially, the computation does not use $\Lambda^{(t^\prime)}_1$.)
				      \item $\ket{\phi^{(t-1).1}}$ is $(1, 2^{\eta/\kappa^2})$-server-side-representable from $P_{pass}\ket{\varphi^{(t-1).1}}$.
				      \item $P_{pass}\ket{\varphi^{(t-1).1}}=P_{pass}\fGdgPrep^{OneRound}_{\fAdv_{(t-1).1}}(K_{in\_(t-1)},\Lambda_1^{(t-1)})\circ \ket{\varphi^{t-1}}$, where:\\
				            The query number of the adversary in this round (which is $\fAdv_{(t-1).1}$) is at most $2^{\sqrt{\kappa}}$.\\
				            In this protocol the computation of the client side messages only use $K_{in\_(t-1)}$, $\Lambda_1^{(t-1)}$ and $K_{out\_(t-1).a}$ (sampled randomly in this round), together with the server's response and the random coins. Especially, it does not use $\Lambda_1^{(t^\prime)}$.
				      \item From the last step we know for any $t^\prime\geq t$, $s\in [\sqrt{\kappa}]$, $j\in [J]$, $\ket{\varphi^{t-1}}$ is $(2^{\eta/7}-(t-1)2^{\eta/100},2^{-\eta/7})$-SC-secure for $\Lambda_1^{(t^\prime)(s)(j)}$ given $K_{in\_(t-1)}$, $\Lambda^{(\geq t-1)}_1-\Lambda_1^{(t^\prime)(s)(j)}$ and $\Lambda_2^{(\geq t-1)}$.
			      \end{enumerate}
			      Thus applying Lemma \ref{lem:basic} completes the proof.

			\item From (almost) the same reasons, $\forall t^\prime\geq t$, $\forall j\in [J]$, $\ket{\varphi^t}$ is $(2^{\eta/7}-t2^{\eta/100},2^{-\eta/7})$-SC-secure for $\Lambda_2^{(t^\prime)(j)}$ given $\Lambda_2^{(\geq t)}-\Lambda_2^{(t^\prime)(j)}$, $K_{in\_t}$ and $\Lambda^{(\geq t)}_1$.\par
			\item From (almost) the same reasons $\ket{\varphi^t}$ is $(1,t\cdot 2^{\eta/100})$-representable from\\ $\ket{\mathfrak{init}}$.
		\end{itemize}\end{mdframed}
	Thus we can use the same technique on $\ket{\varphi^2}$ and decompose $\fPrtl_{\geq 2}\ket{\varphi^2}$ into two states.\par
	And we can repeat this argument to the end without changing anything except the time step index. Recall that the proof pieces in the boxes of ``Details for arguing about the properties of $P_{pass}\ket{\varphi^{t.1}}$'' and the ``state decomposition'' still hold here, since we already write it in a general form. Thus we can repeat the following arguments in an induction-style:
	%
	%
	\begin{enumerate}
		\item Starting from the properties of $\ket{\varphi^t}$, we can prove the general form of (\ref{eq:92r}), which is
		      \begin{equation*}\text{$\forall i$, $P_{pass}\ket{\varphi^{t.1}}$($:=P_{pass}\fPrtl_{=t.1}\ket{\varphi^t}$)}\end{equation*}\begin{equation}\label{eq:94}\text{ is $(2^{\eta/\kappa^2}, 2^{-\sqrt{\kappa}/10})$-SC-secure for $K^{(i)}_{out\_t.a}$ given $K_{out\_t.a}- K_{out\_t.a}^{(i)}$.}\end{equation}
		      This can be proved using the same arguments shown in the box ``Details for arguing about the properties of $P_{pass}\ket{\varphi^{t.1}}$'', and one implicit detail, which is also the reason that this argument can't be applied infinitely, is
		      $$\forall t\leq \log(L/N),2^{\eta/7}-t2^{\eta/100}>2^{\eta/20}$$
		\item We can decompose $P_{pass}\ket{\varphi^{t.1}}$ as $\ket{\phi^{t.1}}+\ket{\chi^{t.1}}$ as discussed in the box of ``state decomposition''.
		\item And we can write
		      $$P_{pass}\fPrtl_{\geq t}\ket{\varphi^t}=\ket{\varphi^\prime}=P_{pass}\fPrtl_{>t.1}\circ\ket{\chi^{t.1}}+P_{pass}\fPrtl_{\geq t+1}\circ Prtl_{=t.2}\circ\ket{\phi^{t.1}}$$
		      $$\text{Denote }\ket{\varphi^{t+1}}:=P_{pass}\fPrtl_{=t.2}\circ\ket{\phi^{t.1}}$$
		      And as discussed in the box of ``Details for arguing about the properties of $\ket{\varphi^{t+1}}$ for $K_{out\_t.b}$'', we can prove
		      \begin{equation}\text{$\forall i$, $\ket{\varphi^{t+1}}$ is $(2^{\eta/\kappa^2},2^{-\eta/\kappa^{2.5}})$-SC-secure for $K_{out\_t.b}^{(i)}$ given $K_{out\_t.b}-K_{out\_t.b}^{(i)}$.}\end{equation}
		\item Finally we can prove ``the properties of $\ket{\varphi^{t+1}}$'' and continue to the next round of this argument.
	\end{enumerate}
	Continue the whole argument to the end, and we know:
	{\small\begin{align}P_{pass}\ket{\varphi^\prime} & =P_{pass}\fPrtl_{\geq 1}\ket{\varphi^1}                                                                                                 \\
		                             & =P_{pass}\fPrtl_{>1.1}\circ\ket{\chi^{1.1}}+P_{pass}\fPrtl_{\geq 2}\circ \fSecurityRefreshing\circ\ket{\phi^{1.1}}                             \\
		                             & =P_{pass}\fPrtl_{>1.1}\circ\ket{\chi^{1.1}}+P_{pass}\fPrtl_{\geq 2}\circ \ket{\varphi^2}                                                \\
		                             & =\cdots                                                                                                                             \\
		                             & =P_{pass}\fPrtl_{>1.1}\circ\ket{\chi^{1.1}}+P_{pass}\fPrtl_{>2.1}\circ\ket{\chi^{2.1}}+\cdots+P_{pass}\fPrtl_{>T.1}\ket{\chi^{T.1}}+\ket{\varphi^{T}}\end{align}}
	where each term about $\ket{\chi}$ has norm $\leq 4.5L2^{-\sqrt{\kappa}/20}$. The last term $\ket{\varphi^{T}}$ is \\$(2^{\eta/\kappa^2},2^{-\eta/\kappa^{2.5}})$-SC-secure for $K_{out}^{(i)}$ given $K_{out}-K_{out}^{(i)}$ (for all $i$). Thus $P_{pass}\ket{\varphi^\prime}$ is $(2^{\eta/\kappa^3},(1+\log L)\cdot 4.5L2^{-\sqrt{\kappa}/20})$-SC-secure for $K_{out}^{(i)}$ given all the other keys. Substitute $\log L\leq \kappa^{1/5}$ inside completes the proof.
\end{proof}
Notice that Protocol \ref{prtl:13} satisfies the correctness and only uses succinct client side quantum computation. Thus by this time we complete the design and proofs of the remote state preparation protocol, thus complete the first step of Outline \ref{ppl:1}.

\cleardoublepage
\chapter{Universal Blind Quantum Computation from Remote Gadget Preparation}\label{cht:8}
In the previous chapter we design a secure remote gadget preparation protocol using succinct client side quantum gates. In this chapter we will show how to make use of the  remote gadget preparation protocol to design a universal blind quantum computation protocol. This will prove Theorem (Theorem \ref{thm:1}).\par
This chapter is organized as follows. In Section \ref{sec:12.1} we give an overview for this final step of our construction, which is to combine the remote gadget preparation protocol from the last chapter, the \emph{8-basis qfactory} protocol, and the BFK's UBQC protocol. In Section \ref{sec:11.2} we give the \emph{8-basis qfactory} protocol that we need, and discuss its security proof.  Finally in Section \ref{sec:12.3} we complete the construction of our universal blind quantum computation protocol (Protocol \ref{prtl:17}), and complete its security proof in Section \ref{sec:11.4} which implies Theorem \ref{thm:1}.
\section{The Roadmap: Our UBQC from Remote Gadget Preparation, 8-basis Qfactory, and BFK's UBQC}\label{sec:12.1}
Before we give the full protocol, let's first review the original UBQC protocol\cite{UBQC}.\par
\begin{prtl}\label{prtl:ubqc}
	(BFK's Universal Blind Quantum Computation)\par
	Input: circuit $C$ to be evaluated.
	\begin{enumerate}
		\item For $i=1,\cdots |C|$, the client samples $\theta^i=\theta_1^i\pi+\theta^i_2\frac{\pi}{2}+\theta^i_3\frac{\pi}{4}$, $\theta^i_c\leftarrow_r\{0,1\}(c=1,2,3)$, prepares the state $\ket{+_{\theta^i}}=\ket{0}+e^{i\theta}\ket{1}$ and sends them to the server.
		\item Run the gadget-assisted circuit evaluation protocol below. (See Protocol \ref{prtl:gaubqc})
		\end{enumerate}
		\end{prtl}
		\begin{prtl}[$\fGAUBQC$, or gadget-assisted UBQC]\label{prtl:gaubqc}
		Input: circuit $C$ to be evaluated; client holds angles $\theta_1\cdots \theta_i, i\in [|C|]$.
		Honest server holds $\ket{+_{\theta^i}}$.
		\begin{enumerate}
		\item The server connects these gadgets as a brickwork state.
		\item For $i=1,\cdots |C|$:
		      \begin{enumerate}
			      \item The client computes the measurement angle $\phi^i$ using:
			            \begin{itemize}\item The circuit description; \item $m_t,t<i$: The measurement results of the previous rounds;\item Angles $\theta$, bits $r$ of the previous rounds (or more formally, $\theta^t$, $r^t$ for $t<i$).\end{itemize}
			      \item The client samples $r^i\leftarrow_r \{0,1\}$, and computes $\delta^i=\theta^i+\phi^i+\pi r^i$. Send it to the server.
			      \item The server makes the measurement corresponding to $\delta^i$ and reports the result.
			    
		      \end{enumerate}
		      \item The client computes the computation result from the output
	\end{enumerate}

\end{prtl}
We omit some details like the form of the brickwork state and how to compute the angles, and only keep what we need.\par
For arguing about the security of our protocol, we need the following two security properties of this $\fGAUBQC$ protocol. The first security property that we need is, as long as the client can hide the two less-significant bits in the bit representation of all the $\theta^i$, the protocol will be secure in the sense of indistinguishability of $C$:
\begin{prop}\label{prop:ubqcsec}
	In protocol \ref{prtl:gaubqc}, for arbitrary server-side initial states, if the client uses $\tilde \theta^i:=\theta_1^i\pi+\tilde\theta_2^i\frac{\pi}{2}+\tilde\theta_3^i\frac{\pi}{4}$ as the input where $\tilde\theta_2^i,\tilde\theta_3^i$ are freshly completely random bits while $\theta_1^i$ could be correlated to the server side states, we have, for any malicious server:
	\begin{equation}
	\fGAUBQC(\theta^i,C)\approx_0\fGAUBQC(\theta^i,0^{|C|})	
	\end{equation}
where $\approx_0$ means these two states are perfectly indistinguishable.
\end{prop}
The second property is, if the replacement from $\theta^i$ to $\tilde\theta^i$ only happens at some rounds, even if the circuit $C$ is public, the adversary is not able to distinguish the angles it gets from completely random angles:
\begin{prop}\label{prop:gaubqcsec0}
	The following two cases are indistinguishable for any malicious server when two parties are executing $\fGAUBQC$, regardless of the circuit $C$ and the initial state, even if $C$ is public:
	\begin{itemize}
	\item For each $t\in [i+1,|C|]$, the client uses $\tilde \theta^i:=\theta_1^i\pi+\tilde\theta_2^i\frac{\pi}{2}+\tilde\theta_3^i\frac{\pi}{4}$ as the input to the protocol where $\tilde\theta_2^i,\tilde\theta_3^i$ are freshly completely random bits.
	\item For each round $t\in [i+1,|C|]$, the client simply sends a completely random angles in $\{0,\cdots,7\}\cdot \frac{\pi}{4}$.
	\end{itemize}

\end{prop}

Notice that in the UBQC protocol \cite{UBQC} the client only needs to prepare some quantum states in the form of $\ket{+_{\theta}}:=\ket{0}+e^{i\theta}\ket{1}$, $\theta=n\pi/4$, $n=0,\cdots 7$, and sends them to the server; then it can instruct the quantum server with classical interactions to do some quantum operations on these gadgets, and it will get the computation result $C\ket{0}$ by decoding the server's measurement results. The problem of this approach is to delegate a circuit $C$, the client has to prepare one such gadget for each gate in $C$, thus the client side quantum operations will be linear in $|C|$. One natural idea is to further delegate the preparation of these states using other protocols. Furthermore, by Proposition \ref{prop:ubqcsec} the client only needs to guarantee the secrecy of the two less-significant bits. An abstraction of this concept is given in paper \cite{qfactory}, which is called the \emph{8-basis qfactory}.\subsubsection{The concept of 8-basis qfactory\cite{qfactory}, and the adaptation to our setting}
The \emph{8-basis qfactory} is defined in \cite{qfactory} using the unpredictability language in the standard model. We revise the definition in the form of indistinguishability in the quantum random oracle model, which is more suitable in our setting. And we further add one revisions that is important in our setting: we add the initial state $\ket{\varphi}$ into the definition. The reason is, in our setting, we can't construct a qfactory protocol ``from scratch''; what the protocol can do is to transform some initial gadget (which in our protocols is part of $\ket{\varphi}$) into the state $\ket{+_\theta}$. We will see, in our protocol, the client and the server can first run the remote gadget preparation protocol in the last chapter to create the gadgets $\ket{\varphi}$ on the server side, then run the 8-basis qfactory protocol using these gadgets as the initial states.
\begin{defn}[8-basis qfactory]\label{def:11.1}
	(Correctness) In an 8-basis qfactory protocol from initial state $\ket{\varphi}$, the client samples $\theta_2,\theta_3\in_r \{0,1\}^2$ in the beginning, and when the protocol completes, the honest server should get $\ket{+_\theta}$ where $\theta=\pi\theta_1+\frac{\pi}{2}\theta_2+\frac{\pi}{4}\theta_3$, $\theta_1$ can be either $0$ or $1$. The client gets $\theta_1$.\par
	(Security) And we say this protocol is secure in the quantum random oracle model against adversaries of query number $\leq 2^\lambda$ with output security $\eta$ for initial state in $\cF$ if:\par
	For any joint purified state $\ket{\varphi}\in \cF$, for any malicious server $\fAdv$ with query number $|\fAdv|\leq 2^\lambda$, for any $\theta_2,\theta_3\in \{0,1\}^2$, suppose the state when the protocol finishes is $\ket{\varphi^\prime}$ and the angles that the client gets are $\theta=\pi\theta_1+\frac{\pi}{2}\theta_2+\frac{\pi}{4}\theta_3$, for any server side distinguisher $\cD$ with query number $|\cD|\leq 2^{\eta}$,
	{\small\begin{equation}|Pr(\cD(P_{pass}\varphi^\prime P_{pass}, (\theta_2,\theta_3))=0)-Pr(\cD(P_{pass}\varphi^\prime P_{pass}, (\theta^\prime_2,\theta^\prime_3)\leftarrow_r \{0,1\}^2)=0)|\leq 2^{-\eta}|\ket{\varphi}|\end{equation}}
	which means the distinguisher can't distinguish the real $(\theta_2,\theta_3)$ from the randomly generated $(\theta^\prime_2,\theta^\prime_3)$ given the final server side state of the protocol.
\end{defn}
An 8-basis qfactory protocol can be used to replace the quantum operations and quantum communications in the UBQC protocol, and its correctness guarantees such replacement won't affect the correctness of the blind quantum computation protocol.\par 
After the replacement, we get our $\fSuccUBQC$ protocol, intuitively as follows. In our $\fSuccUBQC$ protocol, both parties first runs the remote gadget preparation protocol, which gives the server the necessary initial gadgets, and then run the 8-basis qfactory protocol for each gadget to prepare the state $\ket{+_{\theta^i}},i\in [|C|]$. Finally both parties can use BFK to do the remaining work. 
\section{The 8-basis Qfactory Protocol}\label{sec:11.2}
In this section we construct the 8-basis qfactory protocol that we need, and prove its security.
\subsection{Protocol Design and Security Statement}
Let's first formalize the \emph{phase table}, which can allow the server to add an extra phase on the gadget without affecting the security:\par
Suppose a phase gate $R(\theta)=\ket{0}\bra{0}+e^{i\theta}\ket{1}\bra{1}$, $\theta=\frac{n}{D}\pi$ needs to be applied on a gadget, where the key pair on this wire is $K=\{x_0,x_1\}$, then the phase table is defined as follows.
\begin{defn}
	$$\fPhaseLT(K,\theta;\underbrace{ \ell}_{\substack{\text{padding} \\ \text{length}}},\underbrace{ D}_{\substack{\text{Denominator} }})$$ where $K=\{x_0,x_1\}$, $\theta=\frac{n}{D}\pi$ is defined as follows: the client samples $m\leftarrow \{0,\cdots D-1\}$, prepares the table
	$$\fLT(x_0\rightarrow m,x_1\rightarrow m+n;\underbrace{ \ell}_{\substack{\text{padding} \\ \text{length}}},\underbrace{ \ell}_{\substack{\text{tag} \\ \text{length}}})$$
	(Note that we simply choose the tag length to be the same as the pad length, which is enough for usage later.)
\end{defn}
The phase table allows the server to do the following transform, as described in \cite{revgt}:
$$\alpha\ket{x_0}+\beta\ket{x_1}\rightarrow \alpha\ket{x_0}+e^{i\theta}\beta\ket{x_1}$$
Then our 8-basis qfactory protocol is defined as follows.
\begin{mdframed}[style=figstyle]
\begin{prtl}\label{prtl:24}($\fQFac(K;\ell,\kappa)$), where $K=\{y_0,y_1\}$, $\ell$ is the padding length, $\kappa$ is the security parameter.\par
	The honest server should hold the state $(\ket{y_0}+\ket{y_1})\otimes\ket{\phi}$ (where $(\ket{y_0}+\ket{y_1})$ is useful here and $\ket{\phi}$ just means some unrelated state for other purposes.)
	\begin{enumerate}
		\item The client and the server execute $$\fBasisTest(K;\kappa^2,\underbrace{ \ell}_{\substack{\text{padding} \\ \text{length}}},
\underbrace{ \ell+\kappa^2}_{\substack{\text{output} \\ \text{length}}})\text{ (Protocol \ref{prtl:6.2})}$$. The server runs this step with the $\ket{y_0}+\ket{y_1}$ part in the input.
		\item The client chooses $\theta_2,\theta_3\leftarrow_r\{0,1\}^2$.
		\item The client sends $\fPhaseLT(K,\frac{\pi}{2}\theta_2+\frac{\pi}{4}\theta_3; \ell)$ to the server.
		\item The server first prepares the state
		      $$\ket{0}\ket{y_0}+e^{i(\frac{\pi}{2}\theta_2+\frac{\pi}{4}\theta_3)}\ket{1}\ket{y_1}$$
		      using the phase table, then makes Hadamard measurements on all the registers of $y$. Send the results $d$ to the client.
		\item The client computes and stores $\theta_1=d\cdot(y_0+y_1)$. Now an honest server should hold the state $\ket{+_\theta}$ where $\theta=\theta_1\pi+\theta_2\frac{\pi}{2}+\theta_3\frac{\pi}{4}$.
		      %
	\end{enumerate}

\end{prtl}\end{mdframed}

The protocol description already proves the correctness of this protocol. Note that the first $\fBasisTest$ step is non-collapsing and it does not destroy the state in the honest setting. In more details, the step 3 to 4 we make use of the phase table, which is described in Section 2. In step 4 to 5 we make use of the following observation from \cite{MahadevVerification}, where $P_d$ is the projection onto the event that the Hadamard basis measurements output $d$: \par
$$P_{d}(\ket{0}\fH^{\otimes n}\ket{y_0}+\beta\ket{1}\fH^{\otimes n}\ket{y_1})\propto \ket{0}+e^{\mi\pi d\cdot(y_0+y_1)}\beta\ket{1}$$
Now we discuss the security.
\subsubsection{Security statement}
Assuming the initial state $\ket{\varphi}$ satisfies some properties, we can prove this is a secure 8-basis qfactory protocol from $\ket{\varphi}$. The lemma is given below.\par
\begin{lem}\label{lem:11.1}
The following statement holds for sufficiently large security parameter $\kappa$.\par
Protocol
$$\fQFac(K;\underbrace{ \ell}_{\substack{\text{padding} \\ \text{length}}},
\underbrace{ \kappa}_{\substack{\text{security} \\ \text{parameter}}})$$
is secure in the quantum random oracle model against unbounded adversaries of queries number $\leq 2^\kappa$ with output security $\kappa$ for initial states in $\cF$ defined below, when the following inequalities on the various parameters are satisfied:\par
	$\cF$ contains the initial states (denote the joint purified state as $\ket{\varphi^1}$) that satisfy:
	\begin{itemize}
	\item (Security on the inputs) $\ket{\varphi^1}$ is $(2^\eta, 2^{-\eta}|\ket{\varphi^1}|)$-SC-secure for $K$.
	\item (Well-behaveness of the inputs) $\ket{\varphi^1}\in \cWBS(D)$.
	\end{itemize}
Inequalities:
	\begin{itemize}
		\item (Sufficient security on the inputs) , $\eta>5000\kappa^2$
		\item (Well-behaveness of the inputs) $D\leq 2^{\sqrt{\kappa}}$.
		\item (Sufficient padding length) $\ell>6D+10\eta$
	\end{itemize}
\end{lem}
\subsection{Security of our 8-basis Qfactory: the Proof of Lemma \ref{lem:11.1}}\label{sec:11.3}

In this section we prove Lemma \ref{lem:11.1}.\par

First note that the other steps in Protocol \ref{prtl:24} except the $\fBasisTest$ step follow the informal introduction in Section \ref{sec:1.4}. Why do we need the extra $\fBasisTest$ step? The reason is, if we remove the first step of this protocol, the client doesn't have any control on the form of the server's state, and proving the security will be difficult. So we add a $\fBasisTest$ step in the protocol, which gives the client some ability to ``verify'' server's state.\par
However, we note that in the analysis of $\fBasisTest$ in Section \ref{sec:6}, it does not give us ``exponential verifiability'', which means, the adversary may cheat with inverse polynomial probability, and we only know the state is close to the state we want by an inverse-polynomial distance. However, we will show, we can get rid of this problem by giving a different analysis of the $\fBasisTest$ protocol.\par
Simply speaking, for key pair $K=\{x_0,x_1\}$, in Section \ref{sec:6} we try to prove the adversary's state, for an adversary that can pass the test with big probability, after adding some auxiliary information, is close to a state in the form of
\begin{equation}\label{eq:61}\cU(\ket{x_0}\ket{\cdots}+\ket{x_1}\ket{\cdots})\end{equation}
by an inverse polynomial distance. (Note that we are using the natural notation, but in the security proof we purify everything, and the server's state will be entangled with the client's keys.)\par
However the blindness requires exponential security, and such ``verifiability with inverse-polynomial distance'' is not enough. In this section, we will prove, after the $\fBasisTest$ step, we can reduce the server's state in the lemma into a state in the following form, after using some auxiliary-information technique:
\begin{equation}\label{eq:62}\sum_{i\in [\kappa]}\cP_i(\ket{x_0}\ket{\cdots}+\ket{x_1}\ket{\cdots})+\text{some exponentially small value}\end{equation}
Where each $\cP_i$ is an efficient sequence of server-side projection, unitaries and RO queries. Compare to equation (\ref{eq:61}), the main difference is we allow it to be the sum of $\kappa$ terms. And we also need to consider the server-side operations with projections.\par
Thus our proof is divided into two steps. First we can prove the security of our 8-basis qfactory for a restricted case where the initial state can be described in this form, assuming the inner part $\ket{x_0}\ket{\cdots}+\ket{x_1}\ket{\cdots}$ all have some SC-security property. This is done in Lemma \ref{lem:11.3}. Then making use of Lemma \ref{lem:11.3}, we complete the proof of Lemma \ref{lem:11.1} thus prove the security of our 8-basis qfactory protocol.
\subsubsection{Security proof}\label{sec:11.4.1}
We note that the initial state of our 8-basis qfactory protocol should not be understood as the initial state in Protocol \ref{prtl:13}, even if in the security proof some notations are similar. Instead, we will see, when we use this lemma, it is replaced by ``the state after the first step of Protocol \ref{prtl:24} completes, projected onto the passing space'' (which is the post-execution state of Protocol \ref{prtl:13}, projected onto the passing space).\par
We divide the proof into two parts. First we prove the statement when the initial state is in the form of equation (\ref{eq:62}):
\begin{lem}\label{lem:11.3}
The following statement holds for sufficiently large security parameter $\kappa$.\par
Protocol
$$\fQFac^{\geq 2}(K;\underbrace{ \ell}_{\substack{\text{padding} \\ \text{length}}})$$ \begin{center}{where $\geq 2$ means the first step (the $\fBasisTest$ step) is omitted\footnote{ we omit one parameter from the parameter list of the $\fQFac$ protocol since it has no influence on this lemma}}\end{center}
is secure in the quantum random oracle model against unbounded adversaries of query number $\leq 2^\kappa$ with output security $\eta/9$ for initial states in $\cF$ defined below, when the following inequalities on the various parameters are satisfied:\par
	Suppose the key set is $K=\{x_0,x_1\}$, and $\cF$ contains the jointly purified state $\ket{\varphi^1}$ with the form
	$$\ket{\varphi^1}=\sum_{i\in [\kappa]}\cP_i(\ket{\varphi_{i,0}}+\ket{\varphi_{i,1}}), \text{ where $\cP_i$ is a server-side operation (with projections),}$$
	$$\forall i,\text{query number }|\cP_i|\leq 2^\kappa$$
	$$\forall i\in [\kappa],b\in \{0,1\}, P^{S_i}_{x_b}\ket{\varphi_{i,b}}=\ket{\varphi_{i,b}},\text{ where $S_i$ is some server-side system}$$
	and satisfies:
	\begin{itemize}
		\item (Security for the keys) $\ket{\varphi^1}$ is $(2^\eta, 2^{-\eta}|\ket{\varphi^1}|)$-SC-secure for $K$
		\item (Well-behaveness of the inputs) $\ket{\varphi^1}\in \cWBS(D)$. 
		\end{itemize}
		And the inequalities include
		\begin{itemize}
		\item (Security for the keys) $\eta>100\kappa^2$
		\item (Sufficient padding length) $l>6D+10\eta$
	\end{itemize}
\end{lem}
Note that we do not need to project onto the passing space since there is no client-side checking after the first step of the protocol. Further note that we do not consider the ``$Security$'' of the parameter list of $\fQFac$ here since it's not used after the first step.\par
The proof of Lemma \ref{lem:11.3} is relatively simpler, and is given in Appendix \ref{sec:AD}.\par
Below we discuss the overall idea for reducing Lemma \ref{lem:11.1} to Lemma \ref{lem:11.3}. The details are given in Appendix \ref{sec:ap111}.\par 
\paragraph{Ideas for reducing Lemma \ref{lem:11.1} to Lemma \ref{lem:11.3}} 
The keywords are ``elimination of test rounds'' and ``step-by-step reduction''. We will construct a series of statements, and reduce the proof of Lemma \ref{lem:11.1} to these statements step-by-step, and finally reach a statement that has the form of Lemma \ref{lem:11.3}.\par
Notice that in Lemma \ref{lem:11.1} there are $\kappa^2$ rounds of the single-key-pair basis test in the original protocol; we will divide it into $\kappa$ blocks, where each block contains $\kappa$ rounds; and we we will see, we will analyze and these blocks one-by-one, and since each block is itself a multi-round $\fBasisTest$ protocol, we can apply Lemma \ref{lem:6.3} in each block. Every time we do this kind of argument, we eliminate one block of tests and reduce the statement to a new statement whose form is closer to Lemma \ref{lem:11.3}. And we can do this step-by-step until we reach a statement that is basically Lemma \ref{lem:11.3}.\par
The statement we use in each round is as follows. We name it as ``Statement-round-$i$-completed''. Lemma \ref{lem:11.1} is ``Statement-round-$0$-completed''. In the 1st round of reduction, we reduce Lemma \ref{lem:11.1} to ``Statement-round-1-completed'' --- and you can see what the name means. Then we can continue and reduce it to ``Statement-round-2-completed'', and so on, until $s=\kappa$ or hitting some stopping condition in the middle.\par
The ``Statement-round-$i$-completed'' has the similar structure of the ``expanded form'' of Lemma \ref{lem:11.1} (which means, unroll the the security definition of the qFactory). Let's first write down the ``unrolled form'' of Lemma \ref{lem:11.1}.
\begin{mdframed}
\textbf{Unrolled version of Lemma \ref{lem:11.1}}\par
The following statement holds for sufficiently large security parameter $\kappa$.\par
Consider the protocol
$$\fQFac(K;\underbrace{ \ell}_{\substack{\text{padding} \\ \text{length}}},
\underbrace{ \kappa}_{\substack{\text{security} \\ \text{parameter}}})$$
suppose the initial states, described by the purified joint state $\ket{\varphi^1}$, is in $\cF$ defined below, and the following inequalities on the various parameters are satisfied:\par
	\begin{itemize}
	\item (Security on the inputs) $\ket{\varphi^1}$ is $(2^\eta, 2^{-\eta}|\ket{\varphi^1}|)$-SC-secure for $K$.
	\item (Well-behaveness of the inputs) $\ket{\varphi^1}\in \cWBS(D)$.
	\end{itemize}
Inequalities:
	\begin{itemize}
		\item (Sufficient security on the inputs) , $\eta>5000\kappa^2$
		\item (Well-behaveness of the inputs) $D\leq 2^{\sqrt{\kappa}}$.
		\item (Sufficient padding length) $l>6D+10\eta$
	\end{itemize}
	Then for any adversary $\fAdv$ of query number $\leq 2^\kappa$, denote the post-execution state as: 
	$$\ket{\varphi^\prime}=\fQFac(K,\ell,\kappa)\circ \ket{\varphi}$$
	then it satisfies: for any server-side operation $\cD$ with query number $|\cD|\leq 2^\kappa$,
	\begin{align}\label{eq:60}|&Pr(\cD(P_{pass}\varphi^\prime P_{pass}, (\theta_2,\theta_3))=0)-Pr(\cD(P_{pass}\varphi^\prime P_{pass}, (\theta^\prime_2,\theta^\prime_3)\leftarrow_r \{0,1\}^2)=0)|\\\leq& 2^{-\kappa}|\ket{\varphi^1}|\end{align}
\end{mdframed}

\begin{mdframed}
	\textbf{Statement-round-$i$-completed}\par
	The conclusion is the same as Lemma \ref{lem:11.1} , with one difference: the right side of (\ref{eq:60}) is replaced by $(2^{-\kappa}-i2^{-2\eta})|\ket{\varphi^1}|$.\par
	The conditions have the following differences:
	\begin{itemize}
		\item The initial state is
		      \begin{align}\label{eq:100c}
			        & \cP_1(\ket{\psi_0^{1}}+\ket{\psi_1^{1}})+\cP_2(\ket{\psi^{2}_0}+\ket{\psi^{2}_1})+\cdots+\cP_{i}(\ket{\psi^{i}_0}+\ket{\psi^{i}_1})\\
			      + & \ket{\phi^i}
		      \end{align}

		      where $|\ket{\phi^i}|\leq 2^{-i}|\ket{\varphi}|$. Note that some terms in the middle can be zero.\par
		      The notations in equation (\ref{eq:100c}) are as follows:
		      \begin{itemize}
			      \item $\forall i^\prime\in [i],\cP_{i^\prime}$ is a server-side operations with query number $\leq i\cdot 2^{\kappa+3}$.
			      \item $\forall b\in \{0,1\},\forall i^\prime\in [i], P^{S_{i^\prime}}_{x_b}\ket{\psi^{i^\prime}_{b}}=\ket{\psi^{i^\prime}_{b}}$,\\ $\text{ where $S_{i^\prime}$ is some server-side system}$.
			      \item $\forall b\in \{0,1\},\forall i^\prime\in [i],\ket{\psi^{i^\prime}_b}$ is $(1,i^\prime 2^{\kappa+2})$-server-side-representable from
			       \begin{equation}\label{eq:155c}\ket{\phi^{0}}\odot\llbracket\fAuxInf_1^{1\sim i}\rrbracket\odot \llbracket\fBasisTest(K;1\sim i\kappa)\rrbracket\odot \llbracket tag^{1\sim i}\rrbracket\end{equation}
			            where $\ket{\phi^0}$ is just a change of notation for the initial state $\ket{\varphi^1}$, and (note that some notations, for example, the subscript of $\fAuxInf$, are chosen to be consistent to the detailed security proof)
			            \begin{itemize}
			            \item $\llbracket\fBasisTest(K;1\sim i\kappa)\rrbracket$ contains the client-side messages in the $\fBasisTest$ protocol from round $1$ to round $i\kappa$; the parameters are the same as the original protocol. \item $\llbracket tag^{1\sim i}\rrbracket$ is the set of global tags $Tag(r^t)$ where $r^t$ is the output key used in the $t$-th round of $\llbracket\fBasisTest(K;1\sim i\kappa)\rrbracket$; \item $\llbracket\fAuxInf_1^{1\sim i}\rrbracket=\llbracket\fAuxInf^{ 1}_1\rrbracket\odot \llbracket\fAuxInf^{ 2}_1\rrbracket\odot \cdots \llbracket\fAuxInf^{i}_1\rrbracket$, where each one is the same algorithm as the $\llbracket\fAuxInf\rrbracket$ appeared in Lemma \ref{lem:6.3}, but for different superscripts the random coins are independently random.\end{itemize}
			            \item $\ket{\phi^i}$ is also $(1,i 2^{\kappa+2})$-server-side-representable from (\ref{eq:155c}).
		      \end{itemize}
		      
		\item The $\fBasisTest$ step is executed for $(\kappa^2-i\kappa)$ rounds. The parameters of the protocol are the same.
		\item The RO query number of the adversary is $\leq 2^\kappa+i\kappa$.
	\end{itemize}
\end{mdframed}

Thus to complete this reduction, what we we need to do is to prove we can reduce the ``Statement-round-$i$-completed'' to ``Statement-round-$(i+1)$-completed''. We will give the details of such a reduction in Appendix \ref{sec:aj3}. (Note that this is actually the $(i+1)$th round of reductions.)\par
\paragraph{How to understand this statement}Let's explain the statement above.
\begin{enumerate}
\item The conclusion is (almost) the same as Lemma \ref{lem:11.1}, and the extra ``$2^{-2\eta}$'' term is small enough to be ignored if you are trying to get an intuitive understanding.
\item As we do the inductive reduction step-by-step, the conditions will change step-by-step. For the initial state we consider, the number of terms inside will increase, and the norm of the last $\ket{\phi^{\cdots}}$ term will decrease. \item If we ignore the last term, the form of the state is the same as what we want in the beginning of Section \ref{sec:11.3}.\par
And as we can see, the query number in each $\cP_{\cdots}$ is bounded.
\item The different auxiliary information might seem complicated. This comes from the details of the security proof, and they play different roles.\par
One important property is these auxiliary information does not affect the SC-security of each term too much. We can view them as specially designed auxiliary information that both (somewhat) preserves the SC-security and helps the security proof.\par
The reader might get confused when the $\odot$ symbol and the addition of states are used together. This is allowed.
And recall that $\odot$ symbol means ``the client computes this function using its own system and random coins, and sends the result to some fixed place of the read-only buffer as the auxiliary information''. 
\item The number of rounds decreases since one block is removed after one step of this inductive reduction. 
\item And the query number bound increases slightly, which is small enough to be ignored in an intuitive understanding.
\end{enumerate}

See Appendix \ref{sec:ap111} for details.
\section{Final $\fSuccUBQC$ Protocol}\label{sec:12.3}
\subsubsection{Our $\fSuccUBQC$ protocol}
Below we give a top-down description of our protocol. 
\begin{mdframed}[style=figstyle]
\begin{prtl}\label{prtl:17}$\fSuccUBQC(C;\kappa)$ where $\kappa$ is the security parameter is defined as follows
	\begin{enumerate}
		\item The client and the server run $$\fGdgPrep(\underbrace{\kappa^9}_{\text{security}},\underbrace{|C|}_{\substack{\text{output}\\\text{number}}})$$ (Protocol \ref{prtl:13}). Denote the keys got by the client as $K_{out}=\{y^{(i)}_b\}_{i\in [|C|],b\in \{0,1\}}$. The honest server should hold the state $Gadget(K_{out})=\otimes_{i=1}^{|C|}(\ket{y^{(i)}_0}+\ket{y^{(i)}_1})$.
		\item For $i=1,\cdots |C|$:\par
		      Run the protocol $$\fQFac(K_{out}^{(i)};\underbrace{100\kappa^3}_{\substack{\text{padding} \\ \text{length}}},\underbrace{\kappa}_{\text{security}})$$ from initial state $\ket{y_0^{(i)}}+\ket{y_1^{(i)}}$. Suppose the client gets angle $\theta^i$ from this protocol call. Reject if the subprotocol rejects.\par
		      The honest server should hold $\ket{+_{\theta^i}},i\in [|C|].ß$
		\item The client and the server run the gadget-assisted UBQC protocol (Prtl \ref{prtl:gaubqc}) and the client uses $\theta^i,i\in [|C|]$ computed above as the input.
	\end{enumerate}

\end{prtl}\end{mdframed}
Then we can prove our $\fSuccUBQC$ protocol (Protocol \ref{prtl:17}) is secure:
\begin{thm}\label{thm:11.2}
	When $\kappa$ is bigger than some constant, $|C|\leq 2^{\kappa^{1/5}}$, for any adversary $|\fAdv|\leq 2^\kappa$,
	$$|Pr(\fAdv^{\fSuccUBQC}(\kappa,C)=0)-Pr(\fAdv^{\fSuccUBQC}(\kappa,0^{|C|})=0)|\leq 2^{-\kappa/3}$$
	which is the qIND-CPA distinguishing advantage discussed in Section \ref{sec:2.3}.
\end{thm}
Once we prove this theorem, we complete the security proof of the whole protocol and complete the proof of Theorem \ref{thm:1}.\par
\subsubsection*{Some intuitions for the security of these protocols }
Below we will discuss some intuition and subtleness of these protocols and theorems, and why Protocol \ref{prtl:17} is secure when the 8-basis qfactory is defined as above.\par
There is one more thing to care about: as we showed in Protocol \ref{prtl:17}, to prepare the states we need to repeat the qfactory protocol above for many times, so why is Protocol \ref{prtl:24} composable when it's used on different key pairs? In general, it's not (as far as we know). However, in our case, we note that the initial state of the qfactory protocol, which is $\ket{\varphi^1}$, is the output state of a secure remote state preparation protocol, and the security of the remote gadget preparation protocol (Protocol \ref{prtl:13}, Lemma \ref{thm:10.3}) says for all $i\in [L]$, the output state ($\ket{\varphi^1}$ here, $\ket{\varphi^\prime}$ in Lemma \ref{thm:10.3}) is SC-secure for $K_{out}^{(i)}$ even given other keys ($K_{out}-K_{out}^{(i)}$), and the auxiliary information $K_{out}-K_{out}^{(i)}$ here gives us the composability: we can define a series of hybrids, and if an adversary can distinguish two consecutive hybrids, it can \emph{simulate} the extra messages using the auxiliary information ($K_{out}-K_{out}^{(i)}$) and break Lemma \ref{lem:11.1}.\par


\subsection{Security of the $\fSuccUBQC$ Protocol}\label{sec:11.4}

Now we will prove the final security statement of our protocol, which is Theorem \ref{thm:11.2}.

\begin{proof}[Proof of Theorem \ref{thm:11.2}]
	The proof is via a hybrid method. We will replace the bits that the client uses by random bits step-by-step.\par
	For $i=0\cdots L$, define $Hybrid^i(\kappa, C)$ as follows: in $Hybrid^i$, after the second step of Protocol \ref{prtl:17} completes, instead of running $\fGAUBQC(\kappa,(\theta^1,\cdots, \theta^{|C|}),C)$ (recall the client's angles are denoted as $\theta^i$), it runs:
	$$\fGAUBQC(\kappa,(\theta^1,\cdots \theta^i,\tilde\theta^{i+1},\cdots,\tilde\theta^{|C|}),C)$$
	where for each $t$ from $i+1$ to $L$, the client replaces $\theta_{2}^t$ and $\theta_{3}^t$ (recall $\theta^t=\theta_1^t\pi+\theta_2\frac{\pi}{2}+\theta_3\frac{\pi}{4}$) in its stored angles with two new random bits $\tilde\theta^t_2,\tilde\theta_3^t$, and substitute $\tilde\theta^t=\theta_1^t\pi+\tilde\theta_2\frac{\pi}{2}+\tilde\theta_3\frac{\pi}{4}$ as the inputs. \par
	So $Hybrid^L=\fSuccUBQC$, the original protocol. Note that we are doing the hybrids in a backward way, in other words, replace the angles from the last to the first.\par
	Then we will prove for any $i$, under the conditions of Theorem \ref{thm:11.2},
	\begin{equation}\label{eq:10.4.31}|Pr(\fAdv^{Hybrid^i}(\kappa,C)=0)-Pr(\fAdv^{Hybrid^{i-1}}(\kappa,C)=0)|\leq 2^{-\kappa}\end{equation}
	Which means it's hard for the adversary to distinguish whether it's run on $Hybrid^i$ or $Hybrid^{i-1}$.
	\begin{proof}[Proof of equation \eqref{eq:10.4.31}]
	 Here $C$ can be assumed to be public. The only difference of $Hybrid^i$ and $Hybrid^{i-1}$ is whether the second and third bits of the $i$-th angle are replaced by random bits. Note that the $i$-th angle comes from an execution of $\fQFac(K_{out}^{(i)})$, which is in the $i$-th round of the second step of Protocol \ref{prtl:17}.\par
	Suppose there is an adversary $\fAdv$ (with query number $\leq 2^\kappa$) that can break equation \eqref{eq:10.4.31}. Use $\ket{\varphi^1}$ to denote the purified joint state after the first step of Protocol \ref{prtl:17}. First by the security statement (Theorem \ref{thm:10.3}) of remote gadget preparation we know $P_{pass}\ket{\varphi^1}$, the state of projecting $\ket{\varphi^1}$ onto the passing space, either satisfies \begin{itemize}\item (Case 1) $|P_{pass}\ket{\varphi^1}|\leq 2^{-\kappa}$, then (\ref{eq:10.4.31}) is already true;  or \item (Case 2) it is $(2^{\eta},2^{-\eta}|P_{pass}\ket{\varphi^1}|)$-SC-secure for $K_{out}^{(i)}$ given $K_{out}-K_{out}^{(i)}$, where $\eta=10000\kappa^2$. \footnote{Note that when we need to substitute $Security=\kappa^9$ when we apply Theorem \ref{thm:10.3}.}\par
	In this case, the adversary, running from the second step of the protocol with initial state $P_{pass}\ket{\varphi^1}$ can distinguish $Hybrid^{i-1}$ and $Hybrid^i$ with high probability. More formally,
	\begin{equation}\label{eq:10.4.99}|Pr(\fAdv^{Hybrid^i_{\geq 2}}(\kappa,C,P_{pass}\ket{\varphi^1})=0)-Pr(\fAdv^{Hybrid^{i-1}_{\geq 2}}(\kappa,C,P_{pass}\ket{\varphi^1})=0)|\end{equation}\begin{equation*}=p> 2^{-\kappa}\end{equation*}
	where $Pr(\fAdv^{Hybrid^i_{\geq 2}}(\kappa,C,P_{pass}\ket{\varphi^1})=0)$ is the probability of outputting $0$ when the initial state is $P_{pass}\ket{\varphi^1}$, the protocol is executed from the second step of $Hybrid^i$, and the adversary is $\fAdv_{\geq 2}$, defined as the operation in $\fAdv$ from the second step of the protocol.\par
	Starting from equation (\ref{eq:10.4.99}), we can construct an adversary $\fAdv^\prime$ which can break the security (Lemma \ref{lem:11.1}) of the protocol $\fQFac(K_{out}^{(i)})$ as follows.\par
	 The initial state is 
	  \footnote{We can't use $P_{pass}\ket{\varphi^1}$ directly here because in the protocols shown in (\ref{eq:10.4.99}) many new messages are provided besides what are part of $\fQFac(K_{out}^{(i)})$. The adversary $\fAdv^\prime$ should be able to simulate them by itself.}
	$$\ket{\tilde\varphi}:=P_{pass}\ket{\varphi^1}\odot (K_{out}-K_{out}^{(i)})$$
	Which means the adversary gets the keys at all the other index ($K_{out}-K_{out}^{(i)}$) as the auxiliary information.\footnote{We note that the circuit $C$ is already assumed to be public.} This state is $(2^\eta,2^{-\eta}|P_{pass}\ket{\varphi^1}|)$-SC-secure for $K_{out}^{(i)}$ by the security property of the $\fGdgPrep$ protocol. (See in the discussion above equation (\ref{eq:10.4.99}).)\par
	Let's use (\ref{eq:10.4.31}) to derive a contradiction to Lemma \ref{lem:11.1}. by making $\fAdv^\prime$ simulate the operation in (\ref{eq:10.4.31}). Now $\fAdv^\prime$ is defined as follows.\begin{enumerate}\item $\fAdv^\prime$ simulates everything the client sends to the server from the beginning (which is the beginning of the second step of Protocol \ref{prtl:17}) to the $(i-1)$-th round of the second step of Protocol \ref{prtl:17} using $K_{out}-K_{out}^{(i)}$. The simulated client will store the angles $\theta^1,\cdots \theta^i$ and $\fAdv^\prime$ can get access to them.
	\item Then both parties run the $\fQFac(K_{out}^{(i)})$ protocol, which is the same in both the $i$-th round of the UBQC protocol, and the distinguishing game of $\fQFac$ protocol. 
	\item The remaining rounds of the second step of Protocol \ref{prtl:17}, and the first $(i-1)$ rounds of the third step of Protocol \ref{prtl:17} can be simulated using $K_{out}-K_{out}^{(i)}$ and the angles stored on the simulated client.
	\item And for the client-side measurement angle in the $i$-th round of the third step of Protocol \ref{prtl:17}, $\fAdv^\prime$ can simulate it using $\theta_{2},\theta_3$ or $\theta_{2}^\prime,\theta_3^\prime$ from the distinguishing game of $\fQFac$. \item And everything the client sends to the server after the $i$-th round of the third step of Protocol \ref{prtl:17} are just completely random strings by Proposition \ref{prop:gaubqcsec0}. $\fAdv^\prime$ can generate them easily.
	\item Finally $\fAdv^\prime$ uses the answer from equation \ref{eq:10.4.31} as the answer to the distinguishing game of $\fQFac$.
	\end{enumerate}
	  Thus we get an adversary of query number $\leq 2^{\kappa}$ that can break $\fQFac$ on initial state $\ket{\tilde\varphi}$ with distinguishing advantage $p$ and get a contradiction to Lemma \ref{lem:11.1}.\par
	\end{itemize}\end{proof}
	Now we get equation \eqref{eq:10.4.31}. Thus
	\begin{equation}\label{eq:104r}|Pr(\fAdv^{\fSuccUBQC}(\kappa,C)=0)-Pr(\fAdv^{Hybrid^{0}}(\kappa,C)=0)|\leq L\cdot 2^{-\kappa}\leq 2^{-\kappa/2}\end{equation}
	Similarly we can prove
	\begin{equation}\label{eq:105r}|Pr(\fAdv^{\fSuccUBQC}(\kappa,0^{|C|})=0)-Pr(\fAdv^{Hybrid^{0}}(\kappa,0^{|C|})=0)|\leq 2^{-\kappa/2}\end{equation}
	Finally by Proposition \ref{prop:ubqcsec} there is
	\begin{equation}\label{eq:106r}|Pr(\fAdv^{Hybrid^{0}}(\kappa,C)=0)-Pr(\fAdv^{Hybrid^{0}}(\kappa,0^{|C|})=0)|=0\end{equation}
	Summing up (\ref{eq:104r})(\ref{eq:105r})(\ref{eq:106r}) completes the proof.
\end{proof}
At this time we complete all the protocol construction and the security proof. Especially, we prove the main theorem (Theorem \ref{thm:1}).
\newpage
\bibliographystyle{apalike}

\bibliography{main}
\cleardoublepage

\appendix
\chapter{Missing Proofs}\label{cht:b}
\section{Missing Proofs By Section \ref{sec:4.3}}\label{sec:app1}
\begin{proof}[Proof of Lemma \ref{lem:3.4r}]
	First consider the case where $2^{\alpha_1}=1$. Suppose $$\ket{\varphi}=\cP\ket{\varphi_{init}},|\ket{\varphi_{init}}|=1,\cP=\cP_{2^{\alpha_2}}H\cdots \cP_1H\cP_{0}\text{ is a server-side operation}$$
	. Consider the states coming from replacing the last $i$ queries by queries to $H^{pad}$ where $H^{pad}$ is the blinded oracle where $H(\cdots ||pad||\cdots)$ is blinded:
	$$\ket{\varphi^{blind,i}}=\cP^{blind,i} \ket{\varphi_{init}},\cP^{blind,i}=\cP_{2^{\alpha_2}}H^{pad}\cdots H^{pad} \cP_{2^{\alpha_2}-i}H\cdots \cP_1H\cP_{0}$$
	Then
	{\footnotesize $$ |\ket{\varphi^{blind,i}}-\ket{\varphi^{blind,i-1}}|=|\sum_{pad}\frac{1}{\sqrt{2^l}}\ket{pad}\otimes (H-H^{pad})(\cP_{2^{\alpha_2}-i}H\cdots \cP_1H\cP_{0}\ket{\varphi_{init}})|\leq |Pads|\cdot 2^{-l/2+1} $$}
	Thus take $\ket{\varphi^\prime}=\ket{\varphi^{blind,2^{\alpha_2}}}$, then $|\ket{\varphi^\prime}-\ket{\varphi}|\leq |Pads|\cdot 2^{\alpha_2-l/2+1}$.\par
	For the $2^{\alpha_1}>1$ case, by the triangle inequality taking $\ket{\varphi^\prime}=\sum_j\cP_j^{blind,2^{\alpha_2}} \ket{\varphi_{init}}$ completes the proof.
\end{proof}
\begin{proof}[The elementary calculation in Lemma \ref{lem:4.4}]
	The problem is reduced to the following problem: $\omega_0= A$, $\forall i\geq 1,\omega_i\leq \omega_{i-1}/\sqrt{2}$, maximize $\sum_{i=1}^\infty\sqrt{\omega_{i-1}^2-\omega_i^2}$.\par
	Consider it as a function of $\omega_0=A$. Denote it as $f$. Then $f(A)=O(1)A$. And we have $f(A)=\max_{\omega_1\leq A/\sqrt{2}}( f(\omega_1)+\sqrt{A^2-\omega_1^2})$. Substitute and we get $f(A)=(\sqrt{2}+1)A$.
\end{proof}

\section{Basic Lemmas in the SC/ANY-security Framework}\label{sec:basiclemmas}
In this section we give a series of security lemmas for the SC/ANY-security. We refer to Section \ref{sec:techovw} for an overview of this lemmas.
\subsection{Security Influence of Adding Lookup Tables}\label{sec:4.4}
In this section we study how the client providing extra lookup tables to the server will influence the SC-security or ANY-security of the state.\par
Suppose a state $\ket{\varphi}$ is SC-secure for $K$ with parameters being exponential. If an extra lookup table, for example, $\fLT(K\rightarrow K_{out})$, is provided, where $K_{out}$ is sampled randomly, it's still hard for the server to compute two keys simultaneously. Intuitively by the SC-security property the adversary can only get at most one of the keys, thus it can only decrypt one row in the lookup table. And this in turn implies the lookup table does not give the adversary extra power and the state should still be SC-secure for $K$ with similar parameters. The intuition might seem to be a ``circular proof'', but the result itself is indeed true. Formalizing this under different cases (SC/ANY security, lookup table or reversible lookup table, etc) leads to the following lemmas.\par We put the proofs of these lemmas in Appendix \ref{sec:a44}.
\begin{lem}\label{lem:4.8}
	The following statement is true when $\eta$ is bigger than some constant.\par
	Suppose the key pair is denoted as $K=\{x_0,x_1\}$, and the initial state is described by the purified joint state $\ket{\varphi}$. Suppose the following conditions are satisfied:
	\begin{itemize}
		\item (Security of the inputs)	$\ket{\varphi}$ is $(2^{\eta},2^{-\eta}|\ket{\varphi}|)$-SC-secure for $K$.
		\item (Well-behaveness of the inputs) $\ket{\varphi}\in \cWBS(D)$, $D\leq 2^\eta$.
		\item (Sufficient padding length and output length) $l>6D+4\eta$, $\kappa_{out}>l+\eta$.
	\end{itemize}
	A client side algorithm $\fAuxInf$ is defined as follows, where the choices for each round, $b^t$, $p_i^t$ are chosen by the adversary non-adaptively:\\
	\begin{mdframed}
			For $t=1,\cdots n$, the adversary selects one of the followings, and the client adds the result into $\llbracket\fAuxInf\rrbracket$:
			\begin{itemize}
				\item The adversary chooses $b^t\in\{0,1\}$, and $p^t_1,p^t_2,p^t_3$ for this round. The client samples $pad^t\leftarrow_r \{0,1\}^l$, and computes $(pad^t,H(pad^t||p^t_1||x_{b^t}||p^t_2)\oplus p^t_3)$ as the result.
				\item The client computes $\fLT(K\rightarrow K_{out\_t};\underbrace{ \ell}_{\substack{\text{padding} \\ \text{length}}},
\underbrace{ \kappa_{\text{out}}}_{\substack{\text{tag} \\ \text{length}}})$, where $K_{out\_t}$ is sampled randomly from $\{0,1\}^{\kappa_{out}}$.
				\item The client computes $\fRevLT(K\leftrightarrow K_{out\_t};\underbrace{ \ell}_{\substack{\text{padding} \\ \text{length}}})$ and $Tag(K_{out\_t})$, where $K_{out\_t}$ is sampled randomly differently from $\{0,1\}^{\kappa_{out}}$.
			\end{itemize}\end{mdframed}
	Then for all $n<2^{\sqrt{\eta}}$, the choices in each round, and $b^t,p^t_i$ in each round, we have: $\ket{\varphi}\odot \llbracket\fAuxInf\rrbracket$ is $(2^{\eta/6},2^{-\eta/6}|\ket{\varphi}|)$-SC-secure for $K$. What's more, suppose in the $t$-th round the server chooses for the third choice above, we have $\ket{\varphi}\odot \llbracket\fAuxInf\rrbracket$ is $(2^{\eta/6},2^{-\eta/6}|\ket{\varphi}|)$-SC-secure for $K_{out\_t}$.
\end{lem}

Note that everything is chosen non-adaptively, thus we can view $\fAuxInf$ as a client-side algorithm, and use the notation as above.\par
One limit of this lemma is the security decreases multiplicatively. We can prove, when we do not consider the reversible lookup table, the decrease is actually additively:
\begin{lem}\label{lem:4.9}
	The following statement is true when $\eta$ is bigger than some constant.\par
	Consider the key set denoted as $K=\{x^{(i)}_0,x^{(i)}_1\}_{i\in [N]}$. Suppose the initial state is described by the purified joint state $\ket{\varphi}$. Suppose the following conditions are satisfied:
	\begin{itemize}
		\item (Security for the inputs)	$\forall i\in [N]$, $\ket{\varphi}$ is $(2^{\eta},2^{-\eta}|\ket{\varphi}|)$-SC-secure for $K^{(i)}$.
		\item (Well-behaveness of the inputs) $\ket{\varphi}\in \cWBS(D)$, $D\leq 2^\eta$.
		\item (Sufficient padding length, output length) $l>6D+2\eta$, $\kappa_{out}>l+\eta$
	\end{itemize}
	A client side algorithm $\fAuxInf$ is defined as follows, where $i^t$, $b^t$, $p_{\cdots}^t$ in each round are chosen by the adversary non-adaptively:\\
	\begin{mdframed}
			For $t=1,\cdots n$:
			\begin{itemize}
				\item The adversary chooses $i^t\in [N],b^t\in \{0,1\}$, strings $p_0^t,p^t_1,p^t_2,p^t_3$ where each of them has fixed length when $t$ varies. The client samples $pad^t\leftarrow \{0,1\}^l$, computes $(pad^t,H(p_0^t||pad^t||p^t_1||x^{(i^t)}_{b^t}||p^t_2)\oplus p^t_3)$ and adds it to $\llbracket\fAuxInf\rrbracket$.
				      
			\end{itemize}\end{mdframed}
	Then for any $n<\eta^2$, any adversary, for any $i\in [N]$, $\ket{\varphi}\odot \llbracket\fAuxInf\rrbracket$ is $(2^{\eta-4},2^{-\eta+4}|\ket{\varphi}|)$-SC-secure for $K^{(i)}$.
\end{lem}
\paragraph{Note} (1) In the conditions we deal with multi-key setting, and we do not require the initial state to be secure ``given the other keys''; (2)And the condition $n<\eta^2$ might seem unnatural. It could be much looser, but we choose to assume it since the more general version is not used later.\par

The following lemma consider the case where the initial state $\ket{\varphi}$ is $(2^{\eta},C|\ket{\varphi}|)$-SC-secure for $K$, where $C$ might be non-negligible.
\begin{lem}\label{lem:4.10}
	The following statement is true when $\eta$ is bigger than some constant.\par
	Suppose the key pair is denoted as $K=\{x_0,x_1\}$, and the initial state is described the purified joint state $\ket{\varphi}$. Suppose the following conditions are satisfied:
	\begin{itemize}
		\item (Security of the input) $\ket{\varphi}$ is $(2^{\eta},C|\ket{\varphi}|)$-SC-secure for $K$, $C>2^{-\eta/100}$.
		\item (Well-behaveness of the input) $\ket{\varphi}\in \cWBS(D)$, $D\leq 2^\eta$.
		\item (Sufficient padding length, output length) $l>6D+6\eta$, $\kappa_{out}>l+\eta$.
	\end{itemize}
	A client side algorithm $\fAuxInf$ is defined as follows, where the choices for each round, $b^t$, $p_i^t$ are chosen by the adversary non-adaptively:\\
	\begin{mdframed}
			For $t=1,\cdots n$, the adversary chooses one choice below, and the client adds the result into $\llbracket\fAuxInf\rrbracket$:
			\begin{itemize}
				\item The adversary chooses $b^t\in \{0,1\}$, $p^t_1,p^t_2,p^t_3$. The client samples $pad^t\leftarrow \{0,1\}^l$, and computes $(pad^t,H(pad^t||p^t_1||x_{b^t}||p^t_2)\oplus p^t_3)$ as the result.
				\item The client computes $\fLT(K\rightarrow K_{out\_t};\underbrace{ \ell}_{\substack{\text{padding} \\ \text{length}}},
\underbrace{ \kappa_{\text{out}}}_{\substack{\text{tag} \\ \text{length}}})$, where $K_{out}$ is a pair of keys sampled randomly from $\{0,1\}^{\kappa_{out}}$.
				\item The client computes $\fRevLT(K\leftrightarrow K_{out\_t};\underbrace{ \ell}_{\substack{\text{padding} \\ \text{length}}})$, where $K_{out}$ is a pair of different keys sampled randomly from $\{0,1\}^{\kappa_{out}}$.
			\end{itemize}\end{mdframed}

	Then for all $n<\eta^2$, the adversary's choices in each round, and $b^t,p^t_i$ in each round, we have: $\ket{\varphi}\odot \llbracket\fAuxInf\rrbracket$ is $(2^{\eta/37},3C|\ket{\varphi}|)$-SC-secure for $K$.
\end{lem}
\paragraph{Note} The condition $C>2^{-\Theta(\eta)}$ does not mean the lemma cannot be applied to a initial state that is $(2^{\eta},2^{-\eta}|\ket{\varphi}|)$-SC-secure for $K$. We can always relax the parameter before we apply this lemma: if $\ket{\varphi}$ is $(2^{\eta},2^{-\eta}|\ket{\varphi}|)$-SC-secure for $K$, it's certainly $(2^{\eta},C|\ket{\varphi}|)$-SC-secure for $K$ for any bigger $C$. We add this inequality to get rid of the extra exponentially-small terms in the conclusion. Without this condition, the second parameter in the conclusion will be in the form of $O(C)+2^{-\Theta(\eta)}$, and with this condition this become $3C$, which is simpler to understand.\par

\subsection{Oneway-to-hiding and Collapsing Property}\label{sec:4.5}
For the lemmas from Section \ref{sec:4.5} to the end of Section \ref{sec:4}, we postpone their proofs to Appendix \ref{sec:a57}.
\subsubsection{Oneway-to-hiding}
The following lemma is a variant of the famous oneway-to-hiding lemma\cite{onewaytohiding} of the random oracle, described in a way that is more suitable in our setting. This can be proved by a simple hybrid method.
\begin{lem}\label{lem:4.12}
	Suppose the client and the server (adversary) run protocol $\fPrtl$ on initial purified joint state $\ket{\varphi}$. The adversary is $\fAdv$. The number of queries in $\fAdv$ during the protocol is at most $2^\lambda$. Suppose $Set$ is a set of inputs to the random oracle (which might be a deterministic function of the content of some read-only system). Suppose $H^\prime$ is a blinded oracle of $H$ where $Set$ is blinded.\par
	If at any time during the protocol when the adversary is going to make a random oracle query, denote the state as $\ket{\varphi^t}$, there is
	\begin{equation}\label{eq:8}|P_{Set}\ket{\varphi^t}|\leq 2^{-\eta}|\ket{\varphi}|\end{equation}
	where the projection is done on the system that is used for the random oracle query, to the space that the query is in $Set$. And assume the client-side queries does not contain any component in $Set$. Then denote the final state as $\ket{\varphi^\prime}$, and denote the final state when all the oracle queries by the adversary are replaced by queries to the blinded oracle $H^\prime$ as $\ket{\tilde\varphi}$, we have
	\begin{equation}\label{eq:9}
		\ket{\varphi}\approx_{2^{-\eta+\lambda+1}|\ket{\varphi}|}\ket{\tilde\varphi}
	\end{equation}

\end{lem}
The hybrid method can also be done in another direction, and we have:
\begin{lem}\label{lem:4.12n}
	Suppose the client and the server (adversary) run protocol $
\fPrtl$ on initial purified joint state $\ket{\varphi}$. The adversary is $\fAdv$. The number of queries in $\fAdv$ during the protocol is at most $2^\lambda$. Suppose $Set$ is a set of input to the random oracle (that might be a deterministic function of some read-only system). Suppose $H^\prime$ is a blinded oracle of $H$ where $Set$ is blinded.\par
	Suppose $\ket{\varphi^t}$ is defined as follows: replace all the oracle queries in $\fAdv$ before the $t$-th query by queries to $H^\prime$, denote the state just before the $t$-th query as $\ket{\varphi^t}$. If for any $t$, there is
	\begin{equation}\label{eq:10}|P_{Set}\ket{\varphi^t}|\leq 2^{-\eta}|\ket{\varphi}|\end{equation}
	where the projection is done on the system that is used for the random oracle query, to the space that the query is in $Set$. Then denote the final state as $\ket{\varphi^\prime}$, and denote the final state when all the oracle queries are replaced by the queries on the blinded oracle $H^\prime$ as $\ket{\tilde\varphi}$, we have
	\begin{equation}
		\ket{\varphi}\approx_{2^{-\eta+\lambda+1}|\ket{\varphi}|}\ket{\tilde\varphi}
	\end{equation}

\end{lem}
The difference of Lemma \ref{lem:4.12} and \ref{lem:4.12n} is in (\ref{eq:8}) $\ket{\varphi^t}$ comes from queries to the original oracle while in (\ref{eq:10}) it comes from queries to the blinded oracle. Both lemmas are useful later.
\subsubsection{Collapsing property}
And another lemma is the ''collapsing property'' of some special input state. We are not sure where it is first used, one appearance of this technique is in \cite{MahadevVerification}. We describe it in a way that is more suitable in our setting.\par

\begin{lem}\label{lem:4.13}
	Consider a key pair denoted as $K=\{x_0,x_1\}$ . The current state is described by the purified joint state $\ket{\varphi}$. Suppose $P^S_{span\{x_0, x_1\}}\ket{\varphi}=\ket{\varphi}$ where $P^S_{span\{x_0, x_1\}}$ is the server side projection on some system $S$ onto the space of $\{x_0,x_1\}$. And $\ket{\varphi}$ is $(2^{\eta},2^{-\eta}|\ket{\varphi}|)$-SC-secure for $K$.\par
	Suppose a server-side operation $\cD$ satisfies the query number $|\cD|\leq 2^{\eta-4}$. Then
	\begin{equation}
		|\quad|P_0\cD\ket{\varphi}|-|P_0\cD\circ COPY\circ \ket{\varphi}|\quad|\leq 	2^{-\eta/2+2}|\ket{\varphi}|
	\end{equation}
	where $COPY$ is a operation that copies the system $S$ into a system that is not used by $\cD$. $P_0$ is a projection onto $\ket{0}$ on some bit on the server side.
\end{lem}
Intuitively, the first condition means $\ket{\varphi}$ can be written as $\ket{x_0}\ket{\cdots}+\ket{x_1}\ket{\cdots}$ (if we use the natural notation instead of the purified notation). This lemma is similar to saying the state is indistinguishable to the mixed state of $\ket{x_0}\ket{\cdots}$ and $\ket{x_1}\ket{\cdots}$. But here we use $COPY$ operator instead of the standard-basis measurement since this form is useful later.
\subsection{Lemmas about the Blinded Oracle}\label{sec:4.6}
The proof of lemmas in this section is also in Appendix \ref{sec:a57}.\par
\subsubsection{A blinded oracle does not make the adversary more powerful}
We also need to give some lemmas about the blinded oracle. Recall the definition of the blinded oracle. For a pair of keys $K=\{x_0,x_1\}$, when we talk about blinded oracle $H^\prime$ where $\cdots||K||\cdots$ part of the inputs for the oracle $H$ is blinded, we mean an oracle where on the $\cdots||K||\cdots$ part (which contains $H(\cdots ||x_0||\cdots)$ and $H(\cdots ||x_1||\cdots)$) the output of $H$ is independently random from $H$ but on all the other part it's the same as $H$. Note that the paddings before and after $K$ are usually fixed-length, whose values depend on the setting when we use the blinding operation.\par
If we just look at the random oracle itself, $H$ and $H^\prime$ are ``symmetric'', and if the initial state does not depend on the random oracle, giving the adversary $H^\prime$ is not stronger or weaker than $H$; but since when we consider a blinded oracle, the protocol has already be run for some time and the protocol and the adversary all query $H$ instead of $H^\prime$, if during some time in the security proof we need to consider an adversary which can only query the blinded oracle $H^\prime$, it means we temporarily restrict the power of the adversary and do not want it to query the blinded part of $H$. On the other hand, if the adversary has access to the full oracle $H$, it can \emph{simulate} the blinded part with random values, thus giving it access to the blinded oracle does not give it extra power either.
\begin{lem}\label{lem:4.14a}
	$K$ is a pair of keys, stored in some client side read-only system. If $Tag(K)$ is stored in some fixed place of the read-only buffer, an adversary can simulate $H^\prime$ to any finite output length on the space ``$Tag$ is injective on the inputs with the same length as the keys in $K$'', where $H^\prime$ is defined to be a  new blinded oracle of $H$ with $\cdots ||K||\cdots$ part being blinded (the input before and after $K$ are arbitrary but have fixed length). Each query to $H^\prime$ costs four queries to $H$.\par
	(Here ``$\cdots$'' means it's arbitrary strings of a fixed length, but if we replace it with some values that can be computed from the read-only buffer, the statement is still true.)
\end{lem}
We note that the adversary does not need to know $K$ itself to do this simulation. Having $Tag(K)$ is enough. We further emphasize $H^\prime$ is a new blinded oracle; which means, $H^\prime$ does not appear before the time of blinding. Thus this lemma simply says ``having an extra blinded oracle is no more powerful than having the original oracle''. The nontrivial thing here is the adversary may not know $K$.
\begin{proof}
	The simulator samples fresh new random values for the outputs on the $\cdots ||K||\cdots$ part of the inputs of the blinded oracle. On each query to $H^\prime$, the simulator checks whether the query input has the form of $\cdots ||K||\cdots$, (note that this checking does not need the description of $K$; the simulator can use $Tag(K)$ to do it.) stores the checking result (1 for yes and 0 for no) in a separate bit, then behaves as $H$ in the $0$ space and uses the sampled randomness as the output in the $1$ space. Then use another RO queries to disentangle the bit that stores the checking result.
\end{proof}
\subsubsection{The interplay between ANY-security and the blinded oracle}
For these reasons (discussed in the last subsubsection), we can view the blinded oracle as a restricted form of the normal oracle, and the adversary which can only query a (freshly new) blinded oracle is no more powerful than an adversary that can query the original oracle. On the other hand, sometimes we can prove the properties under normal oracle by proving the properties under the blinded oracle, like the lemma below: (before that, we first generalize the SC/ANY-security to the blinded oracle setting.)
\begin{defn}[SC/ANY security under blinded oracles]\label{def:secblind}
	Consider a key pair denoted as $K=\{x_0,x_1\}$. 
	\begin{itemize}\item We say a state $\ket{\varphi}$ is $(2^{\eta},A)$-ANY-secure for $K$ under a blinded oracle $H^\prime$ if for all the server-side operation $\cD$ that only queries $H^\prime$ and the number of oracle queries to $H^\prime$ is at most $2^\eta$, $|P_{span\{x_0, x_1\}}\cD(\ket{\varphi}\odot Tag(K))|\leq A$.
	\item We say a state $\ket{\varphi}$ is $(2^{\eta},A)$-SC-secure for $K$ under a blinded oracle $H^\prime$ if for all the server-side operation $\cD$ that only queries $H^\prime$ and the number of oracle queries to $H^\prime$ is at most $2^\eta$, $|P_{x_0||x_1}\cD(\ket{\varphi}\odot Tag(K))|\leq A$.
	\end{itemize}
\end{defn}

\begin{lem}\label{lem:4.14}
The following is true when $\eta$ is bigger than some constant.\par
	Suppose the key pair is denoted as $K=\{x_0,x_1\}$. The initial state is the purified joint state $\ket{\varphi}$. $Tag(K)$ is stored in some fixed place of the read-only buffer. If:
	\begin{itemize}
		\item $\ket{\varphi}$ is $(2^{\eta},2^{-\eta}|\ket{\varphi}|)$-ANY-secure for $K$ under $H^\prime$ where $H^\prime$ is a new blinded oracle of $H$ with $\cdots||K||\cdots$ being blinded (paddings have fixed length);
		\item $|P_{\text{$Tag$ is not injective on inputs with the same length as $K$}}\ket{\varphi}|<2^{-2\eta}|\ket{\varphi}|$,
	\end{itemize} then $\ket{\varphi}$ is $(2^{\eta/2},2^{-\eta/2+2}|\ket{\varphi}|)$-ANY-secure for $K$.\par
	Here ``$\cdots$'' means it's arbitrary strings of a fixed length. If we replace it with some values that can be computed from the read-only buffer, the statement is still true.
\end{lem}
This lemma is also intuitive: note that we are talking about the ANY security. Informally speaking, since the adversary is hard to compute the keys in $K$, it's hard for it to distinguish $H$ and $H^\prime$, which in turn implies the adversary is hard to compute the keys. This sounds like a circular proof, but we can actually prove it in this way via a hybrid method.\par
The following lemma combines the decomposition lemma (Lemma \ref{lem:4.5}) with a generalization of the previous lemma:
\begin{lem}\label{lem:4.15}
The following is true when $\eta$ is bigger than some constant.\par
	Suppose the key pair is denoted as $K=\{x_0,x_1\}$. The initial state is the purified joint state $\ket{\varphi}$. $Tag(K)$ is stored in some fixed place of the read-only buffer. Suppose:
	\begin{itemize}
	\item $\ket{\varphi}$ is $(2^{\eta},C|\ket{\varphi}|)$-ANY-secure for $K$. $\frac{1}{3}>C>2^{-\eta/24}$.
	\item $|P_{\text{$Tag$ is not injective on inputs with the same length as $K$}}\ket{\varphi}|<2^{-2\eta}|\ket{\varphi}|$.
	\end{itemize}
  Then for any server-side operation $\cD$ with query number $|\cD|\leq 2^{\eta/12}$, consider $\cD^{blind}$ which comes from replacing the oracle queries in $H$ by oracle queries to the blinded oracle $H^\prime$ where $\cdots||K||\cdots$ is blinded (the prefix and suffix ``$\cdots$'' have fixed length), we have
	\begin{equation}
		\cD\ket{\varphi}\approx_{3C|\ket{\varphi}|}\cD^{blind} \ket{\varphi}
	\end{equation}

\end{lem}
Let's compare these two lemmas. Lemma \ref{lem:4.14} is in the form of unpredictability and Lemma \ref{lem:4.15} is in the form of indistinguishability (actually even stronger, in the form of trace-distance). What's more, in Lemma \ref{lem:4.15} $C$ can be inverse-polynomial. These two lemmas are  convenient in different cases.
\subsubsection{The interplay between SC-security and blinded oracle}
In the last subsubsection we know how the blinding can affect the ANY-security. In this subsubsection we consider how it affects the SC-security.\par

When the initial state has SC-security, one technique that we will use commonly is to write the adversary's operation as a sequence of unitaries and queries, insert some projections into them, analyze their properties and combine them back by linearity. Such a method can be used together with the blinded oracle, which leads to the following lemmas that are very useful:
\begin{lem}\label{lem:4.16}
	Consider a pair of keys denoted as $K=\{x_0,x_1\}$, the initial purified joint state is $\ket{\varphi}$. $Tag(K)$ is stored in some fixed place of the read-only buffer. Suppose
	\begin{itemize}
	\item $\ket{\varphi}$ is $(2^\eta,2^{-\eta}|\ket{\varphi}|)$-SC-secure for $K$.
	\item $|P_{\text{$Tag$ is not injective on inputs with the same length as $K$}}\ket{\varphi}|<2^{-2\eta}|\ket{\varphi}|$.
	\end{itemize}
Consider state $\ket{\varphi^\prime}:=P_{x_b}\cU\ket{\varphi}, b\in \{0,1\}$, where $\cU$ is a server-side operation with query number $|\cU|\leq 2^{\eta/3}$. Suppose $\cU^{blind}$ is the blinded version of $\cU$ by replacing the queries in $\cU$ with queries to $H^{\prime}$, which is the blinded oracle of $H$ with $\cdots||x_{1-b}||\cdots$ being blinded (the prefix and suffix ``$\cdots$'' have fixed lengths), correspondingly define $\ket{\tilde\varphi^\prime}=P_{x_b}\cU^{blind}\ket{\varphi}$, then $$\ket{\tilde\varphi^\prime}\approx_{2^{-\eta/3}|\ket{\varphi}|} \ket{\varphi^\prime}$$
\end{lem}
Intuitively, in the end there is a projection onto $P_{x_b}$, thus it should be hard to find a place in the middle of the operation where the adversary can get $x_{1-b}$, otherwise the SC-security of the initial state will be broken.\par
 Note that in many previous lemmas they talk about the blinded oracle where $H(\cdots||x_{b}||\cdots)$ and $H(\cdots||x_{1-b}||\cdots)$ are both blinded; here this lemma only makes one key blinded.\par
Note that Lemma \ref{lem:4.16} is ``backward blinded'', where we replace the oracle queries before the projection onto $x_b$ happens. We can also get a ``forward blinded'' lemma:
\begin{lem}\label{lem:r4.17}

	Suppose the key pair is denoted as $K=\{x_0,x_1\}$, the initial state is the purified joint state $\ket{\varphi}$. $Tag(K)$ is stored in some fixed place of the read-only buffer. Suppose
	\begin{itemize}
	\item $\ket{\varphi}$ is $(2^\eta,2^{-\eta}|\ket{\varphi}|)$-SC-secure for $K$.
	\item $|P_{\text{$Tag$ is not injective on inputs with the same length as $K$}}\ket{\varphi}|<2^{-2\eta}|\ket{\varphi}|$.
	\end{itemize}

	Consider state $\ket{\varphi^\prime}:=\cU P_{x_b}\ket{\varphi}$, $b\in \{0,1\}$, where $\cU$ is a server-side operation and query number $|\cU|\leq 2^{\eta/3}$. Suppose $\cU^{blind}$ is the blinded version of $\cU$ by replacing the queries in $\cU$ with the queries to $H^{\prime}$ which is the blinded oracle with $\cdots||x_{1-b}||\cdots$ being blinded (the prefix and suffix ``$\cdots$'' have fixed length), correspondingly define $\ket{\tilde\varphi^\prime}=\cU^{blind}P_{x_b}\ket{\varphi}$, then $$\ket{\tilde\varphi^\prime}\approx_{2^{-\eta/3}|\ket{\varphi}|} \ket{\varphi^\prime}$$
\end{lem}

\subsection{Indistinguishability of Lookup Tables}\label{sec:4.7}
The following lemma is also intuitively but will be used several times in the later sections. It says if a key is unpredictable then the ciphertexts and lookup tables encrypted under this key is indistinguishable from random strings.
\begin{lem}\label{lem:4.23}
The following statement is true when $\eta$ is bigger than some constant:\par
Suppose the key is denoted as $x_b\in K$ and the initial state is described by the purified joint state $\ket{\varphi}$. If the following conditions are satisfied:
\begin{itemize}
	\item (Security of the input) $\ket{\varphi}$ is $(2^\eta,2^{-\eta}|\ket{\varphi}|)$-unpredictable for $x_b$.
	\item (Well-behaveness of the input) $\ket{\varphi}\in \cWBS(D)$, $D\leq 2^\eta$.
	\item (Sufficient padding length, outut length) $l>6D+4\eta$, $\kappa_{out}>l+\eta$.
\end{itemize}
Consider the following protocol between the client and the adversary:\par
For $t=1,\cdots n<\eta^2$, the adversary chooses to execute one of the followings non-adaptively (which means, the adversary's operation is only to send some pre-computed values to the client and does not do any additional operations between each round):
\begin{itemize}
\item The server chooses and sends $p_0^t$ to the client. The client computes and sends $$\fEn_{x_b}(p_0^t;\underbrace{ \ell}_{\substack{\text{padding} \\ \text{length}}},\underbrace{ \kappa_{out}}_{\substack{\text{output} \\ \text{length}}})$$ to the server.
\item The server chooses and sends $px^t,sx^t,py^t,sy^t$ to the client. Each term has a fixed length when $t$ varies. The client samples $y$ from $\{0,1\}^{\kappa_{out}}$, and if $y$ is already sampled out in some previous round, use the existing $y$. And the client computes and sends the reversible lookup table $$\fRevLT(px^t||x_b||sx^t\leftrightarrow py^t||y||sy^t;\underbrace{ \ell}_{\substack{\text{padding} \\ \text{length}}})$$ to the server.
\end{itemize}	
Denote the post-execution state as $\ket{\varphi^\prime}$. Further denote the post-execution state where the client's responses are all replaced by random strings of the same length as $\ket{\varphi^{\prime\prime}}$. Then for any server-side operation (distinguisher) $\cD$ with query number $|\cD|\leq 2^{\eta/6}$, we have
$$|\quad|P_0\cD\ket{\varphi^\prime}|-|P_0\cD\ket{\varphi^{\prime\prime}}|\quad |\leq 2^{-\eta/6}|\ket{\varphi}|$$
where $P_0$ is some server-side projection.
\end{lem}
Then we can combine (a variant of) this lemma with the decomposition lemma (Lemma \ref{lem:4.7r}) to get:
\begin{lem}\label{lem:4.24}
The following statement is true when $\eta$ is bigger than some constant:\par
Suppose the key is denoted as $x_b\in K$ and the initial state is the purified joint state $\ket{\varphi}$. If the following conditions are satisfied:
\begin{itemize}
	\item (Security of the input) $\ket{\varphi}$ is $(2^\eta,C|\ket{\varphi}|)$-unpredictable for $x_b$. $C>2^{-\eta/38}$.
	\item (Well-behavenss of the input) $\ket{\varphi}\in \cWBS(D)$, $D\leq 2^\eta$.
	\item (Sufficient padding length, output length) $l>6D+4\eta$, $\kappa_{out}>l+\eta$.
\end{itemize}
Consider the following protocol between the client and the adversary:\par
For $t=1,\cdots n<\eta^2$, the adversary chooses to execute one of the followings non-adaptively (which means, the adversary's operation is only to send some pre-computed values to the client and does not do any additional operations):
\begin{itemize}
\item The server chooses and sends $p_0^t$ to the client. The client computes and sends $$\fEn_{x_b}(p_0^t;\underbrace{ \ell}_{\substack{\text{padding} \\ \text{length}}},\underbrace{ \kappa_{out}}_{\substack{\text{output} \\ \text{length}}})$$ to the server.
\item The server chooses and sends $px^t,sx^t,py^t,sy^t$ to the client. The client samples $y$ from $\{0,1\}^{\kappa_{out}}$, and if $y$ is already sampled out in some previous round, use the existing $y$. And the client computes and sends the reversible lookup table $$\fRevLT(px^t||x_b||sx^t\leftrightarrow py^t||y||sy^t;\underbrace{ \ell}_{\substack{\text{padding} \\ \text{length}}})$$ to the server.

\end{itemize}	
Denote the post-execution state as $\ket{\varphi^\prime}$. Further denote the post-execution state where the client's responses are all replaced by random strings of the same length as $\ket{\tilde\varphi^\prime}$. Then for any server-side operation (distinguisher) $\cD$ with query number $|\cD|\leq 2^{\eta/6}$, we have
$$|\quad|P_0\cD\ket{\varphi^\prime}|-|P_0\cD\ket{\tilde\varphi^\prime}|\quad |\leq 2.95C|\ket{\varphi}|$$
where $P_0$ is some server-side projection.
\end{lem}
We note that this lemma still holds if there is a fixed bit-wise permutation on the output key part. (Which means, the rows in the $\fRevLT$ is in the form of $px^t||x||sx^t\leftrightarrow perm(py^t||y||sy^t)$. This case is useful in some later sections.)
\section{Proofs of Lemmas in Section \ref{sec:4.4}}\label{sec:a44}
\begin{proof}[Proof of Lemma \ref{lem:4.8}]	The main technique in this proof is the hybrid method. (This is not the first step below, but it will be used.) However, the backward tables in the reversible lookup tables are one of the obstacles here. To solve this problem, when we define the blinded oracle, we will also make the outputs corresponding to the keys in the backward table blinded. Now let's start the proof.\par
	By Technique \ref{lem:4.2} we can assume $Tag(K)$ is stored in the read-only buffer.\par
	Define $Pads$ as the set of random paddings used in all the rounds, and for the reversible lookup tables, this include the padding used in both forward tables and backward tables.\par
	The paddings in $Pads$ are all sampled independently randomly. By Lemma \ref{lem:3.4r} if we replace all the oracle queries in the representation (see Definition \ref{def:rep}) of $\ket{\varphi}$ by $H\cdot(I-P_{Pads||\cdots})$, denote the final state as $\ket{\tilde\varphi}$, there is
	\begin{equation}\ket{\tilde\varphi}\approx_{2^{-\eta/2}|\ket{\varphi}|}\ket{\varphi}\end{equation}
	Thus
	\begin{equation}\label{eq:179z}
	\text{$\ket{\tilde\varphi}\odot Pads$ is $(2^{\eta},2^{-\eta/2.1}|\ket{\varphi}|)$-SC-secure for $K$.}	
	\end{equation}

	Let's move to study the SC-security of $\ket{\tilde\varphi}\odot \llbracket\fAuxInf\rrbracket$. For any server-side operation $$\cD=\cU_{2^\lambda} H\cU_{2^\lambda-1} H\cdots H\cU_0, \lambda\leq \eta/6$$ applied on $\ket{\tilde\varphi}\odot \llbracket\fAuxInf\rrbracket$, define $\cD_{t,b}$ as the operator that does a projection onto $(I-P_{Pads||K})$ before each of the first $(t-1)$ RO queries and does a projection onto $P_{Pads||x_b}$ before the $t$-th RO queries. \footnote{Recall Notation \ref{nota:2.2} for what $Pads||K$ means.} Define $\cD_0$ as the operator that does a projection onto $(I-P_{Pads||K})$ before each of the RO queries. Then \begin{equation}\cD=\sum_{t\in [2^\lambda],b\in \{0,1\}}\cD_{t,b}+\cD_0\end{equation}
	We will prove \begin{equation}\label{eq:107r}\text{$\forall t\in [2^\lambda],b\in \{0,1\},|P_{x_{1-b}}\cD_{t,b}(\ket{\tilde\varphi}\odot \llbracket\fAuxInf\rrbracket)|\leq 2^{-\eta/3-2}|\ket{\tilde\varphi}|$.}\end{equation}

	Intuitively, this means: once the adversary knows $x_b$, it's hard to know $x_{1-b}$. Let's use the hybrid method to prove it. Denote $K_{out\_rev\_(1-b)}$ is the set of keys used in the reversible lookup tables with subscript $1-b$. Consider a blinded oracle $H^\prime$ where $$\text{$H$ on inputs }Pads||x_{1-b},Pads||K_{out\_rev\_(1-b)};\quad Tag(K_{out\_rev\_(1-b)})$$ are blinded. Denote $\cD^\prime_{t,b}$ as the operation where each oracle query in $\cD_{t,b}$ is replaced by $H^\prime$.\par
	To prove (\ref{eq:107r}), first we can prove \begin{equation}\label{eq:108r}\cD_{t,b}(\ket{\tilde\varphi}\odot \llbracket\fAuxInf\rrbracket)\approx_{2^{-\eta/3-3}|\ket{\tilde\varphi}|}\cD^\prime_{t,b}(\ket{\tilde\varphi}\odot \llbracket\fAuxInf\rrbracket)\end{equation}
	The proof is given below.
	\begin{mdframed}
			\textbf{Proof of (\ref{eq:108r})}\par
			The reason is we can apply the hybrid method and replace $H$ by $H^\prime$ one by one, from the first query to the last query, and bound the difference caused by each step of this replacement. Define the operations by the first $q$ queries in $\cD^\prime_{t,b}$ as ${\cD^\prime}^q_{t,b}$. What we need to prove is:
			\begin{align}\label{eq:169r}
				\forall q\in [2^\lambda],\qquad\qquad |P_{Pads||x_{1-b}}{\cD^\prime}^q_{t,b}(\ket{\tilde\varphi}\odot \llbracket\fAuxInf\rrbracket)| & \leq 2^{-\eta/2-5}|\ket{\tilde\varphi}|,              \\
				|P_{K_{out\_rev\_(1-b)}}{\cD^\prime}^q_{t,b}(\ket{\tilde\varphi}\odot \llbracket\fAuxInf\rrbracket)|                           & \leq 2^{-\eta/2-5}|\ket{\tilde\varphi}|\label{eq:203}
			\end{align}
			\begin{enumerate}
				\item If $q\leq t$, since the queries never contain the input in the form of $Pads||K$ (recall that we add a projection onto $I-P_{Pads||K}$ before each query in $\cD_{t,b}^q$, and add a projection onto $I-P_{Pads||\cdots}$ for each query in $\ket{\tilde\varphi}$) thus (\ref{eq:169r}) holds.\par
				 And by the same reasons above we can get, in the adversary's viewpoint, all the ciphertexts of $K_{out\_rev\_(1-b)}$ can be replaced by random strings without affecting the left hand side of (\ref{eq:203}), thus (\ref{eq:203}) holds since it's the same as guessing a random string.
				\item If $q>t$, (\ref{eq:169r}) holds because otherwise the adversary can break (\ref{eq:179z}). And (\ref{eq:203}) holds by the same reason as before.
			\end{enumerate}
			Then the difference caused by each step of the replacement in the hybrid method can be bounded by (\ref{eq:169r})(\ref{eq:203}), and since $2^{-\eta/2}\cdot 2^{\eta/6}\leq 2^{-\eta/3}$ and we complete the proof of (\ref{eq:108r}).\end{mdframed}
	Then use (\ref{eq:179z}) again we know 
	$|P_{x_{1-b}}\cD^\prime_{t,b}(\ket{\tilde\varphi}\odot \llbracket\fAuxInf\rrbracket)|\leq 2^{-\eta/3-3}|\ket{\tilde\varphi}|$. Thus together with (\ref{eq:108r}) we know (\ref{eq:107r}) is true.\par
	And (\ref{eq:107r}) implies $$|P_{x_{0}||x_1}\cD_{t,b}(\ket{\tilde\varphi}\odot \llbracket\fAuxInf\rrbracket)|\leq 2^{-\eta/3-2}|\ket{\tilde\varphi}|$$
	And by similar reasons $$|P_{x_{0}||x_1}\cD_{0}(\ket{\tilde\varphi}\odot \llbracket\fAuxInf\rrbracket)|\leq |P_{x_{1-b}}\cD_0(\ket{\tilde\varphi}\odot \llbracket\fAuxInf\rrbracket)|\leq 2^{-\eta/2}|\ket{\tilde\varphi}|$$
	. Summing it up for all the $t,b$ we know $|P_{x_0||x_1}\cD(\ket{\tilde\varphi}\odot \llbracket\fAuxInf\rrbracket)|\leq 2^{-\eta/6-0.5}|\ket{\tilde\varphi}|$.\par
	Finally adding back the difference of $\ket{\tilde\varphi}$ and $\ket{\varphi}$ we get $|P_{x_0||x_1}\cD(\ket{\varphi}\odot \llbracket\fAuxInf\rrbracket)|\leq 2^{-\eta/6-0.2}|\ket{\varphi}|$, thus completes the proof. The SC-security of $K_{out\_t}$ holds because otherwise the adversary can compute $K$ through it.
\end{proof}

\begin{proof}[Proof of Lemma \ref{lem:4.9}]
	By Technique \ref{lem:4.2} we can assume $Tag(K)$ is stored in the read-only buffer.\par
	Suppose there exist $i_0\in [N]$, a server-side operation $\cU$ with query number $|\cU|\leq 2^{\eta-4}$ such that
	$$|P_{x_0^{(i_0)}|| x_1^{(i_0)}}\cU(\ket{\varphi}\odot \llbracket \fAuxInf\rrbracket)|>2^{-\eta+4}|\ket{\varphi}|$$
	Suppose the set of random pads used in $\llbracket\fAuxInf\rrbracket$ is $Pads$. $Pads$ is sampled randomly. Applying Lemma \ref{lem:3.4r} we know there exists $\ket{\tilde\varphi}$ such that $\ket{\tilde\varphi}\approx_{2^{-\eta}|\ket{\varphi}|}\ket{\varphi}$ and $\ket{\tilde\varphi}$ does not depend on $H(\cdots||Pads||\cdots)$. (Recall the definition of ``does not depend on'' in Definition \ref{def:ndep}) where the prefix ``$\cdots$'' has length the same with the length of $p_0^t$ and the suffix ``$\cdots$'' has length the same with the length of $p_1^t||x_{0}||p_2^t$. Thus
	\begin{equation}\label{eq:90}|P_{x_0^{(i_0)}|| x_1^{(i_0)}}\cU(\ket{\tilde\varphi}\odot \llbracket\fAuxInf\rrbracket\odot Pads)|>2^{-\eta+3}|\ket{\tilde\varphi}|\end{equation}
	\begin{equation}\label{eq:91}\text{$\forall i$, $\ket{\tilde\varphi}\odot Pads$ is $(2^{\eta},2^{-\eta+1}|\ket{\tilde\varphi}|)$-SC-secure for $K^{(i)}$.}\end{equation}
	Our goal is to find some contradiction between these two. In more details, we will make use of the code of the adversary in (\ref{eq:90}) to construct an adversary that does not satisfy (\ref{eq:91}).\par
	Since $\ket{\tilde\varphi}$ does not depend on $H(\cdots||Pads||\cdots)$, it looks the same no matter whether the randomness of $H(\cdots||Pads||\cdots)$ is sampled beforehand (in the oracle before $\ket{\tilde\varphi}$ appear) or afterwards (after $\ket{\tilde\varphi}$ appear but before the server makes further attack). Thus we can observe that the adversary itself can also sample these randomness by itself and simulate an oracle that looks the same as the original oracle. (When we say ``look the same'', we do not mean the new random oracle should reuse the random coins; instead, it should sample some freshly new randomness, and since the original oracle has not been queried either these two oracles look the same.) And we note that in the purified notation the description of the system $Pads$ itself is also in superposition; the ``does not depend on'' is in the sense of Definition \ref{def:ndep}.\par
	Let's first view the idea 1 below, which does not work completely, but takes us to the main proof.
	\paragraph{Idea 1} Consider the following adversary, denoted by $\cU^\prime$, applied on $\ket{\tilde\varphi}\odot Pads$ (note: it's not $\ket{\tilde\varphi}\odot \llbracket\fAuxInf\rrbracket$), which make use of the unitary-query sequence of $\cU$ and (\ref{eq:90}), to break the (\ref{eq:91}). $\cU^\prime$ is defined as follows:\par
	First $\cU^\prime$ can sample random strings that correspond to all the outputs of \\$H(\cdots||Pads||\cdots)$. (It does not query $H$, but sample strings that have the same size and length needed to ``imitate'' this part.) (Since the output length used by $\cU$ in (\ref{eq:90}) is finite it only needs to sample finite amount of randomness.)\par
	Then it can create a new oracle $\tilde H$ using these random strings:\par
	For query to $\tilde H$ on input $e$:
	\begin{enumerate}
	\item If $e$ has the form $\cdots ||Pads||\cdots$ (which means, the ``$Pads$'' part of the string are contained in the set $pads$ stored in the system $Pads$), return the 	random values sampled just now that are used to ``imitate'' $H(e)$.
	\item Otherwise, return $H(e)$.
	\end{enumerate}
	And replace the oracle queries in $\cU$ by $\tilde H$. This new oracle should look the same (as freshly random as) the original oracle, if the operation is applied on $\ket{\tilde\varphi}$. And we note that for different choices of $Pads$ the new random oracle could be different. In other words, this is not a stand-alone oracle, but an oracle whose construction takes $Pads$ as part of the parameter. However, this does not affect the statement that ``this oracle can be used to replace the original oracle without affecting the adversary's ability of computing the keys in $K^{(i)}$''\par
	 The main problem is, if we compare $\cU^\prime$ and $\cU$, $\cU^\prime$ does not have $\llbracket\fAuxInf\rrbracket$, and since $\cU^\prime$ does not hold the keys, it cannot collect the right values of $\tilde H$ to get a simulated $\llbracket\fAuxInf\rrbracket$. To solve this problem, we consider the following idea:
	 \paragraph{Idea 2:} Let the new oracle $\tilde H$ be actually constructed as follows:\par
	Before the construction of $\tilde H$, the adversary samples random outputs in the following steps:\begin{enumerate}\item First the random oracle entries that corresponds to $H(p_0^t||pad^t||p_1^t||x_{b^t}^{(i^t)}||p_2^t)$ ($\forall t\in [n]$) are sampled randomly and stored separately. Note that the adversary doesn't know the actual values of $x_b^{(i)}$ for any $b,i$, but it knows their tags, so it can still create a \emph{look-up tables} to store these values. \item Then it samples a set of random values on input in the form of $\cdots||Pads||\cdots$ as \emph{``background values''}. This step is the same as the idea 1 above, but these background values are not necessarily the output values of $\tilde H$.\end{enumerate} Then $\tilde H$ is defined as follows. For input $e$:
	\begin{enumerate}
		\item Check if $e$ is in the form of $p_0^t||pad^t||p_1^t||x_{b^t}^{(i^t)}||p_2^t$, $t\in [n]$. If so, return the values from the look-up table. Note that the checking can be done using $Tag(K)$.
		\item Otherwise if $e$ is in the form of $\cdots||Pads||\cdots$ (see Idea 1 for the meaning), return the value from the corresponding background value.
		\item Otherwise return $H(e)$.
	\end{enumerate}
	For each query, $\tilde H$ requires at most 3 queries to $H$ to complete the construction.\par
	Now we can construct the attack $\cU^\prime$:
	\begin{enumerate}
		\item As described in the steps below Idea 2, it samples the ``background values'', and samples the random oracle outputs that correspond to $H(p_0^t||pad^t||p_1^t||x_{b^t}^{(i^t)}||p_2^t)$ ($\forall t\in [n]$), and collects them into the corresponding places of $\llbracket\fAuxInf\rrbracket$. Denote it as $\llbracket\fAuxInf^{fake}\rrbracket$. Fill the system that should store $\llbracket\fAuxInf\rrbracket$ in (\ref{eq:90}) with this fake version.
		\item Construct $\tilde H$ as above.
		\item Run the code of $\cU$, and replace the oracle queries by queries to $\tilde H$. Denote the new operation in this step as $\cU^{fake}$.
	\end{enumerate} 
	Since $\ket{\tilde\varphi}$ does not depend on $H(\cdots||Pads||\cdots)$, $\tilde H$ and $\llbracket\fAuxInf^{fake}\rrbracket$ constructed above should look the same as $H$ and $\llbracket\fAuxInf\rrbracket$ on the ability of outputting the keys in $K$ (on the space that $Tag$ is injective), which means
	\begin{align}|P_{x_0^{(i_0)}|| x_1^{(i_0)}}\cU^\prime(\ket{\tilde\varphi}\odot Pads)|&=|P_{x_0^{(i_0)}|| x_1^{(i_0)}}\cU^{fake}(\ket{\tilde\varphi}\odot \llbracket\fAuxInf^{fake}\rrbracket\odot Pads)|\\&\approx_{2^{-\eta}|\ket{\varphi}|}|P_{x_0^{(i_0)}|| x_1^{(i_0)}}\cU(\ket{\tilde\varphi}\odot \llbracket\fAuxInf\rrbracket\odot Pads)|\end{align}
	(the approximation comes from Fact \ref{fact:injtag}). And the query number to $H$ satisfies $|\cU^\prime|\leq 3|\cU|\leq 2^{\eta-1}$, which contradicts (\ref{eq:90})(\ref{eq:91}). This completes the proof.\par
\end{proof}
\begin{proof}[Proof of Lemma \ref{lem:4.10}]
	By Technique \ref{lem:4.2} we can assume $Tag(K)$ is stored in the read-only buffer.\par
	First use the decomposition lemma (Lemma \ref{lem:4.4})(together with Fact \ref{fact:injtag}) to decompose $\ket{\varphi}$ as $\ket{\phi}+\ket{\chi}$	, where
	\begin{itemize}
		\item $|\ket{\chi}|\leq 2.5C|\ket{\varphi}|$.
		\item $\ket{\phi}$ is $(2^{\eta/6},2^{-\eta/6}C|\ket{\varphi}|)$-SC-secure for $K$ and is $(1,2^{\eta})$-server-side-representable from $\ket{\varphi}$.\\
		      Thus $\ket{\phi}$ is $(2^{D},2^{D}+2^\eta)$-representable from $\ket{\mathfrak{init}}$. What's more, since $|\ket{\phi}|\geq |\ket{\varphi}|-|\ket{\chi}|\geq \frac{1}{6}|\ket{\varphi}|$, we know $\ket{\phi}$ is $(2^{\eta/6},2^{-\eta/6+1}|\ket{\phi}|)$-SC-secure for $K$. 
	\end{itemize}
	Apply Lemma \ref{lem:4.8} on $\ket{\phi}$ we know $\ket{\phi}\odot \llbracket\fAuxInf\rrbracket$ is $(2^{\eta/36-1},2^{-\eta/36+1}|\ket{\phi}|)$-SC-secure for $K$. Add back $\ket{\chi}$ completes the proof.
\end{proof}

\section{Proof of Lemmas from Section \ref{sec:4.5} to the End of Section \ref{sec:4}}\label{sec:a57}
\begin{proof}[Proof of Lemma \ref{lem:4.12}]
	This can be proved by a simple hybrid method. Denote $\fAdv^t$ as the adversary where the RO queries from the $t$-th RO query to the last one are replaced by queries to $H^\prime$, then by (\ref{eq:8})
	$$\fAdv^t\ket{\varphi}\approx_{ 2^{-\eta+1}|\ket{\varphi}|}\fAdv^{t-1}\ket{\varphi}$$
	. Summing up this inequalities for every $t$ completes the proof.
\end{proof}
\begin{proof}[Proof of Lemma \ref{lem:4.12n}]
	This can be proved by a simple hybrid method. Denote $\fAdv^t$ as the adversary where the RO queries from the first to the $t$-th RO query are replaced by queries to $H^\prime$, then $$\fAdv^t\ket{\varphi}\approx_{ 2^{-\eta+1}|\ket{\varphi}|}\fAdv^{t-1}\ket{\varphi}$$. Summing up this inequalities for every $t$ completes the proof.
\end{proof}

\begin{proof}[Proof of Lemma \ref{lem:4.13}]
	Define $\ket{\varphi_b}=P_{x_b}^S\ket{\varphi}$. Then $\ket{\varphi}=\ket{\varphi_0}+\ket{\varphi_1}$. If for some $b\in \{0,1\}$, $|\ket{\varphi_b}|\leq 2^{-\eta/2}|\ket{\varphi}|$, the statement is already true. Otherwise:
	\begin{align}
		                             & |\quad|P_0\cD\ket{\varphi}|^2-|P_0\cD\circ COPY\circ \ket{\varphi}|^2\quad|                                                              \\
		=                            & |tr(P_0\cD\ket{\varphi}\bra{\varphi}\cD^\dagger )-tr(P_0\cD(\ket{\varphi_0}\bra{\varphi_0}+\ket{\varphi_1}\bra{\varphi_1})\cD^\dagger )| \\
		=                            & |\bra{\varphi_1}\cD P_0\cD \ket{\varphi_0}+\bra{\varphi_0}\cD P_0\cD \ket{\varphi_1}|                                                    \\
		\leq                         & (|\bra{x_1}\cD P_0\cD \ket{\varphi_0}|+|\bra{x_0}\cD P_0\cD \ket{\varphi_1}|)\cdot 2^{\eta/2}|\ket{\varphi}|                             \\
		\text{(By SC-security) }\leq & 2^{-\eta/2+1}|\ket{\varphi}|^2
	\end{align}
	This completes the proof.
\end{proof}
\begin{proof}[Proof of Lemma \ref{lem:4.14}]
	Suppose $\cD=\cU_{2^{\lambda}}H\cU_{2^{\lambda}-1}H\cdots HU_0$ where $\lambda\leq\eta/2$. We need to give a bound for $|P_{span\{x_0, x_1\}}\cD\ket{\varphi}|$.\par
	Define $\cD_t$ as the operation where the first $t$ queries in $\cD$ are replaced by queries to $H^\prime$. Since a freshly new blinded oracle can be simulated by the adversary using $Tag(K)$ (on the space that $Tag$ is injective on inputs with length the same as the keys in $K$), by the ANY-security of $\ket{\varphi}$ we know $|P_{span\{x_0, x_1\}}\tilde\cD_t\ket{\varphi}|\leq 2^{-\eta+0.5}|\ket{\varphi}|$ where $\tilde\cD_t$ is the operation in $\cD_t$ from the beginning to the $t$-th queries. Applying Lemma \ref{lem:4.12n} completes the proof.
\end{proof}
\begin{proof}[Proof of Lemma \ref{lem:4.15}]
	By Lemma \ref{lem:4.5} (and adding the space that $Tag$ is not injective) we can decompose $\ket{\varphi}$ together with finite number of server-side ancilla qubits (which are all at state zero), into $\ket{\phi}+\ket{\chi}$ such that \begin{itemize}\item $\ket{\phi}$ is $(2^{\eta/6},2^{-\eta/6+0.5}|\ket{\varphi}|)$-SC-secure for $K$;\item $|\ket{\chi}|\leq 2.9C|\ket{\varphi}|$.\end{itemize} Since $C\leq 1/3$ we know $|\ket{\varphi}|\geq |\ket{\phi}|\geq \frac{1}{6}|\ket{\varphi}|$ thus $\ket{\phi}$ is $(2^{\eta/6},2^{-\eta/6+3}|\ket{\phi}|)$-SC-secure for $K$. Similar to the proof of Lemma \ref{lem:4.14}, applying Lemma \ref{lem:4.12n} on $\cD\ket{\phi}$ we know
	$$\cD\ket{\phi}\approx_{2^{-\eta/12+O(1)}|\ket{\phi}|}\cD^{blind}\ket{\phi}.$$
	Adding $\ket{\chi}$ back completes the proof.
\end{proof}

\begin{proof}[Proof of Lemma \ref{lem:4.16}]
	Define $\cU^t$ as the operation where the first $t$ queries in $\cU$ are replaced by queries to $H^\prime$. We only need to prove for all $t$, $P_{x_b}\cU^t\ket{\varphi}\approx_{2^{-2\eta/3}|\ket{\varphi}|}P_{x_b}\cU^{t-1}\ket{\varphi}$. This is then reduced to prove $|P_{x_b}\cD_1(H-H^\prime)P_{x_{1-b}}\cD_2\ket{\varphi}|\leq 2^{-2\eta/3-1}|\ket{\varphi}|$ where $\cD_1$, $\cD_2$ are the operations after and before the $t$-th oracle queries in $\cU^t$. This comes from the SC-security of the initial state thus we complete the proof.
\end{proof}
\begin{proof}[Proof of Lemma \ref{lem:r4.17}]
	Similar to the proof fo Lemma \ref{lem:4.16}. Define $\cU^t$ as the operation where the first $t$ queries in $\cU$ are replaced by queries to $H^\prime$. We only need to prove for all $t$, $\cU^tP_{x_b}\ket{\varphi}\approx_{2^{-2\eta/3}|\ket{\varphi}|}\cU^{t-1}P_{x_b}\ket{\varphi}$. This is then reduced to prove $|\cD_1(H-H^\prime)P_{x_{1-b}}\cD_2P_{x_b}\ket{\varphi}|\leq 2^{-2\eta/3-1}|\ket{\varphi}|$ where $\cD_1$, $\cD_2$ are the operations after and before the $t$-th oracle queries in $\cU^t$. This comes from the SC-security of the initial state thus we complete the proof.
\end{proof}
\begin{proof}[Proof of Lemma \ref{lem:4.23}]
	Suppose the set of random pads used in the computation of the client's messages is $Set$. First apply Lemma \ref{lem:3.4r} to prove that, if we expand $\ket{\varphi}$ using its representation and replace all the queries to $H$ by $H\cdot (I-P_{Set||\cdots})$ (where $\cdots$ denotes strings of arbitrary length), denote the final state as $\ket{\tilde\varphi}$, we have
	$$\ket{\tilde\varphi}\approx_{2^{-\eta}|\ket{\varphi}|}\ket{\varphi}$$
	And we use the hybrid method to do the remaining steps. Suppose $\tilde H$ is a blinded oracle of $H$ where the following inputs are blinded:
	\begin{equation}
		Set||\cdots||x||\cdots, Set||\cdots||y||\cdots,
	\end{equation}
	where the first $\cdots$ denotes all the possible strings with the same length as $px^t$, the second $\cdots$ denotes all the possible strings with the same length as $sx^t$, the third $\cdots$ denotes all the possible strings with the same length as $py^t$, and the fourth $\cdots$ denotes all the possible strings with the same length as $sy^t$.\par
And define $\ket{\tilde\varphi^\prime}$, $\ket{\tilde\varphi^{\prime\prime}}$ as the result of replacing the initial state by $\ket{\tilde\varphi}$ in the construction of $\ket{\varphi^\prime}$, $\ket{\varphi^{\prime\prime}}$. And use $\cD^{blind}$ to denote the operation that comes from replacing all the queries in $\cD$ by queries to $\tilde H$. Use $\cD^{blind,t}$ to denote the operation coming from replacing the first to the $t$-th queries in $\cD$ by queries to $\tilde H$. And to prove this lemma, we will first prove
\begin{equation}\label{eq:207im}
\cD^{blind,t}\ket{\tilde\varphi^\prime}\approx_{2^{-\eta+4}|\ket{\varphi}|}\cD^{blind,t-1}\ket{\tilde\varphi^\prime}
\end{equation}
This is further reduced to the following two expressions ((\ref{eq:208mm})(\ref{eq:209mm})), where we use $\tilde \cD^{blind,t}$ to denote the operation in $\cD^{blind}$ just before the $t$-th query:
\begin{itemize}
\item \begin{equation}\label{eq:208mm}|P_{x}	\tilde \cD^{blind,t}\ket{\tilde\varphi^\prime}|=|P_{x}	\tilde \cD^{blind,t}\ket{\tilde\varphi^{\prime\prime}}|\leq 2^{-\eta+2}|\ket{\varphi}|\end{equation}
where the inequality comes from the fact that $\ket{\tilde\varphi}$ is $(2^\eta,2^{-\eta+1}|\ket{\varphi}|)$-unpredictable for $x_b$.
\item \begin{equation}\label{eq:209mm}|P_{y}	\tilde \cD^{blind,t}\ket{\tilde\varphi^\prime}|=|P_{y}	\tilde \cD^{blind,t}\ket{\tilde\varphi^{\prime\prime}}|\leq 2^{-\eta}|\ket{\varphi}|\end{equation}
where the inequality comes from the fact that this is predicting a random string.
\end{itemize}
And (\ref{eq:207im}) implies
\begin{equation}
\cD\ket{\tilde\varphi^\prime}\approx_{2^{-\eta/2}|\ket{\varphi}|}\cD^{blind}\ket{\tilde\varphi^\prime}	
\end{equation}
And from (\ref{eq:208mm})(\ref{eq:209mm}) we get
\begin{equation}
\cD\ket{\tilde\varphi^{\prime\prime}}\approx_{2^{-\eta/2}|\ket{\varphi}|}\cD^{blind}\ket{\tilde\varphi^{\prime\prime}}	
\end{equation}
And we also have
\begin{equation}
	|P_0\cD^{blind}\ket{\tilde\varphi^\prime}| = |P_0\cD^{blind}\ket{\tilde\varphi^{\prime\prime}}|
\end{equation}
and these three expressions imply the final lemma.
\end{proof}
\begin{proof}[Proof of Lemma \ref{lem:4.24}]
Apply Lemma \ref{lem:4.7r} we can decompose $\ket{\varphi}$ as $\ket{\phi}+\ket{\chi}$. Apply Lemma \ref{lem:4.23} on $\ket{\phi}$ and then add $\ket{\chi}$ by additivity. This completes the proof.
\end{proof}
\section{Missing Proofs in Section \ref{sec:6.1} to Section \ref{sec:6.2}}\label{sec:ap6p}
\subsubsection{Proof for single round protocol}
The proof of Lemma \ref{lem:6.2} is given below.
\begin{proof}
	Suppose
	\begin{equation}\label{eq:5.2}|P_{pass}\ket{\varphi^\prime}|> (1-\frac{1}{T})|\ket{\varphi}|\end{equation}
	. Otherwise the lemma is already true.\par
	Discuss by cases:
	\begin{itemize}
		\item (Case 1) There exists a server side operation $\cU_1$ with query number $|\cU_1|\leq 2^{\eta/3}$ such that $|P_{span\{x_0, x_1\}}\cU_1\ket{\varphi}|> (1-\frac{4}{T})|\ket{\varphi}|$
		\item (Case 2) $\ket{\varphi}$ is $(2^{\eta/3}, (1-\frac{4}{T})|\ket{\varphi}|)$-ANY-secure for $K$.
	\end{itemize}
	Let's first rule out the Case 2. For Case 2, apply Lemma \ref{lem:4.6} we know $\ket{\varphi}$ can be decomposed as $\ket{\phi}+\ket{\chi}$, where
	\begin{itemize}
		\item $|\ket{\chi}|\leq (1-\frac{2}{T})|\ket{\varphi}|$
		\item $\ket{\phi}$ is $(2^{(\eta/3-150T^2)/6},2^{-(\eta/3-150T^2)/6}|\ket{\varphi}|)$-ANY-secure for $K$ and is $(1,2^{\eta/3})$-representable from $\ket{\varphi}$.\\
		      Then we know $\ket{\phi}$ is $(2^{(\eta/3-150T^2)/6},2^{-(\eta/3-150T^2)/6+\kappa}|\ket{\phi}|)$-ANY-secure for $K$ and it's $(2^{D},2^{D}+2^{\eta/3})$-representable from $\ket{\mathfrak{init}}$. This will be useful below.
	\end{itemize}
	Then (use $\fPrtl$ to denote the protocol)
	\begin{align}\label{eq:5.3}|P_{pass}\ket{\varphi^\prime}| & =|P_{pass}(\fPrtl\circ\ket{\phi}+\fPrtl\circ\ket{\chi})|                           \\
		                                             & \leq |P_{pass}\fPrtl\circ\ket{\phi}|+|P_{pass}\fPrtl\circ\ket{\chi}|\label{eq:31r}
	\end{align}
	By the properties of $\ket{\phi}$, we can prove
	\begin{align}|                                    & P_{pass}\fPrtl\circ \ket{\phi}|                                                                   \\
		\text{(Expand $\fPrtl$)}=              & |P_{r}\fAdv(\ket{\phi}\odot \ket{\fLT(\forall b,x_b\rightarrow r;\ell,\kappa_{out})})| \\
		\text{(By Lemma \ref{lem:4.12})}\leq & 2^{-(\eta/3-150T^2)/6+3\kappa}|\ket{\phi}|                                                      \\
		\leq                                 & 2^{-\kappa}|\ket{\varphi}|\end{align}
	Substitute it back to (\ref{eq:5.3}), together with $|\ket{\chi}|\leq (1-\frac{2}{T})|\ket{\varphi}|$, this leads to a contradiction to (\ref{eq:5.2}).\par
	For Case 1, the only difference of it from the final conclusion (\ref{eq:26r}) is the query number of the server-side operations. We will show the extra $\llbracket\fAuxInf\rrbracket$ will help the adversary prepares the state more easily. First we can write
	$$\cU_1\ket{\varphi}=\ket{\psi_0}+\ket{\psi_1}+\ket{\chi^\prime}$$
	where for some server-side system $S$, \begin{equation}\text{$P^S_{x_0}\ket{\psi_0}=\ket{\psi_0}$, $P^S_{x_1}\ket{\psi_1}=\ket{\psi_1}$, $P^{S}_K\ket{\chi^\prime}=0$,}\end{equation}
	and \begin{equation}|\ket{\psi_0}+\ket{\psi_1}|\geq (1-\frac{4}{T})|\ket{\varphi}|,\quad |\ket{\chi^\prime}|\leq \frac{2\sqrt{2}}{\sqrt{T}}|\ket{\varphi}|\end{equation}
	Expand the protocol and we have
	\begin{align}|P_{pass}\ket{\varphi^\prime}| & =|P_{r}(\fAdv(\ket{\varphi}\odot \llbracket\fLT(\forall b:x_b\rightarrow r)\rrbracket))|\label{eq:18}                                 \\
		                               & =|P_{r}(\fAdv(\cU_1^\dagger(\ket{\psi_0}+\ket{\psi_1}+\ket{\chi^\prime})\odot \llbracket\fLT(\forall b:x_b\rightarrow r)\rrbracket))|\end{align}
	Then we consider the server-side operation $\cU_2$ applied on $\ket{\varphi}\odot \llbracket\fAuxInf\rrbracket$ defined as follows:
	\begin{enumerate}
		\item Take the forward table part of the $\llbracket\fAuxInf\rrbracket$, replace the $\llbracket\fLT(\forall b:x_b\rightarrow r)\rrbracket$ above (right side of equation (\ref{eq:18})) by the forward table $\llbracket\fLT(\forall b:x_b\rightarrow r_b)\rrbracket$, and run $\fAdv$. Instead of outputting $r$ on some system, this operation will output a key in $\{r_b\}_{b\in \{0,1\}}$. Do not do the final projection measurement $P_r$ and keep the register that stores $\{r_b\}_{b\in \{0,1\}}$ in superposition. 
		\item Use the backward table to map $r_b$ back to $x_b$.
	\end{enumerate}
	Let's first analyze the state after the step 1 above. We can prove:
	\begin{align}
		  & \qquad	|\quad |P_{r}(\fAdv(\cU_1^\dagger(\ket{\psi_0}+\ket{\psi_1})\odot \llbracket\fLT(\forall b:x_b\rightarrow r)\rrbracket)|\label{eq:6.2.44}                                                        \\
		  & -|P_{span\{r_0, r_1\}}(\fAdv(\cU_1^\dagger(\ket{\psi_0}+\ket{\psi_1})\odot \llbracket\fLT(\forall b:x_b\rightarrow r_b)\rrbracket)|\quad |\leq 2^{-\eta/24}|\ket{\psi_0}+\ket{\psi_1}|\label{eq:6.2.7}
	\end{align}

	This comes from applying the collapsing property and Lemma \ref{lem:4.12} on both terms. (We put the detailed proof in Appendix \ref{sec:ap6p}.)\par
	Thus we get, after summing back $\cU_1^\dagger\ket{\chi^\prime}$,
	\begin{align}|\quad & |P_{r}(\fAdv(\ket{\varphi}\odot \llbracket\fLT(\forall b:x_b\rightarrow r)\rrbracket)|-|P_{span\{r_0, r_1\}}(\fAdv(\ket{\varphi}\odot \llbracket\fLT(\forall b:x_b\rightarrow r_b)\rrbracket))|\quad |\label{eq:30r} \\
		\leq   & 2^{-\eta/24}|\ket{\psi_0}+\ket{\psi_1}|+\frac{2\sqrt{2}}{\sqrt{T}}|\ket{\varphi}|                                                                        \\
		\leq   & (2^{-\eta/24+2}+\frac{2\sqrt{2}}{\sqrt{T}})|\ket{\varphi}|\end{align}
	From equation (\ref{eq:5.2})(\ref{eq:18}) we know the first term in (\ref{eq:30r}) is at least $(1-\frac{1}{T})|\ket{\varphi}|$. Thus we get
	$$|P_{span\{r_0, r_1\}}(\fAdv(\ket{\varphi}\odot \llbracket\fLT(\forall b:x_b\rightarrow r_b)\rrbracket)) |\geq (1-\frac{3.9}{\sqrt{T}})|\ket{\varphi}|$$
	Note that the server does not really do the projection $P_{span\{r_0, r_1\}}$, but in the second step in $\cU_2$, uses the backward table to map $\{r_b\}_{b\in \{0,1\}}$ back to $K=\{x_b\}_{b\in \{0,1\}}$. Thus
	$$|P_{span\{x_0, x_1\}}\cU_2(\ket{\varphi}\odot \llbracket\fAuxInf\rrbracket)|\geq (1-\frac{4}{\sqrt{T}})|\ket{\varphi}|$$
	Finally we can see the query number $|\cU_2|\leq |\fAdv|+20$. This completes the proof.
\end{proof}
\begin{proof}[Proof of (\ref{eq:6.2.44})(\ref{eq:6.2.7})]
	Let's write $\tilde \cU=\fAdv\circ \cU^\dagger$. The query number is at most $2^{\eta/3+1}$. Then for (\ref{eq:6.2.7}) there is (where the $\approx_\epsilon$ below just means the absolute value of their difference is at most $\epsilon$):
	{\small\begin{align}
	&|P_{span\{r_0, r_1\}}\tilde\cU((\ket{\psi_0}+\ket{\psi_1})\odot \llbracket\fLT(\forall b:x_b\rightarrow r_b)\rrbracket)|\\
	&\approx_{2^{-\eta/3}|\ket{\varphi}|} \\& \sqrt{|P_{span\{r_0, r_1\}}\tilde\cU(\ket{\psi_0}\odot \llbracket\fLT(\forall b:x_b\rightarrow r_b)\rrbracket)|^2+|P_{span\{r_0, r_1\}}\tilde\cU(\ket{\psi_1}\odot \llbracket\fLT(\forall b:x_b\rightarrow r_b)\rrbracket)|^2}
	\end{align}}
	This step is by applying Lemma \ref{lem:4.13}. Note that the projection does not have the form as required in Lemma \ref{lem:4.13} (since the projection is not $P_0$); but we can first strengthen the statement by adding $r_0,r_1$ into the auxiliary information and then apply Lemma \ref{lem:4.13}.\par
Define $Table(x_0\rightarrow r_0,\$)$ as the result of replacing the $x_1$ row in the lookup table by random strings, and $Table(\$,x_1\rightarrow r_1)$ as the result of replacing the $x_0$ row in the lookup table by random strings. Then
	{\small\begin{align}
& \hspace{-60pt}\sqrt{|P_{span\{r_0, r_1\}}\tilde\cU(\ket{\psi_0}\odot \llbracket\fLT(\forall b:x_b\rightarrow r_b)\rrbracket)|^2+|P_{span\{r_0, r_1\}}\tilde\cU(\ket{\psi_1}\odot \llbracket\fLT(\forall b:x_b\rightarrow r_b)\rrbracket)|^2}\\
\approx_{2^{-\eta/3}|\ket{\varphi}|}&\\
 &\hspace{-60pt}\sqrt{|P_{span\{r_0, r_1\}}\tilde\cU(\ket{\psi_0}\odot \llbracket Table(x_0\rightarrow r_0,\$)\rrbracket)|^2+|P_{span\{r_0, r_1\}}\tilde\cU(\ket{\psi_1}\odot \llbracket Table(\$,x_1\rightarrow r_1)\rrbracket)|^2}\\
\approx_{2^{-\eta/3}|\ket{\varphi}|} &\sqrt{|P_{r_0}\tilde\cU(\ket{\psi_0}\odot \llbracket Table(x_0\rightarrow r_0,\$)\rrbracket)|^2+|P_{r_1}\tilde\cU(\ket{\psi_1}\odot \llbracket Table(\$,x_1\rightarrow r_1)\rrbracket)|^2}\label{eq:191n}
	\end{align}}
	where the first step is by Lemma \ref{lem:4.23} and the second step is because it's guessing a random values without giving any information.\par
	Similarly for (\ref{eq:6.2.44}) there is
		\begin{align}
	&|P_r\tilde\cU((\ket{\psi_0}+\ket{\psi_1})\odot \llbracket\fLT(\forall b:x_b\rightarrow r)\rrbracket)|\\
	\approx_{2^{-\eta/3}|\ket{\varphi}|} & \sqrt{|P_r\tilde\cU(\ket{\psi_0}\odot \llbracket\fLT(\forall b:x_b\rightarrow r)\rrbracket)|^2+|P_r\tilde\cU(\ket{\psi_1}\odot \llbracket\fLT(\forall b:x_b\rightarrow r)\rrbracket)|^2}\\
	\approx_{2^{-\eta/3}|\ket{\varphi}|} &\sqrt{|P_r\tilde\cU(\ket{\psi_0}\odot \llbracket Table(x_0\rightarrow r,\$)\rrbracket)|^2+|P_r\tilde\cU(\ket{\psi_1}\odot \llbracket Table(\$,x_1\rightarrow r)\rrbracket)|^2}\label{eq:194n}
	\end{align}
	Notice (\ref{eq:191n}) and (\ref{eq:194n}) are the same since $r,r_0,r_1$ are all sampled randomly. Thus we can combine these expressions and complete the proof.
\end{proof}

\section{Proof of Lemma \ref{lem:6.4}}\label{sec:AB}
Before we prove this lemma, we introduce a notation. We note that this is only a temporary notation needed for this proof.
\begin{nota}
	Suppose $K^{(1)}$, $K^{(3)}$ are the key pairs that we are studying.\par
	We write $\ket{\varphi}=^{st-ind}\ket{\varphi^\prime}$ if for any server-side operation $\cD$ that is unbounded and can make unbounded number of RO queries, $|P_0\cD(\ket{\varphi}\odot K^{(1)}\odot K^{(3)})|=|P_0\cD(\ket{\varphi^\prime}\odot K^{(1)}\odot K^{(3)})|$, where $P_0$ is the projection onto $\ket{0}$ on some server-side system.\par
	And we write $\ket{\varphi}\approx^{st-ind}_\epsilon\ket{\varphi^\prime}$ if for any server-side operation $\cD$ that is unbounded and can make unbounded number of RO queries, $|\quad|P_0\cD(\ket{\varphi}\odot K^{(1)}\odot K^{(3)})|-|P_0\cD(\ket{\varphi^\prime}\odot K^{(1)}\odot K^{(3)})|\quad |\leq \epsilon$.
\end{nota}
\begin{fact}
	If $\ket{\varphi}\approx^{st-ind}_{\epsilon_1}\ket{\varphi^\prime}$, $\ket{\varphi^\prime}\approx^{st-ind}_{\epsilon_2}\ket{\varphi^{\prime\prime}}$, there is $\ket{\varphi}\approx^{st-ind}_{\epsilon_1+\epsilon_2}\ket{\varphi^{\prime\prime}}$.
\end{fact}
Here we add two key pairs as auxiliary information in the definition. Since we are studying Protocol \ref{prtl:r9} we can simply assume they are the two input key pairs $K^{(1)}$ and $K^{(3)}$.\par
We note that we are talking about unbounded adversary with unbounded RO queries. The reader might be confused by wondering what is its difference from saying $\ket{\varphi}$ and $\ket{\varphi^\prime}$ are simply the same. The difference comes from the fact that we are using the purified notation instead of natural notation. $\ket{\varphi}$ and $\ket{\varphi^\prime}$ describe the state of the whole system, including the client and the server. Intuitively this definition is equivalent to saying ``the server-side of these two states are the same'' if the $K^{(1)},K^{(3)}$ terms are removed.\par
Why do we need to add $K^{(1)}$ and $K^{(3)}$ into the auxiliary information? Because this gives us the following property: if $\ket{\varphi}\approx^{st-ind}_\epsilon\ket{\varphi^\prime}$, even if the client sends some messages that only depends on $K^{(1)}$ and $K^{(3)}$, the server-side states of the final states are still the same:
\begin{fact}
	If $\ket{\varphi}\approx^{st-ind}_{\epsilon_1}\ket{\varphi^{\prime}}$, suppose $\fAuxInf$ is a randomized algorithm that only takes $K^{(1)}$, $K^{(3)}$ as inputs where $K^{(1)}$, $K^{(3)}$ are stored in some client-side read-only system, there is $\ket{\varphi}\odot \llbracket\fAuxInf\rrbracket\approx^{st-ind}_{\epsilon_1}\ket{\varphi^{\prime}}\odot \llbracket\fAuxInf\rrbracket$.
\end{fact}

Now we can begin our proof of Lemma \ref{lem:6.4}. The structure of this proof is as follows:
\begin{enumerate}
\item In the beginning we assume the first case in the conclusion of Lemma \ref{lem:6.4} is false; in other words, the norm (or informally and intuitively, probability) of passing the protocol for the adversary is big. And all the proofs after this step are under this assumption.
	\item We first apply the security of single-key-pair basis test on $K^{(3)}$, and then $K^{(1)}$, and get some intermediate results. One subtle thing here is we do not always apply the lemma on the states that really appear during the protocol execution.
	\item Then we will give the construction of the server-side operation that the first case in the conclusion of Lemma \ref{lem:6.4} talks about. This construction will use the intermediate results in the previous step.
	\item Then the remaining work is to prove the server-side operation constructed just now really achieves what we need.
\end{enumerate}
\begin{proof}[Proof of Lemma \ref{lem:6.4}]
	Suppose
	\begin{equation}\label{eq:c40}|P_{pass}\ket{\varphi^\prime}|>(1-C^{12})|\ket{\varphi}|\end{equation}
	Thus we need to prove the second case in the conclusion of Lemma \ref{lem:6.4} assuming (\ref{eq:c40}).\par
	First apply the decomposition lemma for SC-security (Lemma \ref{lem:4.4}) (together with Fact \ref{fact:injtag}) on $K^{(3)}$ we know $\ket{\varphi}$ can be decomposed as $\ket{\phi}+\ket{\chi}$ where
	\begin{itemize}
		\item $|\ket{\chi}|\leq 2.45C|\ket{\varphi}|$
		\item $\ket{\phi}$ is $(2^{\eta/6},2^{-\eta/6+0.1}C|\ket{\varphi}|)$-SC-secure for $K^{(3)}$, and is $(1,2^{\eta})$-server-side representable from $\ket{\varphi}$. We can further get the following for $\ket{\phi}$:
		      \begin{itemize}
			      \item $\ket{\phi}$ is $(2^{D},2^{D}+2^\eta)$-representable from $\ket{\mathfrak{init}}$.
			      \item  $\frac{1}{6}|\ket{\varphi}|\leq |\ket{\phi}|\leq |\ket{\varphi}|$.
			      \item $\ket{\phi}$ is $(2^{\eta/6},2^{-\eta/6+1}|\ket{\phi}|)$-SC-secure for $K^{(3)}$
		      \end{itemize}
	\end{itemize}
	Then consider the result of applying the first step of this protocol, which is \\$\fBasisTest(K^{(3)}; T)$, on $\ket{\phi}$. Denote:
	\begin{equation}\label{eq:118r}\ket{\phi^i}:=\fPrtl_{1.1\sim 1.i}\ket{\phi}:=\fBasisTest_{\fAdv_{1.1\sim 1.i}}(K^{(3)}; i)\circ\ket{\phi}\end{equation}
	Here $\fPrtl_{1.1\sim 1.i}$ is an abbreviated notation for the first $i$ rounds of the first step. In the rightmost expression some parameters of the protocol are implicit.\par
	Apply the security property of $\fBasisTest(K^{(3)}; T)$ (Lemma \ref{lem:6.3}). For the first case in the lemma's conclusion, $|P_{pass}\ket{\phi^T}|\leq \frac{1}{2}|\ket{\phi}|$ thus $|P_{pass}\ket{\varphi^\prime}|\leq |P_{pass}\ket{\phi^T}|+|\ket{\chi}|\leq (1-C^2)|\ket{\varphi}|$, contradiction to (\ref{eq:c40}). Thus we only need to consider the second case in the conclusion of Lemma \ref{lem:6.3}, which is
	\begin{center}\emph{There exist an integer $i\in [0,T)$, a server-side operation $\cU_1$ (which depends deterministically on $\fAdv_{1.1\sim 1.i}$ where $\fAdv_{1.1\sim 1.i}$ is the code of the adversary in the first $i$ rounds of the first step) with query $|\cU_1|\leq |\fAdv|+O(1)$ such that
			\begin{equation}\label{eq:95}|(I-P_{K^{(3)}}^{S_3})(\cU_1 (P_{pass}\ket{\phi^i}\odot \llbracket\fAuxInf_1\rrbracket))|\leq 3T^{-1/4}|P_{pass}\ket{\phi^i}|\leq 3C|\ket{\varphi}|\end{equation} where $P_{K^{(3)}}^{S_3}$ is the projection on $K^{(3)}$ on some server-side system $S_3$.}\end{center}
	Recall that $\llbracket\fAuxInf_1\rrbracket$ is defined in the description of this lemma and has the form we need here.\par
	Similar to (\ref{eq:118r}) define
	$$\ket{\varphi^i}:=\fPrtl_{1.1\sim 1.i}\ket{\varphi}:=\fBasisTest(K^{(3)}; i)\circ\ket{\varphi}$$
	Then since $|\ket{\varphi^i}-\ket{\phi^i}|\leq |\ket{\varphi}-\ket{\phi}|\leq |\ket{\chi}|\leq 2.45C|\ket{\varphi}|$ we have
	\begin{equation}\label{eq:5}|(I-P_{K^{(3)}}^{S_3})(\cU_1 (P_{pass}\ket{\varphi^i}\odot \llbracket\fAuxInf_1\rrbracket))|\leq 3C|\ket{\varphi}|+|\ket{\chi}|\leq 5.45C|\ket{\varphi}|\end{equation}\par
	Then after the $i$-th round of test on $K^{(3)}$ completes, $\fPrtl_{1.(i+1)\sim 1.T}$ is applied on $P_{pass}\ket{\varphi^i}$, where the adversary is $\fAdv_{1.(i+1)\sim 1.T}$. This maps the state to (after projected onto the passing space) $P_{pass}\ket{\varphi^T}$.\par
	Then $\fPrtl_{=2}$ is executed, which is a single round non-collapsing basis test on $K^{(1)}$, and the initial state for this step is $P_{pass}\ket{\varphi^T}$. Since
	\begin{itemize}
		\item $|P_{pass}\ket{\varphi^\prime}|> (1-C^{12})|\ket{\varphi}|>(1-C^{12})|P_{pass}\ket{\varphi^T}|$
		\item $P_{pass}\ket{\varphi^T}$ is $(2^{\eta-10},2^{-\eta+10}|P_{pass}\ket{\varphi^T}|)$-SC-secure for $K^{(1)}$ (This comes from Lemma \ref{lem:4.9}.)
	\end{itemize}
	, applying Lemma \ref{lem:6.2} on $\fPrtl_{=2}$ we know \begin{center}\emph{There exists a server-side operation $\cU_2$ (which depends deterministically on $\fAdv_{=2}$, the code of the adversary on the second step of the protocol) with query number $|\cU_2|\leq |\fAdv|+O(1)$ such that
			\begin{equation}\label{eq:38}\cU_2 (P_{pass}\ket{\varphi^T}\odot \llbracket\fAuxInf_2\rrbracket)=\ket{\psi^{(1)}_0}+\ket{\psi^{(1)}_1}+\ket{\chi^{(1)\prime}}\end{equation}
			{\small\begin{equation}\label{eq:r98}\text{where }\forall b\in \{0,1\},P^{S_1}_{x_b^{(1)}}\ket{\psi_b^{(1)}}=\ket{\psi_b^{(1)}}, P^{S_1}_{K^{(1)}}\ket{\chi^{(1)\prime}}=0, |\ket{\chi^{(1)\prime}}|\leq  3C^3|\ket{\varphi}|\leq \frac{1}{2}C^2|\ket{\varphi}|\end{equation}}
			where $S_1$ is some server-side system.}\end{center}
	Note that in (\ref{eq:5}) and (\ref{eq:38})(\ref{eq:r98}) we express the results in different ways. This is for the convenience of later application.\par
	\textbf{Now we complete the step 1 and 2 in the ``structure of this proof'' given before this proof. What we know by this time is the properties of the states in different time steps --- which are mainly (\ref{eq:5}) and (\ref{eq:38})(\ref{eq:r98}); and next, we will construct the $\cU$ that we claim to exist in the conclusion of this lemma.}\par
	Now we will design a server-side operation $\cU$ which maps the state $$P_{pass}\ket{\varphi^\prime}\odot \llbracket\fAuxInf_1\rrbracket\odot \llbracket\fAuxInf_2\rrbracket$$ into a state of the form of equation (\ref{eq:4}). The operation $\cU$ is defined as follows:
	\begin{enumerate}
		\item Apply $\fAdv^\dagger_{=2}$ on the system corresponding to the server side of $\ket{\varphi^\prime}$ and $\llbracket\fPrtl_{=2}\rrbracket$ in the read-only buffer. Recall that $\fAdv_{=2}$ is the adversary's operation in the second step of the protocol (the adversary's operation after $\llbracket\fPrtl_{=2}\rrbracket$ is received and before $r$ is sent back. Note that we assume the adversary does not do any operation after $r$ is sent out.). This maps the server-side of the state back to the system corresponding to the server side of $\ket{\varphi^T}$.
		\item Suppose the server side of $\ket{\varphi^T}$, $\llbracket\fAuxInf_1\rrbracket$, $\llbracket\fAuxInf_2\rrbracket$ are separately stored in system $S_{\varphi^T}$, $S_{\fAuxInf_1}$, $S_{\fAuxInf_2}$. Then starting from the state coming from step 1 above, apply $\cU_2$ on systems $S_{\varphi^T}$ and $S_{\fAuxInf_2}$, and some extra auxiliary qubits (which are initialized to all zeros). Recall that $\cU_2$ is described in equation (\ref{eq:38}).
		\item Copy the content of the system $S_1$ (defined in (\ref{eq:r98})) that stores $x^{(1)}_0$ and $x^{(1)}_1$ (shown in equation (\ref{eq:38})(\ref{eq:r98})) to a separate empty register $S_1^\prime$. Denote this step as $COPY$.
		\item Apply $\cU_2^\dagger$ on the system that stores the output of the second step. The output of $\cU_2^\dagger$ is in $S_{\varphi^T}$, $S_{\fAuxInf_2}$, and some auxiliary system mentioned in step 2. (Note that these auxiliary system is not necessarily in all zero states now).
		\item Then apply $\fAdv_{1.(i+1)\sim 1.T}^\dagger$ on system $S_{\varphi^T}$, where $\fAdv_{1.(i+1)\sim 1.T}$ is the operations in $\fAdv$ from round $i+1$ to round $T$. Note that in the real execution of the protocol the client needs to send messages to the read-only buffer, and the adversary needs to send messages back. Here we omit these steps and only do the server-side operation in $\fAdv_{1.(i+1)\sim 1.T}^\dagger$. The server-side of the output state is stored in systems that corresponds to the server side of $\ket{\varphi^i}$, and some other systems.
		\item Then apply $\cU_1$ on the server side of $\ket{\varphi^i}$ and $S_{\fAuxInf_1}$. (Note that $\cU_1$ is described in equation (\ref{eq:95})(\ref{eq:5}).) This operation maps the state to systems $S_3$ (defined in (\ref{eq:95})) and some other systems.
		\item Now the systems $S^\prime_1$ and $S_3$ contain the keys $x^{(1)}$ and $x^{(3)}$ we need. $S_1^\prime$ comes from the 3rd step, and $S_3$ comes from the 6th step.
	\end{enumerate}
	First, we analyze the first and second steps in $\cU$. For the second step we can make use of equation (\ref{eq:38}). But note that $\cU_2$ in equation (\ref{eq:38}) is applied on $P_{pass}\ket{\varphi^T}$, while in the second step of $\cU$, $\cU_2$ is applied on $\fAdv^\dagger_{=2}P_{pass}\ket{\varphi^\prime}$, we need to first find the relation between $\fAdv_{=2}^\dagger P_{pass}\ket{\varphi^\prime}$ and $P_{pass}\ket{\varphi^T}$.\par
	In the original protocol, from $P_{pass}\ket{\varphi^T}$ to $P_{pass}\ket{\varphi^\prime}$, the following operations are executed: the client prepares and sends $\llbracket\fPrtl_{=2}\rrbracket$ to the server, which is $\llbracket\fBasisTest(K^{(1)})\rrbracket$; the server applies $\fAdv_{=2}$; the server sends back a copy of the register that stores $r$; and finally a projection on the passing space (``$r$ is correct'') is applied. Expressing these steps with notations, we have
	$$P_{pass}\ket{\varphi^\prime}=P_rSend_{S\rightarrow C} \fAdv_{=2}(P_{pass}\ket{\varphi^T}\odot \llbracket\fPrtl_{=2}\rrbracket)$$
	where we use $Send_{S\rightarrow C}$ to mean the server sends back the response $r$ to the client. Then:
	\begin{align}
		\fAdv_{=2}^\dagger P_{pass}\ket{\varphi^\prime} & =\fAdv_{=2}^\dagger Send_{S\rightarrow C}P_r \fAdv_{=2}(P_{pass}\ket{\varphi^T}\odot \llbracket\fPrtl_{=2}\rrbracket)\label{eq:r101} \\
		                                                & =^{st-ind}\fAdv_{=2}^\dagger P_r \fAdv_{=2}(P_{pass}\ket{\varphi^T}\odot \llbracket\fPrtl_{=2}\rrbracket)\label{eq:r102}             \\
		\text{(By (\ref{eq:c40})) }                     & \approx_{C^6|\ket{\varphi}|} P_{pass}\ket{\varphi^T}\odot \llbracket\fPrtl_{=2}\rrbracket\label{eq:45}
	\end{align}
	Recall that the $=^{st-ind}$ symbols are defined just before this proof. $\approx_\epsilon$ is simply the notation in Section \ref{sec:2.1}. (\ref{eq:r101})(\ref{eq:r102}) is because on the server side, no matter it sends a classical message to the client or not, in the server's viewpoint, the system is exactly the same, even if it gets some auxiliary information that depends on the keys later.\par Here the $=^{st-ind}$ and $\approx_\epsilon$ are all describing the relations between adjacent expressions. In summary, (\ref{eq:r101})(\ref{eq:r102}), (\ref{eq:r102})(\ref{eq:45}) implies
	\begin{equation}\label{eq:133}
		\fAdv_{=2}^\dagger P_{pass}\ket{\varphi^\prime}\approx^{st-ind}_{C^6|\ket{\varphi}|}P_{pass}\ket{\varphi^T}\odot \llbracket\fPrtl_{=2}\rrbracket
	\end{equation}
	The next step in $\cU$ is to apply $\cU_2$. By (\ref{eq:38})(\ref{eq:133}) we have
	\begin{equation}\label{eq:126}\cU_2\fAdv_{=2}^\dagger (P_{pass}\ket{\varphi^\prime}\odot \llbracket\fAuxInf_2\rrbracket)\approx^{st-ind}_{C^6|\ket{\varphi}|} \ket{\psi_0^{(1)}}+\ket{\psi^{(1)}_1}+\ket{\chi^{(1)\prime}}\end{equation}
	where the symbols on the right hand side come from equation (\ref{eq:38}). Thus
	\begin{equation}\label{eq:135}
		(I-P^{S_1}_{K^{(1)}})\cU^{\leq \text{2th}}(P_{pass}\ket{\varphi^\prime}\odot \llbracket\fAuxInf\rrbracket)\leq 	C^6|\ket{\varphi}|+|\ket{\chi^{(1)\prime}}|\leq \frac{2}{3}C^2|\ket{\varphi}|.
	\end{equation}
	where $\cU^{\leq \text{2th}}$ means the first two steps in $\cU$.\par
	Thus the task of bounding the noise on the $K^{(1)}$ part has completed. (which corresponds to (\ref{eq:55r}).) Note that in the 3rd step of $\cU$ we $COPY$ the keys from $S_1$ to $S_1^\prime$, and this system is never used in the operations later. Thus the system that stores $K^{(1)}$ is already prepared. We emphasize we should view $S_1^\prime$ here as ``the $S_1$ in the description of Lemma \ref{lem:6.4}''.\par
	The main thing we need to prove is to prove the remaining system (other than $S_1^\prime$) can be transformed to a form that contains some keys in $K^{(3)}$. Our method is to first analyze the form of the state if the $COPY$ step is not executed, then prove the normal case (where $COPY$ step is executed) does not differ too much from it.\par
	If the $COPY$ step is skipped, the remaining operation in $\cU$  will be $\cU^{\geq \text{4th}}=\cU_1\circ \fAdv^\dagger_{1.(i+1)\sim 1.T}\circ \cU_2^\dagger$, applied on the state $\cU_2\fAdv_{=2}^\dagger (P_{pass}\ket{\varphi^\prime}\odot \llbracket\fAuxInf\rrbracket)$. By (\ref{eq:133}) we know
	\begin{equation}\label{eq:r104}\cU^{\geq \text{4th}}\cU_2 \fAdv_{=2}^\dagger P_{pass}\ket{\varphi^\prime}\approx^{st-ind}_{C^6|\ket{\varphi}|}\cU_1\fAdv_{1.(i+1)\sim 1.T}^\dagger P_{pass}\ket{\varphi^T}\odot \llbracket\fPrtl_{=2}\rrbracket\end{equation}
	Thus to understand the left hand side of (\ref{eq:r104}), we can analyze the right hand side.\par
	By an argument similar to equation (\ref{eq:r101}) to (\ref{eq:45}), we have
	\begin{align}\fAdv_{1.(i+1)\sim 1.T}^\dagger P_{pass}\ket{\varphi^T}\approx^{st-ind}_{C^6|\ket{\varphi}|} & \fAdv^\dagger_{1.(i+1)\sim 1.(T-1)}P_{pass}\ket{\varphi^{T-1}}\odot \llbracket\fPrtl_{=1.T}\rrbracket                        \\
		\approx^{st-ind}_{C^6|\ket{\varphi}|}                                                        & \fAdv^\dagger_{1.(i+1)\sim 1.(T-2)}P_{pass}\ket{\varphi^{T-2}}\odot \llbracket\fPrtl_{=1.(T-1)}\rrbracket\odot \llbracket\fPrtl_{=1.T}\rrbracket \\
		\approx^{st-ind}_{C^6|\ket{\varphi}|}                                                        & \cdots                                                                                                   \\
		\approx^{st-ind}_{C^6|\ket{\varphi}|}                                                        & P_{pass}\ket{\varphi^{i}}\odot \llbracket\fPrtl_{1.(i+1)\sim 1.T}\rrbracket
	\end{align}
	Note that $\approx_{C^6}^{st-ind}$ describes the relation between adjacent terms. In summary we have
	\begin{equation}\label{eq:r108}
		\fAdv_{1.(i+1)\sim 1.T}^\dagger P_{pass}\ket{\varphi^T}\approx^{st-ind}_{(T-i)\cdot C^6|\ket{\varphi}|} P_{pass}\ket{\varphi^{i}}\odot \llbracket\fPrtl_{1.(i+1)\sim 1.T}\rrbracket
	\end{equation}
	where we have $(T-i)\cdot C^6\leq 2C^2$.\par
	Then we have
	{\small\begin{align}\label{eq:79}
		                                          & |(I-P^{S_3}_{K^{(3)}})\cU^{\geq \text{4th}} (\ket{\text{Right of eq (\ref{eq:38})}}\odot \llbracket\fAuxInf_1\rrbracket)|                       \\
		=                                         & |(I-P^{S_3}_{K^{(3)}})(\ket{\text{Right of eq (\ref{eq:r104})}}\odot \llbracket\fAuxInf_1\rrbracket\odot \llbracket\fAuxInf_2\rrbracket)|                           \\
		\text{(By (\ref{eq:r108}))}\leq & |(I-P^{S_3}_{K^{(3)}})\cU_1 (\ket{\text{Right of eq (\ref{eq:r108})}}\odot \llbracket\fAuxInf_1\rrbracket\odot \llbracket\fAuxInf_2\rrbracket)|+2C^2|\ket{\varphi}| \\
		\text{(By (\ref{eq:5}))}\leq               & 5.45C|\ket{\varphi}|+2C^2|\ket{\varphi}|\leq 6C|\ket{\varphi}|\label{eq:206}
	\end{align}}
	Define $$\ket{\tilde\varphi}=P_K^{S_1}\ket{\text{Right of (\ref{eq:38})}}=\ket{\psi_0^{(1)}}+\ket{\psi_1^{(1)}}$$
	. Consider the server-side operation that computes keys in $K^{(3)}$ by applying $\cU^{\geq \text{4th}}$ on $\ket{\tilde\varphi}$. (By now the ``$COPY$'' step is still not considered yet.) Then because the distance of $\ket{\tilde\varphi}$ to $\ket{\text{Right of eq (\ref{eq:38})}}$ is $\leq|\ket{\chi^{(1)\prime}}|\leq \frac{1}{2}C^2|\ket{\varphi}|$, substitute this into (\ref{eq:79}), by (\ref{eq:206}) we have
	\begin{equation}\label{eq:86}|(I-P^{S_3}_{K^{(3)}})\cU^{\geq \text{4th}}(\ket{\tilde\varphi}\odot \llbracket\fAuxInf_1\rrbracket)|\leq 6.5C|\ket{\varphi}|\end{equation}
	Now with the $COPY$ step, by Lemma \ref{lem:4.13} (as the condition for applying this lemma, we can prove $\ket{\tilde\varphi}\odot \llbracket\fAuxInf_1\rrbracket$ is $(2^{\eta/8},2^{-\eta/8}|\ket{\tilde\varphi}|)$-SC-secure for $K^{(1)}$ since we can prove $P_{pass}\ket{\varphi^T}\odot \llbracket\fAuxInf\rrbracket$ is $(2^{\eta/8}+2^{\kappa+1},2^{-\eta/7}|\ket{\varphi}|)$-SC-secure for $K^{(1)}$),
	{\small\begin{align}&|\quad |(I-P^{S_3}_{K^{(3)}})\cU^{\geq \text{4th}}(\ket{\tilde\varphi}\odot \llbracket\fAuxInf_1\rrbracket)|-|(I-P^{S_3}_{K^{(3)}})\cU^{\geq \text{4th}} COPY\circ(\ket{\tilde\varphi}\odot \llbracket\fAuxInf_1\rrbracket)|\quad |\\\leq& 2^{-\eta/30}|\ket{\varphi}|\end{align}}
	Using (\ref{eq:86}), and adding back $\ket{\chi^{(1)\prime}}$ we get
	\begin{equation}
		|(I-P^{S_3}_{K^{(3)}})\cU^{\geq \text{4th}} COPY\circ(\ket{\text{Right of eq (\ref{eq:38})}}\odot \llbracket\fAuxInf_1\rrbracket)|\leq 7C|\ket{\varphi}|
	\end{equation}
	Thus substituting (\ref{eq:38}) it becomes
	\begin{equation}|(I-P^{S_3}_{K^{(3)}})\cU^{\geq 2}(P_{pass}\ket{\varphi^T}\odot \llbracket\fAuxInf\rrbracket)|\leq 7C|\ket{\varphi}|\end{equation}
	Finally using equation (\ref{eq:133}) we get
	\begin{equation}
		|(I-P_{K^{(3)}}^{S_3})\cU(P_{pass}\ket{\varphi^\prime}\odot \llbracket\fAuxInf\rrbracket)|\leq (7C+C^6)|\ket{\varphi}|
	\end{equation}
	This (for $S_3$), together with (\ref{eq:135}) (for $S_1^\prime$), completes the proof.
\end{proof}

\section{Proof of Lemmas in Section \ref{sec:7.2}}\label{sec:AC}
This section contains the proofs for the lemmas about the padded Hadamard test in Section \ref{sec:7.2}. Before we prove these lemma, we first prove the following lemma.
\begin{lem}\label{lem:C1}
	In the padded Hadamard test on keys denoted as $K=\{x_0,x_1\}$, $b\in \{0,1\}$, if the purified joint state $$\ket{\tilde\varphi}=\sum_{pad\in \{0,1\}^l}\frac{1}{\sqrt{2^l}}\underbrace{\ket{pad}}_{\text{client}}\underbrace{\ket{pad}}_{\text{read-only buffer}}\otimes\ket{\tilde\varphi_{pad}}\otimes\underbrace{\ket{pad}}_{\text{environment}}$$ does not depend on (in the sense of Definition \ref{def:ndep}) $H(pad||\cdots)$ where ``$\cdots$'' is the set of all strings of the same length as the keys in $K$ , an adversary $\fAdv^{blind}$ only queries the blinded oracle $H^\prime$ where $H(pad||x_b)$ is blinded, (note the $x_{1-b}$ part is not blinded,) then $$|P_{pass}\fPadHadamard^{\geq 2}_{\fAdv^{blind}}(K)\circ \ket{\tilde\varphi}|\leq \frac{1}{\sqrt{2}}|\ket{\tilde\varphi}|$$
	where the superscript ``$\geq 2$'' means the first step (sampling a random pad) is already and done thus skipped. The $\kappa_{out}$ parameter within the protocol is arbitrary.
\end{lem}
\begin{proof}
	Since $\ket{\tilde\varphi}$ does not depend on $H(pad||\cdots)$ and $\fAdv^{blind}$ does not query $H(pad||x_b)$,\\$\fPadHadamard^{\geq 2}_{\fAdv^{blind}}(K)\circ \ket{\tilde\varphi}$ does not depend on $H(pad||x_b)$. Suppose the server's response for the test is $d=(d_1,d_2)$ where $d_2$ corresponds to the last $\kappa_{out}$ bits, we can expand the left hand as follows:
	\begin{align}     & |P_{pass}\fPadHadamard^{\geq 2}_{\fAdv^{blind}}(K)\circ \ket{\tilde\varphi}|                                                                                                       \\
		=    & \sqrt{\bE_{H(pad||x_b)}|P_{d_1\cdot x_0+d_2\cdot H(pad||x_0)=d_1\cdot x_1+d_2\cdot H(pad||x_1), d_2\neq 0}\sum_{d_1d_2} \ket{d_1}\otimes \ket{d_2}\otimes\ket{\chi_{d_1d_2}}|^2} \\
		\leq & \sqrt{\frac{1}{2}|P_{d_2\neq 0}\sum_{d_1d_2} \ket{d_1}\otimes \ket{d_2}\otimes\ket{\chi_{d_1d_2}}|^2}\leq \frac{1}{\sqrt{2}}|\ket{\tilde\varphi}|\end{align}
\end{proof}
Now we give the proof for the lemmas in Section \ref{sec:7.2}.
\begin{proof}[Proof of Lemma \ref{lem:7.3}]
First we can assume $Tag(K)$ is already stored in the read-only buffer in the beginning of the protocol. We still denote the initial state and the post-execution state as $\ket{\varphi}$, $\ket{\varphi^\prime}$. By the \emph{auxiliary information technique} proving the statement in this setting implies the original lemma. Note that this does not fall into the cases listed in Section \ref{sec:4.2}, but this is still true: the adversary can simply ignore this auxiliary information.\par
	Proof by contradiction. Suppose there exists a server-side operation $\cU$ with query number $|\cU|\leq 2^{\eta/6}$ such that
	\begin{equation}\label{eq:33}|P_{span\{x_0, x_1\}}\cU\ket{\varphi^\prime}|> 2C|\ket{\varphi}|\end{equation}
	where $P_{span\{x_0, x_1\}}$ is a projection onto some server-side system $S$ on outputting either $x_0$ or $x_1$.\par
	Consider the following adversary $\fAdv^\prime$ for $\fPadHadamard$: after it receives the $pad$ from the client, it first runs $\fAdv$ to get a register that stores $d$. Now instead of measuring $d$, it first copies $d$ to some separated register, runs $\cU$ on the remaining registers (which are the server and buffer parts of $\ket{\varphi^\prime}$), and measures to try to get one of $x_0$ or $x_1$, then finally measures the $d$ register and sends it out.\par
	On the one hand, since the measurement on $d$ and the operations on the remaining systems commute, we have
	$$|P_{pass}\fAdv^\prime(\ket{\varphi}\odot pad)|=|P_{pass}\cU\ket{\varphi^\prime}|\geq (1-C^2)|\ket{\varphi}|$$
	On the other hand, we can give an upper bound for $|P_{pass}\cU\ket{\varphi^\prime}|$ by estimating
	\begin{align}|P_{pass}P_{span\{x_0, x_1\}}\cU\ket{\varphi^\prime}|&=|P_{pass}P_{x_0}\cU\ket{\varphi^\prime}+P_{pass}P_{x_1}\cU\ket{\varphi^\prime}|\\&\leq |P_{pass}P_{x_0}\cU\ket{\varphi^\prime}|+|P_{pass}P_{x_1}\cU\ket{\varphi^\prime}|\end{align}
	, where we use $P_{x_0}$ and $P_{x_1}$ to denote the projection onto $x_0$ and $x_1$ on $S$ (defined below (\ref{eq:33})).\par
	Define
	$$p_0=|P_{pass}P_{x_0}\cU\ket{\varphi^\prime}|=|P_{pass}P_{x_0}\cU ( \fPadHadamard_\fAdv\circ\ket{\varphi})|$$
	$$p_1=|P_{pass}P_{x_1}\cU\ket{\varphi^\prime}|=|P_{pass}P_{x_1}\cU ( \fPadHadamard_\fAdv\circ\ket{\varphi})|$$
	Without loss of generality, let's give an upper bound for $p_0$. Assume
	\begin{equation}\label{eq:119i}|P_{x_0}\cU\ket{\varphi^\prime}|\geq 2^{-\eta/4}|\ket{\varphi}|\end{equation}
	(Otherwise we already have a very good upper bound for $p_0$. We will merge this case in the end.)\par
	First note that the first step of $\fPadHadamard$ is to sample a random pad of length $l$. Denote the state after sampling the pad as $\ket{\varphi^1}$. Denote $\ket{\tilde\varphi}$ as the state of replacing all the queries in the representation of $\ket{\varphi}$ by $H\cdot (I-P_{pad||\cdots})$. By Lemma \ref{lem:3.4r} we can get $\ket{\tilde \varphi}\approx_{2^{-\eta+1}|\ket{\varphi}|}\ket{\varphi^1}$ and $\ket{\tilde \varphi}$ does not depend on $H(pad||\cdots)$.\par Then consider the blinded version of the padded Hadamard test, where $\cU$ and $\fAdv$ only query $H^\prime$, where $H(pad||x_1)$ is blinded:
	$$\ket{\psi}=P_{x_0}\cU^{blind} (\fPadHadamard^{\geq 2}_{\fAdv^{blind}}\circ\ket{\tilde \varphi})$$
	Here $\cU^{blind}$ and $\fAdv^{blind}$ are the blinded version of $\cU$ and $\fAdv$ where all the queries to $H$ are replaced by queries to $H^\prime$. The $\geq 2$ superscript means the padding step has already completed.\par
	Since $\ket{\tilde\varphi}$ is $(2^{\eta},2^{-\eta+3}|\ket{\tilde\varphi}|)$-SC-secure for $K$ and the last step is a projection on $x_0$, by Lemma \ref{lem:4.16},
	\begin{align}
		\ket{\psi}           & \approx_{2^{-\eta/3+2}|\ket{\tilde\varphi}|}P_{x_0}\cU (\fPadHadamard_{\fAdv}^{\geq 2}\circ\ket{\tilde\varphi}) \\
		\Rightarrow\ket{\psi} & \approx_{2^{-\eta/3+3}|\ket{\varphi}|}P_{x_0}\cU\ket{\varphi^\prime}\end{align}
	Then by Lemma \ref{lem:C1},
	$$|P_{pass}\ket{\psi}|\leq \frac{1}{\sqrt{2}}|\ket{\psi}|$$
	. Thus
	\begin{equation}p_0\leq \frac{1}{\sqrt{2}}|P_{x_0}\cU\ket{\varphi^\prime}|+2^{-\eta/3+4}|\ket{\varphi}|\end{equation}
	Combining it with the case where (\ref{eq:119i}) does not hold, we get
	\begin{equation}p_0\leq \frac{1}{\sqrt{2}}|P_{x_0}\cU\ket{\varphi^\prime}|+2^{-\eta/4}|\ket{\varphi}|\end{equation}
	And similarly the inequality holds if we replace $p_0$ and $P_{x_0}$ by $p_1$ and $P_{x_1}$. Combining them we know
	\begin{equation}\label{eq:270reh}|P_{pass}P_{span\{x_0, x_1\}}\cU\ket{\varphi^\prime}|\leq \frac{1}{\sqrt{2}}|P_{span\{x_0, x_1\}}\cU\ket{\varphi^\prime}|+2^{-\eta/4+5}|\ket{\varphi}|\end{equation}
	Together with equation (\ref{eq:33}) we know\footnote{The details are as follows: from (\ref{eq:270reh}) we know $|P_{fail}P_{span\{x_0, x_1\}}\cU\ket{\varphi^\prime}|\geq \frac{1}{\sqrt{2}}|P_{span\{x_0, x_1\}}\cU\ket{\varphi^\prime}|-2^{-\eta/4+5}|\ket{\varphi}|$. Thus $|P_{pass}\cU\ket{\varphi^\prime}|=\sqrt{|\ket{\varphi}|^2-|P_{fail}\cU\ket{\varphi^\prime}|^2}<(\sqrt{1-(C/\sqrt{2})^2}+exp(-\kappa))|\ket{\varphi}|<(1-C^2)|\ket{\varphi}|$} $|P_{pass}\cU\ket{\varphi^\prime}|<(1-C^2)|\ket{\varphi}|$, which gives a contradiction.
\end{proof}
Similarly we can prove Lemma \ref{lem:7.4}.
\begin{proof}[Proof of Lemma \ref{lem:7.4}]
	Suppose the conclusion is not true. Consider the following adversary and client: the adversary does not measure $d$, but sends out the $d$ register in quantum state directly. The client holds $d$, and sends $\llbracket\fAuxInf\rrbracket$, then the server runs $\cD$; then the adversary makes a projection on $S$; finally the client measures $d$. Since the two measurements ($d$, and $S$) commute the norm of the passing part should be the same.\par
	Suppose $p>5C$, and without loss of generality, suppose $p_0< p/6$. Thus $p_1<7p/6$.\par
	Use $\fPrtl$ to denote the process of the client providing $\llbracket\fAuxInf\rrbracket$ to the server and the server running $\cD$. We can prove the following lemma using similar argument as proof of Lemma \ref{lem:7.3} above (we omit the subscript $\fAdv$ for simplicity):
	\begin{align}
		\forall b\in \{0,1\},\qquad & |P_{pass}P_S (\fPrtl\circ \fPadHadamard)\ket{\varphi_b}|                                         \\
		\leq                        & \frac{1}{\sqrt{2}}|P_S (\fPrtl\circ \fPadHadamard)\ket{\varphi_b}|+2^{-\eta/4+10}|\ket{\varphi}|
	\end{align}
	Then  we have
	\begin{align}     & |P_{pass}\cD(\ket{\varphi^\prime}\odot \llbracket\fAuxInf\rrbracket)|/|\ket{\varphi}|                         \\
		\leq & \sqrt{1-(p^2-|P_{pass}P_S\cD(\ket{\varphi^\prime}\odot \llbracket\fAuxInf\rrbracket)|/|\ket{\varphi}|^2)}     \\
		<    & \sqrt{1-p^2+(\frac{1}{\sqrt{2}}\cdot (\frac{1}{6}p+\frac{7}{6}p)+2^{-\eta/4+11})^2}<1-C^2\end{align}
	which is a contradiction.
\end{proof}
\section{Properties of $\fRobustRLT$ in Definition \ref{def:rrlt}}\label{sec:afn}
In this section we give some security properties of the robust reversible lookup table. Simply speaking, we can think about the problems in the following form: suppose the initial state is secure for some keys with some parameters, if the client computes and sends a $\fRobustRLT$ to the server, how secure is the final state for these keys? We give detailed study for this problem in this section, and these lemmas will be useful for the proofs in Section \ref{sec:7} and Appendix \ref{sec:ac2}.\par
\paragraph{Organization of this section} This section is organized as follows.
\begin{itemize}\item In Section \ref{sec:ag1} we will give some basic notations and facts, and they will be useful in the proofs later in this section. This can be seen as a preparation for later subsections.\item  In Section \ref{sec:ag2} we study the security for $K^{\text{help}}$ when a $\fRobustRLT$ is provided. This part is relatively easy since $K^{\text{help}}$ is not part of the reversible encoding $\fRobustRLT$ does not introduce complicated structure on it.\item  In Section \ref{sec:ag3} we study how the $\fRobustRLT$ affects the security when $x_1^{\text{help}}$ is unpredictable (which correspond to the identity-style branch).\item In Section \ref{sec:ag4} we study how the $\fRobustRLT$ affects the security when $x_0^{\text{help}}$ is unpredictable (which correspond to the CNOT-style branch).\end{itemize}
\subsection{Some Basic Notations and Facts}\label{sec:ag1}
\subsubsection{Fake keys and $\tilde{tag}$}
First, let's formally define the notation for the fake keys and the $\tilde{tag}$. We have already seen these notations in the informal proof of Lemma \ref{lem:7.6}, and now we repeat and complete the definitions:
\begin{nota}\label{nota:f1}
	Suppose $K_{out}^{(2)}$, $K_{out}^{(3)}$ are the output keys, $K_{out}^{(w)}=\{y_0^{(w)},y_1^{(w)}\}$ ($w\in \{1,2,3\}$). $perm$ and $perm^\prime$ are bit-wise permutations on strings of length $2\kappa_{out}$ where $\kappa_{out}$ is the output key length. $perm$ is called the \emph{real permutation} and $perm^\prime$ is called the \emph{fake permutation}. Define the notations for fake keys as follows:
	\begin{equation}\label{eq:228p}\text{The ``fake keys for $c=0$'', }K_{out}^{fake-0-(w)}=\{y_b^{fake-0-(w)}\}_{b\in \{0,1\}},w\in \{2,3\}\end{equation}
	are defined to be the key pairs of the same length with $K_{out}^{(2)}$, $K_{out}^{(3)}$ that satisfy:
	\begin{equation}\label{eq:228}\forall b\in \{0,1\}, perm^\prime(y_b^{fake-0-(2)}||y_{b}^{fake-0-(3)})=perm (y_b^{(2)}||y_{b}^{(3)}).\end{equation}
	And
	\begin{equation}\label{eq:230p}\text{The ``fake keys for $c=1$'', }K_{out}^{fake-1-(w)}=\{y_b^{fake-1-(w)}\}_{b\in \{0,1\}},w\in \{2,3\}\end{equation}
	are defined to be the key pairs of the same length with $K_{out}^{(2)}$, $K_{out}^{(3)}$ that satisfy:
	\begin{equation}\label{eq:229}\forall b\in \{0,1\}, perm^\prime(y_b^{fake-1-(2)}||y_{1-b}^{fake-1-(3)})=perm (y_b^{(2)}||y_{1-b}^{(3)})\end{equation}
\end{nota}
Recall the intuition behind the definition of fake keys: normally, when the server tries to de-permute $perm (y_b^{(2)}||y_{c+b}^{(3)})$ using the real permutation, it gets the keys in $K^{(2)}_{out}$, $K^{(3)}_{out}$. But if it tries to de-permute them using the fake permutation, it gets the fake keys $K^{fake-c-(2)}_{out}$, $K^{fake-c-(3)}_{out}$. Note that the letter ``$c$'' here will also be used in the same way in the proofs in this section.
\paragraph{Some rules for superscripts and subscripts of fake keys}It might be hard to remember and understand the superscripts and subscripts of fake keys when we meet them in the proofs later, here we make some detailed discussion for it.
\begin{itemize}
\item Note that (\ref{eq:228}) actually contains two equations and thus can define four keys. This four keys form the two fake key pairs for $c=0,w\in \{2,3\}$. (See (\ref{eq:228p}))\par
And similar thing is true for (\ref{eq:228}). (Here $c=1$.)
	\item Notice that (\ref{eq:228})(\ref{eq:229}) do not change the subscripts: the subscripts that come from de-permuting $perm(y_b^{(2)}||y_{b^\prime}^{(3)})$ is still $b,b^\prime$, correspondingly.
	\item The value of ``$c$'' is the xor of the subscripts in the definition equation. (Note that if we view the definition equation as two equations for $w=2$ and $w=3$, we should first fix $w$ then compute the xor of the subscripts.)
\end{itemize}
In the proofs below we will need a special ``global tag'' defined as follows, as the shuffling of three global tags:
\begin{nota}\label{nota:ag2}
Define $\tilde{tag}$ as the random shuffling (a random permutation in $S_3$) of $Tag(K^{(2)}_{out})$, $Tag(K_{out}^{fake-0-(2)})$, $Tag(K_{out}^{fake-1-(2)})$.
\end{nota}
\subsubsection{New notation: $\approx_\epsilon^{\fAdv\in \cA}$, and its variants}
Then let's generalize the notation $\approx$ to contain a set of adversaries as the superscript.
\begin{nota}\label{nota:af2}
	$\ket{\varphi}\approx_\epsilon^{\fAdv\in \cA}\ket{\varphi^\prime}$ if for all $\fAdv\in \cA$, the ``distinguishing norm'', which is defined as
	\begin{equation}\label{eq:108}
		|\quad |P_0\fAdv\ket{\varphi_1}|-|P_0\fAdv\ket{\varphi_2}|\quad |
	\end{equation}
	, is at most $\epsilon$.\par
	And we write $\ket{\varphi}\approx_\epsilon^{\text{no query}}\ket{\varphi^\prime}$ if $\cA$ is the set of adversaries that make $0$ random oracle queries.\par
	And we write $\ket{\varphi}\approx_\epsilon^{\text{st-ind}}\ket{\varphi^\prime}$ if $\cA$ is the set of unbounded adversaries that can make unbounded number of random oracle queries. (Note that this notation appeared temporarily in the proof of Lemma \ref{lem:6.4} but it has different meaning there.) 
\end{nota}
These notations will be useful in the proofs later in this section.
\subsubsection{Basic combinatoric facts}
The following facts can be proved by basic combinatoric and probability calculation, and will be useful later.
\begin{fact}\label{fact:f133}
	The following statement is true when $\kappa_{out}$ is bigger than some constant:\par
	Suppose $IndexSet_1$, $IndexSet_2$ are two sets of size $\kappa_{out}$ which are subsets of $[2\kappa_{out}]$. $perm,perm^\prime$ are sampled independently randomly on the bit-wise permutation on strings of length $2\kappa_{out}$. Denote $IndexSet$ as the sets of elements that are originally in $IndexSet_1$ but mapped to $IndexSet_2$ by ${perm^\prime}^{-1}\circ perm$. Then with probability $>1-2^{-\kappa_{out}/10}$, the number of elements in $IndexSet$ satisfies $|IndexSet|>\frac{1}{10}\kappa_{out}$.
\end{fact}
\begin{fact}\label{fact:f1332}
	Consider the following game: the client samples $K=\{x_0,x_1\}$ and $K^\prime=\{x_0^\prime,x_1^\prime\}$ randomly, which are both pairs of different keys of length $\kappa_{out}$. The client samples $perm$ randomly from the bit-wise permutations on strings of length $2\kappa_{out}$ (but does not reveal it). Then it provides $perm(x_0||x_0^\prime)$ and $perm(x_1||x_1^\prime)$ to the challenger, and the challenger tries to guess either $perm(x_0||x_1^\prime)$ or $perm(x_1||x_0^\prime)$.\par
	The probability that the challenger can win the game above is at most $2^{-\kappa_{out}/10}$.
\end{fact}
\begin{fact}\label{fact:ac6}
In the same setting with Fact \ref{fact:f1332}, the challenger tries to guess $x_0$. Then the probability that the challenger can win the game above is at most $2^{-\kappa_{out}/10}$.
\end{fact}
\begin{fact}\label{fact:ac7}
Consider the following game: the client samples $K=\{x_0,x_1\}$ and $K^\prime=\{x_0^\prime,x_1^\prime\}$ randomly, which are both pairs of different keys of length $\kappa_{out}$, and gives it to the server. The client samples $perm$ randomly from the bit-wise permutations on strings of length $2\kappa_{out}$ (but does not reveal it). The server tries to guess the first half of $perm(x_{b_1}||x_{b_2})$ where $b_1,b_2\in \{0,1\}$ can be arbitrary.\par
	The probability that the challenger can win the game above is at most $2^{-\kappa_{out}/10}$.
\end{fact}

\subsection{SC-security of $K^{\text{help}}$ when a $\fRobustRLT$ is provided}\label{sec:ag2}
The following lemma says giving a robust reversible lookup table to the adversary (and even many other related information) does not affect the SC-security for $K^{\text{help}}$ too much.
\begin{lem}\label{lem:af3}
	The following statement is true when the security parameter $\eta$ is bigger than some constant:\par
	Suppose the input key sets are denoted as $\{K^{\text{help}},K^{(3)}\}$ where each $K^{(w)}$ is a pair of different keys. The initial state is described by the purified joint state $\ket{\varphi}$. Suppose the following conditions are satisfied:
	\begin{itemize}
		\item (Security of the input) $\ket{\varphi}$ is $(2^{\eta},2^{-\eta}|\ket{\varphi}|)$-SC-secure for $K^{\text{help}}$ given $K^{(3)}$.
		\item (Well-behavenss of the input) $\ket{\varphi}\in \cWBS(D)$, $D\leq 2^\eta$.
		\item (Sufficient padding length, output key length) $l>6D+4\eta$, $\kappa_{out}>l+\eta$
	\end{itemize}
	As in the protocol, the client samples $K_{out}=\{K_{out}^{(2)},K_{out}^{(3)}\}$ where each $K_{out}^{(w)}$ is a pair of different keys with key length $\kappa_{out}$, samples $K^{(2)}$ which is a pair of different keys that has the same length as $K^{(3)}$, and samples $perm$ from the bit-wise permutation on strings of length $2\kappa_{out}$, then
	{\small\begin{align}
		\ket{\varphi} & \odot \fRobustRLT(K^{\text{help}},K^{(2,3)}\leftrightarrow K_{out},perm;\underbrace{ \ell}_{\substack{\text{padding} \\ \text{length}}}) \odot K^{(2)}                                   \odot perm\odot K^{(3)}\odot K_{out}^{(2)}\odot K_{out}^{(3)}
	\end{align}}

	is $(2^{\eta/6},2^{-\eta/6}|\ket{\varphi}|)$-SC-secure for $K^{\text{help}}$.
\end{lem}
\begin{proof}
	Since $K^{(2)}$, $K_{out}^{(2)}$, $K_{out}^{(3)}$, $perm$ are all sampled freshly randomly we know $\ket{\varphi}\odot K^{(2)}\odot K_{out}^{(2)}\odot K_{out}^{(3)}\odot perm$ is $(2^{\eta},2^{-\eta}|\ket{\varphi}|)$-SC-secure for $K^{\text{help}}$ given $K^{(3)}$.\par
	When ($K^{(2)}$, $K^{(3)}$, $K_{out}^{(2)}$, $K_{out}^{(3)}$, $perm$) are given to the server as auxiliary information, the $\fRobustRLT$ can be seen as giving some lookup tables (with extra paddings) on $K^{\text{help}}$. (Recall the definition of $\fRobustRLT$ in Section \ref{sec:7.1}.) Thus apply Lemma \ref{lem:4.8} completes the proof. (Or more formally, if this lemma is not true, we can make use of the adversary to break the result in Lemma \ref{lem:4.8}. We just need to compute the corresponding $p^t_1,p^t_2,p^t_3$ in Lemma \ref{lem:4.8} based on $K^{(2)}$, $K^{(3)}$, $K_{out}^{(2)}$, $K_{out}^{(3)}$ and $perm$.)
\end{proof}
Similarly when the auxiliary information is $perm^\prime$ and fake keys, similar statement also holds:
\begin{lem}\label{lem:lrr3}
	Under the same conditions of Lemma \ref{lem:af3}, and the client samples $K_{out}$, $K^{(2)}$, $perm$ similarly, and samples another permutation $perm^\prime$ independently randomly from the bit-wise permutation on strings of length $2\kappa_{out}$, then  $\forall c\in \{0,1\}$,
	\begin{align}
		\ket{\varphi} & \odot \fRobustRLT(K^{\text{help}},K^{(2,3)}\leftrightarrow K_{out},perm;\underbrace{ \ell}_{\substack{\text{padding} \\ \text{length}}})\\& \odot K^{(2)}                                                        \odot perm^\prime\odot K^{(3)}\odot K_{out}^{fake-c-(2)}\odot K_{out}^{fake-c-(3)}
	\end{align}

	is $(2^{\eta/6},2^{-\eta/6}|\ket{\varphi}|)$-SC-secure for $K^{\text{help}}$.
\end{lem}
The proof is similar to the proof of Lemma \ref{lem:af3}: first we know $\ket{\varphi}\odot K^{(2)}\odot K_{out}^{(2)}\odot K_{out}^{(3)}\odot perm\odot perm^\prime$ is $(2^{\eta},2^{-\eta}|\ket{\varphi}|)$-SC-secure for $K^{\text{help}}$ given $K^{(3)}$, then we notice that $K_{out}^{fake-c-(2)}, K_{out}^{fake-c-(3)}$ can be computed from these auxiliary information. Then the remaining proof is similar.\par

Lemma \ref{lem:af3} and \ref{lem:lrr3} discuss the SC-security of $K^{\text{help}}$. These are relatively easier to understand. The following subsections care about the security for keys in the other two wires of the output keys.\par
\subsection{Security effect of $\fRobustRLT$ on Identity-style Branch (when the adversary knows $x^{\text{help}}_0$, but $x_1^{\text{help}}$ is unpredictable)}\label{sec:ag3}
The following lemmas consider the case where $x_1^{\text{help}}$ is unpredictable, which corresponds to the $b_1=0$ branch.
\subsubsection{The usual case (not the ``fake keys'' case)}
 What if the adversary can only know $x_0^{\text{help}}$ and one key in $K^{(3)}$ (while $x_1^{\text{help}}$ and the other key in $K^{(3)}$ are both unpredictable)? As discussed in Section \ref{sec:7.4.3}, in this case intuitively it cannot get both keys in $K_{out}^{(3)}$, even if $K_{out}^{(2)}$ is given. Formalizing this intuition gives us the following lemma.
\begin{lem}\label{lem:lrr2}
	The following statement is true when the security parameter $\eta$ is bigger than some constant:\par
	Suppose the input key sets are denoted as $\{K^{\text{help}},K^{(3)}\}$, $K^{\text{help}}=\{x_b^{\text{help}}\}_{b\in \{0,1\}}$, $K^{(3)}=\{x_b^{(3)}\}_{b\in \{0,1\}}$, the initial state is described by the purified joint state $\ket{\varphi}$, a bit $c\in \{0,1\}$ satisfy:
	\begin{itemize}
		\item (Security of the input) $\ket{\varphi}$ is $(2^\eta,2^{-\eta}|\ket{\varphi}|)$-unpredictable for $x^{\text{help}}_1$ given $x_0^{\text{help}}$ and $K^{(3)}$.
		\item (Security of the input) $\ket{\varphi}$ is $(2^{\eta},C|\ket{\varphi}|)$-unpredictable for $x_{1-c}^{(3)}$ given $x_c^{(3)}$ and $K^{\text{help}}$. $\frac{1}{3}\geq C\geq 2^{-\sqrt{\eta}}$.
		\item (Well-behavenss of the input) $\ket{\varphi}\in \cWBS(D)$, $D\leq 2^\eta$.
		\item (Sufficient padding length, output length) $l>6D+4\eta$, $\kappa_{out}>l+\eta$
	\end{itemize}
	The client samples $K_{out}=\{K_{out}^{(2)},K_{out}^{(3)}\}$ where each $K_{out}^{(w)}$ is a pair of different keys with key length $\kappa_{out}$, samples $K^{(2)}$ which is a pair of different keys with the same key length as $K^{(3)}$, and samples $perm$ from the bit-wise permutations on strings of length $2\kappa_{out}$, then the following conclusion holds:
	\begin{equation}\label{eq:237}\ket{\varphi}\odot \fRobustRLT(K^{\text{help}},K^{(2,3)}\leftrightarrow K_{out},perm;\underbrace{ \ell}_{\substack{\text{padding} \\ \text{length}}})\odot K^{(2)}\odot perm\end{equation}
	is $(2^{\eta/36},3C|\ket{\varphi}|)$-unpredictable for $y_{1-c}^{(3)}$ given $K_{out}^{(2)}$ and $y_c^{(3)}$.
\end{lem}
\begin{proof}[Proof of Lemma \ref{lem:lrr2}]
	For this proof, we can apply Lemma \ref{lem:4.23} and \ref{lem:4.24}.\par
	First by the auxiliary-information technique we can assume $x_0^{\text{help}}$, $x_{c}^{(3)}$ are already stored in the read-only buffer. Then (by the same reason as the proof of Lemma \ref{lem:af3}) we know $\ket{\varphi}\odot x_0^{\text{help}}\odot x_{c}^{(3)}\odot K^{(2)}\odot perm\odot K_{out}^{(2)}\odot y_c^{(3)}$ is $(2^\eta,C|\ket{\varphi}|)$-unpredictable for $x_{1-c}^{(3)}$. And the remaining term in (\ref{eq:237}) is the $\fRobustRLT$.\par
	Define  $\fRobustRLT^{hyb}$ as the ``hybrid'' lookup table where we replace everything encrypted under $x_1^{\text{help}}$ by random strings of the same length. By Lemma \ref{lem:4.23} there is
	\begin{align}&\ket{\varphi}\odot \fRobustRLT\odot K^{(2)}\odot perm\odot K_{out}^{(2)}\odot y_c^{(3)}\\
	\approx_{2^{-\eta/6}|\ket{\varphi}|}^{|\fAdv|\leq 2^{\eta/6}}&\ket{\varphi}\odot \fRobustRLT^{hyb}\odot K^{(2)}\odot perm\odot K_{out}^{(2)}\odot y_c^{(3)}\end{align}
	Further replace the rows in $\fRobustRLT^{hyb}$ that encrypt or are encrypted under $x_{1-c}^{(3)}$ by random strings of the same length, and denote the result as $\fRobustRLT^{hyb2}$. By Lemma \ref{lem:4.24} we have
\begin{align}&\ket{\varphi}\odot \fRobustRLT^{hyb}\odot K^{(2)}\odot perm\odot K_{out}^{(2)}\odot y_c^{(3)}\\
\approx_{2.95C|\ket{\varphi}|}^{|\fAdv|\leq 2^{\eta/6}}&\ket{\varphi}\odot \fRobustRLT^{hyb2}\odot K^{(2)}\odot perm\odot K_{out}^{(2)}\odot y_c^{(3)}\end{align}
	Finally note that in $\fRobustRLT^{hyb2}$ there is no information about $y_{1-c}^{(3)}$. This completes the proof.
\end{proof}
\subsubsection{The \emph{fake keys} case}
And we also have the following lemma, which talks about the unpredictability of the \emph{fake keys}. The reversibility introduces some complicated structure that could not be handle easily using existing lemmas in Section \ref{sec:basiclemmas} so we have to use hybrid methods from scratch.
\begin{lem}\label{lem:lrr2.2}
	Under the same conditions of Lemma \ref{lem:lrr2}, the client samples the same things and additionally samples $perm^\prime$ randomly from the bit-wise permutations on strings of length $2\kappa_{out}$, then
	\begin{align}
		\ket{\varphi} & \odot \fRobustRLT(K^{\text{help}},K^{(2,3)}\leftrightarrow K_{out},perm;\underbrace{ \ell}_{\substack{\text{padding} \\ \text{length}}})\odot K^{(2)}                                   \\
		              & \odot perm^\prime\odot Tag(K_{out}^{(2)})\odot Tag(K_{out}^{fake-0-(2)})\odot Tag(K_{out}^{fake-1-(2)})
	\end{align}

	satisfies: $\forall c^\prime\in \{0,1\}$, it's $(2^{\eta/36},3C|\ket{\varphi}|)$-SC-secure for $K_{out}^{fake-c^\prime-(2)}$ given \\$K_{out}^{fake-0-(3)}$ and  $K_{out}^{fake-1-(3)}$.
\end{lem}
Recall that $K_{out}^{fake-c-(2)}$, $K_{out}^{fake-c-(3)}$ are defined in Notation \ref{nota:f1}. And we expand the content of $\tilde{tag}$ (which means, in the original $\tilde{tag}$ there is a random shuffling, but here we simply provide these global tags without random shuffling.) This makes the argument stronger and cleaner-to-prove.\par
And we note that this lemma is talking about the security of output keys at wire $2$ instead of wire $3$. Thus \textbf{it's actually talking about different thing from Lemma \ref{lem:lrr2}}. Intuitively, it says: if the adversary does not know $x_1^{\text{help}}$, and only knows one of the keys in $K^{(3)}$, even if it knows both $K_{out}^{fake-0-(3)}$ and  $K_{out}^{fake-1-(3)}$, it still cannot compute both keys in $K_{out}^{fake-c^\prime-(2)}$, where $c^\prime$ can be any bit in $\{0,1\}$.\par

\subsubsection*{Proof of Lemma \ref{lem:lrr2.2}}
Note that to prove Lemma \ref{lem:lrr2.2}, it's enough to prove one key in $K_{out}^{fake-c^\prime-(2)}$ is unpredictable. The ``unpredictable key'' has subscript $1+c+c^\prime$ (modular 2). Thus the proof of Lemma \ref{lem:lrr2.2} is reduce to prove:
\begin{lem}[Lemma \ref{lem:lrr2.2}, variant]\label{lem:lrr2.22}
	Under the same conditions of Lemma \ref{lem:lrr2}, the client samples the same things and additionally samples $perm^\prime$ randomly from the bit-wise permutations on strings of length $2\kappa_{out}$, then
	\begin{align}
		\ket{\varphi} & \odot \fRobustRLT(K^{\text{help}},K^{(2,3)}\leftrightarrow K_{out},perm;\underbrace{ \ell}_{\substack{\text{padding} \\ \text{length}}})\odot K^{(2)}                                   \\
		              & \odot perm^\prime\odot K_{out}^{fake-0-(3)}\odot K_{out}^{fake-1-(3)}\\
		              &\odot Tag(K_{out}^{(2)})\odot Tag(K_{out}^{fake-0-(2)})\odot Tag(K_{out}^{fake-1-(2)})
	\end{align}
	satisfies: $\forall c^\prime\in \{0,1\}$, it's $(2^{\eta/36},3C|\ket{\varphi}|)$-unpredictable for $y^{fake-c^\prime-(2)}_{1+c^\prime+c}$.
\end{lem}
The proof of \ref{lem:lrr2.22} is relatively long. To prove it, we make use of a hybrid method (Lemma \ref{lem:4.12n}). To make the proof more readable, we will divide the proofs into several parts, and each part starts with a bold font title. The proof contains Part I and Part II. Part II is mainly a hybrid method, and it's further divided to II.1 and II.2. II.2 further contains II.2.2.
\begin{proof}[Proof of Lemma \ref{lem:lrr2.22}]
	\textbf{Part I: preparation before the hybrid method}\par
	Assume $x_0^{\text{help}}$, $x_c^{(3)}$, $Tag(x_1^{\text{help}})$, $Tag(x_{1-c}^{(3)})$ are given in the read-only buffer. This is reasonable since by Technique \ref{lem:4.2} this only makes the statement stronger.\par
	By the state decomposition lemma (Lemma \ref{lem:4.5}, together with Fact \ref{fact:injtag}) on key $x_{1-c}^{(3)}$, we can decompose $\ket{\varphi}$ as $\ket{\phi}+\ket{\chi}$ such that
	\begin{itemize}\item $\ket{\phi}$ is $(2^{\eta/6-1},2^{-\eta/6+1}|\ket{\varphi}|)$-unpredictable for $x^{(3)}_{1-c}$, and is $(1,2^{\eta-6})$-server-side-representable from $\ket{\varphi}$.
		\item $|\ket{\chi}|\leq 2.5C|\ket{\varphi}|$.
	\end{itemize}
	Thus we can derive that $\ket{\phi}$ satisfies:
	\begin{itemize}
		\item $\frac{1}{6}|\ket{\varphi}|\leq |\ket{\phi}|\leq |\ket{\varphi}|$
		\item $\ket{\phi}$ is $(2^{\eta/2},2^{\eta/2}|\ket{\phi}|)$-unpredictable for $x_1^{\text{help}}$ given $K^{(3)}$
		\item $\ket{\phi}$ is $(2^{\eta/6-1},2^{-\eta/6+4}|\ket{\phi}|)$-unpredictable for $x^{(3)}_{1-c}$
		\item $\ket{\phi}$ is $(2^{D},2^{D}+2^{\eta-6})$-representable from $\ket{\mathfrak{init}}$.
	\end{itemize}
	Then Lemma \ref{lem:lrr2.22} is reduced to prove
	\begin{align}\ket{\phi^\prime}:=\ket{\phi} & \odot \fRobustRLT(K^{\text{help}},K^{(2,3)}\leftrightarrow K_{out},perm;\ell)\odot K^{(2)}\label{eq:245}                  \\
		                              & \odot perm^\prime\odot K_{out}^{fake-0-(3)}\odot K_{out}^{fake-1-(3)}\label{eq:168}                  \\
		                              & \odot Tag(K_{out}^{(2)})\odot Tag(K_{out}^{fake-0-(2)})\odot Tag(K_{out}^{fake-1-(2)})\label{eq:169}
	\end{align}
	\begin{equation}\label{eq:152}\text{is $(2^{\eta/36},2^{-\eta/36}|\ket{\phi}|)$-unpredictable for $y_{1-c+c^\prime}^{fake-c^\prime-(2)}$.}\end{equation}
	Suppose the adversarial server-side operation applied on $\ket{\phi^\prime}$ is $\cD$. Query number $|\cD|\leq 2^{\eta/36}$. Then what we need to prove is
	\begin{equation}\label{eq:249}
		|P_{y_{1-c+c^\prime}^{fake-c^\prime-(2)}}\cD\ket{\phi^\prime}|\leq 2^{-\eta/36}|\ket{\phi}|
	\end{equation}
	where the projection is on some server-side system. Once we prove (\ref{eq:249}), we can add back $\ket{\chi}$ and complete the proof by triangle inequality.\par
	\textbf{Part II: the hybrid method (Lemma \ref{lem:4.12n})}\par
	Suppose the set of random pads used in the computation of $\fRobustRLT$ is $Set$. Consider a blinded random oracle $H^\prime$ where
	\begin{equation}\label{eq:170}H\text{ on inputs in }Set||x_1^{\text{help}}||\cdots,Set||x_0^{\text{help}}||\cdots||x_{1-c}^{(3)}\end{equation}
	\begin{equation}\label{eq:251}
		H\text{ on inputs in }Set||x_0^{\text{help}}||perm(y_0^{(2)}||y_{1-c}^{(3)}),Set||x_0^{\text{help}}||perm(y_1^{(2)}||y_{1-c}^{(3)})
	\end{equation}
	\begin{equation}\label{eq:171}Tag(y_{0}^{fake-(1-c)-(2)}),Tag(y_{1}^{fake-c-(2)})\end{equation}
	are all blinded. The first ``$\cdots$'' means strings of arbitrary length and the second ``$\cdots$'' means all the strings with length the same as $x^{(2)}$.\par
	Denote the blinded inputs as $BI$.\par
	We will replace the oracle queries appeared in (\ref{eq:249}) by $H^\prime$ one-by-one. And finally prove (\ref{eq:249}).\par
	\textbf{Part II.1: replace the queries implicitly hidden in $\ket{\phi}$}\par
	Recall that $\ket{\phi}$ is $(2^{D},2^{D}+2^{\eta-6})$-representable from $\ket{\mathfrak{init}}$. Expand $\ket{\phi}$ using the definition of representability (Definition \ref{def:rep}), and replace the oracle queries within it by queries to $H^\prime$. Denote the new state as $\ket{\tilde\phi}$. By Lemma \ref{lem:3.4r} we have
	\begin{equation}
		\ket{\tilde\phi}\approx_{2^{-\eta}|\ket{\phi}|}\ket{\phi}
	\end{equation}
	Similarly define $\ket{\tilde\phi^\prime}$ as the output state when we use $\ket{\tilde\phi}$ as the initial state in (\ref{eq:245})-(\ref{eq:169}). Thus to prove (\ref{eq:249}), we only need to prove
	\begin{equation}\label{eq:254}
		|P_{y_{1-c+c^\prime}^{fake-c^\prime-(2)}}\cD\ket{\tilde\phi^\prime}|\leq 2^{-\eta/33}|\ket{\tilde\phi}|
	\end{equation}
	\textbf{Part II.2}\par
	Suppose the operation coming from replacing the oracle queries in $\cD$ by $H^\prime$ as $\cD^{blind}$. To prove (\ref{eq:254}), we are going to prove
	\begin{equation}\label{eq:153}\cD\ket{\tilde\phi^\prime}\approx_{2^{-\eta/30}|\ket{\tilde\phi}|}\cD^{blind}\ket{\tilde\phi^\prime}\end{equation}
	To prove (\ref{eq:153}), we can apply Lemma \ref{lem:4.12n}. We only need to prove the norm of ``computing the blinded inputs shown in (\ref{eq:170})-(\ref{eq:171}) using a blinded oracle'' is small. In other words, we want to prove
	\begin{center}
		\emph{For any server-side operation $\cD^{blind-by-time-t}$ which only queries $H^\prime$ and the number of queries is $\leq 2^{\eta/36}$, }\end{center}
	\begin{equation}\label{eq:173r}
		|P_{BI}\cD^{blind-by-time-t}\ket{\tilde\phi^\prime}|\leq 2^{-\eta/15}|\ket{\tilde\phi}|
	\end{equation}
	Denote $\ket{\tilde\phi^{\prime hyb}}$ as the state defined as follows: based on $\ket{\tilde\phi}$, the client provides the following messages to the read-only buffer: the structure of the messages is the same as (\ref{eq:245})(\ref{eq:168})(\ref{eq:169}), but the client replaces the terms that are encrypted under (\ref{eq:170})(\ref{eq:251})(\ref{eq:171}) by random strings of the same length. In other words, it's (where $\$$ means random strings with the same length as the corresponding terms in (\ref{eq:169}))
	\begin{align}\ket{\tilde\phi^{\prime hyb}}:=\ket{\tilde\phi} & \odot \fRobustRLT^{hyb}\odot K^{(2)}\label{eq:257}           \\
		                                                & \odot perm^\prime\odot K_{out}^{fake-0-(3)}\odot K_{out}^{fake-1-(3)}\label{eq:175}                 \\
		                                                & \odot Tag(K_{out}^{(2)})\odot\begin{cases}  Tag(y_0^{fake-0-(2)})\odot \$\odot \$\odot Tag(y_1^{fake-1-(2)})&(c=0)\\\$\odot Tag(y_1^{fake-0-(2)})\odot  Tag(y_0^{fake-1-(2)})\odot \$&(c=1)\end{cases}\label{eq:281rr}
	\end{align}
	where in $\fRobustRLT^{hyb}$, the remaining rows that are \textbf{not} replaced by random strings are the rows that encode the following map:
	{\footnotesize\begin{equation}\label{eq:177}
		\text{(Given $x_0^{\text{help}}$)}:\forall b\in \{0,1\},x_b^{(2)}x_{c}^{(3)}\leftrightarrow perm(y_b^{(2)}||y_c^{(3)})=perm^\prime(y_b^{fake-(b+c)-(2)}||y_c^{fake-(b+c)-(3)})\end{equation}}
	so there are two rows in the forward table and two rows in the backward table. (In (\ref{eq:177}) $b$ varies but $c$ is fixed.) (The equality is from the definition of Notation \ref{nota:f1}.)\par
	And we can observe that, since 
	\begin{enumerate}\item All the oracle queries in $\ket{\tilde\phi}$ have been replaced by queries to the blinded oracle;
	\item In the definition of $\ket{\tilde\phi^{\prime hyb}}$ the auxiliary information does not contain anything about the the random oracle outputs in (\ref{eq:170})-(\ref{eq:171}), except a very small probability that the keys in $K_{out}^{(2)}$, $K_{out}^{fake-0-(2)}$, $K_{out}^{fake-1-(2)}$ have repetitions: this will occupy at most $2^{-\eta}|\ket{\varphi}|$ norm.
	\item All the oracle queries in $\cD^{blind-by-time-t}$ have also been blinded. 
	\end{enumerate}
Thus all the entries encrypted under (\ref{eq:170})-(\ref{eq:171}) look (almost) the same as random strings (if the adversary later does not query $H$, and only queries $H^\prime$.) 
	Thus to prove (\ref{eq:173r}), we only need to prove
	{\small\begin{equation}\label{eq:174r}
		\text{For any $\cD^{blind-by-time-t}$ which only queries $H^\prime$, }|P_{BI}\cD^{blind-by-time-t}\ket{\tilde\phi^{\prime hyb}}|\leq 2^{-\eta/14}|\ket{\tilde\phi}|
	\end{equation}}
	Now we will discuss by different elements in $BI$.\par
	\textbf{Part II.2.2: $BI$ is hard to compute from $\ket{\tilde\phi^{\prime hyb}}$ and the blinded oracle}\par
	The proof of (\ref{eq:174r}) is mostly by existing conditions or some combinatorial arguments. Before we discuss by cases, let's first understand the structure of the auxiliary information in $\ket{\tilde\phi^{\prime hyb}}$.\par
	From (\ref{eq:257})-(\ref{eq:177}) we can see what the adversary in (\ref{eq:174r}) gets is only (or more formally, (\ref{eq:257})(\ref{eq:175})(\ref{eq:281rr}) can be computed on the server side from the followings)
	\begin{equation}\label{eq:264}\text{(Server side of) }\ket{\tilde\phi},x_0^{\text{help}},x_{c}^{(3)},\end{equation}
	\begin{equation}\label{eq:265rr} K^{(2)},perm^\prime, K_{out}^{(2)}\end{equation}
	\begin{equation}\label{eq:266}  K_{out}^{fake-0-(3)}, K_{out}^{fake-1-(3)}\end{equation}
	\begin{equation}\label{eq:284r}
		y_0^{fake-c-(2)},y_1^{fake-(1-c)-(2)}
	\end{equation}
	Note that $\forall b\in \{0,1\},perm(y_b^{(2)}||y_c^{(3)})=perm^\prime(y_b^{fake-(b+c)-(2)}||y_c^{fake-(b+c)-(3)})$, which can be computed from (\ref{eq:266})(\ref{eq:284r}).\par
	Discuss by different classes of inputs in $BI$.
	\begin{itemize}
		\item 	The unpredictability of $x_1^{\text{help}}$ in (\ref{eq:174r}) comes from the condition: since (1)we know $\ket{\tilde\phi}$ is $(2^{\eta/2-2},2^{-\eta/2+2}|\ket{\tilde\phi}|)$-unpredictable for $x_1^{\text{help}}$; (2)(\ref{eq:265rr})(\ref{eq:266})(\ref{eq:284r}) are all freshly sampled randomly according to some distribution, we know
		      \begin{equation}
			      |P_{x_1^{\text{help}}}\cD^{blind-by-time-t}\ket{\tilde\phi^{\prime hyb}}|\leq 2^{-\eta/5}|\ket{\tilde\phi}|
		      \end{equation}
		      (some details: we do not mean (\ref{eq:265rr})(\ref{eq:266})(\ref{eq:284r}) are sampled uniformly randomly. There is some restriction on their distribution, for example, $K_{out}^{fake-0-(3)}$ and $K_{out}^{fake-1-(3)}$ are highly correlated. But here since we only care about the unpredictability of $x_1^{\text{help}}$ we can simplify (\ref{eq:265rr})(\ref{eq:266})(\ref{eq:284r}) by strengthening the power of the adversary: we can assume the adversary gets $K^{(2)}$, $K_{out}^{(2)}$, $K_{out}^{(3)}$, $perm$, $perm^\prime$, from which it can recover everything in (\ref{eq:265rr})(\ref{eq:266})(\ref{eq:284r}), thus the adversary does not become weaker. However, they can also be sampled on the server side thus the server does not get more ability to predict $x_{1}^{\text{help}}$ than what it can do from (\ref{eq:264}).)
		\item Similarly
		      \begin{equation}
			      |P_{x_{1-c}^{(3)}}\cD^{blind-by-time-t}\ket{\tilde\phi^{\prime hyb}}|\leq 2^{-\eta/10}|\ket{\tilde\phi}|
		      \end{equation}
		\item And we can use a combinatorial argument to prove, $\forall c^\prime\in \{0,1\}$,
		      \begin{equation}
			      |P_{y_{1-c+c^\prime}^{fake-c^\prime-(2)}}\cD^{blind-by-time-t}\ket{\tilde\phi^{\prime hyb}}|\leq 2^{-\eta/11}|\ket{\tilde\phi}|
		      \end{equation}
		      or, in other words,
		      \begin{equation}\label{eq:272p}
			      |P_{y_{1}^{fake-c-(2)}}\cD^{blind-by-time-t}\ket{\tilde\phi^{\prime hyb}}|\leq 2^{-\eta/11}|\ket{\tilde\phi}|
		      \end{equation}
		      \begin{equation}\label{eq:273p}
			      |P_{y_{0}^{fake-(1-c)-(2)}}\cD^{blind-by-time-t}\ket{\tilde\phi^{\prime hyb}}|\leq 2^{-\eta/11}|\ket{\tilde\phi}|
		      \end{equation}
		      Let's first write some intuition here since the subscripts and superscripts of the output keys are already complicated.\par
		      \begin{mdframed}
		      \textbf{Intuitions for understanding the subscripts and superscripts of the output keys in (\ref{eq:272p})(\ref{eq:273p}) and (\ref{eq:284r})}\par
		       First note that the keys we consider in (\ref{eq:272p})(\ref{eq:273p}) and (\ref{eq:284r}) form a division of $K_{out}^{fake-0-(2)}$ and $K_{out}^{fake-1-(2)}$. To understand them clearly, first recall how the fake keys are computed: first $K_{out}^{(2)}$, $K_{out}^{(3)}$, $perm$, $perm^\prime$ are sampled, then the fake keys can be computed from them. As what we did in the ``unpredictability of $x_1^{\text{help}}$'' part, we give the adversary some extra information and prove that the adversary still cannot achieve the tasks in (\ref{eq:272p})(\ref{eq:273p}). This can simplify our thinking process.\par
		      The ``extra information'' the client will reveal to the server is $perm$. And $perm^\prime$ is already in the auxiliary information. (This is slightly different from the ``unpredictability of $x_1^{\text{help}}$'' part, where we also reveal all the output keys; we could not reveal so much here.) We will analyze what the server can learn from (\ref{eq:266})(\ref{eq:284r}) and $perm,perm^\prime$ by taking the bit-wise viewpoint on $K_{out}^{(w)}$, $w\in \{2,3\}$.\par
		      ${perm^\prime}^{-1}\circ perm:[2\kappa_{out}]\rightarrow [2\kappa_{out}]$ form a mapping from the bits of output keys to the bits of fake keys. Naturally, use ``$(2)$'' to denote $[\kappa_{out}]$ and use ``$(3)$'' to denote $[\kappa_{out}+1,2\kappa_{out}]$. Define $Bit_{(w)\rightarrow (w^\prime)}$ as the set of bit indexes that is an element of ``$(w)$'', but is mapped to ``$(w^\prime)$'' by ${perm^\prime}^{-1}\circ perm$. (Thus $$Bit_{(w)\rightarrow (2)}\cup Bit_{(w)\rightarrow (3)}=\begin{cases}[\kappa_{out}]&(w=2)\\ [\kappa_{out}+1,2\kappa_{out}]&(w=3)\end{cases}$$.) Then the following is true:
		      \begin{center}
		      If the adversary knows (1)${perm^\prime}^{-1}\circ perm$, (2)the description of $K_{out}^{(2)}$, (3) the values of $K_{out}^{(3)}$ on bit indexes $Bit_{(3)\rightarrow (2)}$, (4) the values of $y_{c}^{(3)}\in K_{out}^{(3)}$ on bit indexes $Bit_{(3)\rightarrow (3)}$, it can recover everything in (\ref{eq:284r}) deterministically.
		      \end{center}
And we can see, the server does not know the bit values of $y_{1-c}^{(3)}\in K_{out}^{(3)}$ on bit indexes $Bit_{(3)\rightarrow (3)}$! In other words, even if the adversary knows (\ref{eq:265rr})(\ref{eq:266})(\ref{eq:284r}), it still does not know the bit values of $y_{1-c}^{(3)}\in K_{out}^{(3)}$ on bit indexes $Bit_{(3)\rightarrow (3)}$, which will imply the unpredictability of $y_1^{fake-c-(2)}$ and $y_0^{fake-(1-c)-(2)}$. (Some more details: notice that \begin{itemize}\item The bit values of $y_{1}^{fake-c-(2)}$ correspond to the bit values of $y_{1}^{(2)}\in K_{out}^{(2)}$ on bit indexes $Bit_{(2)\rightarrow (3)}$ and the bit values of $y_{1-c}^{(3)}\in K_{out}^{(3)}$ on bit indexes $Bit_{(3)\rightarrow (3)}$; \item The bit values of $y_{0}^{fake-(1-c)-(2)}$ correspond to the bit values of $y_{0}^{(2)}\in K_{out}^{(2)}$ on bit indexes $Bit_{(2)\rightarrow (3)}$ and the bit values of $y_{1-c}^{(3)}\in K_{out}^{(3)}$ on bit indexes $Bit_{(3)\rightarrow (3)}$;\end{itemize} where for the $K_{out}^{(3)}$ part both use the bits of $y_{1-c}^{(3)}$.)
		      \end{mdframed}
Now we complete the intuition part and can return to the formal proof. Note that $perm,perm^\prime$ are sampled independently randomly, define a set $Bit_{(3)\rightarrow (2)}$ to be the set of index in $[\kappa_{out}+1,2\kappa_{out}]$ that ${perm^\prime}^{-1}\circ perm$ maps it to $[\kappa_{out}]$, by Fact \ref{fact:f133} with probability $>(1-2^{-\eta/10})$, $|Bit_{(3)\rightarrow (2)}|>\frac{1}{10}\kappa_{out}$. Thus computing $y_{1-c+c^\prime}^{fake-c^\prime-(2)}$ is at least as hard as predicting a random choice on bits with indexes in $Bit_{(3)\rightarrow (2)}$. (With the exception of one string.) Thus 
		      \begin{equation}|P_{y_{1}^{fake-c-(3)}}\cD^{blind-by-time-t}\ket{\tilde\phi^{\prime hyb}}|\leq 2^{-\eta/11}|\ket{\tilde\phi}|
		      \end{equation}
		      \begin{equation}|P_{y_{0}^{fake-(1-c)-(3)}}\cD^{blind-by-time-t}\ket{\tilde\phi^{\prime hyb}}|\leq 2^{-\eta/11}|\ket{\tilde\phi}|
		      \end{equation}
	\end{itemize}
	Thus (\ref{eq:174r}) is true.\par
	Thus we complete the proof of (\ref{eq:173r}). Thus we complete the proof of (\ref{eq:153}). Apply (\ref{eq:173r}) again on $t=|\cD|$, together with (\ref{eq:153}), we complete the proof of (\ref{eq:254}).\par
	Thus we complete the whole proof.

\end{proof}
\subsection{Security effect of $\fRobustRLT$ on CNOT-style Branch (When the Adversary Knows $x_1^{\text{help}}$, and $x_0^{\text{help}}$ is Unpredictable)}\label{sec:ag4}
The following lemma studies the case where the state is unpredictable for $x_0^{\text{help}}$ but the adversary can know $x_1^{\text{help}}$. This corresponds to the CNOT-style branch ($b_1=1$).\par
The lemmas in this subsection will be crucial for the proof of Lemma \ref{lem:7.6} for the $w=2$ case.
\subsubsection{Statement}
\begin{lem}\label{lem:e11}
	The following lemma is true when $\eta$ below is bigger than some constant:\par
	Suppose the input key sets are denoted as $\{K^{\text{help}},K^{(3)}\}$, $K^{(w)}=\{x_0^{(w)},x_1^{(w)}\}(w\in \{1,3\})$ are pairs of different keys, the input state described by the purified joint state $\ket{\varphi}$, some bit $c\in \{0,1\}$ satisfy:
	\begin{itemize}
		\item (Security of the input) $\ket{\varphi}$ is $(2^\eta,2^{-\eta}|\ket{\varphi}|)$-unpredictable for $x^{\text{help}}_0$ given $x_1^{\text{help}}$ and $K^{(3)}$.
		\item (Security of the input) $\ket{\varphi}$ is $(2^\eta,C|\ket{\varphi}|)$-unpredictable for $x_{1-c}^{(3)}$ given $x_c^{(3)}$ and $K^{\text{help}}$. $\frac{1}{3}>C>2^{-\sqrt{\eta}}$.
		\item (Well-behaveness of the input) $\ket{\varphi}\in \cWBS(D)$, $D\leq 2^\eta$.
		\item (Sufficient padding length, output length) $l>6D+4\eta$, $\kappa_{out}>l+\eta$
	\end{itemize}
	The client samples 
	
	\begin{itemize} \item $K_{out}=\{K_{out}^{(2)},K_{out}^{(3)}\}$ where each one is a pair of different keys with key length $\kappa_{out}$; \item  $K^{(2)}$, which is a pair of different keys with key length the same as $K^{(3)}$;\item $perm$ is sampled on the bit-wise permutations of strings of length $2\kappa_{out}$;\item $perm^\prime$ is sampled as the \emph{fake permutation} independently randomly on the bit-wise permutations of strings of length $2\kappa_{out}$\end{itemize}
	 then (meanings of notations are given below)
	\begin{align}\ket{\varphi}                                             & \odot \fRobustRLT(K^{\text{help}},K^{(2,3)}\leftrightarrow K_{out},perm;\underbrace{ \ell}_{\substack{\text{padding} \\ \text{length}}})\odot K^{(2)}\label{eq:263}                                               \\
		                                                          & \odot perm\odot K_{out}^{(3)}\odot Tag(K^{(2)}_{out})\odot Tag(K^{fake-0-(2)}_{out})\odot Tag(K^{fake-1-(2)}_{out})\label{eq:102} \\
		\approx^{\fAdv\in \cA}_{3C|\ket{\varphi}|}  \ket{\varphi} & \odot \fRobustRLT^{Hyb}(K^{\text{help}},K^{(2,3)}\leftrightarrow K_{out},perm;\underbrace{ \ell}_{\substack{\text{padding} \\ \text{length}}})\odot K^{(2)}\label{eq:265}                                         \\
		                                                          & \odot perm\odot K_{out}^{(3)}\odot Tag(K^{(2)}_{out})\odot\$\odot \$\label{eq:103}
	\end{align}
	{\small\begin{align}
		\ket{\varphi}                                             & \odot \fRobustRLT(K^{\text{help}},K^{(2,3)}\leftrightarrow K_{out},perm;\underbrace{ \ell}_{\substack{\text{padding} \\ \text{length}}})\odot K^{(2)}\label{eq:267}                                                             \\
		                                                          & \odot perm^\prime\odot K_{out}^{fake-c-(3)}\odot Tag(K^{(2)}_{out})\odot Tag(K^{fake-0-(2)}_{out})\odot Tag(K^{fake-1-(2)}_{out})\label{eq:143} \\
		\approx^{\fAdv\in \cA}_{3C|\ket{\varphi}|}  \ket{\varphi} & \odot \fRobustRLT^{Hyb}(K^{\text{help}},K^{(2,3)}\leftrightarrow K_{out},perm;\underbrace{ \ell}_{\substack{\text{padding} \\ \text{length}}})\odot K^{(2)}\label{eq:269}                                                       \\
		 & \odot perm^\prime\odot K_{out}^{fake-c-(3)}\odot \begin{cases}\$\odot Tag(K^{fake-0-(2)}_{out})\odot \$&(c=0)\\\$\odot \$\odot Tag(K^{fake-1-(2)}_{out})&(c=1)\end{cases}\label{eq:144}
	\end{align}}
	where 
	\begin{itemize}\item $K^{fake-c-(w)}_{out}$ are defined as in Notation \ref{nota:f1}.
	\item $\$$ denotes random strings with the same length as the corresponding terms. (Compare (\ref{eq:144}) with (\ref{eq:143}), and (\ref{eq:103}) with (\ref{eq:102}).)
	\item $\fRobustRLT^{Hyb}$ is defined as follows. Compared to $\fRobustRLT$, replace all the rows \textbf{other than} the rows $$\text{(Under $x_1^{\text{help}}$)}:x_b^{(2)}||x_c^{(3)}\leftrightarrow perm(y_b^{(2)}||y_{b+c}^{(3)}),b\in \{0,1\}$$ with random strings of the same length. Thus only two rows in the forward table and two rows in the backward table remain the same. ($c$ is fixed, and ``two'' comes from different $b\in \{0,1\}$.)\end{itemize}
	And the notation $\approx^{\fAdv\in \cA}_\epsilon$ is defined in Notation \ref{nota:af2}, where $\cA$ represents the set of adversaries whose attack can be divided into two phases, as follows:
	\begin{enumerate}\item In the first phase, it only operates on the server and read-only buffer part of $\ket{\varphi}$, lookup tables (either $\fRobustRLT$ or $\fRobustRLT^{Hyb}$), and $K^{(2)}$. (In other words, it only operates on (\ref{eq:263})(\ref{eq:265})(\ref{eq:267})(\ref{eq:269}).) And it makes at most $2^{\eta/40}$ queries to $H$ in this phase. \item In the second phase, it operates on all the systems stored in the server and buffer system, including the last five terms shown in the expressions above (we mean (\ref{eq:102})(\ref{eq:103})(\ref{eq:143})(\ref{eq:144})). However, it only queries $H^\prime$ which is a new blinded oracle of $H$ where $\cdots||K^{\text{help}}||\cdots$ part are blinded. The prefix padding has length $l$ and the suffix padding is arbitrary. The adversary makes at most $2^{\eta/40}$ queries to $H^\prime$ in this phase.
	\end{enumerate}
\end{lem}
\subsubsection{Proof of (\ref{eq:263})(\ref{eq:102})(\ref{eq:265})(\ref{eq:103})}
Let's first prove (\ref{eq:263})(\ref{eq:102})(\ref{eq:265})(\ref{eq:103}) and the other one is similar.\par
\begin{proof}[Proof of (\ref{eq:263})(\ref{eq:102})(\ref{eq:265})(\ref{eq:103})]
	Similar to the proof of Lemma \ref{lem:lrr2.2}, we make the proof more readable by dividing it into several phases.\par
	\textbf{Part I: preparation before the hybrid method}\par
	First by Technique \ref{lem:4.2} we can assume $x_1^{\text{help}}$, $x_c^{(3)}$, $Tag(x_0^{\text{help}})$, $Tag(x_{1-c}^{(3)})$ are given to the adversary (stored in some fixed place of the read-only buffer), and this is reasonable since it will only make the adversary more powerful, thus prove the statement under this assumption implies the original statement.\par
	Decompose $\ket{\varphi}$ as $\ket{\phi}+\ket{\chi}$ by applying Lemma \ref{lem:4.7r} (together with Fact \ref{fact:injtag}) for $x_{1-c}^{(3)}$. Then similar to the arguments in the proof of Lemma \ref{lem:lrr2.2}, $|\ket{\chi}|\leq 2.5C|\ket{\varphi}|$ and $\ket{\phi}$ satisfies:
	\begin{itemize}
		\item$ \frac{1}{6}|\ket{\varphi}|\leq |\ket{\phi}|\leq |\ket{\varphi}|$
		\item $\ket{\phi}$ is $(1,2^{\eta/2})$-server-side-representable from $\ket{\varphi}$.
		\item $\ket{\phi}$ is $(2^{\eta/2},2^{-\eta/2+3}|\ket{\phi}|)$-unpredictable for $x_0^{\text{help}}$ given $K^{(3)}$.
		\item $\ket{\phi}$ is $(2^{\eta/12-1},2^{-\eta/12+3}|\ket{\phi}|)$-unpredictable for $x_{1-c}^{(3)}$.
		\item $\ket{\phi}$ is $(2^{D},2^{D}+2^{\eta-6})$-representable from $\ket{\mathfrak{init}}$.
	\end{itemize}
	Then proving (\ref{eq:102})(\ref{eq:103}) is reduced to prove
	\begin{align}
		\ket{\phi}                                                    & \odot \fRobustRLT(K^{\text{help}},K^{(2,3)}\leftrightarrow K_{out},perm;\ell)\odot K^{(2)}\label{eq:281r}                                              \\
		                                                              & \odot perm\odot K_{out}^{(3)}\odot Tag(K^{(2)}_{out})\odot Tag(K^{fake-0-(2)}_{out})\odot Tag(K^{fake-1-(2)}_{out})\label{eq:161} \\
		\approx^{\fAdv\in \cA}_{2^{-\eta/40}|\ket{\phi}|}  \ket{\phi} & \odot \fRobustRLT^{Hyb}(K^{\text{help}},K^{(2,3)}\leftrightarrow K_{out},perm;\ell)\odot K^{(2)}\label{eq:283r}                                        \\
		                                                              & \odot perm\odot K_{out}^{(3)}\odot Tag(K^{(2)}_{out})\odot \$\odot \$\label{eq:162}
	\end{align}
	Once we can prove this, using the fact that $\ket{\phi}\approx_{2.5C|\ket{\varphi}|}\ket{\varphi}$ completes the proof.\par
	\textbf{Part II: hybrid method}\par
	Define $Set$ as the set of random pads used in the computation of $\fRobustRLT$.\par
	Consider a blinded oracle $ \tilde H$ (we use this notation to distinguish it from $H^\prime$ defined in the statement) of $H$ where
	\begin{equation}\label{eq:149}Set||x_0^{\text{help}}||\cdots,Set||x_1^{\text{help}} ||\cdots||x_{1-c}^{(3)}, \end{equation}
	\begin{equation}\label{eq:150} Set||x_1^{\text{help}} ||perm(y_0^{(2)}||y_{1-c}^{(3)}), Set||x_1^{\text{help}} ||perm(y_1^{(2)}||y_{c}^{(3)}),\end{equation}
	\begin{equation}\label{eq:151}Tag(K_{out}^{fake-0-(2)}),Tag(K_{out}^{fake-1-(2)})\end{equation}
	are all blinded. The first ``$\cdots$'' means all the possible strings and the second ``$\cdots$'' in (\ref{eq:149}) means all the possible strings of length equal to the key length in $K^{(2)}$.\par
	Define $BI$ as the set of inputs in (\ref{eq:149})(\ref{eq:150})(\ref{eq:151}).\par
	We will replace the oracle queries in (\ref{eq:281r})(\ref{eq:161})(\ref{eq:283r})(\ref{eq:162}) by $\tilde H$ one by one.
	\begin{enumerate}
		\item Recall that $\ket{\phi}$ is $(2^{D},2^{D}+2^{\eta-6})$-representable from $\ket{\mathfrak{init}}$. Expand $\ket{\phi}$ using the definition of the representability (Definition \ref{def:rep}) and replace the oracle queries to $H$ by queries to $\tilde H$, and denote the resulting state as $\ket{\tilde\phi}$. By Lemma \ref{lem:3.4r} $$\ket{\tilde\phi}\approx_{2^{-\eta}|\ket{\phi}|}\ket{\phi}$$
		      Thus proving (\ref{eq:281r})(\ref{eq:161})(\ref{eq:283r})(\ref{eq:162}) is further reduced to prove
		      {\small\begin{align}\ket{\tilde\phi}                                                          & \odot \fRobustRLT(K^{\text{help}},K^{(2,3)}\leftrightarrow K_{out},perm;\ell)\odot K^{(2)}\label{eq:278}                                               \\
			                                                                                & \odot perm\odot K_{out}^{(3)}\odot Tag(K^{(2)}_{out})\odot Tag(K^{fake-0-(2)}_{out})\odot Tag(K^{fake-1-(2)}_{out})\label{eq:279} \\
			      \approx^{\fAdv\in \cA}_{2^{-\eta/39}|\ket{\tilde\phi}|}  \ket{\tilde\phi} & \odot \fRobustRLT^{Hyb}(K^{\text{help}},K^{(2,3)}\leftrightarrow K_{out},perm;\ell)\odot K^{(2)}\label{eq:280}                                         \\
			                                                                                & \odot perm\odot K_{out}^{(3)}\odot Tag(K^{(2)}_{out})\odot \$\odot \$\label{eq:281}
		      \end{align}}


		\item For $\fAdv\in \cA$, recall that $\fAdv$ can be divided into $\fAdv_1$ and $\fAdv_2$. In this step we will only consider the $\fAdv_1$ part.\par
		      Define $\tilde\fAdv_1$ as the adversary that replaces all the queries to $H$ by queries to $\tilde H$. Our goal is to make use of Lemma \ref{lem:4.12n} to prove
		      \begin{equation}\label{eq:166r}
			      \fAdv_1 \ket{\text{ equation (\ref{eq:278})(\ref{eq:279}) }}\approx_{2^{-\eta/37}|\ket{\tilde\phi}|}\tilde\fAdv_1 \ket{\text{ equation (\ref{eq:278})(\ref{eq:279}) }}
		      \end{equation}
		      Recall that $BI$ as the set of inputs in (\ref{eq:149})(\ref{eq:150})(\ref{eq:151}). Thus by Lemma \ref{lem:4.12n} we only need to prove
		      \begin{equation}\label{eq:315}|P_{\text{BI}}{\tilde\fAdv_1}^t\ket{\text{ equation (\ref{eq:278})(\ref{eq:279}) }}|\leq 2^{-\eta/18}|\ket{\tilde\phi}|\end{equation}
		      where ${\tilde\fAdv_1}^t$ is defined to be the operation in $\tilde\fAdv_1$ before the time of making the $t$-th queries.\par

		      Since (1) $\ket{\tilde\phi}$ does not depend on the RO outputs of $H$ on (\ref{eq:149})(\ref{eq:150})(\ref{eq:151}); (2) all the queries by time $t$ to (\ref{eq:149})(\ref{eq:150})(\ref{eq:151}) of $H$ have already been blinded, we know
		     {\footnotesize \begin{equation}\label{eq:316}|P_{\text{BI}}{\tilde\fAdv_1}^t\ket{\text{ equation (\ref{eq:278})(\ref{eq:279}) }}|\approx_{2^{-\eta/5}|\ket{\tilde\phi}|}|P_{\text{BI}}{\tilde\fAdv_1}^t\ket{\text{ equation (\ref{eq:280})(\ref{eq:281}) }}|\end{equation}}
		      and we need to prove \begin{equation}\label{eq:317}\text{the right side of (\ref{eq:316}) is }\leq 2^{-\eta/15}|\ket{\tilde\phi}|\end{equation}
		      To understand ${\tilde\fAdv_1}^t\ket{\text{ equation (\ref{eq:280})(\ref{eq:281}) }}$ more clearly, recall that the adversary in this phase only operates on the information in (\ref{eq:280}), and what the adversary gets from it can be computed from the followings and random coins:
		      \begin{equation}
			      \text{(Server side of) }\ket{\tilde\phi},x_1^{\text{help}},x_c^{(3)},
		      \end{equation}
		      \begin{equation}\label{eq:303}
			      K^{(2)},
		      \end{equation}
		      \begin{equation}\label{eq:192}perm(y_0^{(2)}||y_c^{(3)}), perm(y_1^{(2)}||y_{1+c}^{(3)})\end{equation}
		      Then the entries in $BI$ are unpredictable if the adversary only queries $\tilde H$, for the following reasons:
		      \begin{itemize}
			      \item We know $\ket{\tilde\phi}$ is $(2^{\eta/2},2^{-\eta/2+6}|\ket{\tilde\phi}|)$-unpredictable for $x_0^{\text{help}}$ and \\$(2^{\eta/12-1},2^{-\eta/12+6}|\ket{\tilde\phi}|)$-unpredictable for $x_{1-c}^{(3)}$. And the output keys, permutation and $K^{(2)}$ in (\ref{eq:303})(\ref{eq:192}) are all sampled randomly (thus can be simulated on the server side), we know
			            \begin{equation}
				            |P_{x_0^{\text{help}}}{\tilde\fAdv_1}^t\ket{\text{ equation (\ref{eq:280})(\ref{eq:281}) }}|\leq 2^{-\eta/2+10}|\ket{\tilde\phi}|
			            \end{equation}
			            \begin{equation}
				            |P_{x_{1-c}^{(3)}}{\tilde\fAdv_1}^t\ket{\text{ equation (\ref{eq:280})(\ref{eq:281}) }}|\leq 2^{-\eta/12+10}|\ket{\tilde\phi}|
			            \end{equation}
			      \item We can also prove
			            \begin{equation}
				            |P_{perm(y_0^{(2)}||y_{1+c}^{(3)})}{\tilde\fAdv_1}^t\ket{\text{ equation (\ref{eq:280})(\ref{eq:281}) }}|\leq 2^{-\eta/14}|\ket{\tilde\phi}|
			            \end{equation} 
			            This is from a combinatoric argument. Applying Fact \ref{fact:f1332} completes the proof. (Intuitively, we can understand it as follows: Because (\ref{eq:192}) looks (almost) the same as two independently sampled strings. Thus for the adversary, computing $perm(y_0^{(2)}||y_{1-c}^{(3)})$ is as difficult as the following task: the client samples a random subset of size $\kappa_{out}$ from $[2\kappa_{out}]$, and swap the corresponding bits in (\ref{eq:192}); the adversary predicting the results. Intuitively the success probability is very small.)\par
			            By similar reason
			            \begin{equation}
				            |P_{perm(y_1^{(2)}||y_c^{(3)})}{\tilde\fAdv_1}^t\ket{\text{ equation (\ref{eq:280})(\ref{eq:281}) }}|\leq 2^{-\eta/14}|\ket{\tilde\phi}|
			            \end{equation}
			            \item For the unpredictability of the \emph{fake keys}, by Fact \ref{fact:ac7} we know
			            \begin{equation*}
				            \forall b\in \{0,1\}, c^\prime\in \{0,1\},\end{equation*}\begin{equation}|P_{y_b^{fake-c^\prime-(2)}}{\tilde\fAdv_1}^t\ket{\text{ equation (\ref{eq:280})(\ref{eq:281}) }}|\leq 2^{-\eta/14}|\ket{\tilde\phi}|
			            \end{equation}
		      \end{itemize}
		      Thus we complete the proof of (\ref{eq:317}). Thus (\ref{eq:315}) is true. Thus (\ref{eq:166r}) is true.
		\item For the $\fAdv_2$ part: \par
		      Note that in the queries in the $\fAdv_2$ part, the oracle queries are done on $H^\prime$, where the queries to $\cdots||K^{\text{help}}||\cdots$ have already been blinded. What we are going to do in this step is to replace the queries to $H^\prime$ by $\tilde H^\prime$, defined as follows:\par
		      For each query on some input:
		      \begin{enumerate}
			      \item If the input falls into the blinded inputs of $H^\prime$, return the values in $H^\prime$.
			      \item If not, and if the input falls into the blinded inputs of $\tilde H$, return the output values in $\tilde H$.
			      \item Otherwise, return the output from $H$.
		      \end{enumerate}
		      Note that $\tilde H^\prime$ can be understood as the result of blinding $\cdots||K^{\text{help}}||\cdots$ part on $\tilde H$.\par
		      Denote the blinded version (where all the queries have been replaced by queries to $\tilde H^\prime$) of the adversary in this phase as $\tilde \fAdv_2$, our goal is to prove:
		      \begin{align}\label{eq:326}&\fAdv_2\tilde\fAdv_1\ket{\text{ equation (\ref{eq:278})(\ref{eq:279}) }}\\
		      \approx_{2^{-\eta/37}|\ket{\tilde\phi}|}&\tilde\fAdv_2\tilde\fAdv_1\ket{\text{ equation (\ref{eq:278})(\ref{eq:279}) }}
		      \end{align}
		      Denote $\tilde\fAdv^t_2$ as the operation in $\tilde\fAdv_2$ by the time just before making the $t$-th query. Denote the blinded entries in (\ref{eq:151}) as $BI_{sub}$. Then this is reduced to prove:
		      \begin{equation}\label{eq:327}|P_{BI_{sub}}\tilde\fAdv_2^t\tilde\fAdv_1\ket{\text{ equation (\ref{eq:278})(\ref{eq:279}) }}|\leq 2^{-\eta/18}|\ket{\tilde\phi}|\end{equation}
		      (Note that this is how the ``blinded part'' helps: the entries corresponding to (\ref{eq:149})(\ref{eq:150}) have already been blinded and the only difference of $H^\prime$ and $\tilde H^\prime$ is on the entries shown in (\ref{eq:151}).)\par
		      Since (1) $\ket{\tilde\phi}$ does not depend on the RO outputs of $H$ on (\ref{eq:149})(\ref{eq:150})(\ref{eq:151}); (2) all the queries by time $t$ to (\ref{eq:149})(\ref{eq:150})(\ref{eq:151}) of $H$ have already been blinded, we know
		      \begin{align}\label{eq:328}&|P_{\text{$BI_{sub}$}}\tilde\fAdv_2^t\tilde\fAdv_1\ket{\text{ equation (\ref{eq:278})(\ref{eq:279}) }}|\\\approx_{2^{-\eta/5}|\ket{\tilde\phi}|}&|P_{\text{$BI_{sub}$}}\tilde\fAdv_2^t\tilde\fAdv_1\ket{\text{ equation (\ref{eq:280})(\ref{eq:281}) }}|\end{align}
		      and what we are going to prove is \begin{equation}\label{eq:329}\text{The right side of (\ref{eq:328}) is $\leq 2^{-\eta/15}|\ket{\tilde\phi}|$.}\end{equation}
		      Notice that, the adversary knows in $\ket{\text{ equation (\ref{eq:280})(\ref{eq:281}) }}$ is (or more formally, what the adversary gets can be computed from the followings and random coins):
		      \begin{equation}
			      \text{(Server side of) }\ket{\tilde\phi},x_1^{\text{help}},x_c^{(3)}
		      \end{equation}
		      \begin{equation}K^{(2)},perm(y_0^{(2)}||y_c^{(3)}),perm(y_1^{(2)}||y_{1+c}^{(3)})\end{equation}
		      \begin{equation}
			      perm, K^{(3)}_{out}
		      \end{equation}
		      \begin{equation}
			      Tag(K_{out}^{(2)})
		      \end{equation}
		      which are simplified to
		      \begin{equation}
			      \text{(Server side of) }\ket{\tilde\phi}, x_1^{\text{help}}, x_c^{(3)}
		      \end{equation}
		      \begin{equation}\label{eq:319}K^{(2)},K_{out}^{(2)},K_{out}^{(3)}\end{equation}
		      \begin{equation}\label{eq:320}
			      perm
		      \end{equation}

		      Then the entries in $BI_{sub}$ are unpredictable if the adversary only queries $H^\prime$, for the following reasons:
		      \begin{itemize}
			      \item Since the adversary's state does not contain any information about $perm^\prime$, nor any fake keys, predicting the keys in $K_{out}^{fake-c-(2)}$ is as hard as winning the game in Fact \ref{fact:ac7}. Thus:
			            {\small\begin{equation}
				            \forall b,c^\prime\in \{0,1\}^2,|P_{y_b^{fake-c^\prime-(2)}}{\tilde\fAdv_2}^t{\tilde\fAdv_1}\ket{\text{ equation (\ref{eq:280})(\ref{eq:281}) }}|\leq 2^{-\eta/14}|\ket{\tilde\phi}|
			            \end{equation}}
		      \end{itemize}
		      Thus we complete the proof of (\ref{eq:329}). Thus we prove (\ref{eq:327}). Thus we prove (\ref{eq:326}).
	
	\item In summary we have
	\begin{align}
		                                         & \fAdv_2\fAdv_1\ket{\text{ equation (\ref{eq:278})(\ref{eq:279}) }}                                           \\
		\approx_{2^{-\eta/37}|\ket{\tilde\phi}|} & \fAdv_2{\tilde\fAdv_1}\ket{\text{ equation (\ref{eq:278})(\ref{eq:279}) }}                                   \\
		\approx_{2^{-\eta/37}|\ket{\tilde\phi}|} & \tilde\fAdv_2{\tilde\fAdv_1}\ket{\text{ equation (\ref{eq:278})(\ref{eq:279}) }}                             \end{align}
		By (\ref{eq:317})(\ref{eq:329}) we also get
		\begin{align}
		                                         & \fAdv_2\fAdv_1\ket{\text{ equation (\ref{eq:280})(\ref{eq:281}) }}                                           \\
		\approx_{2^{-\eta/37}|\ket{\tilde\phi}|} & \fAdv_2{\tilde\fAdv_1}\ket{\text{ equation (\ref{eq:280})(\ref{eq:281}) }}                                   \\
		\approx_{2^{-\eta/37}|\ket{\tilde\phi}|} & \tilde\fAdv_2{\tilde\fAdv_1}\ket{\text{ equation (\ref{eq:280})(\ref{eq:281}) }}                             \end{align}
		And we have
		{\small\begin{align}
		\tilde\fAdv_2{\tilde\fAdv_1}\ket{\text{ equation (\ref{eq:278})(\ref{eq:279}) }}\approx^{st-ind}_0                       & \tilde\fAdv_2{\tilde\fAdv_1}\ket{\text{ equation (\ref{eq:280})(\ref{eq:281})}} 
	\end{align}}
	thus we complete the hybrid method and complete the proof of\\ (\ref{eq:278})(\ref{eq:279})(\ref{eq:280})(\ref{eq:281}).
	\end{enumerate}
	Thus we prove (\ref{eq:281r})(\ref{eq:161})(\ref{eq:283r})(\ref{eq:162}). 
	Thus adding back $\ket{\chi}$ we get\\ (\ref{eq:102})(\ref{eq:103}).
\end{proof}
\subsubsection{Proof of (\ref{eq:267})(\ref{eq:143})(\ref{eq:269})(\ref{eq:144})}
The proof of (\ref{eq:267})(\ref{eq:143})(\ref{eq:269})(\ref{eq:144}) is similar to the proof of\\ (\ref{eq:263})(\ref{eq:102})(\ref{eq:265})(\ref{eq:103}). We will skip the steps that are the same and describe their differences.
\begin{proof}[Proof of (\ref{eq:267})(\ref{eq:143})(\ref{eq:269})(\ref{eq:144})]
	The Part I is the same as the proof of\\ (\ref{eq:263})(\ref{eq:102}) (\ref{eq:265})(\ref{eq:103}). In this step we decompose the state by applying the decomposition lemma for $x_{1-c}^{(3)}$ and reduce the original statement to prove:
	\begin{align}
		\ket{\phi}                                             & \odot \fRobustRLT(K^{\text{help}},K^{(2,3)}\leftrightarrow K_{out},perm;\ell)\odot K^{(2)}\label{eq:267b}                                                             \\
		                                                          & \odot perm^\prime\odot K_{out}^{fake-c-(3)}\odot Tag(K^{(2)}_{out})\odot Tag(K^{fake-0-(2)}_{out})\odot Tag(K^{fake-1-(2)}_{out})\label{eq:143b} \\
		\approx^{\fAdv\in \cA}_{2^{-\eta/40}|\ket{\phi}|}  \ket{\phi} & \odot \fRobustRLT^{Hyb}(K^{\text{help}},K^{(2,3)}\leftrightarrow K_{out},perm;\ell)\odot K^{(2)}\label{eq:269b}                                                       \\
		 & \odot perm^\prime\odot K_{out}^{fake-c-(3)}\odot \begin{cases}\$\odot Tag(K^{fake-0-(2)}_{out})\odot \$&(c=0)\\\$\odot \$\odot Tag(K^{fake-1-(2)}_{out})&(c=1)\end{cases}\label{eq:144b}
	\end{align}
Then in Part II the blinded oracle $\tilde H$ is defined to be the oracle where the followings are blinded:
\begin{equation}\label{eq:149b} Set ||x_0^{\text{help}}||\cdots,Set ||x_1^{\text{help}}||\cdots||x_{1-c}^{(3)}, \end{equation}
	\begin{equation}\label{eq:150b} Set ||x_1^{\text{help}}||perm(y_0^{(2)}||y_{1-c}^{(3)}),Set ||x_1^{\text{help}}||perm(y_1^{(2)}||y_{c}^{(3)}),\end{equation}
	\begin{equation}\label{eq:151b}Tag(K_{out}^{(2)}),Tag(K_{out}^{fake-(1-c)-(2)})\end{equation}
Notice that the only difference is on the third row above.\par
Now the hybrid method starts. 
\begin{enumerate}
	\item The first step is the same, and the problem is reduced to
	{\small\begin{align}
		\ket{\tilde\phi}                                             & \odot \fRobustRLT(K^{\text{help}},K^{(2,3)}\leftrightarrow K_{out},perm;\ell)\odot K^{(2)}\label{eq:267bb}                                                             \\
		                                                           \odot perm^\prime&\odot K_{out}^{fake-c-(3)}\odot Tag(K^{(2)}_{out})\odot Tag(K^{fake-0-(2)}_{out})\odot Tag(K^{fake-1-(2)}_{out})\label{eq:143bb} \\
		\approx^{\fAdv\in \cA}_{2^{-\eta/40}|\ket{\tilde\phi}|}  \ket{\tilde\phi} & \odot \fRobustRLT^{Hyb}(K^{\text{help}},K^{(2,3)}\leftrightarrow K_{out},perm;\ell)\odot K^{(2)}\label{eq:269bb}                                                       \\
		 & \odot perm^\prime\odot K_{out}^{fake-c-(3)}\odot \begin{cases}\$\odot Tag(K^{fake-0-(2)}_{out})\odot \$&(c=0)\\\$\odot \$\odot Tag(K^{fake-1-(2)}_{out})&(c=1)\end{cases}\label{eq:144bb}
	\end{align}}
	\item We replace the oracle queries in $\fAdv_1$ by queries to $\tilde H$.\par
	Similarly we can verify $x_0^{\text{help}}$, $x_{1-c}^{(3)}$ are unpredictable with the same parameters. And the same holds for $perm(y_0^{(2)}||y_{1-c}^{(3)})$, $perm(y_1^{(2)}||y_{c}^{(3)})$ and $y_b^{fake-(1-c)-(2)}$. And we need to additionally prove
	\begin{equation}
				            |P_{y_b^{(2)}}{\tilde\fAdv_1}^t\ket{\text{ equation (\ref{eq:269bb})(\ref{eq:144bb}) }}|\leq 2^{-\eta/14}|\ket{\tilde\phi}|
			            \end{equation}
			            which is true by Fact \ref{fact:ac6}.
			           \item Then we replace the oracle queries in $\fAdv_2$ by queries to $\tilde H^\prime$, which is defined as shown in the ``proof of (\ref{eq:263})(\ref{eq:102})(\ref{eq:265})(\ref{eq:103})''. We notice the auxiliary information in (\ref{eq:269bb})(\ref{eq:144bb}) can be computed from the followings and random coins:
			           \begin{equation}
			      \text{(Server side of) }\ket{\tilde\phi},x_1^{\text{help}},x_c^{(3)}
		      \end{equation}
		      \begin{equation}K^{(2)},perm(y_0^{(2)}||y_c^{(3)}),perm(y_1^{(2)}||y_{1+c}^{(3)})\end{equation}
		      \begin{equation}
			      perm^\prime
		      \end{equation}
		      And we need to (additionally) prove the unpredictability of the keys in $K_{out}^{(2)}$. Formally, we need to prove
		      		            \begin{equation}
				            \forall b\in \{0,1\},|P_{y_b^{(2)}}{\tilde\fAdv_2}^t{\tilde\fAdv_1}\ket{\text{ equation (\ref{eq:280})(\ref{eq:281}) }}|\leq 2^{-\eta/14}|\ket{\tilde\phi}|
			            \end{equation}
			            This is from Fact \ref{fact:ac6}.
			           \item By the same argument as the ``proof of (\ref{eq:263})(\ref{eq:102})(\ref{eq:265})(\ref{eq:103})'' completes the proof.
\end{enumerate}
	 
\end{proof}

\subsubsection{Corollaries of Lemma \ref{lem:e11}}
This lemma leads to the following two corollaries, which are useful in our main proof. Recall that $\tilde{tag}$ is defined in Notation \ref{nota:ag2}.
\begin{cor}\label{cor:e11c1}
	Suppose the initial state described by the purified joint state $\ket{\varphi}$, pad length $l$, output length $\kappa_{out}$, key sets $\{K^{\text{help}},K^{(3)}\}$, bit $c\in \{0,1\}$ satisfy the same conditions given in Lemma \ref{lem:e11}. The client samples $K^{(2)}$ and $K_{out}$, $perm$, $perm^\prime$ similarly. Then define
	$$\ket{\varphi^\prime}:=\ket{\varphi}\odot \fRobustRLT(K^{\text{help}},K^{(2,3)}\leftrightarrow K_{out},perm;\underbrace{ \ell}_{\substack{\text{padding} \\ \text{length}}})\odot K^{(2)}\odot perm^\prime\odot K_{out}^{fake-c-(3)}\odot \tilde{tag}$$
	when the adversary $\fAdv\in \cA$ is as defined in Lemma \ref{lem:e11}, for any key $k\in K_{out}^{fake-(1-c)-(2)}$,  $|P_{k}\fAdv\ket{\varphi^\prime}|\leq 4C|\ket{\varphi}|$.
\end{cor}
\begin{proof}
	Otherwise 	(\ref{eq:143})(\ref{eq:144}) will become distinguishable since the adversary can  compute the keys in $K_{out}^{fake-(1-c)-(2)}$ and check them with the $Tag(K^{fake-(1-c)-(2)}_{out})$.
\end{proof}

\begin{cor}\label{cor:e11c2}
	Suppose the initial state described by the purified joint state $\ket{\varphi}$, pad length $\ell$, key sets $K,K_{out}$ etc satisfy the same conditions given in Lemma \ref{lem:e11}. $\fAdv\in \cA$ as defined in Lemma \ref{lem:e11}. Then
	{\small\begin{align}
		                                           & \ket{\varphi}\odot \fRobustRLT(K^{\text{help}},K^{(2,3)}\leftrightarrow K_{out},perm;\underbrace{ \ell}_{\substack{\text{padding} \\ \text{length}}})\odot K^{(2)}\odot perm\odot K_{out}^{(3)}\odot \tilde{tag}               \label{eq:348v}\\
		\approx^{\fAdv\in \cA}_{6C|\ket{\varphi}|} & \ket{\varphi}\odot \fRobustRLT(K^{\text{help}},K^{(2,3)}\leftrightarrow K_{out},perm;\underbrace{ \ell}_{\substack{\text{padding} \\ \text{length}}})\\&\odot K^{(2)}\odot perm^\prime\odot K_{out}^{fake-c-(3)}\odot \tilde{tag}\label{eq:349v}
	\end{align}}

\end{cor}

\begin{proof}
	Applying Lemma \ref{lem:e11} for both sides, and replace the lookup table with $\fRobustRLT^{hyb}$ and replace $\tilde{tag}$ with (correspondingly) $\mathsf{Shuffle}(K_{out}^{(2)},\$,\$)$ and\\ $\mathsf{Shuffle}(K_{out}^{fake-c-(2)},\$,\$)$, where $\mathsf{Shuffle}$ is a random shuffling on three elements, $\$$ is a random string. This replacement leads to (in total) $6C|\ket{\varphi}|$ error and finally two sides become perfectly indistinguishable.
\end{proof}
Note that if the random shufflings in $\tilde{tag}$ in (\ref{eq:348v})(\ref{eq:349v}) are removed, this statement is not true. And this will be important when we use this corollary.
\section{Proof of the $w=2$ case of Lemma \ref{lem:7.6}}\label{sec:ac2}
Now we give the proof for Lemma \ref{lem:7.6} for the $w=2$ case.\par
 The structure of this proof is as follows: the main body is divided into four steps, step 0 to step 3. During the proof we raise a lemma (Lemma \ref{lem:e1}) and we put the proof of this lemma in the end of this proof, as ``step 4''. Thus there are five steps in total.\par
  To prove this lemma, we need to use the lemmas in the previous section.\par
  The proof is given below.
\begin{proof}
	\textbf{Step 0: Preparation}\par
	Similar to the proof of the $w=3$ case, we assume
	\begin{equation}\label{eq:110n}|P_{pass}\ket{\varphi^\prime}|\geq (1-C^2)|\ket{\varphi}|\end{equation}
	, otherwise the statement is already true.\par

	Suppose there exists a server-side operation $\cU$ with query number $|\cU|\leq 2^{\eta/B}$ such that
	\begin{equation}\label{eq:55}|P_{y_0^{(2)}||y_1^{(2)}}\cU(\ket{\varphi^\prime}\odot K_{out}^{(3)}\odot Tag(K_{out}^{(2)}))|/|\ket{\varphi}|=q\end{equation}
	where the projection is applied on some server-side system, $B$ is a constant chosen to be big enough to make all the arguments below work. (Explicitly, $B=100000$ is enough, but it can be much smaller. We're a little bit lazy here and won't try to get the minimum-possible $B$. What's more, the lemmas later also contain some constants so writing it in this form will make the form of the statement consistent.)\par
	And our goal is to prove $q\leq AC$ for some constant $A$.\par
	Again, we need to make use of Lemma \ref{lem:7.4}. Before that, we first choose a random permutation $perm^\prime$ on the bit-wise permutation on strings of length $2\kappa_{out}$, and define the \emph{fake keys} $K_{out}^{fake-0-(2)}$, $K_{out}^{fake-0-(3)}$ and $K_{out}^{fake-1-(2)}$,$K_{out}^{fake-1-(3)}$, which have the same length as $K_{out}^{(2)}$ and $K_{out}^{(3)}$, as discussed in Notation \ref{nota:f1}. 
	Recall that, intuitively, if the server tries to de-permutation from $perm(y_b^{(2)}||y_{c+b}^{(3)})$ using a \emph{fake permutation} $perm^\prime$, it gets the fake keys $K_{out}^{fake-c-(w)}$.\par
	(Let's add a note on the meaning of $b,c,w$ here. $w\in \{2,3\}$ denote the index of wires. $b$ denotes the bit-value in the second input wire, and $c$ denotes the bit-value in the third input wire. Note that in the lookup table, given $x_1^{\text{help}}$, $x^{(2)}_b||x^{(3)}_c$ is mapped to $perm(y_b^{(2)}||y_{c+b}^{(3)})$.)\par
	Then we will replace the global tags in (\ref{eq:55}) by $\tilde{tag}$. The motivation of this replacement is not clear now, but it's important for later proof, since it allows us to apply Corollary \ref{cor:e11c2}. Then there exists a server-side operation $\cU^\prime$ with query number $|\cU^\prime|\leq |\cU|+O(1)$ such that:
	\begin{equation}\label{eq:56}|P_{y_0^{(2)}||y_1^{(2)}}\cU^\prime(\ket{\varphi^\prime}\odot K_{out}^{(3)}\odot \tilde{tag})|/|\ket{\varphi}|\geq q/\sqrt{3}\end{equation}
	This is because $\cU^\prime$ can just guess the correct tags in the $\tilde{tag}$ and run $\cU$ above.\par
	\textbf{Step 1: make the adversary (after the protocol completes) blind on $K^{\text{help}}$}\par
	Then define a ``blinded'' version of $\cU^\prime$, let's denote it as $\cU^{blind}$. In this operation the RO queries in $\cU$ are replaced by the queries to the blinded oracle $H^\prime$ where $H(\cdots ||K^{\text{help}}||\cdots)$ are blinded. (Which means we blind both $H(\cdots ||x_0^{\text{help}}||\cdots)$ and $H(\cdots ||x_1^{\text{help}}||\cdots)$. And the prefix has length $l$ and the suffix padding is arbitrary.) These blinded parts cover everything in the table thus if we blind these two parts of the random oracle we ``forbid'' the decryption of the whole table.\par
	We define $q^{blind}$ as follows, which replaces the $\cU^\prime$ in (\ref{eq:56}) by the blinded operation:
	\begin{equation}\label{eq:56bb}q^{blind}:=|P_{y_0^{(2)}||y_1^{(2)}}\cU^{blind}(\ket{\varphi^\prime}\odot K_{out}^{(3)}\odot \tilde{tag})|/|\ket{\varphi}|\end{equation}
	Now we are going to use Lemma \ref{lem:7.3}, Lemma \ref{lem:4.15} and (\ref{eq:56}) to get a bound for $q^{blind}$. First by Lemma \ref{lem:7.3} and (\ref{eq:110n}) we know $\ket{\varphi^\prime}\odot K_{out}^{(3)}\odot \tilde{tag}$ is $(2^{\eta/36},2C|\ket{\varphi}|)$-ANY-secure for $K^{\text{help}}$ (the condition for applying Lemma \ref{lem:7.3} is from Lemma \ref{lem:af3}). Then apply Lemma \ref{lem:4.15} we can relate $q^{blind}$ with (\ref{eq:56}) and get:
	\begin{equation}
		q^{blind}\geq q/\sqrt{3}-6C.
	\end{equation}
	\textbf{Step 2: Consider the behavior when the initial state is some branch of $\ket{\varphi}$}\par
	As in the proof of the $w=3$ case, define $\ket{\varphi_0^\prime}$ and $\ket{\varphi_1^\prime}$, the post-execution state of using $\ket{\varphi_b}$ (in equation (\ref{eq:36nn})) as the initial state. Define $q^{blind}_0$ and $q^{blind}_1$ as the quotient when $\ket{\varphi^\prime}$ in equation (\ref{eq:56bb}) is substituted with $\ket{\varphi^\prime_0}$ and $\ket{\varphi^\prime_1}$:
	\begin{equation}\label{eq:56b}\forall b\in \{0,1\},q_b^{blind}:=|P_{y_0^{(2)}||y_1^{(2)}}\cU^{blind}(\ket{\varphi^\prime_b}\odot K_{out}^{(3)}\odot \tilde{tag})|/|\ket{\varphi}|\end{equation}

	Then apply Lemma \ref{lem:7.4} we get (the conditions for applying Lemma \ref{lem:7.3} is proved in Lemma \ref{lem:af3}. And here we only need to consider the case corresponding to (\ref{eq:45r}), since for the case (\ref{eq:44r}) we already get a bound for $q$: $q\leq O(1)C$ for some constant $O(1)$.)
	\begin{equation}\label{eq:58a}q^{blind}_1\geq q^{blind}/6\geq q/6\sqrt{3}-C\end{equation}
	From now on we need to study what the permutation gives us.
	Let's first expand $\ket{\varphi^\prime}$:
	\begin{equation}\label{eq:58}\ket{\varphi^\prime}=(\fPadHadamard_{\fAdv_2}\circ(\fAdv_1(\ket{\varphi}\odot \fRobustRLT(perm)\odot K^{(2)})))\odot perm\end{equation}
	Here we split the operations of $\fAdv$ on different phases of the protocol as $\fAdv_1,\fAdv_2$. And we omit the parameters that are not important here. Further note that this characterization of the adversary is enough: for example, we don't need to add another symbol $\fAdv_3$ in the leftmost, since it can be ``absorbed'' into $\cU$.\par
	\textbf{Step 3} Let's first describe the overall idea of this step. Below \textbf{we will consider what happens when the permutation $perm$ in the fifth step of the protocol is replaced by the \emph{fake permutation}.} This technique is less intuitive but turns out to be a key technique in the proof of this lemma. And we will see some of the unexplained step (for example, replacing $Tag$ with $\tilde{tag}$) is actually the preparation for the proof below.\par
	We define another state $\ket{\phi^\prime}$ as the post-execution state where in the fifth step of the protocol, the client sends $perm^\prime$ to the server instead of $perm$:
	\begin{equation}\label{eq:11}\ket{\phi^\prime}:=(\fPadHadamard_{\fAdv_2}\circ(\fAdv_1(\ket{\varphi}\odot \fRobustRLT(perm)\odot K^{(2)})))\odot perm^\prime\end{equation}
	Then we define $\ket{\phi^\prime_0}$ and $\ket{\phi^\prime_1}$ as the output of replacing $\ket{\varphi}$ above (equation (\ref{eq:11})) with $\ket{\varphi_0}$ and $\ket{\varphi_1}$:
	\begin{equation}\label{eq:359r}\forall b\in \{0,1\},\ket{\phi_b^\prime}:=(\fPadHadamard_{\fAdv_2}\circ(\fAdv_1(\ket{\varphi_b}\odot \fRobustRLT(perm)\odot K^{(2)})))\odot perm^\prime\end{equation}
	So $\ket{\phi^\prime}=\ket{\phi^\prime_0}+\ket{\phi^\prime_1}$\par
	Then we define $q_0^{blind,fake,0}$, $q_1^{blind,fake,0}$ and $q_0^{blind,fake,1}$, $q_1^{blind,fake,1}$, where $q_b^{blind,fake,c}$ stands for the following value: in equation (\ref{eq:56b}), using $\ket{\phi_b^\prime}$ as the initial state, providing $K_{out}^{fake-c-(3)}$ as the revealed keys, and the adversary is trying to compute the fake keys $K_{out}^{fake-c-(2)}$ using $\cU^{blind}$. Formally speaking, they are defined as follows:
	\begin{equation}\label{eq:110}\forall b\in \{0,1\},c\in \{0,1\},q_b^{blind,fake,c}:=\end{equation}\begin{equation*}|P_{y_0^{fake-c-(2)}||y_1^{fake-c-(2)}}\cU^{blind}(\ket{\phi_b^\prime}\odot K_{out}^{fake-c-(3)}\odot \tilde{tag})|/|\ket{\varphi}|\end{equation*}
	We will analyze $q_b^{blind,fake,c}$ below. Note that the $b=0$ and $b=1$ cases are very different here. We can prove (either by existing lemmas or using the lemmas whose proof is postponed):
	\begin{itemize}
		\item For the $b=1$ case, we relate $q_1^{blind,fake,c}$ with $q_1^{blind}$, and get (see Lemma \ref{lem:e1} below):
		      \begin{equation}\label{eq:98}\text{Either }q^{blind,fake,0}_1\geq q_1^{blind}/2-54C\text{ or }q^{blind,fake,1}_1\geq q_1^{blind}/2-54C\end{equation}
		\item For the $b=0$ case, by Lemma \ref{lem:lrr2.2}, we know
		      \begin{equation}\label{eq:99}\forall c\in \{0,1\},q_0^{blind,fake,c}\leq 3\times 4C\times 2+9C=33C\end{equation}
		      The details are as follows: 
		      \begin{enumerate}\item We further make use of $\ket{\varphi_b}\approx_{9C|\ket{\varphi}|}\ket{\varphi_{00}}+\ket{\varphi_{01}}$. Recall (\ref{eq:359r}) and (\ref{eq:110}); now we can study these two branches separately and combine them by the triangle inequality of unpredictability and get the bound for the left side of (\ref{eq:99}). And for each one of these two --- without loss of generality, let's consider $\ket{\varphi_{00}}$ --- we can define $\ket{\phi^\prime_{00}}$ as the result of replacing the $\ket{\varphi_b}$ by $\ket{\varphi_{00}}$ in (\ref{eq:359r}), and define $q_{00}^{blind,fake,c}$ as the result of replacing the $\ket{\phi_b^\prime}$ in (\ref{eq:99}) by $\ket{\phi^\prime_{00}}$. What we need is to prove it's at most $3\times 4C$. (In the right side of (\ref{eq:99}) ``$\times 2$'' and ``$+9C$'' come from this step.)
		      \item One issue for applying Lemma \ref{lem:lrr2.2} here is the ``conditions on $\ket{\varphi_{00}}$'' are described relative to the norm of $\ket{\varphi}$. (For example, we know $\ket{\varphi_{00}}$ is $(2^\eta-2,4C|\ket{\varphi}|)$-unpredictable for $x_1^{(3)}$ given $K^{\text{help}}$, but we need to change the norm $|\ket{\varphi}|$ in this statement to $|\ket{\varphi_{00}}|$.) We can handle this problem as follows:\par
		      (Case 1): If $|\ket{\varphi_{00}}|\leq 2^{-\kappa}|\ket{\varphi}|$ then we automatically have $|\ket{\varphi_{00}}|\leq C|\ket{\varphi}|$.\par
		      (Case 2): Otherwise we know (1)$\ket{\varphi_{00}}$ is $(2^\eta,2^{-\eta+\kappa}|\ket{\varphi_{00}}|)$-unpredictable for $x_1^{\text{help}}$ given $K^{(3)}$ and  (2) $\ket{\varphi_{00}}$ is $(2^\eta,(4C\frac{|\ket{\varphi}|}{|\ket{\varphi_{00}}|})|\ket{\varphi_{00}}|)$-unpredictable for $x_1^{(3)}$ given $K^{\text{help}}$. Then we can apply Lemma \ref{lem:lrr2.2} and choose $C$ (within Lemma \ref{lem:lrr2.2}) to be $4C|\ket{\varphi}|/|\ket{\varphi_{00}}|$ and get $q_{00}^{blind,fake,c}\leq 12C$. Here we implicitly assume $4C|\ket{\varphi}|/|\ket{\varphi_{00}}|<\frac{1}{3}$, since otherwise we still have $q_{00}^{blind,fake,c}\leq |\ket{\varphi_{00}}|/|\ket{\varphi}|\leq 12C$.
		      \end{enumerate}
	\end{itemize}

	Now let's look at (\ref{eq:98}) and (\ref{eq:99}). Note that in (\ref{eq:98}) there are two possible cases. Luckily (\ref{eq:99}) holds for both $c=0$ and $c=1$. Without loss of generality, consider the $c=0$ case in (\ref{eq:98}), and correspondingly take $c=0$ in (\ref{eq:99}), which are
	\begin{align}\label{eq:101}q^{blind,fake,0}_1 & \geq q_1^{blind}/2-54C\geq q/12\sqrt{3}-55C\text{ (by equation (\ref{eq:58a}))} \\
		q_0^{blind,fake,0}               & \leq 33C\label{eq:331}
	\end{align}
	, then notice $q_0^{blind,fake,0}$ and $q_1^{blind,fake,0}$ can also be seen as the output of running some adversary on a post-execution state of padded Hadamard test, we can apply Lemma \ref{lem:7.4} again: we take $\llbracket\fAuxInf\rrbracket$ in Lemma \ref{lem:7.4} to be $\fRobustRLT(perm)\odot K^{(2)}\odot perm^\prime\odot K_{out}^{fake-c-(3)}\odot \tilde{tag}$, and the unitary in the lemma as $\cU^{blind}$, and the conditions for applying Lemma \ref{lem:7.4} come from Lemma \ref{lem:lrr3}. Thus we have
	\begin{equation}\label{eq:120n}\text{ Either }\min\{q^{blind,fake,0}_0,q^{blind,fake,0}_1\}\geq q^{blind, fake,0}/6\geq |q_1^{blind,fake,0}-q_0^{blind,fake,0}|/6\end{equation}
	\begin{equation}\label{eq:121n}\text{, or }5C\geq q^{blind,fake,0}\geq|q_1^{blind,fake,0}-q_0^{blind,fake,0}|\end{equation}

	Both cases imply $q\leq 10000C$. (substitute (\ref{eq:101})(\ref{eq:331}).) Thus we complete the proof of Lemma \ref{lem:7.6}. The remaining work is to fill the missing step:\par

	\textbf{Step 4: Prove equation (\ref{eq:98}).}
	\begin{lem}[A repetition of (\ref{eq:98})]\label{lem:e1}
		$$\text{Either }q^{blind,fake,0}_1\geq q_1^{blind}/2-54C\text{ or }q^{blind,fake,1}_1\geq q_1^{blind}/2-54C$$
	\end{lem}

	\begin{proof}[Proof of (\ref{eq:98})]
		Recall the definition of $\ket{\varphi_{b_1b_2}}$ in equation (\ref{eq:36nn})(\ref{eq:37nn}). Recall that
		\begin{equation}\label{eq:336}
			\ket{\varphi_1}\approx_{9C|\ket{\varphi}|}\ket{\varphi_{10}}+\ket{\varphi_{11}}
		\end{equation}
		Define $\ket{\varphi^\prime_{b_1b_2}}$ as the result of replacing the inner $\ket{\varphi}$ in equation (\ref{eq:58}) with $\ket{\varphi_{b_1b_2}}$:
		\begin{equation}\ket{\varphi^\prime_{b_1b_2}}=(\fPadHadamard_{\fAdv_2}\circ(\fAdv_1(\ket{\varphi_{b_1b_2}}\odot \fRobustRLT(perm)\odot K^{(2)})))\odot perm\end{equation}
		Then define $q^{blind}_{10}$, $q^{blind}_{11}$ as the results when $\ket{\varphi^\prime_{10}}$ and $\ket{\varphi^\prime_{11}}$ are used in equation (\ref{eq:56b}):
		$$q_{1b}^{blind}:=|P_{y_0^{(2)}||y_1^{(2)}}\cU^{blind}(\ket{\varphi^\prime_{1b}}\odot K_{out}^{(3)}\odot \tilde{tag})|/|\ket{\varphi}|$$
		From (\ref{eq:336}), we have
		\begin{equation}\label{eq:119}\text{Either }q^{blind}_{10}\geq q_1^{blind}/2-5C\text{ or }q^{blind}_{11}\geq q_1^{blind}/2-5C\end{equation}
		These two cases correspond to the two cases in the final statement.\par
		Without loss of generality, let's consider the $q^{blind}_{10}\geq q_1^{blind}/2-5C$ case and the other case is similar. Now our task is to find a relation between $q_{10}^{blind}$ and $q_1^{blind,fake,0}$.\par
		Then define $\ket{\phi_{10}^\prime}$ as the result of replacing $\ket{\varphi}$ in equation (\ref{eq:11}) with $\ket{\varphi_{10}}$, and define $\ket{\phi_{11}^\prime}$ as the result of replacing $\ket{\varphi}$ in equation (\ref{eq:11}) with $\ket{\varphi_{11}}$. Then similar to equation (\ref{eq:110}), replacing $\ket{\phi^\prime_{b}}$ with $\ket{\phi^\prime_{10}}$ and $\ket{\phi^\prime_{11}}$, and only considering the $c=0$ case, we define $q_{10}^{blind,fake,0}$, $q_{11}^{blind,fake,0}$ as follows:
		$$q_{10}^{blind,fake,0}:=|P_{y_0^{fake-0-(2)}||y_1^{fake-0-(2)}}\cU^{blind}(\ket{\phi_{10}^\prime}\odot K_{out}^{fake-0-(3)}\odot \tilde{tag})|/|\ket{\varphi^\prime}|$$
		$$q_{11}^{blind,fake,0}:=|P_{y_0^{fake-0-(2)}||y_1^{fake-0-(2)}}\cU^{blind}(\ket{\phi_{11}^\prime}\odot K_{out}^{fake-0-(3)}\odot \tilde{tag})|/|\ket{\varphi^\prime}|$$
		And we have (also by (\ref{eq:336}))
		\begin{equation}\label{eq:120}q_1^{blind,fake,0}\geq |q_{10}^{blind,fake,0}-q_{11}^{blind,fake,0}|-9C\end{equation}
		Now to give a bound for $q_1^{blind,fake,0}$, we only need to give a bound for $q_{10}^{blind,fake,0}$ and $q_{10}^{blind,fake,0}$.\par
		Now it's time to apply the lemmas in Section \ref{sec:afn}.
		\begin{itemize}
			\item For $q_{10}^{blind,fake,0}$, we apply Corollary \ref{cor:e11c2}. From the conditions we know $\ket{\varphi_{10}}$ is $(2^{\eta}-5,4C|\ket{\varphi}|)$-unpredictable for $x_{1}^{(3)}$. Thus
			      \begin{equation}\text{By Corollary \ref{cor:e11c2}: }|q_{10}^{blind,fake,0}-q_{10}^{blind}|\leq 24C\end{equation}
			      (Note that there is one implicit step here, as what we did in the proof of (\ref{eq:99}): we can assume $12C|\ket{\varphi}|\geq|\ket{\varphi_{10}}|\geq 2^{-\sqrt{\kappa}}|\ket{\varphi}|$ because we can discuss the other cases separately. Then when we apply Corollary \ref{cor:e11c2}, with the ``$C$'' in the corollary chosen to be $4C|\ket{\varphi}|/|\ket{\varphi_{10}}|$.)\par 
			      Thus substitute (\ref{eq:119}):
			      \begin{equation}
				      q_{10}^{blind,fake,0}\geq q_1^{blind}/2-29C
			      \end{equation}

			\item For $q_{11}^{blind,fake,0}$, we apply Corollary \ref{cor:e11c1}. The condition is $\ket{\phi_{11}}$ is $(2^{\eta}-5,4C|\ket{\varphi}|)$-unpredictable for $x_{0}^{(3)}$. Thus
			      \begin{equation}
				      \text{By Corollary \ref{cor:e11c1}: } q_{11}^{blind,fake,0}\leq 16C
			      \end{equation}
			      (The implicit step is as above.)
		\end{itemize}
		These two inequalities together with (\ref{eq:120}) complete the proof.
		%

		%
	\end{proof}
	Thus we complete the proof of Lemma \ref{lem:7.6} for the $w=2$ case.
\end{proof}
Let's give a summary on the subtle part of this proof. Notice that we apply Lemma \ref{lem:7.4} twice, by considering different $\llbracket\fAuxInf\rrbracket$, and get (\ref{eq:58a}) and (\ref{eq:120n})(\ref{eq:121n}). And the security provided by the lookup tables gives us the other two inequalities (\ref{eq:98})(\ref{eq:99}). And the whole proof comes from their combinations. The other parts are mostly details when we do hybrid methods on the entries in the lookup tables.
\section{Proof of Lemma \ref{lem:10.1}}\label{sec:AF}
In this section we give the proof of Lemma \ref{lem:10.1}. This section is organized as follows: \begin{enumerate}\item In Section \ref{sec:ah1} we give some lemmas that are useful in the main proof. \item In Section \ref{sec:ah2} we give an overview of the proof.\item And the main proof is divided into four subsections, in Section \ref{sec:ah3} to Section \ref{sec:ah6}.\end{enumerate} So the reader can go to Section \ref{sec:ah2} for an overview of the proof techniques.
\subsection{Preparation}\label{sec:ah1}
Before we prove Lemma \ref{lem:10.1}, first we can prove:
\begin{lem}\label{lem:ah1}
	Under the conditions of Lemma \ref{lem:10.1}, $\ket{\varphi}\odot \llbracket\fSecurityRefreshing\rrbracket$ is $(2^{\eta_1-6},2^{-\eta_1+6}|\ket{\varphi}|)$-SC-secure for $K^{(i)}$, where $\llbracket\fSecurityRefreshing\rrbracket$ are the transcripts of all the messages sent by the client during this protocol.
\end{lem}
\begin{proof}
	The proof is very similar to the proof of Lemma \ref{lem:4.9}. First note \\$\llbracket\fSecurityRefreshing\rrbracket$ contains the followings:
	\begin{itemize}\item The lookup tables sent in the first step of each round of the $\fSecurityRefreshing$ protocol; Denote it as $\llbracket GTs\rrbracket$.\item The random pads for padded Hadamard test in the third step of each round of the $\fSecurityRefreshing$ protocol;\item The random pads in the end of the protocol.\end{itemize} Since in the adversary's viewpoint, these random pads can also be sampled and simulated on the server-side, we only need to prove
	\begin{center}
		\emph{$\ket{\varphi}\odot \llbracket GTs\rrbracket$ is $(2^{\eta_1-6},2^{-\eta_1+6}|\ket{\varphi}|)$-SC-secure for $K^{(i)}$.}
	\end{center}
	Then by Technique \ref{lem:4.2} we can assume the \emph{temporary output keys} \\$K_{temp}=\{K_{temp}^{(i)(j)}\}_{i\in [N],j\in [J]}$ are provided as the auxiliary information. Thus what we will prove is
	\begin{center}
		\emph{$\ket{\varphi}\odot \llbracket GTs\rrbracket\odot K_{temp}$ is $(2^{\eta_1-6},2^{-\eta_1+6}|\ket{\varphi}|)$-SC-secure for $K^{(i)}$.}
	\end{center}

	By proof-by-contradiction, assume there is a server-side operation $\cU$ such that the query number $|\cU|\leq 2^{\eta_1-6}$ and
	\begin{equation}|P_{x_0^{(i)}|| x_1^{(i)}}\cU(\ket{\varphi}\odot \llbracket GTs\rrbracket\odot K_{temp})|>2^{-\eta_1+6}|\ket{\varphi}|\end{equation}
	Suppose the set of random pads used in the computation of $\llbracket GTs\rrbracket$ is $Set$. By Lemma \ref{lem:3.4r} there exists $\ket{\tilde\varphi}\approx_{2^{-\eta}|\ket{\varphi}|}\ket{\varphi}$ that does not depend on $H(Set||\cdots)$ (in other words, the RO output on the entries whose prefixes are in $Set$ does not have influence on the state). Then
	\begin{equation}\label{eq:159}|P_{x_0^{(i)}|| x_1^{(i)}}\cU(\ket{\tilde\varphi}\odot \llbracket GTs\rrbracket\odot K_{temp})|>2^{-\eta_1+5}|\ket{\tilde\varphi}|\end{equation}
	\begin{equation}\label{eq:389r}\text{ (And since $Set,K_{temp}$ are sampled randomly)}\end{equation}\begin{equation} \text{$\ket{\tilde\varphi}\odot K_{temp}\odot Set$ is $(2^{\eta_1},2^{-\eta_1+1}|\ket{\tilde\varphi}|)$-SC-secure for $K^{(i)}$}\end{equation}
	Let's make use of equation (\ref{eq:159}) to construct a unitary $\cU^\prime$ to break the SC-security shown in (\ref{eq:389r}). In more details, we will construct a server-side operation $\cU^\prime$ such that the query number $|\cU^\prime|\leq 2^{\eta_1}$ and
	\begin{equation}|P_{x_0^{(i)}|| x_1^{(i)}}\cU^\prime(\ket{\tilde\varphi}\odot K_{temp}\odot Set)|>2^{-\eta_1+1}|\ket{\tilde\varphi}|\end{equation}
	Starting from (\ref{eq:159}), note that when $K_{temp}$ is given as the auxiliary information, each term in $\llbracket GTs\rrbracket$ can be seen as the tuple of random pads and the hash outputs in the form of (or more formally, if the followings and the random pads are provided to the adversary instead of $\llbracket GTs\rrbracket$, the adversary can compute $\llbracket GTs\rrbracket$ by itself) \begin{equation}\label{eq:394z}H(pad||x_{b_x}^{(i^\prime)}||r_{b_r}^{(j)})\text{ for some }pad\in Set,i^\prime\in [N], j\in [J], b_x,b_r\in \{0,1\}^2\end{equation} to the adversary. (Note that symbol $i$ has been occupied so we have to use $i^\prime$ here. And note for each $i^\prime,j,b_x,b_r$, there are two terms in this form: one for the ciphertext and one for the key tags. See the definition of $\fEn$ in Definition \ref{def:2.13}.)\par
	The remaining steps are very similar to the proof of Lemma \ref{lem:4.9}.\par
	Values of this form fall into the $Set||\cdots$ part of the inputs (the suffix padding has length equal to the length of keys in $K$ plus the length of keys in $\Lambda$), thus the random oracle output values of them do not affect the state $\ket{\tilde\varphi}$ itself. Thus there is no difference whether the outputs of $H$ on $Set||\cdots$ part of the inputs are sampled before the whole protocol, or is sampled just before the operation of $\cU$. Then what $\cU^\prime$ will do is to 
	\begin{enumerate}\item Sample the ``background values'' for the random outputs of $Set||\cdots$ part; and it also samples the random values for the terms in (\ref{eq:394z}), and compute the ``fake version'' of $\llbracket GTs\rrbracket$ from it, denote it as $\llbracket GTs^{fake}\rrbracket$; \item It constructs a ``simulated oracle'' $\tilde H$; 
	The detailed construction is similar to the construction in the proof of Lemma \ref{lem:4.9} in Appendix \ref{sec:a44}. The only difference is in the construction of $\tilde H$, the first step is to check whether the input has the form of $pad||x_{b}^{(i^\prime)}||r_{b^\prime}^{(j)}$, $pad\in Set$. This can be achieved since $Tag(K)$ and $Tag(\Lambda)$ are stored in the read-only buffer. (By Fact \ref{fact:injtag} the space that the $Tag$ is not injective is very small.)\item Suppose the operation that comes from replacing all the queries in $\cU$ by queries to $\tilde H$ as $\cU^{fake}$. $\cU^{\prime}$ will put $\llbracket GTs^{fake}\rrbracket$ onto the place of $\llbracket GTs\rrbracket$ in (\ref{eq:159}), and run $\cU^{fake}$.
	\end{enumerate}
	 Then we have
	 \begin{align}
	 	|P_{x_0^{(i)}|| x_1^{(i)}}\cU^\prime(\ket{\tilde\varphi}\odot K_{temp}\odot Set)|&=|P_{x_0^{(i)}|| x_1^{(i)}}\cU^{fake}(\ket{\tilde\varphi}\odot \llbracket GTs^{fake}\rrbracket\odot K_{temp}\odot Set)|\\\text{(By Fact \ref{fact:injtag} and discussions above)}&\approx_{2^{-\eta}|\ket{\varphi}|} |P_{x_0^{(i)}|| x_1^{(i)}}\cU(\ket{\tilde\varphi}\odot \llbracket GTs\rrbracket\odot K_{temp})|
	 \end{align}
This together with (\ref{eq:159}) completes the proof.
\end{proof}
\subsection{Overview of the proof}\label{sec:ah2}
Now let's prove Lemma \ref{lem:10.1}. We will split this proof into four subsections.
\begin{enumerate}\item In the Section \ref{sec:ah3} we will do a linear decomposition and reduce the statement into two smaller problems, we will call them ``Statement 1'' and ``Statement 2''. We will prove Statement 1 and the proof of Statement 2 is similar. We will list these two statements in Outline \ref{otl:pf}.
	\item For the proof of Statement 1, we first reduce it to a security statement of a \emph{simplified temporary protocol}. We will give this statement in the second subsection (Section \ref{sec:ah4}).
	\item Then the third subsection (Section \ref{sec:ah5}) is for the proof of this security statement of simplified protocol. This is the most difficult part of these four steps. We will give a further overview for this step in the beginning of Section \ref{sec:ah5} (Section \ref{sec:ai52}).
	\item Finally we prove Statement 2 using similar technique (since many steps can be reused the description of the proof is much shorter). Then we combine Statement 1 and 2 via triangle inequality and complete the proof. This is put in the last subsection.\end{enumerate}
\subsection{Part I: Break Lemma \ref{lem:10.1} into Statement 1 and 2 Through Linear Decomposition}\label{sec:ah3}
\begin{proof}[Proof of Lemma \ref{lem:10.1}, Part I]
	Suppose $\fAdv=U_{2^\kappa}HU_{2^\kappa-1}H\cdots U_1HU_0$. (We only write down the adversary's operations. There should be some client side ``computing and sending lookup tables'' operations among these server side operations, thus this expression only has literal meaning. We will give explanations when we use this expression.) \par
	\textbf{ The first step is to decompose the adversary's operation linearly.} Define
	\begin{equation}\label{eq:44n}\fAdv_{t,b}=U_{2^\kappa}HU_{2^\kappa-1}H\cdots HU_tHP_{x^{(i)}_b}U_{t-1}H(I-P_{K^{(i)}})\cdots  U_1H(I-P_{K^{(i)}})U_0\end{equation}
	Here a projection is done before each RO query in the first $t$ server side queries:\begin{itemize}\item $P_{x_b^{(i)}}$ is the server side projection onto the $\cdots||x_{b}^{(i)}||\cdots$ space of the input to the random oracle queries;\item $I-P_{K^{(i)}}=I-P_{x_0^{(i)}}-P_{x_1^{(i)}}$ is a projection onto the space that excludes $\cdots||x_{0}^{(i)}||\cdots$ and $\cdots||x_{1}^{(i)}||\cdots$. \end{itemize}(The prefix padding has length $l$ and the suffix padding has length equal to the key length in $\Lambda$.)\par And (\ref{eq:44n}) means, in $\fAdv_{t,b}$:\begin{itemize}\item In each of the first $(t-1)$ RO queries made by the adversary, the $H(\cdots||x_0^{(i)}||\cdots)$ and $H(\cdots||x_1^{(i)}||\cdots)$ parts of the queries are ``removed'';\item And for the $t$-th query we make a projection and only consider the query on the input $\cdots||x^{(i)}_b||\cdots$.\end{itemize} Then we have:
	\begin{equation}\label{eq:45n}\fAdv=\sum_{t=1}^{2^\kappa}(\fAdv_{t,0}+\fAdv_{t,1})+\fAdv_0\end{equation}
	$$\text{where }\fAdv_0=U_{2^\kappa}H(I-P_{K^{(i)}})U_{2^\kappa-1}H(I-P_{K^{(i)}})\cdots U_TH(I-P_{K^{(i)}})U_{T-1}\cdots  U_1H(I-P_{K^{(i)}})U_0$$
	so in $\fAdv_0$, $(I-P_{K^{(i)}})$ is applied before each query.\par
	We note that (\ref{eq:45n}) is also literal, as discussed above (\ref{eq:44n}): it means when the client-side operations and message transmission operations are inserted suitably into the description of $\fAdv_{\cdots}$, then the equation holds.\par
	We will proceed by proving the following statements one by one:
	\begin{otl}\label{otl:pf}
		\begin{enumerate}
			\item (Statement 1): Suppose the protocol is run against adversary $\fAdv_{t,b}$. Define
			      \begin{equation}\label{eq:348}\ket{\varphi^\prime_{t,b}}=\fSecurityRefreshing_{\fAdv_{t,b}}(K,\Lambda; \underbrace{ \ell}_{\substack{\text{padding} \\ \text{length}}},
\underbrace{ \kappa_{\text{out}}}_{\substack{\text{output} \\ \text{length}}})\circ\ket{\varphi}\end{equation}
			      Then $\forall t\in [2^\kappa],b\in \{0,1\}$, $P_{pass}\ket{\varphi^\prime_{t,b}}$ is $(2^{\eta_2/100\kappa},2^{-\eta_1/4+\kappa}|\ket{\varphi}|)$-SC-secure for $K_{out}^{(i)}$ given $K_{out}-K_{out}^{(i)}$ and $\llbracket\fAuxInf\rrbracket$.
			\item (Statement 2) Similarly, suppose the protocol is run against adversary $\fAdv_{0}$. Define
			      \begin{equation}\ket{\varphi^\prime_{0}}=\fSecurityRefreshing_{\fAdv_{0}}(K,\Lambda; \underbrace{ \ell}_{\substack{\text{padding} \\ \text{length}}},
\underbrace{ \kappa_{\text{out}}}_{\substack{\text{output} \\ \text{length}}})\circ\ket{\varphi}\end{equation}
			      Then $P_{pass}\ket{\varphi^\prime_0}$ is $(2^{\eta_2/100\kappa},2^{-\eta_1/4}|\ket{\varphi}|)$-SC-secure for $K_{out}^{(i)}$ given $K_{out}-K_{out}^{(i)}$ and $\llbracket\fAuxInf\rrbracket$.
			\item Note that $\ket{\varphi^\prime}=\sum_{t\in [2^\kappa],b\in \{0,1\}}\ket{\varphi^\prime_{t,b}}+\ket{\varphi^\prime_0}$. Finally we can combine the Statement 1 and 2 by the triangle inequality of SC-security and draw the conclusion that $P_{pass}\ket{\varphi^\prime}$ is\\ $(2^{\eta_2/100\kappa},2^{-\eta_1/4+2\kappa+2}|\ket{\varphi}|)$-SC-secure for $K_{out}^{(i)}$ given $K_{out}-K_{out}^{(i)}$ and $\llbracket\fAuxInf\rrbracket$, thus complete the proof of Lemma \ref{lem:10.1}.
		\end{enumerate}\end{otl}
		In the next subsection we will reduce Statement 1 to a new statement which is about the security of a temporary protocol. And in Section \ref{sec:ah5} we will see Statement 2 above can also be reduced to it.
\end{proof}

\subsection{Reduce ``Statement 1'' to ``Security of a temporary protocol $\fTempPrtl$''}\label{sec:ah4}
Let's first try to prove the ``Statement 1'' above. To prove it, we will first make use of the technique in Section \ref{sec:4.6} and \ref{sec:4.2} to simplify the protocol and reduce it to the security of a temporary protocol. Then in the next subsection we will prove the security of this temporary protocol and complete the proof of Statement 1.
\begin{proof}[Proof of Statement 1, step 1: reduction]
	First applying Lemma \ref{lem:ah1} we can know $\ket{\varphi}\odot \llbracket\fSecurityRefreshing\rrbracket$ is $(2^{\eta_1-6},2^{-\eta_1+6}|\ket{\varphi}|)$-SC-secure for $K^{(i)}$ given $Tag(K,\Lambda)$.\par
	Note that in $\fAdv_{t,b}$ the adversary gets $x_b^{(i)}$ in the middle of the attack by making a projection. Intuitively by the SC-security proved above the adversary should be hard to compute $x_{1-b}^{(i)}$. To formalize this intuition, we can apply Lemma \ref{lem:r4.17} to switch the oracle queries to $H$ in $\fAdv_{t,b}$ to queries to $H^\prime$ defined as follows:
	\begin{equation}\label{eq:350r}\text{$H^\prime$ is defined to be a new blinded oracle of $H$ where $H(\cdots||x_{1-b}^{(i)}||\cdots)$ is blinded.}\end{equation} (``$\cdots$'' represents arbitrary strings of some length. The prefix padding has length $l$ and the suffix padding has length the same as the keys in $\Lambda$.)\par
	Then define $\fAdv_{t,b}^\prime$ as the adversary that runs the same operations, but queries $H^{\prime}$ instead of $H$. Similar to (\ref{eq:348}), define
	\begin{equation}\label{eq:29}\ket{\varphi^{\prime\prime}_{t,b}}=\fSecurityRefreshing_{\fAdv^\prime_{t,b}}(K,\Lambda; \ell,\kappa_{out})\circ\ket{\varphi}\end{equation}
	The difference of $\fAdv_{t,b}$ and $\fAdv_{t,b}^\prime$ starts after the adversary's $t$-th query. We can prove the state just after the adversary's $t$-th query is $(2^{\eta_1-6},2^{-\eta_1+7}|\ket{\varphi}|)$-SC-secure for $K^{(i)}$ by Lemma \ref{lem:ah1}. (Since the projection can be simulated using $Tag(K)$ assuming $Tag$ is injective on inputs with the same length as the keys in $K$, and the space that $Tag$ is not injective on these inputs has very small norm.) Then by Lemma \ref{lem:r4.17} we have
	\begin{equation}\label{eq:23}\ket{\varphi^{\prime\prime}_{t,b}}\approx_{2^{-\eta_1/3+3}|\ket{\varphi}|}\ket{\varphi^{\prime}_{t,b}}\end{equation}
	(there are some implicit steps here: $\ket{\varphi^{\prime}_{t,b}}$, $\ket{\varphi^{\prime\prime}_{t,b}}$ are defined on real protocol, but Lemma \ref{lem:r4.17} is talking about a server-side operation. But we note that we can assume all the client-side messages are already stored in the read-only buffer but this adversary uses it step-by-step. Then we can apply Lemma \ref{lem:r4.17} and get (\ref{eq:23}).)\par
	\textbf{Thus to prove Statement 1, we can reduce it to ``Statement 3'', which is about the SC-security of $P_{pass}\ket{\varphi^{\prime\prime}_{t,b}}$ for $K_{out}^{(i)}$,} defined as follows:\par
	\begin{center}\emph{(Statement 3) Suppose the security parameter $\kappa$ is bigger than some constant. Suppose the initial state $\ket{\varphi}$ satisfies the conditions listed in Lemma \ref{lem:10.1}. For any adversary $\fAdv^\prime$ that only queries $H^{\prime}$ (see (\ref{eq:350r})) during the protocol, and the total number of queries to $H^\prime$ is at most $2^{\kappa+2}$, the post-execution state,
		$$P_{pass}\fSecurityRefreshing_{\fAdv^\prime}(K,\Lambda; \underbrace{ \ell}_{\substack{\text{padding} \\ \text{length}}},
\underbrace{ \kappa_{\text{out}}}_{\substack{\text{output} \\ \text{length}}})\circ\ket{\varphi}$$
		is $(2^{\eta_2/100\kappa},2^{-\eta_2/100\kappa}|\ket{\varphi}|)$-SC-secure for $K_{out}^{(i)}$ given $K_{out}-K_{out}^{(i)}$ and $\llbracket\fAuxInf\rrbracket$.}\end{center}\par
	Note that $2^{-\eta_2/100\kappa}+2^{-\eta_1/3+O(1)}<2^{-\eta_1/4}$. The choices of the parameters here are for the convenience of later proofs.\par
	Note that the operation of $\fAdv^\prime$ only happens during the protocol, and after the protocol completes, in the definition of SC-security, there is another implicit adversary $\cD$. (Recall that the definition of SC-security says for any server-side operation $|\cD|\leq \cdots$ there is $\cdots$.) In the definition of SC-security the implicit  adversary still queries the original oracle $H$, not $H^\prime$. In other words, we are considering the following setting: the random oracle  is blinded during the protocol, but after the protocol completes, it is not blinded anymore. And we want to prove in this setting the adversary is still hard to output both keys in $K_{out}^{(i)}$.\par
	We will further reduce Statement $3$ to some other statements. During this process we will design a \emph{temporary protocol}, and reduce Statement 3 to the security property of this temporary protocol. In more details, this reduction process is as follows:\begin{enumerate}
		\item First we reduce Statement $3$ to Statement $3^\prime$ by adding auxiliary information. (See Section \ref{sec:4.2} for the auxiliary information technique.)
		\item Then we reduce Statement $3^\prime$ to Statement $3^{\prime\prime}$ by changing the blinded oracle to another oracle that has fewer blinded part. This does not make the adversary weaker since the adversary can also further blind the oracle by itself. This simplifies the later proofs.
		\item Finally we make use of the auxiliary information in step 1 to simplify the protocol and reduce Statement $3^{\prime\prime}$ to the security of a temporary protocol.
	\end{enumerate}
	Let's start the reduction. We will use bold font to divide different steps.\par
	\textbf{First, we apply the auxiliary-information technique (Technique \ref{lem:4.2}) to reduce Statement $3$ to ``Statement $3^\prime$''}, where the adversary is given the following auxiliary information in the beginning:

	\begin{itemize}
		\item $\llbracket\fAuxInf\rrbracket$
		\item $K$
		\item $K_{out}-K_{out}^{(i)}$
		\item $y_{b}^{(i)(j)}$ (recall it's the keys in $K_{temp}^{(i)(j)}$ with subscript $b$) for all $j\in [J]$.
		\item The lookup tables (sent in the step 1 of each round) encrypted under the keys in $K$ that are not at index $i$. (Thus there are $J\times (N-1)$ lookup tables. And the tables that are not provided are the tables encrypted under $K^{(i)}$ and $\Lambda^{(j)}$ for some $j$.) Denote it as $\fLT_{\text{not $K^{(i)}$}}$. 
	\end{itemize}
	(We assume the client has already sampled all the random coins needed thus all of them are well-defined.) In other words, we need to prove,
	\begin{center}\emph{(Statement $3^\prime$) Suppose the security parameter $\kappa$ is bigger than some constant. Suppose the initial state $\ket{\varphi}$ satisfies the conditions listed in Lemma \ref{lem:10.1}. For any adversary $\fAdv^\prime$ that only queries $H^{\prime}$ during the protocol, and the total number of queries to $H^\prime$ is at most $2^{\kappa+3}$, the post-execution state, defined as
		$$P_{pass}\fSecurityRefreshing_{\fAdv^\prime}(K,\Lambda; \ell,\kappa_{out})\circ(\qquad \qquad \qquad \qquad \qquad \qquad \qquad \qquad \qquad \qquad \qquad \qquad$$
		\begin{equation}\label{eq:359}\ket{\varphi}\odot \llbracket\fAuxInf\rrbracket \odot K\odot (K_{out}-K_{out}^{(i)})\odot \{y_{b}^{(i)(j)}\}_{j\in [J]}\odot \fLT_{\text{not $K^{(i)}$}})\end{equation}
		is $(2^{\eta_2/100\kappa},2^{-\eta_2/100\kappa}|\ket{\varphi}|)$-unpredictable for the key in $K_{out}^{(i)}$ with subscript $(1-b)$.}\end{center}
		(In Protocol \ref{prtl:11} we do not have a letter for the final output keys in $K_{out}^{(i)}$, we describe it as ``the key in $K_{out}^{(i)}$ with subscript $(1-b)$''; it is actually ``$pad^i||x_{1-b}^{(i)}||y_{1-b}^{(i)(J)}$'' in Protocol \ref{prtl:11}.)\par
		(Note that in Statement 3 we are talking about SC-security where $Tag(K_{out}^{(i)})$ are provided as auxiliary information; but in Statement $3^\prime$ in the definition of unpredictability only $Tag(pad^i||x_{1-b}^{(i)}||y_{1-b}^{(i)(J)})$ is provided. The reason that we can omit them is: the other global tag (which is $Tag(pad^i||x_{b}^{(i)}||y_{b}^{(i)(J)})$) can be computed on the server-side from the auxiliary information and client-side messages when the protocol completes; thus omitting this part does not make the adversary weaker (if we relax the query number bound on the adversary a little bit).)\par
	Let's talk about the motivation of adding so much auxiliary information. We note that if some client side messages can be deterministically computed from the information in the read-only buffer and some public randomness, these steps in the protocol can be removed (since the server can compute it by itself and does not need the client to send it out), thus we can simplify the protocol. This is why we add so much auxiliary information, and we will use it when we reduce ``Statement $3^{\prime\prime}$'' (not $3^\prime$) to the ``security of a temporary protocol'', in the step after the next step.\par
	\textbf{The next step is to strengthen Statement $3^\prime$ by reducing the blinded part of the random oracle.} Before that, let's give some symbols for the random pads used in the lookup table of this protocol. Notice that the lookup tables sent in the protocol have the following structure:
	\begin{enumerate}\item There is a lookup table for each key in $K^{(i)}$ and each pair of keys in $\Lambda^{(j)}$; \item Each table contains two rows; \item And each row is an output of $\fEn$ (see Definition \ref{def:2.13}), which contains a ciphertext and a key tag.\end{enumerate} Let's use $pad^{(i)(j)}_{b_x,b_r,\text{``ct''}}$ to denote the random pads used in the computation of the ciphertext part of the row of lookup tables encrypted under $x_{b_x}^{(i)}$, $r_{b_r}^{(j)}$; and use $pad^{(i)(j)}_{b_x,b_r,\text{``tg''}}$ to denote the random pads used in the computation of the key tag part of the row of lookup tables encrypted under $x_{b_x}^{(i)}$, $r_{b_r}^{(j)}$. So each row of the lookup table is in the form of $$(pad^{(i)(j)}_{b_x,b_r,\text{``ct''}}, H(pad^{(i)(j)}_{b_x,b_r,\text{``ct''}}||x^{(i)}_{b_x}||r_{b_r}^{(j)})\oplus y_{b_x}^{(i)(j)}),(pad^{(i)(j)}_{b_x,b_r,\text{``tg''}}, H(pad^{(i)(j)}_{b_x,b_r,\text{``tg''}}||x^{(i)}_{b_x}||r_{b_r}^{(j)}))$$
	Suppose $H^{\prime\prime}$ is a freshly new random oracle where \begin{equation}\label{eq:349}H(pad^{(i)(j)}_{1-b,b_r,ct}||x^{(i)}_{1-b}||r_{b_r}^{(j)}),H(pad^{(i)(j)}_{1-b,b_r,tg}||x^{(i)}_{1-b}||r_{b_r}^{(j)}),\forall j\in [J],b_r\in \{0,1\}\end{equation}
	are blinded. (Note that $i$, $1-b$ are fixed. Thus we blind $4J$ entries.)\par
	Since $H^\prime$ can also be understood as the blinded oracle coming from blinding $H^{\prime\prime}$, \textbf{ we can reduce Statement $3^\prime$ to Statement $3^{\prime\prime}$, defined as follows:}
	\begin{center}\emph{(Statement $3^{\prime\prime}$) Suppose the security parameter $\kappa$ is bigger than some constant. Suppose the initial state $\ket{\varphi}$ satisfies the conditions listed in Lemma \ref{lem:10.1}. For any adversary $\fAdv^\prime$ that only queries $H^{\prime\prime}$ and the total number of queries made by the adversary is at most $2^{\kappa+4}$, the post-execution state, defined as
		$$P_{pass}\fSecurityRefreshing_{\fAdv^\prime}(K,\Lambda; \ell,\kappa_{out})\circ(\qquad \qquad \qquad \qquad \qquad \qquad \qquad \qquad \qquad \qquad \qquad \qquad$$
		\begin{equation}\label{eq:399}\ket{\varphi}\odot \llbracket\fAuxInf\rrbracket \odot K\odot (K_{out}-K_{out}^{(i)})\odot \{y_{b}^{(i)(j)}\}_{j\in [J]}\odot \fLT_{\text{not $K^{(i)}$}})\end{equation}
		is $(2^{\eta_2/100\kappa},2^{-\eta_2/100\kappa}|\ket{\varphi}|)$-unpredictable for the key in $K_{out}^{(i)}$ with subscript $(1-b)$.}\end{center}
	(One may get confused by the fact that $pad^{(i)(j)}_{1-b,b_r,ct/tg}$ are sampled during the protocol, but in the security statement we assume the adversary can only query $H^{\prime\prime}$ from the beginning of the protocol. How can we define $H^{\prime\prime}$ if these random pads are not sampled out yet? The answer is here we can assume these pads have already sampled on the client side in the beginning, but they may not have been given to the server. Thus $H^{\prime\prime}$ is still well-defined.)\par
	We note that we give a lot of information to the adversary as public auxiliary information. So what are still not known by the adversary? The following is a list, and these are what are currently protecting the security of the output keys (we mean (\ref{eq:outkeys})):
	\begin{itemize}
		\item During the protocol, the adversary can only query the blinded oracle $H^{\prime\prime}$ where (\ref{eq:349}) are blinded.
		\item $y_{1-b}^{(i)(j)}$, $j\in [J]$ are hidden in the protocol, and they are not given as auxiliary information.
		\item The keys in $\Lambda$ are hidden. They can be seen as the keys that protect the security of $y_{1-b}^{(i)(j)}$, $j\in [J]$.
	\end{itemize}
	And since we already give many auxiliary information to the server in the beginning, we can simplify the protocol (which means, reduce Statement $3^{\prime\prime}$ to the security of a simplified protocol) through the following facts and arguments:
	\begin{enumerate}
	\item First let's review the initial state in equation (\ref{eq:399}):
			\begin{equation*}
				\ket{\varphi}\odot \llbracket\fAuxInf\rrbracket \odot K\odot (K_{out}-K_{out}^{(i)})\odot \{y_{b}^{(i)(j)}\}_{j\in [J]}\odot \fLT_{\text{not $K^{(i)}$}}
			\end{equation*}
		\item If in some steps some the client side messages can be computed from the content of read-only buffer and public random coins, these messages can be removed from the protocol. And we only need to slightly relax the query number bound of the adversary to allow it to compute these information by itself, and it does not become weaker during such simplification.
		\item Note that $i,b$ is already fixed, thus after we complete these simplification, we can define
		$$pad^{(j)}_{b_r,\text{``ct/tg''}}:=pad^{(i)(j)}_{b_x,b_r,\text{``ct/tg''}},y^{(j)}:=y_{1-b}^{(i)(j)},pad_{fixed}:=x^{(i)}_{1-b}$$
		to simplify the notations.
		\item We only care about the security and do not need to discuss the server's honest behavior.
	\end{enumerate}
	  \textbf{Thus to prove Statement $3^{\prime\prime}$, we can reduce it to the security statement of the following temporary simplified protocol}, described as follows: (below we describe not only the protocol itself but also some of the accompanied settings)
	
			\begin{prtl}[A temporary protocol $\fTempPrtl$ for the proof of Lemma \ref{lem:10.1}]\label{prtl:temp}This protocol is run on key set $\Lambda=\{r_{b_r}^{(j)}\}_{b_r\in \{0,1\},j\in [J]}$.\par
			Below $pad_{fixed}$ is stored in some fixed place of the read-only buffer of the initial state.\par 
				      The client samples $\{pad^{(j)}_{b_r,ct},pad^{(j)}_{b_r,tg}\}_{j\in [J]}$ ($b_r\in \{0,1\}$) from $\{0,1\}^l$. 
			      After this is completed, $H^{\prime\prime}$ is well-defined. ($pad_{fixed}:=x_{1-b}^{(i)}$, and see the definition around (\ref{eq:349})).\par
			      For each $j=1,\cdots J$:
			      \begin{enumerate}
				      \item The client samples $y^{(j)}$ from $\{0,1\}^{\kappa_{out}}$ which is different from a string stored in some fixed place of the read-only buffer of the initial state. Then it computes
				            \begin{equation}\label{eq:355r} H(pad^{(j)}_{b_r,ct}||pad_{fixed}||r_{b_r}^{(j)})\oplus y^{(j)},H(pad^{(j)}_{b_r,tg}||pad_{fixed}||r_{b_r}^{(j)}), \forall b_r\in \{0,1\}\end{equation}
				            and sends them together with the random pads $pad^{(j)}_{b_r,ct},pad^{(j)}_{b_r,tg}$ to the server.
				      \item The server runs some server-side operations as the attack, and it is only allowed to query $H^{\prime\prime}$ in this phase.
				      \item The client and the server do a padded Hadamard test on $\Lambda^{(j)}$ with pad length $\ell$ and output length $\kappa_{out}$. As before, the server can only query $H^{\prime\prime}$.
			      \end{enumerate}
			      After all these iterations are completed, the client samples $pad\leftarrow_r \{0,1\}^l$ and sends it to the server. The final output key is
			      \begin{equation}\label{eq:355}pad||pad_{fixed}||y^{(1)}||\cdots y^{(J)}\end{equation}
			      \end{prtl}
			Note that after the protocol completes, the adversary can query $H$.\par
	
	Now let's introduce some notations. Define 
	\begin{equation}\label{eq:371}\ket{\chi^{0}}:=\ket{\text{equation (\ref{eq:399})}}\end{equation}
	. The superscript ``$0$'' is to make it consistent with the notations in the following proofs. And $\ket{\chi^{0}}$ satisfies the followings (note that $\fLT_{\text{not $K^{(i)}$}}$ affects the SC-security, and applying Lemma \ref{lem:4.9} leads to the first condition below):
	\begin{itemize}
	\item $\forall j^\prime\in [J]$, $\ket{\chi^{0}}$ is $(2^{\eta_2-10},2^{-\eta_2+10}|\ket{\varphi}|)$-SC-secure for $\Lambda^{(j^\prime)}$  given $\Lambda-\Lambda^{(j^\prime)}$.
	\item $\ket{\chi^{0}}$ is $(2^{D},2^{D}+2^{\kappa})$-representable under $H$ from $\ket{\mathfrak{init}}$. $D,\log (1/|\ket{\varphi}|)\leq 2^\kappa$. ($|\ket{\chi^0}|=|\ket{\varphi}|$.)
	\end{itemize}

	\textbf{Thus the Statement $3^{\prime\prime}$ can be reduced to the ``Security of $\fTempPrtl$'', as follows} (here we use $\fTempPrtl$ as the notation for Protocol \ref{prtl:temp}):
	\begin{claim}[Security of Protocol \ref{prtl:temp}]
	Suppose the security parameter $\kappa$ is bigger than some constant. Suppose the initial state is $\ket{\chi^{0}}$. And suppose the following conditions are satisfied (the first two are the conditions above):
	\begin{itemize}
	\item (Security of the input) $\forall j^\prime\in [J]$, $\ket{\chi^{0}}$ is $(2^{\eta_2-10},2^{-\eta_2+10}|\ket{\varphi}|)$-SC-secure for $\Lambda^{(j^\prime)}$  given $\Lambda-\Lambda^{(j^\prime)}$. $2^\kappa>\eta_2$. $\eta_2^2>J\geq \eta_2$.
	\item (Well-behaveness of the input) $\ket{\chi^{0}}\in \cWBS(D)$. $D\leq 2^{\kappa}+\kappa$. 
\item (Sufficient padding length, output length) $\ell\geq 6D+7.9\eta_2$, $\kappa_{out}\geq \ell+4\eta_2$.
	\end{itemize}
	 Then if the protocol $\fTempPrtl$ and the adversary behaves as described in Protocol \ref{prtl:temp}, and during the protocol the number of adversary's queries to $H^{\prime\prime}$ is at most $2^{\kappa+5}$, then $P_{pass}\fTempPrtl\circ\ket{\chi^{0}}$ is $(2^{\eta_2/100\kappa},2^{-\eta_2/100\kappa}|\ket{\varphi}|)$-unpredictable for (\ref{eq:355}).
	\end{claim}
	Note that we do not want to define $\ket{\chi^{0}}$ using (\ref{eq:371}) but describe its properties using the two conditions just below (\ref{eq:371}). This is to make this statement more general and later it can be reused to prove the ``Statement 2''. And we also note that we omit the $\fAdv$ subscript in ``$P_{pass}\fTempPrtl_\fAdv\circ\ket{\chi^{0}}$'' since later in the proof we need to use the subscript to denote other things.
\end{proof}

\subsection{Part III: proving the ``Security of Protocol \ref{prtl:temp}''}\label{sec:ah5}
Now the Part III of the proof of Lemma \ref{lem:10.1} will focus on proving ``Security of Protocol \ref{prtl:temp}'', the temporary simplified protocol. Let's first prove some lemmas for preparation, and the outline for this step is given in Section \ref{sec:ai52}. So the reader can go to Section \ref{sec:ai52} for an overview.
\subsubsection{Preparation}
Before that, let's do some preparation. Note that in Protocol \ref{prtl:temp} there is a blinded oracle $H^{\prime\prime}$, and to handle the blinded oracle, we need to generalize some lemmas to this setting.\par
The main task in this subsubsection is to adapt the property of the padded Hadamard test (Corollary \ref{cor:7.3}) to this blinded-oracle setting.\par
First we generalize the Definition \ref{def:rep} to contain both the original oracle and the blinded oracle. This is needed here: recall that there do be two oracles in Protocol \ref{prtl:temp} ($H$ and $H^{\prime\prime}$).
\begin{defn}\label{def:ai1}
We say $\ket{\varphi}$ is $(2^{\alpha_1},2^{\alpha_2})$-representable under $H^{\prime\prime}$ from $\ket{\varphi_{init}}$ if it can be written as $\ket{\varphi}=\sum_{i=1}^{2^{\alpha_1}}\cP_i\ket{\varphi_{init}}$, and $\forall i$, in $\cP_i$ the total query number to $H^{\prime\prime}$ is at most $2^{\alpha_2}$. (Only one oracle is queried here.)\par
We say $\ket{\varphi}$ is $(2^{\alpha_1},2^{\alpha_2})$-representable under both $H,H^{\prime\prime}$ from $\ket{\varphi_{init}}$ if it can be written as $\ket{\varphi}=\sum_{i=1}^{2^{\alpha_1}}\cP_i\ket{\varphi_{init}}$, and $\forall i$, in $\cP_i$ the total query number to $H$ and $H^{\prime\prime}$ is at most $2^{\alpha_2}$. (The queries can contain both oracles.)\par
Define $\mathcal{WBS}^{H,H^{\prime\prime}}(D)$ to be the set of joint purified states (denoted as $\ket{\varphi}$) such that:
$$\text{$\ket{\varphi}$ is $(2^{\alpha_1},2^{\alpha_2})$-representable from $\ket{\mathfrak{init}}$ under both $H,H^{\prime\prime}$. $\alpha_1,\alpha_2,\log(1/|\ket{\varphi}|)\leq D$.} $$
\end{defn}
Similar to Definition \ref{def:rep} we call them the \emph{representation} of $\ket{\varphi}$.\par
And we recall Definition \ref{def:secblind} for the definition of SC-security in this blinded oracle setting, and further generalize it to cover the case where the adversary can query both oracles:
\begin{defn}
	Consider a key pair denoted as $K=\{x_0,x_1\}$. We say a purified joint state $\ket{\varphi}$ is $(2^{\eta},A)$-SC-secure for $K$ under $H$ and $H^{\prime\prime}$ together if:\par For any the server-side operation $\cD$ that queries $H$ and $H^{\prime\prime}$ and the total number of oracle queries to $H$ and $H^{\prime\prime}$ is at most $2^\eta$, $|P_{x_0||x_1}\cD(\ket{\varphi}\odot Tag(K))|\leq A$.
\end{defn}
Then we can describe our adaptation of Corollary \ref{cor:7.3} in the current blinded oracle setting.
\begin{lem}\label{lem:h3}
	Suppose the security parameter $\kappa$ is bigger than some constant. $\Lambda$ is a pair of keys. $Tag(\Lambda)$ is stored in some fixed place of the read-only buffer.\par
	 Suppose $H^{\prime\prime}$ is a blinded oracle of $H$ which comes from blinding $Set$, where $Set$ is a set of entries with size at most $2^\kappa$ and element length bigger than $(l+\text{the length of keys in $\Lambda$})$. (Remark: this is just a simple way to ensure $Set$ is not too ill-behaved and does not have overlaps with some entries we care about. When we apply this lemma these two conditions are satisfied easily.)\par
	  Suppose the initial purified joint state $\ket{\varphi}$ satisfies:
	\begin{itemize}
		\item (Security of the input) $\ket{\varphi}$ is $(2^\eta,2^{-\eta}|\ket{\varphi}|)$-SC-secure for $\Lambda$ under $H^{\prime\prime}$. $\eta>10\kappa$.
		\item (Well-behaveness of the input) $\ket{\varphi}\in \cWBS^{H,H^{\prime\prime}}(D)$, $D\leq 2^\kappa+\kappa$.
		\item (Sufficient padding length and output length) $l>6D+2\eta$. $\kappa_{out}>l+\eta$.
	\end{itemize}
	Then the following conclusion holds:\par
	For any adversary $\fAdv^{\prime}$ that only queries $H^{\prime\prime}$ for less than $2^{\eta/20}$ times, the post-execution state
	$$\ket{\varphi^\prime}=\fPadHadamard_{\fAdv^\prime}(\Lambda;\underbrace{ \ell}_{\substack{\text{padding} \\ \text{length}}},
\underbrace{ \kappa_{\text{out}}}_{\substack{\text{output} \\ \text{length}}})\circ\ket{\varphi}$$
	one of the following two is true:
	\begin{itemize}
		\item 	$|P_{pass}\ket{\varphi^\prime}|\leq \frac{35}{36}|\ket{\varphi}|$
		\item $P_{pass}\ket{\varphi^\prime}$ is $(2^{\eta/6},\frac{1}{3}|\ket{\varphi}|)$-ANY-secure for $\Lambda$ under $H^{\prime\prime}$.
	\end{itemize}
\end{lem}
This statement is an analog of Corollary \ref{cor:7.3} of Lemma \ref{lem:7.3} under a blinded oracle. And the proof is also similar. We just need to adapt some key steps to the current setting.\par
\begin{proof} Recall the proof of Lemma \ref{lem:7.3}. The argument by (\ref{eq:119i}) remains the same. Thus what we need to do is to give a bound for
$$p_0:=|P_{pass}P_{x_0}\cU\ket{\varphi^\prime}|=|P_{pass}P_{x_0}\cU ( \fPadHadamard_\fAdv\circ\ket{\varphi})|$$
The next step is to replace $\ket{\varphi^1}$ (the state after the random pad $pad$ is sampled out, as given in the proof of Lemma \ref{lem:7.3}) by $\ket{\tilde\varphi}$:
\begin{center}
Take the representation (Definition \ref{def:ai1}) of $\ket{\varphi}$, replace $H$ by $H(I-P_{pad||\cdots})$ and replace $H^{\prime\prime}$ by $H^{\prime\prime}(I-P_{pad||\cdots})$, then we get $\ket{\tilde\varphi}$.	
\end{center}
. The setting here is slightly different but we can still apply Lemma \ref{lem:3.4r} since the blinded oracle can be expressed using $H$ and the description of the blinded part. We have $\ket{\tilde \varphi}\approx_{2^{-\eta+1}|\ket{\varphi}|}\ket{\varphi^1}$ and it does not depend on $H(pad||\cdots)$, where ``$\cdots$'' denotes arbitrary strings of length equal to the keys in $\Lambda$.\par
Then we can similarly define
\begin{equation}\label{eq:405z}\ket{\psi}=P_{x_0}\cU^{blind} (\fPadHadamard^{\geq 2}_{\fAdv^{blind}}\circ\ket{\tilde \varphi})\end{equation}
where $\cU^{blind}$ and $\fAdv^{blind}$ mean, each oracle query in them is replaced by the query to the blinded oracle where $pad||x_1$ is blinded. Note that there is a difference: the blinding operation is done on $H^{\prime\prime}$. Then we can similarly apply Lemma \ref{lem:4.16} and get
\begin{align}
		\ket{\psi}           & \approx_{2^{-\eta/3+2}|\ket{\tilde\varphi}|}P_{x_0}\cU (\fPadHadamard_{\fAdv}^{\geq 2}\circ\ket{\tilde\varphi}) \\
		\therefore\ket{\psi} & \approx_{2^{-\eta/3+3}|\ket{\varphi}|}P_{x_0}\cU\ket{\varphi^\prime}\end{align}
Finally similarly we can apply Lemma \ref{lem:C1} on (\ref{eq:405z}) and complete the proof.
\end{proof}

\subsubsection{Overview of the proof of the security of Protocol \ref{prtl:temp}}\label{sec:ai52}
\begin{enumerate}
\item Notice the structure of the protocol can be seen as a loop. First, we use a technique similar to the \emph{multi-round decomposition method} described in Section \ref{sec:4.8.3}. In more details, we need to apply an argument repeatedly. We will iterate a similar argument for many rounds to decompose the state. The argument in each round has a similar form.\par
Each round (for example, analysis of the $j$-th round of the protocol) of argument goes as follows:
\begin{enumerate}
\item[.] The initial state is denoted by $\ket{\chi^{j-1}}$.\par
We note that there is one key difference in our proof here from the technique description in Section \ref{sec:4.8.3}: in each round of decomposition we will use the \emph{auxiliary-information technique} to add the keys $y^{(j-1)}$ as the auxiliary information. This step is necessary for the proofs after this \emph{multi-round decomposition arguments}.\par
\item The post-execution state is decomposed to two states $\ket{\phi^j}$ and $\ket{\chi^j}$. In more details, we will first apply Lemma \ref{lem:h3} to analyze the protocol in this round, and do the decomposition based on a discussion-by-cases on the post-execution state. 
\end{enumerate}
After this iteration-style proof completes, we reduce the ``Security of Protocol \ref{prtl:temp}'' to a list of statements on $\ket{\phi^{j}}$ ($j=1,2\cdots \eta_2/\kappa$), and $\ket{\chi^{\eta_2/\kappa}}$, (see (\ref{eq:427zx})-(\ref{eq:363})) where $\ket{\phi^{\cdots}}$, $\ket{\chi^{\cdots}}$ are defined round-by-round during the proof. And we will see, since the norm of $\ket{\chi^{\eta_2/\kappa}}$ is already exponentially small, we only need to prove the following statement on $\ket{\phi^{j}}$: \begin{center}$\forall j$, $P_{pass}\fTempPrtl_{>j}\circ\ket{\phi^{j}}$ is $(2^{\eta_2/70\kappa},2^{-\eta_2/70\kappa}|\ket{\varphi}|)$-unpredictable for (\ref{eq:355}).\end{center}
We name it as ``Statement 4''
\item The proof of the ``Statement 4'' above can be further divided into the following steps:
\begin{enumerate}
\item In Section \ref{sec:ai54} we slightly simplify the ``Statement 4'' to ``Statement $4^\prime$'' as a preparation for the further proof. ``Statement $4^\prime$'' is ``$\forall j$, $\ket{\phi^{j}}\odot \llbracket\fTempPrtl_{>j}\rrbracket$ is $(2^{\eta_2/70\kappa},2^{-\eta_2/70\kappa}|\ket{\varphi}|)$-unpredictable for (\ref{eq:355})''.
\item We will first prove for each $j$, $\ket{\phi^{j}}\odot \llbracket\fTempPrtl_{>j}\rrbracket\odot\{y^{(j^\prime)}\}_{j^\prime> j}$ is \\$(2^{\eta_2/20\kappa},2^{-\eta_2/20\kappa}|\ket{\varphi}|)$-ANY-secure for $\Lambda^{(j)}$;\item Then based on it, we prove the unpredictability in (\ref{eq:355}).\end{enumerate}
\end{enumerate}
We describe the proof in four subsubsections. The first subsubsection (Section \ref{sec:ai53}) is the first step above. Section \ref{sec:ai54} is the step 2.a above, Section \ref{sec:ai55} is the step 2.b above, and the Section \ref{sec:ai56} is the step 2.c above. 
\subsubsection{Proof of ``Security of Protocol \ref{prtl:temp}'', step 1: decomposition}\label{sec:ai53}
We will organize different steps in a single round of this ``iteration-style'' proof using boxes: the argument in each round can be broken into different pieces, and we put them into boxes. For the argument outside the boxes --- which is, how different pieces are connected together, we will first describe them using the first round as an example, then describe the argument more generally.\par
\begin{proof}[Proof of Lemma \ref{lem:10.1}, part III.Step 1]
	We use $\fTempPrtl_{>i}$ to denote the protocol starting from round $(i+1)$. The index of round counter starts at $1$. Thus the security statement of Protocol \ref{prtl:temp} can be re-written as:
	\begin{center}
		$P_{pass}\fTempPrtl_{>0}\ket{\chi^{0}}$ is $(2^{\eta_2/100\kappa},2^{-\eta_2/100\kappa}|\ket{\varphi}|)$-unpredictable for (\ref{eq:355}).
	\end{center}

	As what we said before, we will first use the first round of the argument as an example, then describe the argument more generally. So the statements outside the boxes use concrete indexes ($0,1$, etc) while the statements inside the boxes use letters to denote the indexes.\par
	Recall that $\ket{\chi^0}$ satisfies:
	\begin{mdframed}
		\textbf{Condition on $\ket{\chi^{0}}$:}
		\begin{center}\emph{$\forall j\in [J]$, $\ket{\chi^{0}}$ is $(2^{\eta_2-10},2^{-\eta_2+10}|\ket{\varphi}|)$-SC-secure for $\Lambda^{(j)}$  given $\Lambda-\Lambda^{(j)}$.}\end{center}
		And apply Lemma \ref{lem:4.9} we know
		\begin{equation}\text{$\ket{\chi^{0}}\odot \llbracket\fTempPrtl_{=1.1}\rrbracket$ is $(2^{\eta_2-20},2^{-\eta_2+20}|\ket{\varphi}|)$-SC-secure for $\Lambda^{(1)}$.}
		\end{equation}
		where $\llbracket\fTempPrtl_{=1.1}\rrbracket$ is the client side message in the first step of the first round. We take it into consideration as the preparation of the following proofs.\par
		Recall $H^{\prime\prime}$ is a freshly new blinded oracle (see (\ref{eq:349})) and the paddings are sampled randomly and can be added into the auxiliary information without affecting the SC-security, apply Lemma \ref{lem:4.14a} (and use Fact \ref{fact:injtag} to show the ``$Tag$ is not injective on these inputs'' has very small norm) we have
				\begin{equation}\text{$\ket{\chi^{0}}\odot \llbracket\fTempPrtl_{=1.1}\rrbracket$ is $(2^{\eta_2-22},2^{-\eta_2+22}|\ket{\varphi}|)$-SC-secure for $\Lambda^{(1)}$ under $H$ and $H^{\prime\prime}$.}
		\end{equation}
		But if we want to apply a statement inductively sometimes we want a statement that has consistent form in each round; one common technique is to loosen the parameters. Here we loosen the parameters from $\eta_2$ to $\eta_2/2$ here to make the statement consistent. The exact meaning of it will be clear later.\par
		Thus we have:\par
		\textbf{Property of $\ket{\chi^{0}}$ for $\Lambda^{(1)}$  after the relaxation:}
						\begin{equation}\label{eq:217}\text{$\ket{\chi^{0}}\odot \llbracket\fTempPrtl_{=1.1}\rrbracket$ is $(2^{\eta_2/2},2^{-\eta_2/2}|\ket{\varphi}|)$-SC-secure for $\Lambda^{(1)}$ under $H$ and $H^{\prime\prime}$.}
		\end{equation}
		\end{mdframed}
	Then we need to study the behavior of the state in the next round. Denote
	\begin{equation}\label{eq:268}\ket{\psi^{1}}:=P_{pass}\fTempPrtl_{=1}\ket{\chi^{0}}\end{equation}
	, the passing part of the post-execution state after the $1$-st round of Protocol \ref{prtl:temp} completes when the initial state is $\ket{\chi^{0}}$.\par 
	We can decompose $\ket{\psi^{1}}$ into $\ket{\phi^{1}}+\ket{\chi^{1}}$ as shown in the following box. Note that (1)  below we describe the decomposition for general $t$ to make it consistent in each round of the iteration-style proof. In the first round we can simply substitute $j=1$. (2) the reader might get confused on why there is ``$\odot y^{(j-1)}$'' below. For $j=1$ this term does not exist, but this is needed for the later round of this iteration-style proof. (We will explain the reason for doing it later.)\\
	\begin{mdframed}
		{\small\textbf{State decomposition at round $j$, which is,} \begin{equation}\label{eq:360r}\ket{\psi^{j}}\quad (:=P_{pass}\fTempPrtl_{=j}(\ket{\chi^{j-1}}\odot y^{(j-1)})),\text{ together with server's ancillas, } =\ket{\phi^{j}}+\ket{\chi^{j}}\end{equation}}

		(Case 0) First, if $|\ket{\chi^{j-1}}|\leq 2^{-\eta_2/10}|\ket{\varphi}|$, take $\ket{\phi^{j}}=0$, $\ket{\chi^{j}}=\ket{\psi^{j}}$ and the decomposition is completed. Go to the summary of this box in the end (``thus for all the cases there are $\cdots$'').\par
		(Case 1) Otherwise, $|\ket{\chi^{j-1}}|> 2^{-\eta_2/10}|\ket{\varphi}|$. 
		Recall that in the $\fTempPrtl_{=j}$ step of the protocol, the following operations are executed:
		\begin{enumerate}
			\item In the first step (denoted as $\fTempPrtl_{=j.1}$) the client computes and sends two ciphertexts (encrypted under $\Lambda^{(j)}$) and two key tags (of $\Lambda^{(j)}$); 
			\item Then a padded Hadamard test on $\Lambda^{(j)}$ is executed.\end{enumerate}
		Let's first analyze the state after the first step. By ``the properties of $\ket{\chi^{j-1}}$'' and the norm of $\ket{\chi^{j-1}}$ in the beginning of this case we know:
		\begin{center}
			$\fTempPrtl_{=j.1}(\ket{\chi^{j-1}}\odot y^{(j-1)})$ is $(2^{\eta_2/2.5},2^{-\eta_2/2.5}|\ket{\chi^{j-1}}|)$-SC-secure for $\Lambda^{(j)}$ under $H$ and $H^{\prime\prime}$.
		\end{center}
		Then we can apply the property of the padded Hadamard Test. 
		Apply Lemma \ref{lem:h3} and discuss by cases:
		\begin{itemize}
			\item (Case 1.1) $|\ket{\psi^{j}}|\leq \frac{35}{36}|\ket{\chi^{j-1}}|$. Then take $\ket{\chi^{j}}=\ket{\psi^{j}}$ and $\ket{\phi^{j}}=0$.
			\item (Case 1.2) $\ket{\psi^{j}}$ is $(2^{\eta_2/15}, \frac{1}{3}|\ket{\chi^{j-1}}|)$-ANY-secure for $\Lambda^{(j)}$ under $H^{\prime\prime}$.\par
			      Then apply the decomposition lemma for ANY-security (Lemma \ref{lem:4.5}) and consider $H^{\prime\prime}$ as the random oracle, we can decompose $\ket{\psi^{j}}$ as $\ket{\phi^{j}}+\ket{\chi^{j}}$ such that
			      \begin{enumerate}
				      \item $|\ket{\chi^{j}}|\leq \frac{2.5}{3}|\ket{\chi^{j-1}}|$, and is $(\eta_2/2, 2^{\eta_2/2})$-server-side representable from $\ket{\psi^{j}}$ under $H^{\prime\prime}$.
				      \item $\ket{\phi^{j}}$ is $(2^{\eta_2/100},2^{-\eta_2/100}|\ket{\varphi}|)$-ANY-secure for $\Lambda^{(j)}$ under $H^{\prime\prime}$, and is $(1,2^{\eta_2/2})$-server-side-representable from $\ket{\psi^{j}}$ under $H^{\prime\prime}$.
			      \end{enumerate}
		\end{itemize}
		Thus for all the cases there are
		\begin{itemize}
			\item $|\ket{\chi^{j}}|\leq \frac{35}{36}|\ket{\chi^{j-1}}|\leq (\frac{35}{36})^j|\ket{\varphi}|$, and is $(\eta_2/2, 2^{\eta_2/2})$-server-side representable from $\ket{\psi^{j}}$ under $H^{\prime\prime}$. 
			\item $\ket{\phi^{j}}$ is $(2^{\eta_2/100},2^{-\eta_2/100}|\ket{\varphi}|)$-ANY-secure for $\Lambda^{(j)}$ under $H^{\prime\prime}$, and is $(1,2^{\eta_2/2})$-server-side-representable from $\ket{\psi^{j}}$ under $H^{\prime\prime}$.
		\end{itemize}

	\end{mdframed}
	After the decomposition, we can reduce ``Security of Protocol \ref{prtl:temp}'' to two statements:
	\begin{equation}\label{eq:363r}\text{
		$	P_{pass}\fTempPrtl_{>1}\ket{\phi^{1}}$ is $(2^{\eta_2/70\kappa},2^{-\eta_2/70\kappa}|\ket{\varphi}|)$-unpredictable for (\ref{eq:355}).}
	\end{equation}
	\begin{equation}\label{eq:360}\text{
		$P_{pass}\fTempPrtl_{>1}\ket{\chi^{1}}$ is $(2^{\eta_2/100\kappa},(2^{-\eta_2/100\kappa}-2^{-\eta_2/70\kappa})|\ket{\varphi}|)$-unpredictable for (\ref{eq:355}).}
	\end{equation}

	For (\ref{eq:363r}), we postpone it and later prove it together with other statements of the similar form. For (\ref{eq:360}), we first apply the \emph{auxiliary-information technique} to add some auxiliary information to (\ref{eq:360}) and reduce (\ref{eq:360}) to the following statement:
	\begin{center}
		$P_{pass}\fTempPrtl_{>1}(\ket{\chi^{1}}\odot y^{(1)})$ is $(2^{\eta_2/100\kappa},(2^{-\eta_2/100\kappa}-2^{-\eta_2/70\kappa})|\ket{\varphi}|)$-unpredictable for (\ref{eq:355}).
	\end{center}
	Why do we want to add the extra $y^{(\cdot)}$ as auxiliary information? The reason is, as the previous cases where we apply the \emph{auxiliary-information technique}, adding more information to the read-only buffer helps us analyze the server's state. On the other hand, to analyze the behavior of the states in the protocol $\fTempPrtl_{>1}$, the conditions we need are only the norm of $\ket{\chi^{1}}$ and its security for $\Lambda^{(j)}$, $j\geq 2$, and its security for $\Lambda^{(1)}$ is not needed any more. Thus providing these information does not affect the proof later.\par
	Note that the reason above does not hold for the $\ket{\phi^{\cdots}}$ branch. We only add this auxiliary information when we do this ``iteration-style proof'' on the $\ket{\chi^{\cdots}}$ branch.\par
	Then we can prove $\ket{\chi^{1}}\odot y^{(1)}$ satisfies a similar statement as ``Properties of $\ket{\chi^{0}}$''. Here we write the arguments below for general $j\geq 0$. Note that we can substitute $j=1$ to get the conditions for this round.
	\begin{mdframed}
		\textbf{Properties of $\ket{\chi^{j}}$:}\par
		We are going to prove:
		\begin{center}\emph{$\forall j^\prime\in [j+1,J]$, $\ket{\chi^{j}}\odot y^{(j)}\odot \llbracket\fTempPrtl_{=(j+1).1}\rrbracket$ is $(2^{\eta_2-22}-j2^{\eta_2/2}-j2^{\kappa+5}, 2^{-\eta_2+22+j\log\eta_2}|\ket{\varphi}|)$-SC-secure for $\Lambda^{(j^\prime)}$ given $\Lambda^{(\geq j+1)}-\Lambda^{(j^\prime)}$ under $H$ and $H^{\prime\prime}$}.\end{center}
		And we additionally have
		\begin{equation}\label{eq:377}
			|\ket{\chi^{j}}|\leq (\frac{35}{36})^j|\ket{\varphi}|
		\end{equation}

		\begin{center}
			$\ket{\chi^{j}}$ is $(2^{D+j\log\eta_2},2^{D}+j2^{\eta_2})$-representable from $\ket{\mathfrak{init}}$ under $H$ and $H^{\prime\prime}$.
		\end{center}
		The proof is given below. Note that (\ref{eq:377}) is already proved inductively in the ``state decomposition'' box. (And we further note that this box is used step-by-step in an upper-level induction-style proof and we could not use it to get the properties of $\ket{\chi^{j}}$ for all $j\in [J]$ once. We need to repeat the argument in this box every time when $j$ increases. If we are currently at round $j$, $\ket{\chi^{j^\prime}}$ for $j^\prime>j$ is not even well-defined yet.)
		\begin{proof}[Proof of the properties of $\ket{\chi^{j}}$]
			Recall how $\ket{\chi^{j}}$ is constructed out.
			\begin{enumerate}
				\item $\ket{\chi^{j}}$ is $(\eta_2/2,2^{\eta_2/2})$-server-side-representable from $\ket{\psi^{j}}$ under $H^{\prime\prime}$; 
				\item $\ket{\psi^{j}}:=P_{pass}\fTempPrtl_{=j}(\ket{\chi^{j-1}}\odot y^{(j-1)})$, where in $\fTempPrtl_{=j}$ the client side messages come from an algorithm on the random paddings, $\Lambda^{(j)}$, $y^{(j)}$ (freshly sampled in this step), and random coins.\\ And the query number of the adversary in this step is at most $ 2^{\kappa+5}$.
				\item And $\ket{\chi^{j-1}}$ is constructed recursively.
			\end{enumerate}

			By Lemma \ref{lem:4.9} and ``the condition of $\ket{\chi^{0}}$'' we know $\ket{\chi^{0}}\odot \llbracket\fTempPrtl_{=(j+1).1}\rrbracket$ is\\ $(2^{\eta_2-22},2^{-\eta_2+22}|\ket{\varphi}|)$-SC-secure for $\Lambda^{(j^\prime)}$ given $\Lambda-\Lambda^{(j^\prime)}$ under $H$ and $H^{\prime\prime}$. (We implicitly apply Lemma \ref{lem:4.14a} to add $H^{\prime\prime}$ to the statement.)  Then by Lemma \ref{lem:basic} and the construction of $\ket{\chi^j}$ listed above we complete the proof of the property on the top of this box. Note that the ``$-j2^{\eta_2/2}$'' and the $j\log \eta_2$ terms come from the decomposition step (step 1 above), and the ``$-j2^{\kappa+5}$'' comes from the adversary's operation.
		\end{proof}
		Then we can relax the parameter and get a statement that has consistent form for any round $t$, and consistent with (\ref{eq:217}):\par
		\textbf{The property after relaxation on $\ket{\chi^{j}}$ for $\Lambda^{(j+1)}$:}
		{\footnotesize\begin{equation}\text{$\ket{\chi^{j}}\odot y^{(j)}\odot \llbracket\fTempPrtl_{=(j+1).1}\rrbracket$ is $(2^{\eta_2/2},2^{-\eta_2/2}|\ket{\varphi}|)$-SC-secure for $\Lambda^{(j+1)}$ under $H$ and $H^{\prime\prime}$.}
		\end{equation}}
	\end{mdframed} 
	Then similar technique can be applied again on $P_{pass}\fTempPrtl_{>1}\ket{\chi^{1}}$, and (\ref{eq:360}) can be further reduced to two statements.\par
	And we can repeat this argument round-by-round. And we can continue to apply the same argument for $\eta_2/\kappa$ rounds (as long as $j\log\eta_2\leq \eta_2/2$, which means we can do the ``relaxation'' within the box of ``properties for $\ket{\chi^{\cdots}}$'') and construct $\ket{\chi^{2}}$, $\ket{\chi^{3}}$, $\ket{\chi^{4}}$ etc, and finally we can reduce the proof of ``Security of Protocol \ref{prtl:temp}'' to the following statements:
	\begin{equation}\label{eq:427zx}\text{
		$	P_{pass}\fTempPrtl_{>1}\ket{\phi^{1}}$ is $(2^{\eta_2/70\kappa},2^{-\eta_2/70\kappa}|\ket{\varphi}|)$-unpredictable for (\ref{eq:355}).}
	\end{equation}
	\begin{center}
		$	P_{pass}\fTempPrtl_{>2}\ket{\phi^{2}}$ is $(2^{\eta_2/70\kappa},2^{-\eta_2/70\kappa}|\ket{\varphi}|)$-unpredictable for (\ref{eq:355}).
	\end{center}
	$$\cdots\cdots $$
	\begin{center}
		$	P_{pass}\fTempPrtl_{>\eta_2/\kappa}\ket{\phi^{\eta_2/\kappa}}$ is $(2^{\eta_2/70\kappa},2^{-\eta_2/70\kappa}|\ket{\varphi}|)$-unpredictable for (\ref{eq:355}).
	\end{center}
	{\scriptsize\begin{equation}\label{eq:363}\text{
		$P_{pass}\fTempPrtl_{>\eta_2/\kappa}\ket{\chi^{\eta_2/\kappa}}$ is $(2^{\eta_2/100\kappa},(2^{-\eta_2/100\kappa}-(\eta_2/\kappa) 2^{-\eta_2/70\kappa})|\ket{\varphi}|)$-unpredictable for (\ref{eq:355}).}
	\end{equation}}
	Note that (\ref{eq:363}) is already proved by the ``properties on $\ket{\chi^{\cdots}}$'':
	\begin{equation}\label{eq:25}|\ket{\chi^{\eta_2/\kappa}}|\leq (\frac{35}{36})^{-\eta_2/\kappa}|\ket{\chi^{0}}|= (\frac{35}{36})^{-\eta_2/\kappa}|\ket{\varphi}|<(2^{-\eta_2/100\kappa}-(\eta_2/\kappa) 2^{-\eta_2/70\kappa})|\ket{\varphi}| \end{equation}
	So the remaining problem is to study the SC-security of $\fTempPrtl_{>j}\circ\ket{\phi^{j}}$.
	Our goal is to prove:
	\begin{center} (Statement 4)\emph{$\forall j$, $P_{pass}\fTempPrtl_{>j}\circ\ket{\phi^{j}}$ is $(2^{\eta_2/70\kappa},2^{-\eta_2/70\kappa}|\ket{\varphi}|)$-unpredictable for (\ref{eq:355}).}\end{center}
\end{proof}
Now we have reduced the proof of ``security of Protocol \ref{prtl:temp}'' to the proof of Statement 4 above. In the next subsubsection we will prove Statement 4.
\subsubsection{A slight simplification of Statement 4: Statement $4^\prime$}\label{sec:ai54}
	By Lemma \ref{lem:basic}, to prove Statement 4, it's enough to prove
	\begin{center}(Statement $4^\prime$)\emph{$\forall j$, $\ket{\phi^{j}}\odot\llbracket\fTempPrtl_{>j}\rrbracket$ is $(2^{\eta_2/70\kappa},2^{-\eta_2/70\kappa}|\ket{\varphi}|)$-unpredictable for (\ref{eq:355}).}\end{center}. 
	Note that in this statement the unpredictability is defined on the original (unblinded) oracle $H$.\par
	
\subsubsection{Step 2b of the overview in Section \ref{sec:ai52}}\label{sec:ai55}
		What conditions do we have on $\ket{\phi^{j}}$? From the ``state decomposition'' box in the previous subsubsection we know
		\begin{equation}\label{eq:282}\text{$\ket{\phi^{j}}$ is $(2^{\eta_2/1000},2^{-\eta_2/1000}|\ket{\varphi}|)$-ANY-secure for $\Lambda^{(j)}$ under $H^{\prime\prime}$}\end{equation}
		To continue, let's write down a summary for the source of state $\ket{\phi^{j}}$:
		\begin{equation}\label{eq:403}
			\ket{\phi^{j}}\text{ is $(1,2^{\eta_2/2})$-server-side representable from $\ket{\psi^{j}}$ under $H^{\prime\prime}$}\end{equation}
		\begin{equation}\label{eq:419}
			\forall j^\prime\leq j, \ket{\psi^{j^\prime}}:=P_{pass}\fTempPrtl_{=j^\prime}(\ket{\chi^{j^\prime-1}}\odot y^{(j^\prime-1)})
		\end{equation}
		\begin{equation*}
			\text{ where the adversary queries $H^{\prime\prime}$ for $\leq 2^{\kappa+5}$ queries, }\end{equation*}\begin{equation*}y^{(j^\prime-1)}\text{ is sampled randomly in }\fTempPrtl_{=j^\prime-1}
		\end{equation*}
		\begin{equation}\label{eq:369}
			\forall j^\prime\leq j, \ket{\chi^{j^\prime}}\text{ is $(\eta_2/2,2^{\eta_2/2})$-server-side representable from $\ket{\psi^{j^\prime}}$ under $H^{\prime\prime}$}\end{equation}
where $\ket{\chi^0}$ is the initial state, whose properties are given in the security statement of $\fTempPrtl$. Thus recall
		\begin{equation}\label{eq:385}\text{$\ket{\chi^0}$ is $(2^{D},2^{D})$-representable from $\ket{\mathfrak{init}}$.}\end{equation}
		And the goal of this subsubsection is to prove
		\begin{equation}\label{eq:283}\text{$\ket{\phi^{j}}\odot\llbracket\fTempPrtl_{>j}\rrbracket\odot\{y^{(j^\prime)}\}_{j^\prime> j}$ is $(2^{\eta_2/20\kappa},2^{-\eta_2/20\kappa}|\ket{\varphi}|)$-ANY-secure for $\Lambda^{(j)}$.}\end{equation}
		What's the difference of (\ref{eq:282}) and (\ref{eq:283})? There are three differences:
		\begin{itemize}\item Whether there is additional auxiliary information $\llbracket\fTempPrtl_{>j}\rrbracket$; note that this part also exists in ``Statement $4^\prime$''; \item We add $\{y^{(j^\prime)}\}_{j^\prime> j}$, the temporary output key in round $[j+1,J]$, into the auxiliary information; this makes this statement stronger, and is useful for the proofs later; \item Whether the ANY-security is under $H$ or $H^{\prime\prime}$.\\
			      Recall that $H^{\prime\prime}$ is the blinded oracle where the entries in the form of $$pad^{j}_{b_r,ct}||pad_{fixed}||r^{(j)}_{b_r},pad^{j}_{b_r,tg}||pad_{fixed}||r^{(j)}_{b_r},\forall b_r\in \{0,1\}, j\in [J]$$ are blinded, and the adversary can only query $H^{\prime\prime}$ during the protocol. However, what we want to prove is the unpredictability under $H$.
		\end{itemize}
		To prove it, we will use a technique that is similar to the proof of Lemma \ref{lem:4.9}.
		\begin{proof}[Proof of (\ref{eq:283})]
			\textbf{We first reduce (\ref{eq:283}) to some simpler-but-stronger statement.}\par
			First note that $\llbracket\fTempPrtl_{>j}\rrbracket$ contains two parts: the client side messages of the third step (the padded Hadamard test) in each round, and the client side messages of the first step in each round, which contain the ``ciphertexts'' and the ``key tags'', as shown in (\ref{eq:355r}). Recall that they are 
			\begin{equation}\label{eq:424}H(pad^{j^\prime}_{b_r,ct}||pad_{fixed}||r_{b_r}^{(j^\prime)})\oplus y^{(j^\prime)},H(pad^{j^\prime}_{b_r,tg}||pad_{fixed}||r_{b_r}^{(j^\prime)}),\forall b_r\in \{0,1\},j^\prime\in [j+1,J]\end{equation}
			Since $\{y^{(j^\prime)}\}_{j^\prime> j}$ is also given as the auxiliary information, (\ref{eq:283}) can be further strengthened and reduced to the following statement:
			\begin{equation}\label{eq:376r}\ket{\phi^{j}}\odot\llbracket HOutput_{>j}\rrbracket\odot \{y^{(j^\prime)}\}_{j^\prime> j}\text{ is $(2^{\eta_2/20\kappa},2^{-\eta_2/20\kappa}|\ket{\varphi}|)$-ANY-secure for $\Lambda^{(j)}$.}\end{equation}
			where $\llbracket HOutput_{>j}\rrbracket$ is defined to be the random oracle outputs in the form of
			\begin{equation}\label{eq:373}H(pad^{j^\prime}_{b_r,ct}||pad_{fixed}||r_{b_r}^{(j^\prime)}), H(pad^{j^\prime}_{b_r,tg}||pad_{fixed}||r_{b_r}^{(j^\prime)}),\forall b_r\in \{0,1\},j^\prime\in [j+1,J]\end{equation}
			which are the oracle outputs of $H$ on the blinded part for $j^\prime\in [j+1,J]$.\par
			Then notice that $y^{(j^\prime)}(j^\prime> j)$ are sampled freshly randomly. Thus (\ref{eq:376r}) is reduced to proving
			\begin{equation}\label{eq:378r}\text{$\ket{\phi^{j}} \odot\llbracket HOutput_{>j}\rrbracket$ is $(2^{\eta_2/20\kappa},2^{-\eta_2/20\kappa}|\ket{\varphi}|)$-ANY-secure for $\Lambda^{(j)}$.}\end{equation}
			\textbf{Now the problem of proving (\ref{eq:283}) boils down to proving (\ref{eq:378r}).}\par
			Corresponding to the definition of $\llbracket HOutput_{>j}\rrbracket$, define $\llbracket HOutput_{<j}\rrbracket$ as the random oracle outputs of $H$ for $j^\prime\in [1,j-1]$:
			\begin{equation}\label{eq:425r}H(pad^{j^\prime}_{b_r,ct}||pad_{fixed}||r_{b_r}^{(j^\prime)}), H(pad^{j^\prime}_{b_r,tg}||pad_{fixed}||r_{b_r}^{(j^\prime)}),\forall b_r\in \{0,1\},j^\prime\in [1,j-1]\end{equation}
			and define $\llbracket HOutput_{=j}\rrbracket$ as the random oracle outputs of $H$ when $j^\prime=j$:
			\begin{equation}\label{eq:381r}H(pad^{j^\prime}_{b_r,ct}||pad_{fixed}||r_{b_r}^{(j^\prime)}), H(pad^{j^\prime}_{b_r,tg}||pad_{fixed}||r_{b_r}^{(j^\prime)}),\forall b_r\in \{0,1\},j^\prime=j\end{equation}
			.  We note that these ``$HOutput$'' describe the random oracle outputs of $H$ (instead of $H^{\prime\prime}$'') on the blinded part. And we notice that
			\begin{itemize}\item $\llbracket HOutput_{<j}\rrbracket$ can be deterministically recovered from the messages stored in the read-only buffer of $\ket{\phi^{j}}$: recall that in each round of the protocol the client sends out (\ref{eq:355r}), and when we do the statement reduction every time (for each $j^\prime\in [2,j]$) we analyze $\fTempPrtl_{=j^\prime}\ket{\chi^{j^\prime-1}}$, we first add $y^{(j^\prime-1)}$ as auxiliary information, as shown in (\ref{eq:360r}). Thus $\llbracket HOutput_{<j}\rrbracket$ can be recovered from the client-side messages and these auxiliary information.
				\item For $\llbracket HOutput_{=j}\rrbracket$, note that we already have (\ref{eq:282}), which means the adversary cannot query the corresponding input entries with high norm.
				\item $\llbracket HOutput_{>j}\rrbracket$ is the main thing that we need to deal with. Notice that the protocol does not use the values of this part before time $j$.
			\end{itemize}
			To ``switch'' the oracle from $H^{\prime\prime}$ to $H$, we will first switch the oracle from $H^{\prime\prime}$ to ``$H^{mid}$'', which the random oracle that 
			\begin{enumerate}
				\item On the entries in the form of (\ref{eq:381r}), it behaves the same as $H^{\prime\prime}$.
				\item Otherwise it behaves the same as $H$.
			\end{enumerate}
And the structure of the following proof is:
\begin{enumerate}
			 \item We will first prove \begin{equation}\label{eq:final461}\text{$\ket{\phi^{j}} \odot\llbracket HOutput_{>j}\rrbracket$ is $(2^{\eta_2/9\kappa},2^{-\eta_2/9\kappa}|\ket{\varphi}|)$-ANY-secure for $\Lambda^{(j)}$ under $H^{mid}$}\end{equation} using a technique similar to the proof of Lemma \ref{lem:4.9};
			 \item  Then we make use of Lemma \ref{lem:4.14} to complete the proof of (\ref{eq:378r}).
			 \end{enumerate}
			To prove (\ref{eq:final461}), by \textbf{proof-by-contradiction} we assume there exists a server-side operation $\cU$ such that the query number to $H^{mid}$ is at most $ 2^{\eta_2/\kappa}$ and
			\begin{equation}\label{eq:411}|P_{\Lambda^{(j)}}\cU(\ket{\phi^{j}}\odot \llbracket HOutput_{>j}\rrbracket )|>2^{-\eta_2/\kappa}|\ket{\varphi}|\end{equation}
			, then assuming (\ref{eq:411}), our goal is to construct a server-side operation $\cU^\prime$, operated on $\ket{\phi^{j}}$ such that $\cU^\prime$ only queries $H^{\prime\prime}$, the number of oracle queries to $H^{\prime\prime}$ is at most $2^{\eta_2/1000}$, and
			\begin{equation}\label{eq:412}|P_{\Lambda^{(j)}}\cU^\prime\ket{\phi^{j}}|>2^{-\eta_2/1000}|\ket{\varphi}|\end{equation}
			Which contradicts (\ref{eq:282}).\par
			Note that $Tag(\Lambda)$ is already in the read-only buffer thus it does not need to be added into (\ref{eq:411})(\ref{eq:412}).\par
			The first step is to prove the initial state (here we mean $\ket{\chi^0}$) is close to a state that does not depend on $H(Pads||\cdots)$ (recall Definition \ref{def:ndep}), where $Pads$ is a set of pads as follows:
			 $$Pads=\{pad^{j^\prime}_{b_r,ct},pad^{j^\prime}_{b_r,tg}:\forall b_r\in \{0,1\}, j^\prime\in [1,J]\}$$
			From (\ref{eq:385}), the pad length $\ell$ in Lemma \ref{lem:10.1} and Lemma \ref{lem:3.4r} we know, define $\ket{\tilde\chi^{0}}$ as the result of replacing each oracle query to $H$ in the representation of $\ket{\chi^{0}}$ by $H\cdot (I-P_{Pads||\cdots})$, (which means, remove the query inputs whose prefixes are contained in the $Pads$ system), there is
			\begin{equation}
				\ket{\tilde\chi^{0}}\approx_{2^{-\eta_2}|\ket{\varphi}|}\ket{\chi^{0}}
			\end{equation}
			Notice that $H\cdot (I-P_{Pads||\cdots})=H^{\prime\prime}\cdot (I-P_{Pads||\cdots})$, thus we can also imagine this state is represented (recall Definition \ref{def:ai1}) under $H^{\prime\prime}$.\par
			%
			Then we view $\ket{\tilde\chi^{0}}$ instead of $\ket{\chi^{0}}$ as the initial state, and based on the same (\ref{eq:403})-(\ref{eq:369}), we can define $\ket{\tilde\phi^{j^\prime}},\ket{\tilde\chi^{j^\prime}}$, etc, inductively and notice that the adversary still only queries $H^{\prime\prime}$. Some of the computation of client side messages will query $H$, but the adversary will not.\par
			And what we want to do (we mean (\ref{eq:412})) is reduced to
			\begin{equation}\label{eq:381}
				|P_{\Lambda^{(j)}}\cU^\prime(\ket{\tilde\phi^{j}}\odot Pads)|>2^{-\eta_2/1000+2}|\ket{\varphi}|
			\end{equation}
			and the condition (we mean (\ref{eq:411})) implies
			\begin{equation}\label{eq:411r}|P_{\Lambda^{(j)}}\cU(\ket{\tilde\phi^{j}}\odot \llbracket HOutput_{>j}\rrbracket)|>2^{-\eta_2/\kappa-2}|\ket{\varphi}|\end{equation}
			Let's explain the intuition of what we are going to do. We know one key difference of (\ref{eq:381}) and (\ref{eq:411r}) is in (\ref{eq:411r}) we ``switch back'' the oracle from $H^{\prime\prime}$ to $H^{mid}$ (in other words, although $\ket{\tilde\phi^{j}}$ comes from $H^{\prime\prime}$, $\cU$ queries $H^{mid}$,) while in (\ref{eq:381}) everything is under $H^{\prime\prime}$. Note that the difference of $H^{mid}$ and $H^{\prime\prime}$ can be described by $\llbracket HOutput_{>j}\rrbracket $ and $\llbracket HOutput_{<j}\rrbracket$, and:\begin{itemize}\item $\llbracket HOutput_{<j}\rrbracket $ could be deterministically recovered from the read-only buffer; \item  $\llbracket  HOutput_{>j}\rrbracket $ is never used (by neither party) when we represent $\ket{{\tilde\phi}^j}$ under $H^{\prime\prime}$.\end{itemize}
			 In other words, $H^{mid}$ can be simulated as follows: first sample a ``fake'' version of $\llbracket HOutput_{>j}\rrbracket $, then make use of $H^{\prime\prime}$, $\llbracket HOutput_{>j}^{fake}\rrbracket $ and $\llbracket HOutput_{<j}\rrbracket$ to simulate it. (Note that we need to be a little bit careful in this simulation to handle the case that some keys in $\Lambda$ are coincidently the same. The probability is small and does not affect the final result.)\par
			We note that there is one difference from this proof to the proof of Lemma \ref{lem:4.9}: here $\llbracket HOutput_{>j}\rrbracket$, together with  $\llbracket HOutput_{<j}\rrbracket$, completely describes the differences of $H^{\prime\prime}$ and $H^{mid}$, thus the operation only needs to sample randomness for the ``fake'' version of $\llbracket HOutput_{>j}\rrbracket$; while in the proof of Lemma \ref{lem:4.9} the operation needs to sample lots of randomness as the ``background values''.\par
			The construction details for $\cU^{\prime}$ are as follows.
			\begin{enumerate}
				\item 
				      $\cU^\prime$ samples
				      $$out^{fake-(j^\prime)}_{b_r,ct}, out^{fake-(j^\prime)}_{b_r,tg},j^\prime\in [j+1,J],b_r\in \{0,1\}$$
				      randomly, with length the same as the corresponding terms shown in (\ref{eq:373}). These are used as the ``$\llbracket HOutput_{>j}^{fake}\rrbracket$''.
				\item The ``fake oracle '' $H^{fake}$ is defined as follows. For each query to $H^{fake}$, suppose the input is $x$, $H^{fake}$ does the following:
				      \begin{enumerate}
					      \item It first check whether $x$ has the form shown in the inputs in (\ref{eq:373})(\ref{eq:425r})(\ref{eq:381r}). This can be achieved with $Tag(\Lambda)$ as long as $Tag$ is injective on inputs with the same length as the keys in $\Lambda$, and the space that $Tag$ is not injective on these inputs are very small by Fact \ref{fact:injtag}.
					      \item If not, or it is in the form of the inputs shown in (\ref{eq:381r}), return $H^{\prime\prime}(x)$.
					      \item Otherwise, if it has the form of (\ref{eq:425r}), return the corresponding values from the $\llbracket HOutput_{<j}\rrbracket$ (recovered in the read-only buffer)
					      \item Otherwise, if it has the form of (\ref{eq:373}), return $out^{fake-(j^\prime)}_{b_r,\text{ct}}$ or $out^{fake-(j^\prime)}_{b_r,\text{tg}}$ (correspondingly).
					      
				      \end{enumerate}
				     Denote the operation where all the queries in $\cU$ are replaced by $H^{fake}$ as $\cU^{fake}$.
				      \item $\cU^{\prime}$ is defined as the the combination of the two steps above: it samples $\llbracket HOutput^{fake}_{>j}\rrbracket$, puts it into the system that are used to store $\llbracket HOutput_{>j}\rrbracket$ in (\ref{eq:411r}), and runs $\cU^{fake}$.
			\end{enumerate}
			When the $\llbracket HOutput_{>j}\rrbracket$ and the oracle queries in (\ref{eq:411r}) are replaced by the fake versions, by the discussion above we will have \begin{align}|P_{\Lambda^{(j)}}\cU^\prime(\ket{\tilde\phi^{j}}\odot Pads)|&=|P_{\Lambda^{(j)}}\cU^{fake}(\ket{\tilde\phi^{j}}\odot \llbracket HOutput_{>j}^{fake}\rrbracket\odot Pads))|\\\text{(By discussions above and Fact \ref{fact:injtag})}&\approx_{2^{-\eta_2}|\ket{\varphi}|}|P_{\Lambda^{(j)}}\cU(\ket{\tilde\phi^{j}}\odot \llbracket HOutput_{>j}\rrbracket)|\end{align}
			Thus (\ref{eq:381}) is satisfied and we get a contradiction. Thus we complete the proof of (\ref{eq:final461}).\par
			Finally apply Lemma \ref{lem:4.14} we can switch the oracle from $H^{mid}$ to $H$ and the proof of (\ref{eq:378r}) is completed. Thus we complete this step ((\ref{eq:283}), the first part of the proof of Statement $4^\prime$).
		\end{proof}
	\subsubsection{Step 2c of the overview in Section \ref{sec:ai52}}\label{sec:ai56}
	\textbf{Now we can proceed to the step 2c of the overview in Section \ref{sec:ai52}, which is the second part of the proof of Statement $4^\prime$.} We will move from the security for keys in $\Lambda$ to the unpredictability of the output keys.\par
	The idea is intuitively as follows:\begin{enumerate}\item  $y^{(j)}$ is encrypted under $pad_{fixed}||r_0^{(j)}$ and $pad_{fixed}||r_1^{(j)}$. And intuitively we can reduce the unpredictability of plaintext (here it's $y^{(j)}$) to the unpredictability of keys (ignoring $pad_{fixed}$, it's $\Lambda^{(j)}=\{r_{b_r}^{(j)}\}_{b_r\in \{0,1\}}$).\par In more details, $y^{(j)}$ is unpredictable because it's encrypted as follows:
	\begin{enumerate}
		\item The four random pads $pad_{b_r,\text{``ct''/``tg''}}^{(j)}, b_r\in \{0,1\}$ are sampled randomly. (``Four'' corresponds to two possible choices of $b_r$ and two choices for ``ct''/``tg''.)
		\item 	During the protocol the adversary queries the blinded oracle $H^{\prime\prime}$ where $pad_{b_r,\text{``ct''/``tg''}}^{(j)}||pad_{fixed}||r_{b_r}^{(j)}$ are blinded.
		\item And after the protocol by (\ref{eq:283}) the state is $(2^{\eta_2/20\kappa},2^{-\eta_2/20\kappa}|\ket{\varphi}|)$-unpredictable for both $r_0^{(j)}$ and $r_1^{(j)}$, by the conclusion in Section \ref{sec:ai55}.
	\end{enumerate}
	\item Note that during Protocol \ref{prtl:temp}, $y^{(j)}$ is used as a part of (\ref{eq:355}). Intuitively we can reduce the unpredictability of (\ref{eq:355}) to the unpredictability of $y^{(j)}$;
	\end{enumerate}
	Formalizing the intuition above gives us the second step of the proof of Statement $4^\prime$.\par
	Let's first study the unpredictability for $y^{(j)}$, the output key in the $j$-th round of $\fTempPrtl$.\par 
		The statement we will prove is
		\begin{equation}\label{eq:383}\ket{\phi^{j}}\odot\llbracket\fTempPrtl_{>j}\rrbracket\text{ is $(2^{\eta_2/70\kappa},2^{-\eta_2/70\kappa}|\ket{\varphi}|)$-unpredictable for $y^{(j)}$ given $\{y^{(j^\prime)}\}_{j^\prime\in [J], j^\prime\neq j}$.}\end{equation}
		Note that one thing we need to be careful of is: in the definition of the unpredictability we need to add $Tag(y^{(j)})$ as the auxiliary information. The proof is given below.
		\begin{proof}[Proof of (\ref{eq:383})]
			Since $\{y^{(j^\prime)}\}_{j^\prime\in [J], j^\prime< j}$ is already stored in the read-only buffer, we can assume what we want to prove is the unpredictability for $y^{(j)}$ given $\{y^{(j^\prime)}\}_{j^\prime\in [J], j^\prime> j}$.\par
			Suppose $\cD$ is a server-side operation that tries to compute $y^{(j)}$ in (\ref{eq:383}). Note that it only queries $H$ and the query number $|\cD|\leq 2^{\eta_2/30\kappa}$. (But note that within the definition of $\ket{\phi^j}$, which is (\ref{eq:403})-(\ref{eq:385}), there do be queries to $H^{\prime\prime}$.)\par 
			Let's further define two blinded oracles. 
			\begin{itemize}\item Denote $\tilde H$ as a blinded oracle of $H$ where the followings are blinded:
			\begin{equation}\label{eq:382}H(pad_{0}||pad_{fixed}||r_0^{(j)}), H(pad_{1}||pad_{fixed}||r_1^{(j)}),Tag(y^{(j)})\end{equation}
			 where $pad_0,pad_1$ are the abbreviation of $pad_{b_r,\text{``ct''}}^{(j)}, b_r\in \{0,1\}$, the random pads used in the computation of ciphertexts that encrypt $y^{(j)}$.\par
			 \item Then define $\tilde H^{\prime\prime}$ as the blinded oracle of $H^{\prime\prime}$ where $Tag(y^{(j)})$ is blinded using the same output values as $\tilde H$. (Recall that $H^{\prime\prime}$ itself is a blinded oracle; and we further blind $Tag(y^{(j)})$ on it.)\end{itemize}
			  Now we do the followings step by step:
			\begin{enumerate}
				\item Expand $\ket{\phi^{j}}$ using (\ref{eq:403})-(\ref{eq:385}). Then in the sense of Definition \ref{def:ai1}:\begin{enumerate}
					\item $\ket{\phi^j}$ is $(2^{j\log\eta_2},j2^{\eta_2})$-representable from $\ket{\chi^0}$ under $H,H^{\prime\prime}$ (where the queries to $H$ are only from the computation of the client side messages);
					\item $\ket{\chi^0}$, as (\ref{eq:385}) says, is $(2^{D},2^{D})$-representable from $\ket{\mathfrak{init}}$.
				\end{enumerate} 
				 Let's start from the representation of $\ket{\chi^0}$ above. Within the representation of $\ket{\chi^0}$, before each query to $H$, do a projection onto $I-P_{span(pad_0||\cdots,\quad pad_1||\cdots)}$ (removing the queries which have prefix in $\{pad_0,pad_1\}$). Denote the result state as $\ket{\tilde\chi^0}$. Define $\ket{\tilde\phi^j}$ (similar to $\ket{\phi^j}$) through (\ref{eq:403})-(\ref{eq:385}), using $\ket{\tilde\chi^0}$ (instead of $\ket{\chi^0}$) above as the initial state. By Lemma \ref{lem:3.4r},
				      \begin{equation}\label{eq:385nn}\ket{\tilde\phi^{j}}\approx_{2^{-\eta_2/2}|\ket{\varphi}|}\ket{\phi^{j}}\end{equation}
				\item Starting from this step, we will replace the oracle queries by queries to $\tilde H$ or $\tilde H^{\prime\prime}$ step-by-step. 
				The replacement in this step will consider the queries within the definition of $\ket{\tilde\phi^j}$, and in the next step we will consider the query in $\cD$. What we will do in this step is as follows:\begin{enumerate}\item Within the ``representation'' of $\ket{\tilde\chi^0}$, we will replace the queries to $H$ by queries to $\tilde H$.\\ In the previous step we have already ``removed'' the queries that contain prefixes in $\{pad_0,pad_1\}$, which covers the first two in (\ref{eq:382}). If we compare $H\cdot(I-P_{span(pad_0||\cdots,\quad pad_1||\cdots)})$ and $\tilde H\cdot(I-P_{span(pad_0||\cdots,\quad pad_1||\cdots)})$, we still need to use the hybrid method to blind the oracle output on $Tag(y^{(j)})$.\item In the representation of $\ket{\tilde\phi^j}$ from $\ket{\tilde\chi^0}$, we will replace the queries to $H^{\prime\prime}$ by queries to $\tilde H^{\prime\prime}$. (Here client side queries to $H$ remain the same.)\par And when we replace $H^{\prime\prime}$ by $\tilde H^{\prime\prime}$, we only need to blind the $Tag(y^{(j)})$ part.\end{enumerate}
				       Since $y^{(j)}$ is sampled randomly, if we do this replacement step by step, we have the following fact on the difference caused by each step of the replacement and the total number of steps of replacement:
				      \begin{itemize}\item At some step of this replacement, since (1)all the queries to $Tag(y^{(j)})$ by this time have been replaced, (2) if the server only queries $\tilde H^{\prime\prime}$, the client side message in the $j$-th round looks the same as random strings, and on the server side, predicting $y^{(j)}$ is as hard as predicting a random string of the same length. Thus we have, each step of the replacement makes at most a difference of $2^{-\kappa_{out}/2+D+(\eta_2/\kappa)\cdot \log\eta_2}|\ket{\varphi}|$ on the output state. ($D$ comes from (\ref{eq:385}) and $\log\eta_2$ comes from (\ref{eq:369}). And notice that $t<\eta_2/\kappa$.)
					      \item The total number of steps in this replacement is bounded by $2^{D}+2^{\eta_2/\kappa}+2^{\kappa+5}$.
				      \end{itemize}
				      Thus if we denote the final state after this step completes as $\ket{\tilde{\tilde\phi}^{j}}$, there is
				      \begin{equation}\label{eq:387}\ket{\tilde{\tilde\phi}^{j}}\approx_{2^{-\eta_2/2}|\ket{\varphi}|}\ket{\tilde\phi^{j}}\end{equation}
				\item In this step we replace the oracle queries in $\cD$ by queries to the blinded oracle. Suppose $\cD^{blind}$ is the operation coming from replacing the oracle queries in $\cD$ by queries to $\tilde H$. Suppose $\cD^s$ is the operation in $\cD^{blind}$ from the beginning to the time just before the $s$-th query. By (\ref{eq:283})(\ref{eq:385nn})(\ref{eq:387}) we have\footnote{Details: First note that since all the queries to $Tag(y^{(j)})$ has been blinded, if we replace the $Tag(y^{(j)})$ term in (\ref{eq:387r}) by random strings it makes no difference. Then we can simply replace all the queries to the blinded version of $Tag(y^{(j)})$ by the original $Tag(y^{(j)})$ since they are both just random strings and there is no difference. This step replace the queries to $\tilde H$ in $\cD$ by queries to a blinded oracle where only the first two terms in (\ref{eq:382}) are blinded, and replace $\ket{\tilde{\tilde\phi}^{j}}$  by $\ket{\tilde\phi^j}$. Then we can apply (\ref{eq:385nn}) to replace $\ket{{\tilde\phi}^{j}}$ by $\ket{{\phi}^{j}}$, which introduce a little bit extra noise; finally we can apply (\ref{eq:283}).}
				      \begin{align}\forall s, \quad  |P_{\Lambda^{(j)}}\cD^s(\ket{\tilde{\tilde\phi}^{j}}\odot \llbracket\fTempPrtl_{>j}\rrbracket\odot \{y^{(j^\prime)}\}_{j^\prime>j}\odot Tag(y^{(j)}))|\label{eq:387r} 
					      \leq             & 2^{-\eta_2/22\kappa}|\ket{\varphi}|
				      \end{align}
				      And we also have
				      \begin{equation}\forall s, \quad  |P_{y^{(j)}}\cD^s(\ket{\tilde{\tilde\phi}^{j}}\odot \llbracket\fTempPrtl_{>j}\rrbracket\odot \{y^{(j^\prime)}\}_{j^\prime>j}\odot Tag(y^{(j)}))|  \leq              2^{-\eta_2/22\kappa}|\ket{\varphi}|\label{eq:390r}
				      \end{equation}
				      The reason is, after all the queries to (\ref{eq:382}) have been replaced, the ciphertexts of  $y^{(j)}$, which is
				      $$H(pad_{b_r}||pad_{fixed}||r_{b_r}^{(j)})\oplus y^{(j)}$$
				      become random strings, and predicting $y^{(j)}$ is as hard as predicting a freshly-new random string.\par
				      By Lemma \ref{lem:4.12n}
				      \begin{align}                                              & \cD(\ket{\tilde{\tilde\phi}^{j}}\odot \llbracket\fTempPrtl_{>j}\rrbracket\odot\{y^{(j^\prime)}\}_{j^\prime>j}\odot Tag(y^{(j)}))                       \\
					      \approx_{2^{-\eta_2/60\kappa}|\ket{\varphi}|} & \cD^{blind}(\ket{\tilde{\tilde\phi}^{j}}\odot \llbracket\fTempPrtl_{>j}\rrbracket\odot\{y^{(j^\prime)}\}_{j^\prime>j}\odot Tag(y^{(j)}))\label{eq:389}\end{align}
			\end{enumerate}


			Note that (\ref{eq:390r}) can also be applied on $\cD^{blind}$. Then summing up the error term in (\ref{eq:385nn})(\ref{eq:387})(\ref{eq:389}) completes the proof of (\ref{eq:383}). 
	\end{proof}
	\textbf{Now we can complete the proof of Statement $4^\prime$: }
	\begin{proof}[Proof of Statement $4^\prime$] Then we need to reduce the unpredictability of (\ref{eq:355}) to the unpredictability of $y^{(j)}$. We need to be a little bit careful here, since in the definition of the unpredictability of (\ref{eq:355}), $Tag(\text{ (\ref{eq:355}) })$ is provided to the adversary, which does not exist in the unpredictability of $y^{(j)}$.\par
	Recall that $$\text{(\ref{eq:355})}=pad||pad_{fixed}||y^{(1)}||\cdots || y^{(j)}||\cdots ||y^{(J)}$$
	Here $pad$ is sampled randomly, and all the other parts (other than $y^{(j)}$) have already been stored in the read-only buffer or given as the auxiliary information, $Tag(\text{ (\ref{eq:355}) })$ can be viewed as a hash value of $y^{(j)}$ with some paddings. Then applying Lemma \ref{lem:4.9} completes the proof of Statement $4^\prime$.\par 
\end{proof}
Then combine Statement 4 and equation (\ref{eq:25}) and apply the triangle inequality of SC-security (Lemma \ref{lem:4.1}), the proof of the Security of Protocol \ref{prtl:temp} is completed.\par

\subsection{Remaining Steps}\label{sec:ah6}
Now we combine everything and complete the proof of Lemma \ref{lem:10.1}.\par
First we can complete the proof of Statement 1.
\begin{proof}[Proof of Lemma 10.1, Part III]
	Recall that we reduce Statement 3 to ``Security of Protocol \ref{prtl:temp}''. Since ``Security of Protocol \ref{prtl:temp}'' is proved in the last subsection, the proof of Statement 3 has been completed.\par
	Summing everything up by the triangle inequality of the SC-security, the $2^{-\eta_1/3+3}|\ket{\varphi}|$ term (in equation (\ref{eq:23})) dominates ($\eta_2$ is much bigger than $\eta_1$). Thus we complete the proof of Statement 1.\end{proof}
	Then we prove the ``Statement 2'' in Outline \ref{otl:pf}, Section \ref{sec:ah3}.
\begin{proof}[Proof of Statement 2]The Statement 2 (about $\fAdv_0$) can be proved in a similar way. 
	Let's describe the proof in more details. Similar to the proof in Section \ref{sec:ah4}, we can reduce ``Statement 2'' to the ``Statement $3^{\prime\prime}$'', as follows:\par
		Note that one difference of ``Statement 2'' from ``Statement 1'' is both the $x_0^{(i)}$ and $x_1^{(i)}$ parts of the queries are ``removed''. Thus what the adversary can do during the attack is less than what it can do in the setting of Statement 1, thus proving it is actually easier. To reuse the proof of Statement 1, we strengthen the statement. We choose $b\in \{0,1\}$ arbitrarily and assume during the protocol execution the adversary can actually query $H^{\prime\prime}$ defined below, where only $x_{1-b}^{(i)}$ part is blinded.\par
		Same as the proof in Section \ref{sec:ah4}, define $H^{\prime\prime}$ as the blinded oracle where
		\begin{equation}H(pad^{(i)(j)}_{1-b,b_r,ct}||x^{(i)}_{1-b}||r_{b_r}^{(j)}),H(pad^{(i)(j)}_{1-b,b_r,tg}||x^{(i)}_{1-b}||r_{b_r}^{(j)}),\forall j\in [1,J],b_r\in \{0,1\}\end{equation}
		are blinded. (Note that $i$, $1-b$ are fixed. Thus we blind $4J$ entries.) Then we can reduce ``Statement 2'' to the statement below by adding auxiliary information and narrowing the blinded part of the random oracle:
		\begin{center}\emph{(``Statement $3^{\prime\prime}$'', repeated) Suppose $\kappa$ is bigger than some constant and the initial state $\ket{\varphi}$ satisfies the conditions listed in Lemma \ref{lem:10.1}. For any adversary $\fAdv^\prime$ that only queries $H^{\prime\prime}$ during the protocol, and the total number of queries to $H^{\prime\prime}$ is at most $2^{\kappa+4}$, the post-execution state, defined as
			$$P_{pass}\fSecurityRefreshing_{\fAdv^\prime}(K,\Lambda; \ell,\kappa_{out})\circ(\qquad \qquad \qquad \qquad \qquad \qquad \qquad \qquad \qquad \qquad \qquad \qquad$$
			\begin{equation}\label{eq:412c}\ket{\varphi}\odot \llbracket\fAuxInf\rrbracket \odot K\odot (K_{out}-K_{out}^{(i)})\odot  \{y_{b}^{(i)(j)}\}_{j\in [1,J]}\odot \fLT_{\text{not $K^{(i)}$}})\end{equation}
			is $(2^{\eta_2/100\kappa},2^{-\eta_2/100\kappa}|\ket{\varphi}|)$-unpredictable for (\ref{eq:355}).}\end{center}
	And the statement can be further reduced to the ``security of Protocol \ref{prtl:temp}''. 
	Thus we can use the same ``security of $\fTempPrtl$'' to prove the ``Statement 2''.
\end{proof}
Finally we can put everything together and complete the third step of the Outline \ref{otl:pf} thus complete the whole proof.
\begin{proof}
	By triangle inequality of the SC-security we can combine Statement 1 and Statement 2 to complete the proof of Lemma \ref{lem:10.1}. See Outline \ref{otl:pf} for more details.
\end{proof}

\section{Proof of Lemma \ref{lem:11.3}}\label{sec:AD}
\begin{proof}
	The idea is to expand everything. It's a little bit similar to the proof of Lemma \ref{lem:4.13}.\par
	Expand the expression of the distinguishing advantage, we need to prove for any distinguisher $\cD$ run by the server with queries $|\cD|\leq 2^\kappa$, $\tilde\theta=\theta_2\pi/2+\theta_3\pi/4$, there is:
	\begin{align}
		     & |tr(P_0(\cD(\varphi^1 \odot \fPhaseLT(\tilde\theta)\odot (\theta_2,\theta_3))\cD^\dagger))\\ &-tr(P_0\cD(\varphi^1\odot \fPhaseLT(\tilde\theta)\odot (\theta^\prime_2,\theta^\prime_3\leftarrow_r \{0,1\}^2))\cD^\dagger)|\label{eq:31} \\
		\leq & 2^{-\eta/9}|\ket{\varphi^1}|                                                                                                                                                                                                    
	\end{align}
	Where $\varphi^1:=\ket{\varphi^1}\bra{\varphi^1}$, $\ket{\varphi^1}=\sum_{i\in [\kappa]}\cP_i\sum_b\ket{\varphi_{i,b}}$. When we substitute it into the trace operations above, each term above has the following form, where $\theta^?$ can be replaced by $\theta_2,\theta_3$ or $\theta_2^\prime,\theta_3^\prime$:
	\begin{align}
		  & \tr(P_0\cD(\varphi^1\odot \fPhaseLT(\tilde\theta)\odot \theta^?_{2,3})\cD^\dagger)\label{eq:68}                                                                       \\
		= & \sum_i\sum_j\sum_b\tr(P_0\cD(\cP_i(\ket{\varphi_{i,b}}\bra{\varphi_{j,b}})\cP_j^\dagger\odot \fPhaseLT(\tilde\theta)\odot \theta_{2,3}^?)\cD^\dagger)\label{eq:34}    \\
		  & +\sum_i\sum_j\sum_b\tr(P_0\cD(\cP_i(\ket{\varphi_{i,b}}\bra{\varphi_{j,1-b}}))\cP_j^\dagger\odot \fPhaseLT(\tilde\theta)\odot\theta_{2,3}^?)\cD^\dagger)\label{eq:70}
	\end{align}
	We will call (\ref{eq:34}) ``the first term'' and (\ref{eq:70}) ``the second term''. Notice that the first term is
	{\small\begin{equation}\label{eq:36}\tr(P_0\cD(\ket{\varphi^1_0}\bra{\varphi^1_0} \odot \fPhaseLT(\tilde\theta)\odot (\theta^?_2,\theta^?_3))\cD^\dagger))+\tr(P_0\cD(\ket{\varphi^1_1}\bra{\varphi^1_1}\odot \fPhaseLT(\tilde\theta)\odot (\theta^?_2,\theta^?_3)\cD^\dagger)\end{equation}}
	where
	$$\ket{\varphi_b^1}:=\sum_{i\in [\kappa]}\cP_i\ket{\varphi_{i,b}}$$
	First we can prove each of the second term is exponentially small. That's because, for all $i\in [\kappa],j\in [\kappa],b\in \{0,1\}$:
	\begin{align}
		     & \tr(P_0\cD(\cP_i(\ket{\varphi_{i,b}}\bra{\varphi_{j,1-b}}))\cP_j^\dagger\odot \fPhaseLT(\tilde\theta)\odot\theta_{2,3}^?)\cD^\dagger) \\
		\leq & |P_{x_{1-b}}\cP_j\cD P_0\cD \cP_i(\ket{\varphi_{i,b}}\ket{\fPhaseLT(\tilde\theta)}\ket{\theta_{2,3}^{?}})|\label{eq:163}
	\end{align}

	Note that all the operations in $P_{x_{1-b}}\cP_j\cD P_0\cD \cP_i$ are server-side. Since $\ket{\varphi_{i,b}}\odot \fPhaseLT(\tilde\theta)$ is $(2^{\eta/8},2^{-\eta/8}|\ket{\varphi^1}|)$-SC-secure for $K=\{x_0,x_1\}$ (by Lemma \ref{lem:4.8}), and $P_{x_b}^{S_i}\ket{\varphi_{i,b}}=\ket{\varphi_{i,b}}$, we know $\ket{\varphi_{i,b}}\odot \fPhaseLT(\tilde\theta)$ is $(2^{\eta/8},2^{-\eta/8}|\ket{\varphi^1}|)$-unpredictable for $x_{1-b}$. And we have the query number $|\cP_j\cD P_0\cD \cP_i|\leq 2^{\eta/8}$, thus we know (\ref{eq:163}) is $\leq 2^{-\eta/8}|\ket{\varphi^1}|$.\par
	Note that equation (\ref{eq:70}) contains $2\kappa^2$ such terms thus the norm of (\ref{eq:70}) is at most $2^{-\eta/8.5}|\ket{\varphi^1}|$.\par
	Thus to bound the distinguishing advantage, we only need to consider the first term in the expansion of (\ref{eq:68}), which is (\ref{eq:36}). Substitute (\ref{eq:36}) into (\ref{eq:31}), we only need to give a bound for:
	{\small\begin{equation}\label{eq:39}\sum_b|\tr(P_0\cD(\ket{\varphi^1_b}\bra{\varphi^1_b} \odot \fPhaseLT(\tilde\theta)\odot (\theta_2,\theta_3))\cD^\dagger))-\tr(P_0\cD(\ket{\varphi^1_b}\bra{\varphi^1_b}\odot \fPhaseLT(\tilde\theta)\odot (\theta^\prime_2,\theta^\prime_3)\cD^\dagger)|\end{equation}}t
	First we know $\ket{\varphi^1_b}$ is $(2^{\eta/2},2^{-\eta/2}|\ket{\varphi^1}|)$-unpredictable for $x_{1-b}$. Thus consider 
	$$\tr(P_0\cD(\ket{\varphi^1_b}\bra{\varphi^1_b} \odot \fPhaseLT^{hyb}(\tilde\theta)\odot (\theta^?_2,\theta^?_3))\cD^\dagger))$$
	where $\fPhaseLT^{hyb}$ comes from replacing the entries in $\fPhaseLT$ that are encrypted under $x_{1-b}$ by random values, by Lemma \ref{lem:4.23} this causes at most $2^{-\eta/8}|\ket{\varphi^1}|$ difference:
	\begin{align}\label{eq:75} & |\tr(P_0\cD(\ket{\varphi^1_b}\bra{\varphi^1_b} \odot \fPhaseLT^{hyb}(\tilde\theta)\odot (\theta^?_2,\theta^?_3))\cD^\dagger))\\&-\tr(P_0\cD(\ket{\varphi^1_b}\bra{\varphi^1_b} \odot \fPhaseLT(\tilde\theta)\odot (\theta^?_2,\theta^?_3))\cD^\dagger))| \\
		\leq          & 2^{-\eta/8}|\ket{\varphi^1}|
	\end{align}
	. And when the phase tables are replaced, $\theta$ and $\theta^\prime$ look completely the same:
	\begin{align}\label{eq:76} & \tr(P_0\cD(\ket{\varphi^1_b}\bra{\varphi^1_b} \odot \fPhaseLT^{hyb}(\tilde\theta)\odot (\theta_2,\theta_3))\cD^\dagger))               \\
		=             & \tr(P_0\cD(\ket{\varphi^1_b}\bra{\varphi^1_b} \odot \fPhaseLT^{hyb}(\tilde\theta)\odot (\theta^\prime_2,\theta^\prime_3))\cD^\dagger))
	\end{align}
	Summing up (\ref{eq:75})(\ref{eq:76}) for $b\in\{0,1\}$ and $?\in \{\text{empty},^\prime\}$ gives a bound for (\ref{eq:39}). Thus we complete the proof.
\end{proof}

\section{Proof of Lemma \ref{lem:11.1}}\label{sec:ap111}
\subsection{An Overview of Proof of Lemma \ref{lem:11.1}}
Recall that an overview of this proof is also given in Section \ref{sec:11.4.1}.\par To make it easier to understand, we give a list for the meaning of different characters:
\begin{itemize}
	\item $i\in [\kappa]$ is the index of blocks; (recall that we divide the $\kappa^2$ rounds of tests to $\kappa$ blocks where each block has $\kappa$ rounds of tests.) Another viewpoint is it denotes the round count of the induction-style reduction proof. In each round of the reduction we analyze $\kappa$ rounds of basis test. $i$-th blocks ($i$-th round of the reduction) corresponds the $(i-1)\kappa+1\sim i\kappa$ rounds of the test in the protocol. 
	\item $t$ denotes the index of round within some block. $t\in [\kappa]$.\\
	      So the $t$-th test in the $i$-th block corresponds to the $((i-1)\kappa+t)$-th round of test in the original protocol.
\end{itemize}

\paragraph{Ideas of the proof, repeated} Note that our proof of Lemma \ref{lem:11.1} is as follows: we reduce Lemma \ref{lem:11.1} to the ``Statement-round-1-completed'', then reduce it to ``Statement-round-2-completed'', etc, where the statements are described in Section \ref{sec:11.4.1}.\par
Note that the description above is simplified; during each round of the reduction, we need to add an assumption on the norm of the state; and we will stop this reduction if this assumption does not hold. In Section \ref{sec:aj3} we will describe the reduction from ``Statement-round-$i$-completed'' to ``Statement-round-$(i+1)$-completed'', where $i\in [\kappa]$ is arbitrary. And we will describe the overall proof in Section \ref{sec:aj4}.\par
In each round of this reduction, the structure of the argument is as below. We use the $(i+1)$-th round of the argument as an example. (Note that $i$ starts from $0$ here, thus what we mean is the reduction from ``Statement-round-$i$-completed'' to ``Statement-round-$(i+1)$-completed''):\begin{enumerate}\item Starting from the ``Statement-round-$i$-completed'', we consider a fixed but arbitrary adversary and apply Lemma \ref{lem:6.3} (assuming the norm of the state we consider is not too small). This leads to two cases. 
\begin{itemize}
\item Case 1: there exists a time $t$ in the middle such that the state at this time can be controlled. For this case, we will do the following two step reduction:\begin{enumerate}\item Eliminate the test after time $t$ in this block of test and reduce the statement to ``Statement-round-$(i+1)$.Case1.1'', \item Eliminate the test before time $t$ in this block of test and reduce it further to ``Statement-round-$(i+1)$.Case1.2'';\end{enumerate} \item Case 2: the norm of passing space of the final state can be bounded. For this case the analogous two step reduction becomes: \begin{enumerate}\item Write down a statement that has an analogous form to the statement in Case 1, and name it as ``Statement-round-$(i+1)$.Case2.1'', \item Eliminate the test in this block and reduce the statement to ``Statement-round-$(i+1)$.Case2.2''.\end{enumerate}
\end{itemize} \item Finally we show how the statements coming out of these two cases can both be covered by a single statement, and this statement exactly has the form of ``Statement-round-$(i+1)$-completed''.\end{enumerate}
\subsection{Statement Reduction: from ``Statement-round-$i$-completed'' to \\``Statement-round-$(i+1)$-completed''}\label{sec:aj3}
Let's consider the $(i+1)$-th round, $0\leq i\leq \kappa-1$. This corresponds to the reduction from ``Statement-round-$i$-completed'' to ``Statement-round-$(i+1)$-completed''. And we need to analyze the block of tests with index starting from round $i\kappa+1$. (Which is the first block, or round, in the post-elimination protocol. Here we use the corresponding index before the elimination.)
\begin{proof}[Reduction from ``Statement-round-i-completed'' to ``Statement-round-$(i+1)$-completed'']
	Now we want to reduce the ``Statement-round-i-completed'' to some other statement. Recall the description of the ``Statement-round-i'' in Section \ref{sec:11.4.1}. (Substitute $s=i$.) Note that in the beginning ``Statement-round-0-completed'' is defined to be Lemma \ref{lem:11.1} itself.\par
	There are $\kappa^2-i\kappa$ rounds of $\fBasisTest$. We can represent the post-execution state after all the $\kappa^2-i\kappa$ rounds of tests as 
	\begin{align}\label{eq:93} & P_{pass}\fBasisTest_{\fAdv}(K;i\kappa+1\sim\kappa^2)\circ(\cP_1(\ket{\psi^{1}_0}+\ket{\psi^{1}_1})+\cdots +\cP_i(\ket{\psi^{i}_0}+\ket{\psi^{i}_1})) \\
		+             & P_{pass}\fBasisTest_{\fAdv}(K;i\kappa+1\sim\kappa^2)\circ\ket{\phi^i}\label{eq:425}\end{align}
	Here we use $i\kappa+1\sim\kappa^2$ to mean these remaining basis tests correspond to the $i\kappa+1$-th to $\kappa^2$-th rounds of tests in the original protocol (we mean the ``$\fBasisTest(K;\kappa^2)$'' in Lemma \ref{lem:11.1}.) Note that since each round of the basis test is the same operation, it doesn't matter whether it's the $i\kappa+1$-th to $\kappa^2$-th rounds or $1$st to $(\kappa^2-i\kappa)$-th rounds; we choose this notation because it shows how this single-round reduction is used in the upper-level proof: recall that we need to first analyze the first block, which corresponds to the $1\sim \kappa$ rounds in the tests, and this reduces Lemma \ref{lem:11.1} to ``Statement-round-1-completed''; and we continue this process to analyze the second block, third block, etc. And currently we are focusing on the $(i+1)$-th block, which corresponds to the $i\kappa+1\sim (i+1)\kappa$ rounds in the original protocol.\par
	In other words, here \textbf{the test round with index $i\kappa+1$ is actually the first round}.\par
	Further note that when $i=0$ there is no (\ref{eq:93}); and $\ket{\phi^0}$ appeared in (\ref{eq:425}) is defined to be $\ket{\varphi^1}$ in Lemma \ref{lem:11.1}.\par
	As we discussed above, consider the first block of $\kappa$ rounds of tests here, which can be denoted as $\fBasisTest(K;i\kappa+1\sim (i+1)\kappa)$. Now we use $\ket{\phi^{i,t}}$ to denote the post-execution state after time $t\in [0,\kappa]$ (which means, after the basis tests in round $i\kappa+1\sim i\kappa+t$) when the protocol is applied on $\ket{\phi^{i}}$:
	$$\ket{\phi^{i,t}}=\fBasisTest_{\fAdv_{BT:i\kappa+1\sim i\kappa+t}}(K;i\kappa+1\sim i\kappa+t)\circ \ket{\phi^i}$$
	Here we use $\fAdv_{BT:i\kappa+1\sim i\kappa+t}$ to denote the code of the adversary between round $i\kappa+1\sim i\kappa+t$.\par
	We can apply Lemma \ref{lem:6.3}. Let's first give the conditions for applying this lemma. From ``Statement-round-$i$-completed'' (see Section \ref{sec:11.4.1}) we can prove the followings on $\ket{\phi^i}$: 
	\begin{mdframed}
		\textbf{SC-security of $\ket{\phi^i}$ for $K$}\\
		From ``Statement-round-$i$-completed'' we know 
			$\ket{\phi^{i}}$ is \\$(1,i2^{\kappa+2})$-server-side-representable from 
			\begin{equation}\text{$\ket{\phi^{0}}\odot \llbracket \fAuxInf^{1\sim i}_1\rrbracket\odot \llbracket\fBasisTest(K;1\sim i\kappa)\rrbracket\odot \llbracket tag^{1\sim i}\rrbracket$}
		\end{equation}
		And we know $\ket{\phi^0}$ is $(2^{\eta},2^{-\eta}|\ket{\varphi^1}|)$-SC-secure for $K$. By Lemma \ref{lem:4.8} we know
		\begin{equation}
			\text{$\ket{\phi^i}$ is $(2^{\eta/12},2^{-\eta/12}|\ket{\varphi^1}|)$-SC-secure for $K$.}
		\end{equation}
		Which implies, (loosen the parameters to make the statement consistent in each round),
		\begin{equation}
			\text{Either $|\ket{\phi^i}|< 2^{-\kappa}|\ket{\varphi^1}|$, or $\ket{\phi^i}$ is $(2^{\eta/20},2^{-\eta/20}|\ket{\phi^i}|)$-SC-secure for $K$.}
		\end{equation}
	\end{mdframed}
	thus we get, if \begin{equation}\label{eq:430}|\ket{\phi^i}|\geq 2^{-\kappa}|\ket{\varphi^1}|\end{equation}, there is (by applying Lemma \ref{lem:6.3})
	\begin{itemize}
		\item (Case 1): There exists $t_{i+1}\in [0,\kappa-1]$, a server-side operation $\cU_{i+1}$ with query number $\leq 2^\kappa+i\kappa+20$ (whose form might depend on the code of $\fAdv$ on the $(i\kappa+t_{i+1}+1)$-th round) such that
		      \begin{equation}|(I-P_K^{S_{i+1}})\cU_{i+1}(P_{pass}\ket{\phi^{i,t_{i+1}}}\odot \llbracket \fAuxInf_1^{i+1}\rrbracket )|\leq \frac{1}{10}|\ket{\phi^{i}}|\label{eq:114}\end{equation}
		      \begin{equation} S_{i+1}\text{ is a server-side system,}\end{equation}
		      \begin{center}the algorithm of $\fAuxInf_1^{i+1}$ is the same as the $\llbracket\fAuxInf\rrbracket$ in Lemma \ref{lem:6.3}, but run on freshly new random coins.\end{center}
		\item (Case 2): \begin{equation}|P_{pass}\ket{\phi^{i,\kappa}}|\leq \frac{1}{2}|\ket{\phi^i}|\label{eq:118}\end{equation}
	\end{itemize}
	Discuss by cases. 
	 In both cases the reductions contain two steps. The ``first step'' of these reductions are to ``remove the tests after the time $t_{i+1}$'', while the ``second step'' of these reductions are to ``remove the tests by the time $t_{i+1}$''. (The meaning of these descriptions will become clear later.) 
	\begin{itemize}
		\item Case 1: 
		      \textbf{The first step} is to remove the $\fBasisTest$ from round $i\kappa+t_{i+1}+1$ to $(i+1)\kappa$. The idea is to replace the real protocol by auxiliary information that contain the client-side messages (and some other things) in the protocol, and prove that this does not make the adversary weaker. Define
		      \begin{equation}\label{eq:181rr}\llbracket\fAuxInf^{i+1}_>\rrbracket=\llbracket\fBasisTest(K;(i\kappa+t_{i+1}+1)\sim (i+1)\kappa)\rrbracket\odot \llbracket tag_{(i\kappa+t_{i+1}+1)\sim (i+1)\kappa}^{i+1}\rrbracket\end{equation}
		      , where
		      \begin{itemize}
			      \item 	$\llbracket\fBasisTest(K;(i\kappa+t_{i+1}+1)\sim (i+1)\kappa)\rrbracket$ is the client side messages from round $(i\kappa+t_{i+1}+1)\sim (i+1)\kappa$.
			      \item $\llbracket tag_{(i\kappa+t_{i+1}+1)\sim (i+1)\kappa}^{i+1}\rrbracket$ contains all the $Tag(r^{t^\prime}),t^\prime\in [i\kappa+t_{i+1}+1,(i+1)\kappa]$ where $r^{t^\prime}$ denotes the output key used in the computation of the $t^\prime$-th round of the $\fBasisTest$ protocol. (See protocols in Section \ref{sec:6.3}, where we use the same symbol $r$ to denote the output keys.)
		      \end{itemize}
		      (The subscript ``$>$'' in (\ref{eq:181rr}) means ``$>t_{i+1}$''. Later we will see similar notations with subscript $\leq$.)\par
		      Now an adversary can simulate the execution of $\fBasisTest$ from round $i\kappa+t_{i+1}+1$ to round $(i+1)\kappa$ with $\llbracket\fAuxInf^{i+1}_>\rrbracket$. Note that it can also simulate $P_{pass}$ with $TAG_{(i\kappa+t_{i+1}+1)\sim (i+1)\kappa}^{(i+1)}$, as long as $Tag$ is injective on inputs with that length (by Fact \ref{fact:injtag} the norm that it's not injective is very small). This implies that, if an adversary can get some distinguishing advantage (defined as equation (\ref{eq:60})) in the original statement, where \begin{itemize}\item the tests from round $(i\kappa+t_{i+1}+1)$ to $(i+1)\kappa$ are really executed; \item the number of RO queries by the adversary is at most $2^\kappa+i\kappa$\end{itemize} there exists an adversary which can distinguish with at least the same advantage minus a very small value (which is the norm that $Tag$ is not injective on inputs with length $\kappa_{out}$), where \begin{itemize} \item the tests from round $i\kappa+t_{i+1}+1$ to round $(i+1)\kappa$ are not executed, but $\llbracket\fAuxInf_>^{i+1}\rrbracket$ is provided instead; \item the number of RO queries by the adversary is at most $2^\kappa+i\kappa+(\kappa-t_{i+1})$, where $(\kappa-t_{i+1})$ is for simulating $P_{pass}$. \end{itemize}
		      \textbf{Thus to prove the ``statement-round-$i$-completed'', for Case 1, we only need to prove the similar statement: }
		      \begin{mdframed}
			      \textbf{Statement-round-$(i+1)$.Case1.1}\par
			      The conclusion is the same as Lemma \ref{lem:11.1} (we mean the statement below ``then the following conclusion holds''), with one difference: the right side of (\ref{eq:60}) is replaced by $(2^{-\kappa}-i2^{-2\eta}-2^{-4\eta})|\ket{\varphi^1}|$.\par
			      The conditions have the following differences:
			      \begin{itemize}
				      \item The initial state is
				            \begin{align}
					              & P_{pass}\ket{\phi^{i,t_{i+1}}}\odot \llbracket\fAuxInf^{i+1}_{1}\rrbracket\odot \llbracket\fAuxInf^{i+1}_{>}\rrbracket                                                                                                   \\
					            = & P_{pass}(\fBasisTest_{\fAdv_{BT:i\kappa+1\sim i\kappa+t_{i+1}}}(K;i\kappa+1\sim i\kappa+t_{i+1})\circ\label{eq:437}                                          \\
					              & \quad\quad\cP_1(\ket{\psi^{1}_0}+\ket{\psi^{1}_1})+\cdots +\cP_i(\ket{\psi^{i}_0}+\ket{\psi^{i}_1})                                                        \\
					              & \quad\quad\quad)\odot \llbracket\fAuxInf^{i+1}_{1}\rrbracket\odot \llbracket\fAuxInf^{i+1}_{>}\rrbracket\label{eq:117}                                                                                                   \\
					              & +P_{pass}(\fBasisTest_{\fAdv_{BT:i\kappa+1\sim i\kappa+t_{i+1}}}(K;i\kappa+1\sim i\kappa+t_{i+1})\circ\ket{\phi^i}\label{eq:119r}\\
					              &\quad\quad\quad)\odot \llbracket\fAuxInf^{i+1}_{1}\rrbracket\odot \llbracket\fAuxInf^{i+1}_{>}\rrbracket
				            \end{align}
				            where
				            \begin{itemize}
					            \item $\fAdv_{BT:i\kappa+1\sim i\kappa+t_{i+1}}$ is the code of the adversary  from round $i\kappa+1\sim i\kappa+t_{i+1}$ of the basis test part, which has query number at most $2^\kappa+i\kappa$.
					            \item $\cP_1,\cdots \cP_i$, $\ket{\psi_0^1},\ket{\psi_1^1},\cdots \ket{\psi_0^i},\ket{\psi_1^i}$,$\ket{\phi^i}$ satisfy the conditions listed in the \\``statement-round-$i$-completed'';
					            \item $\llbracket\fAuxInf^{i+1}_{1}\rrbracket, \llbracket\fAuxInf^{i+1}_{>}\rrbracket$ are defined in (\ref{eq:114})(\ref{eq:181rr}).
				            \end{itemize}
				            Note that the (\ref{eq:119r}) is actually $P_{pass}\ket{\phi^{i,t_{i+1}}}$ and it satisfies (\ref{eq:114}).
				      \item The protocol is as follows: in the $\fBasisTest$ step, the tests are executed for $\kappa^2-(i+1)\kappa$ rounds. The parameters of the protocol (pad length, output key length) are the same.
				      \item The adversary is $\fAdv^{\prime}$, and it satisfies $|\fAdv^{\prime}|\leq 2^\kappa+(i+1)\kappa$
			      \end{itemize}
		      \end{mdframed}
		      \textbf{The second step} is: we can do something to ``flatten'' the
		      \begin{equation}\label{eq:432}P_{pass}\fBasisTest_{\fAdv_{BT:i\kappa+1\sim i\kappa+t_{i+1}}}(K;i\kappa+1\sim i\kappa+t_{i+1})\end{equation}
		       in (\ref{eq:437})(\ref{eq:119r}). This will lead us to a further different statement.\par
		      Define
		      \begin{equation}\label{eq:433}\llbracket\fAuxInf^{i+1}_\leq\rrbracket=\llbracket\fBasisTest(K;(i\kappa+1)\sim i\kappa+t_{i+1})\rrbracket\odot \llbracket tag^{i+1}_{i\kappa+1\sim i\kappa+t_{i+1}}\rrbracket\end{equation} where $\llbracket tag^{i+1}_{\kappa+1\sim \kappa+t_{i+1}}\rrbracket$ contains all the $Tag(r^{t})$ where $r^t$ denotes the output keys used in the computation of $\fBasisTest(K;(i\kappa+1)\sim i\kappa+t_{i+1})$.\par
		      Then if the server holds $\llbracket\fAuxInf_\leq^{i+1}\rrbracket$, it can simulate the protocol execution, the adversary's operation and $P_{pass}$ in (\ref{eq:432}) and get a ``simulated $P_{pass}\ket{\phi^{i,t_{i+1}}}$''. This simulation works as follows: the adversary goes through each step of the protocol in (\ref{eq:432}), and:
		      \begin{itemize}
		      \item It can run the code contained in $\fAdv_{BT:i\kappa+1\sim i\kappa+t_{i+1}}$ itself;
		      \item It already gets all the client-side messages from the $\llbracket\fAuxInf^{i+1}_{\leq}\rrbracket$, and does not need to wait for the client to send the message;
		      \item And it can simulate the projection onto the passing space with the global tags of the output keys, as long as $Tag$ is injective on inputs with length $\kappa_{out}$; (by Fact \ref{fact:injtag} the norm that this is not true is very small.)
		      \item Finally for the ``sending messages to the client'' step, it simply initializes some empty qubits and stores the response in it.
		      \end{itemize}
		       Denote the server's operation in this simulation as $\cP_{i+1}$. Note the query number of $\cP_{i+1}$ is upper bounded by $2^{\kappa+0.1}$. We will use $\cP_{i+1}$ to replace (\ref{eq:432}) in (\ref{eq:437})(\ref{eq:119r}). The reason that the adversary does not become weaker is as follows: let's think about what is the difference of this ``simulated $P_{pass}\ket{\phi^{i,t_{i+1}}}$'' from the real $P_{pass}\ket{\phi^{i,t_{i+1}}}$. The difference is when the server needs to reply, whether it writes on the client side empty register or it writes on some empty qubits on its own. Thus these two states are actually the same one if we ignore the position of this part of system! And in the simulated case the server holds more system than the real case. Thus replacing the initial state in this way does not make the adversary weaker.\par
		      We also notice that an inequality similar to (\ref{eq:114}) still holds for this ``simulated state'':
		      \begin{equation}
			      |(I-P_K^{S_{i+1}})\cU_{i+1}(\cP_{i+1}(\ket{\phi^i}\odot \llbracket\fAuxInf^{i+1}_{\leq}\rrbracket)\odot \llbracket\fAuxInf_1^{i+1}\rrbracket)|\leq \frac{1}{10}|\ket{\phi^i}|
		      \end{equation}
		      Thus to prove the ``Statement-round-$(i+1)$.Case1.1'', we can prove the following statement:
		      \begin{mdframed}
			      \textbf{Statement-round-$(i+1)$.Case1.2}\par
			      The conclusion is the same as Lemma \ref{lem:11.1} (we mean the statement below ``then the following conclusion holds''), with one difference: the right side of (\ref{eq:60}) is replaced by $(2^{-\kappa}-i2^{-2\eta}-2\cdot 2^{-4\eta})|\ket{\varphi^1}|$ \par The conditions have the following differences:
			      \begin{itemize}
				      \item The initial state has the following form:
				            \begin{align}
					              & \cP_{i+1}((\cP_1(\ket{\psi_0^{1}}+\ket{\psi_1^{1}})+\cdots +\cP_i(\ket{\psi_0^{i}}+\ket{\psi_1^{i}})+\ket{\phi^i})\odot \llbracket\fAuxInf^{i+1}\rrbracket)\label{eq:190r} \\
					            = & \cP_{i+1}((\cP_1(\ket{\psi_0^{1}}+\ket{\psi_1^{1}})+\cdots +\cP_i(\ket{\psi_0^{i}}+\ket{\psi_1^{i}}))\odot \llbracket\fAuxInf^{i+1}\rrbracket)\label{eq:146}               \\
					              & +\cP_{i+1}(\ket{\phi^i}\odot \llbracket\fAuxInf^{i+1}\rrbracket)\label{eq:147}
				            \end{align}
				            where 
				            \begin{center}
					            $\cP_{i+1}$ is a fixed server-side operation (with projection) with query number $\leq 2^{\kappa+0.1}$,
				            \end{center}\begin{center}$\cP_1,\cdots \cP_i$, $\ket{\psi_b^{1}},\cdots \ket{\psi_b^{i}}$, $\ket{\phi^i}$ satisfy the conditions in ``statement-round-i-completed'',\end{center}
				            \begin{equation}\label{eq:448}\llbracket\fAuxInf^{i+1}\rrbracket :=\llbracket \fAuxInf^{i+1}_1\rrbracket\odot \llbracket\fAuxInf^{i+1}_>\rrbracket\odot \llbracket\fAuxInf^{i+1}_\leq\rrbracket,\end{equation}
				            (see (\ref{eq:114})(\ref{eq:181rr})(\ref{eq:433})),\par
				            and there exists a fixed server-side operation $\cU_{i+1}$ with query number $\leq 2^{\kappa+0.1}$, a server-side system $S_{i+1}$, define
				            \begin{equation}\label{eq:191r}
					            \ket{{\chi}^{i+1}}:=(I-P_K^{S_{i+1}})\cU_{i+1}\ket{\text{ equation (\ref{eq:147}) }}
				            \end{equation}
				            there is

				            $$|\ket{{\chi}^{i+1}}|\leq \frac{1}{10}|\ket{\phi^i}|$$
				      \item The protocol is as follows: in the $\fBasisTest$ step, the basis tests on $K$ are executed for $\kappa^2-(i+1)\kappa$ rounds. The parameters of the protocol (pad length, output key length) are the same.
				      \item The adversary is $\fAdv^{\prime}$, and it satisfies $|\fAdv^{\prime}|\leq 2^\kappa+(i+1)\kappa$.
			      \end{itemize}
		      \end{mdframed}
		      Now define \begin{equation}\label{eq:192rr}\ket{\phi^{i+1}}:=\cU_{i+1}^\dagger\ket{{\chi}^{i+1}}=\cU_{i+1}^\dagger(I-P_K^{S_{i+1}})\cU_{i+1}\cP_{i+1}(\ket{\phi^i}\odot \llbracket\fAuxInf^{i+1}\rrbracket)\end{equation}
		      Note that we have
		      \begin{equation}\label{eq:451}
					            \ket{\text{ equation (\ref{eq:147}) }}=\cU_{i+1}^\dagger (\ket{\psi_0^{i+1}}+\ket{\psi_1^{i+1}})+\ket{\phi^{i+1}}
				            \end{equation}
\begin{equation}\label{eq:190}
					            \text{where }\forall b\in \{0,1\},\ket{{\psi}^{i+1}_b}:=P^{S_{i+1}}_{x_b}\cU_{i+1}\ket{\text{ equation (\ref{eq:147}) }}
				            \end{equation}
				            We will use (\ref{eq:451}) when we further reduce this statement to ``Statement-round-$(i+1)$-completed''.
		\item Case 2 (\ref{eq:118}):
		      In this case, we can reduce ``Statement-round-$i$-completed'' to a new statement as follows. The analog of the first step of Case 1 can be skipped, but we still write down a similar statement for consistency.
		      \begin{mdframed}
			      \textbf{Statement-round-$(i+1)$.Case2.1}\par
			      The conclusion is the same as Lemma \ref{lem:11.1} (we mean the statement below ``then the following conclusion holds''), but the conditions have the following differences:
			      \begin{itemize}\item The initial state is
				            \begin{align}  & P_{pass}\fBasisTest_{\fAdv_{BT:i\kappa+1\sim (i+1)\kappa}}(K;i\kappa+1\sim (i+1)\kappa)\circ(            \\
					              & \qquad\cP_1(\ket{\psi^{1}_0}+\ket{\psi^{1}_1})+\cdots +\cP_i(\ket{\psi^{i}_0}+\ket{\psi^{i}_1}))       \\
					            + & P_{pass}\fBasisTest_{\fAdv_{BT:i\kappa+1\sim (i+1)\kappa}}(K;i\kappa+1\sim (i+1)\kappa)\circ\ket{\phi^i}\end{align}
					            where \begin{itemize}\item $\fAdv_{BT:i\kappa+1\sim i\kappa+t_{i+1}}$ is the code of the adversary  from round $i\kappa+1\sim i\kappa+t_{i+1}$ of the basis test part, which has query number at most $2^\kappa+i\kappa$.\item $\cP_1\cdots \cP_i$, $\ket{\psi_0^1},\ket{\psi_1^1}\cdots \ket{\psi_0^i},\ket{\psi_1^i}$, $\ket{\phi^i}$ satisfy the conditions in ``Statement-round-$i$-completed''.\end{itemize}
				      \item the $\fBasisTest$ is executed for $\kappa^2-(i+1)\kappa$ rounds; the parameters of the protocol are the same.
				      \item The adversary $\fAdv^\prime$ is slightly more powerful than the original adversary: $|\fAdv^\prime|\leq 2^\kappa+(i+1)\kappa$.
			      \end{itemize}\end{mdframed}
		      Then we do similar thing as the step 2 in Case 1. We reduce the statement-round-$(i+1)$.Case2.1 further to the following statement:
		      \begin{mdframed}
			      \textbf{Statement-round-$(i+1)$.Case2.2}\par
			      The conclusion is the same as Lemma \ref{lem:11.1} (we mean the statement below ``then the following conclusion holds''), with one difference: the right side of (\ref{eq:60}) is replaced by $(2^{-\kappa}-i2^{-2\eta}-2^{-4\eta})|\ket{\varphi^1}|$.\par The conditions have the following differences:
			      \begin{itemize}\item The initial state can be written as
				            \begin{align}
					            \cP_{i+1}(\quad & (\cP_1(\ket{\psi^{1}_0}+\ket{\psi^{1}_1})+\cdots +\cP_i(\ket{\psi^{i}_0}+\ket{\psi^{i}_1}))+\label{eq:456} \\
					                            & \ket{\phi^i})\quad\odot \llbracket\fBasisTest(K;i\kappa+1\sim (i+1)\kappa)\rrbracket\odot \llbracket tag^{i+1}\rrbracket)\label{eq:453}
				            \end{align}
				            where \begin{itemize}\item $\cP_1\cdots \cP_i$, $\ket{\psi_b^1}\cdots \ket{\psi_b^i}$, $\ket{\phi^i}$ satisfy the conditions in\\ ``Statement-round-$i$-completed''.\item The query number of $\cP_{i+1}$ is at most $2^{\kappa+0.1}$.\item  $\llbracket tag^{i+1}\rrbracket$ contains all the $Tag(r^t)$ where $r^t$ is the output key used in the $t$-th round of the tests in $\fBasisTest(K;i\kappa+1\sim (i+1)\kappa)$.\end{itemize}\item And the following is satisfied: $|\cP_{i+1}\ket{\text{equation (\ref{eq:453})}}|\leq \frac{1}{2}|\ket{\phi^i}|$.
				      \item the $\fBasisTest$ is executed for $\kappa^2-(i+1)\kappa$ rounds; the parameters of the protocol are the same.
				      \item The adversary $\fAdv^{\prime}$ is slightly more powerful than the original adversary: $|\fAdv^{\prime}|\leq 2^\kappa+(i+1)\kappa$.
			      \end{itemize}
		      \end{mdframed}
		      Let's define some symbols that match (\ref{eq:190})(\ref{eq:192rr}): just choose \begin{equation}\label{eq:502}\ket{{\psi}_0^{i+1}}=\ket{{\psi}_1^{i+1}}=0,\quad \ket{\phi^{i+1}}=\cP_{i+1}\ket{\text{equation (\ref{eq:453})}}\end{equation}
	\end{itemize}
	To summarize, \textbf{the third step} is to combine these two cases: we can reduce both ``Statement-round-$(i+1)$.Case1.2'' and ``Statement-round-$(i+1)$.Case2.2'' to a new statement (which is the same for both cases). This new statement is called ``Statement-round-$(i+1)$-completed''.\par \textbf{And we note that the symbols in this statement may have different meanings as the symbols in the previous statement, and we will explain the difference after giving the statement:}
	\begin{mdframed}
		\textbf{Statement-round-$(i+1)$-completed}\par
		The conclusion is the same as Lemma \ref{lem:11.1} (we mean the statement below ``then the following conclusion holds''), with one difference: the right side of (\ref{eq:60}) is replaced by $(2^{-\kappa}-(i+1)2^{-2\eta})|\ket{\varphi^1}|$.\par The conditions have the following differences:
		\begin{itemize}
			\item The initial state is in the form of
			      \begin{equation}\label{eq:458}
				      \cP_1(\ket{\psi_0^{1}}+\ket{\psi_1^{1}})+\cP_2(\ket{\psi^{2}_0}+\ket{\psi^{2}_1})+\cdots+\cP_{i+1}(\ket{\psi^{i+1}_0}+\ket{\psi^{i+1}_1})+\ket{\phi^{i+1}}
			      \end{equation}
			      where 
			      \begin{itemize}
			      	
			      \item $|\ket{\phi^{i+1}}|\leq (\frac{1}{2})^{i+1}|\ket{\varphi^1}|$
				      \item $\cP_1$, $\cP_2$, $\cdots\cP_{i+1}$ are all server-side operations with query number $\leq (i+1)2^{\kappa+3}$.
				      \item $\forall b\in \{0,1\},\forall i^\prime\in [1,i+1], P^{S_{i^\prime}}_{x_b}\ket{\psi^{i^\prime}_{b}}=\ket{\psi^{i^\prime}_{b}}$, $\text{ $S_{i^\prime}$ is some server-side system}$.
				      \item $\forall b\in \{0,1\},\forall i^\prime\in [1,i+1],\ket{\psi^{ i^\prime}_b}$ is $(1,i^\prime 2^{\kappa+2})$-server-side-representable from
				       \begin{equation}\label{eq:108n}\ket{\phi^0}\odot\llbracket \fAuxInf^{1\sim i+1}_1\rrbracket\odot \llbracket\fBasisTest(K;1\sim (i+1)\kappa)\rrbracket\odot \llbracket tag^{1\sim i+1}\rrbracket \end{equation}
				            where
				            \begin{itemize}
					            \item 
					            $\llbracket\fAuxInf^{1\sim i+1}_1\rrbracket=\llbracket\fAuxInf^{ 1}_1\rrbracket\odot \llbracket\fAuxInf^{ 2}_1\rrbracket\odot \cdots \odot \llbracket\fAuxInf^{i+1}_1\rrbracket$
					            \item $\llbracket\fBasisTest(K;1\sim (i+1)\kappa)\rrbracket$ contains the client side messages from round $1\sim (i+1)\kappa$ of the tests. 
					            \item $\llbracket tag^{1\sim i+1}\rrbracket$ are the set of $Tag(r^t)$ where $r^t$ is the output key used in the $t$-th round of $\llbracket\fBasisTest(K;1\sim (i+1)\kappa)\rrbracket$.
				            \end{itemize}
				     \item $\ket{\phi^{i}}$ is $(1,i 2^{\kappa+2})$-server-side-representable from (\ref{eq:108n}).
			      \end{itemize}
			\item The protocol is as follows: in the $\fBasisTest$ step, the basis tests are executed for $\kappa^2-(i+1)\kappa$ rounds. The parameters of the protocol (pad length, output key length) are the same.
			\item The adversary is $\fAdv$, and it satisfies the query number $|\fAdv|\leq 2^\kappa+(i+1)\kappa$.
		\end{itemize}\end{mdframed}
	The reduction that we do in this step is as follows:
	\begin{itemize}
		\item Simplify the notation: First notice that in (\ref{eq:448}) the notation can be simplified as follows: $$\llbracket\fAuxInf_>^{i+1}\rrbracket\odot \llbracket\fAuxInf_\leq^{i+1}\rrbracket=\llbracket\fBasisTest(K;i\kappa+1\sim (i+1)\kappa)\rrbracket\odot \llbracket tag^{i+1}\rrbracket$$
		      Thus in the final statement we can combine the auxiliary information in the form of the left hand side above and use the notation in the right hand side.
		\item Merge the auxiliary information: the auxiliary information in ``Statement-round-$(i+1)$.Case2.2'' is a subset of the auxiliary information in ``Statement-round-$(i+1)$.Case1.2''. (See (\ref{eq:448}) and (\ref{eq:453}).) In the final statement we simply consider (\ref{eq:448}). The reason for it is, in the Case 2, by Technique \ref{lem:4.2} adding more auxiliary information does not make the adversary weaker. Thus merging the ``auxiliary information'' will not make the statement weaker. 
		\item Change the symbol for the server-side operators: We note that $\cP_1,\cdots \cP_{i+1}$ in (\ref{eq:458}) should not be understood as the same things as the operations with the same symbols in ``Statement-round-i-completed'', or ``Statement-round-$(i+1)$-Case1.2'', or ``Statement-round-$(i+1)$-Case2.2''. Instead:\begin{itemize}\item In Case 1, $\cP_1,\cdots \cP_{i}$ here should be understood as $\cP_{i+1}\cP_1,\cdots\cP_{i+1}\cP_i$ in (\ref{eq:146}) and $\cP_{i+1}$ here is $\cU_{i+1}^\dagger$ in (\ref{eq:451}). \item In Case 2 it can be understood as the operations in (\ref{eq:456})(\ref{eq:453}).\end{itemize} Note that in ``Statement-round-$(i+1)$-completed'' there is no restriction on the form of these server-side operations, thus this statement can cover both cases.
		\item Change the meaning of the states: Note that $\ket{\psi_b^{i^\prime}}$ in the ``Statement-round-$(i+1)$-completed'' have different meaning from the intermediate statements in the two cases and the previous rounds. The difference is it contains extra auxiliary information. Note that we add $\llbracket\fAuxInf^{i+1}_1\rrbracket\odot \llbracket\fBasisTest(K;i\kappa+1\sim (i+1)\kappa)\rrbracket\odot \llbracket tag^{i+1}\rrbracket$ as the auxiliary information in this step. And when we describe the form of the state, this auxiliary information should be appended into all the ``basis states'' shown in (\ref{eq:93}). Even if the symbol is the same, the ``basis state'' in the ``Statement-round-$(i+1)$'' contains more auxiliary information.\par 
		And we note that $\ket{\psi_0^{i+1}},\ket{\psi_1^{i+1}}$ come from (\ref{eq:190})(\ref{eq:502}).
		\item We still use $\fAdv$ (instead of $\fAdv^\prime$ or $\fAdv^{\prime\prime}$) to denote the adversary.
		\item We loosen some parameters to make the statement looks simpler.
	\end{itemize}
	Thus we complete the reduction in this round.
\end{proof}

\subsection{Overall Reduction}\label{sec:aj4}
Finally we put the argument in Section \ref{sec:aj3} to a bigger picture and prove Lemma \ref{lem:11.1}. Recall Section \ref{sec:11.3} for the  ``Statement-round-$i$-completed'' used in this induction-style reduction.
\begin{proof}[Proof of Lemma \ref{lem:11.1}]
	In the last section we show how to reduce ``Statement-round-$i$-completed'' to ``Statement-round-$(i+1)$-completed''. And the form of the initial statement is Lemma \ref{lem:11.1}. Intuitively we can just repeat this reduction process. However, note that in each round of this reduction there is an implicit assumption: $|\ket{\phi^i}|\geq 2^{-\kappa}|\ket{\varphi^1}|$, which is assumed in (\ref{eq:430}) as a condition of this reduction. If this condition is not satisfied at some step, we need to stop this ``statement reduction'' process.\par
	Thus the overall reduction process is done as follows. Note that here we also give an explicit explanation on how we deal with the quantifier on the adversary.
	\begin{enumerate}
		\item We want to prove Lemma \ref{lem:11.1}. We only need to prove the conclusion in Lemma \ref{lem:11.1} for a specific but arbitrarily-chosen adversary $\fAdv$.
		\item To prove it, we can turn to prove ``Statement-round-1-completed''. Note that in this new statement there is a $\forall$ quantifier before the adversary in it; and it certainly covers the case where the adversary is $\fAdv$ that we consider in the last step.
		\item To prove ``Statement-round-1-completed'', we only need to prove the conclusion for a specific but arbitrarily-chosen adversary $\fAdv$. (We use the same notation since $\fAdv$ in the step above is already useless here.) Then we consider whether $|\ket{\phi^1}|\geq 2^{-\kappa}|\ket{\varphi^1}|$ is satisfied. If not, stop. Otherwise, reduce it to ``Statement-round-2-completed''.
		\item Similarly consider whether $|\ket{\phi^2}|\geq 2^{-\kappa}|\ket{\varphi^1}|$ is satisfied. If not, stop. Otherwise, reduce it to ``Statement-round-3-completed''.
		\item $\cdots\cdots$
	\end{enumerate}

	This argument can be repeated as long as $|\ket{\phi^i}|\geq 2^{-\kappa}|\ket{\varphi^1}|$ and  $i\in [\kappa]$ hold. Then when this repetition of reduction stops, (suppose it stops at round $s$,) no matter it stops because which one is violated, we always have $|\ket{\phi^s}|\leq 2^{-\kappa}|\ket{\varphi^1}|$. 
	Then the final state has the form we want, except an exponential small additional term, and we can apply Lemma \ref{lem:11.3}. The conditions for applying Lemma \ref{lem:11.3} hold because:
	\begin{itemize}
		\item The state form in ``statement-round-$s$-completed'' is the same as the requirement in Lemma \ref{lem:11.3}. And $|\ket{\phi^s}|\leq 2^{-\kappa}|\ket{\varphi^1}|$.
		\item As we say in the ``statement-round-s-completed'', 
		      $\ket{\psi^{i}_b}$ is $(1,\kappa 2^{\kappa+3})$-server-side representable from
		      \begin{equation}\label{eq:191}\ket{\phi^{0}}\odot \llbracket\fAuxInf^{1\sim s}_1\rrbracket\odot \llbracket\fBasisTest(K;1\sim s\kappa)\rrbracket\odot \llbracket tag^{1\sim s}\rrbracket\end{equation}
		      Since (\ref{eq:191}) is $(2^{\eta/7},2^{-\eta/7}|\ket{\varphi}|)$-SC-secure for $K$ (by Lemma \ref{lem:4.8}), we know $\ket{\psi^{i}_b}$ is $(2^{\eta/7.5},2^{-\eta/7}|\ket{\varphi}|)$-SC-secure for $K$.
	\end{itemize}
	Thus applying Lemma \ref{lem:11.3} we complete the proof.

\end{proof}
\end{document}